\documentclass[12pt,a4paper, twoside]{report}

\usepackage{latexsym}
\usepackage{fancyhdr}
\usepackage{amsfonts}
\usepackage{amssymb}
\usepackage{amsmath}
\usepackage{amsthm}

\usepackage{graphicx} 
\usepackage{indentfirst}
\usepackage{bbm}
\usepackage{enumerate}
\usepackage{appendix}
\usepackage{chngcntr}
\usepackage{etoolbox}

\usepackage{verbatim}
\usepackage{mathrsfs}

\usepackage{dsfont}
\usepackage{cite}
\usepackage{xcolor}
\usepackage{enumitem}
\usepackage{ytableau}
\usepackage{tikz}
\usepackage[colorlinks=true,citecolor=blue, urlcolor=blue ]{hyperref}

\allowdisplaybreaks

\AtBeginEnvironment{subappendices}{%
\chapter*{Appendix}
\addcontentsline{toc}{chapter}{Appendices}
\counterwithin{figure}{section}
\counterwithin{table}{section}
}

\graphicspath{{Images/}}

\parskip=\medskipamount

\def\a{\alpha}
\def\b{\beta}
\def\c{\chi}
\def\d{\delta}
\def\e{\epsilon}
\def\f{\phi}
\def\g{\gamma}

\def\j{\psi}
\def\k{\kappa}
\def\l{\lambda}
\def\m{\mu}
\def\mub{\bar{\mu}}

\def\o{\omega}

\def\q{\theta}
\newcommand{\qb}{{\bar{\theta}}}
\def\r{\rho}
\def\s{\sigma}
\def\t{\tau}

\def\x{\xi}
\def\z{\zeta}
\def\D{\Delta}
\def\F{\Phi}
\def\J{\Psi}
\def\L{\Lambda}
\def\O{\Omega}
\def\P{\Pi}
\def\Q{\Theta}
\def\S{\Sigma}
\def\U{\Upsilon}

\newcommand {\cA}{{\cal A}}

\newcommand {\cD}{{\cal D}}
\newcommand {\cE}{{\cal E}}

\newcommand {\cG}{{\cal G}}
\newcommand {\cH}{{\cal H}}

\newcommand {\cL}{{\cal L}}
\newcommand {\cM}{{\cal M}}
\newcommand {\cN}{{\cal N}}

\newcommand {\cQ}{{\cal Q}}
\newcommand {\cR}{{\cal R}}
\newcommand {\cS}{{\cal S}}
\newcommand {\cT}{{\cal T}}

\newcommand{\sSp}{\mathsf{Sp}}
\newcommand{\sSU}{\mathsf{SU}}
\newcommand{\sSL}{\mathsf{SL}}
\newcommand{\sGL}{\mathsf{GL}}
\newcommand{\sSO}{\mathsf{SO}}
\newcommand{\sU}{\mathsf{U}}

\newcommand{\sOSp}{\mathsf{OSp}}

\newcommand{\ad}{{\dot{\alpha}}}                           
\newcommand{\bd}{{\dot{\beta}}}   
\newcommand{\mud}{{\dot{\mu}}} 
\newcommand{\qB}{{\bar{\theta}}} 
\newcommand{\gd}{{\dot\g}}
\newcommand{\dd}{{\dot\d}}                        
\newcommand{\ve}{\varepsilon}                          
\newcommand{\cDB}{{\bar\cD}}                        
\newcommand{\DB}{\bar{D}}

\newcommand{\ab}{{\a\b}}

\renewcommand{\aa}{{\a\ad}}
\newcommand{\bb}{{\b\bd}}
\newcommand{\pa}{\partial}                      
\newcommand{\hf}{\frac12}

\def\rd{{\rm d}}
\def\ri{{\rm i}}

\newcommand{\Nabla}{\bm{\nabla}}
\newcommand{\bNabla}{\bar{\bm{\nabla}}}
\newcommand{\vf}{\varphi}

\newcommand{\be}{\begin{equation}}
\newcommand{\ee}{\end{equation}}
\newcommand{\bsubeq}{\begin{subequations}}
\newcommand{\esubeq}{\end{subequations}}
\newcommand{\ba}{\begin{align}}
\newcommand{\ea}{\end{align}}
\newcommand{\bea}{\begin{eqnarray}}
\newcommand{\eea}{\end{eqnarray}}
\newcommand{\non}{\nonumber}

\newcommand{\mc}{\mathcal}
\newcommand{\mf}{\mathfrak}
\newcommand{\ms}{\mathscr}
\newcommand{\mb}{\mathbb}

\newcommand{\bm}[1]{\mbox{\boldmath$#1$}}

\def \un{\underline}

\begin{document}

\pagestyle{fancy}
\pagenumbering{Roman}

\begin{titlepage}
\begin{center}

{\LARGE \bf Models for (super)conformal higher-spin\\[8pt] fields  on curved backgrounds}

        \vspace{1.5cm}

       {\Large{\textbf{Michael Ponds}}}\\
\large  
  \vspace{1cm}
Supervisor: ~~~~~~~~~~~~~~~~Prof. Sergei M. Kuzenko\\
Co-supervisor:   ~~~~~~~~~~~~A/Prof. Evgeny I. Buchbinder

  \vspace{2cm}

        \includegraphics[width=0.3\textwidth]{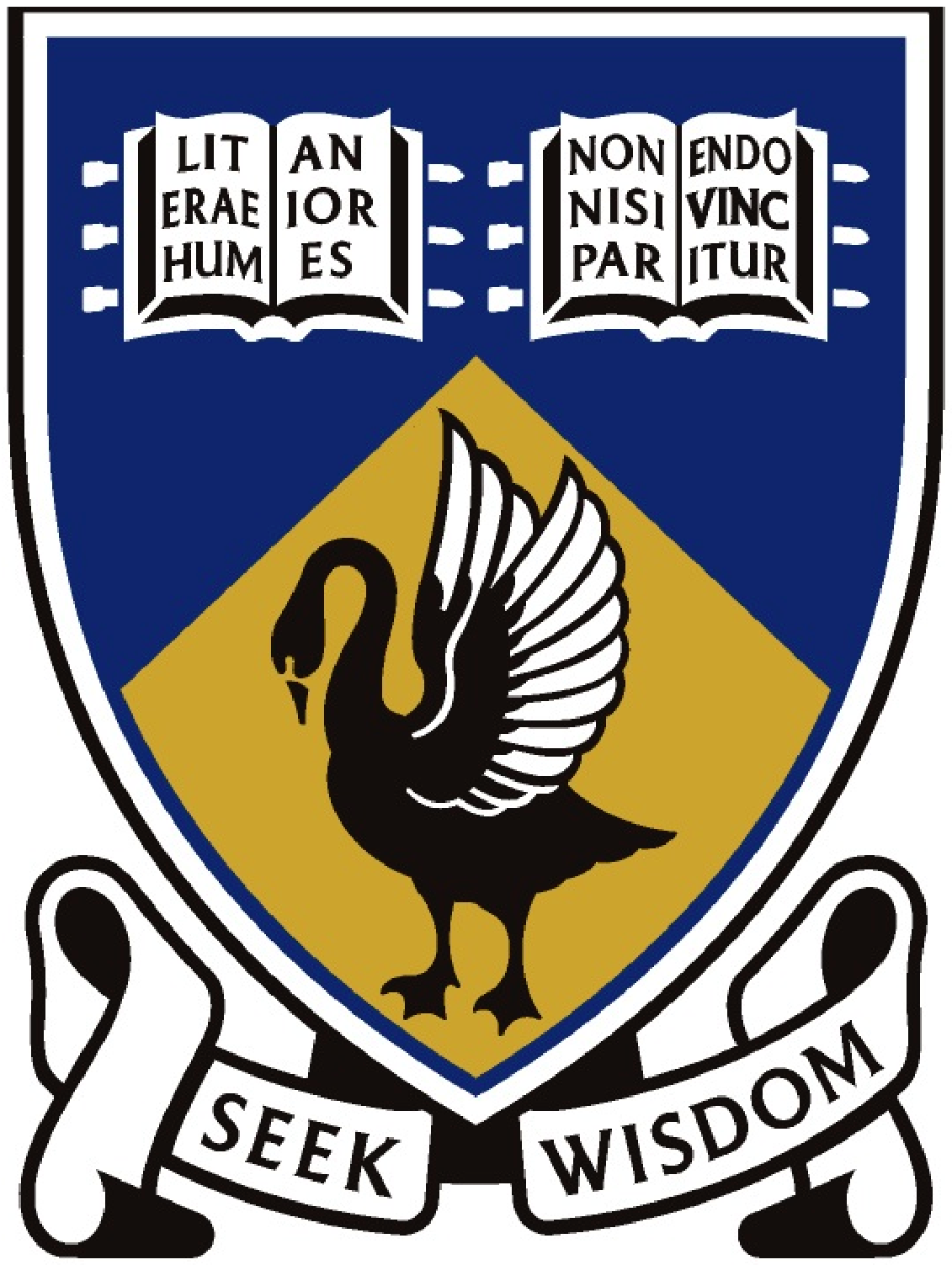}

  \vspace{1.5cm}

This thesis is presented for the degree of Doctor of Philosophy\\
The University of Western Australia\\
Department of Physics\\
September 2021

\vspace{0.7cm}

\flushleft{
Examiners:\\
Prof. Dmitri Sorokin \hfill (I.N.F.N. Sezione di Padova, Italy)\\
Prof. Rikard von Unge ~\hfill ~(Masaryk University, Czech Republic)
}

\end{center}
\end{titlepage}
\chapter*{Abstract}
\newenvironment{changemargin}[2]{%
\begin{list}{}{%
\setlength{\topsep}{0pt}%
\setlength{\leftmargin}{#1}%
\setlength{\rightmargin}{#2}%
\setlength{\listparindent}{\parindent}%
\setlength{\itemindent}{\parindent}%
\setlength{\parsep}{\parskip}%
}%
\item[]}{\end{list}}

\begin{changemargin}{-0.1cm}{-0.1cm}

This thesis is devoted to the construction of theories describing the consistent propagation of (super)conformal higher-spin fields on curved three- and four-dimensional (super)spaces.
Gauge-invariant actions for free conformal higher-spin fields in Minkowski space were proposed  over thirty years ago. 
Since then there have been many attempts to promote these models to curved backgrounds.
However, due to the presence of higher-derivative operators, preservation of the higher-spin gauge symmetry has turned out to be a major technical obstacle.

In the first half of this thesis we systematically derive models for conformal fields of arbitrary rank on various types of curved spacetimes. 
On generic conformally-flat backgrounds in three $(3d)$ and four $(4d)$ dimensions, we obtain closed-form expressions for the actions which are manifestly gauge and Weyl invariant.
Similar results are provided for generalised conformal fields, which have higher-depth gauge transformations. 
In three dimensions, conformally-flat spacetimes are the most general backgrounds allowing consistent propagation.
 In four dimensions, it is widely expected that gauge invariance can be extended to Bach-flat backgrounds, although no complete models for spin greater than two exist. 
We confirm these expectations for the first time by constructing a number of complete gauge-invariant models for conformal fields with higher spin. 
In the second half of this thesis we employ superspace techniques to extend the above results to conformal higher-spin theories possessing off-shell supersymmetry.  
It is shown that the existence of such theories has interesting implications on the conformal models embedded within them.

Several novel applications of our results are also provided.  
In particular, transverse projection operators are constructed in $4d$ anti-de Sitter (AdS$_4$) space, and their poles are shown to be associated with partially-massless fields. 
This allows us to demonstrate that on such backgrounds, the (super)conformal higher-spin kinetic operator factorises into products of second order operators. 
Similar conclusions are drawn in AdS$_3$ (super)space. 
Finally, we make use of the (super)conformal higher-spin models in $3d$ Minkowski and AdS (super)space to build topologically massive gauge theories.

\end{changemargin}

\chapter*{Authorship Declaration}
This thesis is based on eight published papers \cite{Topological, 3Dprojectors, Confgeo, AdSprojectors, spin3depth3, SCHS, SCHSgen, CottonAdS}. Their details are as follows:

\begin{enumerate}

\item S.~M.~Kuzenko and M.~Ponds,  \\
{\it Topologically massive higher spin gauge theories,}\\
JHEP {\bf 1810}, 160 (2018) 
\href{https://arxiv.org/abs/1806.06643}{[arXiv:1806.06643 [hep-th]]}.\\
{\bf Location in thesis:} Chapters \ref{Chapter3D} and \ref{Chapter3Dsuperspace}.


\item E. I. Buchbinder, S. M. Kuzenko, J. La Fontaine and M. Ponds, \\
{\it Spin projection operators and  higher-spin Cotton tensors in three dimensions,} \\
Phys.\ Lett.\ B {\bf 790}, 389 (2019)
 \href{https://arxiv.org/abs/1812.05331}{ [arXiv:1812.05331 [hep-th]]}.\\
{\bf Location in thesis:} Chapter \ref{Chapter3D}.


\item S.~M.~Kuzenko and M.~Ponds,\\
{\it Conformal geometry and (super)conformal higher-spin gauge theories,}\\
JHEP {\bf 1905},  113 (2019) 
\href{https://arxiv.org/abs/1902.08010}{[arXiv:1902.08010 [hep-th]]}.\\
{\bf Location in thesis:}  Chapters \ref{Chapter3D}, \ref{Chapter4D}, \ref{Chapter3Dsuperspace} and \ref{Chapter4Dsuperspace}.


\item S.~M.~Kuzenko and M.~Ponds,\\
{\it Spin projection operators in (A)dS and partial masslessness,}\\
Phys. Lett. B \textbf{800}, 135128 (2020)
\href{https://arxiv.org/pdf/1910.10440.pdf}{[arXiv:1910.10440 [hep-th]]}.\\
{\bf Location in thesis:} Chapter \ref{Chapter4D}.


\item S.~M.~Kuzenko and M.~Ponds,\\
 {\it Generalised conformal higher-spin fields in curved backgrounds,}\\
  JHEP {\bf 2004}, 021 (2020)
 \href{https://arxiv.org/abs/1912.00652}{[arXiv:1912.00652 [hep-th]]}.\\
{\bf Location in thesis:} Chapter \ref{Chapter4D}.


\item S.~M.~Kuzenko, M.~Ponds and E.~S.~N. Raptakis,\\ 
{\it New locally (super)conformal gauge models in Bach-flat backgrounds,}\\
JHEP \textbf{2008}, 068 (2020)
\href{https://arxiv.org/abs/2005.08657}{[arXiv:2005.08657 [hep-th]]}.\\
{\bf Location in thesis:} Chapters \ref{Chapter4D} and \ref{Chapter4Dsuperspace}.


\item S.~M.~Kuzenko, M.~Ponds and E.~S.~N.~Raptakis,\\
{\it Generalised superconformal higher-spin multiplets,}\\
JHEP \textbf{03}, 183 (2021)
\href{https://arxiv.org/abs/2011.11300}{[arXiv:2011.11300 [hep-th]]}.\\
{\bf Location in thesis:} Chapters \ref{Chapter4D} and \ref{Chapter4Dsuperspace}.


\item S.~M.~Kuzenko and M.~Ponds,\\
{\it Higher-spin Cotton tensors and massive gauge-invariant actions in AdS$_3$,}\\ 
JHEP \textbf{05}, 275 (2021)
\href{https://arxiv.org/abs/2103.11673}{[arXiv:2103.11673 [hep-th]]}.\\
{\bf Location in thesis:} Chapters \ref{Chapter3D} and \ref{Chapter3Dsuperspace}.


\end{enumerate}

The published papers \cite{AdSuperprojectors, AdS3(super)projectors} were also
produced during the candidature of my PhD, 
but only partial results from them contribute to the thesis. Their details are as follows:


\begin{enumerate}

\setcounter{enumi}{8}

%
%

\item E.~I.~Buchbinder, D.~Hutchings, S.~M.~Kuzenko and M.~Ponds, \\
{ \it AdS superprojectors,}\\
 JHEP {\bf 2104}, 074 (2021)
\href{https://arxiv.org/abs/2101.05524}{[arXiv:2101.05524 [hep-th]]}.\\
{\bf Location in thesis:} Chapter \ref{Chapter4Dsuperspace}.


\item D.~Hutchings, S.~M.~Kuzenko and M.~Ponds,\\
{ \it AdS (super)projectors in three dimensions and partial masslessness,}\\
JHEP \textbf{2110}, 090 (2021), \href{https://arxiv.org/abs/2107.12201}{[arXiv:2107.12201 [hep-th]]}.\\
{\bf Location in thesis:}  Chapter \ref{Chapter4D}.


\end{enumerate}

\hspace{5pt}

Permission has been granted to use the above work.

\hspace{5pt}

Evgeny Buchbinder

\hspace{5pt}

Daniel Hutchings

\hspace{5pt}

Sergei Kuzenko

\hspace{5pt}

James La Fontaine

\hspace{5pt}

Emmanuel Raptakis

\hspace{5pt}
\chapter*{Acknowledgements}

Completing a PhD is something that cannot be achieved alone and there are many people to whom I am indebted. 
First of all, I would like to thank  my supervisor, Sergei Kuzenko, for his endless guidance, unwavering patience and boundless insight.
Special thanks also goes to my co-supervisor, Evgeny Buchbinder,  for providing many valuable comments on my thesis. 
Their world-class expertise and desire to give everyone a fair go has seen them become beacons of inspiration for many students in the isolated land down under, including myself.  Your efforts are much appreciated!

Of course, the past four years would not have been possible, or nearly as enjoyable, without the limitless support of my family and friends. 
Thank you all for the invaluable company, being so genuine and allowing me to appreciate the smaller things in life. 
 You make it impossible and unimaginable to forget who I am and where I've come from.  

To the UWA crew -- Ben, Conway, Daniel, Darren, Emmanuel, James and Jess -- thanks for all of the amusing, wacky and hilarious moments. You'd think that after so many years of eating lunch and sharing offices with the same people, the conversations would get stale, but it has been quite the opposite! 
To the non-UWA crew -- Dom, Nicole, Sam, Stefano, Tarnee and the Clifton Street gang (past and present) -- thank you for your friendship, the unforgettable experiences, and providing me with an escape from the world of academia when it was needed.
To the Ponds and Chase tribe -- I won't attempt to name you all, but you know who you are -- thanks for always having my back, being beautifully weird and bringing out the (perhaps not so beautiful) weird in me.  To me you are not just family, but also some of my greatest friends, and I'm proud to be part of the tribe. 

This research was supported by a Hackett Postgraduate Scholarship, under the Australian Government Research Training Program.


{\hypersetup{hidelinks}
\tableofcontents
}
\clearpage\pagenumbering{arabic}


\chapter{Introduction}

Currently, our understanding of the universe is divided into two regimes.
At large scales, its structure is accurately described by Einstein's theory of general relativity, which is a classical model of gravity.
At small scales, where the other fundamental forces become dominant, our best description of nature is the Standard Model of particle physics, which is a quantum field theory. 
Although the two models describe physics at different scales and are formulated within different frameworks, they have several principles in common which underpin their foundation. 
In particular, both theories may be understood as gauge theories of an underlying symmetry group.
 The symmetry groups commonly encountered in physics fall roughly into two categories -- internal and spacetime groups.  
 The Standard Model is a gauge theory of the internal symmetry group $\sSU(3)\times\sSU(2)\times\sU(1)$.
 In contrast,  Einstein gravity may be formulated as a gauge theory of the Poincar\'e group, which is an example of a spacetime symmetry group.

The Poincar\'e group is a symmetry group of any closed relativistic system.
As such, one can classify elementary particles according to its unitary irreducible representations, which in turn are specified by mass $m$ and spin $s$ \cite{Wigner}.
Theories describing particles with spin $s>2$ are traditionally known as higher-spin theories. Such theories have been the subject of intense study for over eighty years. Within this period, an enormous amount of research has been invested into their development, and we will not attempt to provide a complete overview. For this purpose we refer the reader to the review articles \cite{VasilievReview1, SorokinHS, BekaertVasilievReview, VasilievReview2, BekaertReview, DidenkoSkvortsov,RahmanTaronna} and references therein.

The gauge principle plays a pivotal role in higher-spin theories.
Indeed, in order to guarantee the correct number of physical degrees of freedom,  massless higher-spin theories are necessarily described by gauge fields.
Preservation of the gauge symmetry then serves as a guiding principle when building models for such fields.
However, the problem of constructing consistent (i.e. gauge invariant) theories of interacting massless higher-spin fields has turned out to be extraordinarily difficult.

Free massless spin-$s$ actions were constructed by Fang and Fronsdal in both Minkowski \cite{Fronsdal1, FF} and (anti-)de Sitter $\big($(A)dS$\big)$ \cite{Fronsdal2, FF2}   spaces.\footnote{Models for free massless higher-spin fields posessing off-shell $\mc{N}=1$  supersymmetry were derived in four-dimensional Minkowski and anti-de Sitter superspace in \cite{KPS,KS,KS94}.
Models with on-shell supersymmetry in four-dimensional Minkowski space were proposed earlier in \cite{Curtright:1979uz, Vasiliev:1980as}. Various cubic vertices for massless higher-spin supermultiplets have been constructed, for example, in\cite{SHSI1, SHSI2, SHSI3, SHSI4, BHK, SHSI5}.} 
Around the same time, it was shown by Aragone and Deser \cite{AragoneDeser} that a consistent covariantisation of these models on general gravitational backgrounds does not exist.
Their analysis led to the belief that gravitational interactions of massless higher-spin fields were impossible to formulate.
While the results of \cite{AragoneDeser} are true, it was later shown by Fradkin and Vasiliev \cite{FV-vertices87a, FV-vertices87b} that the proper setting for consistent introduction of such interactions is on an (A)dS background. 

 Since the work of \cite{FV-vertices87a, FV-vertices87b}, a full non-linear theory of massless higher-spin fields has been established at the level of equations of motion  \cite{VasilievFull, VasilievFulld}. 
 However, the study of its quantum aspects has been hampered by the lack of a corresponding Lagrangian formulation.
The theory of conformal higher-spin (CHS) fields, pioneered in \cite{FT}, is a younger and  more tractable example of an interacting higher-spin theory as compared to the massless case. 
Nevertheless, the construction of consistent models describing free CHS fields on curved backgrounds is an important open problem. 
This is the research problem that we aim to address in this thesis, with a focus on spacetimes of three and four dimensions. 
In addition, we will also be interested in CHS theories possessing off-shell $\mc{N}=1$ supersymmetry, known as superconformal higher-spin theories. 
Before describing these specific problems in more detail, we first provide a brief overview of CHS theory. 

\begin{flushleft}
{\bf Conformal higher-spin theory}
\end{flushleft}


The study of CHS theory was initiated in 1985 by Fradkin and Tseytlin \cite{FT}, where they introduced  free CHS models in four-dimensional Minkowski space $\mb{M}^4$.
The latter are generalisations of electrodynamics ($s=1$) and linearised conformal gravity ($s=2$) to conformal fields with spin $s>2$.  
Among the attractive features of CHS theories are the following: (i) maximal spin-$s$ gauge symmetry consistent with locality \cite{FT}; (ii) Lagrangian formulation for a complete interacting bosonic CHS theory \cite{Tseytlin, Segal, BJM1, BJM2, Bonezzi}; (iii) interesting quantum properties \cite{Tseytlin13, TseytlinSixSphere, Beccaria:2014jxa, BT2015, Joung:2015eny, Beccaria:2016syk, BeccariaTseytlin9000, BeccariaTseyltinSphere, AABD, Adamo1, Adamo2, Beccaria:2018rxp, ABD}; and (iv) natural connection to the AdS/CFT correspondence \cite{Tseytlin, Metsaev10, AdS/CFT3, AdS/CFT4, AdS/CFT5}. 
Let us elaborate on some of these points. 

In the case of integer spin $s$, the free CHS model of \cite{FT} is described by a totally symmetric rank-$s$ tensor field $\bm h_{a_1 \dots a_s}= \bm h_{(a_1\dots a_s)} \equiv \bm h_{a(s)}$
with the action functional
\begin{align}
S_{\text{CHS}}^{(s)}[{\bm h}_{(s)}] \propto \int \text{d}^4x {\bm h}^{a(s)}\Box^s \Pi^{\perp}_{(s)}{\bm h}_{a(s)}~.\label{intro.1}
\end{align}
Here $\Pi^{\perp}_{(s)}$ is the spin-$s$ projection operator, originally proposed by Behrends and Fronsdal \cite{BF, Fronsdal58} in the 1950s, which projects $\bm h_{a(s)}$ onto its transverse and traceless (TT) component
\begin{align}
\pa^b \Pi^{\perp}_{(s)} \bm h_{ba(s-1)}=0~,\qquad \eta^{bc}\Pi^{\perp}_{(s)} \bm h_{bca(s-2)}=0~.\label{intro.2}
\end{align}
For further discussion on the projectors $\Pi^{\perp}_{(s)}$, we refer the reader to section \ref{secFTCHSMd}. A field $\bm h_{a(s)}$ satisfying \eqref{intro.2} is said to describe a pure spin-$s$ field off the mass shell. The corresponding equation of motion may be shown to be equivalent to $\Box^s\bm h_{a(s)} =0$, for TT $\bm h_{a(s)} $. As shown in \cite{FT}, this system encodes $s(s+1)$ dynamical degrees of freedom. 

As a consequence of the properties \eqref{intro.2}, the action \eqref{intro.1} is invariant under the following differential and algebraic gauge transformations
\begin{align}
\delta_{\xi,\l}\bm h_{a(s)} = \pa_{(a_1}\xi_{a_2\dots a_s)}+\eta_{(a_1a_2}\l_{a_3\dots a_s)}~, \label{intro.3}
\end{align}
with $\xi_{a(s-1)}$ traceless.
They can be considered as higher-spin generalisations of linearised diffeomorphisms and Weyl transformations. 
 In order for the CHS action to be local and invariant under \eqref{intro.3}, the corresponding Lagrangian must be of a higher-derivative nature\footnote{There exists an `ordinary derivative' formulation of CHS theories, which removes higher-derivatives through the introduction of auxilliary fields \cite{MetsaevOrd1,MetsaevOrd2}.} (the $\Box^s$ in \eqref{intro.1} compensates for the non-locality of  $\Pi^{\perp}_{(s)}$), which spoils unitarity.

In this thesis we work 
 exclusively with the totally symmetric and traceless field $h_{a(s)}$. This is achieved by making use of the algebraic gauge symmetry in \eqref{intro.3} to gauge away the trace of $\bm h_{a(s)}$
\begin{align}
\bm h_{a(s)}&\equiv h_{a(s)}~,\qquad \eta^{bc}h_{bc a(s-2)}=0~. \label{intro.4}
\end{align}
From general principles, it follows that the action \eqref{intro.1} is invariant under rigid conformal transformations, where $h_{a(s)}$ is a primary field with conformal weight $\Delta_s=2-s$.
We refer the reader to section \ref{SectionCHSprep} for further clarification on these points. 


Since the work of \cite{FT} there has been much interest in CHS theories in diverse spacetime dimensions $d$. 
 For $d>4$, free CHS actions were proposed in \cite{Segal}  (see also \cite{Marnelius, Marnelius2, Vasiliev2009, Metsaev:2012hr}), although in odd dimensions they are non-local and do not describe genuine field theories.
In $d=3$, consistent non-linear and local actions for CHS fields were constructed by Pope and Townsend \cite{PopeTownsend}, and independently by Fradkin and Linetsky \cite{FL-3D}.\footnote{In three dimensions, CHS fields do not generally carry any dynamical degrees of freedom. }   
In both cases the models are formulated as gauge theories of certain conformal higher-spin (super)algebras with a corresponding Chern-Simons action. 
The free actions for $3d$ bosonic CHS fields were deduced in \cite{PopeTownsend}, where they arise from a weak field expansion of the Chern-Simons functional. The free actions for $3d$ fermionic CHS fields were derived some $30$ years later by Kuzenko \cite{K16} using superspace techniques. 
In both cases, higher-spin generalisations of the linearised Cotton tensor, which lie at the heart of many recent studies (see e.g. \cite{DD, BHT,BKRTY,Nilsson1, Nilsson2,HHL, LN, KO,  KT17, BBB, HKO, HLLMP, BHHK, Grigoriev:2019xmp}), are of central importance.


At the non-linear level, $4d$ CHS superalgebras
and associated gauge theories in the cubic approximation were constructed in \cite{FL-algebras} and \cite{FL1} respectively. These results built upon the seminal work by Fradkin and Vasiliev on higher-spin superalgebras \cite{FV1,FV2,Vasiliev88} and interacting massless higher-spin theories \cite{FV-vertices87a,FV-vertices87b}.
Finally, the Lagrangian formulation for a complete interacting bosonic CHS theory 
was sketched by Tseytlin \cite{Tseytlin} and fully developed by Segal \cite{Segal} in 2002. 
The models proposed in \cite{Segal} are valid for any even dimension.
Further understanding of the interacting CHS theory was obtained in the important work \cite{BJM2} (see also \cite{BJM1,Bonezzi}).

Conformal higher-spin theory is the first (non-stringy, 4$d$) 
example of a complete non-linear theory of higher-spin fields admitting an action.\footnote{The so-called chiral higher-spin gravity  \cite{ChiralHS1, ChiralHS2, ChiralHS3, ChiralHS4, ChiralHS5, ChiralHS6, ChiralHS7, ChiralHS8, ChiralHS9} is a recent example of another consistent theory of fully interacting $4d$ higher-spin fields admitting an action.}
In even dimensions, the latter may be understood as an induced action. 
More specifically, it corresponds to the logarithmically divergent sector of the effective action obtained after integrating out a complex conformal scalar field 
 coupled to an infinite tower of background CHS fields. 
 With its establishment came the possibility to study its corresponding properties at the quantum level, leading to many interesting results.


For instance, in four dimensions it has been shown  \cite{Joung:2015eny,Beccaria:2016syk, Adamo1, Adamo2}  that various tree-level three- and four-point scattering amplitudes with massless external states, mediated by the tower of CHS fields, vanish. 
 These results suggest that when restricted to massless higher-spin (i.e. unitary) external states, the S-matrix of CHS theory is trivial.   
 Another interesting feature is the vanishing of the free one-loop partition function\cite{Tseytlin13, Beccaria:2014jxa, BT2015}, and of the $a$ coefficent of the associated Weyl anomaly \cite{Tseytlin13,TseytlinSixSphere,AdS/CFT3,AdS/CFT4} (see also \cite{AABD,ABD}).
In each case, all such cancellations are non-trivial, since the corresponding computations involve the summation of an infinite number of contributions (with suitable regularisation) 
from each spin-$s$ CHS field. 
These cancellations are believed to be a constraint imposed by the large underlying global symmetry of the interacting CHS theory, 
and is a good indication towards consistency as a quantum theory.


%


\begin{flushleft}
{\bf Superconformal higher-spin theory}
\end{flushleft}

Off-shell\footnote{A supermultiplet is called off-shell if the the algebra of supersymmetry transformations closes without having to impose the equations of motion.} $\cN=1$ superconformal higher-spin (SCHS) multiplets in four dimensions were briefly discussed by Howe, Stelle and Townsend  \cite{HST81} in 1981, a few years before 
Fradkin and Tseytlin \cite{FT} 
constructed the free CHS actions.
The authors of \cite{HST81} were interested only in the structure of the gauge supermultiplets, and as such no actions were proposed.
It was only until 2017 that the higher-spin gauge prepotentials
(describing the superspin-$(s+\hf)$ multiplet, with $s=2,3,\dots$) introduced in \cite{HST81} 
and more general off-shell gauge supermultiplets were finally used to construct 
free $\cN=1$ SCHS actions \cite{KMT}. Recently, the free actions for $\mc{N}=2$ superconformal gauge multiplets were constructed in \cite{KRN=2SCHS}, and duality invariant (S)CHS models with $0\leq \mc{N} \leq 2$ were investigated in \cite{KRDuality}. 
Parallel studies in three dimensions describing SCHS gauge multiplets and the corresponding Chern-Simons actions were conducted in \cite{K16,KT17}, \cite{KO,HKO}  and \cite{BHHK} for the cases 
$\cN=1$, $\cN=2$ and $3\leq \mc{N} \leq 6$  respectively.

Within the frame-like component approach, gauge-invariant actions for superconformal higher-spin multiplets have been constructed by Fradkin and Linetsky at the cubic level in 
both three \cite{FL-3D} and four \cite{FL-4D} dimensions. 
The $3d$ and $4d$ off-shell constructions 
described above open the possibility to develop a manifestly supersymmetric setting for the cubic SCHS theories of \cite{FL-3D,FL-4D}. 
Not much is known about an interacting SCHS theory beyond the cubic level. 
However, by virtue of the off-shell formulation of \cite{KMT}, it becomes feasible, as was briefly discussed in \cite{KMT}, to formulate a complete non-linear $4d$ $\mc{N}=1$ SCHS theory by developing a superfield analogue of Segal's bosonic CHS theory \cite{Segal}.

An important feature of the approach advocated in \cite{KMT} is that it provides a new avenue to study the problem of consistent propagation of CHS fields on curved backgrounds. 
Specifically, we are referring to the fact that SCHS superfields contain CHS fields at the component level. Hence, an effective method of studying various CHS models can be to study the corresponding SCHS models which generate them. This sentiment will be echoed throughout this thesis.
 Indeed at various points we will exploit this fact to gain useful insight into
CHS models via the associated SCHS ones (and vice versa).

\begin{flushleft}
{\bf (Super)conformal higher-spin models on curved backgrounds}
\end{flushleft}



The problem of consistently deforming the free CHS actions from flat to curved gravitational backgrounds is non-trivial. 
By a consistent deformation it is meant that both the gauge and conformal symmetry of the action is preserved. 
The conformal symmetry is held intact by ensuring invariance under Weyl transformations of the background metric, which on its own is relatively straightforward.  
The difficulty lies within securing the gauge symmetry which, in turn, spoils the manifest Weyl symmetry. This can be attributed to the higher-derivative nature of the action:
 for every positive (half-)integer $s \geq 1$, the action for a conformal spin-$s$ field involves $2s+d-4$ derivatives.


As a consequence, gravitational interactions of conformal higher-spin fields still remain quite mysterious. 
For example, since the original 1985 paper \cite{FT} it was reasonably clear that there should exist a consistent formulation for all CHS models on arbitrary conformally flat backgrounds.\footnote{Their existence on such backgrounds is guaranteed due to the conformal symmetry. The non-trivial problem is obtaining the actions in a manifestly covariant form. } 
However, explicit expressions for the gauge invariant actions describing CHS fields with arbitrary integer spin have been derived only in the case of (anti-)de Sitter space \cite{Metsaev2014} (see also \cite{NTCHS}). 
Similarly, the procedure to couple off-shell SCHS multiplets to conformal supergravity was explained in \cite{KMT}, though explicit calculations were carried out only for the superconformal gravitino (i.e. superspin 1) multiplet.

The principal goal of this thesis is to fill this gap in the literature.
We will construct concrete gauge invariant models for (super)conformal higher-spin fields propagating on various (super)gravity backgrounds. 
As a by-product of this overarching objective, we will derive many other novel results in closely related areas, and applications thereof. 

As previously mentioned, our attention will be restricted to (super)spaces of dimension three and four.  
In three dimensions the Weyl tensor is identically zero and all information about the conformal structure of spacetime is encoded in the Cotton tensor. In particular, spacetime is conformally-flat if and only if the Cotton tensor vanishes \cite{Eisen}.
It is well known that the dynamics of the conformal graviton ($s=2$) can be consistently defined at most on conformally-flat backgrounds.
This is because the corresponding gauge-invariant action may be obtained by linearising the action of $3d$ conformal gravity \cite{DJT1,DJT2,vN, HW}  about its stationary point, and the equation of motion is the vanishing of the Cotton tensor.
Similarly, in off-shell $\cN$-extended conformal supergravity,
the equation of motion states that the super-Cotton tensor vanishes \cite{BKNT-M2,KNT-M13}, hence curved superspace is conformally-flat. 
It is therefore natural to expect that the most general type of background on which (S)CHS fields can consistently propagate are conformally-flat ones.
These are the precisely the types of backgrounds for which we will derive unique closed-form models.

  The situation in $d=4$ is much more intricate
   as far as the issue of consistent  propagation of (S)CHS fields on curved backgrounds is concerned. This is because the equation of motion for Weyl's conformal gravity is the vanishing of the Bach tensor, which does not neccesarily correspond to a conformally-flat background. Likewise, the equation of motion for $\mc{N}=1$ conformal supergravity  \cite{KTvN1,KTvN2} is the vanishing of the so-called super-Bach tensor \cite{BK88} (see e.g. \cite{BK} for a review).
   It follows that the conformal graviton, and its corresponding supermultiplet (i.e. superspin 3/2), may be consistently defined only if the background (super)space is Bach-flat.

It would therefore seem that to ensure the consistent propagation of a single conformal  field with spin $s>2$, the vanishing of the background Bach tensor is a necessary condition.   
For quite some time it was  believed that this condition was also sufficient.
However, recent studies of the spin-3 theory
\cite{NTCHS, GrigorievT, BeccariaT, Manvelyan2} have demonstrated \cite{GrigorievT, BeccariaT} that this is in fact not the case. 
In particular, it has been conjectured by Grigoriev and Tseytlin \cite{GrigorievT} that in order to maintain gauge invariance of the conformal spin-$3$ field on a Bach-flat background, it is necessary to introduce a non-minimal coupling to a conformal spin-$1$ field.\footnote{The presence of such non-diagonal terms may introduce corrections into the corresponding quantum theory. This is the case, for example, with the $c$ coefficient of the Weyl anomaly \cite{BeccariaT}.}


A more detailed account 
of the conformal spin-3 model can be found in section \ref{secCHSBach}. For now, it suffices to mention that a complete gauge invariant model is still lacking, and its existence (or uniqueness) is not guaranteed. 
Though we do not provide a resolution to this specific story, in this thesis we will provide various complete models of (super)conformal higher-spin fields propagating on Bach-flat backgrounds. Gauge invariance of many of these models require couplings to subsidiary conformal fields, and are the first complete examples exhibiting this mechanism. This result supports the conjecture of \cite{GrigorievT}.

\begin{flushleft}
{\bf Conformal (super)space}
\end{flushleft}

In order to keep the (super-)Weyl symmetry under control, we will develop a formalism with manifest local (super)conformal symmetry. This will be achieved by employing a well-known (super)conformal tensor calculus created by supergravity practitioners \cite{MacDowellMansouri, KTvN1,  KTvN2, vN} (see \cite{ButterN=1, ButterN=2, BKNT-M1, BKNT-M5D, BKNT17} for superspace formulations).\footnote{Another modern formulation of conformal geometry was developed by mathematicians and is often referred to as conformal tractor calculus \cite{BEG,Gover}. Its roots go back to the work of Thomas \cite{Thomas}; see \cite{Gover:2008sw, Gover:2008pt, Bonezzi:2010jr, Grigoriev:2011gp, Joung:2013doa} for some of its interesting applications in physics. Conformal geometry was recently studied in \cite{JoungConfgeo} using the unfolded formulation.}
In this approach, which we will refer to as conformal (super)space, the geometry of conformal (super)gravity is formulated as a gauge theory of the (super)conformal group (upon imposing suitable constraints).

This framework will serve as one of our main technical tools. 
In the non-supersymmetric case, the corresponding geometry is encoded within the conformally covariant derivative, denoted $\nabla_a$. 
Its utility can be summarised by the following equations
\begin{subequations}
\begin{align}
d=3:~~~~~~~~~~~~~~~\big[\nabla_a,\nabla_b\big]&=\phantom{-}\frac{1}{2}C_{ab}{}^{c}K_c~,\\
d>3:~~~~~~~~~~~~~~~\big[\nabla_a,\nabla_b\big]&=-\frac{1}{2}W_{ab}{}^{cd}M_{cd}+\frac{1}{2(d-3)}\nabla^cW_{abc}{}^{d}K_d~.
\end{align}
\end{subequations}
Here, $C_{abc}$ and $W_{abcd}$ are the Cotton and Weyl tensors respectively, which completely control the algebra of $\nabla_a$.
The operators $M_{ab}$ and $K_a$ are the Lorentz and special conformal generators of the conformal group. 
From these equations, we see that if spacetime is conformally-flat, then $\nabla_a$ satisfies a trivial algebra. Therefore, in practice, working on such backgrounds is akin to working in flat space. 

\begin{flushleft}
{\bf Generalised (S)CHS gauge fields}
\end{flushleft}



As pointed out earlier, the conformal spin-$s$ field $h_{a(s)}$ (see eq. \eqref{intro.4}) carries conformal weight $\big(2-s\big)$. This value is uniquely fixed by its gauge transformation law
\begin{align}
\delta_{\xi}h_{a(s)}=\pa_{(a_1}\xi_{a_2\dots a_s)} - (\text{traces})~.
\end{align}
It is because of their low conformal weights and consequently high derivative Lagrangians that CHS fields are difficult to work with in curved space. 
However, there exists a broader class of conformal gauge fields, the so-called generalised ones $h^{(t)}_{a(s)}$. The latter carry the conformal weight $\big( t+1-s\big)$ and are characterised by gauge transformations with depth-$t$
\begin{align}
\delta_{\xi}h^{(t)}_{a(s)}=\pa_{(a_1}\cdots\pa_{a_t}\xi_{a_{t+1}\dots a_s)}- (\text{traces})~,\qquad 1\leq t \leq s~. \label{intro.7}
\end{align}
 By virtue of their relatively higher conformal weights and consequently lower derivative Lagrangians, they can provide a much friendlier environment in which to study CHS theory. Indeed, we will construct gauge invariant actions on Bach-flat backgrounds for the conformal spin $s=5/2$ and $s=3$ gauge fields with maximal depth.
 

Historically, the first generalised conformal field appeared in the seminal work by Deser and Nepomechie \cite{DeserN1, DeserN2}, where they discussed the maximal-depth conformal graviton. This concept was later extended to tensors of a generic symmetry type by Vasiliev in \cite{Vasiliev2009}, where the corresponding conformal and gauge invariant actions in $\mathbb{M}^d$  were also given (see \cite{BG2013, BT2015, Barnich, GrigorievH, Grigoriev:2019xmp, Grigoriev:2020lzu} for more recent related studies). 
 The supersymmetric multiplets containing generalised CHS fields have not been studied, neither at the superspace nor component level. 
We will rectify this issue, and provide the corresponding gauge and super-Weyl invariant actions.

Another reason for the interest drawn by generalised CHS fields is their role in the context of the AdS/CFT correspondence. More specifically, it is known \cite{BG2013} that generalised CHS fields in $\mathbb{M}^d$ may be identified with the boundary values of partially massless fields 
 propagating in the bulk of AdS$_{d+1}$. 
Partially massless fields are naturally defined in (A)dS, and are another example of fields with gauge transformations of higher depth. However, unlike generalised CHS fields, they are not usually conformal. Their study was also initiated in \cite{DeserN1, DeserN2}, with early developments in \cite{Higuchi1, Higuchi2, Higuchi3, Metsaev2006, DeserW1, DeserW2, DeserW3, DeserW4, DeserW5, Zinoviev, DNW, SV, BMVpm} (see \cite{Brust} for a review).\footnote{Publications  on supersymmetric partially massless theory have appeared recently \cite{G-SHR, BKSZ, BG-SHR, AdSuperprojectors, AdS3(super)projectors}.} 

 Partially massless fields are ubiquitous in the higher-spin literature. 
For example, in AdS$_d$ with $d$ even, it has been shown\footnote{This had already been noticed for some lower-spin values in \cite{DeserN1, DeserN2, Tseytlin5, Tseytlin6 }, and was conjectured for generic spin in \cite{Tseytlin13, Karapet1} (see also \cite{Karapet2}).} by several groups \cite{Metsaev2014, NTCHS, GrigorievH} that the integer spin-$s$ CHS kinetic operator factorises into $s$ products of second-order operators associated with (partially-)massless fields of all depths (plus $(d-4)/2$ massive modes). 
We will extend this result to CHS operators associated with fermionic and mixed symmetry fields in $d=4$,  and to arbitrary (half-)integer spin fields in $d=3$ (being the first derivation in odd dimensions). 
To establish these factorisation properties, we will construct and make use of AdS covariant extensions of the spin projection operators \eqref{intro.2}.\footnote{Superspin projection operators on AdS$_d$ superspace have been recently constructed in $d=4$ \cite{AdSuperprojectors} and in $d=3$ \cite{AdS3(super)projectors}. They will be used to prove analogous factorisation properties of SCHS operators.  } 
We show that the poles of these projectors are associated with (partially-)massless fields, providing a new way to understand the latter.

\begin{flushleft}
{\bf Thesis structure}
\end{flushleft}

 This thesis is organised as follows. In chapter \ref{Chapter2} we will establish our notations and conventions by reviewing the various background materials necessary for the subsequent chapters. This will include a summary of the vielbein formulation of gravity and conformal gravity in sections \ref{SectionVielbein} and \ref{SectionCFT},  as well as a discussion of conformal symmetry in Minkowski and curved spacetimes. In section \ref{SectionCHSprep} we discuss in detail CHS fields on $\mb{M}^d$, whilst in section \ref{GCA} we give a thorough review of conformal geometry as a gauge theory.

 Chapters \ref{Chapter3D} -- \ref{Chapter4Dsuperspace} constitute the bulk of the original material in this thesis, and are based on the publications  \cite{Topological, 3Dprojectors, Confgeo, AdSprojectors, spin3depth3, SCHS, SCHSgen, CottonAdS, AdSuperprojectors, AdS3(super)projectors}. A more detailed description regarding the locations of each publication can be found in the authorship declaration section and, where applicable, at the beginning of each main section. The first half of these chapters are based strictly on non-supersymmetric CHS theories in three (chapter \ref{Chapter3D}) and four (chapter \ref{Chapter4D}) dimensions. The second half are based on $\mc{N}=1$ SCHS theories in superspaces of dimension three (chapter \ref{Chapter3Dsuperspace}) and four (chapter \ref{Chapter4Dsuperspace}). 
 A concluding discussion and future outlook is provided in chapter \ref{ChapterFinito}.
 A detailed outline of the content and plan of chapters \ref{Chapter3D} -- \ref{Chapter4Dsuperspace} is given on the first page of each chapter. We do not repeat this information here, however, below we provide some comments on their general structure. 
 
 Individually, each chapter is organised according to the same scheme. Specifically, the first section of each chapter discusses the salient features and the main ingredients for (S)CHS models on a generic (super)gravity background.  Each middle section is dedicated to the explicit construction of (S)CHS models on a specific background, ordered by increasing calculational difficulty. In three dimensions, this order is Minkowski, conformally-flat and  AdS (super)space.\footnote{In AdS we will be interested in obtaining the CHS actions in terms of the Lorentz covariant derivative rather than the conformally covariant one. In $3d$, the former description proves to be more challenging.} In four dimensions, the order is Minkowski, AdS, conformally-flat and then Bach-flat (super)spaces. 
 
 We will present the original results from the papers \cite{Topological, 3Dprojectors, Confgeo, AdSprojectors, spin3depth3, SCHS, SCHSgen, CottonAdS, AdSuperprojectors, AdS3(super)projectors}, which contribute to this thesis, as if it were the first time they have appeared. We will therefore not usually provide citations for these results in the main body. 
 However, in the last section of each chapter \ref{Chapter3D} -- \ref{Chapter4Dsuperspace}, a summary of the main original results obtained (with relevant citations) is given.
 As stated in the authorship declaration, some results from \cite{AdSuperprojectors, AdS3(super)projectors} are not strictly part of this thesis (e.g. the superprojectors in AdS superspace), and will be cited accordingly. 

 
\chapter{Background material} \label{Chapter2}

In  order to establish the notation and conventions used throughout this thesis, below we review some well-known facts on selected topics in gravity and classical field theory. Our presentation is not intended to be mathematically rigorous or complete,    
and is instead skewed towards a more streamlined approach.
We do not provide any background material on supersymmetry or supergravity in this chapter; the relevant information on these subjects will be supplied in later chapters as it becomes necessary. 

\section{Vielbein formulation of gravity} \label{SectionVielbein}

In this section we review the vielbein formulation of (Einstein) gravity. We begin by describing the gauge group of gravity which, in this approach, consists of general coordinate transformations and local Lorentz rotations. 
 Our presentation is largely influenced by \cite{BK, SKsymm, ButterN=1, FVP}.

\subsection{Local Lorentz symmetry}



Spacetime is described by a Lorentzian manifold $\big(\mc{M}^d,g\big)$, where $\mc{M}^d$ is a connected and differentiable $d$-dimensional manifold and $g$ is a non-degenerate metric with Lorentzian signature $(-+\cdots +)$. 
Let $x^m$ be local coordinates on $\mc{M}^d$, where $m=0,1,\dots,d-1$. For the tangent space $T_p\mc{M}^d$ at $p\in\mc{M}^d$, we take the holonomic coordinate basis $\{\pa_m := \frac{\pa}{\pa x^m}\}$ with corresponding dual basis $\{\text{d}x^m\}$.
 The components of $g$ in this basis are $g_{mn}(x)=g_{nm}(x)$, which are non-singular when considered as a matrix; $\text{det}(g_{mn})\neq 0$. 
 In the metric approach to gravity, $g_{mn}$ is the basic dynamical variable.


In the vielbein formulation of gravity, the basic dynamical field is the vielbein $e_{a}{}^{m}(x)$.\footnote{There is no consensus in the literature on whether to call $e_{a}{}^{m}$ or $e_m{}^{a}$ the vielbein, see e.g. \cite{Eguchi, Ortin}.  } 
The set of vector fields $\{e_a:=e_a{}^{m}(x)\pa_m\}$, known as a frame, constitutes a basis in $T_p\mc{M}^d$  which is orthonormal at each point $p\in\mc{M}^d$ in spacetime,
\begin{align}
e_{a}{}^{m}(x)e_{b}{}^{n}(x)g_{mn}(x)=\eta_{ab} \qquad \Longleftrightarrow \qquad  g_{mn}(x)=e_m{}^{a}(x)e_n{}^{b}(x)\eta_{ab}~. \label{MetricFromVeil}
\end{align}  
This means that $e:=\text{det}(e_m{}^{a})\neq 0$, where we have denoted the inverse vielbein by $e_{m}{}^{a}(x)$:
\begin{align}
e_m{}^{a}(x)e_a{}^{n}&(x)=\delta_{m}{}^{n}~,\qquad e_{a}{}^{m}(x)e_{m}{}^{b}(x)=\delta_{a}{}^{b}~.\label{VielInverse}
\end{align} 
Here $a=0,1,\dots, d-1$ and $\eta_{ab}=\text{diag}(-+\dots+)$ is the Minkowski metric.  The corresponding dual basis of one-forms, or co-frame, is denoted $\{E^a:=\text{d}x^m e_m{}^{a}(x)\}$.

From \eqref{MetricFromVeil} we see that the vielbein $e_{a}{}^{m}$  provides the change in basis of $T_p\mc{M}^d$ which diagonalises the metric at $p$.
 The price paid for this is that the basis $\{e_a\}$ is anholonomic. In other words, it satisfies the non-vanishing commutation relations
\begin{align}
[e_a,e_b]=\ms{C}_{ab}{}^{c}e_c~,\qquad \ms{C}_{ab}{}^{c}:=\big(e_ae_b{}^{m}-e_be_a{}^{m}\big)e_m{}^{c}~, \label{anholonomy}
\end{align}
which is to be contrasted with the holonomic basis $\{\pa_m\}$. The coefficients $\ms{C}_{ab}{}^{c}=-\ms{C}_{ba}{}^{c}$ are known as the anholonomy coefficients.

The vielbein appears to carry $\frac{1}{2}d(d-1)$ extra degrees of freedom as compared to the metric. 
However, these correspond to pure gauge degrees of freedom, since the orthonormal frame $e_a$ belongs to an equivalence class defined up to local Lorentz transformations. 
Let us illustrate this point in the infinitesimal case.
Let $K_{a}{}^{b}(x)$ be some infinitesimal local transformation matrix, and consider the conditions under which  the deformed frame $e'_a=(\delta_{a}{}^{b}+K_a{}^{b})e_b$ remains orthonormal. We see that if $K_{ab}:=K_a{}^{c}\eta_{bc}=-K_{ba}$, then $e'_a{}^{m}$ also satisfies \eqref{MetricFromVeil}. Therefore, this deformation represents a local Lorentz transformation (LLT) in $T_p\mathcal{M}^d$. It follows that any two orthonormal frames may be related by an LLT. 

Under such a transformation, the vielbein and its inverse transform covariantly in the vector representation of the Lorentz group\footnote{The variation $\delta\Phi(x)$ of a field $\Phi(x)$ is defined according to $\delta\Phi(x):=\Phi'(x)-\Phi(x)$. Subscripts belonging to $\delta$ denote the transformation parameter, whilst superscripts indicate the type of transformation.  }
\begin{align}
\delta^{(\text{llt})}_{K}e_a{}^{m}=K_a{}^{b}e_{b}{}^{m}~,\qquad \delta^{(\text{llt})}_{K}e_{m}{}^{a}=K^a{}_{b}e_{m}{}^{b}~. \label{vielLLT}
\end{align}
The Lorentz algebra is spanned by the generators $M_{ab}=-M_{ba}$ of the Lorentz group. The latter satisfy the commutation relations
\begin{align}
\big[M_{ab},M_{cd}\big]=\eta_{ac}M_{bd}-\eta_{bc}M_{ad}-\eta_{ad}M_{bc}+\eta_{bd}M_{ac}~,\label{LorentzAlgebra}
\end{align}
and their action on a vector $v_a$ is defined through 
\begin{align}
M_{ab}v_c=2\eta_{c[a}v_{b]}:=\eta_{ca}v_b-\eta_{cb}v_a~.
\end{align}
In this thesis, (anti-)symmetrisation of $n$ indices is accompanied by a factor of $1/n!$.

Under an invertible general coordinate transformation (GCT) $x^m \rightarrow x'^m=f^m(x)$, the vielbein and its inverse transform as a vector and a covector respectively
 \begin{align}
 e'_a{}^m(x')=\frac{\pa x'^m}{\pa x^n}e_a{}^n(x)~,\qquad e'_m{}^{a}(x')=\frac{\pa x^n}{\pa x'^m}e_n{}^{a}(x)~.
 \end{align}
 Infinitesimally we may write the GCT 
 as $x^m\rightarrow x'^m=x^m-\xi^m$, 
 and to first order in $\xi^m$ the above transformation rules are equivalent to
\begin{subequations}\label{vielgct}
\begin{align}
\delta_{\xi}^{(\text{gct})}e_{a}{}^{m}&= \mc{L}_{\xi}e_{a}{}^{m} =\xi^n\pa_n e_{a}{}^{m} - (\pa_n\xi^m)e_{a}{}^{n}~,\label{vielgct1}\\
\delta_{\xi}^{(\text{gct})}e_{m}{}^{a}&= \mc{L}_{\xi}e_{m}{}^{a} =\xi^n\pa_n e_{m}{}^{a} + (\pa_m\xi^n)e_{n}{}^{a}~.\label{vielgct2}
\end{align}
\end{subequations}
Here $\mc{L}_{\xi}$ denotes the Lie derivative along the vector field $\xi=\xi^m\pa_m$.

Any tensor field defined on spacetime can be expressed with respect to either of the bases $(\text{d}x^m,~\pa_m)$ or $(E^a,~e_a)$. As an example, the vector field $v$ may be written as either $v=v^m(x)\pa_m$ or $v=v^a(x)e_a=v^a(x)e_a{}^{m}(x)\pa_m$. Thus the components are related via $v^m=v^ae_{a}{}^{m}~\implies ~ v^a=v^me_m{}^{a}$, and we see that it is the vielbein (and its inverse) which provides the map between the two bases. The components $v^m(x)$ transform as a vector field under GCTs and are inert under LLTs. On the other-hand, the components $v^a(x)$ transform as a scalar field under GCTs and as a vector field under LLTs.


In this thesis we use indices from the middle of the Latin alphabet to denote world indices (those which transform under the general linear group) and indices from the beginning of the Latin alphabet to denote Lorentz/frame indices (those which transform under the local Lorentz group).
Where possible, we will always express a tensor field with respect to the (co-)frame. This is done by using the vielbein to convert any world indices it may possess into Lorentz ones. 

The composition of general coordinate and local Lorentz transformations comprises the gauge group $\mc{G}$ of (Einstein) gravity. A tensor field $\Phi(x)$ with only Lorentz indices (which are suppressed), is said to be $\mc{G}$-covariant if under $\mc{G}$ it transforms according to
\begin{align}
\delta_{\L}^{(\mc{G})}\Phi=\Big(\delta_{\xi}^{(\text{gct})}+\delta_{K}^{(\text{llt})}\Big)\Phi=\L\Phi~,\qquad \L:=\xi^be_b+\frac{1}{2}K^{bc}M_{bc}~, \label{LorentzG}
\end{align}
for local gauge parameters $\xi^{a}(x)=\xi^m e_m{}^{a}$ and $K_{ab}(x)=-K_{ba}(x)$.  Similarly, in Einstein gravity the gauge freedom of the vielbein is described by the transformation law
\begin{align}
\delta_{\L}^{(\mc{G})}e_{a}{}^{m}=\Big(\delta_{\xi}^{(\text{gct})}+\delta_{K}^{(\text{llt})}\Big)e_{a}{}^{m}=\Big(\mc{L}_{\xi}+\frac{1}{2}K^{bc}M_{bc}\Big)e_{a}{}^{m}~. \label{vielPGgt}
\end{align}
A $\mc{G}$-covariant field does not transform with a derivative on its gauge parameter.

The notation $\mc{G}$ will be used throughout this thesis to denote the relevant gauge group of spacetime. The group $\mc{G}$ will change depending on the gravitational theory under consideration (for example Einstein or conformal gravity). The symbol $\L$ is reserved for the gauge parameter of a full group transformation.

\subsection{The Lorentz covariant derivative} \label{SectionLCD}

If the field $\Phi$ behaves as a scalar under a GCT, then so too does its derivative $e_a\Phi$. However, under an LLT the latter is not $\mc{G}$-covariant because it transforms with a derivative on the gauge parameter. This necessitates the introduction of a Lorentz connection, which maps derivatives of $\mc{G}$-covariant tensors to $\mc{G}$-covariant tensors. The corresponding Lorentz covariant derivative takes the form
\begin{align}
\mc{D}_m:=\pa_m-\frac{1}{2}\omega_{m}{}^{ab}M_{ab}~,\qquad \mathcal{D}_a:=e_a{}^{m}\mc{D}_m~,\label{LCDeriv}
\end{align}
where $\omega_{a}{}^{bc}(x):= e_a{}^{m}\o_{m}{}^{bc}=-\omega_{a}{}^{cb}(x)$ is the Lorentz spin connection. Under an LLT, $\omega_{abc}$ transforms in such a way as to render $\mc{D}_a\Phi$ covariant, whilst under a GCT it transforms as a scalar: $\delta_{\xi}^{(\text{gct})}\o_{abc}=\xi^de_d\o_{abc}$. In particular, for an infinitesimal\footnote{For a finite GCT and LLT, the Lorentz spin connection transforms according to $\omega'_{abc}(x')=\Lambda_{a}{}^{d}\Lambda_{b}{}^{e}\Lambda_{c}{}^{f}\omega_{def}(x)-\eta_{ef}\Lambda_a{}^{d}\Lambda_b{}^{e}e_{d}\Lambda_{c}{}^{f}$ where $\Lambda=\text{exp}\big(\frac{1}{2}K^{bc}M_{bc}\big)$ with $M_{bc}$ in the vector representation. } LLT  parametrised by $K_{ab}$, its transformation law is
\begin{align}
\delta_{K}^{(\text{llt})}\omega_{abc}= e_aK_{bc}+K_{a}{}^{f}\omega_{fbc}+K_{b}{}^{f}\omega_{afc}+K_{c}{}^{f}\omega_{abf}=\mc{D}_{a}K_{bc}+K_{a}{}^{f}\o_{fbc}~. \label{LConLLT}
\end{align}
 Using \eqref{LorentzG} and \eqref{LConLLT}, one can show that $\mc{D}_a\Phi$ transforms covariantly (i.e. without a derivative on the gauge parameter) under a $\mc{G}$-transformation generated by $\L$
\begin{align}
\delta_{\L}^{(\mc{G})} \mc{D}_a \Phi = \xi^b e_b \mc{D}_a \Phi +K_{a}{}^{b}\mc{D}_b\Phi +\frac{1}{2}K^{bc}\mc{D}_a M_{bc}\Phi~. \label{DPhiTrans}
\end{align}
Writing this as $\delta^{(\mc{G})}_{\L}\big(\mc{D}_a\Phi\big)=\L\mc{D}_a\Phi$ allows us to read off the commutator
\begin{align}
\big[M_{ab},\mc{D}_c\big]=2\eta_{c[a}\mc{D}_{b]} \label{LCDMCom}
\end{align}
and the transformation law of $\mc{D}_a$
\begin{align}
\delta_{\L}^{(\mc{G})}\mc{D}_a=\big[\L,\mc{D}_a\big]~, \qquad \L=\xi^ae_a+\frac{1}{2}K^{ab}M_{ab}~. \label{LCDgt}
\end{align}

The Lorentz covariant derivative may be shown to satisfy the commutation relation
\begin{align}
\big[\mc{D}_a, \mc{D}_b\big] = -T_{ab}{}^{c}\mc{D}_c -\frac{1}{2}R_{ab}{}^{cd}M_{cd}~, \label{LCDCommutator}
\end{align}
where the torsion and Lorentz curvature are defined, respectively,  according to
\begin{subequations}
\begin{align}
T_{ab}{}^{c}&= -\ms{C}_{ab}{}^{c}+2\omega_{[ab]}{}^{c}~,\label{Torsion}\\
R_{ab}{}^{cd} &= 2e_{[a}\omega_{b]}{}^{cd}-\ms{C}_{ab}{}^{f}\omega_{f}{}^{cd}+2\omega_{[a}{}^{fc}\omega_{b]f}{}^{d}~.\label{Riemann}
\end{align}
\end{subequations}
Here $\ms{C}_{abc}$ are the anholonomy coefficients \eqref{anholonomy}. 
Both \eqref{Torsion} and \eqref{Riemann} are $\mc{G}$-covariant tensor fields and transform as prescribed by eq. \eqref{LorentzG}.

In Einstein (and conformal) gravity, the vielbein (or equivalently the metric) is the only independent geometric object. To ensure that this is the case, one imposes the torsion-free constraint on the algebra \eqref{LCDCommutator}
\begin{align}
T_{ab}{}^{c}=0~. \label{TFC}
\end{align}
This off-shell constraint determines the spin-connection in terms of the vielbein
\begin{align}
\omega_{abc}\equiv \omega_{abc}(e) = \frac{1}{2}\big(\ms{C}_{abc}-\ms{C}_{acb}-\ms{C}_{bca}\big)~. \label{TFLorCon}
\end{align}
Using \eqref{TFC}, one can then show that the vielbein is covariantly constant,\footnote{When computing $\mc{D}_ae_b{}^{m}$ one must include a compensating term for the world index, which involves the usual Levi-Civita Christoffel connection.} $\mc{D}_ae_b{}^{m}=0$. 

In this thesis we will always work with the torsion-free Lorentz covariant derivative. In this case, on account of the Jacobi identity
\begin{align}
0=\big[\mc{D}_a,[\mc{D}_b,\mc{D}_c]\big]+\big[\mc{D}_b,[\mc{D}_c,\mc{D}_a]\big]+\big[\mc{D}_c,[\mc{D}_a,\mc{D}_b]\big]~,
\end{align}
 the Lorentz curvature satisfies the algebraic Bianchi identity
\begin{align}
0=R_{[abc]d} \quad \implies \quad R_{abcd}=R_{cdab} \label{Bianchi1}
\end{align}
and the differential Bianchi identity
\begin{align}
\mc{D}_{[a}R_{bc]fg}=0~. \label{Bianchi2}
\end{align}
As a consequence, the Ricci tensor $R_{ab}:=\eta^{cd}R_{acbd}$ is symmetric, $R_{ab}=R_{ba}$, and satisfies
\begin{align}
\mathcal{D}^{d}R_{abcd}=-2\mc{D}_{[a}R_{b]c} \quad \implies \quad \mc{D}^bR_{ab}=\frac{1}{2}\mc{D}_{a}R~, \label{Bianchi3}
\end{align}
where $R:=\eta^{ab}R_{ab}$ is the scalar curvature.


For various computations, the definition of a general coordinate transformation used above may not be the most economical one. One way to see this is as follows. Let $\Phi$ be a $\mc{G}$-covariant tensor field with only  (suppressed) Lorentz indices, and perform a GCT followed by an LLT:\footnote{We assume that under a gauge transformation, gauge parameters themselves do not transform.}
\begin{align}
\delta^{(\text{llt})}_{K}\delta^{(\text{gct})}_{\xi}\Phi = \delta^{(\text{llt})}_{K}\big(\xi^{m}\pa_m\Phi\big)=\frac{1}{2}\xi^m\pa_m\big(K^{ab}M_{ab} \Phi\big)~. \label{CGCTMotivation}
\end{align} 
We see that under this combination, $\delta^{(\text{gct})}_{\xi}\Phi$ transforms with a derivative on the LLT gauge parameter, and hence $\delta^{(\text{gct})}_{\xi}\Phi$ is non-covariant. We can fix this issue by accompanying every GCT with an $\xi$ dependent LLT according to the rule
\begin{align}
\delta^{(\text{cgct})}_{\xi}:=\delta^{(\text{gct})}_{\xi}+\delta^{(\text{llt})}_{K[\xi]}~,\qquad K[\xi]_{ab}:=-\xi^c\o_{cab}~.\label{cgct}
\end{align}
 It follows that under \eqref{cgct},  $\F$ transforms according to
\begin{align}
\delta^{(\text{cgct})}_{\xi}\Phi=\xi^a\big(e_a-\frac{1}{2}\o_a{}^{bc}M_{bc}\big)\Phi=\xi^a\mc{D}_a\Phi~,
\end{align}
From \eqref{DPhiTrans} we see that $\delta^{(\text{llt})}_{K}\delta^{(\text{cgct})}_{\xi}\Phi=\frac{1}{2}\xi^a K^{bc}M_{bc}\mc{D}_{a}\Phi$ and the GCT has been `covariantised' with respect to Lorentz rotations. For this reason we will refer to a transformation of the form \eqref{cgct} as a covariantised general coordinate transformation (CGCT). 

The definition \eqref{cgct} may be thought of as a change in the basis of transformations of the gravity gauge group $\mc{G}$. In this thesis we will always use covariantised diffeomorphisms to implement the translational part of a spacetime symmetry group $\mc{G}$ (see section \ref{GCA} for further comments on this). In the current case, a drawback is that under a CGCT the Lorentz gauge field no longer transforms as a covariant scalar, but rather as
\begin{align}
\delta_{\xi}^{(\text{cgct})}\omega_{mab}=\xi^ce_m{}^{d}R_{abcd} \quad \Longleftrightarrow \quad \delta_{\xi}^{(\text{cgct})}\omega_{abc}=\xi^dR_{dabc}-\big(\mc{D}_a\xi^d\big)\o_{dbc}~.
\end{align}
On the other hand, the translational gauge field and its inverse transform according to 
\begin{subequations}\label{VielCGCT}
\begin{align}
-\delta_{\xi}^{(\text{cgct})}e_{a}{}^{m}&=\mc{D}_{a}\xi^{m}=\big(\mc{D}_{a}\xi^{b}\big)e_b{}^{m}~,\label{VielCGCT2} \\
\delta_{\xi}^{(\text{cgct})}e_{m}{}^{a}&=\mc{D}_{m}\xi^{a}=\big(\mc{D}_{b}\xi^a\big)e_m{}^{b}~,\label{VielCGCT1}
\end{align}
\end{subequations}
where we have made use of the torsion-free condition. 

Under a transformation of the full gravity gauge group (in the new basis) with gauge parameter $\L$, we find that the transformation law for $\Phi$ is
\begin{align}
\delta^{(\mc{G})}_{\L}\Phi=\Big(\delta^{(\text{cgct})}_{\xi}+\delta^{(\text{llt})}_{K}\Big)\Phi=\L\Phi ~, \qquad \L:=\xi^a\mc{D}_a+\frac{1}{2}K^{ab}M_{ab}~.\label{PoinPhi}
\end{align}
 The corresponding gauge freedom in the vielbein is described by
 \begin{align}
\delta_{\L}^{(\mc{G})}e_{a}{}^{m}=\Big(\delta_{\xi}^{(\text{cgct})}+\delta_{K}^{(\text{llt})}\Big)e_{a}{}^{m}=\Big(-\mc{D}_{a}\xi^{b}+K_a{}^{b}\Big)e_{b}{}^{m}~, \label{vielPGcgt}
\end{align}
whilst the Lorentz covariant derivative obeys by the rule
\begin{align}
\delta_{\L}^{(\mc{G})}\mc{D}_{a}=\big[\L,\mc{D}_a\big]~, \label{PoinDer}
\end{align}
with $\L$ defined as in \eqref{PoinPhi}.

\subsection{Perturbing the vielbein} \label{SectionVielPert}


Given an action functional $S[\Phi,e_{a}{}^{m}]$ describing the dynamics of some collection of matter fields $\Phi=\{\Phi^i\}$ coupled to the gravitational field $e_a{}^m$,  two important issues are: 
\begin{enumerate}[label=(\roman*)]
\item Deriving the equations of motion for the vielbein; and
\item `Linearising' this action about some gravitational background.
\end{enumerate}
In order to address these points, it is necessary to study what effect a deformation of the vielbein has on the geometrical objects which are constructed from it (e.g. curvature, connection, covariant derivative etc.). 

Let us perform a finite deformation of the background vielbein $e_{a}{}^{m}$  according to
\begin{align}
\tilde{e}_{a}{}^{m}=F_{a}{}^{b}e_{b}{}^{m}~,\qquad \tilde{e}_{m}{}^{a}=e_{m}{}^{b}\big(F^{-1}\big)_{b}{}^{a} \label{vieldeform}
\end{align}
for some invertible matrix $F_{ab}$. As a result of \eqref{vieldeform}, the background torsion-free covariant derivative $\mc{D}_a$ suffers a corresponding deformation
\begin{align}
\tilde{\mc{D}}_a=\tilde{e}_{a}{}^{m}\pa_m-\frac{1}{2}\tilde{\omega}_{a}{}^{bc}M_{bc}\equiv F_{a}{}^{b}\mc{D}_a-\frac{1}{2}\Xi_{a}{}^{bc}M_{bc}~.
\end{align} 
Here $\Xi_{abc}=-\Xi_{acb}$ represents the deformation of the spin-connection
\begin{align}
\Xi_{abc}= \tilde{\omega}_{abc}-F_{a}{}^{d}\omega_{dbc}~,
\end{align}
which, by imposing the torsion-free constraint on $\tilde{\mc{D}}_a$, may be determined as follows 
\begin{align}
\Xi_{abc}=F_{a}{}^{g}\mc{D}_{g}F_{[b|}{}^{d}F^{-1}_{d|c]}-F_{[b|}{}^{g}\mc{D}_{g}F_{a}{}^{d}F^{-1}_{d|c]}-F_{[b|}{}^{g}\mc{D}_{g}F_{|c]}{}^{d}F^{-1}_{da}~.
\end{align}
Using this one may express the curvature tensor for $\tilde{\mc{D}}_a$,
\begin{align}
\tilde{R}_{abcd}=F_{a}{}^{g}F_{b}{}^{h}R_{ghcd}+2F_{[a|}{}^{g}\mc{D}_{g}\Xi_{|b]cd}+2\Xi_{[ab]}{}^{g}\Xi_{gcd}+2\Xi_{[a|c}{}^{g}\Xi_{|b]gd}
\end{align}
solely in terms of the background covariant derivative and the deformation matrix. 

In order to answer the queries (i) and (ii) above, it is sufficient to study only the first and second order deformations (respectively) to geometric objects caused by some infinitesimal perturbation to the vielbein. Therefore, it suffices to perturbatively expand the finite deformation matrix $F_{ab}$ and its inverse $F^{-1}_{ab}$ to second order in some infinitesimal (dimensionless) fluctuation $h_{ab}$ defined according to\footnote{Of course, there exist alternative pertubative schemes, each of which amounts to a redefinition of $h_{ab}$. For example, in the metric theory one often uses the expansion $\tilde{g}_{mn}=g_{mn}+H_{mn}$ with corresponding inverse $\tilde{g}^{mn}=g^{mn}-H^{mn}+H^{mp}H_{p}{}^{n}+\cdots$. In this case, using \eqref{MetricFromVeil} and \eqref{vieldeform} we find that this scheme and \eqref{DefMatrix} are related by $H_{mn}=-2h_{mn}-3h_{m}{}^{p}h_{np}$ where $h_{mn}:=e_m{}^{a}e_n{}^{b}h_{ab}$. Yet another scheme corresponds to the expansion $F_{ab}=\eta_{ab}+h_{ab}+\frac{1}{2}h_{a}{}^{c}h_{bc}+\dots$ with inverse $F^{-1}_{ab}= \eta_{ab}-h_{ab}+\frac{1}{2}h_{a}{}^{c}h_{bc}+\cdots$. In this case one must make the replacement $h_{ab}\rightarrow h_{ab}+\frac{1}{2}h_{a}{}^{c}h_{bc}$ in all of the formulae below. Each of these redefinitions is valid only up to (and including) second order.  }
\begin{subequations} \label{DefMatrix}
\begin{align}
F_{ab}&=\eta_{ab}+h_{ab}~,\label{DefMatrix1}\\
F^{-1}_{ab}&= \eta_{ab}-h_{ab}+h_{a}{}^{c}h_{bc}~.\label{DefMatrix2}
\end{align}
\end{subequations}
It is clear that the antisymmetric part of the deformation, $F_{[ab]}=h_{[ab]}$, corresponds to an infinitesimal local Lorentz transformation (cf. eq. \eqref{vielLLT}). The effect of an infinitesimal LLT on all geometrical objects was discussed in the previous section. Therefore, to isolate the effect of the symmetric part of the perturbation, we hereby make the restriction
\begin{align}
 h_{ab}=h_{ba}~.
\end{align}
Having chosen $F_{ab}$ according to \eqref{DefMatrix1}, it is clear from the relation \eqref{vieldeform},
\begin{align}
\tilde{e}_{a}{}^{m}=e_{a}{}^{m}+h_{a}{}^{b}e_{b}{}^{m}~\quad \implies \quad \delta^{(\text{dfm})}_{h} e_{a}{}^{m}= h_{a}{}^{b}e_{b}{}^{m}~,\label{vielExp}
\end{align}
(here `dfm' stands for deformation) that we are expanding the vielbein $\tilde{e}_a{}^{m}$ about some background vielbein $e_{a}{}^{m}$, with $h_{ab}$ representing an infinitesimal perturbation. Linearisation of the gravity-matter system $S[\Phi, \tilde{e}_a{}^{m}]$ then consists of expanding $\tilde{e}_a{}^{m}$ about the background $e_{a}{}^{m}$ according to \eqref{vielExp} and keeping terms of second order in the perturbation.


With these remarks in mind we find that, to second order in $h_{ab}$, the deformation in the spin-connection is given by
\begin{align}
\Xi_{abc}={}^{(0)}\Xi_{abc}+{}^{(1)}\Xi_{abc}+{}^{(2)}\Xi_{abc}
\end{align}
where we have denoted
\begin{subequations}
\begin{align}
{}^{(0)}\Xi_{abc}&=0~, \label{SpinDeform0}\\
{}^{(1)}\Xi_{abc}&=-2\mc{D}_{[b}h_{c]a}~, \label{SpinDeform1}\\
{}^{(2)}\Xi_{abc}&=h_{[b|}{}^{d}\mc{D}_{a}h_{|c]d}-2h_{[b|}{}^{d}\mc{D}_{d}h_{|c]a}+h^{d}{}_{a}\mc{D}_{[b}h_{c]d}-h^d{}_{[b}\mc{D}_{c]}h_{ad}~. \label{SpinDeform2}
\end{align}
\end{subequations}
It follows that, to first order, the Lorentz spin-connection and covariant derivative suffer the disturbances 
\begin{subequations}\label{LC(D)Deform}
\begin{align}
\delta^{(\text{dfm})}_{h} \omega_{abc}&=h_{a}{}^{d}\o_{dbc}-2\mc{D}_{[b}h_{c]a}~,\label{LCDeform}\\
\delta^{(\text{dfm})}_{h} \mc{D}_{a}&=h_{a}{}^{b}\mc{D}_{b}+\big(\mc{D}^{b}h_{a}{}^{c}\big)M_{bc}\label{LCDDeform}~.
\end{align} 
\end{subequations}

Expanding the curvature to second order and expressing it as 
\begin{align}
\tilde{R}_{abcd}={}^{(0)}\hspace{-1mm}R_{abcd}+{}^{(1)}\hspace{-1mm}R_{abcd}+{}^{(2)}\hspace{-1mm}R_{abcd}~,
\end{align}
we find, upon grouping terms to make the algebraic symmetries of the Riemann tensor manifest, the following fluctuations\footnote{Here we use both square brackets $[\cdots]$ and braces $\{\cdots\}$ to denote anti-symmetrisation.}
\begin{subequations}
\begin{align}
{}^{(0)}\hspace{-1mm}R_{abcd}&=R_{abcd}~, \label{RiemannDeform0}\\
{}^{(1)}\hspace{-1mm}R_{abcd}&=-h^{f}{}_{[a}R_{b]fcd}-h^{f}{}_{[c}R_{d]fab}-2\mc{D}_{[a}\mc{D}_{\{c}h_{d\}b]}-2\mc{D}_{[c}\mc{D}_{\{a}h_{b\}d]}~, \label{RiemannDeform1}\\
{}^{(2)}\hspace{-1mm}R_{abcd}&=h_{a}{}^{f}h_{b}{}^{g}R_{fgcd}+h_{c}{}^{f}h_{d}{}^{g}R_{abfg}-4h_{[a|}{}^{f}h_{\{c}{}^{g}R_{d\}f|b]g}-2h_{[a}{}^{f}h_{\{c}{}^{g}R_{d\}b]fg} \non \\
&\phantom{=}-4h_{[a|}{}^{f}\mc{D}_{f}\mc{D}_{\{c}h_{d\}|b]}-4h^{f}{}_{[c|}\mc{D}_{f}\mc{D}_{\{a}h_{b\}|d]} -2h^f{}_{[a}\mc{D}_{b]}\mc{D}_{\{c}h_{d\}f}-2h^f{}_{[c}\mc{D}_{d]}\mc{D}_{\{a}h_{b\}f}\non\\
&\phantom{=}+2\mc{D}^{f}h_{[a\{c|}\mc{D}_{f}h_{|d\}b]}+6\mc{D}_{[a}h_{b]}{}^{f}\mc{D}_{[c}h_{d]f}+2\mc{D}_{[a}h^{f}{}_{\{c}\mc{D}_{b]}h_{d\}f}+2\mc{D}_{[c}h^{f}{}_{\{a}\mc{D}_{d]}h_{b\}f}\non\\
&\phantom{=}-2\mc{D}_{[c}h^{f}{}_{\{a}\mc{D}_{d]}h_{b\}f}-4\mc{D}^fh_{[a\{c}\mc{D}_{d\}}h_{b]f}-4\mc{D}^fh_{[c\{a}\mc{D}_{b\}}h_{d]f}~.\label{RiemannDeform2}
\end{align}
\end{subequations}
The corresponding deformations to the Ricci tensor and scalar curvature may be easily obtained by taking appropriate traces of the above formulae.  In particular, their first-order deformations are given by 
\begin{subequations}\label{RiemannDescDeform}
\begin{align}
{}^{(1)}\hspace{-1mm}R_{ab}&= 2h^{c}{}_{(a}R_{b)c}+2\mc{D}^c\mc{D}_{(a}h_{b)c}-\Box h_{ab}-\mc{D}_{a}\mc{D}_{b}h   ~,\label{RicciDeform}\\
{}^{(1)}\hspace{-1mm}R&=2h^{ab}R_{ab}+2\mc{D}^a\mc{D}^bh_{ab}-2\Box h \label{ScalarDeform}
\end{align}
\end{subequations}
where $\Box :=\mc{D}^a\mc{D}_a$ and $h:=\eta^{ab}h_{ab}$. Finally, to obtain a perturbative expansion of the vielbein determinant, we make use of the relation $  \delta_{a}{}^{b}+M_{a}{}^{b} \approx \text{exp}\big[M_{a}{}^{b}-\frac{1}{2}M_a{}^{c}M_{c}{}^{b}\big]$, which is true at second order in some matrix $M$. It follows that
\begin{align}
\text{det}\big(\delta_{a}{}^{b}+M_{a}{}^{b}\big)\approx \text{exp}\big[\text{Tr}\big(M-\frac{1}{2}M^2\big)\big]\approx 1+\text{Tr}(M)+\frac{1}{2}\text{Tr}^2(M)-\frac{1}{2}\text{Tr}\big(M^2\big)~,
\end{align}
where $\text{Tr}$ denotes the trace operation. Hence, to second order in $h_{ab}$, we have
\begin{align}
\text{det}\big(\tilde{e}_{m}{}^{a}\big)= \text{det}\big(e_{m}{}^{b}\big)\text{det}^{-1}\big(F_{b}{}^{a}\big)=\text{det}\big(e_{m}{}^{b}\big)\big(1-h+\frac{1}{2}h^2+\frac{1}{2}h^{ab}h_{ab}\big)~.\label{DetDeform}
\end{align}


\subsection{Covariantised diffeomorphisms and Weyl transformations}

We have already seen that the antisymmetric part of the deformation $h_{ab}$ corresponds to a Lorentz transformation. 
There are two other types of deformations which are of particular interest:
those generated by a vector field $\xi_{a}(x)$ 
\begin{align}
h_{ab}=-\mc{D}_{(a}\xi_{b)} \equiv h[\xi]_{ab}\label{DiffDeform}~,
\end{align}
and those generated by a scalar field $\sigma(x)$
\begin{align}
h_{ab}=\eta_{ab}\sigma \equiv h[\s]_{ab}\label{WeylDeform}~.
\end{align}
 The following discussions will be limited to first order in $h_{ab}$. 
 
 The  deformation \eqref{DiffDeform} is closely related to the covariantised diffeomorphisms introduced in the previous section. This can be seen through comparison with \eqref{VielCGCT2}, which may be expressed as\footnote{One should bear in mind that there are two different $K[\xi]_{ab}$, one in \eqref{ConformalDeform} and the other in \eqref{cgct}.}
\begin{align}
\delta_{\xi}^{(\text{cgct})}e_{a}{}^{m}=\Big(h[\xi]_{a}{}^{b}-K[\xi]_{a}{}^{b}\Big)e_{b}{}^{m}~,\qquad K[\xi]_{ab}:=\mc{D}_{[a}\xi_{b]} \label{ConformalDeform}
\end{align}
 Therefore the deformation \eqref{DiffDeform}
 is seen to correspond to a combination of a CGCT and a particular LLT
 (this should be compared with \eqref{vielPGcgt})
\begin{align}
\delta^{(\text{dfm})}_{h[\xi]}e_{a}{}^{m}= h[\xi]_a{}^{b}e_{b}{}^{m} =\big(\delta_{\xi}^{(\text{cgct})}+\delta_{K[\xi]}^{(\text{llt})}\big)e_{a}{}^{m}~. \label{GravDeform}
\end{align}
Indeed, one may insert \eqref{DiffDeform} into (for example) the formulae \eqref{RiemannDeform1} and verify that the induced transformation on $R_{abcd}$ is 
\begin{align}
\delta^{(\text{dfm})}_{h[\xi]}R_{abcd}=\Big(\xi^f\mc{D}_{f}+\frac{1}{2}K[\xi]^{fg}M_{fg}\Big)R_{abcd}~.
\end{align}
 This should be compared with the rule \eqref{PoinPhi} which applies to all covariant fields. By virtue of \eqref{GravDeform} and \eqref{DetDeform}, we see that to first order, the vielbein determinant transforms under a CGCT according to $\delta^{(\text{cgct})}_{\xi} e = e\mc{D}^a\xi_a$. 

It is interesting to note that if $\xi=\xi^ae_a$ is a Killing vector field, then by definition $h[\xi]_{ab}=0$ and hence $\delta^{(\text{dfm})}_{h[\xi]}e_{a}{}^{m}=0$. In other words, if $\xi^a$ is a Killing vector then the particular combination of transformations on the right-hand side (RHS) of \eqref{GravDeform} leaves the vielbein invariant. This is a well known property of Killing vectors. Similarly, if $\xi=\xi^ae_a$ is a conformal Killing vector field (see section \ref{SectionCKV}), then $h[\xi]_{ab}=-\frac{1}{d}\eta_{ab}\mc{D}^c\xi_c$ and \eqref{ConformalDeform} is precisely \eqref{CKVviel}.

Next let us consider the deformation \eqref{WeylDeform}, for which \eqref{vieldeform}  takes the form
\begin{align}
\delta^{(\text{weyl})}_{\s}e_{a}{}^{m}=\s e_{a}{}^{m}~, \qquad \delta^{(\text{weyl})}_{\s}e_{m}{}^{a}=-\s e_{m}{}^{a}~. \label{vielWeyl}
\end{align}
This is known as a Weyl transformation. From \eqref{LC(D)Deform}, we find that the spin connection and covariant derivative suffer the following deformations
\begin{subequations}
\begin{align}
\delta^{(\text{weyl})}_{\s}\omega_{abc}&=\s \omega_{abc}+2\eta_{a[b}\mc{D}_{c]}\s~, \label{LCWeyl}\\
\delta^{(\text{weyl})}_{\s}\mc{D}_a&= \s \mc{D}_a - \big(\mc{D}^b\s \big)M_{ab}~. \label{LCDWeyl}
\end{align}
\end{subequations}
 Under this variation, eqs. \eqref{RiemannDeform1} and \eqref{RiemannDescDeform} imply that the curvature tensor and its traces transform according to 
 \begin{subequations}\label{WeylRDesc}
\begin{align}
\delta^{(\text{weyl})}_{\s}R_{abcd}&=2\s R_{abcd }-4\eta_{[a\{c}\mc{D}_{d\}}\mc{D}_{b]}\s~,\label{RiemannWeyl}\\
\delta^{(\text{weyl})}_{\s}R_{ab}&= 2\s R_{ab}-(d-2)\mc{D}_{a}\mc{D}_{b}\s-\eta_{ab}\Box\s~\label{RicciWeyl},\\
\delta^{(\text{weyl})}_{\s}R &= 2\s R -2(d-1)\Box\s~, \label{ScalarWeyl}
\end{align} 
\end{subequations}
whilst the vielbein determinant transforms as
\begin{align}
\delta^{(\text{weyl})}_{\s} e = -d\s e~. \label{DetWeyl}
\end{align}
We note that a finite Weyl transformation of the vielbein takes the form $\tilde{e}_a{}^{m}=e^\s e_a{}^{m}$.
 \subsection{Conformal descendants of the curvature} \label{SecConfCurve}

In this thesis, there are four descendants of the Riemann tensor which play a prominent role and are ubiquitous throughout. They are the Schouten tensor\footnote{To denote these curvatures, we prefer to use the first letter of their founder's surname. }
\begin{align}
S_{ab}:=\frac{1}{(d-2)}\bigg(R_{ab}-\frac{1}{2(d-1)}\eta_{ab}R\bigg)~, \label{SchoutenT}
\end{align}
the Weyl tensor
\begin{align}
W_{abcd}:=R_{abcd}-2\eta_{a[c}S_{d]b}+2\eta_{b[c}S_{d]a}~, \label{WeylT}
\end{align}
the Bach tensor 
\begin{align}
 B_{ab} := \Big(\mc{D}^c\mc{D}^d-(d-3)S^{cd}\Big)W_{cabd}~, \label{BachT}
\end{align}
and the Cotton tensor
\begin{align}
C_{abc}:=2\mc{D}_{[a}S_{b]c} ~. \label{CottonT}
\end{align}
As we will see, they control the conformal geometry of spacetime. 

Under a Weyl transformation, the Schouten tensor transforms according to the rule 
\begin{align}
\delta^{(\text{weyl})}_{\s}S_{ab}=2\s S_{ab}-\mc{D}_a\mc{D}_b\s~. \label{SchoutenWeyl}
\end{align}
The inhomogeneous term in \eqref{SchoutenWeyl} 
can be used to kill the inhomogeneous term in the Weyl transformation of the Riemann tensor \eqref{RiemannWeyl}. This means that the Weyl tensor transforms homogeneously with conformal (or Weyl) weight two,
\begin{align}
\delta^{(\text{weyl})}_{\s}W_{abcd}=2\s W_{abcd}~. \label{WeylWeyl}
\end{align}
In addition to possessing all of the symmetries of the Riemann tensor, the Weyl tensor is also traceless in any pair of indices,
\begin{subequations}
\begin{align}
W_{abcd}=-W_{bacd}&=-W_{abdc}=W_{cdab}~, \\
 W_{[abc]d}=0~,&\qquad W^{c}{}_{abc}=0~.
\end{align}
\end{subequations}
On account of \eqref{Bianchi3} it satisfies the following Bianchi identity\footnote{Although the first equation in \eqref{Bianchi4} applies only for $d>3$, the second equation holds for all $d>2$. }
\begin{align}
\mc{D}^{d}W_{abcd}=-(d-3)C_{abc}~ \quad \implies \quad \mc{D}^bS_{ab}=\mc{D}_{a}S~, \label{Bianchi4}
\end{align}
where $S:=\eta^{ab}S_{ab}=\frac{1}{2(d-1)}R$. From the Weyl and Schouten tensors, one may construct the two derivative descendant \eqref{BachT} known as the  Bach tensor. By virtue of \eqref{Bianchi4}, \eqref{CottonAlgebraic} and \eqref{CottonDifferBianchi}, it is both symmetric and traceless,
\begin{align}
B_{ab}=B_{ba}, \qquad \eta^{ab}B_{ab}=0
\end{align}
and has divergence equal to
\begin{align}
\mc{D}^bB_{ab}=(d-3)(d-4)S^{bc}C_{abc}~. \label{BachTransverse}
\end{align}
Under an infinitesimal Weyl transformation it transforms according to the rule
\begin{align}
\delta^{(\text{weyl})}_{\s}B_{ab}=4\s B_{ab}-2(d-4)\mc{D}^c\s\mc{D}^dW_{c(ab)d}~. \label{BachWeyl}
\end{align}
In the case of four-dimensional spacetimes, $d=4$, the right hand side of \eqref{BachTransverse} vanishes and $B_{ab}$ is said to be covariantly conserved. The second term in \eqref{BachWeyl} also vanishes and $B_{ab}$ transforms homogeneously under Weyl transformations with conformal weight four. In four dimensions, the Bach tensor arises as the first functional derivative of the action for conformal gravity (see section \ref{sectionCG4}). From this point of view the above properties of $B_{ab}$ are to be expected. For example, its transversality \eqref{BachTransverse} may be understood as a consequence of the invariance of this action under infinitesimal diffeomorphisms.



Two spacetimes $\big(\mc{M}^d,e_a{}^{m}\big)$  and $\big(\mc{M}^d, \tilde{e}_{a}{}^{m}\big)$ are said to be (locally) conformally related if their vielbein are related by a finite Weyl transformation $\tilde{e}_a{}^{m}(x)=e^{\s} e_{a}{}^{m}(x)$. The latter condition may be equivalently stated as $\widetilde{\mc{D}}_a=e^{\s}\big( \mc{D}_a - \mc{D}^b\s M_{ab}\big)$. If a spacetime is conformally related to Minkowski space then it is said to be conformally flat.  It is a well known fact that spacetimes with $d\geq 4$ are conformally flat if and only if their  Weyl tensor vanishes,
\begin{align}
W_{abcd}=0 \quad \Leftrightarrow \quad e_{a}{}^{m}=e^{\s}\delta_{a}{}^{m} \quad \Leftrightarrow \quad g_{mn}(x) = e^{-2\s(x)}\eta_{mn}~.
\end{align} 
This emphasises the importance of the Weyl tensor in the context of conformal geometry. 


In three dimensions, $d=3$, the Weyl tensor vanishes exactly and the conformal geometry of spacetime is controlled by the Cotton tensor \eqref{CottonT}. In particular, spacetime is conformally flat if and only if the Cotton tensor vanishes \cite{Eisen} (see \cite{BKNT-M1} for a modern proof). For generic $d$, under a Weyl transformation the Cotton tensor transforms via
\begin{align}
\delta^{(\text{weyl})}_{\s}C_{abc}=3\s C_{abc} +W_{abcd}\mc{D}^d\s~.
\end{align}
The algebraic properties of the Cotton tensor are 
\begin{align}
C_{abc}=-C_{bac}~,\qquad C_{[abc]}=0~, \qquad C_{ab}{}^b =0~, \label{CottonAlgebraic}
\end{align}
and is covariantly conserved on account of the Bianchi identity \eqref{Bianchi4}
\begin{align}
\mc{D}^cC_{abc}=0~.\label{CottonDifferBianchi}
\end{align}
In $d=3$, under a Weyl transformation $C_{abc}$ transforms homogeneously with conformal weight three, since there the Weyl tensor vanishes identically. Introducing the Levi-Civita tensor\footnote{In this thesis we adopt the normalisation $\ve_{01\dots d-1}=-\ve^{01\dots d-1}=-1$. Furthermore, the frame version $\ve_{ab\dots d}$ of the world tensor $\ve_{mn\dots p}:=\sqrt{-g}\hat{\ve}_{mn\dots p}$, obeys the relation $\ve_{ab\dots d}:=e_a{}^{m}e_b{}^{n}\cdots e_{d}{}^{p}\ve_{mn\dots p}=\hat{\ve}_{ab\dots d}$. Here $\hat{\ve}_{mn\dots p}$ is the totally antisymmetric symbol, which is a world tensor density. } $\ve_{abc}$, we may define the dual $C_{ab}$ of the Cotton tensor $C_{abc}$:
\begin{align}
C_{ab}:= \hf \ve_{acd} C^{cd}{}_b ~ \qquad \Longleftrightarrow \qquad C_{abc}=-\ve_{ab}{}^{d}C_{cd}~. \label{CottonTDual}
\end{align}
The latter is symmetric and traceless,
\bea
C_{ba} =C_{ab}~, \qquad C^a{}_a =0~,
\eea
and also transverse, 
\bea
\mc{D}^b C_{ab}=0~.
\eea
The above properties of the Cotton tensor are not surprising given that $C_{ab}$ arises as the first functional derivative of the action for conformal gravity in $3d$ (see section \ref{sectionCG3}).


It will be of interest to find the infinitesimal variation of the Schouten, Weyl and Cotton tensors under a generic symmetric perturbation in the vielbein. Of course the latter may always be decomposed according to
\begin{align}
h_{ab}=\hat{h}_{ab}+\eta_{ab}\s[h]
\end{align}
where $\hat{h}^{ab}$ is the symmetric traceless part and $\s[h]$ is the pure trace part,
\begin{align}
\hat{h}_{ab}:=h_{ab}-\frac{1}{d}\eta_{ab}h~,\qquad \s[h]:=\frac{1}{d}h~.
\end{align}
It is clear that the trace sector corresponds to a Weyl transformation with parameter $\s=\s[h]$, and all of the corresponding variations for this have been given above. Therefore we may concern ourselves only with the variations produced by a symmetric and traceless perturbation, which are given by 
\begin{subequations}
\begin{align}
\delta^{(\text{dfm})}_{\hat{h}}S_{ab}&=2\hat{h}^c{}_{(a}S_{b)c}-\frac{1}{(d-1)}\eta_{ab}\hat{h}^{cd}S_{cd}+\frac{2}{(d-2)}\Big[\hat{h}_{ab}S +\mc{D}^c\mc{D}_{(a}\hat{h}_{b)c}-\frac{1}{2}\Box\hat{h}_{ab}\non\\
&\phantom{=}-\frac{1}{2(d-1)}\eta_{ab}\mc{D}^c\mc{D}^d\hat{h}_{cd}\Big]~,\label{SchoutenDeform}\\
\delta^{(\text{dfm})}_{\hat{h}}W_{abcd}&=-\hat{h}^f{}_{[a}W_{b]fcd}-\hat{h}^f{}_{[c}W_{d]fab}+2\hat{h}^f{}_{[a}\eta_{b]\{c}S_{d\}f}+2\hat{h}^f{}_{[c}\eta_{d]\{a}S_{b\}f}+4\hat{h}_{[a\{c}S_{d\}b]} \non\\
&\phantom{=}+\frac{4}{(d-1)}\eta_{a[c}\eta_{d]b}\hat{h}^{fg}S_{fg}-\frac{8}{(d-2)}\eta_{[a\{c}\hat{h}_{d\}b]}S+\frac{4}{(d-2)}\eta_{[a\{c}\Box \hat{h}_{d\}b]} \non\\
&\phantom{=}-2\mc{D}_{[a}\mc{D}_{\{c}\hat{h}_{d\}b]}-2\mc{D}_{[c}\mc{D}_{\{a}\hat{h}_{b\}d]}-\frac{4}{(d-1)(d-2)}\eta_{a[c}\eta_{d]b}\mc{D}^f\mc{D}^g\hat{h}_{fg} \non\\
&\phantom{=} -\frac{4}{(d-2)}\eta_{[a\{c}\mc{D}^f\mc{D}_{d\}}\hat{h}_{b]f}-\frac{4}{(d-2)}\eta_{[c\{a}\mc{D}^f\mc{D}_{b\}}\hat{h}_{d]f}~. \label{WeylDeform1}
\end{align}
\end{subequations}
 We postpone the calculation of the variation in the Cotton tensor to section \ref{sectionCG3}, which will be done for $d=3$ using the two component spinor formalism.


\section{Elements of classical conformal field theory} \label{SectionCFT}

In this section we will review aspects of classical conformal field theory, with a particular emphasis on the gauge group of conformal gravity.

\subsection{Conformal Killing vectors} \label{SectionCKV}
 
In terms of the vielbein, the line element of spacetime takes the form
\begin{align}
\text{d}s^2=g_{mn}\text{d}x^m\text{d}x^n=\eta_{ab}E^aE^b~.
\end{align}
Making use of equation \eqref{VielCGCT1} we see that, under an infinitesimal covariantised general coordinate transformation in the direction $-\xi^a$,\footnote{By this we mean an infinitesimal GCT $x^m\rightarrow x'^m=x^m+\xi^m$ supplemented by an LLT with gauge parameter $K[\xi]_{ab}=\xi^{c}\omega_{cab}$, in accordance with \eqref{vielgct2} and \eqref{cgct}.} the line element changes according to
\begin{align}
\text{d}s^2\rightarrow \text{d}s'^2=\text{d}s^2 -2\mc{D}_{(a}\xi_{b)}E^aE^b~.
\end{align}
  Let us now consider a particular family of vector fields, whose components we hereby denote by $\xi^a\equiv \z^a$, which satisfy the equation
 \begin{align}
 \mc{D}_{(a}\z_{b)}=\eta_{ab}\s[\z]~,\qquad \s[\z]:=\frac{1}{d}\mc{D}^a\z_{a}~.  \label{CKE}
 \end{align}
 Then, it is clear that under such a CGCT, $\text{d}s^2$ gets locally scaled,
 \begin{align}
 \text{d}s^2\rightarrow \text{d}s'^2=(1-2\s[\z])\text{d}s^2~,
 \end{align}
and hence induces a Weyl transformation on the metric. 
 
The equation \eqref{CKE} is called the conformal Killing equation, and any vector field $\z=\z^ae_a$ satisfying it is called a conformal Killing vector.  Under an infinitesimal CGCT generated by $-\z$, the inverse vielbein transforms according to \eqref{VielCGCT1},
 \begin{align}
 \delta_{-\z}^{(\text{cgct})}e_{m}{}^{a}=-\s[\z] e_m{}^{a}-e_m{}^{b}K[\z]_{b}{}^{a}= \Big(\delta^{(\text{weyl})}_{\s[\z]}+\delta_{K[\z]}^{(\text{llt})}\Big)e_m{}^{a}~, \label{CKCGCT}
 \end{align}
where we have denoted $K[\z]_{ab}:=\mc{D}_{[a}\z_{b]}$. It is beneficial to view equation \eqref{CKCGCT} from slightly different point of view. Specifically, a simple  rearrangement indicates that the combination of a CGCT along a conformal Killing vector $-\z$, a Weyl transformation with scale parameter $\s[\z]$ and an LLT parametrised by $K[\z]_{ab}$ does not change the vielbein
\begin{align}
0=\Big(\delta^{(\text{cgct})}_{\z}+\delta_{K[\z]}^{(\text{llt})}+\delta^{(\text{weyl})}_{\s[\z]}\Big)e_m{}^{a}= e_m{}^{b}\Big(\mc{D}_{b}\z^a-K[\z]_{b}{}^{a}-\s[\z]\delta_{b}{}^{a}\Big) \label{CKVviel}
\end{align}
and hence the metric. 

An equivalent definition of a conformal Killing vector is any vector $\z^a$ which satisfies
\begin{align}
0=
\big[\L[\z],\mc{D}_a\big]+\delta^{(\text{weyl})}_{\s[\z]}\mc{D}_a~,\qquad \L[\z]:=\z^a\mc{D}_a+\frac{1}{2}K[\z]^{ab}M_{ab}~,
\end{align}
where we have used \eqref{PoinDer}. In this sense they correspond to the conformal isometries of spacetime which preserve $\mc{D}_a$. This notion of a conformal killing vector has a straightforward generalisation to superspace, and is the one we will use in later chapters. 

Let $\z_{(1)}=\z^a_{(1)}e_a$ and $\z_{(2)}=\z^a_{(2)}e_a$ be two distinct conformal Killing vectors  on $\mc{M}^d$. It is not difficult to show that their Lie bracket,
\begin{align}
\big[\z_{(1)},\z_{(2)}\big]= \z^a_{(3)}e_a~,\qquad \z_{(3)}^a:=\z_{(1)}^b\mc{D}_b\z^a_{(2)}-\z_{(2)}^b\mc{D}_b\z^a_{(1)} \label{CKVLiebracket}
\end{align}
is also a conformal Killing vector field (i.e. $\z^a_{(3)}$ satisfies \eqref{CKE}). Therefore, the linear space of conformal Killing vector fields, endowed with the Lie bracket \eqref{CKVLiebracket}, forms a Lie algebra. The latter is known as the conformal algebra associated with the spacetime $\big(\mc{M}^d,e_{a}{}^{m}\big)$. It may be shown that if two spacetimes are conformally related
then they share the same conformal Killing vectors and hence conformal algebras. In particular, any conformally flat spacetime has the same conformal algebra as Minkowski space $\mb{M}^d$.
 
\subsection{The conformal algebra of Minkowski space} \label{secConfalMd}

 The conformal algebra of Minkowski space is isomorphic to $\mf{so}(d,2)$, which we will hereby simply refer to as the conformal algebra. In an arbitrary representation, $\mf{so}(d,2)$ is characterised by the commutation relations 
  \begin{subequations} \label{confal}
\begin{align}
[M_{ab},M_{cd}]&=2\eta_{c[a}M_{b]d}-2\eta_{d[a}M_{b]c}~, \phantom{inserting blank space inserting}\label{confalLor}\\
[M_{ab},P_c]&=2\eta_{c[a}P_{b]}~, \qquad \qquad \qquad \qquad ~ [\mathbb{D},P_a]=P_a~,\\
[M_{ab},K_c]&=2\eta_{c[a}K_{b]}~, \qquad \qquad \qquad \qquad [\mathbb{D},K_a]=-K_a~,\\
[K_a,P_b]&=2\eta_{ab}\mathbb{D}+2M_{ab}~,
\end{align}
\end{subequations}
 with all other commutators vanishing. This algebra is spanned by the translation $P_a$, Lorentz $M_{ab}$, dilatation $\mathbb{D}$ and the special conformal $K_a$ generators.  
 
 In Cartesian coordinates $x^a$, the vielbein of Minkowski space may be globally chosen to take the form $e_m{}^{a}=\delta_{m}{}^{a}$, and the general solution to \eqref{CKE} is given by
 \begin{align}
 \z^a(x)=\xi^a  - K^{a}{}_{b}x^b+\s x^a -\tau^a x^bx_b+2x^a x^b\tau_b~, \label{CKVflat}
 \end{align}
 where $\xi^a$, $K_{ab}=-K_{ba}$, $\s$ and $\tau^a$ are arbitrary real constant parameters. 
 We can express this conformal Killing vector in the equivalent form
 \begin{align}
 \z=\z^a\pa_a=\xi^ap_a+\frac{1}{2}K^{ab}m_{ab}+\s D + \tau^ak_a
 \end{align}  
 where we have defined the differential operators $\{p_a,m_{ab}=-m_{ba},D,k_a\}$ 
 \begin{align}
 p_a:=\pa_a~,\quad m_{ab}:=x_a\pa_b-x_b\pa_a~,\quad D:=x^a\pa_a~,\quad k_a:=2x_ax^b\pa_b-x^bx_b\pa_a~. \label{CAdiffop}
 \end{align}
They satisfy the commutation relations \eqref{confal}  of $\mf{so}(d,2)$ (up to an overall minus sign).


Let us denote the space of tensors fields $\F(x)$ (with indices suppressed) defined on $\mb{M}^d$ by $\ms{L}(\mb{M}^d)$.
In order to obtain a representation of $\mf{g}\equiv \mf{so}(d,2)$ on $\ms{L}(\mb{M}^d)$, whose generators we denote $\{\mf{P}_a, \mf{M}_{ab}=-\mf{M}_{ba}, \mf{D}, \mf{K}_a\}$,  one makes use of the theory of induced representations \cite{Mackey, MackSalam}. 

 The infinitesimal action of the conformal group $\mc{G}$ on a point $x^a\in \mb{M}^d$ is given by\footnote{In order to have a well defined rigid action of finite special conformal transformations (SCTs) on $\mb{M}^d$, one must instead consider the compactification of the latter.  In this thesis we will not concern ourselves with such issues, since our attention will be restricted to infinitesimal conformal transformations. }
\begin{align}
x^a\mapsto x'^a=x^a+\z^a=(x^a+\xi^a ) - K^{a}{}_{b}x^b+\s x^a -\tau^a x^bx_b+2x^a x^b\tau_b~.
\end{align}  
Clearly, the subgroup of transformations leaving the origin $x_0^m=0$ invariant is generated by $\{M_{ab},\mb{D},K_a\}$. This subgroup is called the stability  subgroup (or little group) of $x_0^m$, and is denoted by $\mc{H}$, with corresponding Lie algebra $\mf{h}=\text{span}\{M_{ab},\mb{D},K_a\}$. Let us define a representation of $\mf{h}$ on the space $\ms{L}(\mb{M}^d)$ at $x_0^m$ according to
\begin{align}
\mf{M}_{ab}\Phi(0)=M_{ab}\Phi(0)~,&\qquad \mf{D}\Phi(0)=\Delta \Phi(0)~, \qquad \mf{K}_a \Phi(0)=0~. \label{primary13}
\end{align}
A field transforming in this way, with $\mf{K}_a$ acting trivially, is said to be primary with scaling dimension $\Delta \in \mb{R}$ \cite{MackSalam}. In this thesis, we reserve the notation $M_{ab}$ for the Lorentz generators which, when acting on any tensor field, return only the spin contribution (and not the `orbital' part described by $m_{ab}$ in \eqref{CAdiffop}). In particular, $M_{ab}\Phi(0)=\mc{S}_{ab}\Phi(0)$ where $\mc{S}_{ab}$ is the spin matrix\footnote{The matrices $\mc{S}_{ab}$ satisfy the commutation relations $\big[\mc{S}_{ab},\mc{S}_{cd}\big]=-\eta_{ac}\mc{S}_{bd}+\cdots$, which have opposite sign to \eqref{confalLor}. This is because $M_{ab}$ passes through $\mc{S}_{cd}$: $M_{ab}M_{cd}\Phi(0)=\mc{S}_{cd}M_{ab}\Phi(0)=\mc{S}_{cd}\mc{S}_{ab}\Phi(0)$. } appropriate for the Lorentz representation of $\Phi$.

To obtain a representation of $\mf{so}(d,2)$ at any point away from the origin, we define the translation operator $\mf{P}_a$ to act on $\F(x)$ according to the rule
\begin{align}
\mf{P}_a\F(x')\big|_{x'=0}=\pa'_a\F(x')\big|_{x'=0} \quad  \implies \quad\mf{P}_a\Phi(x)&=\pa_a\Phi(x)~.
\end{align}
Then, using the identity $\Phi(x)=\big(e^{x^a \cdot \mf{P}_a}\Phi(x')\big)\big|_{x'=0}$ and 
 the Baker-Campbell-Hausdorff formula, it may be shown that
\begin{subequations} \label{ConfalDiffRel}
\begin{align}
\mf{P}_a \Phi(x)&= p_a\Phi(x)~,\\
\mf{M}_{ab}\Phi(x) &= \big(m_{ab}+M_{ab}\big)\Phi(x)~,\\
\mf{D}\Phi(x) &= \big(D+\Delta\big) \Phi(x)~,\\
\mf{K}_a \Phi(x) &= \big(k_a+2x^bM_{ab}+2\Delta x_a\big)\Phi(x)~,\label{ConfalDiffReld}
\end{align} 
\end{subequations}
 where we have made use of the commutation relations \eqref{confal} and the operators \eqref{CAdiffop}. 
 
 In deriving the above results, we have assumed that the generators $\{\mf{P}_a, \mf{M}_{ab}, \mf{D}, \mf{K}_a\}$ act only on the field, and not on the coordinates, derivatives or transformation parameters. This corresponds to an active interpretation and, consequently, the (infinitesimal) action of the conformal group $\mc{G}$ that we have discussed is given by $x^a\rightarrow x'^a=x^a$ and\footnote{This may be rearranged as $\delta_{\L}^{(\mc{G})}\Phi(x)=\big(\z^b\pa_b+\frac{1}{2}K[\z]^{ab}M_{ab}+\Delta \s[\z]\big)\Phi(x)$, where $\z^a$ is a conformal Killing vector \eqref{CKVflat} whilst $K[\z]_{ab}=\pa_{[a}\z_{b]}$ and $\s[\z]=\frac{1}{d}\pa^a\z_a$. See the discussion surrounding \eqref{Iwanttorefertothis}. }
\begin{align}
\qquad\delta_{\L}^{(\mc{G})}\Phi(x)=\L\Phi(x)~,\qquad \L:=\xi^a\mf{P}_a+\frac{1}{2}K^{ab}\mf{M}_{ab}+\s\mf{D}+\tau^a\mf{K}_a~. \label{ConfGPhiTrans}
\end{align} 
A field transforming in this way is said to be primary with conformal (Weyl) weight $\Delta$.
 
\subsection{Conformal and Weyl symmetry: there and back again} \label{SectionHobbit}

Consider an action $S[\Phi, e_a{}^{m}=\delta_{a}{}^{m}]\equiv S[\Phi] =\int \text{d}^dx \ms{L}\big(\Phi,\pa_a\Phi,\cdots\big)$ describing the dynamics of some matter field $\Phi$ (with supressed Lorentz indices) in flat Minkowski space. This system may be minimally coupled to a non-trivial gravitational background $e_{a}{}^{m}$ through the usual prescription: one simply makes the replacements (i) $\delta_{a}{}^{m}\rightarrow e_{a}{}^{m}$; (ii) $\pa_a \rightarrow \mc{D}_a$; and (iii) $\ms{L}\rightarrow e \ms{L}$. If, in addition, one includes extra terms which vanish in the flat-limit (e.g. curvature dependent terms), then the coupling is said to be non-minimal.

Suppose that through this prescription it is possible to promote the action $S[\Phi]$ to the gravitational background $e_{a}{}^{m}$ in such a way\footnote{The coupling may be minimal or non-minimal; the important point is that $S[\Phi, e_a{}^{m}]$ reduces to $S[\Phi]$ in the flat limit.} that under an infinitesimal Weyl transformation of the vielbein and matter field,
\begin{align}
\delta_{\s}^{(\text{weyl})}e_{a}{}^{m}=\s e_{a}{}^{m}~,\qquad \delta_{\s}^{(\text{weyl})}\Phi = \Delta \s \Phi~, \qquad \Delta \in \mathbb{R}~, \label{VariousWeylT}
\end{align}
the resulting system 
is invariant $\delta_{\s}^{(\text{weyl})}S[\Phi, e_a{}^{m}]=0$.
  Then, by virtue of general covariance under CGCTs and local Lorentz transformations, it follows that $S[\Phi, e_a{}^{m}]$ is invariant under the transformations
\begin{subequations}
\begin{align}
\delta \Phi &= \Big( \xi^a\mc{D}_a +\frac{1}{2}K^{ab}M_{ab}+\Delta\s \Big) \Phi~,\label{CGphigauge}\\
\delta e_{a}{}^{m} &= \Big( -\mc{D}_{a}\xi^b+K_a{}^{b}+\s\delta_{a}{}^{b}\Big)e_b{}^{m}~, \label{CGvielgauge}
\end{align}
\end{subequations}
for arbitrary local parameters $\xi^a$, $K_{ab}=-K_{ba}$ and $\s$. A field $\Phi$ transforming as in \eqref{CGphigauge} is also said to be primary with conformal weight $\Delta$. Now, let us further suppose that $\xi^a\equiv \z^a$ is a conformal Killing vector and that $K_{ab}=K[\z]_{ab}=\mc{D}_{[a}\z_{b]}$ and $\s=\s[\z]=\frac{1}{d}\mc{D}^a\z_a$. Then, in accordance with section \ref{SectionCKV}, the combination \eqref{CGvielgauge} does not change the vielbein, $\delta e_{a}{}^{m}=0$. Therefore, it follows that  under the transformations \eqref{CGphigauge},
\begin{align}
\delta_{\z} e_{a}{}^{m}=0~,\qquad \delta_{\z} \Phi = \Big( \z^a\mc{D}_a +\frac{1}{2}K[\z]^{ab}M_{ab}+\Delta\s[\z] \Big) \Phi ~, \label{CGphiCKV}
\end{align}
the action  is invariant, $\delta_{\z} S[\Phi, e_a{}^{m}]=0$. In this work we will say that such actions are conformal. In the future (see section \ref{SectionActionCS}), $S[\Phi, e_a{}^{m}]$ will also be said to be primary.  

Justification for this name is provided by the fact that in the flat limit the action $S[\Phi]$, with which we began, is invariant under the infinitesimal conformal transformations \eqref{CGphiCKV}, with $\z^a$ given by \eqref{CKVflat}. 
 Indeed, inserting the explicit form \eqref{CKVflat} of $\z^a$  into \eqref{CGphiCKV} and using $K[\z]_{ab}=K_{ab}+4\tau_{[a}x_{b]}$ and $\s[\z]=\s+2\tau^ax_a$, the transformation \eqref{CGphiCKV} may be seen to be equivalent to 
\begin{align}
\delta_{\z} \Phi = \Big(\z^b\pa_b+\frac{1}{2}K[\z]^{ab}M_{ab}+\Delta \s[\z]\Big)\Phi = \Big( \xi^a\mf{P}_a + \frac{1}{2}K^{ab}\mf{M}_{ab}+\s \mf{D}+ \tau^a\mf{K}_a\Big)\Phi ~,\label{Iwanttorefertothis}
\end{align}
where we have made use of \eqref{ConfalDiffRel}. This rigid action coincides with \eqref{ConfGPhiTrans} and hence furnishes a representation of the conformal algebra $\mf{so}(d,2)$. Therefore, we see that if a model $S[\Phi]$ can be lifted to a curved background in a Weyl invariant way, then $S[\Phi]$ is conformally invariant.

\subsection{Conformal gravity action and its linearisation} \label{SectionCGpeasant}

In section \ref{SectionVielbein} we discussed in detail the gauge group of Einstein gravity, which corresponds to (covariantised) general coordinate and local Lorentz transformations. Below we review the models for conformal gravity in three and four dimensions, whose gauge group is enlarged with respect to the Einstein case by the inclusion of Weyl symmetry. 

\subsubsection{Four dimensions} \label{sectionCG4}

The action for conformal gravity in four dimensions is the square of the Weyl tensor
\begin{align}
S^{(d=4)}_{\text{CG}}[e]=\int \text{d}^4x \, e\, W^{abcd}W_{abcd}~. \label{CGA4}
\end{align}
Like the Einstein-Hilbert action, it is manifestly invariant under general coordinate and local Lorentz transformations (and hence CGCTs). However, unlike Einstein gravity,  this model is also invariant under Weyl transformations of the vielbein
\begin{align}
\delta_{\s}^{(\text{weyl})}e_{a}{}^{m}=\s e_{a}{}^{m} \qquad \implies \qquad \delta_{\s}^{(\text{weyl})}S^{(d=4)}_{\text{CG}}[e]=0~,
\end{align}
which follows directly from \eqref{DetWeyl} and \eqref{WeylWeyl}. Therefore the gauge group $\mc{G}$ of conformal gravity is characterised by the following symmetries of the vielbein
\begin{align}
\delta^{(\mc{G})}_{\L}e_{a}{}^{m}=\Big(\delta_{\xi}^{(\text{cgct})}+\delta_{K}^{(\text{llt})}+\delta_{\s}^{(\text{weyl})}\Big)e_{a}{}^{m}=\Big(-\mc{D}_a\x^b+K_a{}^{b}+\s\delta_a{}^{b}\Big)e_b{}^{m}~. \label{CGgaugetrans1}
\end{align}

By virtue of Weyl and Lorentz invariance, when deriving the equations of motion we may perturb the vielbein according to $\delta_{\hat{h}}e_a{}^{m}=\hat{h}_a{}^{b}e_b{}^{m}$, where $\hat{h}_{ab}$ is traceless and symmetric. Making use of the variational formula  \eqref{WeylDeform1}, we find the field equation 
\begin{align}
B_{ab}=\big(\mc{D}^c\mc{D}^d-S^{cd}\big)W_{cabd}=0~, \label{BachFlat}
\end{align}
where $B_{ab}$ is the Bach tensor \eqref{BachT}.
 Spacetimes satisfying \eqref{BachFlat} are said to be Bach-flat. Some obvious examples are (i) conformally-flat spacetimes; and (ii) Einstein spaces $R_{ab}=\lambda\eta_{ab}$ with $\l$ constant (bearing in mind the Bianchi identity \eqref{Bianchi4}).

 This action may be linearised about a background vielbein $e_a{}^{m}$ which solves the equation of motion \eqref{BachFlat}, i.e. is Bach-flat. To do this, we perturb the vielbein $\tilde{e}_a{}^{m}$ in accordance with the prescription outlined in section \ref{SectionVielPert}, except with $h_{ab}$ replaced by the symmetric and traceless $\hat{h}_{ab}$, 
\begin{align}
\tilde{e}_{a}{}^{m}=\Big(\delta_{a}{}^{b}+\hat{h}_{a}{}^{b}\Big)e_b{}^{m}~,\qquad \tilde{e}_m{}^{a}= e_{m}{}^{b}\Big(\delta_{b}{}^{a}-\hat{h}_{b}{}^{a}+\hat{h}_{b}{}^{c}\hat{h}_{c}{}^{a}\Big)~. \label{CGpert}
\end{align} 
 Then the action may be expanded to second order in $\hat{h}_{ab}$ as follows
 \begin{align}
 S_{\text{CG}}^{(d=4)}[\tilde{e}]=\int \text{d}^4 x \, \tilde{e} \, \widetilde{W}^{abcd}\widetilde{W}_{abcd}={}^{(0)}\hspace{-1mm}S_{\text{CG}}^{(d=4)}[e]+{}^{(1)}\hspace{-1mm}S_{\text{CG}}^{(d=4)}[e,\hat{h}]+{}^{(2)}\hspace{-1mm}S_{\text{CG}}^{(d=4)}[e,\hat{h}]~. \label{CG4expanded}
 \end{align} 
 To zeroth order we simply have ${}^{(0)}\hspace{-1mm}S_{\text{CG}}^{(d=4)}[e] = S_{\text{CG}}^{(d=4)}[e] $. The sector linear in $\hat{h}_{ab}$ is 
\begin{align}
{}^{(1)}\hspace{-1mm}S_{\text{CG}}^{(d=4)}[e,\hat{h}]=2\int \text{d}^4x \, e\, W^{abcd}~{}^{(1)}\hspace{-0.7mm}W_{abcd} =\Big|_{B_{ab}=0} ~~0
\end{align}
 where $W_{abcd}$ is the background Weyl tensor and ${}^{(1)}\hspace{-0.7mm}W_{abcd}\equiv \delta_{\hat{h}}^{(\text{dfm})}W_{abcd}$ is given by \eqref{WeylDeform1}. Of course, modulo a total derivative, the linear part vanishes when the background is Bach-flat. The sector which is quadratic in $\hat{h}_{ab}$ takes the form 
\begin{align}
{}^{(2)}\hspace{-1mm}S_{\text{CG}}^{(d=4)}[e,\hat{h}]=\int \text{d}^4x \, e \, \Big\{ \frac{1}{2}\hat{h}^{fg}\hat{h}_{fg}W_{abcd}W^{abcd} + {}^{(1)}\hspace{-0.7mm}W_{abcd}{}^{(1)}\hspace{-0.7mm}W^{abcd}+2W_{abcd}{}^{(2)}\hspace{-1mm}R^{abcd}\Big\} \label{CGquad}
\end{align}
 where ${}^{(2)}\hspace{-1mm}R_{abcd}$ is given by \eqref{RiemannDeform2} and we have used \eqref{DetDeform}. 
 
 At this point it is instructive to recall that the action $S_{\text{CG}}^{(d=4)}[\tilde{e}]$ is invariant under the gauge transformations \eqref{CGgaugetrans1}, which for $\tilde{e}_{a}{}^{m}$ read\footnote{Here and below we have used the notation $\tilde{\xi}^a:=\xi^m\tilde{e}_m{}^{a}$ and $\xi^a:=\xi^me_m{}^{a}$, where $\xi^m$ is the vector generating the diffeomorphism. }
\begin{align}
 \delta \tilde{e}_{a}{}^{m}= \Big(-\tilde{\mc{D}}_{a}\tilde{\xi}^b+K_{a}{}^{b}+\s\delta_a{}^{b}\Big)\tilde{e}_{b}{}^{m} \label{2.104}
 \end{align}
  Upon expanding around the background $e_{a}{}^{m}$, the latter induce corresponding invariance transformations on the linearised action \eqref{CG4expanded}, which may be interpreted in one of two ways. First there is the passive point of view, where both the background and perturbation transform according to 
\begin{subequations} 
\begin{align}
\delta e_{a}{}^{m}&=\Big(-\mc{D}_a\xi^{b}+K_{a}{}^{b}+\s\delta_a{}^{b}\Big)e_b{}^{m}~,\\
\delta \hat{h}_{ab}&=\Big( \xi^c\mc{D}_c+\frac{1}{2}K^{cd}M_{cd}\Big)\hat{h}_{ab}~. \label{CgravitonPrimary}
\end{align}
\end{subequations}
This reflects the fact that our background belongs to an equivalence class defined up to general coordinate, local Lorentz and local Weyl transformations, each of which leave the Bach-flat condition invariant. Furthermore, in accordance with previous discussions, the transformation rule \eqref{CgravitonPrimary} means that $\hat{h}_{ab}$ is a primary field with zero conformal weight. Hence, under a Weyl transformation of the background vielbein and $\hat{h}_{ab}$, 
\begin{align}
\delta_{\s}^{(\text{weyl})}e_a{}^{m}=\s e_a{}^{m}~,\qquad \delta_{\s}^{(\text{weyl})}\hat{h}_{ab}=0~,
\end{align} 
the linearised action, and independently the quadratic sector \eqref{CGquad}, is invariant. The action \eqref{CGquad} is therefore conformal in the sense of section \ref{SectionHobbit}.

On the other hand, we may adopt an active point of view, where the background is considered to be fixed, $\delta e_{a}{}^{m}=0$, and the whole burden of transformation falls onto the perturbation. To examine the consequences of this view, it is useful to rewrite the diffeomorphism sector of \eqref{2.104} as 
\begin{align}
 \delta \tilde{e}_{a}{}^{m}= \Big(\tilde{\mc{D}}_{(a}\tilde{\xi}_{b)}-\frac{1}{4}\eta_{ab}\tilde{\mc{D}}^c\tilde{\xi}_c\Big)\tilde{e}^{bm} ~,\label{2.105a}
 \end{align}
 where we have absorbed some $\tilde{\xi}$ dependent terms into the definitions of $K_{ab}$ and $\s$ and redefined $\xi^m \rightarrow -\xi^m$ for simplicity. To the lowest order in small quantities this means that $\hat{h}_{ab}$ is defined modulo the gauge transformations
\begin{align}
\delta_{\x}\hat{h}_{ab}=\mc{D}_{(a}\x_{b)}-\frac{1}{4}\eta_{ab}\mc{D}^c\xi_c~. \label{LinDiff}
\end{align}
Therefore, when restricted to Bach-flat backgrounds, the quadratic action \eqref{CGquad} is invariant under these gauge transformations 
 \begin{align}
\delta_{\x}\hat{h}_{ab}=\mc{D}_{(a}\x_{b)}-\frac{1}{4}\eta_{ab}\mc{D}^c\xi_c~ \quad \implies \quad  \delta_{\x}{}^{(2)}\hspace{-1mm}S_{\text{CG}}^{(d=4)}[e,\hat{h}]=\Big|_{B_{ab}=0} ~~0~. \label{QuadInvariance}
\end{align}
The perturbation $\hat{h}_{ab}$ is called the conformal graviton. When viewed as a primary field with zero conformal weight that is also defined modulo the gauge transformations \eqref{LinDiff}, the gauge parameter $\xi_a$ must also be primary but with conformal weight $-1$.

\subsubsection{Three dimensions} \label{sectionCG3}
In three dimensions, the Einstein-Hilbert action $S_{\text{EH}}$ is known to propagate no local degrees of freedom.\footnote{The usual Einstein-Hilbert action for $3d$ gravity with a cosmological term can also be interpreted as the Chern-Simons action for the AdS group \cite{AT, Witten}.} However, non-trivial dynamics emerge in topologically massive gravity \cite{DJT1,DJT2}, obtained by combining $S_{\text{EH}}$ with a Lorentz Chern-Simons term.
 The latter may be understood as the action for $3d$ conformal gravity \cite{vN, HW} and is given by 
\begin{align}
S^{(d=3)}_{\text{CG}}[e]=\int \text{d}^3x \, e\, \ve^{abc}\Big\{R_{ab}{}^{fg}\omega_{cfg}-\frac{2}{3}\o_{ad}{}^{f}\o_{bf}{}^{g}\o_{cg}{}^{d}\Big\}~. \label{CGA3}
\end{align}
This action is manifestly invariant under general coordinate transformations. That it is invariant under local Lorentz transformations is not obvious and must be verified using \eqref{LConLLT}. 
In addition to these, it is also invariant under local Weyl transformations 
\begin{align}
\delta_{\s}^{(\text{weyl})}e_{a}{}^{m}=\s e_{a}{}^{m} \qquad \implies \qquad \delta_{\s}^{(\text{weyl})}S^{(d=3)}_{\text{CG}}[e]=0~,
\end{align}
which can be checked using the relations \eqref{LCWeyl} and \eqref{RiemannWeyl}.

 The equation of motion corresponding to an infinitesimal displacement in the vielbein $\delta_{\hat{h}}^{(\text{dfm})}  e_{a}{}^{m}=\hat{h}_{a}{}^{b}e_{b}{}^{m}$, with $\hat{h}_{ab}$ symmetric and traceless, may be obtained using the variational relations \eqref{LCDeform} and \eqref{RiemannDeform1}. The resulting field equation is the vanishing of the (dual of the) Cotton tensor \eqref{CottonT}
 \begin{align}
 C_{ab}=0~. \label{CottonZero}
 \end{align}
 As discussed earlier, in three dimensions, solutions to \eqref{CottonZero} correspond to conformally flat spacetimes. Such a background should be used to linearise the conformal gravity action \eqref{CGA3}. This is done in complete analogy with the linearisation of four-dimensional conformal gravity from the previous section (though the calculation is longer).  The sector quadratic in the perturbation $\hat{h}_{ab}$ proves to be given by 
 \begin{align}
 {}^{(2)}\hspace{-1mm}S_{\text{CG}}^{(d=3)}[e,\hat{h}]=  4 \int \text{d}^3x\, e \, \hat{h}_{ab}{}^{(1)}C^{ab} ~,\label{CG3Lin1}
 \end{align}
 where ${}^{(1)}C_{ab}$ is the linearised Cotton tensor.  The latter may be determined by extracting the first order terms from $\widetilde{C}_{ab}=\ve_{a}{}^{cd}\widetilde{\mc{D}}_{c}\widetilde{S}_{db}$ using the relations \eqref{LCDDeform} and \eqref{SchoutenDeform}. Below we follow an alternative approach, and instead derive the corresponding expression in the language of two component spinors. A summary  of our spinor conventions is given in 
 appendix \ref{AppSpinor} to which the reader is referred for the technical details.
 
 Associated with the traceless part of the Ricci tensor, $R_{ab}-\frac{1}{3}\eta_{ab}R$, and  
 the Cotton tensor are symmetric rank-four spinors defined by 
 \begin{subequations}
 \bea
  R_{\a\b\g\d}&=&\big(\gamma^a\big)_{\a\b}\big(\gamma^b\big)_{\g\d}\Big(R_{ab}-\frac{1}{3}\eta_{ab}R\Big)=R_{(\a\b\g\d)}~, \\
 C_{\a\b\g\d}&=&(\g^a)_{\a\b}(\g^b)_{\g\d}C_{ab}=C_{(\a\b\g\d)}~.
 \eea
 \end{subequations}
 It is possible to show that the latter may be expressed in terms of $R_{\a(4)}$ via
 \bea 
 C_{\a\b\g\d}=\mathcal{D}^{\sigma}{}_{(\a}R_{\b\g\d)\sigma}~, \label{CottonRicci0}
 \eea
 where 
 $\mathcal{D}_{\a\b}=(\g^a)_{\a\b}\mathcal{D}_a$. Using \eqref{RicciDeform} and \eqref{ScalarDeform}, the variation in the traceless Ricci tensor is given by 
 \begin{align}
 \delta_{h}^{(\text{dfm})} R_{\a(4)}=-\mathcal{D}_{(\a_1}{}^{\b_1}\mathcal{D}_{\a_2}{}^{\b_2}h_{\a_3\a_4)\b(2)}+\frac{1}{2}R^{\b(2)}{}_{(\a_1\a_2}h_{\a_3\a_4)\b(2)}+\frac{1}{6}Rh_{\a(4)}~,
 \end{align}
 where we have defined $h_{\a\b\g\d}:=(\g^a)_{\a\b}(\g^b)_{\g\d}\hat{h}_{ab}=h_{(\a\b\g\d)}$. The deformation of the covariant derivative \eqref{LCDDeform} is given by 
 \begin{align}
 \delta_{h}^{(\text{dfm})}  \mc{D}_{\a\b}=-\frac{1}{2}h_{\a\b}{}^{\g\d}\mc{D}_{\g\d}+\frac{1}{2}\mc{D}^{\s\g}h_{\s\a\b}{}^{\d}M_{\g\d}~.\label{LCDSpinorDeform}
 \end{align} 
Then, using \eqref{CottonRicci} and \eqref{LCDSpinorDeform}, the deformation of the Cotton tensor proves to be 
 \begin{align}
  \delta_{h}^{(\text{dfm})} C_{\a(4)}&=\frac{1}{2}C^{\b(2)}{}_{(\a_1\a_2}h_{\a_3\a_4)\b(2)}-\frac{1}{2}\mathcal{D}_{(\a_1}{}^{\b_1}\mathcal{D}_{\a_2}{}^{\b_2}\mathcal{D}_{\a_3}{}^{\b_3} h_{\a_4)\b(3)}-\frac{1}{2}\Box\mathcal{D}_{(\a_1}{}^{\b_1} h_{\a_2\a_3\a_4)\b_1}
\notag\\
&+\big(\mathcal{D}_{(\a_1}{}^{\b_1}R_{\a_2\a_3}{}^{\b_2\b_3}\big) 
 h_{\a_4)\b(3)}+\frac{1}{12}\big(\mathcal{D}_{(\a_1}{}^{\b_1}R\big) h_{\a_2\a_3\a_4)\b_1}
-\frac{1}{12}R\mathcal{D}_{(\a_1}{}^{\b_1} h_{\a_2\a_3\a_4)\b_1}\notag\\
&+2R^{\b_1\b_2}{}_{(\a_1\a_2}\mathcal{D}_{\a_3}{}^{\b_3} h_{\a_4)\b(3)}-\frac{3}{4}R^{\b_1}{}_{\d(\a_1\a_2}\mathcal{D}^{\d\b_2}  h_{\a_3\a_4)\b(2)}~. \label{lincot}
 \end{align}
 Finally, the action \eqref{CG3Lin1} quadratic in $h_{\a(4)}$ is given by
 \begin{align}
 {}^{(2)}\hspace{-1mm}S_{\text{CG}}^{(d=3)}[e,h]=   \int \text{d}^3x\, e \, h^{\a(4)}\mc{C}_{\a(4)}(h) ~,\label{CG3Lin2}
 \end{align}
 where $\mc{C}_{\a(4)}(h)=\delta_{h}^{(\text{dfm})}  C_{\a(4)}$ is obtained from \eqref{lincot} by setting $C_{\a(4)}=0$.
 
Similar to the four dimensional case, upon expanding around a background, the symmetries \eqref{2.104} of the full action $S^{(d=3)}_{\text{CG}}[\tilde{e}]$ give rise to corresponding symmetries in the linearised action. In particular, the quadratic action \eqref{CG3Lin2} is invariant under linearised diffeomorphisms (on a conformally-flat background) under which $h_{\a(4)}$ transforms as
\begin{align}
\delta_{\xi}\hat{h}_{ab}= \mc{D}_{(a}\xi_{b)}-\frac{1}{3}\eta_{ab}\mc{D}^c\xi_c    \qquad \Longleftrightarrow\qquad \d_\x {h}_{\a(4) } =\cD_{(\a_1 \a_2 } \x_{\a_3 \a_4) }~.
\end{align}
Once again, the quadratic action \eqref{CG3Lin2} proves to be conformal when $h_{\a(4)}$
is  interpreted as a primary field of dimension zero.

\subsubsection{Higher-spin generalisations} 

 The linearised actions \eqref{CGquad} and \eqref{CG3Lin2} describe the gauge and Weyl invariant dynamics of the spin-2 conformal graviton $\hat{h}_{ab}$ in $d=4$ and $d=3$. By construction, gauge invariance can be achieved at most on Bach-flat and conformally-flat backgrounds respectively. Therefore, the latter constitute the most general type of backgrounds on which $\hat{h}_{ab}$ can consistently propagate. 

The main goal of this thesis is to investigate extensions of these quadratic actions to conformal higher-spin (CHS) fields on various types of curved gravitational backgrounds in three and four dimensions. Like the conformal graviton, it is widely thought that gauge invariance of the conformal higher-spin models can only be achieved at most on  conformally-flat and Bach-flat backgrounds respectively. 
The results of our later chapters support this statement. 
 Indeed, we will see many examples of complete CHS models where this statement holds true (and none where it does not). 

Rather than obtaining the quadratic CHS actions by linearising the models of \cite{Segal} (see \cite{GrigorievT, BeccariaT} for attempts in this direction),
 we will instead take a more straightforward and kinematic approach. 
More specifically, we will minimally couple the CHS models in Minkowski space to a non-trivial gravitational background and introduce non-minimal curvature dependent terms, using gauge invariance as our guiding principle. Of course, this must all be done in such a way as to preserve the Weyl symmetry. To keep the latter symmetry under control we will make use of a conformal calculus known as conformal space, which we review in section \ref{GCA}.


\section{Conformal higher-spin gauge fields on $\mb{M}^d$} \label{SectionCHSprep}

It is pertinent to discuss the generic structure and properties of conformal higher-spin gauge fields and their corresponding actions in Minkowski space. Below we review some of their general features in $\mathbb{M}^d$, leaving a detailed description of the specific cases $d=3$ and $d=4$ to later chapters.  

\subsection{The Fradkin-Tseytlin CHS model} \label{secFTCHSMd}

Models describing the dynamics of free conformal higher-spin fields were first proposed in four dimensional Minkowski space $\mathbb{M}^4$ by Fradkin and Tseytlin \cite{FT}. They were later generalised to $\mathbb{M}^d$ by Segal \cite{Segal}, where their  non-linear extensions were also described. According to the original formulation \cite{FT}, 
a conformal field with integer spin $s\geq 1$ is described by a real rank-$s$ tensor field ${\bm h}_{a_1\dots a_s}$ which is totally symmetric and traceful,
\begin{align}
{\bm h}_{a_1\dots a_s}={\bm h}_{(a_1\dots a_s)}\equiv {\bm h}_{a(s)}~, \qquad {\bm h}^b{}_{b a(s-2)}\neq 0~. \label{tracefulprep}
\end{align}
It is defined modulo the differential gauge transformation\footnote{The gauge transformation law \eqref{FlatDGT} is often generalised by removing the condition  $ \x^{b}{}_{b a (s-3)} =0$. However the resulting transformation law is equivalent to \eqref{FlatDGT} modulo a trace-shift transformation.} 
\begin{align}
\delta^{(\text{hsgt})}_{\xi}{\bm h}_{a(s)}=\pa_{(a_1}\xi_{a_2 \dots a_s)}~,\qquad \xi^{b}{}_{ba(s-3)}=0~,\label{FlatDGT}
\end{align}
 and the algebraic gauge transformation\footnote{The algebraic gauge symmetry is present only for $s\geq 2$.}
\begin{align}
\delta^{(\text{trs})}_{\lambda}{\bm h}_{a(s)}=\eta_{(a_1 a_2}\lambda_{a_3 \dots a_s)}~.\label{FlatAGT}
\end{align}
for totally symmetric gauge parameters $\xi_{a(s-1)}$ and $\l_{a(s-2)}$. Respectively, they are higher-spin generalisations of (i) diffeomorphisms (known as higher-spin gauge transformations); and (ii) Weyl transformations (known as generalised Weyl or trace-shift transformations).

Traditionally, the action for the conformal higher-spin gauge field in $\mathbb{M}^d$ is expressed in the following schematic form
\begin{align}
S_{\text{CHS}}^{(s,d)}[{\bm h}]=\frac{1}{2}\int \text{d}^dx {\bm h}^{a(s)}\pa^{(2s+d-4)}\Pi^{\perp}_{a(s)}({\bm h})~.\label{DdimCHSA}
\end{align}
 The operator $\Pi^{\perp}_{[s]}$, where $\Pi^{\perp}_{a(s)}({\bm h})=\Pi^{\perp}_{a(s)}{}^{b(s)}{\bm h}_{b(s)}\equiv\Pi^{\perp}_{[s]}{\bm h}_{a(s)}$, is known as the traceless and transverse (TT) spin-$s$ projector (or alternatively, the spin-projection operator).
The operator $\Pi^{\perp}_{[s]}$ acts on the space of totally symmetric and traceful fields $\Phi_{a(s)}$, and possesses the following defining properties:
\begin{enumerate}[label=(\roman*)]
\item \textbf{Idempotent:} $\Pi^{\perp}_{[s]}$ squares to itself, 
\begin{subequations}\label{TTProjPropDDim}
\begin{align}
\Pi^{\perp}_{[s]}\Pi^{\perp}_{[s]}\Phi_{a(s)}=\Pi^{\perp}_{[s]}\Phi_{a(s)}~. 
\end{align}
\item \textbf{Traceless:} $\Pi^{\perp}_{[s]}$ maps $\Phi_{a(s)}$ to a totally symmetric and traceless field 
\begin{align}
\Pi^{\perp}_{a_1\dots a_s}(\Phi)=\Pi^{\perp}_{a(s)}(\Phi)~,\qquad \eta^{bc}\Pi^{\perp}_{a(s-2)bc}(\Phi)=0~. \label{TTsymmetry}
\end{align}
\item   \textbf{Transverse:} $\Pi^{\perp}_{[s]}$ maps 
  $\Phi_{a(s)}$ to a transverse field
\begin{align}
\pa^b\Pi^{\perp}_{ba(s-1)}(\Phi)=0~. \label{TTproperty}
\end{align}
\item  \textbf{Surjective:} Every transverse and traceless field $\Psi_{a(s)}$ belongs to the image of $\Pi^{\perp}_{[s]}$, 
\begin{align}
\pa^{b}\Psi_{ba(s-1)}=\eta^{bc}\Psi_{bca(s-2)}=0~\quad \implies \quad \Pi^{\perp}_{[s]}\Psi_{a(s)}=\Psi_{a(s)}~.
\end{align}
\end{subequations}
In other words, $\Pi^{\perp}_{[s]}$ acts as the identity operator on the space of TT fields.
\end{enumerate}
By virtue of the properties \eqref{TTproperty} and \eqref{TTsymmetry}, the CHS action \eqref{DdimCHSA} is invariant under both types of gauge transformations \eqref{FlatDGT} and \eqref{FlatAGT},
\begin{align}
\delta^{(\text{hsgt})}_{\xi}S_{\text{CHS}}^{(s,d)}[{\bm h}]=0~,\qquad \delta^{(\text{trs})}_{\l}S_{\text{CHS}}^{(s,d)}[{\bm h}]=0~. \label{CHSinv}
\end{align}

Explicit expressions for the TT projectors for integer and half-integer spin fields were first derived in $d=4$ by Behrends and Fronsdal \cite{BF, Fronsdal58}. They were later generalised to $d$-dimensions in \cite{Segal}  in the case of integer spin (see also \cite{FMS, PoTs, Bonezzi, Isaev:2017nud}) and in \cite{Isaev:2017nud} for half-integer spin (see also \cite{3Dprojectors} for $d=3$). In chapters \ref{Chapter3D} and \ref{Chapter4D} we will derive explicit expressions for the spin-projection operators in $d=3$ and $d=4$ respectively. 

 In contrast to the free massless higher-spin actions of Fang and Fronsdal \cite{Fronsdal1,FF},  for fixed $s$ the CHS model \eqref{DdimCHSA} does not contain any lower-spin fields (or compensators) at the level of the action. The latter is said to describe pure spin-$s$ states and possesses a maximal degree of gauge invariance and irreducibility off-shell. 
 These properties can be attributed to the spin-$s$ TT projector in \eqref{DdimCHSA}, the presence of which threatens the locality of the model. In order to recover locality, one is forced to insert  higher-derivatives into the Lagrangian. From the point of view of curved space extensions, the higher-derivative nature of the CHS action is problematic, and presents the largest technical obstacle in this thesis.

\subsection{Equivalent realisations of CHS prepotentials}

There exist different ways to describe conformal higher-spin fields. They differ only in the sector 
of purely gauge degrees of freedom (compensators) that can be eliminated 
algebraically by applying local symmetry transformations without derivatives. 
For example, by virtue of \eqref{CHSinv} the generalised Weyl symmetry \eqref{FlatAGT} may be used to make the gauge field ${\bm h}_{a(s)}$ traceless by requiring
\bea
  {\bm h}_{a(s)} =h_{a(s)}~, \qquad 
  \eta^{bc} {h}_{bc a (s-2) } =0~. 
 \eea 
In order to preserve this gauge, the higher-spin gauge transformations must be accompanied by a $\xi$-dependent trace shift, yielding the new rule
\begin{align}
\delta_{\xi}h_{a(s)}=\pa_{(a_1}\xi_{a_2\dots a_s)}-\frac{(s-1)}{(2s+d-4)}\eta_{(a_1 a_2} \pa^b\xi_{a_3 \dots a_s)b}\equiv \pa_{(a_1}\xi_{a_2\dots a_s)}- (\text{traces})~.\label{TracelessGT}
\end{align} 
  
In principle, one may instead use Fronsdal's doubly traceless
spin-$s$ field 
\cite{Fronsdal1,Fronsdal2}
\bea
{\mathfrak h}_{a_1 \dots a_s} =h_{a_1\dots a_s} 
+ \eta_{(a_1 a_2} \vf_{a_3 \dots a_s) }~, \qquad
\eta^{bc} \vf_{bc a(s-4)} =0~,
\eea
to describe conformal spin-$s$ dynamics. In such an approach $\vf_{a(s-2)}$ is a compensator.
The gauge transformation law of ${\mathfrak h}_{a(s)}$ is given by 
\bea
\d {\mathfrak h}_{a(s)} = \pa_{(a_1} \x_{a_2 \dots a_s)} 
+ \eta_{(a_1 a_2} \tilde{\l}_{a_3 \dots a_s)}~, \qquad 
\eta^{bc} \x_{bc a(s-3)} =0~,\quad
\eta^{bc} \tilde{\l}_{bc a(s-4)} =0~.
\label{1.6}
\eea 
It is clear that the compensator $\vf_{a(s-2)}$  may be gauged away by applying 
a  $\tilde \l$-transformation, 
and then we are back to the formulation in terms of $h_{a(s)}$.

Another description of conformal spin-$s$ dynamics is obtained by employing 
 Vasiliev's frame field \cite{Vasiliev1980,Vasiliev87}
\bea
{\bm e}_{m, \, a_1 \dots a_{s-1}} = {\bm e}_{m, \, (a_1 \dots a_{s-1})} ~,
\qquad \eta^{bc} {\bm e}_{m, \, bc a (s-3)} =0~.
\eea 
In addition to a higher-spin $\x$-transformation, $\d {\bm e}_{m, a(s-1) } = \pa_m \x_{a(s-1)}$,
there are two additional local symmetries in this setting. These are 
 generalised Lorentz and Weyl transformations,  which do not involve 
 derivatives and allow one to gauge away compensating degrees of freedom 
 contained in ${\bm e}_{m, a(s-1) }$ by imposing the gauge condition 
 that ${\bm e}_{m, a(s-1) }$ is completely symmetric and traceless, 
 ${\bm e}_{m, a(s-1) } = h_{m a(s-1)}$.

It is natural to think of  $h_{a(s)}$ as the  genuine conformal spin-$s$ gauge field, due to several reasons. Firstly, one can consistently  define $h_{a(s) }$ to be a conformal primary field, see eq. \eqref{PrimaryConsistency} below. Secondly, the other degrees of freedom contained in ${\bm h}_{a(s)}$ are purely gauge ones, and as such they may become essential only at the nonlinear level. Finally, the nonlinear  conformal higher-spin theory of \cite{Segal} is formulated in terms of the fields $h_{a(s)}$.

Furthermore, in $d=3$ and $d=4$, the traceless and totally symmetric prepotential $h_{a(s)}$ has a straightforward translation into the language of two component spinors. The advantage of the two component spinor formalism is (at least) threefold: (i) it allows for a unified treatment of fields with integer (bosonic) and half-integer (fermionic) spins; (ii) it is the most natural language for superspace techniques; and (iii) it leads to remarkably simple and compact expressions as compared to their counterparts in vector and four component notation. Therefore, this will be the description favoured in this thesis.

\subsection{Conversion to $d=3,4$ two component spinor notation}

Below we recall the dictionary between the CHS field in vector notation and its two component spinor counterpart in $d=3$ and $d=4$. A detailed account of our specific two component spinor conventions in both dimensions may be found in appendix \ref{AppSpinor}.  

\subsubsection{Three dimensions}

The $d=3$ spin group\footnote{${\sSL} (2, {\mathbb R})$ is 
a double covering of the connected Lorentz group $\sSO_0(2,1)$.
The universal covering group of ${\sSL} (2, {\mathbb R})$ is not a matrix group 
and cannot be embedded in any group
$\sGL (n, {\mathbb R})$, see \cite{Novikov} for the proof.}
is ${\sSL} (2, {\mathbb R})$, 
and its irreducible field representations may be realised on the space of real totally symmetric rank-$n$ spinor fields
\begin{align} 
\Phi_{\a_1\dots\a_{n}}=\Phi_{(\a_1\dots\a_{n})}\equiv \Phi_{\a(n)}~.
\end{align}
 As a slight abuse of terminology, we may sometimes say that $\Phi_{\a(n)}$ carries spin $s=\frac{1}{2}n$.

In the current case, the symmetric and traceless rank-$s$ real tensor $h_{a(s)}$ is in one-to-one correspondence with the real\footnote{We make use of real gamma matrices, which means that complex conjugation amounts only to the exchange in the position of fields. For example, $(f_{\a(m)}g_{\b(n)})^*=(g_{\b(n)})^*(f_{\a(m)})^* = g_{\b(n)}f_{\a(m)}=(-1)^{\ve_f+\ve_g}f_{\a(m)}g_{\b(n)}$ where $\ve_f=m$ and $\ve_g=n$ are the Grassmann parities of $f_{\a(m)}$ and $g_{\b(n)}$.} and totally symmetric rank-$(2s)$ bosonic spinor field $h_{\a(2s)}$ via the relations 
\begin{subequations}
\begin{align}
h_{\a(2s)}&=\big(\g^{a_1}\big)_{\a_1\a_2}\cdots\big(\g^{a_s}\big)_{\a_{2s-1}\a_{2s}}h_{a_1\dots a_s}~,\\
h_{a(s)}&= \left(-\hf \right)^s(\g_{a_1})^{\a_1 \a_2 } \cdots  (\g_{a_s})^{\a_{2s-1} \a_{2s} } 
h_{\a_1 \dots \a_{2s}} ~.
\end{align}
\end{subequations}
Similarly, the real and totally symmetric rank-$s$ tensor-spinor $\psi_{a_1 \dots a_s \g}$, which is gamma traceless (and hence traceless in its Latin indices)
\begin{align}
(\g^b)^{\a\b}\psi_{ba(s-1)\b}=0~,
\end{align}
is in one-to-one correspondence with the real totally symmetric rank-$(2s+1)$ fermionic spinor field $\psi_{\a(2s+1)}$ via the relations
\begin{subequations}
\begin{align}
\psi_{\a(2s+1)}&=\big(\g^{a_1}\big)_{\a_1\a_2}\cdots\big(\g^{a_s}\big)_{\a_{2s-1}\a_{2s}}\psi_{a_1\dots a_s\a_{2s+1}}~,\\
\psi_{a_1 \dots a_s \a}&= \left(-\hf \right)^s(\g_{a_1})^{\a_1 \a_2 } \dots  (\g_{a_s})^{\a_{2s-1} \a_{2s} } \psi_{\a_1  \dots \a_{2s}\a} ~.
\end{align}
\end{subequations}
We combine both into the field $h_{\a(n)}$, for which the corresponding higher-spin gauge transformations take the form
\begin{align}
\delta_{\xi}h_{\a(n)}=\pa_{(\a_1\a_2}\xi_{\a_3\dots\a_n)}~. \label{CHS3gtFlat1}
\end{align}
The gauge parameter $\xi_{\a(n-2)}$ is also totally symmetric and $\pa_{\a\b}:=(\g^a)_{\a\b}\pa_a$. 

Finally, sometimes it will be very convenient to make use of the convention whereby spinor indices denoted by the same Greek letter are to be symmetrised over.
 Thus, for example, given two totally symmetric spinors $U_{\a(m)}$ and $V_{\b(n)}$, we have
\begin{align}
U_{\a(m)}V_{\a(n)}\equiv U_{(\a_1\dots\a_m}V_{\a_{m+1}\dots\a_{m+n})}=\frac{1}{(m+n)!}\big(U_{\a_1\dots\a_m}V_{\a_{m+1}\dots\a_{m+n}} + \dots\big)~. \label{SymCon1}
\end{align}
This convention will be used frequently in chapters \ref{Chapter3D} and \ref{Chapter3Dsuperspace}.

\subsubsection{Four dimensions}

The $d=4$ spin group is ${\sSL} (2, {\mathbb C})$, the double cover of the connected Lorentz group ${\sSO}_0 (3,1)$, and its irreducible representations may be labelled by $(m/2,n/2)$ for integers $m\geq 0$ and $n\geq 0$. The latter may be realised on (in general complex) rank-$(m,n)$ spinor fields $\Phi_{\a_1\dots\a_m\ad_1\dots\ad_n}$ which are symmetric in their undotted indices and, independently, in their dotted indices:  
\begin{align}
\Phi_{\a_1\dots\a_m\ad_1\dots\ad_n}=\Phi_{(\a_1\dots\a_m)(\ad_1\dots\ad_n)}\equiv \Phi_{\a(m)\ad(n)}~.
\end{align}
We may sometimes say that a field $\Phi_{\a(m)\ad(n)}$ carries spin $s=\frac{1}{2}(m+n)$.

The real symmetric and traceless rank-$s$ tensor $h_{a(s)}$ is in one-to-one correspondence with the real totally symmetric rank-$(s,s)$ spinor field $h_{\a(s)\ad(s)}$ via 
\begin{subequations}
\begin{align}
h_{\a(s)\ad(s)}&=\big(\s^{a_1}\big)_{\a_1\ad_1}\cdots\big(\s^{a_s}\big)_{\a_{s}\ad_{s}}h_{a_1\dots a_s}~,\\
h_{a(s)}&= \left(-\hf \right)^s(\tilde{\s}_{a_1})^{\ad_1 \a_1 } \cdots  (\tilde{\s}_{a_s})^{\ad_{s} \a_{s} } 
h_{\a_1 \dots \a_{s}\ad_1\dots\ad_s} ~.
\end{align}
\end{subequations}

A spin-$(s+\frac{1}{2})$ conformal field, for integer $s>0$, is usually described by the totally symmetric rank-$s$ Majorana tensor-(four)spinor ${\bm \Psi}_{a(s)}$ which is gamma-traceless 
\begin{align}
\Gamma^b {\bm \Psi}_{b a(s-1)}=0~,\qquad \Gamma^b=\begin{pmatrix} 0 && \s^b \\ \tilde{\s}^b && 0\end{pmatrix}~. \label{GammaT4CHS}
\end{align}
It may be represented as a column matrix consisting of a pair of tensor-(two)spinors,
\begin{align}
{\bm \Psi}_{a(s)}=\begin{pmatrix} \psi_{a(s)\a} \\ \bar{\psi}_{a(s)}{}^{\ad}\end{pmatrix}
\end{align}
where $\psi_{a(s)\a}$ and $\bar{\psi}_{a(s)\ad}$ are complex conjugates of eachother, $(\psi_{a(s)\a})^*=\bar{\psi}_{a(s)\ad}$.\footnote{Complex conjugation in the $d=4$ two component spinor formalism interchanges both (i) the dotted and undotted indices; and (ii) the relative positions of fields in composite expressions. For example $(f_{\a(m)\ad(n)}g_{\b(p)\bd(q)})^*=(g_{\b(p)\bd(q)})^*(f_{\a(m)\ad(n)})^*\equiv \bar{g}_{\b(q)\bd(p)}\bar{f}_{\a(n)\ad(m)}=(-1)^{\ve_f+\ve_g}\bar{f}_{\a(n)\ad(m)}\bar{g}_{\b(q)\bd(p)}$, where $\ve_f=m+n$ and $\ve_g=p+q$ are the Grassmann parities of $f_{\a(m)\ad(n)}$ and $g_{\b(p)\bd(q)}$.} On account of \eqref{GammaT4CHS} they are each $\s$-traceless,
\begin{align}
(\tilde{\s}^b)^{\ad\a}\psi_{b a(s-1)\a}=0~, \qquad (\tilde{\s}^b)^{\ad\a}\bar{\psi}_{b a(s-1)\ad}=0~,
\end{align}
and hence traceless in their Latin indices. The tensor-spinor $\psi_{a(s)\a}$ is in one-to-one correspondence with the complex totally-symmetric rank-$(s+1,s)$ fermionic spinor field $\psi_{\a(s+1)\ad(s)}$ via
\begin{subequations}
\begin{align}
\psi_{\a(s+1)\ad(s)}&=\big(\s^{a_1}\big)_{\a_1\ad_1}\cdots\big(\s^{a_s}\big)_{\a_{s}\ad_{s}}\psi_{a_1\dots a_s\a_{s+1}}~,\\
\psi_{a(s)\a}&= \left(-\hf \right)^s(\tilde{\s}_{a_1})^{\ad_1 \a_1 } \cdots  (\tilde{\s}_{a_s})^{\ad_{s} \a_{s} } 
\psi_{\a\a_1 \dots \a_{s}\ad_1\dots\ad_s} ~.
\end{align}
\end{subequations}
Similar relations between $\bar{\psi}_{a(s)\ad}$ and $\bar{\psi}_{\a(s)\ad(s+1)}$ may be obtained by complex conjugation. 

In chapter \ref{Chapter4D} of this thesis, we will consider the more general type of CHS field $h_{\a(m)\ad(n)}$ (with complex conjugate $\bar{h}_{\a(n)\ad(m)}$), of which the two described above are special cases. For such fields the higher-spin gauge transformations take the form
\begin{align}
\delta_{\xi}h_{\a(m)\ad(n)}=\pa_{(\a_1(\ad_1}\xi_{\a_2\dots\a_m)\ad_2\dots\ad_n)}~,
\end{align}
where the gauge parameter $\xi_{\a(m-1)\ad(n-1)}$ is totally symmetric and $\pa_{\a\ad}:=(\s^a)_{\a\ad}\pa_a$. In the case when $m+n$ is even and $m\geq n$, the pair $(h_{\a(m)\ad(n)},\bar{h}_{\a(n)\ad(m)})$ is in one-to-one correspondence with the mixed symmetry field $h_{a(p), b(q)}$ associated with the traceless two row Young tableau \scalebox{0.55}{
\begin{tikzpicture}
\draw (0,0) -- (2,0) -- (2,0.5) -- (0,0.5) -- cycle;
\draw (0,0) -- (1,0) -- (1,-0.5) -- (0,-0.5) -- cycle;
\filldraw[black] (1,0.25) circle (0pt) node {$p$};
\filldraw[black] (0.5,-0.25) circle (0pt) node {$q$};
\end{tikzpicture}
}, where $p=\frac{1}{2}(m+n)$ and $q=\frac{1}{2}(m-n)$ (see e.g. \cite{DidenkoSkvortsov}).

We extend the convention \eqref{SymCon1} to dotted and, independently, to undotted indices:
\begin{align}
U_{\a(m)\ad(p)}V_{\a(n)\ad(q)}\equiv U_{(\a_1\dots\a_m(\ad_1\dots\ad_p}V_{\a_{m+1}\dots\a_{m+n})\ad_{p+1}\dots\ad_{p+q})}~. \label{SymCon2}
\end{align}
This convention will be used frequently in chapters \ref{Chapter4D} and \ref{Chapter4Dsuperspace}.

\subsection{Conformal properties of the prepotentials} \label{secConfPropCHSbosMd}

In order to realise a rigid action of the conformal group $\mc{G}$, we define the gauge field $h_{a(s)}$ to be primary with some conformal weight $\Delta_{h_{(s)}}$ (see eq. \eqref{ConfGPhiTrans} or \eqref{Iwanttorefertothis})
\begin{align}
\delta_{\L}^{(\mc{G})}h_{a(s)}&=\Big(\xi^b\mf{P}_b+\frac{1}{2}K^{bc}\mf{M}_{bc}+\s\mf{D}+\tau^b\mf{K}_b\Big)h_{a(s)}~  \non\\
&=\Big(\xi^b\pa_b~+\frac{1}{2}K^{bc}\big[2x_b\pa_c+M_{bc}\big]+\s\big[x^b\pa_b+\Delta_{h_{(s)}}\big]\non\\
&\phantom{\Big(}~~+\tau^b\big[2x_bx^c\pa_c-x^2\pa_b+2x^cM_{bc}+2\Delta_{h_{(s)}}x_b\big]\Big)h_{a(s)}~.\label{CHSprepPrimary}
\end{align}
We also define the gauge parameter $\xi_{a(s-1)}$ in \eqref{TracelessGT} to be primary with conformal weight $\Delta_{\xi_{(s-1)}}$. However, these definitions must be made in such a way that after performing a higher-spin gauge transformation \eqref{TracelessGT}, $h_{a(s)}\rightarrow h'_{a(s)}=h_{a(s)}+\delta_{\xi}h_{a(s)}$, the prepotential remains primary. In particular, under a dilatation transformation, one finds that $\delta_{\xi}h_{a(s)}$ transforms according to 
\begin{align}
\delta_{\s}^{(\text{dil})}\big(\delta_{\xi}h_{a(s)}\big)=\s\Big(x^b\pa_b+(\Delta_{\xi_{(s-1)}}+1)\Big)\pa_{(a_1}\xi_{a_2\dots a_s)}~.
\end{align}
This tells us that the dimension of $h_{a(s)}$ is one unit greater than that of $\xi_{a(s-1)}$; $\Delta_{h_{(s)}}=\Delta_{\xi_{(s-1)}}+1$. Next, under an SCT one may show that $\delta_{\xi}h_{a(s)}$ transforms as (cf. \eqref{CHSprepPrimary})
\begin{align}
\delta_{\tau}^{(\text{sct})}&\big(\delta_{\xi}h_{a(s)}\big)= \tau^b\big[2x_bx^c\pa_c-x^2\pa_b+2x^cM_{bc}+2\big(\Delta_{\xi_{(s-1)}}+1\big)x_b\big]\delta_{\xi}h_{a(s)} \non\\
&+2\big[\Delta_{\xi_{(s-1)}}+(s-1)\big]\tau^b\Big(\eta_{b(a_1}\xi_{a_2\dots a_s)}-\frac{(s-1)}{(2s+d-4)}\eta_{(a_1 a_2}\xi_{a_3\dots a_s)b}\Big)~. \label{PrimaryConsistency}
\end{align}
Therefore, it follows that in order for $h_{a(s)}$ to be primary and also be defined modulo the gauge transformations \eqref{TracelessGT}, its conformal weight must be fixed to take the value
\begin{align}
\Delta_{h_{(s)}}=2-s~. \label{WeightFix}
\end{align}
By performing a similar calculation for the traceful prepotential ${\bm h}_{a(s)}$ (see eq. \eqref{tracefulprep}), it may be shown that one cannot consistently define ${\bm h}_{a(s)}$ to be both primary and possess the gauge freedom \eqref{FlatDGT}.

The spin-projection operators $\Pi^{\perp}_{[s]}$ have only been constructed on maximally-symmetric (super)spaces recently \cite{AdSprojectors, AdSuperprojectors, AdS3(super)projectors} (see also sections \ref{secCHSAdS4} and \ref{secAdS4|4}). Extending them to more generic gravitational backgrounds is a very challenging problem. Therefore, in order to describe CHS models on curved backgrounds, we will abandon the description \eqref{DdimCHSA} in favour  
of one which uses manifestly local, gauge and Weyl invariant field strengths. 

In particular, all of the conformal higher-spin actions encountered in this thesis that are invariant under the gauge transformations \eqref{TracelessGT}, may be expressed in the form
\begin{align}
S_{\text{CHS}}^{(s,d)}[h]=\int\text{d}^dx \, h^{a(s)} \mc{F}_{a(s)}(h)~.
\end{align}
Here the tensor $\mc{F}_{a(s)}(h)$ represents some (as of yet) unspecified transverse, traceless and  gauge invariant field strength which is a primary descendent of $h_{a(s)}$ with conformal weight $\Delta_{\mc{F}_{(s)}}=s+d-2$. By making use of the transformation law \eqref{CHSprepPrimary} for $h_{a(s)}$, and a similar rule for $\mc{F}_{a(s)}(h)$ (with $\Delta_{h_{(s)}}$ replaced by $\Delta_{\mc{F}_{(s)}}$), one can show that the above action is invariant under infinitesimal rigid conformal transformations, 
\begin{align}
\delta_{\L}^{(\mc{G})}h_{a(s)}=\L h_{a(s)} \qquad \implies \qquad \delta_{\L}^{(\mc{G})}S_{\text{CHS}}^{(s,d)}[h]=0~. \label{ConvInvAct}
\end{align}
Here $\L$ is defined as in \eqref{ConfGPhiTrans} (see also \eqref{Iwanttorefertothis}). Thus, obtaining the CHS action amounts to deriving the gauge invariant and primary descendant $\mc{F}_{a(s)}(h)$.



\section{Conformal gravity as a gauge theory} \label{GCA}





In the vielbein formulation of conformal gravity, we have seen that the gauge group consists of general coordinate, local Lorentz and Weyl transformations.  In this section we will review an alternative approach, where conformal gravity arises naturally as a gauge theory of the full conformal group \cite{KTvN1} (see also \cite{MacDowellMansouri}). The resulting geometry is referred to as conformal space, which is not to be confused with the ambient conformal space of Dirac \cite{DiracCS} (see however \cite{VasilievCS} ). The symmetry group in the vielbein formulation proves to correspond to a gauge fixed version of the structure group of conformal space. Our presentation follows that of \cite{ButterN=1, BKNT-M1} (see also \cite{IvanovCG, Lord1, Heyl1, FVP}).

\subsection{General remarks on the gauging procedure}
Let $\mc{G}$ be a spacetime symmetry group of $\mb{M}^d$, such as the conformal group $\mathsf{SO}_{0}(d,2)$ or the Poincar\'e group $\mathsf{ISO}_{0}(d-1,1)$ (a similar discussion holds for the corresponding supergroups).
The corresponding Lie algebra $\mf{g}$ is generated by the operators $X_{\tilde{a}}$, which can be split into two distinct classes. 
The first consists of those operators $P_a$ which generate spacetime translations,
 whilst the second class is formed by the remaining generators $X_{\un{a}}$. The latter form a subalgebra (known as the internal or homogeneous part) of $\mf{g}$ which we denote by $\mf{h}$ and the corresponding subgroup it generates is denoted by $\mc{H}$.

Consider again our prototypical dynamical matter system $S[\Phi]=\int \text{d}^dx \ms{L}\big(\Phi,\pa_a\Phi\big)$ defined on a Minkowski background (we assume Cartesian coordinates), and suppose that it is invariant under the rigid action of $\mc{G}$; $S[\Phi_{g}]=S[\Phi]$ for $g\in \mc{G}$. We adopt the active point of view, whereby the symmetry group is defined to act only on the matter fields and not on the coordinates themselves.  Infinitesimally this reads
\begin{align}
x^a\rightarrow x'^a=x^a~,\qquad \Phi(x) \rightarrow \Phi'(x)=\big(\mathds{1}+\L\big)\Phi(x)~,
\end{align}
which may alternatively be written  as
\begin{align}
\delta_{\L}^{(\mc{G})}x^a=0~,\qquad \delta_{\L}^{(\mc{G})}\Phi(x)=\L\Phi(x)~. \label{GlobalGAction}
\end{align}
Here $\L=\L^{\tilde{a}}X_{\tilde{a}}$ is an element of $\mf{g}$ with constant parameters $\L^{\tilde{a}}$ and with the generators $X_{\tilde{a}}$ in an appropriate representation (for example, in the case of the conformal algebra, this is \eqref{ConfGPhiTrans}). As a consequence, derivatives of the field transform according to 
\begin{align}
\delta_{\L}^{(\mc{G})}\big(\pa_a\Phi\big)=\pa_a\big(\L\Phi\big)=\L^{\tilde{b}}\pa_a\big(X_{\tilde{b}}\Phi\big)~. \label{GlobalDerivPhi}
\end{align}

We wish to gauge this symmetry by promoting the parameters to be local functions on spacetime $\L^{\tilde{a}}\equiv \L^{\tilde{a}}(x)$ in such a way that the action remains invariant. However, when passing to the local theory, it is clear that $\pa_a\Phi$ will transform with non-covariant terms involving derivatives of the parameter. Thus \eqref{GlobalDerivPhi} no longer holds and the action is not locally $\mc{G}$-invariant. 
There are typically two routes to go about rectifying this, and they each turn out to be equivalent to the other. 

The most direct approach is to start with $S[\Phi]$ on $\mb{M}^d$, localise the gauge parameters and then work out how to compensate for the non-covariant terms by introducing a gauge covariant derivative $\nabla_a$ involving appropriate gauge fields. The latter are defined to transform under $\mc{G}$ such that $\nabla_a\Phi$ transforms according to
\begin{align}
\delta_{\L}^{(\mc{G})}\big(\nabla_a\Phi\big)=\delta_{\L}^{(\mc{G})}\big(\nabla_a\big)\Phi+\nabla_a\big(\L\Phi\big)=\L(\nabla_a\Phi)~, \label{LocalDerivPhi}
\end{align}
 which is the local version of the rigid rule \eqref{GlobalDerivPhi}. Eventually, the gauge field for the translational symmetry may be identified with the vielbein and the translations are seen to correspond roughly to general coordinate transformations.  
Furthermore, one finds that gauging the full symmetry group $\mc{G}$ has the effect of endowing spacetime with: (i) non-zero torsion and curvature; (ii) local $\mc{H}$-symmetry in the tangent space at each point; and (iii) local diffeomorphism covariance. 
Therefore, by replacing $\pa_a \rightarrow \nabla_a$ and $\ms{L}\rightarrow e \ms{L}$,  the minimally coupled action $S[\Phi]$ is invariant under diffeomorphisms and local $\mc{H}$-transformations, and we may interpret $S[\Phi]$ to be defined on a curved spacetime.

On the other hand one can take an alternative (but equivalent) approach:  start with a non-trivial spacetime $\big(\mathcal{M}^d,e_a{}^{m}\big)$ from the beginning (with general coordinate covariance built in) and endow it with a local $\mc{H}$-structure. The latter is accomplished as follows; for each generator $X_{\un{a}}$ we associate a connection one form $B^{\un{a}}=\text{d}x^mB_{m}{}^{\un{a}}$ and introduce the corresponding $\mc{H}$-covariant derivative $\nabla_m:=\pa_m-B_m{}^{\un{a}}X_{\un{a}}$. The $\mc{H}$-transformation rules for the gauge fields may be determined either by requiring \eqref{LocalDerivPhi} to hold or by demanding that they furnish a representation of $\mf{h}$. Modulo internal gauge transformations, translations may then be shown to correspond to GCTs. From these ingredients one may easily construct models $S[\Phi]$ with diffeomorphism and local $\mc{H}$ invariance. Then, in the limiting case when the geometry degenerates to Minkowski space, one can show that the rigid $\mc{G}$-invariance is recovered. 

In the next section we will employ the second approach described above to gauge the conformal algebra, which corresponds to the case $\mf{g}=\mathfrak{so}(d,2)$. 
Before moving on, we point out some important conventions. Firstly, when performing successive gauge transformations, the gauge parameters themselves do not transform. Secondly, objects like $X_{\tilde{a}}\Phi$ are treated as if they are a single covariant field, so that in particular  
\begin{align}
\delta^{(\mc{G})}_{\L_1}\delta^{(\mc{G})}_{\L_2}\Phi=\L_1^{\tilde{a}}\L_2^{\tilde{b}}X_{\tilde{a}}X_{\tilde{b}}\Phi~.
\end{align}

\subsection{Conformal space: gauging the conformal algebra}

The conformal algebra $\mathfrak{so}(d,2)$, for $d>2$ dimensions,  is spanned by the translation $(P_a)$, Lorentz $(M_{ab})$, special conformal $(K_a)$ and dilatation $(\mathbb{D})$ generators. The non-zero commutators are given by \eqref{confal}, which we reiterate below for convenience
\begin{subequations} \label{confal0}
\begin{align}
[M_{ab},M_{cd}]&=2\eta_{c[a}M_{b]d}-2\eta_{d[a}M_{b]c}~, \phantom{inserting blank space inserting}\\
[M_{ab},P_c]&=2\eta_{c[a}P_{b]}~, \qquad \qquad \qquad \qquad ~ [\mathbb{D},P_a]=P_a~,\\
[M_{ab},K_c]&=2\eta_{c[a}K_{b]}~, \qquad \qquad \qquad \qquad [\mathbb{D},K_a]=-K_a~,\\
[K_a,P_b]&=2\eta_{ab}\mathbb{D}+2M_{ab}~.
\end{align}
\end{subequations}
The generators $M_{ab},K_a$ and $\mathbb{D}$ span a subalgebra $\mf{h}$
of $\mathfrak{so}(d,2)$
and are hereby collectively referred to as $X_{\un{a}}$. In contrast, we denote the generators of the full algebra by $X_{\tilde{a}}=\big(P_a,X_{\un{a}}\big)$. Then, the commutation relations \eqref{confal0} may be rewritten as follows\footnote{We adopt the convention whereby a factor of 1/2 is inserted when summing over pairs of (anti-)symmetric indices. For example, $f_{\un{a}\un{b}}{}^{\un{c}}X_{\un{c}}=f_{\un{a}\un{b}}{}^{K_c}K_{c}+\frac{1}{2}f_{\un{a}\un{b}}{}^{M_{cd}}M_{cd}+\dots$.}
\begin{subequations}\label{confal2}
\begin{align}
[X_{\tilde{a}},X_{\tilde{b}}]&= -f_{\tilde{a}\tilde{b}}{}^{\tilde{c}}X_{\tilde{c}}~,\label{confal20}
\end{align}
or, explicitly separating the translation generator,\footnote{We prefer to explicitly indicate which generator an index belongs to in the structure constants. Otherwise it is difficult, for example, to distinguish between $f_{\mathbb{D}, P_a}{}^{P_b}$ and $f_{\mathbb{D}, K_a}{}^{K_b}$.}
\begin{align}
[X_{\un{a}},X_{\un{b}}]&=-f_{\un{a}\un{b}}{}^{\un{c}}X_{\un{c}}~,\label{confal2a}\\
[X_{\un{a}},P_b]&=-f_{\un{a}b}{}^{\un{c}}X_{\un{c}}-f_{\un{a}P_b}{}^{P_c}P_c~. \label{confal2b}
\end{align}
\end{subequations}
Here $f_{\tilde{a}\tilde{b}}{}^{\tilde{c}}$ are
the structure constants whose non-vanishing components are 
\begin{subequations} \label{strcon1}
\begin{align}
f_{M_{ab},M_{cd}}{}^{M_{fg}}&=8\eta_{[a\{c}\delta_{d\}}^{[f}\delta_{b]}^{g]}~,\qquad ~~~ f_{K_a,P_b}{}^{M_{cd}}=-4\delta_{a}^{[c}\delta_{b}^{d]}~,\\
f_{M_{ab},P_c}{}^{P_d}&=-2\eta_{c[a}\delta_{b]}^d~, \qquad \qquad ~~ f_{\mathbb{D}, P_a}{}^{P_b}=-\delta_a^b~, \\
f_{M_{ab},K_c}{}^{K_d}&=-2\eta_{c[a}\delta_{b]}^d~, \qquad \qquad ~ f_{\mathbb{D}, K_a}{}^{K_b}=\delta_a^b~,\\
f_{K_a,P_b}{}^{\mathbb{D}}&=-2\eta_{ab}~,
\end{align}
\end{subequations} 
and which satisfy the Jacobi identities
\begin{align} \label{Jacobi}
0=f_{[\tilde{a}\tilde{b}}{}^{\tilde{d}}f_{\tilde{c}]\tilde{d}}{}^{\tilde{e}}=f_{[\tilde{a}\tilde{b}}{}^{P_d}f_{\tilde{c}]P_d}{}^{\tilde{e}}+f_{[\tilde{a}\tilde{b}}{}^{\un{d}}f_{\tilde{c}]\un{d}}{}^{\tilde{e}}~.
\end{align}

Let us denote by $\mc{H}$ the subgroup generated by the operators $X_{\un{a}}$, and let us consider a field $\Phi(x)$ (with indices suppressed) transforming linearly under some tensor representation of $\mc{H}$. We say that $\F$ is $\cH$-covariant if, under an $\cH$-transformation parametrised by the local gauge parameter $\L^{\un{a}}(x)$, it transforms with no derivative on $\L^{\un{a}}$:
\begin{align}
\delta^{(\mc{H})}_{\L}\Phi=\Lambda^{\un{a}}X_{\un{a}}\Phi~.
\end{align}
In addition, if $\Phi$ satisfies
\begin{align}
K_{a}\Phi=0~,\qquad \mathbb{D}\Phi=\Delta\Phi~, \label{PrimaryDef}
\end{align}
then it is said to be a primary field of dimension (or Weyl/conformal weight) $\D$. Such fields will be of primary interest in the future, but for the current discussion we will not place this restriction on $\Phi$.

It is clear that $\pa_m \F$ is not $\cH$-covariant:
\begin{align}
\delta^{(\mc{H})}_{\L}\pa_m\Phi= \pa_m\big(\L^{\un{a}}\big)X_{\un{a}}\Phi +  \L^{\un{a}}\pa_m\big(X_{\un{a}}\Phi\big)~.
\end{align}
To rectify this, with each generator $X_{\un{a}}$ of $\mc{H}$ we associate a connection one-form $B^{\un{a}}=\text{d}x^mB_{m}{}^{\un{a}}$ and with the translation generator $P_a$ we associate the vielbein $E^a=\text{d}x^me_m{}^{a}$. We then introduce the conformally covariant derivative $\nabla_m$:
\begin{align}\label{2.5}
\nabla_m=\partial_m-B_m{}^{\un{a}}X_{\un{a}}~, \qquad \nabla_a:=e_a{}^{m}\nabla_m~.
\end{align} 
If we postulate that the connections and vielbein transform under $\mc{H}$ according to
\begin{subequations} \label{GeometricHTrans}
\begin{align}
\delta^{(\mc{H})}_{\L}B_{m}{}^{\un{a}}&=\pa_m\L^{\un{a}}+e_m{}^{b}\L^{\un{c}}f_{\un{c}P_b}{}^{\un{a}}+B_{m}{}^{\un{b}}\L^{\un{c}}f_{\un{c}\un{b}}{}^{\un{a}}~,\label{ConnHTrans} \\
\delta^{(\mc{H})}_{\L}e_{m}{}^{a}&=e_m{}^{b}\L^{\un{c}}f_{\un{c}P_b}{}^{P_a}~, \label{VielHTrans}
\end{align}
\end{subequations}
then it is not difficult to show that $\nabla_a \Phi$ is $\mc{H}$-covariant,
\begin{align}
 \delta^{(\mc{H})}_{\L}\big(\nabla_a\Phi\big)=\Lambda^{\un{b}}\nabla_aX_{\un{b}}\Phi-\Lambda^{\un{b}}f_{\un{b}P_a}{}^{P_c}\nabla_c\Phi-\Lambda^{\un{b}}f_{\un{b}P_a}{}^{\un{c}}X_{\un{c}}\Phi~.
 \end{align}
Expressing this in the form $\delta^{(\mc{H})}_{\L}\big(\nabla_a \Phi\big) =\Lambda^{\un{b}}X_{\un{b}}\nabla_a\Phi$ allows us to deduce to relation
\begin{align}
 [X_{\un{a}},\nabla_b]=-f_{\un{a}P_b}{}^{\un{c}}X_{\un{c}}-f_{\un{a}P_b}{}^{P_c}\nabla_c~.\label{confal3}
 \end{align}
 It is useful to combine \eqref{ConnHTrans} and \eqref{VielHTrans} to obtain the transformation rule for the connection $B_{a}{}^{\un{b}}:=e_a{}^{m}B_{m}{}^{\un{b}}$ with only frame indices,
 \begin{align}
\delta^{(\mc{H})}_{\L} B_{a}{}^{\un{b}}=e_a\L^{\un{b}}+\L^{\un{c}}f_{\un{c}P_a}{}^{\un{b}}+B_{a}{}^{\un{c}}\L^{\un{d}}f_{\un{d}\un{c}}{}^{\un{b}}-\L^{\un{d}}f_{\un{d}P_a}{}^{P_c}B_{c}{}^{\un{b}}~.\label{ConnHTransFrame}
\end{align}

Comparing \eqref{confal3} with \eqref{confal2b}, we see that $X_{\un{a}}$ satisfies the same commutation relations with $\nabla_{{b}}$ as it does with $P_b$. This is a consequence of the fact that the transformation rules \eqref{GeometricHTrans} furnish a representation of $\mf{h}$ in the sense that
\begin{align}
\big[\delta^{(\mc{H})}_{\L_1} ,\delta^{(\mc{H})}_{\L_2} \big]B_m{}^{\un{a}}=\delta^{(\mc{H})}_{\L_3} B_{m}{}^{\un{a}}~,\qquad \big[\delta^{(\mc{H})}_{\L_1} ,\delta^{(\mc{H})}_{\L_2} \big]e_m{}^{a}=\delta^{(\mc{H})}_{\L_3} e_{m}{}^{a}~,
\end{align}
where $\L_3^{\un{a}}:=-\L^{\un{b}}_1\L^{\un{c}}_2f_{\un{b}\un{c}}{}^{\un{a}}$.
Indeed, this is the motivation for the ans\"atze \eqref{GeometricHTrans}.

 However, unlike the translation generators $P_a$, 
the commutator of two covariant derivatives is not zero but is instead given by 
\begin{align} 
\big[\nabla_a,\nabla_b\big]=-\mathcal{T}_{ab}{}^{c}\nabla_c-\mathcal{R}_{ab}{}^{\un{c}}X_{\un{c}}~.\label{nablacom1}
\end{align}
In equation \eqref{nablacom1}, $\mathcal{T}_{ab}{}^{c}$ and $\mathcal{R}_{ab}{}^{\un{c}}:=\big\{\mc{R}(M)_{ab}{}^{cd}, ~\mc{R}(\mb{D})_{ab},~\mc{R}(K)_{ab}{}^{c}\big\}$ are the torsion and curvature tensors respectively,
\begin{subequations} \label{fieldstr}
\begin{align}
\mathcal{T}_{ab}{}^{c}&=-\mathscr{C}_{ab}{}^{c}+2B_{[a}{}^{\un{d}}f_{P_b]\un{d}}{}^{P_c}~,\\
\mathcal{R}_{ab}{}^{\un{c}}&=-\mathscr{C}_{ab}{}^{d}B_{d}{}^{\un{c}}+2B_{[a}{}^{\un{d}}f_{P_b]\un{d}}{}^{\un{c}}+B_{[a}{}^{\un{e}}B_{b]}{}^{\un{d}}f_{\un{de}}{}^{\un{c}}+2e_{[a}B_{b]}{}^{\un{c}} ~,
\end{align}
\end{subequations}
where $e_a=e_a{}^{m}\partial_m$ and $\mathscr{C}_{ab}{}^{c}$ are the anholonomy coefficients \eqref{anholonomy}.
Using the transformation rules \eqref{ConnHTrans} and the Jacobi identities \eqref{Jacobi}, we find that the torsion and curvature tensors \eqref{fieldstr} transform 
covariantly under $\mathcal{H}$ according to
\begin{subequations}\label{2.11}
\begin{align}
\delta^{(\mc{H})}_{\L}\mathcal{T}_{ab}{}^{c}&=\mathcal{T}_{ab}{}^{d}\Lambda^{\un{e}}f_{\un{e}P_d}{}^{P_c}
-2\Lambda^{\un{e}}f_{\un{e}[P_a}{}^{P_d}\mathcal{T}_{b]d}{}^c~,\\
\delta^{(\mc{H})}_{\L}\mathcal{R}_{ab}{}^{\un{c}}&=\mathcal{R}_{ab}{}^{\un{e}}\Lambda^{\un{d}}f_{\un{de}}{}^{\un{c}}+2\Lambda^{\un{d}}f_{\un{d}[P_a}{}^{P_e}\mathcal{R}_{b]e}{}^{\un{c}}+\mathcal{T}_{ab}{}^{e}\Lambda^{\un{f}}f_{\un{f}P_e}{}^{\un{c}}~.
\end{align}
\end{subequations}
 On account of the Jacobi identity 
\begin{align}\label{BianchiCCDCS}
0=\big[\nabla_a,[\nabla_b,\nabla_c]\big]+\big[\nabla_b,[\nabla_c,\nabla_a]\big]+\big[\nabla_c,[\nabla_a,\nabla_b]\big]~,
\end{align}  
the tensors \eqref{fieldstr} satisfy the Bianchi identities
\begin{subequations} \label{BianchiCA}
\begin{align}
0&= \nabla_{[a}\mc{T}_{bc]}{}^{d}-\mc{T}_{[ab}{}^{f}\mc{T}_{c]f}{}^{d}-\mc{R}_{[ab}{}^{\un{f}}f_{P_c]\un{f}}{}^{P_d} ~,\label{BianchiCA1}\\
0&=\nabla_{[a}\mc{R}_{bc]}{}^{\un{d}}-\mc{T}_{[ab}{}^{f}\mc{R}_{c]f}{}^{\un{d}}-\mc{R}_{[ab}{}^{\un{f}}f_{P_c]\un{f}}{}^{\un{d}} ~.\label{BianchiCA2}
\end{align}
\end{subequations}

In this formulation infinitesimal general coordinate transformations, generated by a local parameter $\xi^a$, are not covariant with respect to $\mathcal{H}$ (in the same sense as \eqref{CGCTMotivation}). To remedy this, they must be supplemented by an additional $\mathcal{H}$-transformation with gauge parameter $\Lambda[\xi]^{\un{a}}$,\footnote{We recall that the GCT in \eqref{CovDiffConf} is to be interpreted actively, which means that $\delta^{(\text{gct})}_{\xi} x^m =0$ whilst all other objects transform according to their Lie derivative $\mc{L}_{\xi}$ along $\xi^a$. In particular, $\delta^{(\text{gct})}_{\xi} \F=\xi^ae_a\F$.}
\begin{align}
\delta^{(\text{cgct})}_{\xi}:=\delta^{(\text{gct})}_{\xi}+\delta^{(\mc{H})}_{\L[\xi]}~,\qquad \L[\xi]^{\un{a}}:=-\xi^bB_{b}{}^{\un{a}}~.\label{CovDiffConf}
\end{align}
 It follows that such transformations act on fields $\Phi$ (with all indices Lorentz) as $\delta^{(\text{cgct})}_{\xi}\Phi=\xi^a\nabla_a\Phi$. The resulting gauge group, denoted by $\mathcal{G}$, is then generated by the set of operators $\{\nabla_a,X_{\un{a}}\}$ under which $\Phi$ transforms as
 \begin{align}
 \delta^{(\mathcal{G})}_{\L}\Phi=\L\Phi,\qquad \L=\xi^b\nabla_b+\L^{\un{b}}X_{\un{b}}~. \label{GtransPhi}
 \end{align}

Using the rules \eqref{ConnHTrans} and \eqref{VielHTrans}, it may be shown that under a covariantised general coordinate transformation \eqref{CovDiffConf}, the connections transform according to  
 \begin{subequations} \label{ConnCGCT}
 \begin{align}
 \delta^{(\text{cgct})}_{\xi}B_m{}^{\un{a}}&= \xi^be_m{}^{c}\mc{R}_{bc}{}^{\un{a}}+\xi^{b}B_{m}{}^{\un{c}}f_{P_b\un{c}}{}^{\un{a}}~,\\
 \delta^{(\text{cgct})}_{\xi} e_m{}^{a}&= \pa_m\xi^a+\xi^bB_{m}{}^{\un{c}}f_{P_b\un{c}}{}^{P_a}+\xi^be_m{}^{c}\mc{T}_{bc}{}^{a}~. \label{VielCGCTCA}
 \end{align}
 \end{subequations}
 Finally, the gauge transformation of $\nabla_a$ under $\mathcal{G}$ proves to obey the relation
 \begin{subequations} \label{CCDGaugeT}
 \begin{align}
 \delta^{(\mathcal{G})}_{\L}\nabla_a=[\L,\nabla_a]~,\qquad \L=\xi^b\nabla_b+\L^{\un{b}}X_{\un{b}} 
 \end{align}
 provided we interpret\footnote{Care must be taken when applying these identities since, for example, one can have $\L^{\un{b}}=0$ but $\nabla_a\L^{\un{b}}\neq 0$ if $\xi^a\neq 0$. \label{GaugeWarning}}
 \begin{align}
 \nabla_a\xi^b:=e_a\xi^b+\xi^cB_a{}^{\un{d}}f_{P_c\un{d}}{}^{P_b}~,\qquad \nabla_{a}\Lambda^{\un{b}}:=e_a\Lambda^{\un{b}}+\xi^cB_a{}^{\un{d}}f_{P_c\un{d}}{}^{\un b}+\Lambda^{\un{c}}B_{a}{}^{\un{d}}f_{\un{cd}}{}^{\un{b}}~.
 \end{align}
 \end{subequations}
 

The rigid limit corresponds to the scenario when the $\mc{H}$-connections vanish, $B_{m}{}^{\un{a}}=0$, and the vielbein can (globally) be chosen to take the form  $e_{m}{}^{a}=\delta_m{}^{a}$. Consequently, the torsion and curvatures vanish and our spacetime degenerates to flat Minkowski space $\mb{M}^d$ in Cartesian coordinates.  The parameters of $\mc{G}$-transformations preserving this rigid configuration are constrained by the equations (cf. \eqref{ConnHTrans}, \eqref{VielHTrans} and \eqref{ConnCGCT})
\begin{align}
0&=\pa_mK^{ab}-4\tau^{[a}\delta_m{}^{b]}~, \qquad 0=\pa_m\s-2\tau_m~,\non\\
0&=\pa_m\tau^a~,\qquad 0=\pa_m\xi^a-\s\delta_m{}^{a}+\delta_{m}{}^{b}K^{a}{}_{b}~.
\end{align}
Here $\xi^a$, $K^{ab}=-K^{ba}$, $\s$  and  $\tau^a$ parametrise translations, Lorentz rotations, dilatations and special conformal boosts respectively. These equations may be integrated to give 
\begin{align}
\tau^a=\tau_{0}^a~,\qquad \s&=\s_0+2\tau_0^ax_a~,\qquad K^{ab}=K_0^{ab}+4\tau_0^{[a}x^{b]}~,\non\\
\xi^a=\xi^a_0+&\s_0x^a+K_0^{ba}x_b-\tau_0^{a}x^bx_b+2\tau^b_0x_bx^a~,
\end{align}
for arbitrary real constants $\xi_0^a,~ K_0^{ab}=-K^{ba}_0,~\s_0,~\tau_0^a$. Then, given a primary field $\Phi$ (see \eqref{PrimaryDef}), the transformation law $\delta_{\L}^{(\mc{G})}\Phi=\L\Phi$ where $\L=\xi^a\pa_a+\frac{1}{2}K^{ab}M_{ab}+\s\mb{D}+\tau^aK_a$, precisely reproduces the infinitesimal rigid action \eqref{ConfGPhiTrans} of the conformal group in $\mb{M}^d$.

\subsection{Conformal gravity from conformal space} \label{secCGfromCS}

Through the above gauging procedure we have managed to localise the conformal symmetry, which has effectively endowed our spacetime with the local structure group $\mf{so}(d,2)$. So far, the vielbein $e_m{}^{a}$ and each of the connections $B_a{}^{\un{b}}$ have remained completely independent of one another. However, we will be interested in constructing theories coupled to conformal gravity as it was described in section \ref{SectionCFT} using the vielbein approach. There, the vielbein was the only independent geometric field, and the corresponding gauge symmetry is captured by \eqref{CGgaugetrans1}.  Therefore, in this section we describe how one can constrain the gauge group $\mc{G}$ to recover the vielbein approach to conformal gravity. 

The explicit form of the conformally covariant derivative, given by eq. \eqref{2.5}, is
\begin{align}
\nabla_a=e_a{}^{m}\pa_m-\frac{1}{2}\hat{\o}_{a}{}^{bc}M_{bc}-\mathfrak{b}_a\mathbb{D}-\mathfrak{f}_{a}{}^{b}K_b~,
\label{2.177}
\end{align}
where $\hat{\o}_{a}{}^{bc}$, $\mathfrak{b}_a$  and $\mathfrak{f}_a{}^{b}$ are the Lorentz, dilatation and special conformal connections respectively. They satisfy the commutation relations
\begin{align}
[\nabla_a,\nabla_b]=-\mathcal{T}_{ab}{}^{c}\nabla_c-\frac{1}{2}\mathcal{R}(M)_{ab}{}^{cd}M_{cd}-\mathcal{R}(K)_{ab}{}^{c}K_c-\mathcal{R}(\mathbb{D})_{ab}\mathbb{D}
\end{align}
where the torsion and curvatures are given by \eqref{fieldstr},
\begin{subequations}
 \begin{align}
 \mathcal{T}_{ab}{}^c&=-\mathscr{C}_{ab}{}^{c}+2\hat{\o}_{[ab]}{}^c+2\mathfrak{b}_{[a}\delta_{b]}{}^{c}~,\label{CATorsion}\\
 \mathcal{R}(M)_{ab}{}^{cd}&=\hat{R}_{ab}{}^{cd}+8\mathfrak{f}_{[a}{}^{[c}\delta_{b]}{}^{d]}~,\label{LorCur}\\ 
 \mathcal{R}(K)_{ab}{}^{c}&=-\mathscr{C}_{ab}{}^d\mathfrak{f}_d{}^{c}-2\hat{\o}_{[a}{}^{cd}\mathfrak{f}_{b]d}-2\mathfrak{b}_{[a}\mathfrak{f}_{b]}{}^{c}+2e_{[a}\mathfrak{f}_{b]}{}^{c}~, \label{2.19cc}\\
 \mathcal{R}(\mathbb{D})_{ab}&= -\mathscr{C}_{ab}{}^{c}\mathfrak{b}_c+4\mathfrak{f}_{[ab]}+2e_{[a}\mathfrak{b}_{b]}~.
 \end{align}
 \end{subequations}
In eq. \eqref{LorCur} we have defined the Riemann tensor $\hat{R}_{abcd}$,
\begin{align}
\hat{R}_{ab}{}^{cd}:=-\mathscr{C}_{ab}{}^{f}\hat{\o}_{f}{}^{cd}+2e_{[a}\hat{\o}_{b]}{}^{cd}-2\hat{\o}_{[a}{}^{cf}\hat{\o}_{b]f}{}^{d}~, \label{RiemannB}
\end{align}
corresponding to the Lorentz connection  $\hat{\o}_{abc}$. At this stage, neither $\hat{R}_{abcd}$ nor $\hat{\o}_{abc}$ can be identified with their regular gravity counterpart from section \ref{SectionLCD}. 

In order to ensure that the vielbein is the only independent field in the theory
(modulo purely gauge degrees of freedom; see section \ref{SectionDegauging}), 
we have to impose covariant constraints on the algebra.
 It turns out that the correct constraints are
 \begin{subequations}\label{constraints}
 \begin{align}
 \mathcal{T}_{ab}{}^c=0~,\label{constraintTF}\\
 \eta^{bd}\mathcal{R}(M)_{abcd}=0~.\label{constraintCG}
 \end{align}
 \end{subequations}
These conditions are covariant in the sense that they are preserved by $\mathcal{H}$-transformations, which may be verified through \eqref{2.11}.\footnote{The torsion and curvatures (with frame indices) transform as scalars under CGCTs, and hence the constraints \eqref{constraints} are invariant under these too. } The first constraint is analogous to the torsion-free constraint \eqref{TFC} and determines the Lorentz connection in terms of the vielbein and the dilatation connection, 
   \begin{align}\label{spincon}
  \hat{\omega}_{abc}&=\o_{abc}-2\eta_{a[b}\mathfrak{b}_{c]}~.
  \end{align}
 Here $\o_{abc} \equiv \o_{abc} (e)
 =\frac{1}{2}\big(\mathscr{C}_{abc}-\mathscr{C}_{acb}-\mathscr{C}_{bca}\big)$ 
 coincides with the standard torsion-free Lorentz connection \eqref{TFLorCon} of section \ref{SectionLCD}. This allows us to decompose the Lorentz covariant derivative with connection $\hat{\o}_{abc}$ as follows
\begin{align}\label{2.100}
 \hat{\mathcal{D}}_{a}&=e_a-\frac{1}{2}\hat{\o}_{a}{}^{bc}M_{bc}=\mathcal{D}_a+\mathfrak{b}^cM_{ac}
 \end{align}
 where $\mathcal{D}_a=e_a-\frac{1}{2}\o_{a}{}^{bc}M_{bc}$ is the torsion-free 
 Lorentz covariant derivative of section \ref{SectionLCD}.

 The second constraint \eqref{constraintCG} has no analogy in the vielbein formulation, and is known as the conformal gravity constraint. It algebraically resolves the special conformal connection in terms of the vielbein and the dilatation connection, 
 \begin{align}
 \mathfrak{f}_{ab}=-\frac{1}{2}\hat{S}_{ab}~,\qquad \hat{S}_{ab}:=\frac{1}{(d-2)}\Big(\hat{R}_{ab}-\frac{1}{2(d-1)}\eta_{ab}\hat{R}\Big)~, \label{2.16}
 \end{align}
 where $\hat{S}_{ab}$ is the (non-symmetric) Schouten tensor, $\hat{R}_{ab}=\eta^{cd}\hat{R}_{acbd}$ is the (non-symmetric) Ricci tensor and $\hat{R}=\eta^{ab}\hat{R}_{ab}$ is the scalar curvature. 
 
Making use of  \eqref{RiemannB} and \eqref{spincon}
 allows us to decompose $\hat{R}_{abcd}$ 
  into those terms which depend solely on the vielbein
 and those involving the dilatation connection,
\begin{subequations} \label{2.99}
 \bea
 \hat{R}_{abcd}&=&R_{abcd}-4e_{[a}\eta_{b][c}\mathfrak{b}_{d]}
 -4\eta_{\{c[a}\omega_{b]d\}}{}^{g}\mathfrak{b}_g
 +4\mathfrak{b}_{[c}\eta_{d][a}\mathfrak{b}_{b]}
 +2\eta_{c[a}\eta_{b]d}\mathfrak{b}^f\mathfrak{b}_f~,
\label{2.999}\\
 \hat{R}_{ab}&=&R_{ab}
+ (d-2) \Big\{ e_a\mathfrak{b}_b - \o_{ab}{}^{c}\mathfrak{b}_c - \mathfrak{b}_a\mathfrak{b}_b
 \Big\}
\non\\
&&
\phantom{R_{ab}}
+\eta_{ab} \Big\{ e_c\mathfrak{b}^c -\o^c{}_{cd}\mathfrak{b}^d
+  (d-2)  \mathfrak{b}_c  \mathfrak{b}^c\Big\}
~,\\
\hat{R}&=&R+2 (d-1) \Big\{ e_a\mathfrak{b}^a-\o^a{}_{ab}\mathfrak{b}^b
+\hf (d-2)\mathfrak{b}^a\mathfrak{b}_a \Big\}
~.
 \eea
 \end{subequations}
Here $R_{abcd}$ is the (usual) Riemann tensor \eqref{Riemann} associated with  
the (usual) Lorentz connection $\o_{abc}$ \eqref{TFLorCon},
and $R_{ab}$ and $R$  stand for  the corresponding (symmetric) Ricci tensor 
and scalar curvature, respectively. 
Inserting the relations \eqref{2.99} into the solution 
of the conformal gravity constraint \eqref{2.16} yields
 \begin{align}\label{2.96}
 \mathfrak{f}_{ab}=-\frac{1}{2}S_{ab}+\frac{1}{2}\mathfrak{b}_a\mathfrak{b}_b-\frac{1}{4}\eta_{ab}\mathfrak{b}^c
\mathfrak{b}_c+\frac{1}{2}\o_{ab}{}^{c}\mathfrak{b}_c-\frac{1}{2}e_{a}\mathfrak{b}_b~,
 \end{align} 
 where $S_{ab}$ is the (usual) Schouten tensor \eqref{SchoutenT}.

We now turn to extracting all possible information from the Bianchi identities \eqref{BianchiCA}. Making use of the constraint \eqref{constraintTF}, the Bianchi identity \eqref{BianchiCA1} takes the form 
\begin{align}
0=\mc{R}(M)_{[abc]d}+\mc{R}(\mb{D})_{[ab}\eta_{c]d}~.
\end{align}
Contracting with $\eta^{cd}$ and using \eqref{constraintCG}, we find that the dilatation curvature vanishes,
\begin{align}
\mathcal{R}(\mathbb{D})_{ab}&=0~.
\end{align}
From \eqref{BianchiCA2} and the above relations, one may derive the following Bianchi identities
\begin{subequations}
\begin{align}
\mathcal{R}(K)_{[abc]}&=0~,\label{Bianchi54}\\
\mathcal{R}(M)_{[abc]d}&=0~,\label{Bianchi55}\\
\nabla_{[a}\mathcal{R}(K)_{bc]d}&=0~,\\
\nabla_{[a}\mathcal{R}(M)_{bc]}{}^{de}-4\mathcal{R}(K)_{[ab}{}^{[d}\delta_{c]}{}^{e]}&=0~.\label{2.87}
\end{align}
\end{subequations} 
 
 Using \eqref{2.999} and \eqref{2.96}, one can show that 
 the dependence on the dilatation connection drops out of \eqref{LorCur}
 and we obtain
 \begin{align}
\mathcal{R}(M)_{abcd}=W_{abcd}~. \label{2.98}
 \end{align}
Here $W_{abcd}$ is the (usual) Weyl tensor \eqref{WeylT},
which is a primary field of dimension +2,
\begin{align}
K_e W_{abcd}=0~, \qquad \mathbb{D} W_{abcd}=2W_{abcd}~. \label{WeylTPrimary}
\end{align}
The above relations may be deduced from \eqref{2.11} by stripping off the gauge parameter.  The fact that the Lorentz curvature is equal to the Weyl tensor is consistent with \eqref{Bianchi55}.

So far our considerations have been valid for all spacetime dimensions with $d\geq 3$. However, for further analysis of the constraints, 
it is necessary to consider separately the choices $d=3$ and $d>3$. 
 
\subsubsection{Conformal gravity in $d>3$}

In the  $d>3$ case, it follows from \eqref{2.87} and \eqref{2.98} that the special conformal curvature is given by 
 \begin{align}\label{2.60}
 \mathcal{R}(K)_{abc}=\frac{1}{2(d-3)}\nabla^d W_{abcd}~.
 \end{align}
 As a result, the algebra of conformal covariant derivatives is
\begin{align}\label{2.61}
[\nabla_a,\nabla_b]=-\frac{1}{2}W_{abcd}M^{cd}-\frac{1}{2(d-3)}\nabla^dW_{abcd}K^c~.
\end{align}
It is determined by a single primary tensor field, 
the Weyl tensor. In particular, from \eqref{2.61} it follows that if the spacetime under consideration is conformally flat, 
then the conformal covariant derivatives commute,
\begin{align}\label{2.62}
W_{abcd}=0 \quad \implies \quad [\nabla_a,\nabla_b]=0~.
\end{align}
This observation will be important for our subsequent analysis in chapter \ref{Chapter4D} of this thesis.

\subsubsection{Conformal gravity in $d=3$}

The Weyl tensor vanishes identically in three dimensions.
As a result, the Lorentz curvature \eqref{2.98} also vanishes and the algebra of conformal covariant derivatives takes the form 
\begin{align}\label{2.21}
[\nabla_a,\nabla_b]=-\mathcal{R}(K)_{ab}{}^cK_c~.
\end{align}
The Lorentz covariant derivative \eqref{2.100} allows us to represent the special conformal curvature \eqref{2.19cc} as
\begin{align}\label{2.23}
 \mathcal{R}(K)_{abc}&= 2\mathcal{D}_{[a}\mathfrak{f}_{b]c}-2\mathfrak{b}_{[a}\mathfrak{f}_{b]c}+2\mathfrak{b}_c\mathfrak{f}_{[ab]}+2\eta_{c[a}\mathfrak{f}_{b]d}\mathfrak{b}^d~.
 \end{align}
  We note that $\mc{D}_a\mf{f}_{bc}=\big(e_a-\frac{1}{2}\o_a{}^{bc}M_{bc}\big)\mf{f}_{bc}$ is well defined since $\mf{f}_{ab}$ is covariant under LLTs. Using \eqref{2.96}, after some algebra one may show that the dependence on $\mathfrak{b}_a$ in eq. \eqref{2.23} drops out such that 
 \begin{align}\label{2.25}
 \mathcal{R}(K)_{abc}=-\frac{1}{2}C_{abc}~.
 \end{align}
Here  $C_{abc}$ is the Cotton tensor \eqref{CottonT}, which in $d=3$ is a primary field of dimension $+3$,
\begin{align}
K_dC_{abc}=0~,\qquad \mathbb{D}C_{abc}=3C_{abc}~.
\end{align} 
The above relations may be deduced from \eqref{2.11} by stripping off the gauge parameter.
 
We find that in $d=3$ the algebra of conformally covariant derivatives is 
\begin{align}
[\nabla_a,\nabla_b]=\frac{1}{2}C_{ab}{}^cK_c~. \label{CCDA3}
\end{align}
Therefore, all information about conformal geometry 
is encoded in 
a single primary field, the Cotton tensor. In particular,  as follows from   \eqref{CCDA3}, 
  the commutator of conformal covariant derivatives vanishes
  in the conformally flat case,
 \begin{align} \label{2.33}
 C_{abc}=0\quad \implies \quad [\nabla_a,\nabla_b]=0~.
 \end{align}
This observation will be important for our subsequent analysis in chapter \ref{Chapter3D} of this thesis.

  \subsection{Degauging and the emergence of Weyl symmetry} \label{SectionDegauging}

In the previous section we imposed constraints on the gauge algebra such that only the vielbein $e_m{}^{a}$ and the dilatation connection $\mf{b}_a$ remained as the independent geometric fields. However, in both Einstein and conformal gravity, only the vielbein (or equivalently the metric) carries independent physical degrees of freedom. Rather than imposing an extra constraint to fix $\mathfrak{b}_a$ in terms of the vielbein, we observe that under a special conformal transformation (SCT) with local parameter $\tau_a$, $\mathfrak{b}_a$ transforms as
 \begin{align}\label{2.17}
 \delta_{\tau}^{(\text{sct})}\mathfrak{b}_a=-2\tau_a~.
 \end{align}
 It follows that we may impose the gauge condition 
 \begin{align}\label{degauge}
 \mathfrak{b}_a=0~.
 \end{align}

After this choice, only the vielbein remains as an independent field. We note that whenever the gauge \eqref{degauge} is chosen, all hatted objects from the previous sections coincide with their non-hatted counterparts.  In particular
 \begin{align}
 \hat{\mathcal{D}}_a\big|_{\mathfrak{b}_a=0}=\mathcal{D}_a~,\qquad \hat{R}_{abcd}\big|_{\mathfrak{b}_a=0}=R_{abcd}~,
 \end{align}
 and in this gauge we may therefore abandon the hat notation without any ambiguity. Furthermore, in this case it is clear that the special conformal connection \eqref{2.96} (see also \eqref{2.16}) reduces to 
 \begin{align}
 \mathfrak{f}_{ab}=-\frac{1}{2}S_{ab}~.  \label{DegaugedSCCon}
 \end{align} 
It follows that the conformal covariant derivative takes the form  
\be
\nabla_a=\mathcal{D}_a+\frac 12 S_{a}{}^{b}K_b~. \label{DegaugedCCD}
\ee 
The process of imposing the gauge \eqref{degauge}, exchanging all occurrences of $\nabla_a$ in favour of $\mc{D}_a$ using \eqref{DegaugedCCD} and evaluating the action of $K_a$ is called `degauging'. For our purposes, it is desirable to keep this  symmetry intact throughout calculations and degauge only at the end when we wish to extract physically meaningful results.

The gauge \eqref{degauge} breaks the special conformal symmetry, in the sense that it is no longer an independent symmetry.  To see this, it is instructive to recall the full $\mc{H}$-transformation rule \eqref{ConnHTransFrame} for $\mf{b}_a$,
\begin{align}
\delta_{\Lambda}^{(\mc{H})} \mf{b}_a=e_a\s -2\tau_a +\s\mf{b}_a+K_a{}^{b}\mf{b}_a~.
\end{align}
Evidently, the gauge choice \eqref{degauge} is not preserved under dilatations. To remedy this, we must supplement all dilatation transformations with a $\s$-dependent SCT according to 
\begin{align}
\delta_{\s}^{(\text{weyl})}:= \delta_{\s}^{(\text{dil})}+\delta_{\tau[\s]}^{(\text{sct})}~,\qquad \tau[\s]_a:=\frac{1}{2}e_a\s~. \label{BrokenDil}
\end{align}
To justify the notation $\delta_{\s}^{(\text{weyl})}$ on the left hand side we observe that when implemented on the vielbein and a primary field $\Phi$ with weight $\Delta$, the outcome agrees with \eqref{VariousWeylT}. Furthermore, using \eqref{CCDGaugeT}  (bearing in mind the cautionary remarks in footnote \ref{GaugeWarning}) to examine the variation of $\nabla_a$ under \eqref{BrokenDil}, we find
\begin{align}
\delta_{\s}^{(\text{weyl})}\nabla_a &=\s\nabla_a-2\tau[\s]^bM_{ab}-\Big(\s\mf{f}_{ab}+e_a\tau[\s]_b-\o_{ab}{}^{c}\tau[\s]_c\Big)K^b \non\\
&=\Big(\s\mc{D}_a-\mc{D}^b\s M_{ab}\Big)+\frac{1}{2}\Big(2\s S_{ab}-\mc{D}_{(a}\mc{D}_{b)}\s\Big)K^b~, \non\\
&=\delta_{\s}^{(\text{weyl})}\mc{D}_a+\frac{1}{2}\Big(\delta_{\s}^{(\text{weyl})}S_{ab}\Big)K^b~,
\end{align}
where we have used \eqref{DegaugedSCCon} and \eqref{DegaugedCCD}. This allows us to read off the Weyl transformation of the Lorentz covariant derivative and the Schouten tensor, and both agree with the expressions \eqref{LCDWeyl} and \eqref{SchoutenWeyl} derived earlier.

 
Therefore, we see that Weyl symmetry emerges as the residual dilatation symmetry left after fixing the gauge \eqref{degauge}. In particular, after degauging, the gauge symmetry of the vielbein reduces precisely to that which is present in the vielbein formulation of conformal gravity \eqref{CGgaugetrans1}.

\subsection{On the action principle in conformal space} \label{SectionActionCS}

To finish this section, we give some general remarks concerning the action principle within the formulation described above, and describe how the actions \eqref{CGA4} and \eqref{CGA3} for conformal gravity in $d=4$ and $d=3$ dimensions arise.  

\subsubsection{Local $\mc{G}$-invariance }

We consider a functional of the form
\begin{align}
S[\Phi, e_{m}{}^{a},B_a{}^{\un{b}}]=\int \text{d}^dx \, e \, \ms{L}\big(e_m{}^{a},\Phi, \nabla_a\Phi,\dots\big)~. \label{CAmodelaction}
\end{align}
 The Lagrangian may be constructed solely from the vielbein, or it may involve matter fields $\Phi$, though we only allow for dependence on the $\mc{H}$-connections through $\nabla_a$ (with the exception of Chern-Simons type theories, to be discussed briefly in the next section). The latter restriction allows us to assume that $\ms{L}$ is $\mc{H}$-covariant; $\delta_{\L}^{(\mc{H})}\ms{L}=\L^{\un{a}}X_{\un{a}}\ms{L}$. Then, requiring \eqref{CAmodelaction} to be invariant under the local gauge group $\mc{H}$ (i.e. $\delta_{\L}^{(\mc{H})}S=0$ ) implies 
\begin{align}
0&=\int \text{d}^dx\, e \, \Big\{ \delta_{\L}^{(\mc{H})}\ms{L}+e_a{}^{m}\delta_{\L}^{(\mc{H})}\big(e_m{}^{a}\big)\ms{L}\Big\}= \int \text{d}^dx\, e \, \Big\{\frac{1}{2}K^{ab}M_{ab}+\s\big(\mb{D}-d\big)+\tau^a K_a \Big\}\ms{L} ~,\non 
\end{align}
where we have used \eqref{VielHTrans} and the variational formula $\delta e = e e_a{}^{m}\delta e_m{}^{a}$. Therefore, for $\mc{H}$ gauge invariance we see that the Lagrangian must be a primary scalar field with conformal weight equal to $d$,
\begin{align}
M_{ab}\ms{L}=0~,\qquad K_a\ms{L}=0~,\qquad \mb{D}\ms{L}=d\ms{L}~. \label{HcovLag}
\end{align}
Furthermore, using \eqref{VielCGCTCA} and \eqref{CATorsion}, imposing invariance under infinitesimal CGCTs yields the constraint
\begin{align}
0=\int \text{d}^d x \, e \, \Big\{ \delta_{\xi}^{(\text{cgct})}\ms{L}+\Big[e_a\xi^a-\xi^a\ms{C}_{ab}{}^{b}+d\xi^a\mf{b}_a\Big]\ms{L}\Big\}~.
\end{align} 
This is satisfied modulo the boundary term $\int \text{d}^d x \pa_m\big(e \xi^m \ms{L}\big)$ provided the Lagrangian transforms as a covariant scalar under a CGCT,
\begin{align}
\delta_{\xi}^{(\text{cgct})}\ms{L} = \xi^a \nabla_a\ms{L}~, \label{CGCTcovLag}
\end{align}
where we have used \eqref{HcovLag}.  

Any action \eqref{CAmodelaction} whose Lagrangian $\ms{L}$ possesses the properties \eqref{HcovLag} and \eqref{CGCTcovLag} is invariant under the full gauge group $\mc{G}$. In particular, this means that upon imposing \eqref{constraints} and degauging \eqref{degauge}, the action \eqref{CAmodelaction} is invariant under Weyl transformations,
\begin{align}
\mf{b}_a=0\qquad \implies \qquad \delta_{\s}^{(\text{weyl})}S =0~.
\end{align} 
In this thesis such action functionals will be said to be primary.

Finally, we note that in the gauge \eqref{degauge}, the transformation of the vielbein under $\mc{G}$ (cf. \eqref{VielHTrans} and \eqref{VielCGCTCA}) is given by 
\begin{align}
\delta_{\L}^{(\mc{G})}e_{m}{}^{a}=\mc{D}_{m}\xi^a-\s e_m{}^{a}-e_m{}^{b}K_{b}{}^{a}~.
\end{align}
It is clear the $\mc{G}$-transformations which preserve the vielbein, $\delta_{\L}^{(\mc{G})}e_{m}{}^{a}=0$, are generated by the conformal Killing vectors $\xi^a\equiv \z^a$ of the manifold (with $\s\equiv \s[\z]=\frac{1}{d}\mc{D}^a\z_a$ and $K_{ab}\equiv K_{ab}[\z]=\mc{D}_{[a}\z_{b]}$). From the invariance of $S$ under $\mc{G}$, it also follows that $S$ is invariant under the conformal transformations \eqref{CGphiCKV}.

\subsubsection{Integration by parts} \label{secIBPCS}

Consider a functional of the form \eqref{CAmodelaction}, with $\ms{L}$ satisfying \eqref{HcovLag} and \eqref{CGCTcovLag}. Let us further suppose that $\ms{L}$ takes the form
\begin{align}
\ms{L}=g^J\mathcal{A}_J(h) \label{Z.0}
\end{align}
where $g_J$ and $h_J$ are primary fields with abstract index structure and $\mathcal{A}$ is a linear differential operator such that $\mathcal{A}_J(h)\equiv \mathcal{A}_J{}^{I}h_I$ is also primary.  We define the transpose of the operator $\mathcal{A}$ by
\begin{align}
\int\text{d}^dx\, e \, g^J\mathcal{A}_J(h)
=\int\text{d}^dx\, e \, h^J\mathcal{A}^T_J(g)+\int\text{d}^dx\, e \, \Omega \label{Z.1}
\end{align}
where $\Omega$ is a total conformal derivative and may be written as $\Omega=\nabla_aV^a$ for some composite vector field $V^a=V^{a}(g,h)$ with Weyl weight $(d-1)$. The first term on the right hand side of \eqref{Z.1} is the result of integrating the left hand side by parts in the usual way. 

In general we cannot conclude that the second term on the right hand side of \eqref{Z.1} vanishes. However, under the condition that $\mathcal{A}^T_J(g)$ is primary then $\Omega$ must also be primary. It follows that 
\begin{align}
0=K_a\Omega=[K_a,\nabla_b]V^b+\nabla^bK_aV_b= \nabla^bK_aV_b~.\label{Y.0}
\end{align}
 It is clear that the condition $\nabla^bK_aV_b=0$ is satisfied if $V_a$ is primary. What is not so clear is that any solution $V_a$ to this equation is necessarily primary. However, for all cases known to us this is true, and we shall make this assumption in this thesis. 
 
Since the Lagrangian in \eqref{CAmodelaction} is primary, all dependence on the dilatation connection $\mathfrak{b}_a$ drops out, which means that the conformal covariant derivative takes the form \eqref{DegaugedCCD}. Consequently, the total conformal derivative arising in \eqref{Z.1} vanishes,
 \begin{align}
\int\text{d}^dx\, e \, \Omega=\int\text{d}^dx\, e \, \bigg(\mc{D}^aV_a+\frac{1}{2}S^{ab}K_bV_a\bigg)= 0~, \label{Z.111}
\end{align} 
where we have ignored the total derivative arising from torsion-free Lorentz covariant derivative $\mc{D}_a$ and used that $V_{a}$ is primary. 

 Therefore, we arrive at the following rule for integration by parts:
\begin{align}
\int\text{d}^dx\, e \, g^J\mathcal{A}_J(h)=\int\text{d}^dx\, e \, h^J\mathcal{A}^T_J(g) \label{Y.10}
\end{align}
if $K_ag_I=K_ah_I=K_a\big(\mathcal{A}_I(h)\big)=K_a\big(\mathcal{A}^T_I(g)\big)=0$. 

It may happen that we wish to integrate $ g^J\mathcal{A}_J(h)$ by parts, but the resulting $h^J\mathcal{A}^T_J(g)$ is not primary. In this case we cannot assume that the $\Omega$ on the right hand side of \eqref{Z.1} is primary.  In particular this means that 
\begin{align}
\int\text{d}^dx\, e \, \Omega=\frac{1}{2}\int\text{d}^dx\, e \, S^{ab}K_bV_a~, 
\end{align}
and the best we can do is evaluate the action of $K_a$ on $V_b$ with a non-zero result.  Since $S_{ab}$ appears explicitly, the manifest conformal symmetry may be lost. However, in some cases it may be restored through the use of the Bianchi identity 
\be 
\nabla^d W_{abcd}=-2(d-3)\mathcal{D}_{[a}S_{b]c}~. \label{BianchiIBP}
\ee 
Indeed, we will see an example of this in appendix \ref{AppIBPS3}. Within the setting of conformal space, this Bianchi identity may be derived by imposing the gauge \eqref{degauge} and equating the two equivalent representations \eqref{2.19cc} and \eqref{2.60} for the special conformal curvature. 

\subsubsection{Action for conformal gravity in $d=4$}

In the case $d=4$, the action for conformal gravity \eqref{sectionCG4} may be obtained \`{a} la MacDowell and Mansouri \cite{MacDowellMansouri} as the square of the Lorentz curvature,
\begin{align}
S_{\text{CG}}^{(d=4)}=-\frac{1}{4}\int\text{d}^4x \, e \, \ve^{abcd}\ve^{efgh}\mc{R}(M)_{abef}\mc{R}(M)_{cdgh} = \int\text{d}^4x \, e \, W^{abcd}W_{abcd}~. \label{ACGCA4}
\end{align}
Here we have made use of the constraints \eqref{constraints} and their consequences. It is dependent only on the vielbein and, by virtue of \eqref{WeylTPrimary}, the Lagrangian clearly possesses the properties \eqref{HcovLag} and \eqref{CGCTcovLag}. The action \eqref{ACGCA4} is therefore invariant under the full conformal gravity gauge group $\mc{G}$. It is also the unique action, of the form proposed in \cite{MacDowellMansouri}, with this property. This is the sense by which it is meant that conformal gravity is the gauge theory of the conformal group. 


\subsubsection{Action for conformal gravity in $d=3$}

As we now demonstrate, the action \eqref{CGA3} for conformal gravity in three dimensions  coincides with the Chern-Simons action for the gauged conformal algebra. To write down the latter, it is advantageous to use the language of differential forms, for which we use the super-form conventions of \cite{WB} (in particular, the exterior derivative acts from the right). 

Let us employ a unified notation whereby the vielbein and connection one-forms are packaged according to $B^{\tilde{a}}=\text{d}x^m B_m{}^{\tilde{a}}=(E^a,B^{\un{a}})$. We also package the torsion and curvature two-forms as $\mathcal{R}^{\tilde{a}}:=\frac{1}{2}E^c\wedge E^b \mathcal{R}_{bc}{}^{\tilde{a}}=(\mathcal{T}^a,\mathcal{R}^{\un{a}})$, they prove to be equal to
\begin{align}
\mc{R}^{\tilde{a}}=\text{d}B^{\tilde{a}}-\frac{1}{2}B^{\tilde{b}}\wedge B^{\tilde{c}}f_{\tilde{c}\tilde{b}}{}^{\tilde{a}}~.
\end{align}
In order to construct the relevant Chern-Simons action, we make use of the symmetric non-degenerate Cartan-Killing metric $\Gamma_{\tilde{a}\tilde{b}}=\Gamma_{\tilde{b}\tilde{a}}$ on $\mathfrak{so}(3,2)$. The latter is defined according to $\Gamma_{\tilde{a}\tilde{b}}:=f_{\tilde{a}\tilde{d}}{}^{\tilde{c}}f_{\tilde{b}\tilde{c}}{}^{\tilde{d}}$ and its non-zero components are given by 
\begin{align}
\Gamma_{M_{ab},M_{cd}}=-12\eta_{a[c}\eta_{d]b}~,\qquad \Gamma_{K_a,P_b}=-12\eta_{ab}~,\qquad \Gamma_{\mathbb{D},\mathbb{D}}=6~.
\end{align}
Using $\Gamma_{\tilde{a}\tilde{b}}$ we may define the structure constants with all indices downstairs $f_{\tilde{a}\tilde{b}\tilde{c}}:=f_{\tilde{a}\tilde{b}}{}^{\tilde{d}}\Gamma_{\tilde{d}\tilde{c}}$, which prove to be totally antisymmetric $f_{\tilde{a}\tilde{b}\tilde{c}}=f_{[\tilde{a}\tilde{b}\tilde{c}]}$.

Then the Chern-Simons action with the gauge algebra $\mf{so}(3,2)$ is \cite{vN} 
\begin{align}\label{CG}
S^{(d=3)}_{\text{CG}}= \frac{2}{3}
 \int\Sigma_{\text{CS}}~,\qquad \Sigma_{\text{CS}}:=\mathcal{R}^{\tilde{b}}\wedge B^{\tilde{a}}\Gamma_{\tilde{a}\tilde{b}}+\frac{1}{6}B^{\tilde{c}}\wedge B^{\tilde{b}}\wedge B^{\tilde{a}}f_{\tilde{a}\tilde{b}\tilde{c}}~.
\end{align} 
Under an infinitesimal $\mathcal{H}$-transformation, the three form proves to vary by an exact form,
\begin{align}\label{exact}
\delta^{(\mc{H})}_{\L}\Sigma_{\text{CS}}=\text{d}\big(\text{d} B^{\tilde{b}}\Lambda^{\tilde{a}}\Gamma_{\tilde{a}\tilde{b}}\big)~,\qquad \Lambda^{\tilde{a}}=\big(0,~\Lambda^{\un{a}}\big)~,
\end{align} 
It follows immediately that \eqref{CG} is invariant under the gauge group $\mc{G}$. To derive \eqref{exact}, the following transformation rules are useful
\begin{align}
\delta_{\L}^{(\mc{H})}B^{\tilde{a}}=\text{d}\L^{\tilde{a}}+B^{\tilde{c}}\L^{\tilde{b}}f_{\tilde{b}\tilde{c}}{}^{\tilde{a}}~,\qquad \delta_{\L}^{(\mc{H})}\mc{R}^{\tilde{a}}=\mc{R}^{\tilde{c}}\L^{\tilde{b}}f_{\tilde{b}\tilde{c}}{}^{\tilde{a}}~,
\end{align}
 as well as the Bianchi identities,
 \begin{align}
 0=\text{d}\mc{R}^{\tilde{a}}+\mc{R}^{\tilde{c}}\wedge B^{\tilde{b}}f_{\tilde{b}\tilde{c}}{}^{\tilde{a}}~.
 \end{align}
These are equivalent to \eqref{ConnHTrans}, \eqref{VielHTrans}, \eqref{2.11} and \eqref{BianchiCA} respectively. 
By making use of the relevant component expressions and the constraints \eqref{constraints}, the action \eqref{CG} may be show to be equal to
 \begin{align}\label{CG1}
 S^{(d=3)}_{\text{CG}}=
 \int\text{d}^3x\,e\,\ve^{abc}\bigg\{\hat{R}_{ab}{}^{fg}\hat{\o}_{cfg}
 -\frac{2}{3}\hat{\o}_{ad}{}^{e}\hat{\o}_{be}{}^{f}\hat{\o}_{cf}{}^{d}+8\mathfrak{f}_{ab}\mathfrak{b}_c\bigg\}~.
 \end{align}
 Since \eqref{CG} is inert under SCTs (i.e. it is primary) up to a total derivative, 
 the dependence on $\mathfrak{b}_a$ once again drops out,\footnote{This may be shown explicitly using the relations \eqref{spincon}, \eqref{2.999} and \eqref{2.96}.} and it simplifies to
 \begin{align}\label{CG2}
 S^{(d=3)}_{\text{CG}}=
 \int\text{d}^3x\,e\,\ve^{abc}\bigg\{R_{ab}{}^{fg}\o_{cfg}-\frac{2}{3}\o_{ad}{}^{e}\o_{be}{}^{f}\o_{cf}{}^{d}\bigg\}~.
 \end{align}
 This depends solely on the vielbein, which is itself a primary field, so that the special conformal symmetry trivialises. 

Equivalently, one may arrive at \eqref{CG2} from \eqref{CG1} by making use of the special conformal symmetry to impose the gauge \eqref{degauge}. As mentioned earlier, in this gauge the residual dilatation symmetry becomes the usual Weyl symmetry, and the actions \eqref{CG} and \eqref{CGA3} are seen to be equivalent.

\begin{subappendices}

\section{Two component spinor conventions} \label{AppSpinor}


In this appendix we summarise the conventions used throughout this thesis for the two component spinor formalism in three and four dimensions. 

\subsection{Three dimensions}

Our two-component spinor notation and conventions in three dimensions follow \cite{KLT-M11}.
In particular, the Minkowski metric is
$\eta_{ab}=\mbox{diag}(-1,1,1)$.
The spinor indices are  raised and lowered using
the $\rm SL(2,{\mathbb R})$ invariant tensors
\bea
\ve_{\a\b}=\left(\begin{array}{cc}0~&-1\\1~&0\end{array}\right)~,\qquad
\ve^{\a\b}=\left(\begin{array}{cc}0~&1\\-1~&0\end{array}\right)~,\qquad
\ve^{\a\g}\ve_{\g\b}=\d^\a_\b
\eea
by the standard rule:
\bea
\psi^{\a}=\ve^{\a\b}\psi_\b~, \qquad \psi_{\a}=\ve_{\a\b}\psi^\b~.
\label{A2}
\eea

We make use of real gamma-matrices,  $\g_a := \big( (\g_a)_\a{}^\b \big)$, 
which obey the algebra
\be
\gamma_a \gamma_b=\eta_{ab}{\mathbbm 1} + \varepsilon_{abc}
\gamma^c~,
\label{A3}
\ee
where the Levi-Civita tensor is normalised as
$\varepsilon^{012}=-\varepsilon_{012}=1$. The completeness
relation for the gamma-matrices reads
\be
(\gamma^a)_{\alpha\beta}(\gamma_a)^{\rho\sigma}
=-(\delta_\alpha^\rho\delta_\beta^\sigma
+\delta_\alpha^\sigma\delta_\beta^\rho)~.
\label{A4}
\ee
Here the symmetric matrices 
$(\gamma_a)^{\alpha\beta}$ and $(\gamma_a)_{\alpha\beta}$
are obtained from $(\g_a)_\a{}^{\b}$ by the rules (\ref{A2}).
Some useful relations involving $\g$-matrices are 
\begin{subequations}
\bea
\ve_{abc}(\g^b)_{\a\b}(\g^c)_{\g\d}&=&
\ve_{\g(\a}(\g_a)_{\b)\d}
+\ve_{\d(\a}(\g_a)_{\b)\g}
~,
\\
\text{tr}[\g_a\g_b\g_{c}\g_d]&=&
2\eta_{ab}\eta_{cd}
-2\eta_{ac}\eta_{db}
+2\eta_{ad}\eta_{bc}
~.
\eea
\end{subequations}
Given a three-vector $V_a$,
it  can be equivalently described by a symmetric second-rank spinor $V_{\a\b}$
defined as
\bea
V_{\a\b}:=(\g^a)_{\a\b}V_a=V_{\b\a}~,\qquad
V_a=-\hf(\g_a)^{\a\b}V_{\a\b}~.
\eea
In the $d=3$ case,  an
antisymmetric tensor $F_{ab}=-F_{ba}$ is Hodge-dual to a three-vector $F_a$, 
specifically
\bea
F_a=\hf\ve_{abc}F^{bc}~,\qquad
F_{ab}=-\ve_{abc}F^c~.
\label{hodge-1}
\eea
Then, the symmetric spinor $F_{\a\b} =F_{\b\a}$, which is associated with $F_a$, can 
equivalently be defined in terms of  $F_{ab}$: 
\bea
F_{\a\b}:=(\g^a)_{\a\b}F_a=\hf(\g^a)_{\a\b}\ve_{abc}F^{bc}
~.
\label{hodge-2}
\eea
These three algebraic objects, $F_a$, $F_{ab}$ and $F_{\a \b}$, 
are in one-to-one correspondence with each other, 
$F_a \leftrightarrow F_{ab} \leftrightarrow F_{\a\b}$.
The corresponding inner products are related to each other as follows:
\bea
-F^aG_a=
\hf F^{ab}G_{ab}=\hf F^{\a\b}G_{\a\b}
~.
\eea

The Lorentz generators with two vector indices ($M_{ab} =-M_{ba}$),  one vector index ($M_a$)
and two spinor indices ($M_{\a\b} =M_{\b\a}$) are related to each other by the rules:
$M_a=\hf \ve_{abc}M^{bc}$ and $M_{\a\b}=(\g^a)_{\a\b}M_a$.
These generators 
act on a vector $V_c$ 
and a spinor $\J_\g$ 
as follows:
\bea
M_{ab}V_c=2\eta_{c[a}V_{b]}~, ~~~~~~
M_{\a\b}\J_{\g}
=\ve_{\g(\a}\J_{\b)}~.
\label{generators}
\eea

\subsection{Four dimensions}\label{App4dSpinorconventions}

Our two-component spinor notation and conventions in four dimensions follow \cite{BK}. In particular the Minkowski metric is $\eta_{ab}=\text{diag}(-1,1,1,1)$. The spinor indices are raised and lowered using the $\sSL(2,\mb{C})$ invariant tensors
\begin{subequations}
\begin{align}
\ve_{\a\b}=\left(\begin{array}{cc}0~&-1\\1~&0\end{array}\right)~,\qquad
\ve^{\a\b}=\left(\begin{array}{cc}0~&1\\-1~&0\end{array}\right)~,\qquad
\ve^{\a\g}\ve_{\g\b}=\d^\a_\b~,\\
\ve_{\ad\bd}=\left(\begin{array}{cc}0~&-1\\1~&0\end{array}\right)~,\qquad
\ve^{\ad\bd}=\left(\begin{array}{cc}0~&1\\-1~&0\end{array}\right)~,\qquad
\ve^{\ad\gd}\ve_{\gd\bd}=\d^{\ad}_{\bd}~,
\end{align}
\end{subequations}
by the standard rules:
\begin{align}
\psi^{\a}=\ve^{\a\b}\psi_{\b}~,\qquad \psi_{\a}=\ve_{\a\b}\psi^{\b}~,\qquad \psi^{\ad}=\ve^{\ad\bd}\psi_{\bd}~,\qquad \psi_{\ad}=\ve_{\ad\bd}\psi^{\bd}~.
\end{align}

A four-vector $V_{a}$ is in one-to-one correspondence with the rank-(1,1) spinor $V_{\a\ad}$ via
\begin{align}
V_{\a\ad}=(\s^a)_{\a\ad}V_{a}~,\qquad V_{a}=-\frac{1}{2}(\tilde{\s}_a)^{\ad\a}V_{\a\ad}~.
\end{align}
Here $(\tilde{\s}_{a})^{\ad\a}:=\ve^{\a\b}\ve^{\ad\bd}(\s_a)_{\b\bd}$ and $\s_a=(\mathds{1},\vec{\s})$ where $\vec{\s}$ are the Pauli matrices.\footnote{If the spinor indices of $\s^a$ (or $\tilde{\s}^a$) are not indicated, then it is assumed that they are all downstairs (or upstairs): $\s^a\equiv (\s^a)_{\a\ad}$ and $\tilde{\s}^a\equiv (\tilde{\s}^a)^{\ad\a}$.} It is common to define the double-$\s$ matrices according to
\begin{align}
\big(\s^{ab}\big)_{\a}{}^{\b}:=-\frac{1}{4}\big(\s^a\tilde{\s}^b-\s^b\tilde{\s}^a\big)_{\a}{}^{\b}~,\qquad \big(\tilde{\s}^{ab}\big)^{\ad}{}_{\bd}:=-\frac{1}{4}\big(\tilde{\s}^a\s^b-\tilde{\s}^b\s^a\big)^{\ad}{}_{\bd}
\end{align}
In addition to the standard $\s$ identities, which may be found e.g. in \cite{BK}, one may derive the following useful identities
\begin{align}
(\s^a)_{\a\ad}(\s_{ab})_{\g\d}&=\ve_{\a(\g}(\s_b)_{\d)\ad}~,\qquad (\s^a)_{\a\ad}(\tilde{\s}_{ab})_{\gd\dd}=(\s_b)_{\a(\gd} \ve_{\dd)\ad}~,\non\\
(\s^{ab})_{\a\b}(\s_{ab})_{\g\d}=-2\ve_{\a(\g}&\ve_{\d)\b}~,\quad(\s^{ab})_{\a\b}(\tilde{\s}_{ab})_{\ad\bd}=0~, \quad (\tilde{\s}^{ab})_{\ad\bd}(\tilde{\s}_{ab})_{\gd\dd}=-2\ve_{\ad(\gd}\ve_{\dd)\bd}~,\non\\
\text{Tr}\big(\s^{ab}\s^{cd}\big)=-\eta^{c[a}&\eta^{b]d}-\frac{\ri}{2} \ve^{abcd}~,\qquad \text{Tr}\big(\tilde{\s}^{ab}\tilde{\s}^{cd}\big)=-\eta^{c[a}\eta^{b]d}+\frac{\ri}{2} \ve^{abcd}~,\non\\
\frac{1}{2}\ve^{abcd}&\s_{cd}=-\ri \s^{ab}~,\qquad  \frac{1}{2}\ve^{abcd}\tilde{\s}_{cd}=\ri\tilde{\s}^{ab}~,
\end{align}
where we have denoted $\text{Tr}\big(\s^{ab}\s^{cd}\big)=(\s^{ab})_{\a}{}^{\b}(\s^{cd})_{\b}{}^{\a}$ and $\text{Tr}\big(\tilde{\s}^{ab}\tilde{\s}^{cd}\big)=(\tilde{\s}^{ab})^{\ad}{}_{\bd}(\tilde{\s}^{cd})^{\bd}{}_{\ad}$.


The double-$\s$ matrices may be used to relate a real antisymmetric tensor to a pair of symmetric two-spinors related via complex conjugation. For example, the antisymmetric Lorentz generators $M_{ab}$ are related to the pair of symmetric spinors $\big(M_{\a\b},\bar{M}_{\ad\bd}\big)$ via
\begin{align}
M_{ab}=&~(\s_{ab})^{\a\b}M_{\a\b}-(\tilde{\s}_{ab})^{\ad\bd}\bar{M}_{\ad\bd}~,\non\\
M_{\a\b}=\frac{1}{2}(\s^{ab})_{\a\b}&M_{ab}~,\qquad \bar{M}_{\ad\bd}=-\frac{1}{2}(\tilde{\s}^{ab})_{\ad\bd}M_{ab}~.
\end{align}
The Lorentz generators $M_{\a\b} $ and ${\bar M}_{\ad \bd}$ 
act on two-component spinors as follows: 
\begin{subequations} 
\bea
M_{\a\b} \,\j_\g=
\hf(\ve_{\g\a}\j_{\b}+\ve_{\g\b}\j_{\a})
~,\quad&&\qquad M_{\a\b}\, {\bar \j}_{\gd}=0~,\\
{\bar M}_{\ad\bd} \,{\bar \j}_{\gd}=
\hf(\ve_{\gd\ad}{\bar \j}_{\bd}+\ve_{\gd\bd}{\bar \j}_{\ad})
~,\quad&&\qquad {\bar M}_{\ad\bd}\, \j_{\g}=0~.
\eea
\end{subequations} 
In this thesis we usually work with irreducible $\sSL(2,\mb{C})$ tensor fields which are totally symmetric in their undotted indices and, separately, in the dotted ones:
$\F_{\a(m)\ad(n)} = \F_{(\a_1 ... \a_m)(\ad_1 ... \ad_n)} = \F_{\a_1 ... \a_m\ad_1 ... \ad_n} $.
In this case, the following identities hold:
\begin{subequations}
	\bea
	M_{\a_1}{}^{\b}\F_{\b \a_2 ... \a_{m}\ad(n)} &=& - \hf (m+2)\F_{\a(m)\ad(n)}~,\\
	\bar{M}_{\ad_1}{}^{\bd}\F_{\a(m)\bd \ad_2 ... \ad_{n}} &=& - \hf (n+2)\F_{\a(m)\ad(n)}~, \\
	M^{\b\g}M_{\b\g}  \F_{\a(m)\ad(n)} &=& -\hf m(m+2) \F_{\a(m)\ad(n)}~, \\
	\bar{M}^{\bd\gd} \bar{M}_{\bd\gd} \F_{\a(m)\ad(n)} &=& - \hf n(n+2)\F_{\a(m)\ad(n)}~.
	\eea
\end{subequations}

\end{subappendices}


\chapter{CHS models in three dimensions} \label{Chapter3D}

In this chapter we elaborate on the models describing the gauge invariant dynamics of conformal higher-spin fields on various types of curved three-dimensional backgrounds. For every conformal gauge field $h_{\a(n)}$, with $n\geq 2$ an integer, we show that there corresponds a primary field strength $\mc{C}_{\a(n)}(h)$ known as the higher-spin Cotton tensor. Its properties are such that knowledge of $\mc{C}_{\a(n)}(h)$ amounts to full knowledge of the spin $s=\frac{n}{2}$ CHS action at the linearised level. Accordingly, they will be the main object of interest.

 The models presented are higher-spin generalisations of the linearised action for conformal gravity in three dimensions, which was discussed in section \ref{sectionCG3}.
It is widely believed that a gauge-invariant action for  conformal fields of spin $s> 2$ may be defined only if the background metric is a solution to the equation of motion for conformal gravity. 
In three dimensions, this means that the vielbein must have vanishing Cotton tensor.
Therefore, the most general type of background on which CHS fields can consistently propagate are conformally-flat ones. These are the precisely the types of backgrounds for which we are able to obtain unique closed-form expressions for $\mc{C}_{\a(n)}(h)$.
 
This chapter is based on the publications \cite{Topological, 3Dprojectors, Confgeo,CottonAdS} and is organised as follows. 
In section \ref{sec3DCHSgen} we give some general remarks regarding models for CHS fields on an arbitrary three dimensional background. In section \ref{sec3DCHSflat} we specialise to a flat Minkowski background and derive a novel representation of the higher-spin Cotton tensors in terms of the $3d$ spin-projection operators. The corresponding CHS actions are then used to generate topologically massive higher-spin gauge models.  In section \ref{sec3DCHSCF} we make use of  conformal space to obtain closed-form expressions for CHS models on arbitrary conformally-flat backgrounds. Extensions of these models to higher-depth CHS fields are also provided. In section \ref{sec3DCHSAdS} we derive an explicit expression for $\mc{C}_{\a(n)}(h)$ on an AdS$_3$ background which allows us to (i) demonstrate the factorisation of the CHS kinetic operator; and (ii) analyse the dynamics of (new) topologically massive higher-spin models in AdS$_3$. 
A summary of the results obtained is given in section \ref{sec3DCHSdis}.

\section{Conformal higher-spin models in $\mc{M}^3$ }  \label{sec3DCHSgen}


 
As alluded to earlier, the Cotton tensor $\mc{C}_{\a(n)}(h)$ of the conformal higher-spin gauge field $h_{\a(n)}$ is of central importance when constructing the gauge and conformally invariant quadratic action for $h_{\a(n)}$.
In this section we give a general discussion regarding the key features of $\mc{C}_{\a(n)}(h)$ which are common to all background spacetimes. The material in this section is based on our paper \cite{Topological}.

We first review some basic geometric facts. 
Given a three dimensional spacetime $\big(\cM^3,e_{a}{}^{m}\big)$, its geometry may be described in terms of  the torsion-free Lorentz covariant derivative \eqref{LCDeriv}
\begin{align}
\mc{D}_{a}
=e_a{}^{m}\pa_m-\frac{1}{2}\omega_{a}{}^{bc}M_{bc}
~, \qquad \big[ \cD_a , \cD_b \big] = -\hf R_{ab}{}^{cd} M_{cd}~. \label{VecDer}
\end{align}
The condition of vanishing torsion is invariant under Weyl transformations which, together with (covariantised) diffeomorphisms and local Lorentz rotations, constitute the gauge group of conformal gravity $\mc{G}$,
\begin{align}
\delta_{\L}^{(\mc{G})} e_{a}{}^{m} = \Big( -\mc{D}_{a}\xi^b+K_a{}^{b}+\s\delta_{a}{}^{b}\Big)e_b{}^{m}~.
\end{align}
In three dimensions the Weyl tensor vanishes exactly. This means that the Riemann curvature is completely determined by the Schouten tensor, 
\begin{align}
R_{abcd}=2\eta_{a[c}S_{d]b}-2\eta_{b[c}S_{d]a}~, \qquad S_{ab}=R_{ab}-\frac{1}{4}\eta_{ab}R~, \label{Riemann3D}
\end{align}
as can be seen from eq. \eqref{WeylT}. Consequently, the algebra \eqref{VecDer} 
reduces to 
\begin{align}
\big[\mc{D}_{a},\mc{D}_{b}\big]=2S^{c}{}_{[a}M_{b]c}~. \label{3dAlgebro}
\end{align}

As mentioned in section \ref{SecConfCurve}, in $d=3$ the role of the Weyl tensor is played by the Cotton tensor \eqref{CottonT}; spacetime is conformally flat if and only if the Cotton tensor  vanishes. The latter may be represented by its dual $C_{ab}$ \eqref{CottonTDual}, which is symmetric, traceless and covariantly conserved.
We recall that in two component spinor notation,  $C_{ab}$ and the symmetric traceless part of the Ricci tensor are in one-to-one correspondence with the following completely symmetric rank-4 spinors:
\begin{subequations}
\begin{align}
 C_{\a\b\g\d}&:=\big(\gamma^a\big)_{\a\b}\big(\gamma^b\big)_{\g\d}C_{ab}  =C_{(\a\b\g\d)}~,\\
 R_{\a\b\g\d}&:=\big(\gamma^a\big)_{\a\b}\big(\gamma^b\big)_{\g\d}\Big(R_{ab}-\frac{1}{3}\eta_{ab}R\Big)=R_{(\a\b\g\d)}~. 
 \end{align}
 \end{subequations}
 They prove to be related via the useful identity
 \begin{align}
 C_{\a\b\g\d}=\mc{D}^\r{}_{(\a} R_{\b\g \d) \r}~. \label{CottonRicci}
\end{align}
From \eqref{WeylRDesc}, one may show that the Weyl transformation of $R_{\a\b\g\d}$ is 
\begin{align}
\d^{(\text{weyl})}_\s R_{\a\b\g\d} = 2\s R_{\a\b\g\d} -\mc{D}_{(\a\b} \mc{D}_{\g\d)} \s~.\label{RspinorWeyl}
\end{align}
Using this in conjunction with the spinor version of \eqref{LCDWeyl},
\begin{align}
\delta_{\s}^{(\text{weyl})}\mc{D}_{\a\b}=\s\mc{D}_{\a\b}+\big(\mc{D}^{\g}{}_{(\a}\s \big)M_{\b)\g}~,
\end{align}
allows one to show that $C_{\a(4)}$ is a primary tensor field with Weyl weight $+3$,
\begin{align}
\d^{(\text{weyl})}_\s C_{\a(4)} = 3\s C_{\a(4)}~.
\end{align}

Finally, we note that the algebra \eqref{3dAlgebro} is equivalent to
\begin{align}
\big[ \mc{D}_{\a(2)},\mc{D}_{\b(2)}\big]=R^{\g}{}_{\b(2)\a}M_{\a\g}-R^{\g}{}_{\a(2)\b}M_{\b\g}+\frac{1}{3}R\ve_{\b\a}M_{\a\b}~,
\end{align}
where we are employing convention \eqref{SymCon1}. 

\subsection{Conformal higher-spin gauge fields} \label{secCHS3prep}

Let us fix a positive integer $n \geq 2$.
In two-component spinor notation, 
a real tensor field $h_{\a(n) } = h_{\a_1 \dots \a_n } 
=h_{(\a_1 \dots \a_n)} $  is said to be a conformal spin-$\frac n2$ gauge field on $\mc{M}^3$ if: 
\begin{enumerate}[label=(\roman*)]
\item   $h_{\a(n) }$ is a primary tensor field with conformal weight $\Delta_{h_{(n)}}=(2-\frac{n}{2})$, 
\bea
\d^{(\text{weyl})}_\s h_{\a(n)} = \Big(2-\frac{n}{2}\Big)\s h_{\a(n)}~. \label{4777577398.4}
\eea
\item  $h_{\a(n) }$  is defined modulo gauge transformations of the form
\begin{align}
\d_{\xi} h_{\a(n) } =\mc{D}_{(\a_1 \a_2 } \xi_{\a_3 \dots \a_n) }~, \label{GTidc}
\end{align}
where the real gauge parameter $\xi_{\a(n-2)}$ is a weight 
$\Delta_{\x_{(n-2)}}=(1-\frac{n}{2})$ primary field.
\end{enumerate}
The consistency of these two properties is what uniquely fixes $\Delta_{h_{(n)}}$.

By virtue of property (i), under $\mc{G}$ the prepotential transforms according to the rule
\begin{align}
\d_{\L}^{(\mc{G})} h_{\a(n)} =\Big( \xi^a\mc{D}_a +\frac{1}{2}K^{ab}M_{ab}+\Delta_{h_{(n)}}\s \Big) h_{\a(n)} ~. \label{transrule6676}
\end{align}
Let us consider the background spacetime $\big(\cM^3,e_{a}{}^{m}\big)$  to be fixed, and suppose that it admits a conformal Killing vector field $\z = \z^m \pa_m = \z^a e_a$, 
\bea
 \cD^a \z^b + \cD^b \z^a = 2 \eta^{ab} \s[\z] ~, \qquad \s[\z] = \frac{1}{3} \cD_b \z^b  ~.
\label{CCVF1.3}
 \eea
 Then, in accordance with the discussion in section \ref{SectionHobbit}, the primary tensor field $h_{\a(n)}$  possesses the following conformal transformation law
\begin{align}
\d^{(\mc{G})}_{\L[\z]} h_{\a(n)} = \Big(\z^b \cD_b +\hf K^{bc}[\z] M_{bc} + \Delta_{h_{(n)}} \s[\z]\Big) h_{\a(n)} ~, \qquad 
K^{bc} [\z] &=  \cD^{[b} \z^{c]} ~. \label{ConfT3D}
\end{align}
In the case of a Minkowski background,  
$\big(\mc{M}^3,e_a{}^{m}\big)=\big(\mb{M}^3,\delta_a{}^{m}\big)$,
 \eqref{ConfT3D} reduces to the rigid conformal transformation rule \eqref{CHSprepPrimary} upon 
 substituting \eqref{CKVflat} for $\z^a$. 


\subsection{Linearised higher-spin Cotton tensors} \label{secLinCotVA}

Starting  with $h_{\a(n) } $ one can construct its descendant, denoted by
$\mc{C}_{\a(n)}(h)$, known as the linearised higher-spin Cotton tensor. It is 
defined 
by the following the properties: 
\begin{enumerate}[label=(\roman*)]

\item
$\mc{C}_{\a(n)}(h)$ is a primary tensor field with conformal weight $\Delta_{\mc{C}_{(n)}}=(1+\frac{n}{2})$, 
\begin{align}
\d_\s^{(\text{weyl})} \mc{C}_{\a(n)}(h) = \Big(1+\frac{n}{2} \Big)  \s \mc{C}_{\a(n)}(h)~.
\label{2.13}
\end{align}

\item  
$\mc{C}_{\a(n)}(h)$ is of the form $\mc{C}_{\a(n)}(h)=\cA h_{\a(n)}$, 
where $\cA$ is a linear differential operator of order $(n-1)$ involving 
the Lorentz covariant derivative $\mc{D}_a$, the curvature tensors
 $R_{\a(4)}$ and $R$, 
and their covariant derivatives.

\item
$\mc{C}_{\a(n)}(h)$ has vanishing gauge variation under \eqref{GTidc} if spacetime is conformally flat,
\begin{align}
\d_{\xi} \mc{C}_{\a(n)} (h)= \mc{O}\big( C\big)~.
\label{3188}
\end{align}

\item
 $\mc{C}_{\a(n)}(h)$ is divergenceless if spacetime is conformally flat,
\begin{align}
\mc{D}^{\b(2)} \mc{C}_{\b(2)\a(n-2)}(h) = \mc{O}\big( C\big)~.
\end{align} 
Here and in \eqref{3188}, 
$\mc{O}\big( C\big)$ stands for contributions involving the Cotton tensor and
its covariant derivatives.
\end{enumerate}


We now consider several examples on an arbitrary background. 
Given a conformal spin-1 gauge field
$h_{\a\b } = h_{\b\a}$, 
\bea
\d^{(\text{weyl})}_\s h_{\a\b} = \s h_{\a\b} ~,
\eea
the required Weyl primary descendant is 
\begin{align}
\mc{C}_{\a\b} (h)= \mc{D}^\g{}_{(\a} h_{\b)\g}~. \label{MaxwellianFS}
\end{align}
This corresponds to the Maxwell field strength, 
$\mc{C}_{ab}(h) = \mc{D}_a h_b - \mc{D}_b h_a$,
of the gauge potential $h_a$. The latter is both gauge invariant and conserved 
\bea
\delta_{\xi}\mc{C}_{\a(2)}(h)=0~,\qquad \mc{D}^{\b(2)} \mc{C}_{\b(2)}(h)=0~.
\eea

Next consider a conformal spin-$\frac 32$ gauge field $h_{\a(3)}$
(i.e. conformal gravitino),
\bea
\d^{(\text{weyl})}_\s h_{\a(3)} = \hf \s h_{\a(3)}~.
\eea
The required Weyl primary descendant is 
\bea
\mc{C}_{\a(3)}(h)
=\frac{3}{4}\mc{D}_{(\a_1}{}^{\b_1} \mc{D}_{\a_2}{}^{\b_2} 
h_{\a_3)\b(2)}
+\frac{1}{4}\Box h_{\a(3)}
-\frac{3}{4}R^{\b(2)}{}_{(\a_1\a_2}h_{\a_3)\b(2)}
+\frac{1}{16}R h_{\a(3)}~. \label{linCgravitinoM3}
\eea
Its  gauge transformation is 
\bea
\d_{\xi} \mc{C}_{\a(3)}(h)=-\frac{1}{2}C_{\a(3)}{}^{\b}\xi_{\b}~.
\eea
Computing its divergence gives
\bea
\mc{D}^{\b(2)} \mc{C}_{\b(2)\a}(h)=-\frac{1}{2}C_{\a}{}^{\b(3)}h_{\b(3)}~.
\eea

Our last example is a conformal spin-2 gauge field $h_{\a(4)}$
(i.e. conformal graviton),
\bea
\d^{(\text{weyl})}_{\sigma} h_{\a(4)}=0~.
\eea
The required Weyl primary descendant of $h_{\a(4)}$ is
\begin{align}
\mc{C}_{\a(4)}(h)&
=\frac{1}{2}\mc{D}_{(\a_1}{}^{\b_1}\mc{D}_{\a_2}{}^{\b_2}
\mc{D}_{\a_3}{}^{\b_3} h_{\a_4)\b(3)}
+\frac{1}{2}\Box\mc{D}_{(\a_1}{}^{\b} h_{\a_2\a_3\a_4)\b}
-\big(\mc{D}_{(\a_1}{}^{\b_1}R_{\a_2\a_3}{}^{\b_2\b_3}\big) 
h_{\a_4)\b(3)}\notag\\
&-\frac{1}{12}\big(\mc{D}_{(\a_1}{}^{\b}R\big)h_{\a_2\a_3\a_4)\b}
+\frac{1}{12}R\mc{D}_{(\a_1}{}^{\b}h_{\a_2\a_3\a_4)\b}-2R^{\b_1\b_2}{}_{(\a_1\a_2}\mc{D}_{\a_3}{}^{\b_3}h_{\a_4)\b(3)}\notag\\
&+\frac{3}{4}R^{\b_1}{}_{\d(\a_1\a_2}\mc{D}^{\d\b_2}h_{\a_3\a_4)\b(2)}~.\label{NotQuiteLinCot}
\end{align}
Its gauge transformation is
\begin{align}
\d_{\xi} \mc{C}_{\a(4)}(h)=&
-\big(\mc{D}^{\g\d}C_{\g(\a_1\a_2\a_3}\big)\xi_{\a_4)\delta}
-\frac{1}{2}\big(\mc{D}_{(\a_1\a_2}C_{\a_3\a_4)}{}^{\b(2)}\big)\xi_{\b(2)}+C_{\g_1(\a_1\a_2\a_3}\mc{D}^{\g(2)}\xi_{\a_4)\g_2}\notag\\
&-\frac{11}{12}C_{\a(4)}\mc{D}^{\b(2)}\xi_{\b(2)}
-\frac{1}{2}C^{\b}{}_{\g(\a_1\a_2}\mc{D}_{\a_3}{}^{\g}\xi_{\a_4)\b}~.
\end{align}
The divergence of $\mc{C}_{\a(4)}(h)$ may be shown to be
\begin{align}
\mc{D}^{\b(2)} \mc{C}_{\b(2)\a(2)}(h)&=
\frac{1}{2}\big(\mc{D}_{\g(\a_1}C^{\g\b(3)}\big) h_{\a_2)\b(3)}
-\frac{5}{12}\big(\mc{D}_{\a(2)}C^{\b(4)}\big) h_{\b(4)}
+\frac{1}{12}C^{\b(4)}\mc{D}_{\a(2)}h_{\b(4)}\notag\\
&+\frac{3}{2}C_{\g_1(\a_1}{}^{\b(2)}\mc{D}^{\g(2)}h_{\a_2)\g_2\b(2)}
-C_{\a(2)}{}^{\b(2)}  \mc{D}^{\g(2)} h_{\b(2)\g(2)}~.
\end{align}

Notice that \eqref{NotQuiteLinCot} coincides with \eqref{lincot} modulo the term involving the background Cotton tensor (and an overall normalisation). This is because the above properties do not determine the higher-spin Cotton tensor  uniquely on a non-conformally-flat background.

\subsection{Conformal higher-spin action }
Suppose that the spacetime under consideration is conformally flat, 
\bea
C_{\a(4)}=0~.
\label{3.30}
\eea
Then the tensor $\mc{C}_{\a(n)}(h)$ 
is gauge invariant and covariantly conserved,\footnote{It follows from the Weyl transformation law \eqref{2.13} that 
$\mc{D}^{\b(2)} \mc{C}_{\b(2)\a(n-2)}(h)$ is a primary field,
$
\d^{(\text{weyl})}_\s \big( \mc{D}^{\b(2)} \mc{C}_{\b(2)\a(n-2)}(h) \big)
= \big(2+\frac{n}{2} \big)  \s 
\mc{D}^{\b(2)} \mc{C}_{\b (2) \a(n-2)}(h).
\label{2.26}
$
This property means that the conservation equation \eqref{CotPropCFMb} is Weyl invariant.}
\begin{subequations} \label{CotPropCFM}
\bea
\d_\xi \mc{C}_{\a(n)}(h)&=&0~,  \label{CotPropCFMa}\\ 
\mc{D}^{\b(2)} \mc{C}_{\b(2)\a(n-2)}(h) &=&0~. \label{CotPropCFMb}
\eea
\end{subequations}
and proves to be unique. These properties, and the Weyl transformation law \eqref{2.13}, tell us 
that the following action 
\bea
S_{\rm{CHS}}^{(n)} [ h] 
=\frac{\text{i}^n}{2^{\left \lfloor{n/2}\right \rfloor +1}
} \int \rd^3 x \, e\, h^{\a(n)} 
\mc{C}_{\a(n) } (h)~,
\label{CHSActionM3}
\eea
 is gauge and Weyl invariant, 
\bea
\d_\xi S_{\rm{CHS}}^{(n)} [h] =0~, \qquad 
\d^{(\text{weyl})}_\s S_{\rm{CHS}}^{(n)} [ h ] =0~.
\eea
Here $\left \lfloor{x}\right \rfloor$ denotes the floor  function; it
coincides with the integer part of a real number $x\geq 0$. This is the conformal higher-spin action in three dimensions.
Condition \eqref{3.30} is required to guarantee the gauge invariance 
of $S_{\rm{CHS}}^{(n)} [h] $ for $n>2$. 
The above action is actually Weyl invariant in an arbitrary curved space. 
By virtue of the invariance of \eqref{CHSActionM3} under the spacetime gauge group $\mc{G}$, it follows that it is invariant under the conformal transformations \eqref{ConfT3D}, 
 \begin{align}
 \d^{(\mc{G})}_{\L[\z]}S_{\rm{CHS}}^{(n)} [ h] =0~,
 \end{align}
where the background is held fixed, $\d^{(\mc{G})}_{\L[\z]} e_{a}{}^{m}=0$.

\section{Conformal higher-spin models in $\mb{M}^3$} \label{sec3DCHSflat}
The expressions \eqref{MaxwellianFS}, \eqref{linCgravitinoM3}  and \eqref{NotQuiteLinCot} for the $n=2,3,4$ linearised Cotton tensors were derived by considering all possible structures with the correct Weyl weight and imposing invariance under Weyl transformations. Though straightforward, this method is very tedious for $n>4$, even if the background is conformally-flat.  This is because as $n$ increases, so too does the number of covariant derivatives contained in $\mc{C}_{\a(n)}(h)$, and consequently the number of possible curvature dependent terms. 


In section \ref{sec3DCHSCF} we will utilize the power of conformal space to side-step these issues. 
Of central importance to the approach advocated there are the models for conformal higher-spin fields in Minkowski space $\mb{M}^3$. It is therefore pertinent to recall the main features of these models. 
Our presentation follows the spirit of the original papers \cite{FT, Segal} (see section \ref{secFTCHSMd}), where a key role is played by the transverse-traceless projectors and the spin-$\frac{n}{2}$ CHS action takes the schematic form \eqref{DdimCHSA}
\begin{align}
S_{\text{CHS}}^{(n)}[h]\propto \int \text{d}^3x h^{\a(n)}\pa^{(n-1)}\Pi_{\perp}^{(n)}h_{\a(n)}~.\label{CHS3Aproj}
\end{align}
 This allows us to provide a novel reformulation of the higher-spin Cotton tensors in terms of the $d=3$ spin projection operators $\Pi_{\perp}^{(n)}$.
From this point of view, the defining features of the higher-spin Cotton tensor, which are otherwise obscured, are made manifest. This section is based on our papers \cite{Topological, 3Dprojectors}.

\subsection{On-shell massive fields in $\mb{M}^3$} \label{SectionIrrepPoin3}

We start by discussing tensor fields realising irreducible massive 
representations of the Poincar\'e group in three dimensions \cite{Binegar}. We restrict our attention to the case of integer and half-integer spin values; for a discussion of the anyon representations see, e.g., \cite{JN}.


For $n>1$, an on-shell  field $\Phi_{\a (n)}(x) $ of mass $m$
is subject to satisfy the following differential equations 
 \cite{GKL} (see also \cite{TV,BHT}):
\begin{subequations}\label{1.1.12}
\bea
\pa^{\b \g} \Phi_{\b \g \a (n-2)}&=&0\,, 
\label{1.1.1}
\\
\pa^\b{} _{(\a_1} 
\Phi_{\a_2 \dots \a_n) \b} &=& m \s \Phi_{\a (n)}\,, \qquad \s =\pm 1 \,. 
\label{1.1.2}
\eea
\end{subequations}
In the spinor case, $n=1$, eq.  \eqref{1.1.1} is absent, and it is the Dirac equation 
\eqref{1.1.2} which defines a massive field. 
The constraints \eqref{1.1.1} and  \eqref{1.1.2}
imply the mass-shell equation\bea
(\Box -m^2 ) \F_{\a (n)} =0~.
\label{mass-shell}
\eea
Eqs. \eqref{1.1.1} and 
\eqref{mass-shell} prove to be equivalent to the $3d$ Fierz-Pauli field equations \cite{FP}.

Let $P_a$ and $J_{ab}= -J_{ba}$ be the generators of the $3d$ Poincar\'e group.
The Pauli-Lubanski pseudo-scalar 
\bea
\mb{W}:= \hf \ve^{abc}P_a J_{bc} = -\hf P^{\a\b} J_{\a\b}
\label{PauliL}
\eea
commutes with the generators $P_a$ and $J_{ab}$.
Irreducible unitary representations of the Poincar\'e group 
are labelled by two parameters,  mass $m$ and helicity $\l$, 
which are associated with the Casimir operators, 
\bea
 P^a P_a = -m^2 {\mathbbm 1} ~, \qquad \mb{W}=m \l {\mathbbm 1}~.
 \label{Casimirs}
 \eea
 The parameter $|\l|$ is identified with spin. 
 
In the case of field representations, we have
\bea
\mb{W}= \hf \pa^{\ab} M_{\a\b}~, \label{PLscalar}
\eea
where the action of $M_{\a\b}=M_{\b\a}$ on a field 
$\F_{\g(n)}$ is defined by 
\bea
M_{\a\b} \F_{\g_1 \cdots \g_n} = \sum_{i=1}^n
\ve_{\g_i (\a} \F_{\b) \g_1 \cdots \widehat{\g_i} \dots\g_n}~,
\eea
where the hatted index of $\F_{\b \g_1 \cdots \widehat{\g_i} \dots\g_n}$  is omitted.
It follows from \eqref{1.1.2} and the second relation in \eqref{Casimirs} that 
the helicity of the on-shell massive field $\F_{\a(n)}$ is 
\bea
\l = \frac{n}{2} \s~.
\eea 

\subsection{Spin-projection operators} \label{SectionSPO3}

Having described the irreducible tensor fields carrying definite helicity, we can now construct the operators which project onto these states.  To this end, we introduce the off-shell projection operators
\be 
\P^{(\pm)}_{\ \a}{}^{\b} :=\frac{1}{2} \Big( \d_{\a}{}^{ \b} \pm \frac{1}{\sqrt{\Box}} \pa_{\a}{}^{ \b}\Big)\,. 
\label{2.12}
\ee
and their higher-rank extensions 
\begin{subequations} \label{3.13}
\bea
&&
 \Pi^{(+n)}_{\ \a(n)}{}^{\b(n)} :=  \Pi^{(+)}_{\ (\a_1}{}^{\b_1} \cdots 
\Pi^{(+)}_{\ \a_n)}{}^{\b_n}\,, 
 \\
&&
\Pi^{(-n)}_{\ \a(n)}{}^{\b(n)} :=  \Pi^{(-)}_{\ (\a_1}{}^{\b_1} \cdots \Pi^{(-)}_{\ \a_n)}{}^{\b_n}\,. 
\eea
\end{subequations}
Given an off-shell field $h_{\a(n)}$ (which need not be a CHS field), the action of $ \Pi^{(\pm n) } $ on $h_{\a(n)}$ 
is defined by 
\begin{subequations} \label{3.14}
\bea
\Pi^{(+n)} h_{\a(n)} &:=&  \P^{(+)}_{\ \a_1}{}^{\b_1} \cdots  \P^{(+)}_{\ \a_n}{}^{\b_n}
h_{\b_1 \dots \b_n}  \equiv h^{(+)}_{\a(n)}~, \\
 \Pi^{(-n)} h_{\a(n)} &:=& \P^{(-)}_{\ \a_1}{}^{\b_1} \cdots  \P^{(-)}_{\ \a_n}{}^{\b_n}
h_{\b_1 \dots \b_n}  \equiv h^{(-)}_{\a(n)}~.
\eea
\end{subequations}
Expanding out each factor, it is not difficult to show that the following relation holds
 \bea
h^{(\pm)}_{\a(n)} &=&\frac{1}{2^n}\sum_{j=0}^{n}\binom{n}{j}
\frac{(\pm 1)^j}{\Box^{j/2}}\partial_{(\a_1}{}^{\b_1}\dots\partial_{\a_j}{}^{\b_j}
h_{\a_{j+1}\dots\a_n)\b_1\dots\b_j}~. \label{HelicityPrepExp}
 \eea

The operators $ \Pi^{(+ n) } $  and $\Pi^{(-n) }$ are orthogonal projectors since 
\begin{subequations}
\begin{align}
 \Pi^{(+n)}  \Pi^{(+n)} = \Pi^{(+n)} ~, \qquad &
 \Pi^{(-n)}  \Pi^{(-n)} = \Pi^{(-n)} ~, \\
 \Pi^{(+n)}  \Pi^{(-n)} &=0~.
\end{align}
\end{subequations}
Furthermore, one may also check that the following relations hold
\bea
\pa^{\a_1 \a_2}  \P^{(\pm)}_{\ \a_1}{}^{\b_1}  \P^{(\pm)}_{\ \a_2}{}^{\b_2} =0~,
\qquad
\P^{(\pm)}_{~ \a_1}{}^{\b_1}  \P^{(\pm)}_{\ \a_2}{}^{\b_2} \pa_{\b_1 \b_2}=0~.
\label{3.16}
\eea
 The first identity in \eqref{3.16}  implies that the field $ h^{(\pm)}_{\a(n)}$ is transverse, 
\begin{align}
\partial^{\b \g}h^{(\pm)}_{\b \g \a(n-2)}=0~. \label{2.13.5}
\end{align}
The second identity implies that $h^{(\pm )}_{\a(n)}$ is invariant under 
the higher-spin gauge transformations, 
\bea
\d_{\xi} h_{\a(n)} = \pa_{(\a_1\a_2} \xi_{\a_3 \dots \a_n)} \qquad \implies \qquad \delta_{\xi}h^{(\pm )}_{\a(n)}=0~. \label{2.13.6}
\eea
In addition to these, one may show that $h^{(\pm )}_{\a(n)}$ satisfies the identity
\begin{align}\label{2.12.5}
\partial^{\b}{}_{(\a_1}h^{(\pm )}_{\a_2\dots\a_n)\b}=\pm\sqrt{\Box}h^{(\pm )}_{\a(n)}~.
\end{align}

As a consequence of the above analysis, it follows that if $h_{\a(n)}$ satisfies the Klein-Gordon equation \eqref{mass-shell}, 
\begin{align}
\big(\Box-m^2\big)h_{\a(n)}=0~, \label{CHSKGE}
\end{align}
then $h^{(\pm)}_{\a(n)}=\Pi^{(\pm n) } h_{\a(n)}$ is a solution of the equations \eqref{1.1.1} and \eqref{1.1.2},
\begin{align}
\pa^{\b\g}h^{(\pm)}_{\b\g\a(n-2)}=0~,\qquad \partial^{\b}{}_{(\a_1}h^{(\pm)}_{\a_2\dots\a_n)\b}=\pm m h^{(\pm)}_{\a(n)}~.
\end{align} Therefore, we see that the operator $\Pi^{(\pm n) }$ projects onto the space of transverse fields carrying definite helicity $\pm s$. The operators $\Pi^{(\pm n)}$ contain terms involving the non-local operator  ${\Box}^{-1/2}$, which requires a special definition. Using either of these projectors in \eqref{CHS3Aproj} would lead to a gauge invariant, but inevitably non-local action.

However, let us define the following combinations
\begin{subequations}\label{3DMVPProj}
\begin{align}
\Pi_{\perp}^{[n]}&:=\Pi^{(+n)}+\Pi^{(-n)}~,\label{3DTTProj}\\
\Pi_{\parallel}^{[n]}&:=\mathds{1}-\Pi_{\perp}^{(n)}~. \label{3DLongProj}
\end{align}
\end{subequations}
They are well defined as all terms involving odd powers of ${\Box}^{-1/2}$ cancel out and consequently they contain only inverse powers of $\Box$.  From the properties of their constituents, it is clear that $\Pi_{\perp}^{[n]}$ and $\Pi_{\parallel}^{[n]}$ are orthogonal projectors
\begin{subequations}
\begin{align}
\Pi_{\perp}^{[n]}  \Pi_{\perp}^{[n]}=\Pi_{\perp}^{[n]} ~, \qquad &
 \Pi_{\parallel}^{[n]}  \Pi_{\parallel}^{[n]}= \Pi_{\parallel}^{[n]} ~, \\
\Pi_{\perp}^{[n]}  \Pi_{\parallel}^{[n]} &= 0~,
\end{align}
\end{subequations}
and that they resolve the identity, $\mathds{1}=\Pi_{\perp}^{[n]}+ \Pi_{\parallel}^{[n]}$. An important observation is that the map
$h_{\a (n)} \to  \Pi_{\perp}^{[n]} h_{\a (n)}$ projects the space of  
symmetric fields $h_{\a(n)}$
onto the space of divergence-free fields, 
\begin{align}
h^{\perp}_{\a(n)}:= \Pi_{\perp}^{[n]} h_{\a (n)}~, \qquad \pa^{\b(2)}h^{\perp}_{\b(2)\a(n-2)}=0~, \label{3DTTCHSflat}
\end{align}
in accordance with \eqref{2.13.5}. It may also be shown that $ \Pi_{\perp}^{[n]} $ acts as the identity on the space of transverse fields $\psi_{\a(n)}$
\begin{align}
\pa^{\b(2)}\psi_{\b(2)\a(n-2)}=0~\quad \implies \quad  \Pi_{\perp}^{[n]}\psi_{\a(n)}=\psi_{\a(n)}~.
\end{align}
Therefore, the projectors $ \Pi_{\perp}^{[n]} $  satisfy the properties \eqref{TTProjPropDDim} and can be considered as $3d$ analogues of the Behrends-Fronsdal projection operators \cite{BF,Fronsdal58}.\footnote{The helicity projectors $\Pi^{(\pm n)}$ given by eq. \eqref{3.13} do not satisfy the surjectivity property in \eqref{TTProjPropDDim}.} As a consequence of the above analysis, it follows that if $h_{\a(n)}$ satisfies the Klein-Gordon equation \eqref{CHSKGE}, 
then $ \Pi_{\perp}^{[n] }h_{\a(n)} $ is a solution of 
the 3$d$ Fierz-Pauli field equations \eqref{1.1.1} and \eqref{mass-shell} (it contains both helicity modes).

Furthermore, it may be shown that $\Pi_{\parallel}^{[n]}$ projects $h_{\a(n)}$ onto its longitudinal (pure gauge) component
\begin{align}
h_{\a(n)}^{\parallel}:=\Pi_{\parallel}^{[n]}h_{\a(n)}~,\qquad h_{\a(n)}^{\parallel}=\pa_{(\a_1\a_2}h_{\a_3\dots\a_n)}~,
\end{align}
for some unconstrained totally symmetric field $h_{\a(n-2)}$. It follows that any off-shell rank-$n$ field $h_{\a(n)}$ may be decomposed according to (we make use of the convention \eqref{SymCon1})
\begin{subequations} \label{3dDecompFlat1}
\begin{align}
h_{\a(2s)}&= h^{\perp}_{\a(2s)}+\pa_{\a(2)}h^{\perp}_{\a(2s-2)}+\pa_{\a(2)}\pa_{\a(2)}h^{\perp}_{\a(2s-4)}+\cdots + \underbrace{\pa_{\a(2)}\cdots\pa_{\a(2)}}_{s-\text{times}}h^{\perp}~,\\
h_{\a(2s+1)}&= h^{\perp}_{\a(2s+1)}+\pa_{\a(2)}h^{\perp}_{\a(2s-1)}+\pa_{\a(2)}\pa_{\a(2)}h^{\perp}_{\a(2s-3)}+\cdots + \underbrace{\pa_{\a(2)}\cdots\pa_{\a(2)}}_{s-\text{times}}h^{\perp}_{\a}~,
\end{align}
\end{subequations}
for integer and half-integer spin respectively. This may be written more compactly as 
\begin{align}
h_{\a(n)}=\sum_{t=0}^{\lfloor n/2 \rfloor}\big(\pa_{\a(2)}\big)^{t} h^{\perp}_{\a(n-2t)}~. \label{3dDecompFlat2}
\end{align}
In eqs. \eqref{3dDecompFlat1} and \eqref{3dDecompFlat2}, each of the fields $ h^{\perp}_{\a(n-2t)}$ with $0 \leq t \leq \lfloor n/2 \rfloor -1$ are transverse,
while both $ h^{\perp}_{\a}$ and $ h^{\perp}$ are unconstrained. Using the identity \eqref{3DTTProj}, one can take this decomposition a step further and split each term into positive and negative helicity components, each of which are transverse:
\begin{align}
h_{\a(n)}=\sum_{t=0}^{\lfloor n/2 \rfloor}\big(\pa_{\a(2)}\big)^{t} \big(h^{(+)}_{\a(n-2t)}+h^{(-)}_{\a(n-2t)}\big)~. \label{3dDecompFlat52}
\end{align}


We now give several examples of the spin projectors \eqref{3DTTCHSflat}:
 \begin{subequations} \label{3.21abc}
 \begin{align}
  \Pi_{\perp}^{[2] } h_{\a(2) } &= \frac{1}{2}\frac{1}{\Box}\bigg(\partial_{(\a_1}{}^{\b_1}\pa_{\a_2)}{}^{\b_2}h_{\b(2)}+\Box h_{\a(2)}\bigg)~,\\
  \Pi_{\perp}^{[3] } h_{\a(3) } &= \frac{1}{2^2}\frac{1}{\Box}\bigg(3\partial_{(\a_1}{}^{\b_1}\pa_{\a_2}{}^{\b_2}h_{\a_3)\b(2)}+\Box h_{\a(3)}\bigg)~, \\
  \Pi_{\perp}^{[4] } h_{\a(4) } &= \frac{1}{2^3}\frac{1}{\Box^2}\bigg(\pa_{(\a_1}{}^{\b_1}\cdots\pa_{\a_4)}{}^{\b_4}h_{\b(4)}+6\Box\pa_{(\a_1}{}^{\b_1}\pa_{\a_2}{}^{\b_2}h_{\a_3\a_4)\b(2)}+\Box^2 h_{\a(4)}\bigg)~,\\
  \Pi_{\perp}^{[5] } h_{\a(5) } &= \frac{1}{2^4}\frac{1}{\Box^2}\bigg(5\pa_{(\a_1}{}^{\b_1}\cdots\pa_{\a_4}{}^{\b_4}h_{\a_5)\b(4)}+10\Box\pa_{(\a_1}{}^{\b_1}\pa_{\a_2}{}^{\b_2}h_{\a3\a_4\a_5)\b(2)}~~~~~~~~~~~~~~~~~~~~~~~~ \notag \\
 & \phantom{blank~~~~~~}+\Box^2h_{\a(5)}\bigg)~,\\
  \Pi_{\perp}^{[6] } h_{\a(6) } &= \frac{1}{2^5}\frac{1}{\Box^3}\bigg(\pa_{(\a_1}{}^{\b_1}\cdots\pa_{\a_6)}{}^{\b_6}h_{\b(6)}+15\Box\pa_{(\a_1}{}^{\b_1}\cdots \pa_{\a_4}{}^{\b_4}h_{\a_5\a_6)\b(4)} \notag \\
& \phantom{blank~~~~~~}+15\Box^2\pa_{(\a_1}{}^{\b_1}\pa_{\a_2}{}^{\b_2}h_{\a_3\dots\a_6)\b(2)}+\Box^3h_{\a(6)}  \bigg)~. 
 \end{align}
 \end{subequations} 
In vector notation, the examples \eqref{3.21abc} are equivalent to 
\begin{subequations}
\begin{align}
\Pi_{\perp}^{[2] } h_{a\phantom{b\a} }&=\frac{1}{\Box}\bigg(\Box h_a-\pa_a\pa^b h_b \bigg)~,\\
\Pi_{\perp}^{[3] } h_{a \g \phantom{b} }&=\frac{1}{\Box}\bigg( \Box h_{a\g}
-\pa_a\pa^bh_{b\g}-\frac{1}{2}\ve_{abc}(\g^b)_{\g}{}^{\d}\pa^c\pa^dh_{d\d}\bigg)~,\\
\Pi_{\perp}^{[4] } h_{ab\phantom{\a}}&=\frac{1}{\Box^2}\bigg(\Box^2 h_{ab}-2\Box\pa^c\pa_{(a}h_{b)c}+\frac{1}{2}\Box\eta_{ab}\pa^c\pa^dh_{cd}+\frac{1}{2}\pa_a\pa_b\pa^c\pa^dh_{cd}\bigg)~,\\
\Pi_{\perp}^{[5] } h_{ab\g}&=\frac{1}{\Box^2}\bigg(\Box^2h_{ab\g}-2\Box\pa^c\pa_{(a}h_{b)c\g}+\frac{1}{4}\Box\eta_{ab}\pa^c\pa^dh_{cd\g}+\frac{3}{4}\pa_a\pa_b\pa^c\pa^dh_{cd\g} \phantom{~~~~~~~~~~~~~~~~~~~~~~~~~~~}\notag\\
&\phantom{blank sp}-\frac{1}{2}(\g^c)_{\g}{}^{\d}\ve_{cd(a}\big[\Box\pa^d\pa^fh_{b)f\d}-\pa_{b)}\pa^d\pa^f\pa^gh_{fg\d}\big]\bigg)~,\\
\Pi_{\perp}^{[6] } h_{abc}&=\frac{1}{\Box^3}\bigg(\Box^3h_{abc}-3\Box^2 \pa^d\pa_{(a}h_{bc)d}+\frac{3}{4}\Box^2\pa^d\pa^f\eta_{(ab}h_{c)df}+\frac{9}{4}\Box \pa^d\pa^f \pa_{(a}\pa_bh_{c)df}\notag \\
&\phantom{blank sp}-\frac{3}{4}\Box\eta_{(ab}\pa_{c)}\pa^d\pa^f\pa^gh_{dfg}-\frac{1}{4}\pa_a\pa_b\pa_c\pa^d\pa^f\pa^g h_{dfg}\bigg)~.
\end{align}
\end{subequations}
One may check that they are each traceless and/or $\g$-traceless. 

Finally, we note that new realisations of the spin-projection operators $\Pi^{[n]}_{\perp}$ and the helicity projectors $\Pi^{(\pm n)}$ were obtained in \cite{AdS3(super)projectors}. The corresponding operators are expressed in terms of only the Casimir operators $\mb{W}$ (see eq. \eqref{PLscalar}) and $\Box$.

\subsection{Conformal higher-spin action } \label{SectionD3CotFlat}

With the rank-$n$ Behrends-Fronsdal projectors $\Pi^{[n]}_{\perp}$ at our disposal, we now turn to the construction of the $d=3$ gauge invariant action for the CHS prepotential $h_{\a(n)}$.

Using the expressions \eqref{HelicityPrepExp}, one may show that the explicit form of $\Pi_{\perp}^{[n] } h_{\a(n) }$ is
\begin{align}
\Pi_{\perp}^{[n] } h_{\a(n) }=\frac{1}{2^{n-1}}\Box^{-\lfloor n/2 \rfloor}\sum_{j=0}^{\lfloor n/2 \rfloor}&\binom{n}{2\lfloor n/2 \rfloor-2j}\Box^{j}\pa_{(\a_1}{}^{\b_1}\cdots\pa_{\a_{2\lfloor n/2 \rfloor-2j}}{}^{\b_{2\lfloor n/2 \rfloor-2j}}\non\\
&\times h_{\a_{2\lfloor n/2 \rfloor-2j+1}\dots\a_{n})\b(2\lfloor n/2 \rfloor-2j)}~. \label{UglyTTExplicit}
\end{align}
Here $\left \lceil{x}\right \rceil$ denotes the ceiling function; it returns the smallest integer which is greater than or equal to the real number $x\geq 0$. 
To arrive at \eqref{UglyTTExplicit}, we have split the sum over $j$ in \eqref{HelicityPrepExp} into even and odd parts. At this point we separate the analysis into the cases when $n$ is even, $n=2s$, and when $n$ is odd, $n=2s+1$. In particular, we find
\begin{subequations} \label{3DTTAll}
\begin{align}
\Pi_{\perp}^{[2s] } h_{\a(2s) }&=\frac{1}{2^{2s-1}}\frac{1}{\Box^{s}}\sum_{j=0}^{s}\binom{2s}{2j}\Box^{j}\pa_{(\a_1}{}^{\b_1}\cdots\pa_{\a_{2s-2j}}{}^{\b_{2s-2j}}h_{\a_{2s-2j+1}\dots\a_{2s})\b(2s-2j)}~,\label{3DTTBos}\\
\Pi_{\perp}^{[2s+1] } h_{\a(2s+1) }&=\frac{1}{2^{2s}}\frac{1}{\Box^{s}}\sum_{j=0}^{s}\binom{2s+1}{2j+1}\Box^{j}\pa_{(\a_1}{}^{\b_1}\cdots\pa_{\a_{2s-2j}}{}^{\b_{2s-2j}}h_{\a_{2s-2j+1}\dots\a_{2s})\b(2s-2j)}~.\label{3DTTFerm}
\end{align}
\end{subequations}

From \eqref{3DTTBos}, it is clear that for integer spin the Lagrangian $h^{\a(2s)}\Box^s\Pi_{\perp}^{[2s] } h_{\a(2s) }$ will lead to an action which is both gauge invariant and local. Hence it is a candidate Lagrangian for the CHS action \eqref{CHS3Aproj}. However, we recall that the conformal weight of the CHS field $h_{\a(n)}$ is fixed to be $\Delta_{h_{(n)}}=2-\frac{n}{2}$, see equation \eqref{WeightFix}.\footnote{Though the calculation leading to the conclusion \eqref{WeightFix} was done explicitly only for integer spin, it may be easily extended to the case of half-integer spin using the two component spinor formalism. } 
We must therefore reject this candidate, 
on the grounds that the resulting action has no chance to be invariant under conformal transformations. The only other possibility, which is consistent with locality, leads to the gauge invariant action
\begin{align}
S_{\text{CHS}}^{(2s)}[h]=\frac{1}{2}\left(-\hf \right)^s\int \text{d}^3x \, h^{\a(2s)}\Box^{s-1}\pa_{\a}{}^{\b}\Pi^{[2s]}_{\perp}h_{\a(2s-1)\b}~.\label{CHS3BosFlat}
\end{align}
In the fermionic case, the simplest candidate Lagrangian for a gauge invariant and local action 
is the correct one, and it is given by
\begin{align}
S_{\text{CHS}}^{(2s+1)}[h]=\frac{\ri}{2}\left(-\hf \right)^s\int \text{d}^3x \, h^{\a(2s+1)}\Box^{s}\Pi^{[2s+1]}_{\perp}h_{\a(2s+1)}~.\label{CHS3FermFlat}
\end{align}
We may unify both actions by expressing them in the form
\begin{align}
S_{\text{CHS}}^{(n)}[h]=\frac{\ri^n}{2^{\lfloor n/2 \rfloor + 1}}\int \text{d}^3x \, h^{\a(n)}\mc{C}_{\a(n)}(h)~,\label{CHS3ActCot}
\end{align}
where we have made use of the definitions
\begin{subequations} \label{HSCotProj}
\begin{align}
\mc{C}_{\a(2s)}(h)&:=\Box^{s-1}\pa_{\a}{}^{\b}\Pi^{[2s]}_{\perp}h_{\a(2s-1)\b}~,\\
\mc{C}_{\a(2s+1)}(h)&:= \Box^{s}\Pi^{[2s+1]}_{\perp}h_{\a(2s+1)}~.
\end{align}
\end{subequations}

The local field strength $\mc{C}_{\a(n)}(h)$ is known as the linearised higher-spin Cotton tensor of $h_{\a(n)}$. Though the formulation \eqref{HSCotProj} of $\mc{C}_{\a(n)}(h)$ in terms of the spin-projection operators is novel, higher-spin Cotton tensors in $\mb{M}^3$ have been studied for over thirty years. We now review their main features.
  
 

\subsection{Linearised higher-spin Cotton tensors}



The higher-spin Cotton tensor $\mc{C}_{\a(n)}(h)$ arises naturally when constructing a local action for the conformal higher-spin gauge field $h_{\a(n)}$. 
Given an integer $n\geq 2$, we recall that the conformal gauge field $h_{\a(n)}$ is defined modulo the gauge transformations 
\bea
\d_{\xi} h_{\a(n)} = \pa_{(\a_1\a_2} \xi_{\a_3 \dots \a_n)}~.
\label{3.32}
\eea
and is a primary field with conformal weight $\Delta_{h_{(n)}}=(2-\frac{n}{2})$, see eq. \eqref{CHSprepPrimary}.
Then, given an $h_{\a(n)}$, its Cotton tensor $\mc{C}_{\a(n)}(h)$ is defined uniquely 
by the following properties: %
\begin{enumerate}[label=(\roman*)]
\item $\mc{C}_{\a(n)}  (h)$ is a primary field with conformal weight $\Delta_{\mc{C}_{(n)}}=(1+\frac{n}{2})$.
\item $\mc{C}_{\a(n)}  (h)$ is invariant under the gauge transformations \eqref{3.32} 
\bea
\d_{\xi} \mc{C}_{\a(n)}(h) =0~. \label{CotDefPropFlat1}
\eea
\item $\mc{C}_{\a(n)}  (h)$ is conserved (i.e. transverse)
\bea
\pa^{\b(2)} \mc{C}_{\b(2) \a(n-2)}(h) =0~. \label{CotDefPropFlat2}
\label{3.334}
\eea
\end{enumerate}
These properties ensure that the corresponding action \eqref{CHS3ActCot} is invariant under conformal transformations aswell as higher-spin gauge transformations.

 From the property (i) it follows that $\mc{C}_{\a(n)}(h)$ is a descendent of $h_{\a(n)}$ of the form $\mc{C}_{\a(n)}(h)=\mc{A} h_{\a(n)}$, where $\cA$ is a linear differential operator of order $(n-1)$. Then, modulo an overall normalisation, the unique solution  to the properties (i)--(iii) is the following
\begin{align}
\mc{C}_{\a(n)}  (h)
&=\frac{1}{2^{n-1}} \sum_{j=0}^{\lceil n/2 \rceil-1} \binom{n}{2j+1}\Box^{j}\pa_{(\a_{1}}{}^{\b_{1}}\dots\pa_{\a_{n-2j -1}}{}^{\b_{n-2j -1}}h_{\a_{n-2j}\dots\a_{n})\b(n-2j -1) }~ \non\\
&=\frac{1}{2^{n-1}} \sum_{j=0}^{\lceil n/2 \rceil-1} \binom{n}{2j+1}\Box^{j}\big(\pa_{\a}{}^{\b}\big)^{n-2j-1}h_{\a(2j+1)\b(n-2j-1)}~.\label{2.31}
\end{align}
This expression for the linearised higher-spin Cotton tensor, and the associated CHS action \eqref{CHS3ActCot}, was first derived by Pope and Townsend \cite{PopeTownsend} in the case of integer spin, $n=2s$, and by Kuzenko \cite{K16}\footnote{Relation \eqref{2.31} was obtained 
in \cite{K16} via component reduction of the HS $\cN=1$ super-Cotton tensor, which has a remarkably simple form (see eq. \eqref{HSscotMink3a}). This is an example of the power of supersymmetry.} in the case of half-integer spin, $n=2s+1$. 
 It reduces to the linearised Cotton tensor for $n=4$ (cf. the flat limit of eq. \eqref{lincot}), 
 and to the Maxwell field strength for  $n=2$.
It should be pointed out that the conformal spin-3 case, $n=6$, 
was studied for the first time in \cite{DD}.
 The spin-3/2 case, $n=3$,  
was considered in \cite{ABdeRST} and the field strength $\mc{C}_{\a(3)}(h)$ is the linearised version of the
Cottino vector spinor \cite{DK,GPS}.


Although it is not difficult to directly check that the expression \eqref{2.31} satisfies \eqref{CotDefPropFlat1} and \eqref{CotDefPropFlat2}, these two properties are far from obvious. 
However, using the identity \eqref{3DTTAll}, one may show that \eqref{2.31} is exactly equivalent to \eqref{HSCotProj}, which expresses the Cotton tensor in terms of the spin-projection operators.
Furthermore, on account of the identity \eqref{2.12.5}, we may rewrite $\mc{C}_{\a(n)}(h)$ in terms of the positive and negative helicity components of $h_{\a(n)}$ as follows 
\begin{align} \label{HelicityCotton}
\mc{C}_{\a(n)}(h)=\Box^{\frac{1}{2}(n-1)}\Big(h_{\a(n)}^{(+)}-(-1)^nh_{\a(n)}^{(-)}\Big)~.
\end{align}
By virtue of eqs. \eqref{2.13.5} and \eqref{2.13.6}, this form renders both the gauge invariance and the transversality of the higher-spin Cotton tensor manifest. It is also straightforward, albeit tedious, to show that  $\mc{C}_{\a(n)}(h)$ transforms as a primary field with weight $(1+\frac{n}{2})$ under the conformal group. 
We will not derive this statement explicitly, since it follows directly from the fact that $\mc{C}_{\a(n)}(h)$ admits a Weyl covariant extension to curved spacetime, which is shown in section \ref{sec3DCHSCF}. 

An alternative derivation of both the fermionic and bosonic higher-spin Cotton tensors
was given in \cite{HHL,HLLMP} (see also \cite{BBB}), where they played an integral role in establishing a conformal geometry of higher-spin gauge fields.\footnote{By conformal geometry we mean the construction of all invariants under the conformal higher-spin gauge transformations and the study of their properties. } Using cohomological techniques, the authors showed that the expression \eqref{2.31} for the Cotton tensor is actually the most general solution to the conservation equation \eqref{CotDefPropFlat2}. 
An alternative proof of this property, which is  based on supersymmetry considerations, was given in \cite{K16}. 

\subsubsection{Gauge completeness} \label{SecGaugeCom}

Another important result established in \cite{HHL,HLLMP} is that the Cotton tensor $\mc{C}_{\a(n)}(h)$ of a field $h_{\a(n)}$ (which need not be CHS) vanishes if and only if the prepotential is pure gauge:
 \bea
\mc{C}_{\a(n)}(h) =0 \quad \Longleftrightarrow \quad 
h_{\a(n)}=\pa_{\a(2)}\x_{\a(n-2)}~,
\label{GaugeCompleteness}
\eea
for some $\xi_{\a(n-2)}$. Below we sketch an alternate proof for this result.

The Cotton tensor is invariant under the gauge transformations \eqref{CotDefPropFlat1}, which allows us to impose the gauge condition
\bea
\pa^{\b(2) } h_{\b(2) \a(n-2)} = 0~.
\label{GFcondCot}
\eea
In this gauge the field strength \eqref{2.31} reduces to the form 
\begin{subequations} \label{GaugeFixedCotFlat}
\bea
\mc{C}_{\alpha(2s)}(h)&=&\Box^{s-1}\partial^{\beta}{}_{(\a_1}h_{\a_2\dots\a_{2s})\beta}
=\Box^{s-1}\partial^{\beta}{}_{\a_1}h_{\a_2\dots\a_{2s}\beta}
~, 
\label{3.41a}
\\
\mc{C}_{\alpha(2s+1)}(h)&=&\Box^s h_{\alpha(2s+1)}~, 
\label{3.41b}
\eea
\end{subequations}
as a consequence of the identity\footnote{Alternatively, since the transverse projector $\Pi_{\perp}^{[n]}$ acts as the identity on fields satisfying \eqref{GFcondCot}, the relations \eqref{GaugeFixedCotFlat} follow immediately from \eqref{HSCotProj}. }
\bea
 \sum\limits_{j=0}^{ \left \lfloor{n/2}\right \rfloor }\binom{n}{2j+1} = 2^{n-1}~.
\eea
Now suppose $\mc{C}_{\a(n)}(h) =0 $. We assume $h_{\a(n)}$ to decrease at infinity, 
such that its Fourier transform is well defined. It follows that, for $n>2$, $h_{\a(n) }$ obeys the wave equation 
\bea
\Box h_{\a(n) } =0~,
\label{KGCHS3D}
\eea
in addition to the transverse condition \eqref{GFcondCot}. The equations \eqref{GFcondCot} and \eqref{KGCHS3D} are invariant under a restricted class of gauge transformations \eqref{CotDefPropFlat1}
such that the gauge parameter $\x_{\a(n-2)} $ is constrained by 
\bea
\pa^{\b(2) } \x_{\b(2) \a(n-4)} = 0~, \qquad \Box \x_{\a(n-2)} = 0~.
\eea
This residual gauge freedom suffices to gauge $h_{\a(n)}$ away, which implies 
\eqref{GaugeCompleteness}. The case $n=2$ is special since the condition $\mc{C}_{\alpha(2)}(h)=0$
gives the first-order equation
\bea
\partial^{\beta}{}_{(\a_1} h_{\a_2) \beta}
=\partial^{\beta}{}_{\a_1}h_{\a_2\beta}=0 \quad \implies \quad \Box h_{\a(2) } =0~,
\eea
and thus the vector field $h_{\a(2)}$ has a single independent massless component. 
The residual gauge freedom is described by a parameter $\x$ constrained by 
$\Box \x =0$. This gauge freedom allows us to gauge away $h_{\a(2)}$. 

Yet another way to arrive at the \eqref{GaugeCompleteness} is by using the results of the previous section. In particular, we observe that the expressions \eqref{HSCotProj} for the HS Cotton tensor in terms of the spin-projection operators allow us to express the transverse component of $h_{\a(n)}$ as 
\bea
h^{\perp}_{\a(2s)}=\Box^{-s} \pa_{\a}{}^{\b}\mc{C}_{\a(2s-1)\b}(h)~,\qquad 
h^{\perp}_{\a (2s+1)}=\Box^{-s} \mc{C}_{\a(2s+1)}(h)~.
\eea
Finally, it remains to recall that $h_{\a(n)}$ may be decomposed according to \eqref{3dDecompFlat2}, so that if $\mc{C}_{\a(n)}(h) =0 $, only the longitudinal (pure gauge) component remains. 

Actually, a statement analogous to \eqref{GaugeCompleteness} holds in any conformally flat spacetime $\cM^3$ if the field $h_{\a(n)}$ is CHS. Namely, the Cotton tensor $\mc{C}_{\a(n)}(h) $ vanishes if and only if the gauge field $h_{\a(n)}$ is pure gauge, 
\bea
\mc{C}_{\a(n)}(h) =0 \quad \Longleftrightarrow \quad 
h_{\a(n)}=\mc{D}_{\a(2)}\x_{\a(n-2)}~,
\label{GaugeCompletenessCF}
\eea
for some $\xi_{\a(n-2)}$. This follows from the fact that both equations in \eqref{GaugeCompletenessCF} are invariant under Weyl transformations in curved space 
provided $h_{\a(n)} $ and $\x_{\a(n-2)} $ are primary fields of dimension
$(2-n/2 )$ and $(1-n/2)$, as described in section \ref{secCHS3prep}. 

\subsection{Topologically massive higher-spin gauge models} \label{secMassiveActFlat}

A unique feature of three spacetime dimensions is the existence of topologically 
massive Yang-Mills and gravity theories  \cite{Siegel, JT, Schonfeld, DJT1,DJT2}. 
These theories  are obtained by augmenting the usual Yang-Mills 
or gravitational action by a gauge-invariant topological term of the Chern-Simons type.
This section is devoted to the construction and analysis of higher-spin extensions 
of the linearised actions for these topologically massive theories.\footnote{It should be pointed out that the problem of constructing topologically massive 
higher-spin theories was considered in  \cite{Chen1,Chen2}. However, 
the non-linear action proposed possesses only a restricted gauge freedom 
in the presence of the Lagrange multiplier $\b$ that enforces the torsion-free conditions
on the spin connections. Alternative approaches are worth pursuing. }

\subsubsection{Topologically massive actions} \label{SecTMHS}

Topologically massive models for higher-spin gauge fields are higher-spin extensions of linearised topologically massive gravity \cite{DJT1,DJT2}. They are obtained by coupling the massless Fronsdal \cite{Fronsdal2} and  
Fang-Fronsdal \cite{FF2} actions to the bosonic (even $n$)  and fermionic 
(odd $n$) CHS action \eqref{CHS3ActCot},\footnote{The CHS action plays the role of the Chern-Simons term.} respectively. 
This is possible because both the CHS and massless actions may be formulated in terms of the same gauge field $h_{\a(n)}$. To begin the construction, we describe the massless (non-conformal) higher-spin gauge actions in $\mb{M}^3$.

 There are two types of massless higher-spin actions, first-order and second-order ones. 
Given an integer $n\geq 4$, the first-order model is described by the set of real fields 
$\{ h_{\a(n)},y_{\a(n-2)}, z_{\a(n-4)} \}$ 
which are defined modulo gauge transformations of the form 
\begin{subequations}\label{MasslessFOFlatGT}
\bea
\d_{\xi} h_{\a(n)}&=&\pa_{(\a_1\a_2}\xi_{\a_3\dots\a_n)} ~,\label{MasslessFOFlatGTa}\\
\d_{\xi} y_{\a(n-2)}&=&\frac{1}{n}\pa^\b{}_{(\a_1} \xi_{\a_2\dots\a_{n-2})\b }~,\label{MasslessFOFlatGTb}\\
\d_{\xi} z_{\a(n-4)}&=&\pa^{\b(2)}\xi_{\a(n-4)\b(2)}~. \label{MasslessFOFlatGTc}
\eea
\end{subequations}
The Fang-Fronsdal-type gauge-invariant action,\footnote{It is worth pointing out 
that the $4d$ Fang-Fronsdal action for a massless spin-$(s+\hf)$ field  \cite{FF}
is also described in terms of a triplet of fermionic gauge fields, 
$\{\J_{\a(s+1) \ad(s)},~\J_{\a(s-1) \ad(s)},~\J_{\a(s-1) \ad(s-2)}\}$
and their conjugates, if one makes use of the two-component spinor notation,
see section 6.9 of \cite{BK}. 
More generally, there exist bosonic and fermionic higher-spin triplet 
models in higher dimensions \cite{FS,ST,FPT,Sorokin:2008tf,AAS}.
On-shell supersymmetric formulations for the generalised 
triplets in diverse dimensions have recently been given in 
\cite{Sorokin:2018djm}. 
} 
$S_{\rm{FF}}^{(n)} 
= S_{\rm{FF}}^{(n)} [h,y,z]$,
is
\begin{align}
S_{\rm{FF}}^{(n)}=\frac{\text{i}^n}{2^{\lceil n/2\rceil}}&\int\text{d}^3x\,
\bigg\{ h^{\a(n-1)\g}\pa_{\g}{}^{\d}h_{\d\a(n-1)}
+2(n-2)y^{\a(n-2)}\pa^{\b(2)}h_{\a(n-2)\b(2)} \non\\
&+4(n-2) y^{\a(n-3)\g}\pa_{\g}{}^{\d}
y_{\d\a(n-3)}
+2n\frac{(n-3)}{(n-1)}z^{\a(n-4)}\pa^{\b(2)}y_{\a(n-4)\b(2)}
~~~~~\notag\\
&-\frac{(n-3)(n-4)}{(n-1)(n-2)}z^{\a(n-5)\g}
\pa_{\g}{}^{\d} z_{\d\a(n-5)} \bigg\}~. \label{MasslessFOFlat}
\end{align}
This model can be reformulated in terms of a single reducible tensor field $\boldsymbol{h}_{\b(2),\a(n-2)}$, which corresponds to the massless first order model introduced by Tyutin and Vasiliev \cite{TV} (see appendix B of \cite{Topological} for a review).

Given an integer $n\geq 4$, the second-order model is described by the set of real fields 
$\{ h_{\a(n)}, y_{\a(n-4)} \}$ which are 
defined modulo gauge transformations of the form 
\begin{subequations}\label{MasslessSOFlatGT}
\bea
\d_{\xi} h_{\a(n)}&=&\pa_{(\a_1\a_2}\xi_{\a_3\dots\a_n)}~,\label{MasslessSOFlatGTa}\\
\d_{\xi} y_{\a(n-4)}&=&\frac{(n-2)}{(n-1)}\pa^{\b(2)}\xi_{\a(n-4)\b(2)}~.\label{MasslessSOFlatGTb}
\eea
\end{subequations}
The Fronsdal-type gauge-invariant action, 
$S_{\rm{F}}^{(n)} = S_{\rm{F}}^{(n)} [h,y]$,
is
\begin{align}
S_{\rm{F}}^{(n)}=&\frac{\text{i}^n}{2^{\lfloor n/2 \rfloor +1}}
\int\text{d}^3x\, e \, \bigg\{
h^{\a(n)}\Box h_{\a(n)}
-\frac{n}{4}\pa_{\g(2)}h^{\g(2)\a(n-2)}\pa^{\b(2)}h_{\a(n-2)\b(2)}\notag\\
&-\frac{(n-3)}{2}y^{\a(n-4)}\pa^{\b(2)}\pa^{\g(2)}h_{\a(n-4)\b(2)\g(2)}
-\frac{2(n-3)}{n}y^{\a(n-4)}\Box y_{\a(n-4)} \notag\\
& -\frac{(n-3)(n-4)(n-5)}{4n(n-2)}\pa_{\g(2)}y^{\g(2)\a(n-6)}
\pa^{\b(2)}y_{\b(2)\a(n-6)}\bigg\}~.\label{MasslessSOFlat}
\end{align}

Separately, each of the gauge-invariant actions  \eqref{CHS3ActCot}, \eqref{MasslessFOFlat}
and \eqref{MasslessSOFlat} prove to describe no local propagating degrees of freedom.
However, on-shell the models
\begin{subequations} \label{HSTMGFlat}
\bea
S_{\rm TM}^{(2s+1)}[h,y,z] &=& S_{\rm{CHS}}^{(2s+1)}[h]-m^{2s-1}S_{\rm{FF}}^{(2s+1)}[h,y,z]\label{HSTMGFlatF} \\
S_{\rm TM}^{(2s)}[h,y] &=& S_{\rm{CHS}}^{(2s)}[h]-m^{2s-3}S_{\rm{F}}^{(2s)} [h,y] \label{HSTMGFlatB}
\eea
\end{subequations}
describe irreducible fields on $\mb{M}^3$ with mass $|m|$ and helicities $\s (s+\frac{1}{2})$ and $\s s$ respectively, where $\s= m/|m|$. 



\subsubsection{Topologically massive models: the fermionic case} 

We now demonstrate that  on-shell, the model \eqref{HSTMGFlatF} indeed describes an 
irreducible massive spin-$(s+\frac{1}{2})$ field with a single propagating degree of freedom.

The equations of motion corresponding to \eqref{HSTMGFlatF} are
\begin{subequations}
\bea
0&=& m^{2s-1}\big(\partial^{\beta}{}_{(\alpha_1}h_{\a_2\dots\a_{2s+1})\beta}
-(2s-1)\partial_{(\a_1\a_2}y_{\a_3\dots\a_{2s+1})}\big)
- \mc{C}_{\alpha(2s+1)}(h)~,\label{3.9a}\\
0&=&\partial^{\beta(2)}h_{\alpha(2s-1)\beta(2)}
+4\partial^{\beta}{}_{(\a_1}y_{\a_2\dots\a_{2s-1})\beta}
-\frac{(s-1)(2s+1)}{s(2s-1)}\partial_{(\a_1\a_2}z_{\a_3\dots\a_{2s-1})}~,\label{3.9b}\\
0&=&(2s-1)\partial^{\beta(2)}y_{\alpha(2s-3)\beta(2)}
-\frac{(2s-3)}{(2s+1)}\partial^{\beta}{}_{(\a_1}z_{\a_2\dots\a_{2s-3})\beta}~.
\label{3.9c}
\eea
\end{subequations}
The gauge transformation \eqref{MasslessFOFlatGTc} tells us that  
$z_{\alpha(2s-3)}$ can be  completely gauged away, that is, we are able to impose 
the gauge condition
\begin{align}
z_{\alpha(2s-3)}=0~. \label{3.10}
\end{align} 
Then, the residual gauge freedom is described by $\xi_{\alpha(2s-1)}$ constrained by 
\bea \label{3.11}
\partial^{\beta(2)}\xi_{\alpha(2s-3)\beta(2)}=0 \quad
\implies  \quad
\partial^{\b}{}_{(\alpha_1}\xi_{\alpha_2\dots\alpha_{2s-1})\b}
=\partial^{\b}{}_{\alpha_1}\xi_{\alpha_2\dots\alpha_{2s-1}\b}~. 
\eea
In the gauge \eqref{3.10}, the equation of motion \eqref{3.9c} 
becomes the condition for $y_{\a(2s-1)}$ to be divergenceless, 
\bea\label{3.12}
\partial^{\beta(2)}y_{\alpha(2s-3)\beta(2)}=0 \quad
\implies \quad
\partial^{\b}{}_{(\alpha_1}y_{\alpha_2\dots\alpha_{2s-1})\b}
=\partial^{\b}{}_{\alpha_1}y_{\alpha_2\dots\alpha_{2s-1}\b}~.
\eea
Due to \eqref{3.12}, the gauge transformation \eqref{MasslessFOFlatGTb} becomes
\bea
\delta_\xi y_{\alpha(2s-1)}=\frac{1}{2s+1}
\partial^{\b}{}_{\alpha_1}\xi_{\alpha_2\dots\alpha_{2s+1}\b} ~.
\label{3.144}
\eea
Since $y_{\alpha(2s-1)}$ and $\xi_{\alpha(2s-1)}$ have the same functional type, 
we are able 
to completely gauge away the $y_{\alpha(2s-1)}$ field,
\begin{align}
y_{\alpha(2s-1)}=0~. \label{3.18}
\end{align}
In accordance with \eqref{3.144} and \eqref{3.18}, 
the residual gauge freedom is described by the parameter $\xi_{\alpha(2s-1)}$ 
constrained by 
\bea
\partial^{\b}{}_{~\alpha_1}\xi_{\a_2\dots\a_{2s-1}\b}=0
\quad \implies  \quad \Box  \xi_{\alpha(2s-1)}=0~.
\label{3.18.5}
\eea
In the gauge \eqref{3.18}, the equation of motion \eqref{3.9b} tells us that $h_{\a(2s+1)}$ is divergenceless, 
\begin{align}
\partial^{\beta(2)}h_{\alpha(2s-1)\beta(2)}=0 
\quad \implies  \quad
\partial^{\b}{}_{(\alpha_1}h_{\alpha_2\dots\alpha_{2s+1})\b}
=\partial^{\b}{}_{\alpha_1}h_{\alpha_2\dots\alpha_{2s+1}\b}~.
\label{3.19}
\end{align}
Until now, the above analysis has been identical to that of the massless model, given in appendix B of \cite{KO}.

Due to \eqref{3.19},
the Cotton tensor reduces
to the expression \eqref{3.41b}.
Therefore, in the gauge \eqref{3.18}, 
the equation of motion \eqref{3.9a} becomes
\begin{align}
m^{2s-1}\partial^{\beta}{}_{\a_1}h_{\a_2\dots\a_{2s+1}\beta}
-\Box^s h_{\alpha(2s+1)}=0~.\label{3.21}
\end{align}
This equation has two types of solutions, massless and massive ones,
\begin{subequations}
\bea
\partial^{\b}{}_{\alpha_1}h_{\alpha_2\dots\alpha_{2s+1}\b}=0
\quad \implies \quad  \Box h_{\alpha(2s+1)}=0~;
\label{4.14a}\\
m^{2s-1} h_{\a (2s+1)}- \Box^{s-1} \partial^{\beta}{}_{\a_1} h_{\alpha_2 \dots \a_{2s+1} \b}=0~.
\label{4.14b}
\eea
\end{subequations}
We point out that $\J_{\a_1 \dots \a_{2s+1}} :=\partial^{\beta}{}_{\a_1} h_{\alpha_2 \dots \a_{2s+1} \b}$ 
is completely symmetric and divergenceless, 
$\J_{\a_1 \dots \a_{2s+1}} = \J_{(\a_1 \dots \a_{2s+1})}$ and 
$ \pa^{\b\g} \J_{\b\g \a_1 \dots \a_{2s-1}} =0$.

Let us show that the massless solution \eqref{4.14a} is a pure gauge degree of freedom.\footnote{Of course, since $\Box h_{\a(2s+1)}=0$ and $\pa^{\b(2)}h_{\b(2)\a(2s-1)}=0$ implies $\mc{C}_{\a(n)}(h)=0$, this also follows from the gauge completeness property \eqref{GaugeCompleteness}.}
Since both the gauge field $h_{\alpha(2s+1)}$ and 
the residual gauge parameter $\xi_{\alpha(2s-1)}$ are on-shell massless, 
 it is useful to switch to momentum space, by replacing 
$h_{\a(2s+1)} (x) \to h_{\a(2s+1)} (p)  $ and $\xi_{\a(2s-1)}(x) \to\xi_{\a(2s-1)}(p)$, where
the three-momentum $p^a$ is light-like, $p^{\a\b}p_{\a\b}=0$. 
For a given three-momentum, 
we can choose a frame in which the only non-zero component of 
$p^{\a\b}= (p^{11}, p^{12} =p^{21}, p^{22}) $ is $p^{22}=p_{11}$. 
Then, the conditions $p^\b{}_{\a_1} h_{\a_2 \dots \a_{2s+1} \b}(p)=0$ and 
$p^\b{}_{\a_1} \xi_{\a_2 \dots \a_{2s-1} \b}(p)=0$ are equivalent to 
\bea
h_{\a(2s)2 }(p)=0~, \qquad \xi_{\a(2s-2)2 }(p)=0~.
\eea
Thus the only non-zero components of $h_{\a(2s+1)} (p)$ and  $\xi_{\a(2s-1)} (p)$ 
are $h_{1\dots 1}(p) $ and  $\xi_{1\dots 1}(p) $. 
The residual gauge freedom, 
$\d_{\xi} h_{1 \dots 1} (p) \propto p_{11 } \xi_{1 \dots 1}$, 
allows us to gauge away the field $h_{\a(2s+1) } $ completely.

Thus, it remains to analyse the general solution of the equation \eqref{4.14b}, 
which implies 
\begin{align}
\Big(\Box^{2s-1}-(m^2)^{2s-1}\Big) h_{\alpha(2s+1)}=0~.
 \label{3.22}
\end{align}
This equation 
in momentum space yields
\begin{align}
\bigg(1-\bigg(\frac{-p^2}{m^2}\bigg)^{2s-1}\bigg)h_{\alpha(2s+1)}(p)=0~. \label{3.25}
\end{align}
Since the polynomial equation $z^{2s-1}-1=0$ has only one real root, 
$z=1$, 
the only  real solution to \eqref{3.25} is $p^2=-m^2$, 
from which it follows that $h_{\alpha(2s+1)}$ satisfies the ordinary  Klein-Gordon equation,
\begin{align}
\big(\Box-m^2\big)h_{\alpha(2s+1)}=0~. \label{3.26}
\end{align}
Applying \eqref{3.26} to \eqref{3.21} reveals that $h_{\alpha(2s+1)}$ satisfies the equation of motion corresponding to a massive spin $(s+\frac{1}{2})$-field with mass 
$|m|$ and helicity $\s (s+\frac{1}{2})$,
\begin{align}
\partial^{\beta}{}_{\alpha_1}h_{\a_2\dots\a_{2s+1}\beta}=\s |m|
h_{\alpha(2s+1)}~,\qquad \s:=\frac{m}{|m|}~. 
\label{3.27}
\end{align}

Finally, for completeness let us recall  the proof of the fact that equation \eqref{3.27}
describes 
a single propagating degree of freedom. 
 The field $h_{\alpha(2s+1)}$ is on-shell with momentum satisfying 
 $p^2=-m^2$, we can therefore transform equation \eqref{3.27} into momentum space and boost into the rest frame where $p^a=(|m|,0,0) \implies p^1{}_{1}=p^2{}_{2}=0,~ p^1{}_{2}=-p^2{}_{1}=|m|$, 
\begin{align}
\text{i}h_{\alpha(2s)1}(p)-\s h_{\alpha(2s)2}(p)=0~. \label{3.28}
\end{align}
Due to the symmetry of the field $h_{\alpha(2s+1)}$, equation \eqref{3.28} states that there is only a single degree of freedom. Taking the independent field component to be $h_{11\dots 1}(p)$ allows us to express all other components 
in terms of it.

Along with the fermionic model \eqref{HSTMGFlatF}, which corresponds to $n=2s+1$,
we could consider a bosonic one described by the action
\bea
S_{\rm TM}^{(2s)}[h,y,z]
&=& S_{\rm{CHS}}^{(2s)}[h]
-m^{2s-2}S_{\rm{FF}}^{(2s)} [h,y, z]~,
\label{4.25}
\eea
which corresponds to $n=2s$. Most of the above analysis would remain 
valid in this case as well. However, in place of eq. \eqref{3.25} we would have 
\begin{align}
\bigg(1-\bigg(\frac{-p^2}{m^2}\bigg)^{2s-2}\bigg)h_{\alpha(2s)}(p)=0~.
\end{align}
This equation has both physical ($p^2=-m^2$) and tachyonic ($p^2=m^2$)
solutions. Therefore, the model \eqref{4.25} is unphysical.
This may be interpreted  as a manifestation of the spin-statistics theorem.


\subsubsection{Topologically massive models: the bosonic case} \label{secTMHSbos}

We now demonstrate that on-shell, the model \eqref{HSTMGFlatB} describes an 
irreducible massive spin-$s$ field which propagates a single degree of freedom.

The equations of motion corresponding to \eqref{HSTMGFlatB} are
\begin{subequations}
\begin{align}
&0=m^{2s-3}\Big(\Box h_{\alpha(2s)}+\frac{1}{2}s\partial^{\beta(2)}\partial_{(\a_1\a_2}h_{\a_3\dots\a_{2s})\beta(2)}+\notag\\
&\phantom{BBBBBBBBB}
-\frac{1}{4}(2s-3)\partial_{(\a_1\a_2}\partial_{\a_3\a_4}y_{\a_5\dots\a_{2s})}\Big)
-\mc{C}_{\alpha(2s)}(h)~,\label{3.32a}\\
&0=\partial^{\beta(2)}\partial^{\gamma(2)}h_{\alpha(2s-4)\beta(2)\gamma(2)}
+\frac{4}{s}\Box y_{\alpha(2s-4)}+\notag\\
&\phantom{How much wood can a woodchop chop}-\frac{(s-2)(2s-5)}{2s(s-1)}\partial^{\beta(2)}\partial_{(\a_1\a_2}y_{\a_3\dots\a_{2s-4})\beta(2)}~.
\label{3.32b}
\end{align}
\end{subequations}
As follows from the gauge transformation law \eqref{MasslessSOFlatGTb}, 
it is possible
 to completely gauge away the field $y_{\alpha(2s-4)}$,
\begin{align}
y_{\alpha(2s-4)}=0~.
\label{3.33}
\end{align}
Then, the residual gauge freedom is described by a parameter $\xi_{\alpha(2s-2)}$
constrained by
\bea \label{3.34}
\partial^{\beta(2)}\xi_{\alpha(2s-4)\beta(2)}=0 \quad
\implies \quad
\partial^{\b}{}_{(\alpha_1}\xi_{\alpha_2\dots\alpha_{2s-2})\b}
=\partial^{\b}{}_{\alpha_1}\xi_{\alpha_2\dots\alpha_{2s-2}\b}~.
\eea
In the gauge \eqref{3.33}, the equation of motion \eqref{3.32b} becomes 
\begin{align}
\partial^{\gamma(2)}\partial^{\beta(2)}h_{\alpha(2s-4)\beta(2)\gamma(2)}=0~.
\label{3.36}
\end{align}
According to \eqref{MasslessSOFlatGTa}, the divergence of $h_{\alpha(2s)}$ transforms as
\begin{align}
\delta_\xi\big(\partial^{\beta(2)}h_{\alpha(2s-2)\beta(2)}\big)&=\partial^{\beta_1\beta_2}\partial_{(\alpha_1\alpha_2}\xi_{\alpha_3\dots\alpha_{2s-2}\beta_1\beta_2)} 
=-\frac{2}{s}\Box \xi_{\alpha(2s-2)} \label{3.37}
\end{align}
where we have made use of \eqref{3.34}.
 Since $\xi_{\alpha(2s-2)}$ and $\partial^{\beta(2)}h_{\alpha(2s-2)\beta(2)}$
 have the same functional type,  it is possible  to completely gauge away 
 the divergence of $h_{\alpha(2s)}$,
\bea\label{3.38}
\partial^{\beta(2)}h_{\alpha(2s-2)\beta(2)}=0 \quad 
\implies \quad
\partial^{\b}{}_{(\alpha_1}h_{\alpha_2\dots\alpha_{2s})\b}
=\partial^{\b}{}_{\alpha_1}h_{\alpha_2\dots\alpha_{2s}\b}~.
\eea
Under the gauge conditions imposed, there still remains 
some residual gauge freedom described by the gauge parameter $\xi_{\a(2s-2)}$ constrained by 
\eqref{3.34} and 
 $\Box \xi_{\alpha(2s-2)}=0$. 
So far the above analysis has been identical to that of the massless model given in appendix B of \cite{KO}.

As a consequence of \eqref{3.38}, the Cotton tensor \eqref{2.31} 
reduces to the simple form \eqref{3.41a}.
Making use of the gauge conditions \eqref{3.33} and \eqref{3.38} 
in conjunction with eq.  \eqref{3.41a}, the equation of motion \eqref{3.32a} becomes
\begin{align}
\Big( m^{2s-3} \d^\b{}_{\a_1}
-\Box^{s-2}\partial^{\beta}{}_{\a_1}
\Big)
\Box h_{\a_2\dots\a_{2s}\beta}=0 ~.
\label{3.40}
\end{align}
This equation has two types of solutions, massless and massive ones,
\begin{subequations}
\bea
\Box h_{\a (2s)}&=&0 ~;  \label{4.30a}\\ 
m^{2s-3} h_{\a(2s)} 
-\Box^{s-2}\partial^{\beta}{}_{\a_1}
 h_{\a_2\dots\a_{2s}\beta}&=&0 ~. \label{4.30b}
\eea
\end{subequations}

Let us show that the massless solution \eqref{4.30a} is a pure gauge degree of freedom.
Since both the gauge field $h_{\alpha(2s)}$ 
and the gauge parameter  $\xi_{\alpha(2s-2)}$ are on-shell 
massless, it is useful to switch to momentum space 
 by replacing 
$h_{\a(2s)} (x) \to h_{\a(2s)} (p)  $ and $\z_{\a(2s-2)}(x) \to\z_{\a(2s-2)}(p)$, where
the three-momentum $p^a$ is light-like, $p^{\a\b}p_{\a\b}=0$. 
As in the fermionic case studied in the previous subsection,  
we can choose a frame in which the only non-zero component of 
$p^{\a\b}= (p^{11}, p^{12} =p^{21}, p^{22}) $ is $p^{22}=p_{11}$. 
In this frame, the equations \eqref{3.34} and \eqref{3.38} are equivalent to
\begin{align}
h_{\alpha(2s-2)22}(p)=0 ~, \qquad  \xi_{\alpha(2s-4)22}(p)=0~.\label{3.43}
\end{align}
These conditions tell us 
 that the only non-zero components in this frame are $h_{1\dots 1}(p)$, $h_{1\dots 12}(p)$ and $\xi_{1\dots 1}(p)$, $\xi_{1\dots 12}(p)$. However, the gauge transformation \eqref{MasslessSOFlatGTa} is equivalent to $\delta_{\xi} h_{1\dots 1}(p)\propto \xi_{1\dots1}(p)$ and $\delta_{\xi} h_{1\dots 12}(p)\propto \xi_{1\dots12}(p)$, allowing us to completely gauge away the $h_{\alpha(2s)}$ field. 

Let us turn to the other equation  \eqref{4.30b}, which implies 
\begin{align}
\Big(\Box^{2s-3}-(m^2)^{2s-3}\Big) h_{\alpha(2s)}=0~. 
\label{3.44}
\end{align}
Here the mass parameter has the same form  
as in the fermionic case, eq. \eqref{3.22}.
Transforming eq. \eqref{3.44} to momentum space gives
\begin{align}
\bigg(1-\bigg(\frac{-p^2}{m^2}\bigg)^{2s-3}\bigg) h_{\alpha(2s)}(p)=0~.
\end{align}
In complete analogy with the fermionic 
case considered in the previous subsection, 
this equation has the unique real solution
$p^2=-m^2$. 

 It follows that $h_{\alpha(2s)}$ satisfies the Klein-Gordon equation, 
\begin{align}
(\Box-m^2)h_{\alpha(2s)}=0~.\label{KGBaby}
\end{align}
As a consequence, 
the  equation of motion \eqref{4.30b} leads to
\begin{align}
\partial^{\beta}{}_{\alpha_1}h_{\a_2\dots\alpha_{2s}\beta}=\s |m| h_{\alpha(2s)}~,
\qquad  \s:=\frac{m}{|m|}~. 
\label{3.47}
\end{align}
Therefore $h_{\a(2s)}$ is an irreducible on-shell massive field 
with mass $m$ and helicity $\l =\s s$.
Equation \eqref{3.47} implies that $h_{\a(2s)}$ describes a single 
propagating degree of freedom.

%


\subsubsection{New topologically massive actions}

The so-called `new topologically massive' (NTM) models for bosonic fields were first introduced in \cite{BKRTY} in Minkowski space. Here we propose extensions of these models to fields with half-integer spin. 

These models are formulated solely in terms of the gauge prepotentials $h_{\a(n)}$ and 
the associated Cotton tensors $\mc{C}_{\a(n)}(h)$.  
In particular, given an integer $n\geq 2$, the gauge-invariant NTM action for the field $h_{\a(n)}$ is  
\begin{align}
S_{\text{NTM}}^{(n)}[h]=\frac{1}{m}\int \text{d}^3 x \, h^{\a(n)}\Big\{\pa_{\a}{}^{\b}\mc{C}_{\a(n-1)\b}(h)-m\mc{C}_{\a(n)}(h)\Big\}~.\label{HSNTMGflat}
\end{align}
where $m$ is some arbitrary non-zero mass parameter. The equation of motion obtained by varying \eqref{HSNTMGflat} with respect to the field $h_{\a(n)}$ is 
\begin{align}
\pa^{\b}{}_{(\a_1}\mc{C}_{\a_2\dots\a_n)\b}(h)=\s |m|\mc{C}_{\a(n)}(h)~,\qquad \s:=\frac{m}{|m|}~.
\label{NTMGeom1Flat} \end{align}
 In accordance with section \ref{SectionIrrepPoin3} this equation, in conjunction with the off-shell conservation identity \eqref{3.334}, means that the field strength $\mc{C}_{\a(n)}(h)$ itself describes a spin-$n/2$ field with mass $|m|$ and helicity $\s n/2$.
 
 On the other hand, we may choose to analyse solutions to \eqref{NTMGeom1Flat} in terms of the prepotential. In this case, the gauge freedom allows us to impose the  gauge condition 
 \begin{align}
 h_{\a(n)}\equiv h_{\a(n)}^{\perp} ~,\qquad \pa^{\b(2)}h_{\b(2)\a(n-2)}^{\perp}=0~.
 \end{align}
From \eqref{GaugeFixedCotFlat}, it follows that in this gauge, the field equation \eqref{NTMGeom1Flat} implies
 \begin{align}
 0=\Box^{\lfloor n/2 \rfloor} \Big(\pa_{(\a_1}{}^{\b}-\s |m| \delta_{(\a_1}{}^{\b}\Big)h^{\perp}_{\a_2\dots\a_n)\b}~. \label{NTMGeom2Flat}
 \end{align}
There are two types of independent solutions to this equation. The first are those satisfying $\Box^{\lfloor n/2 \rfloor} h^{\perp}_{\a(n)}=0$, which have vanishing Cotton tensor and are hence pure gauge by property \eqref{GaugeCompleteness}. The second satisfy
\begin{align}
 \pa_{(\a_1}{}^{\b}h^{\perp}_{\a_2\dots\a_n)\b}=\s |m| h^{\perp}_{\a(n)}~,
\end{align}
and describe a spin-$n/2$ field with mass $|m|$ and helicity $\s n/2$.

The massive spin $s=\frac{n}{2}$ gauge actions described above are of order $2s$ in derivatives and do not involve auxilliary fields. Recently, models describing massive integer spin $s$ (even $n$) fields which are of order $2s-1$ and also do not require auxilliary fields were proposed \cite{Dalmazi}. The corresponding models were lifted to AdS$_3$ in \cite{AdS3(super)projectors}.


\section{Conformal higher-spin models on conformally-flat backgrounds} \label{sec3DCHSCF}


In section \ref{GCA} we developed the formalism of conformal space, which is an extremely efficient tool for constructing field theories with local conformal symmetry on curved manifolds. In this section we make use of this framework to construct gauge and Weyl invariant models for conformal higher-spin fields on arbitrary conformally-flat backgrounds. Some comments on more general backgrounds will also be given.  This section is based on our paper \cite{Confgeo}.

We recall that in  $d=3$ conformal space, after imposing appropriate constraints on the algebra, the commutator of two conformally covariant derivatives $\nabla_a$ takes the form 
\begin{align}
\big[\nabla_a,\nabla_b\big]=\frac{1}{2}C_{ab}{}^{c}K_c~. \label{ComCom54}
\end{align}
Here $C_{abc}$ is the Cotton tensor and $K_{c}$ is the special conformal generator, see section \ref{GCA} for more details.  This algebra showcases one of the biggest advantages of the formalism:    the only curvature tensor which enters the geometry is the Cotton tensor $C_{\a(4)}$. The (traceless) Ricci tensor and scalar curvature, $R_{\a(4)}$ and $R$, appear only after degauging and only through the special conformal connection (see section \ref{sec3DDegauging} below). 

From a computational standpoint, this is a very desirable feature for a number of reasons. Firstly, it means that there are less curvatures and consequently less structures that one has to consider when constructing a Lagrangian (e.g. one can ignore all non-minimal curvature dependent terms in \eqref{linCgravitinoM3}  and \eqref{NotQuiteLinCot}). Secondly, since $\nabla_{\a(2)}$ and $C_{\a(4)}$ are conformally covariant (in contrast to $\mc{D}_{\a(2)}$, $R_{\a(4)}$ and $R$), it is much easier to build primary descendants -- one just needs to use the algebra \eqref{3DKNcom} to ensure that they are annihilated by $K_a$. Finally, for any conformally-flat spacetime the Cotton tensor vanishes and the conformally covariant derivatives commute
\begin{align}
C_{abc}=0\quad \implies \quad \big[\nabla_{a},\nabla_{b}\big]=0~. \label{CommutingDerivs}
\end{align}
Therefore, in practice, working on such backgrounds is akin to working in flat space. 


In two component spinor notation, the commutator \eqref{ComCom54} is equivalent to 
\begin{subequations}\label{confal4}
\bea
[\nabla_{\a\b},\nabla_{\g\d}]=\frac{1}{4}\ve_{\g(\a}C_{\b)\d}{}^{\r(2)}K_{\r(2)}
+\frac{1}{4}\ve_{\d(\a}C_{\b)\g}{}^{\r(2)}K_{\r(2)}
~.
\eea
The commutation relations of the other generators of the conformal group 
with  the covariant derivatives are as follows:
\bea \label{confal579}
\big[M_{\a\b},\nabla_{\g\d}\big]&=&\ve_{\g(\a}\nabla_{\b)\d}+\ve_{\d(\a}\nabla_{\b)\g}~,\\
\big[\mathbb{D},\nabla_{\a\b} \big]&=&\nabla_{\a\b}~,\\
\big[K_{\a\b},\nabla_{\g\d} \big]&=&4\ve_{\g(\a}\ve_{\b)\d}\mathbb{D}
-2\ve_{\g(\a}M_{\b)\d}-2\ve_{\d(\a}M_{\b)\g} \label{3DKNcom}
~.
\eea
\end{subequations}
For the sake of completeness, below we provide the conformal algebra $\mf{so}(3,2)$ in two component spinor notation (cf. \eqref{confal0}),
\begin{subequations}\label{confal5}
\begin{align}
[M_{\a\b},M_{\g\d}]&=\ve_{\g(\a}M_{\b)\d}+\ve_{\d(\a}M_{\b)\g}~, \phantom{inserting blank space inserting}\\
[M_{\a\b},P_{\g\d}]&=\ve_{\g(\a}P_{\b)\d}+\ve_{\d(\a}P_{\b)\g}~, \qquad \qquad \qquad \qquad ~ ~[\mathbb{D},P_{\a\b}]=P_{\a\b}~,\\
[M_{\a\b},K_{\g\d}]&=\ve_{\g(\a}K_{\b)\d}+\ve_{\d(\a}K_{\b)\g}~, \qquad \qquad \qquad \qquad [\mathbb{D},K_{\a\b}]=-K_{\a\b}~,\\
[K_{\a\b},P_{\g\d}]&=4\ve_{\g(\a}\ve_{\b)\d}\mathbb{D}-4\ve_{(\g(\a}M_{\b)\d)}~,
\end{align}
\end{subequations}
where $M_{\a\b}:=(\g^a)_{\a\b}M_a=\frac{1}{2}(\g^a)_{\a\b}\ve_{abc}M^{bc},~ P_{\a\b}:=(\g^a)_{\a\b}P_a$ and $K_{\a\b}:=(\g^a)_{\a\b}K_a$.

\subsection{Conformal higher-spin gauge fields} 

Within the setting of conformal space, a spin-$\frac{n}{2}$ conformal gauge field is also realised by the real totally symmetric rank-$n$ spinor field $h_{\a(n)}$. However, in place of those listed in section \ref{secCHS3prep}, we now define $h_{\a(n)}$  to satisfy the following properties: 
\begin{enumerate}[label=(\roman*)]

\item $h_{\a(n)}$ is a primary tensor field
 with conformal weight $\Delta_{h_{(n)}}=(2-n/2)$
 \begin{align}
 K_{\b(2)}h_{\a(n)}=0~,\qquad \mathbb{D}h_{\a(n)}=\left(2-\frac{n}{2}\right)h_{\a(n)}~. \label{CHS3Dprepprop}
 \end{align}
 
 \item $h_{\a(n)}$ is defined modulo gauge transformations of the form
 \begin{align}
\d_\x h_{\a(n) } = \nabla_{(\a_1 \a_2 } \x_{\a_3 \dots \a_n) }~, \label{CHSprepGT3}
\end{align}
where the real gauge parameter $\xi_{\a(n-2)}$ is a primary tensor field with conformal weight $\Delta_{\x_{(n-2)}}=(1-\frac{n}{2})$,
\begin{align}
 K_{\b(2)}\x_{\a(n-2)}=0~,\qquad \mathbb{D}\xi_{\a(n-2)}=\left(1-\frac{n}{2}\right)\x_{\a(n-2)}~.
\end{align}

\end{enumerate}

In this context, the conformal weight of $h_{\a(n)}$ is fixed by consistency of the above properties with the conformal algebra \eqref{confal579}. From \eqref{GtransPhi}, it follows that under the conformal gravity gauge group $\mc{G}$, the prepotential transforms according to the rule 
\begin{align}
\delta_{\L}^{(\mc{G})}h_{\a(n)}=\Big(\xi^a\nabla_a + \L^{\un{a}}X_{\un{a}}\Big)h_{\a(n)}=\Big(\xi^a\nabla_a +\frac{1}{2}K^{ab}M_{ab}+\Delta_{h_{(n)}}\s\Big)h_{\a(n)}~, \label{transrule579}
\end{align}
where we have used the properties \eqref{CHS3Dprepprop}.
We note that after degauging, i.e. imposing the gauge condition \eqref{degauge}, the transformation rules \eqref{CHSprepGT3} and \eqref{transrule579} reduce to \eqref{GTidc} and \eqref{transrule6676} respectively. Therefore, we see that (upon degauging) the definition of a conformal gauge field  in conformal space is equivalent to the one given in the vielbein approach. 

\subsection{Linearised higher-spin Cotton tensors}  

In conformal space we can construct the higher-spin Cotton tensor $\mc{C}_{\a(n)}(h)$ of $h_{\a(n)}$ in complete analogy with the guiding principles set out in section \ref{secLinCotVA}. In particular, the analogue of its defining features are as follows:

\begin{enumerate}[label=(\roman*)]

\item 
$\mc{C}_{\a(n)}(h)$ is a primary field of dimension $\Delta_{\mc{C}_{(n)}}=(1+\frac{n}{2})~,$
\begin{align}
K_{\b(2)}\mc{C}_{\a(n)}(h)=0~,\qquad \mathbb{D}\mc{C}_{\a(n)}(h)=\left(1+\frac{n}{2} \right)\mc{C}_{\a(n)}(h)~. \label{3.444}
\end{align}

\item 
$\mc{C}_{\a(n)}(h)$ is of the form $\mc{C}_{\a(n)}(h)=\mathcal{A}h_{\a(n)}$, where $\mathcal{A}$ is a linear differential operator of order $(n-1)$ involving the conformally covariant derivative $\nabla_a$, the Cotton tensor $C_{\a(4)}$, and its covariant derivatives.

\item
$\mc{C}_{\a(n)}(h)$ has vanishing gauge variation under \eqref{CHSprepGT3} if spacetime is conformally flat,
\begin{align}
\d_{\xi} \mc{C}_{\a(n)} (h)= \mc{O}\big( C\big)~.
\label{3188.5}
\end{align}

\item
 $\mc{C}_{\a(n)}(h)$ is divergenceless if spacetime is conformally flat,
\begin{align}
\nabla^{\b(2)} \mc{C}_{\b(2)\a(n-2)}(h) = \mc{O}\big( C\big)~.
\end{align} 
\end{enumerate}

 As mentioned in the beginning of this section, an attractive feature of this formulation occurs when the spacetime under consideration is conformally flat,
 \bea
 C_{\a(4)} =0~,
 \eea
  and therefore the conformal covariant derivatives commute \eqref{CommutingDerivs}. Let us  exclusively consider such backgrounds for the moment. 
  By virtue of property (ii) above, the only possible expression for $\mc{C}_{\a(n)}(h)$, which reduces to \eqref{2.31} in the flat limit, is the minimal lift of \eqref{2.31}, obtained by replacing $\pa_a \rightarrow \nabla_a$. Therefore,
 the unique expression  (up to an overall normalisation) for the spin-$\frac{n}{2}$ Cotton tensor is 
\begin{align}\label{3.7}
 \mc{C}_{\a(n)}(h)=\frac{1}{2^{n-1}}\sum_{j=0}^{\lceil n/2\rceil -1}\binom{n}{2j+1}(\Box_c)^j\big(\nabla_{\a}{}^{\b}\big)^{n-2j-1}h_{\a(2j+1)\b(n-2j-1)}~,
 \end{align}
 where $\Box_c=\nabla^a\nabla_a$ is the conformal d'Alembertian. 

It is clear that \eqref{3.7} is of the form $\mc{C}_{\a(n)}(h)=\mathcal{A}h_{\a(n)}$ and has Weyl weight equal to $(1+n/2)$. From \eqref{CommutingDerivs} and the transversality and gauge invariance of the higher-spin Cotton tensors in Minkowski space, it also follows that \eqref{3.7} satisfies both properties (iii) and (iv). It remains to show that it is primary.

 Using the algebra \eqref{confal4} it is possible to show, by induction on $j$, that the following two identities hold true\footnote{We note that \eqref{40.40a} and \eqref{40.40b} hold only in conformally-flat spacetimes, however \eqref{40.40b} holds in any spacetime when acting on a primary field. }
\begin{subequations}\label{40.40}
\begin{align}
\big[K_{\g(2)},(\Box_c)^j\big]&=2j(\Box_c)^{j-1}\nabla_{\g(2)}\big(2\mathbb{D}+2j-3\big)-4j(\Box_c)^{j-1}\nabla_{\g}{}^{\d}M_{\g\d} ~,\label{40.40a}\\[10pt]
\big[K_{\g(2)},\big(\nabla_{\a\b}\big)^j\big]&=4j\ve_{\a\g}\ve_{\g\b}\big(\nabla_{\a\b}\big)^{j-1}\big(\mb{D}+j-1\big)+4j\big(\nabla_{\a\b}\big)^{j-1}\ve_{\g(\a}M_{\b)\g}~\non\\
&-j(j-1)\ve_{\a\b}\ve_{\a\b}\big(\nabla_{\a\b}\big)^{j-2}\nabla_{\g(2)}~. \label{40.40b}
\end{align}
\end{subequations}
It is important to note that in the above identities we are using the convention \eqref{SymCon2}. Therefore, under a special conformal transformation we have
\begin{align}
&2^{n-1} K_{\g(2)}\mc{C}_{\a(n)}(h)=\sum_{j=0}^{\lceil n/2\rceil -1}\binom{n}{2j+1}\bigg\{\big[K_{\g(2)},(\Box_c)^j\big]\big(\nabla_{\a}{}^{\b}\big)^{n-2j-1}\phantom{BLANKSP}\notag\\ 
&\phantom{BLANK}+(\Box_c)^j\big[K_{\g(2)},\big(\nabla_{\a}{}^{\b}\big)^{n-2j-1}\big]\bigg\}h_{\a(2j+1)\b(n-2j-1)} \notag\\
&=\sum_{j=0}^{\lceil n/2\rceil -1}\binom{n}{2j+1}\bigg\{2j(2j+1)(\Box_c)^{j-1}\bigg[\nabla_{\g(2)}\big(\nabla_{\a}{}^{\b}\big)^{n-2j-1}h_{\a(2j+1)\b(n-2j-1)} \notag\\
&\phantom{BLA}-2\nabla_{\g\a}\big(\nabla_{\a}{}^{\b}\big)^{n-2j-1}h_{\a(2j+1)\b(n-2j-1)\g}\bigg] -(n-2j-1)(n-2j-2)\notag\\
&\phantom{BLA}\times(\Box_c)^j\bigg[\nabla_{\g(2)}\big(\nabla_{\a}{}^{\b}\big)^{n-2j-3}h_{\a(2j+3)\b(n-2j-3)} \notag\\
&\phantom{BLA}-2\nabla_{\g\a}\big(\nabla_{\a}{}^{\b}\big)^{n-2j-3}h_{\a(2j+2)\b(n-2j-3)\g}\bigg] \bigg\}=0~.\notag
\end{align}
In the last line we have used the fact that the second and third terms vanish for $j=\lfloor n/2 \rfloor$ and the first and last terms vanish for $j=0$ to shift the summation variable. This shows that in any conformally flat space \eqref{3.7} is the unique tensor satisfying the properties listed at the beginning of this section.

In a conformally-flat background, the first two properties determine $\mc{C}_{\a(n)}(h)$ uniquely. However,  for  $n\geq 4$, this is not true in a general curved background, since we can always add appropriate terms involving $C_{\a(4)}$.
For example, below we give the most general expressions for $\mc{C}_{\a(n)}(h)$,
for $n=2,3,4,5$, which satisfy properties (i) and (ii) above:\footnote{In a general curved space, $\mc{C}_{\a(n)}(h)$ must reduce to the expression \eqref{3.7} in the conformally-flat limit. Therefore, in such spaces the right hand side of eq. \eqref{3.7} constitutes the skeleton of $\mc{C}_{\a(n)}(h)$. It immediately follows that its divergence and 
 gauge variation under \eqref{CHSprepGT3} are  proportional to terms involving $C_{\a(4)}$ and its covariant derivatives. Therefore they also satisfy property (iii) and (iv). }
\begin{subequations}\label{58.9}
\begin{align}
\mc{C}_{\a(2)}(h)&=\frac{1}{2^{~}}\Big(2\nabla_{(\a_1}{}^{\b}h_{\a_2)\b}\Big)~,\\
\mc{C}_{\a(3)}(h)&=\frac{1}{2^2}\bigg(3\nabla_{(\a_1}{}^{\b_1}\nabla_{\a_2}{}^{\b_2}h_{\a_3)\b(2)}+\Box_ch_{\a(3)} \bigg)~,\\
\mc{C}_{\a(4)}(h)&=\frac{1}{2^3}\bigg(4\nabla_{(\a_1}{}^{\b_1}\nabla_{\a_2}{}^{\b_2}\nabla_{\a_3}{}^{\b_3}h_{\a_4)\b(3)}+4\Box_c\nabla_{(\a_1}{}^{\b}h_{\a_2\a_3\a_4)\b} \non\\
&\phantom{BLAN}+a_0C_{(\a_1\a_2}{}^{\b(2)}h_{\a_3\a_4)\b(2)} \bigg)~,\label{3.6c} \\
\mc{C}_{\a(5)}(h)&=\frac{1}{2^4}\bigg(5\nabla_{(\a_1}{}^{\b_1}\nabla_{\a_2}{}^{\b_2}\nabla_{\a_3}{}^{\b_3}\nabla_{\a_4}{}^{\b_4}h_{\a_5)\b(4)} +10\Box_c\nabla_{(\a_1}{}^{\b_1}\nabla_{\a_2}{}^{\b_2}h_{\a_3\a_4\a_5)\b(2)}+(\Box_c)^2h_{\a(5)}\notag\\
&\phantom{BLAN}+\Big(\frac{745}{16}-\frac{1}{2}a_1+\frac{5}{2}a_2+3a_3\Big)C^{\b_1\b_2}{}_{(\a_1\a_2}\nabla_{\a_3}{}^{\b_3}h_{\a_4\a_5)\b(3)}\notag\\
&\phantom{BLAN}+\Big(\frac{564}{48}-\frac{1}{2}a_1+\frac{1}{2}a_2+a_3\Big)C_{\d(\a_1\a_2}{}^{\b_1}\nabla^{\d\b_2}h_{\a_3\a_4\a_5)\b(2)}\notag\\
&\phantom{BLAN}+\Big(-40+\frac{1}{2}a_1-\frac{1}{2}a_2-a_3\Big)C^{\b(3)}{}_{(\a_1}\nabla_{\a_2\a_3}h_{\a_4\a_5)\b(3)}\notag\\
&\phantom{BLAN}+a_1\nabla^{\d\b_1}C^{\b_2}{}_{\d(\a_1\a_2}h_{\a_3\a_4\a_5)\b(2)}+a_2\nabla_{(\a_1}{}^{\b_1}C_{\a_2\a_3}{}^{\b_2\b_3}h_{\a_4\a_5)\b(3)}\notag\\
&\phantom{BLAN}+a_3\nabla_{(\a_1\a_2}C_{\a_3}{}^{\b(3)}h_{\a_4\a_5)\b(3)}\bigg)~,\label{spin52Baby}
\end{align}
\end{subequations}
 where the $a_i$ are arbitrary constants reflecting the non-uniqueness.\footnote{However, in the  $n=4$ case, one way to  fix the constant to $a_0=4$ is by explicitly linearising $C_{\a(4)}$ around an arbitrary background as in \eqref{lincot}.} 
These tensors, and more generally the higher-spin Cotton tensors, are not gauge invariant and the above ambiguity associated with their definition cannot rectify this. This supports the widespread belief that in $d=3$, conformal higher-spin fields can consistently propagate only on conformally-flat backgrounds.

\subsection{Conformal higher-spin action}
 The properties given in eqs. \eqref{CHS3Dprepprop} and \eqref{3.444} ensure that the linearised conformal higher-spin action  
\begin{align}\label{CHS3DActCF}
S_{\rm{CHS}}^{(n)} [ h] 
=\frac{\text{i}^n}{2^{\left \lfloor{n/2}\right \rfloor +1}
} \int \rd^3 x \, e\, h^{\a(n)} 
\mc{C}_{\a(n)}(h)~,
\end{align}
is invariant under the conformal gravity gauge group $\mathcal{G}$, $\delta_{\L}^{(\mc{G})}S_{\rm{CHS}}^{(n)} [ h]=0$. Upon degauging, this means that \eqref{CHS3DActCF} is invariant under Weyl transformations. 
Moreover, for every conformally flat spacetime, the  tensor \eqref{3.7}
has the following properties:
\begin{enumerate}

\item $\mc{C}_{\a(n)}(h)$ is conserved,
 \begin{align} \label{3.99}
\nabla^{\b(2)}\mc{C}_{\a(n-2)\b(2)}(h)=0~.
 \end{align}
 
 \item $\mc{C}_{\a(n)}(h)$ is 
 invariant under the gauge transformations \eqref{CHSprepGT3},
 \begin{align}\label{3.9}
 \delta_{\xi}h_{\a(n)}=\nabla_{(\a_1\a_2}\xi_{\a_3\dots\a_n)}
 \quad \implies
 \quad  \delta_{\xi}\mc{C}_{\a(n)}(h)=0~. \
 \end{align}

\end{enumerate}
Therefore, in any conformally flat space \eqref{CHS3DActCF} is invariant under the gauge transformations \eqref{CHSprepGT3},
\begin{align}
C_{ab}=0 \quad \implies \quad \delta_{\xi}S_{\text{CHS}}^{(n)}[h]=0~.
\end{align}
In the flat limit, the action \eqref{CHS3DActCF} coincides with \eqref{CHS3ActCot}. 

Finally, we would like to point out that the spin projection operators derived in section \ref{SectionSPO3} may be trivially extended to any conformally-flat space within this formalism. In particular, they are obtained by minimally lifting \eqref{2.12} and \eqref{3.13}:
\begin{align}
\Pi^{(\pm)}_{~\a}{}^{\b}=\frac{1}{2}\bigg(\delta_{\a}{}^{\b}\pm\frac{\nabla_{\a}{}^{\b}}{\sqrt{\Box_c}}\bigg)~,\qquad \Pi^{(\pm n)}_{~\a(n)}{}^{\b(n)}=\Pi^{(\pm )}_{~(\a_1}{}^{\b_1}\dots \Pi^{(\pm)}_{~\a_n)}{}^{\b_n}~.
\end{align}
All of their properties follow through on account of \eqref{CommutingDerivs}.
Then, in any conformally-flat spacetime, one can express the higher-spin Cotton tensor (and hence the CHS action) in terms of the projectors, as was done in section \ref{SectionD3CotFlat}. Of course, such operators are well defined only on conformal fields.


\subsection{Generalised CHS models}

As an extension of the previous constructions, we now consider  a conformal higher-spin gauge field $h^{(t)}_{\a(n)}$ which is defined modulo gauge transformations with $t$ derivatives,\footnote{Similar gauge transformations occur in the 
description of partially massless fields in AdS$_3$.} 
\begin{align}
\delta_{\xi}h^{(t)}_{\a(n)}=\nabla_{(\a_1\a_2}\cdots\nabla_{\a_{2t-1}\a_{2t}}\xi_{\a_{2t+1}\dots\a_n)}=\big(\nabla_{\a(2)}\big)^t\xi_{\a(n-2t)}~. \label{50.39}
\end{align}
Here $t$ is some integer $1\leq t \leq \lfloor \frac n2 \rfloor$ which is referred to as the depth. We require both the prepotential and the gauge parameter $\x_{\a(n-2t)}$ to be primary which, using the identity 
\begin{align}
\big[K_{\g(2)},\big(\nabla_{\a(2)}\big)^t\big]&=4t\ve_{\a\g}\ve_{\g\a}\big(\nabla_{\a(2)}\big)^{t-1}(\mathbb{D}+t-1)-4t\big(\nabla_{\a(2)}\big)^{t-1}\ve_{\a\g}M_{\g\a}~,
\end{align}
 proves to fix the dimension of $h^{(t)}_{\a(n)}$ uniquely to $(t+1)-\frac n2$. The conformal properties of $h^{(t)}_{\a(n)}$ and its gauge parameter may then be summarised by 
\begin{subequations}
\begin{align}
K_{\b(2)}h^{(t)}_{\a(n)}&=0~,\qquad 
\mathbb{D}h^{(t)}_{\a(n)}=\Big(t+1-\frac{n}{2}\Big)h^{(l)}_{\a(n)}~,\label{50.41}\\
K_{\b(2)}\xi_{\a(n-2t)}&=0~,\qquad 
\mathbb{D}\xi_{\a(n-2t)}=\Big(1-\frac{n}{2}\Big)\xi_{\a(n-2t)}~.
\end{align}
\end{subequations}

The following analysis will be restricted to the case of a conformally-flat background, 
\begin{align}
C_{\a(4)}=0~.
\end{align}
As was done earlier in the case $t=1$, from $h^{(t)}_{\a(n)}$ we may construct a generalised higher-spin Cotton tensor $\mc{C}_{(n,t)}(h)$ which possesses the following properties:

\begin{enumerate}[label=(\roman*)]

\item 
$\mc{C}^{(t)}_{\a(n)}(h)$ is a primary field of dimension $\Delta_{\mc{C}^{(t)}_{(n)}}=(2-t+\frac{n}{2})~,$
\begin{align}
K_{\b(2)}\mc{C}^{(t)}_{\a(n)}(h)&=0~,\qquad 
\mathbb{D}\mc{C}^{(t)}_{\a(n)}(h)=\Big(2-t+\frac{n}{2}\Big)\mc{C}^{(t)}_{\a(n)}(h)~. \label{50.42}
\end{align}

\item 
$\mc{C}^{(t)}_{\a(n)}(h)$ is of the form $\mc{C}^{(t)}_{\a(n)}(h)=\mathcal{A}h^{(t)}_{\a(n)}$, where $\mathcal{A}$ is a linear differential operator of order $(1+n-2t)$ involving the conformally covariant derivative $\nabla_a$.

\item
$\mc{C}^{(t)}_{\a(n)}(h)$ has vanishing gauge variation under \eqref{50.39},
\begin{subequations}\label{50.45}
\begin{align}
 \delta_{\xi}\mc{C}^{(t)}_{\a(n)}(h)=0~.\label{50.45b}
\end{align}
\item
 $\mc{C}^{(t)}_{\a(n)}(h)$ is `partially conserved',
\begin{align}
\big(\nabla^{\b(2)}\big)^{t}\mc{C}^{(t)}_{\a(n-2t)\b(2t)}(h)=0~.\label{50.45a}
\end{align}
\end{subequations}

\end{enumerate}

The properties (i) and (ii) determine $\mc{C}^{(t)}_{\a(n)}(h)$ uniquely
to be 
\begin{align}\label{50.43}
 \mc{C}^{(t)}_{\a(n)}(h)&=\frac{1}{2^{n-2t+1}}\sum_{j=t-1}^{\lceil n/2\rceil -1}\binom{n}{2j+1}\binom{j}{t-1}(\Box_c)^{j-t+1}\big(\nabla_{\a}{}^{\b}\big)^{n-2j-1}h^{(t)}_{\a(2j+1)\b(n-2j-1)}~.
 \end{align} 
To derive \eqref{50.43} we have made use of the identities \eqref{40.40}.
The proofs that \eqref{50.43} satisfies the properties \eqref{50.45} are non-trivial and are given in appendix \ref{AppGenCotProp}.\footnote{It would be of interest to apply the methods of 
\cite{HHL,HLLMP} to demonstrate that \eqref{50.43} is the most general solution 
of the $t$-folded conservation equation \eqref{50.45a} 
in the case of Minkowski spacetime.} 
The properties \eqref{50.41} and \eqref{50.42} mean that the generalised conformal higher-spin action,
\begin{align}\label{50.44}
S_{\rm{CHS}}^{(n,t)} [ {h}] 
=\frac{\text{i}^n}{2^{\left \lfloor{n/2}\right \rfloor +1 }
} \int \rd^3 x \, e\, { h}_{(t)}^{\a(n)} 
{\mc{C}}^{(t)}_{\a(n) } (h)~,
\end{align}
is invariant under the conformal gravity gauge group $\mathcal{G}$. Moreover, 
as a consequence of the properties \eqref{50.45b} and \eqref{50.45a}, the action \eqref{50.44} is also invariant under the gauge transformations \eqref{50.39},
\begin{align}
\delta_{\xi}S_{\rm{CHS}}^{(n,t)}[h] =0~.\label{50.46}
\end{align}
Like the minimal depth ($t=1$) case, the above action does not admit gauge invariant extensions to non-conformally-flat backgrounds. 

An interesting question to ask is the following. For a given spin, which values of $t$ yield first and second-order Lagrangians in the action \eqref{50.44}? To answer this question, we observe that the number of covariant derivatives in \eqref{50.44} is $(n-2t+1)$ so that $t=\frac 12 n$ and $t=\frac 12 (n-1)$, respectively. Since $t$ must be an integer it immediately follows that first-order conformal models exist only for bosonic spin whilst second-order models must be fermionic. 
These models are said to have `maximal depth' since $t$ assumes its maximal value of $t=\lfloor \frac n2 \rfloor$.  Our conclusions regarding second-order models are in agreement with those drawn long ago in \cite{Sachs}. 


\subsection{Degauging} \label{sec3DDegauging}
In the gauge \eqref{degauge}, the spinor conformally covariant derivative assumes the form 
 \begin{align}\label{3.1313}
 \nabla_{\a(2)}=\mathcal{D}_{\a(2)}-\frac{1}{4}S_{\a(2),}{}^{\b(2)}K_{\b(2)}~,
 \end{align}
where $S_{\a(2),\b(2)}:=\big(\g^a\big)_{\a(2)}\big(\g^b\big)_{\b(2)}S_{ab}$. We may decompose $S_{\a(2),\b(2)}$ into irreducible ${\sSL} (2, {\mathbb R})$ components as
\begin{align}
S_{\a(2),\b(2)}=R_{\a(2)\b(2)}+\frac{1}{6}\ve_{\a_1(\b_1}\ve_{\b_2)\a_2}R~,
\end{align} 
where $R_{\a(4)}$ is the background traceless Ricci tensor and $R$ is the scalar curvature. Then $\nabla_{\a(2)}$ becomes
\begin{align}
 \nabla_{\a(2)}=\mc{D}_{\a(2)}-\frac{1}{4}R_{\a(2)}{}^{\b(2)}K_{\b(2)}+\frac{1}{24}RK_{\a(2)}~.\label{DegaugedCCDM3}
\end{align}
  As described in section \ref{SectionDegauging}, the aim of the degauging procedure is to replace all occurrences of $\nabla_{\a(2)}$ with $\mc{D}_{\a(2)}$ through \eqref{DegaugedCCDM3}  and use the algebra \eqref{confal4} to eliminate the special conformal generator $K_a$.

It turns out that, even in the case of a maximally symmetric spacetime, the implementation of the degauging procedure for the higher-spin Cotton tensor is extremely non-trivial.
For example, in AdS$_3$, the conformally covariant derivative is 
\begin{align} \label{CCDads}
\nabla_{\a(2)}=\mathcal{D}_{\a(2)}+\mathcal{S}^2K_{\a(2)}
\end{align}
 whilst the conformal d'Alembertian is
\begin{align} 
\Box_c=\square+6\mathcal{S}^2\mathbb{D}-\mathcal{S}^2\mathcal{D}^{\a(2)}K_{\a(2)}-\frac{1}{2}\mathcal{S}^4K^{\a(2)}K_{\a(2)}~.
\end{align}
 Here and in \eqref{CCDads}, the parameter $\mathcal{S}$ is related to the AdS scalar curvature via $R=24\mc{S}^2$, see section \ref{secAdS3Geom} below.  Making use of the above relations, one may show that the degauged version of \eqref{3.7}, for  $3\leq n \leq 6$ is given by 
 \begin{subequations}\label{LowCot}
\begin{align}
\mc{C}_{\a(3)}(h)=&\frac{1}{2^2}\bigg(3\big(\cD_{\a}{}^{\b}\big)^2
{h}_{\a\b(2)}+\big(\cQ -9\mc{S}^2\big){h}_{\a(3)}\bigg)~,\\
\mc{C}_{\a(4)}(h)=&\frac{1}{2^3}\bigg(4\big(\cD_{\a}{}^{\b}\big)^3h_{\a\b(3)}
+4( \cQ  - 20 \cS^2) \cD_{\a}{}^{\b}{h}_{\a(3)\b}
\bigg)~,\\
\mc{C}_{\a(5)}(h)=&\frac{1}{2^4}\bigg(5\big(\cD_{\a}{}^{\b}\big)^4{h}_{\a\b(4)}
+10\big(\cQ-33\mc{S}^2\big)\big(\cD_{\a}{}^{\b}\big)^2{h}_{\a(3)\b(2)}
\notag\\
&\phantom{extra}
+\big(\cQ-57\mc{S}^2\big)\big(\mc{Q}-25\mc{S}^2\big) {h}_{\a(5)}\bigg)~,\\
\mc{C}_{\a(6)}(h)=&\frac{1}{2^5}\bigg(6\big(\cD_{\a}{}^{\b}\big)^5{h}_{\a\b(5)}
+20\big(\cQ-48\mc{S}^2\big)\big(\cD_{\a}{}^{\b}\big)^3{h}_{\a(3)\b(3)}\notag\\
&\phantom{extra}+\big(6\cQ^2-704\mathcal{S}^2+18432\mathcal{S}^4\cQ\big)\cD_{\a}{}^{\b}h_{\a(5)\b}\bigg)~,
\end{align}
\end{subequations}
where $\mc{Q}$ is defined in \eqref{Q}  and we have made use of the convention \eqref{SymCon1}.

\section{Conformal higher-spin models in AdS$_3$} \label{sec3DCHSAdS}



%
The description \eqref{3.7} of $\mc{C}_{\a(n)}(h)$  in terms of $\nabla_a$ suffices to describe 
the conformal higher-spin action \eqref{CHS3DActCF}. However, in order to formulate non-conformal massive actions, such as those derived in section \ref{secMassiveActFlat}, one has to express 
$\mc{C}_{\a(n)}(h)$ in terms of the Lorentz covariant derivatives $\cD_a$.
In the previous section this was done in AdS$_3$ for $\mc{C}_{\a(n)}(h)$ with $3\leq n \leq 6$, however there does not seem to be a simple systematic method to degauge $\mc{C}_{\a(n)}(h)$ for generic $n$. 

 In this section we solve this problem and present a closed-form expression for the higher-spin Cotton tensor  in terms of the AdS$_3$ Lorentz covariant derivative. We also give several novel applications of this result, including a proof of the factorisation of the CHS action and the construction of massive gauge actions in AdS$_3$. This section is based on our paper \cite{CottonAdS}.

\subsection{AdS$_3$ geometry}  \label{secAdS3Geom}

 The Lorentz covariant derivatives of AdS$_3$ satisfy the commutation relations\footnote{In \cite{CottonAdS} we used the convention $\big[\mc{D}_a,\mc{D}_b\big]=\frac{1}{2}R_{ab}{}^{cd}M_{cd}$ and $R=-24\mc{S}^2$. In order to maintain the same commutator algebra, $\big[\mc{D}_{a},\mc{D}_{b}\big]=-4\mc{S}^2M_{ab}$, in this thesis we take $R$ to be positive, $R=24\mc{S}^2$. }
\begin{align}
\big[\mc{D}_a,\mc{D}_b\big]=-4\mc{S}^2M_{ab}\qquad\Longleftrightarrow \qquad
\big[\mc{D}_{\a\b},\mc{D}_{\g\d}\big]=4\mc{S}^2\big(\ve_{\g(\a}M_{\b)\d}+\ve_{\d(\a}M_{\b)\g}\big) ~.
\label{alg2.1}
\end{align}
Here the parameter $\mc{S}$ is related to the AdS radius $\ell$ and
the scalar curvature $R$ via $\ell^{-1} = 2 \cS$ and  
$R=24\mc{S}^2$. 
We parametrise the curvature in terms of $\cS$ in order for the notation to be consistent with that used in $\cN=1$ AdS superspace \eqref{algN1}.

Isometries of AdS$_3$ are generated by those solutions of the conformal Killing equation \eqref{CCVF1.3}, which obey the additional restriction $\s[\z]=0$. 
They are called the Killing vector fields on AdS$_3$. Given a  tensor field $h_{\a(n)}$ on AdS$_3$, its AdS transformation law is obtained from \eqref{ConfT3D} by setting $\s[\z]=0$.  

There are two quadratic Casimir operators of the AdS$_3$ isometry algebra 
$\mf{so}(2,2)\cong \mf{sl}(2,{\mathbb R}) \oplus \mf{sl}(2,{\mathbb R})$, and they can be chosen as follows  (see, e.g., \cite{BPSS}):\footnote{In the flat limit, the casimir \eqref{F} is proportional to the Pauli-Lubanski scalar \eqref{PLscalar}. }
\begin{subequations}
\bea
\mc{F}&:=&\mc{D}^{\a\b}M_{\a\b}~,\qquad ~~~~~~~~~~\phantom{..}\big[\mc{F}, \mc{D}_{\a(2)}\big]=0~, \label{F}
\\
\mc{Q}&:=&\Box - 2\mc{S}^2M^{\a\b}M_{\a\b}~,\qquad \big[\mc{Q},\mc{D}_{\a(2)}\big]=0~, \label{Q}
\eea
\end{subequations}
where $\Box= \mc{D}^a\mc{D}_{a}=-\frac{1}{2}\mc{D}^{\a\b}\mc{D}_{\a\b}$. These Casimir operators are independent of one another, 
which may be seen by the action of $\mc{F}^2$ on the (unconstrained) field $h_{\a(n)}=h_{(\a_1\dots\a_n)}$:
\begin{align}
\mc{F}^2h_{\a(n)}=n^2\big[\mc{Q}-(n-2)(n+2)\mc{S}^2\big]h_{\a(n)}+n(n-1)\mc{D}_{\a(2)}\mc{D}^{\b(2)}h_{\a(n-2)\b(2)}~, \label{ID0}
\end{align}
for any integer $n\geq 0$. It should be pointed out that the divergence in the second term of \eqref{ID0} is not defined for $n=0 $ and 1, however the corresponding numerical coefficient is equal to zero in both cases. 


\subsection{On-shell massive and (partially-)massless fields} \label{section 2.2}

We define an on-shell (real) field $h_{\a(n)}$, with $n\geq 2 $, to be one which satisfies the first-order constraints\footnote{For $n=0$ only \eqref{OS1c} is present, whilst for $n=1$ there is only \eqref{OS1b}.}
\begin{subequations}\label{OS1}
\begin{align}
0&=\mc{D}^{\b\g}h_{\a(n-2)\b\g}~,\label{OS1a}\\
0&=\big(\mc{F}-\rho\big)h_{\a(n)} \label{OS1b}~, 
\end{align}
\end{subequations}
for some mass parameter  
$\rho \in \mb{R}$.
An equivalent form of the second irreducibility condition \eqref{OS1b} is (cf. \eqref{1.1.2})
\begin{align}
\mc{D}_{(\a_1}{}^{\b}h_{\a_2\dots\a_n)\b}=\frac{\rho}{n} h_{\a(n)}~.
\label{OS1b+}
\end{align}
The on-shell field $h_{\a(n)}$ is said to be transverse with pseudo-mass $\rho$, spin $n/2$ and helicity $\s n/2$, where $\s=\rho/|\rho|$. 
The equations \eqref{OS1a} and \eqref{OS1b+} were introduced in 
\cite{BHRST} (see also \cite{BPSS}).\footnote{In the flat-space limit, these equations 
reduce to those considered in \cite{GKL,TV}.} 

From the relation \eqref{ID0} we see that when restricted to the space of transverse fields, the quadratic Casimirs \eqref{F} and \eqref{Q} are related via
\begin{align}
\mc{F}^2h_{\a(n)}=n^2\Big(\mc{Q}-(n-2)(n+2)\mc{S}^2\Big)h_{\a(n)}~.
\end{align}
Applying $\mc{F}$ to \eqref{OS1b}, one finds that the on-shell conditions \eqref{OS1} lead to the following second-order mass-shell equation
\begin{align}
0=\Big(\mc{Q}-\big[\big(\rho/n\big)^2+(n-2)(n+2)\mc{S}^2\big]\Big)h_{\a(n)}~.\label{OS1c}
\end{align}
In terms of the AdS d'Alembertian, $\Box$, this reads
\begin{align}
0=\Big(\Box - \big[(\rho/n)^2-2(n+2)\mc{S}^2\big]\Big)h_{\a(n)}~.\label{OS1d}
\end{align}
The transverse projectors on AdS$_3$, which take a field satisfying \eqref{OS1c} and select the component satisfying \eqref{OS1}, were derived recently in \cite{AdS3(super)projectors} (which is not part of this thesis).

An on-shell field $h^{(t)}_{\a(n)}$ is said to be partially massless\footnote{Partially massless fields have been studied in diverse dimensions for over 35 years, see \cite{DeserN1, DeserN2,Higuchi1, Higuchi2, Higuchi3, Metsaev2006, DeserW1, DeserW2, DeserW3, DeserW4, DeserW5, Zinoviev, DNW, SV, BMVpm} for some of the earlier works.
 Lagrangian models for partially massless fields in AdS$_d$ were constructed in \cite{Zinoviev, Metsaev2006, SV}, and in \cite{BSZ1,BSZ2} 
 for the specific case of AdS$_3$.  } 
with depth $t$ if, in addition to the conditions \eqref{OS1}, the pseudo-mass satisfies
\begin{align}
\rho  \equiv \rho^{(\pm)}_{(t,n)} = \pm n(n-2t)\mc{S}~,\qquad 1\leq t \leq \lfloor  n/2 \rfloor ~. \label{PM1}
\end{align}
We will refer to $\rho^{(\pm)}_{(t,n)}$ as the depth-$t$ pseudo-mass values.\footnote{We note that the $\pm$ in \eqref{PM1} coincides the sign of the helicity, since $\s= \rho^{(\pm)}_{(t,n)}/ |\rho^{(\pm)}_{(t,n)}|=\pm1$. The positive and negative branches are related via  the identity $\rho^{(+)}_{(\lfloor n/2 \rfloor+t,n)}=\rho^{(-)}_{(\lceil n/2 \rceil-t,n)}$ for arbitrary $t$.} The corresponding mass-shell equation \eqref{OS1c} reduces to 
\begin{align}
0=\big(\mc{Q}-\tau_{(t,n)}\mc{S}^2\big)h^{(t)}_{\a(n)} \label{PMWE}
\end{align}
where the constants $\tau_{(t,n)}$ are the partially massless values defined by
\begin{align}
\tau_{(t,n)}= 2n(n-2t) + 4 (t-1)(t+1)~. \label{PM2}
\end{align}
The following identity holds for any transverse field $h_{\a(n)}$
\begin{align}
\big(\mc{F}-\rho^{(+)}_{(t,n)}\big)\big(\mc{F}-\rho^{(-)}_{(t,n)}\big)h_{\a(n)}=n^2\big(\mc{Q}-\tau_{(t,n)}\mc{S}^2\big)h_{\a(n)}~, \label{ID6}
\end{align}
and can be used to factorise second order operators involving the partially massless values into first order ones.

Using the formalism developed in appendix \ref{AppGenForm}, it may be shown that at the partially massless points \eqref{PM1}, the system of equations \eqref{OS1} admits a depth-$t$ gauge symmetry of the form
\begin{align}
\delta_{\xi}h^{(t)}_{\a(n)}=\mc{D}_{(\a_1\a_2}\cdots\mc{D}_{\a_{2t-1}\a_{2t}}\xi_{\a_{2t+1}\dots\a_n)}\equiv \big(\mc{D}_{\a(2)}\big)^t\xi_{\a(n-2t)}~. \label{PMgt}
\end{align}
This is true if the gauge parameter $\xi_{\a(n-2t)}$ is also on-shell with the same pseudo-mass\footnote{The system of equations \eqref{GPC1a} and \eqref{GPC1b} is defined only if $n-2t\geq 2$. For  $n-2t =1 $ only eq.  \eqref{GPC1b} is present. If $n-2t=0$, then these conditions must be replaced with $0=\big(\mc{Q}-(n-2)(n+2)\mc{S}^2\big)\xi$. }
\begin{subequations} \label{GPC1}
\begin{align}
0&= \mc{D}^{\b\g}\xi_{\a(n-2t-2)\b\g}~, \label{GPC1a}\\
0&=\big(\mc{F}-\rho^{(\pm)}_{(t,n)} \big)\xi_{\a(n-2t)}~. \label{GPC1b}
\end{align}
\end{subequations}
 We note that strictly massless fields $h_{\a(n)}$ correspond to those partially massless fields with the minimal depth $t=1$ and therefore have pseudo-mass equal to $\rho^{(\pm)}_{(1,n)}$. They are defined modulo the standard first order gauge transformations,
\begin{align}
\delta_{\xi}h_{\a(n)}=\mc{D}_{(\a_1\a_2}\xi_{\a_3\dots\a_n)}~.
\end{align} 

In the interest of making contact with the representation theory of the AdS$_3$ isometry group $\sSO(2,2)$, it is useful to recast eq. \eqref{OS1d} into the notation of \cite{BHRST}, 
\begin{align}
0=\Big(\Box+ \ell^{-2}\big[n/2+1-\eta^{-2}\big]\Big)h_{\a(n)}~,\qquad \ell^{-1}:= 2\mc{S}~,\qquad \eta^{-1}:=\frac{\rho }{2n\mc{S}}~.
\end{align}
In this context, the minimal energy of the corresponding $\mf{so}(2,2)$ irreducible representation\footnote{The unitary irreducible representations of $\mf{so}(2,2)$
are denoted $D(E_0,s)$, where $E_0$ is the minimal energy, $s$ the helicity, and $|s|$ the spin. See \cite{DKSS} and references therein. The unitary bound is $E_0 \geq |s|$.}
 is $E_0=1+|\eta^{-1}|$, and the unitarity bound is $E_0\geq n/2$, or equivalently $|\rho|\geq n(n-2)\mc{S}$. The minimal energy of the  partially massless depth-$t$ field $h_{\a(n)}^{(t)}$ is $E^{(t,n)}_0=n/2+(1-t)$. Therefore, strictly massless fields saturate the unitarity bound, whilst true partially massless fields violate it (they are non-unitary). Finally, we note that the physical mass (see e.g. \cite{DeserW}) is related to the pseudo-mass via 
\begin{align}
\rho_{\text{phys}}^2=(\rho/n)^2-(\rho^{(\pm)}_{(1,n)}/n)^2~.
\end{align}
 In terms of $\rho_{\text{phys}}$, the unitarity  bound is $\rho_{\text{phys}}^2\geq 0$. Partially massless fields are thus seen to have negative physical mass-squared: $(\rho_{\text{phys}}^{(t,n)})^2=-4(t-1)(n-t-1)\mc{S}^2$. 

\subsection{Linearised higher-spin Cotton tensors}

It is a difficult technical problem to obtain an explicit expression for $\mc{C}_{\a(n)}(h)$ in terms of the AdS covariant derivative. In particular, the minimal uplift of \eqref{2.31} is not gauge invariant or transverse. To restore these properties, \eqref{2.31} must be supplemented with curvature dependent terms, as in the low spin cases \eqref{LowCot}.  However, for generic spin, previous attempts to do this in \cite{Topological, Confgeo} were unsuccessful. In order to derive the Cotton tensors in AdS$_3$, it is advantageous to recast the problem into the framework of homogeneous polynomials (or generating functions). Some basic identities in this formalism are given in appendix \ref{AppGenForm}. 

We begin by making an ansatz for the higher-spin Cotton tensor, which must be done separately for the bosonic ($n=2s$) and fermionic ($n=2s+1$) cases. In terms of homogeneous polynomials, our ans\"atze are as follows
\begin{subequations}\label{ansatz}
\begin{align}
\mc{C}_{(2s)}(h)&=\sum_{j=0}^{s-1}b_{j}\prod_{t=1}^{j}\bigg(\mc{Q}-\tau_{(s-t,2s)}\mc{S}^2\bigg) \mc{D}_{(2)}^{s-j-1}\mc{D}_{(0)}\mc{D}_{(-2)}^{s-j-1}h_{(2s)}~,\label{ansatzb}\\
\mc{C}_{(2s+1)}(h)&=\sum_{j=0}^{s}f_{j}\prod_{t=1}^{j}\bigg(\mc{Q}-\tau_{(s-t+1,2s+1)}\mc{S}^2\bigg) \mc{D}_{(2)}^{s-j}\mc{D}_{(-2)}^{s-j}h_{(2s+1)}~, \label{ansatzf}
\end{align}
\end{subequations}
for undetermined real coefficients $b_j$ and $f_j$, and where $\tau_{(t,n)}$ are the partially massless values \eqref{PM2}. We refer the reader to appendix \ref{AppGenForm} for an explanation of the notation. The main motivation for  
\eqref{ansatz}
is the expectation\footnote{Strictly speaking, to the best of our knowledge, most results regarding the factorisation of the conformal higher-spin kinetic operators apply only in even dimensions and only in the bosonic case.} that in an AdS background, the gauge invariant action \eqref{CSAdS3} for a conformal higher-spin field $h_{\a(n)}$ should factorise (in the transverse gauge) into products of minimal second-order operators involving all partial mass values. The factorisation of the Cotton tensors and their corresponding CHS action will be elaborated on in the next section. 
 
 To fix the coefficients in \eqref{ansatz}, we require that the expressions in \eqref{ansatz} are transverse \eqref{CotPropCFMb} and gauge invariant \eqref{CotPropCFMa}. Condition \eqref{CotPropCFMb} takes the form $0=\mc{D}_{(-2)}\mc{C}_{(n)}(h)$ which, upon employing the identity \eqref{ID15}, yields the recurrence relations
\begin{subequations}\label{ansatzT}
\begin{align}
0&=b_j-4(s-j)(s+j)b_{j-1}~,\qquad ~~~~\phantom{..}1\leq j \leq s-1~,\label{ansatzTb}\\
0&=f_j-4(s-j+1)(s+j)f_{j-1}~, \qquad 1\leq j\leq s ~.\label{ansatzTf}
\end{align}
\end{subequations}
 These two systems determine the coefficients $b_j$ and $f_j$ up to an overall normalisation:
 \begin{subequations}\label{coeffSol}
\begin{align}
b_j&=2^{2j}\binom{s+j}{2j+1}\frac{(2j+1)!}{s}b_0~,\\
f_j&=2^{2j}\binom{s+j}{2j}(2j)! f_0~.
\end{align} 
\end{subequations}
To ensure that $\mc{C}_{\a(n)}(h)$ reduces to \eqref{2.31} in the flat limit, we choose $b_0$ and $f_0$ as follows
\begin{align}
b_0=\frac{1}{2^{2s-1}(2s-1)!}~,\qquad f_0=\frac{1}{2^{2s}(2s)!}~.
\end{align}
It is interesting to note that even though the imposition of transversality completely fixes the coefficients in \eqref{ansatz}, the resulting descendent is automatically gauge invariant (and vice versa). This can be attributed to the symmetry property \eqref{sym} and the Noether identity which follows as a consequence. Gauge invariance of \eqref{ansatz}, with the coefficients given by \eqref{coeffSol}, can also be checked explicitly by expressing the gauge transformations in the form $\delta_{\xi}h_{(n)} \propto\mc{D}_{(2)}\xi_{(n-2)}$ and using the identity \eqref{ID14}.
 
To conclude this section, we provide the 
final expressions for the higher-spin Cotton tensors  
with explicit indices.
They are given by  
\begin{subequations}\label{cotE}
\begin{align}
\mc{C}_{\a(2s)}(h)&=\frac{1}{2^{2s-1}}\sum_{j=0}^{s-1}2^{2j+1}\binom{s+j}{2j+1}\prod_{t=1}^{j}\bigg(\mc{Q}-\tau_{(s-t,2s)}\mc{S}^2\bigg) \non\\
&\phantom{\frac{1}{2^{2s-1}}\sum_{j=0}^{s-1}2^{2j+1}\binom{s+j}{2j+1}}\times\mc{D}_{\a(2)}^{s-j-1}\mc{D}_{\a}{}^{\b}\big(\mc{D}^{\b(2)}\big)^{s-j-1}h_{\a(2j+1)\b(2s-2j-1)}~, \label{cotF}\\
\mc{C}_{\a(2s+1)}(h)&=\frac{1}{2^{2s}}\sum_{j=0}^{s}2^{2j}\binom{s+j}{2j}\frac{(2s+1)}{(2j+1)}\prod_{t=1}^{j}\bigg(\mc{Q}-\tau_{(s-t+1,2s+1)}\mc{S}^2\bigg) ~~~~~~~~~~~~~~~~~~~~~~~~~~~\non\\
&\phantom{\frac{1}{2^{2s}}\sum_{j=0}^{s}\binom{s+j}{2j}\frac{(2s+1)}{(2j+1)}}\times\mc{D}_{\a(2)}^{s-j}\big(\mc{D}^{\b(2)}\big)^{s-j}h_{\a(2j+1)\b(2s-2j)}~, \label{cotB}
\end{align}
\end{subequations}
in the bosonic and fermionic cases respectively. We note that new expressions for the Cotton tensors in terms of the transverse projectors of AdS$_3$, and in terms of (only) the Casimir operators $\mc{F}$ and $\mc{Q}$, were obtained recently in \cite{AdS3(super)projectors}.


\subsection{Factorisation of the Cotton tensors and CHS action} \label{factor1}

The higher-spin Cotton tensors \eqref{cotE} are both gauge invariant and transverse.
This means that the action \eqref{CHSActionM3} for the conformal higher-spin field $h_{\a(n)}$,
\begin{align}
S_{\text{CHS}}^{(n)}[h]=\frac{\text{i}^n}{2^{\lfloor n/2 \rfloor+1}}\int\text{d}^3x\, e \, h^{\a(n)}\mc{C}_{\a(n)}(h)~,
\label{CSAdS3}
\end{align}
 is gauge invariant. Since $h_{\a(n)}$ and $\mc{C}_{\a(n)}(h)$ are primary fields with dimension \eqref{4777577398.4} and \eqref{2.13} respectively, 
the action is invariant under the conformal transformations \eqref{ConfT3D}. 
Furthermore, \eqref{CSAdS3} is symmetric in the sense that, modulo a total derivative, the relation
\begin{align} 
\int\text{d}^3x\, e \, g^{\a(n)}\mc{C}_{\a(n)}(h) = \int\text{d}^3x\, e \, h^{\a(n)}\mc{C}_{\a(n)}(g) \label{sym}
\end{align}
holds for arbitrary fields $g_{\a(n)}$ and $h_{\a(n)}$. 

Since the action is gauge invariant, we may impose the transverse gauge condition
\begin{align}
h_{\a(n)}\equiv h^{\text{T}}_{\a(n)}~,\qquad 0=\mc{D}^{\b(2)}h^{\text{T}}_{\b(2)\a(n-2)}~. \label{Tcond}
\end{align}
When the prepotential is transverse, only the $j=\lceil n/2 \rceil -1$ term contributes to \eqref{cotE}, and the corresponding Cotton tensors reduce to
\begin{subequations}\label{CF}
\begin{align}
\mc{C}_{\a(2s)}(h^{\text{T}})&=\prod_{t=1}^{ s-1 }\big(\mc{Q}-\tau_{(t,2s)}\mc{S}^2\big) \mc{D}_{\a}{}^{\b}h^{\text{T}}_{\a(2s-1)\b}~,\label{CFB}\\
\mc{C}_{\a(2s+1)}(h^{\text{T}})&=\prod_{t=1}^{ s}\big(\mc{Q}-\tau_{(t,2s+1)}\mc{S}^2\big) h^{\text{T}}_{\a(2s+1)}~. \label{CFF}
\end{align}
\end{subequations}
As a result, in this gauge the fermionic conformal higher-spin action \eqref{CSAdS3} fully factorises into products of minimal second order operators involving partial mass values of all possible depths. From \eqref{CFB} it is clear that the bosonic Cotton tensor does not wholly factorise and does not include the maximal depth partial mass $\tau_{(s,2s)}$ as a factor. However, in this case one may show that the following descendent of $\mc{C}_{\a(2s)}(h)$ fully factorises
\begin{align}
 \mc{D}_{\a}{}^{\b}\mc{C}_{\a(2s-1)\b}(h^{\text{T}})=\prod_{t=1}^{ s }\big(\mc{Q}-\tau_{(t,2s)}\mc{S}^2\big) h^{\text{T}}_{\a(2s)}~,\label{FE}
\end{align}
and its factors include all possible partial mass values. Using the identity \eqref{ID6}, from \eqref{CF} one may derive the following useful alternative forms for the Cotton tensors,
\begin{subequations} \label{cotFO}
\begin{align}
\mc{C}_{\a(2s)}(h^{\text{T}})&=\frac{1}{(2s)^{2s-1}}\mc{F}\prod_{t=1}^{ s-1 }\big(\mc{F}-\rho^{(-)}_{(t,2s)}\big) \big(\mc{F}-\rho^{(+)}_{(t,2s)}\big)h^{\text{T}}_{\a(2s)}~, \label{cotFOB}\\
\mc{C}_{\a(2s+1)}(h^{\text{T}})&=\frac{1}{(2s+1)^{2s}}\prod_{t=1}^{ s }\big(\mc{F}-\rho^{(-)}_{(t,2s+1)}\big) \big(\mc{F}-\rho^{(+)}_{(t,2s+1)}\big)h^{\text{T}}_{\a(2s+1)}~. \label{cotFOF}
\end{align}
\end{subequations}
We see that $\mc{C}_{\a(n)}(h)$ factorises into products of first-order differential operators involving all partial pseudo-mass values.\footnote{We note that $\rho_{(s,2s)}^{(\pm)}=0$, so that the maximal-depth bosonic pseudo-mass also appears in \eqref{cotFOB}.} 


\subsection{Partially massless prepotentials}

In this subsection we do not assume that $h_{\a(n)}$ is a conformal field.  From the above expressions for $\mc{C}_{\a(n)}(h)$, it is clear that the Cotton tensor of any on-shell partially massless field vanishes. More specifically, if $h_{\a(n)}\equiv h_{\a(n)}^{(t)}$ satisfies the on-shell conditions \eqref{OS1} and has depth-$t$, i.e. $\rho$ is given by \eqref{PM1}, then 
\begin{align}
\mc{C}_{\a(n)}(h)=0~.
\end{align}
In section \ref{SecGaugeCom} we saw that the vanishing of the Cotton tensor associated with the field $h_{\a(n)}$, on any conformally-flat background such as AdS$_3$, is a necessary and sufficient condition for $h_{\a(n)}$ to be pure gauge:\footnote{The derivation of \eqref{PG} in section \ref{SecGaugeCom} relied on $h_{\a(n)}$ being a CHS field. However, eq. \eqref{PG} may be shown to hold for a generic field using the AdS$_3$ spin projection operators of \cite{AdS3(super)projectors}. }
\begin{align}
\mc{C}_{\a(n)}(h)=0 \qquad \Longleftrightarrow \qquad h_{\a(n)}=\mc{D}_{\a(2)}\xi_{\a(n-2)}~, \label{PG}
\end{align}
 for some $\xi_{\a(n-2)}$. This provides another way to understand why partially massless fields in AdS$_3$ do not have any local dynamical degrees of freedom.

In fact, for on-shell partially massless fields $h_{\a(n)}^{(t)}$, the statement \eqref{PG} can be taken a step further: Writing  $h_{\a(n)}^{(t)}=\mc{D}_{\a(2)}\xi_{\a(n-2)}$ and requiring this to be transverse yields the equation
\begin{align}
0=\big(\tau_{(t,n)}-\tau_{(1,n)}\big)\mc{S}^2\xi_{\a(n-2)}-\frac{(n-2)(n-3)}{4(n-1)}\mc{D}_{\a(2)}\mc{D}^{\b(2)}\xi_{\a(n-4)\b(2)}~. \label{PGDT}
\end{align}
 Here we have made use of the 
 identities \eqref{ID14} and \eqref{PMWE}. Therefore, if $t=1$ (i.e. $h$ is massless) then $\xi_{\a(n-2)}$ is transverse. On the other hand, if $2\leq t \leq \lfloor n/2 \rfloor$ then \eqref{PGDT} implies
 \begin{align}
\xi_{\a(n-2)}=\mc{D}_{\a(2)}\xi_{\a(n-4)}\qquad \implies \qquad h_{\a(n)}^{(t)}=\mc{D}_{\a(2)}\mc{D}_{\a(2)}\xi_{\a(n-4)}~,
 \end{align}
 for some non-zero $\xi_{\a(n-4)}$. Once again, requiring $h_{\a(n)}^{(t)}$ to be transverse yields the equation
 \begin{align}
 0=\big(\tau_{(t,n)}-\tau_{(2,n)}\big)\mc{S}^2\mc{D}_{\a(2)}\xi_{\a(n-4)} - \frac{(n-4)(n-5)}{8(n-2)}\mc{D}_{\a(2)}\mc{D}_{\a(2)}\mc{D}^{\b(2)}\xi_{\a(n-6)\b(2)}~,
 \end{align} 
 and similar remarks hold. This procedure terminates when $t$ gradients have been extracted, whereupon $h_{\a(n)}^{(t)}$ takes the form
\begin{align}
h_{\a(n)}^{(t)}=\underbrace{\mc{D}_{\a(2)}\cdots\mc{D}_{\a(2)}}_{t-\text{times}}\xi_{\a(n-2t)}~,
\end{align}
 for some non-zero $\xi_{\a(n-2t)}$ being also on-shell (i.e. satisfying \eqref{GPC1}). The partially massless field $h_{\a(n)}^{(t)}$ will be said to be `pure depth-$t$ gauge'.


\subsection{Topologically massive higher-spin gauge models}

In this section we demonstrate that the properties of the higher-spin Cotton tensor allow us to analyse the on-shell dynamics of massive higher-spin gauge-invariant actions in AdS$_3$. 
The models presented are AdS$_3$ extensions of those proposed in section \ref{secMassiveActFlat}.

\subsubsection{New topologically massive higher-spin gauge models}

We recall that the models for new topologically massive (NTM) higher-spin fields are 
 formulated solely in terms of the gauge prepotentials $h_{\a(n)}$ and 
the associated Cotton tensors $\mc{C}_{\a(n)}(h)$.  
Having obtained explicit expressions for the latter in AdS, we now take a closer look at these models.

Given an integer $n\geq 2$, the gauge-invariant NTM action for the field $h_{\a(n)}$  is  
\begin{align}
S_{\text{NTM}}^{(n)}[h]=\frac{\text{i}^n}{2^{\lfloor n/2 \rfloor+1}}\frac{1}{\rho}\int\text{d}^3x\, e \, \mc{C}^{\a(n)}(h) \big(\mc{F}-\rho\big)h_{\a(n)}~, \label{HSNTMG}
\end{align}
where $\rho$ is some arbitrary mass parameter. The equation of motion obtained by varying \eqref{HSNTMG} with respect to the field $h_{\a(n)}$ is 
\begin{align}
0=\big(\mc{F}-\rho\big)\mc{C}_{\a(n)}(h)~. \label{EOM1}
\end{align}
 We may analyse the solutions to \eqref{EOM1} in terms of (i) the gauge prepotential $h_{\a(n)}$; or (ii) its field strength  $\mc{C}_{\a(n)}(h)$.

Let us first work in terms of the prepotential. Since the action \eqref{HSNTMG} is invariant under the gauge transformations \eqref{GTidc}, we may impose the transverse gauge condition \eqref{Tcond}, whereupon the field equation \eqref{EOM1} takes the form
\begin{subequations}\label{EOM2}
 \begin{align}
 0&=\big(\mc{F}-\rho\big)\mc{F}\prod_{t=1}^{ s-1 }\big(\mc{F}-\rho^{(-)}_{(t,2s)}\big) \big(\mc{F}-\rho^{(+)}_{(t,2s)}\big)h^{\text{T}}_{\a(2s)}~,\label{EOM2a}\\ 
 0&=\big(\mc{F}-\rho\big)\prod_{t=1}^{ s }\big(\mc{F}-\rho^{(-)}_{(t,2s+1)}\big) \big(\mc{F}-\rho^{(+)}_{(t,2s+1)}\big)h^{\text{T}}_{\a(2s+1)}~, \label{EOM2b}
 \end{align}
 \end{subequations}
in the bosonic and fermionic cases respectively. For simplicity, here we will only seek particular solutions to \eqref{EOM2}, rather than general ones. Clearly, any $h_{\a(n)}^{\text{T}}$ satisfying
\begin{align}
0=\big(\mc{F}-\rho\big)h_{\a(n)}^{\text{T}}~,\qquad \rho\neq \rho_{(t,n)}^{(\pm)}~,\label{EOM3}
\end{align}
 is a solution to \eqref{EOM2}. In addition, any $h_{\a(n)}^{\text{T}}\equiv h_{\a(n)}^{(t)}$ satisfying 
 \begin{align}
0=\big(\mc{F}-\rho_{(t,n)}^{(\pm)}\big) h_{\a(n)}^{(t)}~,\label{EOM4}
\end{align}
 for some $1\leq t \leq \lfloor n/2 \rfloor$, is also a solution. However, in this case its Cotton tensor vanishes
\begin{align}
0=\mc{C}_{\a(n)}(h^{(t)})~. \label{HSCF}
\end{align}
As discussed section \ref{factor1}, this condition is necessary and sufficient to conclude that $h_{\a(n)}^{(t)}$ is pure depth-$t$ gauge. Hence, of these solutions, the only non-trivial one is \eqref{EOM3}, which propagates a single degree of freedom with pseudo-mass $\rho$, spin $n/2$ and helicity $ \text{sgn}(\rho) n/2$.

If instead we wish to analyse solutions to \eqref{EOM1} in terms of the field strength $\mc{C}_{\a(n)}(h)$, then the analysis should be split into two cases. First, if $\mc{C}_{\a(n)}(h)$ satisfies 
\begin{align}
0=\big(\mc{F}-\rho\big)\mc{C}_{\a(n)}(h)~, \qquad \rho\neq \rho_{(t,n)}^{(\pm)}~, \label{EOM58}
\end{align}
then, in accordance with section \ref{section 2.2}, eq. \eqref{EOM58} together with the conservation identity \eqref{CotPropCFMb} means that the field strength $\mc{C}_{\a(n)}(h)$ itself defines an on-shell  spin $n/2$ field with pseudo-mass $\rho$, and helicity $ \text{sgn}(\rho) n/2$. Next we consider the case when $\mc{C}_{\a(n)}(h)$ satisfies 
\begin{align}
0=\big(\mc{F}-\rho_{(t,n)}^{(\pm)}\big)\mc{C}_{\a(n)}(h)~, \label{EOM59}
\end{align}
for some $1\leq t \leq \lfloor n/2 \rfloor$. 
 Now let us examine the Cotton tensor of the field strength $\mc{C}_{\a(n)}(h)$, which for ease of notation we denote $\mc{G}_{\a(n)}(h)$. Since $\mc{C}_{\a(n)}(h)$ is identically conserved, eq. \eqref{CotPropCFMb}, its Cotton tensor factorises according to \eqref{cotFO}, 
\begin{subequations} \label{cotFOG}
\begin{align}
\mc{G}_{\a(2s)}(h)&=\frac{1}{(2s)^{2s-1}}\mc{F}\prod_{t=1}^{ s-1 }\big(\mc{F}-\rho^{(-)}_{(t,2s)}\big) \big(\mc{F}-\rho^{(+)}_{(t,2s)}\big)\mc{C}_{\a(2s)}(h)~, \label{cotFOGB}\\
\mc{G}_{\a(2s+1)}(h)&=\frac{1}{(2s+1)^{2s}}\prod_{t=1}^{ s }\big(\mc{F}-\rho^{(-)}_{(t,2s+1)}\big) \big(\mc{F}-\rho^{(+)}_{(t,2s+1)}\big)\mc{C}_{\a(2s+1)}(h)~. \label{cotFOGF}
\end{align}
\end{subequations}
On account of \eqref{EOM59}, we conclude that $\mc{G}_{\a(n)}(h)=0$. Therefore, $\mc{C}_{\a(n)}(h)$ is itself pure gauge: 
\begin{align}
\mc{C}_{\a(n)}(h)= \mc{D}_{\a(2)}\widehat{\mc{C}}_{\a(n-2)}(h)~,\label{EOM60}
\end{align}
 for some $\widehat{\mc{C}}_{\a(n-2)}(h)$. Since $\mc{C}_{\a(n)}(h)$ is gauge invariant, we can take $h_{\a(n)}$ in \eqref{EOM60} to be transverse, whereupon the left hand side of \eqref{EOM60} factorises. Consequently, the only way that $\mc{C}_{\a(n)}(h)$ can have the form \eqref{EOM60}, is if $h_{\a(n)}$ is pure gauge, and hence $\mc{C}_{\a(n)}(h)=0$. 
 
 Therefore, from both points of view, the conclusion is that if the pseudo-mass $\rho$ in \eqref{HSNTMG} takes on one of the partially massless values, then it describes only pure gauge degrees of freedom. If however $\rho$ satisfies $\rho\neq \rho_{(t,n)}^{(\pm)}$, but is otherwise arbitrary,\footnote{For unitarity one should restrict $\rho$ to satisfy $|\rho|> n(n-2)\mc{S}$.} then on-shell the model describes a spin $n/2$ mode with pseudo-mass $\rho$ and helicity $ \text{sgn}(\rho) n/2$.


\subsubsection{Topologically massive higher-spin gauge models}

In this section we construct AdS$_3$ extensions to the Topologically massive higher-spin gauge models presented in section \ref{SecTMHS} and analyse their on-shell content. Such models
are obtained by coupling the AdS$_3$ counterparts of 
the massless Fronsdal \eqref{MasslessSOFlat} and  
Fang-Fronsdal \eqref{MasslessFOFlat} actions to the bosonic (even $n$)  and fermionic 
(odd $n$) Chern-Simons action \eqref{CSAdS3}, respectively. 

There are two types of the higher-spin massless actions, first-order and second-order ones. 
Given an integer $s\geq 2$, the first-order model is described by real fields 
$h_{\a(2s+1)},\,y_{\a(2s-1)}$ and $z_{\a(2s-3)}$, which are defined modulo gauge transformations of the form 
\begin{subequations} \label{FOGT}
\begin{align}
\d_{\xi} h_{\a(2s+1)}&=\mc{D}_{\a(2)}\xi_{\a(2s-1)} ~,\\
\d_{\xi} y_{\a(2s-1)}&=\frac{1}{(2s+1)(2s-1)}\big(\mc{F}-\rho^{(-)}_{(1,2s+1)}\big)\xi_{\a(2s-1)}~,\\
\d_{\xi} z_{\a(2s-3)}&=\mc{D}^{\b(2)}\xi_{\a(2s-3)\b(2)}~.
\end{align}
\end{subequations}
The three-dimensional counterpart of the  Fang-Fronsdal action is\footnote{There are actually two gauge invariant first-order Fang-Fronsdal actions in AdS$_3$ \cite{HHK}. The other is obtained by interchanging $\rho^{(+)}_{(1,2s+1)}$ with $\rho^{(-)}_{(1,2s+1)}$ everywhere in \eqref{FOGT} and \eqref{FOA}, or equivalently, sending $\mc{S}\rightarrow - \mc{S}$.}
\begin{align}
S_{\text{FF}}&^{(2s+1)}[h,y,z]=\bigg(-\frac{1}{2}\bigg)^s\frac{\ri}{2}\int \text{d}^3x\, e\, \bigg\{\frac{1}{(2s+1)}h^{\a(2s+1)}\big(\mc{F}-\rho^{(-)}_{(1,2s+1)}\big)h_{\a(2s+1)}  \non\\
&+4y^{\a(2s-1)}\big(\mc{F}-\rho^{(+)}_{(1,2s+1)}\big)y_{\a(2s-1)}-\frac{(s-1)}{s(2s-1)}z^{\a(2s-3)}\big(\mc{F}-\rho^{(-)}_{(1,2s+1)}\big)z_{\a(2s-3)} \non\\
&+2y^{\a(2s-1)}\bigg[(2s-1)\mc{D}^{\b(2)}h_{\a(2s-1)\b(2)}-\frac{(2s+1)(s-1)}{s}\mc{D}_{\a(2)}z_{\a(2s-3)}\bigg]\bigg\}~. \label{FOA}
\end{align}

Given an integer $s\geq 2$, the second-order model is described by the real fields 
$h_{\a(2s)}$ and $y_{\a(2s-4)}$,  which are
defined modulo gauge transformations of the form 
\begin{subequations} \label{SOGT}
\begin{align}
\d_{\x} h_{\a(2s)}&=\mc{D}_{\a(2)}\xi_{\a(2s-2)}~, \label{SOGT1}\\
\d_{\x} y_{\a(2s-4)}&=\frac{2s-2}{2s-1}\mc{D}^{\b(2)}\xi_{\a(2s-4)\b(2)}~.
\end{align}
\end{subequations}
The corresponding gauge-invariant Fronsdal-type action is
\begin{align}
S_{\text{F}}^{(2s)}[h,y]&=\frac{1}{2}\bigg(-\frac{1}{2}\bigg)^s\int\text{d}^3x \, e \,\bigg\{h^{\a(2s)}\big(\mc{Q}-\tau_{(1,2s)}\mc{S}^2\big)h_{\a(2s)} +\frac{s}{2}h^{\a(2s)}\mc{D}_{\a(2)}\mc{D}^{\b(2)}h_{\a(2s-2)\b(2)} \non\\
&-\frac{(2s-3)}{2}y^{\a(2s-4)}\mc{D}^{\b(2)}\mc{D}^{\b(2)}h_{\a(2s-4)\b(4)}-\frac{(2s-3)}{s}\bigg[y^{\a(2s-4)}\big(\mc{Q}-\tau_{(2,2s)}\mc{S}^2\big)y_{\a(2s-4)}\non\\
&-\frac{(s-2)(2s-5)}{8(s-1)}y^{\a(2s-4)}\mc{D}_{\a(2)}\mc{D}^{\b(2)}y_{\a(2s-6)\b(2)}\bigg]\bigg\}~. \label{SOA}
\end{align}
For an extension of the above first and second order massless actions to the case of general $n$ (not neccasarily odd or even), see \cite{Topological}.

When $n$ is even, $n=2s$, our action \eqref{SOA}
is the unique gauge-invariant 3D counterpart 
to the Fronsdal action in AdS${}_4$ \cite{Fronsdal2}.
The Fronsdal action \cite{Fronsdal2}
can also be generalised to $d$-dimensional AdS backgrounds
\cite{Segal01,Buchbinder:2001bs}. Such an action in AdS${}_d$ is formulated in terms 
of a symmetric double-traceless field 
and it is fixed  by the condition of gauge invariance.\footnote{The dynamical equations
for massless higher-spin fields in AdS${}_d$ were  studied by Metsaev
\cite{M1,M2,M3,M4}. For alternative descriptions of massless higher-spin dynamics
in AdS${}_d$, see \cite{CF,FMM}.
}

For spin $s+1/2$ and $s$ respectively, the corresponding topologically massive models are described by the gauge-invariant actions
\begin{subequations}\label{TM}
\begin{align}
S_{\text{TM}}^{(2s+1)}[h,y,z]&= S^{(2s+1)}_{\text{CHS}}[h]-\mu(\rho,s)S_{\text{FF}}^{(2s+1)}[h,y,z]~,\label{TMF}\\
S_{\text{TM}}^{(2s)}[h,y]&= S^{(2s)}_{\text{CHS}}[h]-\nu(\rho,s)S_{\text{F}}^{(2s)}[h,y]~.\label{TMB}
\end{align}
\end{subequations}
The functions $\mu(\rho,s)$ and $\nu(\rho,s)$ are defined by 
\begin{subequations} \label{coupleC}
\begin{align}
\mu(\rho,s)&=\frac{1}{(2s+1)^{2s-1}}\big(\rho-\rho_{(1,2s+1)}^{(+)}\big)\prod_{t=2}^{s}\big[\rho^2-\big(\rho_{(t,2s+1)}^{(\pm)}\big)^2\big]~,\label{muF}\\
\nu(\rho,s)&=\frac{1}{(2s)^{2s-3}}\big(\rho-\rho_{(s,2s)}^{(\pm)}\big)\prod_{t=2}^{s-1}\big[\rho^2-\big(\rho_{(t,2s)}^{(\pm)}\big)^2\big]~,\label{muB}
\end{align}
\end{subequations}
where $\rho\in \mb{R}$ is some arbitrary real constant with mass dimension one. For unitarity, the mass parameter $\rho$ must satisfy $|\rho|> n(n-2)\mc{S}$. For such values the functions \eqref{muF} and \eqref{muB} are always non-vanishing. For the (non-unitary) true partially massless values $\rho=\rho_{(t,n)}^{(\pm)}$, with $2\leq t \leq \lfloor n/2 \rfloor$, the coupling constants \eqref{coupleC} vanish.

It may be shown that on-shell, the model $S_{\text{TM}}^{(n)}$ describes an irreducible spin-$n/2$ field with pseudo-mass $\rho$ and helicity whose sign is equal to $\rho/|\rho|$. Let us sketch how this can be seen in the bosonic case with $n=2s$. 

In analogy with the analysis in section \ref{secTMHSbos},  using the equation of motion obtained by varying \eqref{TMB} with respect to $y_{\a(2s-4)}$, it follows that we may impose the gauge
\begin{align}
y_{\a(2s-4)}=0~,\qquad \mc{D}^{\b(2)}h_{\b(2)\a(2s-2)}=0~. \label{GF1}
\end{align} 
In this gauge the Cotton tensor factorises in accordance with section \ref{factor1}, and the equation of motion obtained by varying \eqref{TMB} with respect to $h_{\a(2s)}$ is
\begin{align}
0=\bigg(\mc{F}\prod_{t=2}^{s-1}\big(\mc{Q}-\tau_{(t,2s)}\mc{S}^2\big)+2s\nu(\rho,s)\bigg)\bigg(\mc{Q}-\tau_{(1,2s)}\mc{S}^2\bigg)h_{\a(2s)}~. \label{EOM5}
\end{align}
There are two types of solutions to \eqref{EOM5}. The first type solves $0=\big(\mc{Q}-\tau_{(1,2s)}\mc{S}^2\big)h_{\a(2s)}$, in which case $h_{\a(2s)}$ is massless with only pure gauge degrees of freedom. The second type of solutions are those satisfying
\begin{align}
0=\bigg(\mc{F}\prod_{t=2}^{s-1}\big(\mc{Q}-\tau_{(t,2s)}\mc{S}^2\big)+2s\nu(\rho,s)\bigg)h_{\a(2s)}~. \label{EOM6}
\end{align}
Once again, here we will only seek particular solutions to \eqref{EOM6}. For $\nu(\rho,s)$ defined as in \eqref{muB}, one can show that any $h_{\a(2s)}$ satisfying
\begin{align}
0=\big(\mc{F}-\rho\big)h_{\a(2s)} \quad \implies \quad 0=\big(\mc{Q}-[\rho^2/(2s)^2+4(s-1)(s+1)\mc{S}^2]\big)h_{\a(2s)}~,\label{sol1}
\end{align}
where $\rho$ satisfies $\rho \neq \rho_{(t,2s)}^{(\pm)}$, but is otherwise arbitrary, solves \eqref{EOM6}. We note that the residual gauge symmetry preserving the gauge fixing condition \eqref{GF1} is described by \eqref{SOGT1}, where the gauge parameter satisfies
\begin{align}
0=\mc{D}^{\b(2)}\xi_{\b(2)\a(2s-4)}~,\qquad 0=\big(\mc{F}-\rho_{(1,2s)}^{(\pm)}\big)\xi_{\a(2s-2)}~. \label{ResGS}
\end{align} 
For the massive solutions \eqref{sol1}, the residual gauge symmetry is exhausted since both \eqref{sol1} and \eqref{ResGS} imply $\xi_{\a(2s-2)}=0$ for $\rho \neq \rho_{(t,2s)}^{(\pm)}$.

 In accordance with section \ref{section 2.2}, the field $h_{\a(2s)}$ satisfying the second equation in \eqref{GF1} and the first equation in \eqref{sol1} defines an on-shell spin-$s$ field with pseudo-mass $\rho$ and helicity $s\rho/|\rho|$.  It is important to note that the above discussion assumes that the pseudo-mass does not take on any of the true partially massless values. If this is the case then the coupling constant  $\nu$ vanishes, and we are left with the Chern-Simons sector, which describes only pure gauge degrees of freedom.  The analysis in the fermionic case proceeds in a similar fashion and we do not repeat it here.


\section{Summary of results} \label{sec3DCHSdis}

This chapter was dedicated to the construction of CHS models in curved $3d$ spacetimes and applications thereof. We began in section \ref{sec3DCHSgen} by describing the salient features of the CHS field $h_{\a(n)}$, and its higher-spin Cotton tensor $\mc{C}_{\a(n)}(h)$, on a generic background. 
The descendent $\mc{C}_{\a(n)}(h)$ is defined \cite{Topological} such that if spacetime is conformally-flat, then it has the characteristic features of a covariantly conserved conformal current, leading to a gauge and Weyl invariant CHS action \eqref{CHSActionM3}. 
We then argued for the existence of $\mc{C}_{\a(n)}(h)$ by explicit construction in the cases $2\leq n \leq 4$ (eqs. \eqref{MaxwellianFS}, \eqref{linCgravitinoM3} and \eqref{NotQuiteLinCot}) \cite{Topological}. 

These calculations were done within the vielbein formulation, which we found to be too computationally expensive for derivation of arbitrary $n$ expressions. To bypass some of the associated technical difficulties, we employed the framework of conformal space in section \ref{sec3DCHSCF}. This allowed us to construct the gauge and Weyl invariant CHS action $S_{\text{CHS}}^{(n)}$ \eqref{CHS3DActCF} for all $n$ on arbitrary conformally-flat backgrounds \cite{Confgeo}. Gauge and Weyl actions \eqref{50.44} were also obtained for generalised depth-$t$ CHS fields on conformally-flat backgrounds \cite{Confgeo}.  
Finally, we deduced that it is not possible to extend the gauge invariance of $\mc{C}_{\a(n)}(h)$ to non-conformally-flat backgrounds through the addition of non-minimal terms \cite{Confgeo}. This was demonstrated by explicit computation in the spin-$5/2$ case \eqref{spin52Baby}. 

The actions \eqref{CHS3DActCF} were obtained by minimally lifting the expressions for $\mc{C}_{\a(n)}(h)$ in $\mb{M}^3$ to conformal space. In section \ref{sec3DCHSflat} we gave a novel derivation of $\mc{C}_{\a(n)}(h)$ in Minkowski space. In particular, their defining features (gauge invariance, transversality) were made manifest by expressing them in terms of the $3d$ spin-projection operators $\Pi_{\perp}^{[n]}$ \cite{3Dprojectors}, see \eqref{HSCotProj}.  The projectors $\Pi_{\perp}^{[n]}$, eq. \eqref{3DTTProj}, were constructed in terms of the operators $\Pi^{(\pm n)}$, eq. \eqref{3.14}, which project onto a state of definite helicity. The latter were derived in \cite{3Dprojectors}. This allows one to express $\mc{C}_{\a(n)}(h)$ as a sum of the positive and negative helicity modes of the prepotential, eq. \eqref{HelicityCotton}. 
  
The actions \eqref{CHS3DActCF} are constructed in terms of the conformally covariant derivative $\nabla_a$. For some purposes it is desirable to have expressions in terms of the Lorentz covariant derivative $\mc{D}_a$. However, even in the case of AdS$_3$, the process of degauging the actions \eqref{CHS3DActCF} is highly non-trivial. To avoid this issue, in section \ref{sec3DCHSAdS} we constructed $\mc{C}_{\a(n)}(h)$ directly in terms of the AdS$_3$ covariant derivative \cite{CottonAdS}, see eq. \eqref{cotE}. This allowed us to show, for the first time in a spacetime of odd dimension, that the AdS$_3$ CHS action factorises into second order operators associated with (partially-)massless fields of all depths \cite{CottonAdS}, see eq. \eqref{CF}.

Having the CHS action in terms of $\mc{D}_a$ allowed us to construct fermionic \eqref{TMF} and bosonic \eqref{TMB} topologically massive higher-spin actions in AdS$_3$, and their new variant \eqref{HSNTMG} \cite{CottonAdS}.\footnote{These models were first proposed in \cite{Topological}. However, in \cite{Topological} we were not able to obtain explicit expressions for $\mc{C}_{\a(n)}(h)$ with $n$ arbitrary in AdS$_3$.} The aforementioned factorisation properties facilitated their on-shell analysis. These topologically massive gauge models reduce to \eqref{HSTMGFlatF}, \eqref{HSTMGFlatB} and \eqref{HSNTMGflat} respectively in the flat limit.  The latter were first given in \cite{Topological}, although in principle they may be derived from $\mc{N}=1 \rightarrow \mc{N}=0 $  reduction of the models in \cite{KT17}, which in turn may be obtained from $\mc{N}=2 \rightarrow \mc{N}=1$ superspace reduction of the models in \cite{KO}.

%
%

\begin{subappendices}

\section{Higher-spin Cotton tensor as a descendant of the Fang-Fronsdal field strengths}

The Cotton tensor is defined in terms of the Ricci tensor according to \eqref{CottonRicci}. The latter determines the equations of motion corresponding to the Einstein-Hilbert action. In this appendix we show that a similar relation holds 
between the linearised higher-spin Cotton tensor \eqref{2.31} and the (Fang-)Fronsdal field strengths.


\subsubsection{The first-order case}

We begin by demonstrating that 
the higher-spin Cotton tensor  \eqref{2.31}
is a descendant of the gauge-invariant field strengths which determine 
the equations of motion in the first-order model \eqref{MasslessFOFlat}.
Associated with the dynamical variables $h_{\a(n)},y_{\a(n-2)}$ and $z_{\a(n-4)}$
are the following gauge-invariant field strengths:
\begin{subequations}
\bea
F_{\a(n)}(h)&:=&\partial_{(\a_1}{}^{\b}h_{\a_2\dots\a_{n})\b}-(n-2)\partial_{(\a_1\a_2}y_{\a_3\dots\a_{n})}~, \label{C.1a}\\
G_{\a(n-2)}(h,y,z)&:=&\partial^{\b(2)}h_{\a(n-2)\b(2)}+4\partial_{(\a_1}{}^{\b}y_{\a_2\dots\a_{n-2})\b}\non\\
&-&\frac{n(n-3)}{(n-1)(n-2)}\partial_{(\a_1\a_2}z_{\a_3\dots \a_{n-2})}~,~~~~~~ \label{C.1b}\\
H_{\a(n-4)}(y,z)&:=&(n-2)\partial^{\b(2)}y_{\a(n-4)\b(2)}-\frac{n-4}{n}\partial_{(\a_1}{}^{\b}z_{\a_2\dots\a_{n-4})\b}~. \label{C.1c}
\eea
\end{subequations}
The equations of motion corresponding to \eqref{MasslessFOFlat} are the conditions 
that these field strengths vanish. Furthermore, the gauge symmetry implies that  $F_{\a(n)},~G_{\a(n-2)}$ and $ H_{\a(n-4)}$ are related to each other via the  
Noether identity
\bea
0=\partial^{\b(2)}F_{\a(n-2)\b(2)}-\frac{n-2}{n}\partial_{(\a_1}{}^{\b}G_{\a_2\dots\a_{n-2})\b}+\frac{n(n-3)}{(n-1)(n-2)}\partial_{(\a_1\a_2}H_{\a_3\dots\a_{n-2})}~. \label{C.2}
\eea

We claim that the Cotton tensor $\mc{C}_{\a(n)} (h)$
may be expressed as 
$\mc{C}_{\alpha(n)}=(\mathcal{A}_1F)_{\a(n)}+(\mathcal{A}_2G)_{\a(n)}+(\mathcal{A}_3H)_{\a(n)}$, for some linear differential operators $\mathcal{A}_i$ of order $n-2$. A suitable ansatz for such an expression is
\begin{align}
\mc{C}_{\a(n)}&=\sum_{j=0}^{\lfloor\frac{n}{2}\rfloor-1}a_j\square^j\partial_{(\a_1}{}^{\b_1}\cdots\partial_{\a_{n-2j-2}}{}^{\b_{n-2j-2}}F_{\a_{n-2j-1}\dots\a_{n})\b_1\dots\b_{n-2j-2}} \notag\\
&+\sum_{k=0}^{\lceil \frac{n}{2} \rceil-2}b_k\square^k\partial_{(\a_1}{}^{\b_1}\cdots\partial_{\a_{n-2k-3}}{}^{\b_{n-2k-3}}\partial_{\a_{n-2k-2}\a_{n-2k-1}}G_{\a_{n-2k}\dots\a_{n})\b_1\dots\b_{n-2k-3}}\label{C.3}\\
&+\sum_{l=0}^{\lfloor\frac{n}{2}\rfloor-2}c_l\square^l\partial_{(\a_1}{}^{\b_1}\cdots\partial_{\a_{n-2l-4}}{}^{\b_{n-2l-4}}\partial_{\a_{n-2l-3}\a_{n-2l-2}}
\notag\\
&
\qquad \qquad 
\times 
\partial_{\a_{n-2l-1}\a_{n-2l}}
H_{\a_{n-2l+1}\dots\a_{n})\b_1\dots\b_{n-2l-4}} \notag
\end{align}
for some coefficients $a_j,b_k$ and $c_l$. It may be shown that the values of these coefficients are not unique and that there are $\lfloor\frac{n}{2}\rfloor-1$ free parameters.  For example, when $n=5$ one may show that the general solution is 
\bea
\begin{pmatrix}
a_0\\a_1\\b_0\\b_1\\c_0
\end{pmatrix} =\begin{pmatrix}
\frac{1}{2}+\frac{18}{5}c_0\\\frac{1}{2}-\frac{18}{5}c_0\\\frac{9}{80}-\frac{36}{25}c_0\\\frac{3}{80}-\frac{18}{25}c_0\\c_0
\end{pmatrix}~. \notag
\eea
We may use this freedom to eliminate the $\lfloor\frac{n}{2}\rfloor-1$ coefficients $c_l$ so that only the 
field strengths $F_{\a(n)}$ and $G_{\a(n-2)}$ appear in \eqref{C.3}. This fixes the solution uniquely to
\begin{subequations}
\bea
a_j&=&\frac{1}{2^{n-2}}\frac{(n-1)}{(2j+1)}\binom{n-2}{2j} \qquad \qquad ~~~~ \text{for}~~ 0\leq j \leq \bigg\lfloor\frac{n}{2}\bigg\rfloor-1~, \label{C.4a}\\[2 ex]
b_k&=&\frac{1}{2^{n-1}}\frac{(n-2)^2}{n(2k+1)}\binom{n-3}{2k} \qquad\qquad ~~\text{for}~~0\leq k \leq \bigg\lceil \frac{n}{2} \bigg\rceil-2~, \label{C.4b} \\[2ex]
c_l&=&0 \phantom{BLANK SPACEEEEEEEEE}~ \text{for}~~ 0\leq l \leq \bigg\lfloor\frac{n}{2}\bigg\rfloor-2~.
\eea
\end{subequations}
The fact that there are $\lfloor\frac{n}{2}\rfloor-1$ free parameters may be understood as a consequence of the Noether identity \eqref{C.2}. To see this, observe that, in principle, we may use \eqref{C.2} to replace all occurrences of $H_{\a(n-4)}$ with $F_{\a(n)}$ and $G_{\a(n-2)}$ in the ansatz \eqref{C.3}. There will then be only two sets of independent coefficients, say $\tilde{a}_j$ and $\tilde{b}_k$, 
whose unique  values coincide with those of \eqref{C.4a} and \eqref{C.4b}.  


\subsubsection{The second-order case}

We now consider the the second-order model \eqref{MasslessFOFlat}, which is described by the real fields $h_{\a(n)}$ and $y_{\a(n-4)}$. 
Associated with these two fields are the following gauge-invariant field strengths:
\begin{subequations}
 \begin{align}
 F_{\a(n)}(h,y)&:=\square h_{\a(n)}+\frac{n}{4}\partial^{\b(2)}\partial_{(\a_1\a_2}h_{\a_3\dots\a_{n})\b(2)}-\frac{n-3}{4}\partial_{(\a_1\a_2}\partial_{\a_3\a_4}y_{\a_5\dots\a_{n})}~,\label{C.5a}\\
 G_{\a(n-4)}(h,y)&:= \partial^{\b(2)}\partial^{\b(2)}h_{\a(n-4)\b(4)}+\frac{8}{n}\square y_{\a(n-4)} \non\\
 &-\frac{(n-4)(n-5)}{n(n-2)}\partial^{\b(2)}\partial_{(\a_1\a_2}y_{\a_3\dots\a_{n-4})\b(2)}~.~~~~~~~~~~~\label{C.5b}
 \end{align}
 \end{subequations}
 The equations of motion for the model
   are $F_{\a(n)}=0$ and $G_{\a(n-4)}=0$. The two field strengths are related by the Noether identity
 \bea
 \partial^{\b(2)}F_{\a(n-2)\b(2)}=\frac{(n-3)(n-2)}{4(n-1)}\partial_{(\a_1\a_2}G_{\a_3\dots\a_{n-2})}~.
 \eea
 We claim that the Cotton tensor $\mc{C}_{\a(n)} (h)$
 may be written as 
 $\mc{C}_{\alpha(n)}=(\mathcal{A}_1F)_{\a(n)}+(\mathcal{A}_2G)_{\a(n)}$ where the $\mathcal{A}_i$ are linear differential operators of order $n-3$. A suitable ansatz for such an expression is
\begin{align}
\mc{C}_{\a(n)}=&\sum_{j=0}^{\lceil \frac{n}{2} \rceil-2}a_j\square^j\partial_{(\a_1}{}^{\b_1}\cdots\partial_{\a_{n-2j-3}}{}^{\b_{n-2j-3}}F_{\a_{n-2j-2}\dots\a_{n})\b_1\dots\b_{n-2j-3}} \label{C.7}\\
&+\sum_{k=0}^{\lceil \frac{n}{2} \rceil-3}b_k\square^k\partial_{(\a_1}{}^{\b_1}\cdots\partial_{\a_{n-2k-5}}{}^{\b_{n-2k-5}}
\non \\
& \qquad 
\times
\partial_{\a_{n-2k-4}\a_{n-2k-3}}\partial_{\a_{n-2k-2}\a_{n-2k-1}}
G_{\a_{n-2k}\dots\a_{n})\b_1\dots\b_{n-2k-5}}~,
\notag~
\end{align}
for some coefficients $a_j$ and $b_k$. It may be shown that the choice of these coefficients is not unique, and that there are $\lceil \frac{n}{2} \rceil-2$ free parameters. For example, when $n=6$ one may show that the general solution is 
\bea
\begin{pmatrix}
a_0\\a_1\\b_0
\end{pmatrix} =
\begin{pmatrix}
\frac{5}{8}-\frac{10}{3}b_0 \\ \frac{3}{8}+\frac{10}{3}b_0\\b_0
\end{pmatrix}~. \notag
\eea
We can use this freedom to completely eliminate the $\lceil \frac{n}{2} \rceil-2$ coefficients $b_k$ so that only the top field strength, $F_{\a(n)}$, appears in \eqref{C.7}. This gives the unique solution
\begin{subequations}
\bea
a_j&=&(j+1)\frac{\binom{n-3}{2j}}{\binom{2j+3}{3}}\frac{n(n-1)}{3\cdot 2^{n-2}}~\phantom{BLANK SPA} \text{for}~~0\leq j\leq \bigg\lceil \frac{n}{2} \bigg\rceil-2~,\\
b_k&=&0 \phantom{BLAN SPACEEEEEEEEE}~~~~~~~ \text{for}~~ 0\leq k\leq \bigg\lceil \frac{n}{2} \bigg\rceil-3~.
\eea
\end{subequations}

\section{Properties of the generalised HS Cotton tensor}  \label{AppGenCotProp}
In this appendix we present the main steps that are needed in order to prove the two properties \eqref{50.45} of the generalised higher-spin Cotton tensor $\mc{C}^{(t)}_{\a(n)}(h)$. Namely, that in any conformally flat spacetime it is partially conserved and gauge invariant.

It may be shown that the $t^{\text{th}}$ divergence of $\mc{C}^{(t)}_{\a(n)}(h)$ is given by
\begin{align}
&2^{n-2t+1}\nabla^{\b_1\b_2}\cdots\nabla^{\b_{2t-1}\b_{2t}}\mc{C}^{(t)}_{\a(n-2t)\b(2t)}\notag\\
&=\sum_{j=t-1}^{\lceil \frac n2 \rceil -1}\binom{n}{2j+1}\binom{j}{t-1}(\Box_c)^{j-t+1}\nabla^{\b_1\b_2}\cdots\nabla^{\b_{2t-1}\b_{2t}}\notag\\
&\phantom{Spa}\times\frac{1}{n!}\sum_{k=0}^{t}\binom{t}{k}\binom{n-2j-1}{2k}(2k)!\binom{n-2t}{n-2j-2k-1}(n-2j-2k-1)!(2j+1)! \notag\\
&\phantom{Spa}\times\nabla_{\b_1}{}^{\g_1}\cdots\nabla_{\b_{2k}}{}^{\g_{2k}}\nabla_{(\a_1}{}^{\g_{2k+1}}\cdots\nabla_{\a_{n-2j-2k-1}}{}^{\g_{n-2j-1}}h^{(t)}_{\a_{n-2j-2k}\dots\a_{n-2t})\b_{2k+1}\dots\b_{2t}\g_1\dots\g_{n-2j-1}} \notag\\
&=\sum_{k=0}^{t}\sum_{j=t-k}^{\lceil \frac n2 \rceil -k-1}(-1)^k\binom{n}{2j+1}\binom{j}{t-1}\binom{t}{k}\binom{n-2j-1}{2k}\binom{n-2t}{n-2j-2k-1}\frac{(2k)!(2j+1)!}{n!}\notag\\
&\phantom{Spa}\times(n-2j-2k-1)!(\Box_c)^{j+k-t+1}\nabla^{\g_1\g_2}\cdots\nabla^{\g_{2t-1}\g_{2t}}\nabla_{(\a_1}{}^{\g_{2t+1}}\cdots\nabla_{\a_{n-2j-2k-1}}{}^{\g_{n-2j-2k+2t-1}}\notag\\
&\phantom{Spa}\times h^{(t)}_{\a_{n-2j-2k}\dots\a_{n-2t})\g_1\dots\g_{n-2j-2k+2t-1}} \notag\\
&=\sum_{j=t}^{\lceil \frac n2 \rceil -1}\binom{n-2t}{n-2j-1}(\Box_c)^{j-t+1}\nabla^{\g_1\g_2}\cdots\nabla^{\g_{2t-1}\g_{2t}}\nabla_{(\a_1}{}^{\g_{2t+1}}\cdots\nabla_{\a_{n-2j-1}}{}^{\g_{n-2j+2t-1}}\notag\\
&\phantom{Spa}\times h^{(t)}_{\a_{n-2j}\dots\a_{n-2t})\g_1\dots\g_{n-2j+2t-1}}\bigg\{\sum_{k=0}^{t}(-1)^k\binom{j-k}{t-1}\binom{t}{k}\bigg\} ~.\notag
\end{align}
Making use of the combinatoric identity
 \begin{align}
 \sum_{k=0}^{t}(-1)^k \binom{j-k}{t-1}\binom{t}{k}=0 \qquad \forall ~j\geq t~,\label{G.1}
 \end{align}
 which may be proved by induction on $t$, it follows that the last line in the above is equal to zero. In the second line we have used the combinatoric factors to shift the upper and lower bounds of the summation over $j$. Then, in the third line we have shifted the dummy variable $j\mapsto j-k$.

 Under the gauge transformations \eqref{50.39}, one may show that $\mc{C}^{(t)}_{\a(n)}(h)$ transforms as 
 \begin{align}
 \delta_{\x}&\bigg(2^{n-2t+1}\mc{C}^{(t)}_{\a(n)}(h)\bigg)\notag\\
 =&\sum_{j=t-1}^{\lceil \frac n2 \rceil-1}\binom{n}{2j+1}\binom{j}{t-1}(\Box_c)^{j-t+1}\nabla_{(\a_1}{}^{\b_1}\cdots\nabla_{\a_{n-2j-1}}{}^{\b_{n-2j-1}}\notag\\
 &\phantom{Spa}\times\frac{1}{n!}\sum_{k=0}^{t}\binom{n-2j-1}{2k}\binom{t}{k}(2k)!\binom{2j+1}{2t-2k}(2t-2k)!(n-2t)! \notag\\
 &\phantom{Spa}\times\nabla_{|\b_1\b_2}\cdots\nabla_{\b_{2k-1}\b_{2k}|} \nabla_{\a_{n-2j}\a_{n-2j+1}}\cdots\nabla_{\a_{n-2j-2k+2t-2}\a_{n-2j-2k+2t-1}}\notag\\
 &\phantom{Spa}\times\x_{\a_{n-2j-2k+2t}\dots\a_n)\b_{2k+1}\dots\b_{n-2j-1}} \notag\\
 =&\sum_{k=0}^{t}\sum_{j=t-k}^{\lceil \frac n2 \rceil-k-1}(-1)^k\binom{n}{2j+1}\binom{j}{t-1}\binom{n-2j-1}{2k}\binom{t}{k}\binom{2j+1}{2t-2k}\frac{(2k)!(2t-2k)!(n-2t)!}{n!} \notag\\
 &\phantom{Spa}\times (\Box_c)^{j+k-t+1}\nabla_{(\a_1\a_2}\cdots\nabla_{\a_{2t-1}\a_{2t}}\nabla_{\a_{2t+1}}{}^{\b_{2t+1}}\cdots\nabla_{\a_{n-2j-2k+2t+1}}{}^{\b_{n-2j-2k+2t+1}}\notag\\
 &\phantom{Spa}\times\x_{\a_{n-2j-2k+2t+2}\dots\a_n)\b_{2t+1}\dots\b_{n-2j-2k+2t+1}} \notag\\
 =&\sum_{j=t}^{\lceil \frac n2 \rceil-1}\binom{n-2t}{2j-2t+1}(\Box_c)^{j-t+1}\nabla_{(\a_1\a_2}\cdots\nabla_{\a_{2t-1}\a_{2t}}\nabla_{\a_{2t+1}}{}^{\b_{2t+1}}\cdots\nabla_{\a_{n-2j+2t+1}}{}^{\b_{n-2j+2t+1}} \notag\\
 &\phantom{Spa}\times\x_{\a_{n-2j+2t+2}\dots\a_n)\b_{2t+1}\dots\b_{n-2j+2t+1}}\bigg\{\sum_{k=0}^{t}(-1)^k \binom{j-k}{t-1}\binom{t}{k}\bigg\}\notag.
 \end{align}
It follows that the gauge variation vanishes after making use of \eqref{G.1} once more. In the second line of the above we have used the combinatoric factors to shift the upper and lower bounds of the summation over $j$. Then, in the third line we have shifted the dummy variable $j\mapsto j-k$.

\section{Generating function identities in AdS$_3$} \label{AppGenForm}

In order to easily prove the properties of the higher-spin Cotton tensors in AdS$_3$, it will be useful to reformulate the problem into one in terms of homogeneous polynomials. This framework is also sometimes referred to as the generating function formalism. 

Associated with a
symmetric 
rank-$n$ spinor field $\phi_{\a(n)}:=\phi_{\a_1\dots\a_n}=\phi_{(\a_1\dots\a_n)}$ 
is 
 a homogeneous polynomial $\phi_{(n)}(\U)$ of degree $n$ defined by 
\begin{align}
\phi_{(n)}:=\U^{\a_1}\cdots\U^{\a_n}\phi_{\a_1 \dots \a_n}~,
\end{align}
where the  auxiliary variables $\U^{\a}$ are chosen to be commuting,
hence 
$\U^{\a}\U_{\a}:=\U^{\a}\ve_{\a\b}\U^{\b}=0$.
The correspondence $\phi_{\a(n)} \to \phi_{(n)}$ is one-to-one.
The linear space of such polynomials will be denoted $\mc{H}_{(n)}$.

Introducing the auxiliary derivative $\pa_{\a}:=\frac{\pa}{\pa\U^{\a}}$, whose index is raised according to the usual rule $\pa^{\a}:=\ve^{\a\b}\pa_{\b}$, we define the AdS differential operators
\begin{align}
\mc{D}_{(2)}:=\U^{\a}\U^{\b}\mc{D}_{\a\b}~,\qquad \mc{D}_{(0)}:=\U^{\a}\mc{D}_{\a}{}^{\b}\pa_{\b}~,\qquad \mc{D}_{(-2)}:=\mc{D}^{\a\b}\pa_{\a}\pa_{\b}~.
\end{align}
They raise the degree of homogeneity of any polynomial on which they act by 2, 0 and -2 respectively. Amongst themselves, they may be shown to satisfy the algebra\footnote{The auxiliary variables $\U^{\a}$ are defined to be inert with respect to the Lorentz generators. Alternatively, one could define their action on any $\phi_{(n)}\in \mc{H}_{(n)}$ to be $M_{\a\b}\phi_{(n)}:=-\U_{(\a}\pa_{\b)}\phi_{(n)}$.}
\begin{subequations}\label{ID2}
\begin{align}
\big[\mc{D}_{(2)}, \mc{D}_{(-2)}\big]&=4\big(\bm{\U}_{(0)} +1\big)\Box +8\mc{S}^2 \bm{M}_{(0)}\big(\bm{\U}_{(0)}+1\big)~,\label{ID2a}\\
\big[\mc{D}_{(2)}, \mc{D}_{(0)}\big]&=4\mc{S}^2\bm{M}_{(2)}\big(\bm{\U}_{(0)}+2\big)~,\label{ID2b}\\
\big[\mc{D}_{(-2)}, \mc{D}_{(0)}\big]&=-4\mc{S}^2\bm{M}_{(-2)}\big(\bm{\U}_{(0)}+2\big)~.\label{ID2c}
\end{align}
\end{subequations}
where we have made use of the definitions\footnote{The identities \eqref{ID3c} and \eqref{ID3d} imply that the commutators \eqref{ID2b} and \eqref{ID2c} vanish on $\mc{H}_{(n)}$. This is not surprising given that on $\mc{H}_{(n)}$, the operator $\mc{D}_{(0)}$ can be identified with the quadratic Casimir $\mc{F}$.}
\begin{subequations}\label{ID3}
\begin{align}
\bm{\U}_{(0)}&:=\U^{\a}\pa_{\a}~,\qquad \qquad ~~~~~~~~\bm{\U}_{(0)}\phi_{(n)}=n\phi_{(n)}~,\label{ID3a}\\
\bm{M}_{(0)}&:=\U^{\a}M_{\a}{}^{\b}\pa_{\b}~,\qquad \qquad~\bm{M}_{(0)}\phi_{(n)}=-\frac{1}{2}n(n+2)\phi_{(n)}~,\label{ID3b}\\
\bm{M}_{(2)}&:=\U^{\a}\U^{\b}M_{\a\b}~, \qquad \qquad \bm{M}_{(2)}\phi_{(n)}=0~,\label{ID3c}\\
\bm{M}_{(-2)}&:=M^{\a\b}\pa_{\a}\pa_{\b}~, \qquad \qquad \bm{M}_{(-2)}\phi_{(n)}=0~.\label{ID3d}
\end{align}
\end{subequations}
Using the above identities it is possible to show, via induction on $t$, that for any $\phi_{(n)}\in \mc{H}_{(n)}$ the following relations hold
\begin{subequations}
\begin{align}
\big[\mc{D}_{(2)}, \mc{D}^{\phantom{.}t}_{(-2)}\big]\phi_{(n)}&=4t(n-t+2)\big(\mc{Q}-\tau_{(t,n+2)}\mc{S}^2\big)\mc{D}_{(-2)}^{t-1}\phi_{(n)}~,\label{ID14}\\
\big[\mc{D}_{(-2)}, \mc{D}^{\phantom{.}t}_{(2)}\big]\phi_{(n)}&=-4t(n+t)\big(\mc{Q}-\tau_{(t,n+2t)}\mc{S}^2\big)\mc{D}_{(2)}^{t-1}\phi_{(n)}~.\label{ID15}
\end{align}
\end{subequations}
Here $\mc{Q}$ is the Casimir operator \eqref{Q} and $\tau_{(t,n)}$ are the partially massless values \eqref{PM2}.

\end{subappendices}


\chapter{CHS models in four dimensions} \label{Chapter4D}

In this chapter we elaborate on models which describe the gauge invariant dynamics of conformal higher-spin fields on various types of curved four dimensional backgrounds. The models presented are higher-spin generalisations of the linearised action for conformal gravity in $d=4$, which was discussed in section \ref{sectionCG4}. Gauge invariance of the latter holds at most on Bach-flat backgrounds, and as such, this is suspected to be the most general type of background on which CHS fields can consistently propagate. 

For every conformal gauge field $h_{(m,n)}$ of Lorentz type $(\frac{m}{2}, \frac{n}{2})$, with $m\geq 1$ and $n\geq 1$ integers, 
we construct a primary field strength $\mc{W}_{(m,n)}(h)$ known as the linearised higher-spin Weyl tensor. Its properties are such that knowledge of $\mc{W}_{(m,n)}(h)$ amounts to full knowledge of the CHS action on conformally-flat backgrounds. 
For quite some time it was believed that on Bach-flat backgrounds, it was possible to construct another primary field strength $\mc{B}_{(m,n)}(h)$, known as the linearised higher-spin Bach tensor. Its properties are such that it completely determines the corresponding CHS action on Bach-flat backgrounds. However, recent studies indicate that, in general, such a descendent does not exist and one is required to couple $h_{(m,n)}$ to additional fields to achieve gauge invariance. Indeed, below we present the first few examples of complete models exhibiting this mechanism. 
 
This chapter is based on the publications \cite{Confgeo, AdSprojectors, spin3depth3, SCHS, SCHSgen, AdS3(super)projectors} and is organised as follows. 
Remarks on the general structure of CHS models on an arbitrary background are given in section \ref{secCHSM4}. In section \ref{secCHSMink4} we describe the gauge invariant model for $h_{(m,n)}$ in $\mb{M}^4$ and provide the corresponding transverse projectors. These results are extended to AdS$_4$ in section \ref{secCHSAdS4} and are used to show that (i)  the poles of the projectors correspond to partially-massless fields; and (ii) 
the CHS kinetic operator factorises into products of second-order operators. In section \ref{secCHSCF4} we derive all gauge invariant CHS models on conformally-flat backgrounds, including their generalised cousins with higher-derivative gauge transformations.
In section \ref{secCHSBach} we derive gauge-invariant models for various CHS fields on Bach-flat backgrounds. 
A summary of the results obtained is given in section \ref{sec4DCHSdis}.

\section{Conformal higher-spin models in $\mc{M}^{4}$} \label{secCHSM4}



Given a four dimensional spacetime $\big(\cM^4,e_{a}{}^{m}\big)$, its geometry may be described in terms of the torsion-free Lorentz covariant derivative $\mc{D}_{a}$ satisfying the algebra \eqref{LCDCommutator}. The latter may be rewritten using \eqref{WeylT} as follows
\begin{align}
 \big[ \cD_a , \cD_b \big] = -\hf W_{ab}{}^{cd} M_{cd}+\hat{R}^c{}_{[b}M_{b]c}-\frac{1}{12}RM_{ab}~, \label{VecDer4}
\end{align}
where $\hat{R}_{ab}:=R_{ab}-\frac{1}{4}\eta_{ab}R$ is the traceless Ricci tensor. 

In order to transition smoothly to $d=4$ two component spinor notation, it is useful to recall the decomposition 
\begin{align}
W_{abcd}=W^{(+)}_{abcd}~+~W_{abcd}^{(-)}~,
\end{align}
of the Weyl tensor into its self-dual and anti-self-dual parts respectively,
\begin{align}
W^{(\pm)}_{abcd}:=\frac{1}{2}W_{abcd}\mp\frac{\ri}{4}\varepsilon_{ab}{}^{fg}W_{cd fg}~,\qquad \frac{1}{2}\varepsilon_{ab}{}^{fg}W_{fg cd}^{(\pm)}=\pm\ri W_{abcd}^{(\pm)}~.
\end{align}
The latter possess the same algebraic properties as the Weyl tensor,
\begin{align}
W^{(\pm)}_{abcd}=-W^{(\pm)}_{bacd}&=-W^{(\pm)}_{abdc}=W^{(\pm)}_{cdab}~, \qquad  W^{(\pm)}_{[abc]d}=0~,\qquad W^{(\pm)}_{cab}{}^{c}=0~.
\end{align}
It may be shown that $W_{abcd}$ is in one-to-one correspondence with the two totally symmetric rank four spinor fields $W_{\a\b\g\d}=W_{(\a\b\g\d)}$ and $\bar{W}_{\ad\bd\gd\dd}=\bar{W}_{(\ad\bd\gd\dd)}$ via the relations
\begin{subequations} \label{SpinorWeylTensor}
\begin{align}
W_{\a\b\g\d}&= \frac{1}{2}(\s^{ab})_{\a\b}(\s^{cd})_{\g\d}W^{(-)}_{abcd}=\frac{1}{2}(\s^{ab})_{\a\b}(\s^{cd})_{\g\d}W_{abcd}~,\\
\bar{W}_{\ad\bd\gd\dd}&= \frac{1}{2}(\tilde{\s}^{ab})_{\ad\bd}(\tilde{\s}^{cd})_{\gd\dd}W^{(+)}_{abcd}= \frac{1}{2}(\tilde{\s}^{ab})_{\ad\bd}(\tilde{\s}^{cd})_{\gd\dd}W_{abcd}~,\\
W_{abcd}&=\frac{1}{2}(\s_{ab})^{\a\b}(\s_{cd})^{\g\d}W_{\a\b\g\d}+\frac{1}{2}(\tilde{\s}_{ab})^{\ad\bd}(\tilde{\s}_{cd})^{\gd\dd}\bar{W}_{\ad\bd\gd\dd}~.
\end{align}
\end{subequations}
Accordingly, the tensors $\bar{W}_{\ad(4)}$ and $W_{\a(4)}$ describe the self-dual and anti-self-dual parts of the Weyl tensor respectively. Since they encode all information about the Weyl tensor and are the complex conjugates of each other, $\bar{W}_{\ad(4)}=\big(W_{\a(4)}\big)^*$, we will simply refer to both of them as the Weyl tensor.  We note that the following relation also holds
\begin{align}
W_{\a\ad\b\bd\g\gd\d\dd}:=\big(\s^a\big)_{\a\ad}\big(\s^b\big)_{\b\bd}\big(\s^c\big)_{\g\gd}\big(\s^d \big)_{\d\dd}W_{abcd}=2\ve_{\ad\bd}\ve_{\gd\dd}W_{\a\b\g\d}+2\ve_{\a\b}\ve_{\g\d}\bar{W}_{\ad\bd\gd\dd}~.
\end{align}  

Using these definitions, it may be shown that in two component notation the algebra of covariant derivatives \eqref{VecDer4} takes the form
\begin{align}
-\big[\mc{D}_{\a\ad},\mc{D}_{\b\bd}\big]=\ve_{\ad\bd}&\Big[W_{\a\b}{}^{\g(2)}M_{\g(2)}+\frac{1}{2}R_{\a\b}{}^{\gd(2)}\bar{M}_{\gd(2)}+\frac{1}{6}RM_{\a\b}\Big] \non\\
+\ve_{\a\b}&\Big[\bar{W}_{\ad\bd}{}^{\gd(2)}\bar{M}_{\gd(2)}+\frac{1}{2}R^{\g(2)}{}_{\ad\bd}M_{\g(2)}+\frac{1}{6}R\bar{M}_{\ad\bd}\Big]~.
\end{align}
Finally, we note that the Bach tensor \eqref{BachT} corresponds to the real rank-(2,2) spinor
\begin{align}
B_{\a\b\ad\bd}:=\big(\s^a\big)_{\a\ad}\big(\s^b\big)_{\b\bd}B_{ab}=\Big(\mc{D}_{(\ad}{}^{\g}\mc{D}_{\bd)}{}^{\g}-\frac{1}{2}R^{\g(2)}{}_{\ad\bd}\Big)W_{\a\b\g(2)}=\bar{B}_{\a\b\ad\bd}~.\label{BachSpinor}
\end{align}
Above we have defined the real rank-$(2,2)$ spinor $R_{\a\b\ad\bd}:=\big(\s^a\big)_{\a\ad}\big(\s^b\big)_{\b\bd}\hat{R}_{ab}=R_{(\a\b)(\ad\bd)}$.

\subsection{Conformal higher-spin gauge fields} \label{secCHS4prep}

Let us fix two positive integers $m \geq 1$ and $ n \geq 1 $.
In two-component spinor notation,  
a complex tensor field $h_{\a(m)\ad(n) } = h_{\a_1 \dots \a_m \ad_1\dots\ad_n } 
=h_{(\a_1 \dots \a_m)(\ad_1\dots\ad_n)} $  is said to be a conformal spin-$\frac{1}{2}(m+n)$ gauge field on $\mc{M}^4$ if 
\begin{enumerate}[label=(\roman*)]
\item   $h_{\a(m)\ad(n)}$ is a primary tensor field with conformal weight $\Delta_{h_{(m,n)}}=\big(2-\frac{1}{2}(m+n)\big)$, 
\bea
\d^{(\text{weyl})}_\s h_{\a(m)\ad(n)} = \Big(2-\frac{1}{2}(m+n)\Big)\s h_{\a(m)\ad(n)}~. \label{4777577398.44}
\eea
\item  $h_{\a(m)\ad(n)}$  is defined modulo gauge transformations of the form
\begin{align}
\d_{\xi} h_{\a(m)\ad(n) } =\mc{D}_{(\a_1 (\ad_1 } \xi_{\a_2 \dots \a_m)\ad_2\dots\ad_n) }~, \label{GTCHSprepM4}
\end{align}
where the complex gauge parameter $\xi_{\a(m-1)\ad(n-1)}$ is a primary field with conformal weight $\Delta_{\xi_{(m-1,n-1)}}=\big(1-\frac{1}{2}(m+n)\big)$ .

\end{enumerate}
The consistency of these two properties is what uniquely fixes $\Delta_{h_{(m,n)}}$.

The corresponding properties of the field $\bar{h}_{\a(m)\ad(n)}$ may be obtained from those above via complex conjugation. In the bosonic spin-$s$ case, specifically when $m=n=s$, it is possible to impose the reality condition
\begin{align}
h_{\a(s)\ad(s)}=\bar{h}_{\a(s)\ad(s)}~. \label{RealityCHS}
\end{align}
In this case the prepotential $h_{\a(s)\ad(s)}$ is equivalent to the one described in section \ref{SectionCHSprep}, which in vector notation was denoted by $h_{a(s)}$.

By virtue of property (i), under $\mc{G}$ the prepotential transforms according to the rule
\begin{align}
\d_{\L}^{(\mc{G})} h_{\a(m)\ad(n)} =\Big( \xi^a\mc{D}_a +\frac{1}{2}K^{ab}M_{ab}+\Delta_{h_{(m,n)}}\s \Big) h_{\a(m)\ad(n)} ~. \label{transruleM444}
\end{align}
Let us consider the background spacetime $\big(\cM^4,e_{a}{}^{m}\big)$  to be fixed, and suppose that it admits a conformal Killing vector field $\z = \z^m \pa_m = \z^a e_a$, 
\bea
 \cD^a \z^b + \cD^b \z^a = 2 \eta^{ab} \s[\z] ~, \qquad \s[\z] = \frac{1}{4} \cD_b \z^b  ~.
\label{CCVF1.31}
 \eea
 Then, in accordance with the discussion in section \ref{SectionHobbit}, the primary tensor field $h_{\a(m)\ad(n)}$  possesses the following conformal transformation law
\begin{align}
\d^{(\mc{G})}_{\L[\z]} h_{\a(m)\ad(n)} = \Big(\z^b \cD_b +\hf K^{bc}[\z] M_{bc} + \Delta_{h_{(m,n)}} \s[\z]\Big) h_{\a(m)\ad(n)} ~, \label{ConfT4D} 
\end{align}
where $K^{bc} [\z] =  \cD^{[b} \z^{c]}$. In the case of a Minkowski background,  $\big(\mc{M}^4,e_a{}^{m}\big)=\big(\mb{M}^4,\delta_a{}^{m}\big)$,
\eqref{ConfT4D} reduces to the rigid conformal transformation rule \eqref{CHSprepPrimary} upon 
substituting \eqref{CKVflat} for $\z^a$. 


\subsection{Linearised higher-spin Weyl and Bach tensors} \label{secLinWeylBachVA}

Starting  with $h_{\a(m)\ad(n) } $, one can construct two important descendants known as the linearised higher-spin Weyl tensors, which we denote by $\mc{W}^{(m,n)}_{\a(m+n)}(h)$ and $\overline{\mc{W}}^{(m,n)}_{\ad(m+n)}(h)$. The labels $(m,n)$ indicate the Lorentz type of the field on which they act. We will usually only be dealing with fixed (but arbitrary) values of $m$ and $n$, or in other words a single CHS field $h_{\a(m)\ad(n) } $ (along with its conjugate) at a time. Therefore, when it is clear which gauge field we are talking about, we will often drop these labels. 

Let us denote by $V_{(m,n)}$ the linear space of tensor fields with Lorentz type $(\frac{m}{2},\frac{n}{2})$. From an operator point of view, $\mc{W}^{(m,n)}$ converts all dotted indices of its argument into undotted ones, $\mc{W}^{(m,n)}:V_{(m,n)}\rightarrow V_{(m+n,0)}$. On the other hand, $\overline{\mc{W}}^{(m,n)}$ converts all undotted indices into dotted ones,  $\overline{\mc{W}}^{(m,n)}:V_{(m,n)}\rightarrow V_{(0,m+n)}$.
Given a CHS field $h_{\a(m)\ad(n)}$, its higher-spin Weyl tensors are defined by the following  properties: 
\begin{enumerate}[label=(\roman*)]

\item
$\mc{W}_{\a(m+n)}(h)$ and $\overline{\mc{W}}_{\ad(m+n)}(h)$ are primary tensors with conformal weights $\Delta_{\mc{W}^{(m,n)}}=\big(2-\frac{1}{2}(m-n)\big)$ and $\Delta_{\overline{\mc{W}}^{(m,n)}}=\big(2+\frac{1}{2}(m-n)\big)$ respectively, 
\begin{subequations} \label{HSWeylWeylT}
\begin{align}
\d_\s^{(\text{weyl})} \mc{W}_{\a(m+n)}(h) &= \Big(2-\frac{1}{2}(m-n)\Big)  \s \mc{W}_{\a(m+n)}(h)~,\\
\d_\s^{(\text{weyl})} \overline{\mc{W}}_{\ad(m+n)}(h) &= \Big(2+\frac{1}{2}(m-n)\Big)  \s \overline{\mc{W}}_{\ad(m+n)}(h)~.\label{ThisIsNotImportant}
\end{align}
\end{subequations}

\item  
$\mc{W}_{\a(m+n)}(h)$ and $\overline{\mc{W}}_{\ad(m+n)}(h)$  are descendants of the form $\mc{W}_{\a(m+n)}(h)=(\hat{\cA} h)_{\a(m+n)}$ and $\overline{\mc{W}}_{\ad(m+n)}(h)=(\check{\cA} h)_{\ad(m+n)}$. Here $\hat{\cA}$ and $\check{\cA}$ are some linear differential operators of order $n$ and $m$ respectively, each involving 
the Lorentz covariant derivative $\mc{D}_a$, the curvature tensors
 $\big(W_{\a(4)},\bar{W}_{\ad(4)}\big)$, $R_{\a(2)\ad(2)}$ and $R$, 
and their covariant derivatives.

\item
$\mc{W}_{\a(m+n)}(h)$ and $\overline{\mc{W}}_{\ad(m+n)}(h)$ have vanishing gauge variation under \eqref{GTCHSprepM4} if spacetime is conformally-flat,
\begin{subequations} \label{HSWeylGIM4}
\begin{align}
\d_{\xi} \mc{W}_{\a(m+n)} (h)&= \mc{O}\big( W\big)~,\\
\d_{\xi} \overline{\mc{W}}_{\ad(m+n)} (h)&= \mc{O}\big( W\big)~.
\end{align}
\end{subequations}
Here $\mc{O}\big( W\big)$ stands for contributions involving the Weyl tensor and
its covariant derivatives.
\end{enumerate}

The operators $\mc{W}^{(m,n)}$ and $\overline{\mc{W}}^{(m,n)}$ act naturally on the space $V_{(m,n)}$, with $m$ and $n$ being arbitrary positive integers. As the notation suggests, we will see soon that the two types of HS Weyl tensors are related through complex conjugation as follows
\begin{align}
\Big(\mc{W}^{(m,n)}_{\a(m+n)}(h)\Big)^*=\overline{\mc{W}}^{(n,m)}_{\ad(m+n)}(\bar{h})~,\qquad \Big(\mc{W}^{(n,m)}_{\a(m+n)}(\bar{h})\Big)^*=\overline{\mc{W}}^{(m,n)}_{\ad(m+n)}(h)~.\label{HSWeylCC}
\end{align}
Therefore, we will usually choose to formulate models using only  $\mc{W}^{(m,n)}$.

Another important conformal field strength that may be constructed from $h_{\a(m)\ad(n)}$ is the linearised higher-spin Bach tensor $\mc{B}^{(m,n)}_{\a(n)\ad(m)}(h)$. As an operator, it interchanges the number of dotted and undotted indices carried by its argument, $\mc{B}^{(m,n)}:V_{(m,n)}\rightarrow V_{(n,m)}$. For a given CHS field $h_{\a(m)\ad(n)}$, it is defined to possess the following properties:
\begin{enumerate}[label=(\roman*)]
\item
$\mc{B}_{\a(n)\ad(m)}(h)$ is a primary tensor with conformal weight $\Delta_{\mc{B}^{(m,n)}}=\big(2+\frac{1}{2}(m+n)\big)$, 
\begin{align}
\d_\s^{(\text{weyl})} \mc{B}_{\a(n)\ad(m)}(h) = \Big(2+\frac{1}{2}(m+n) \Big)  \s \mc{B}_{\a(n)\ad(m)}(h)~.
\label{HSBachWeylT}
\end{align}

\item  
$\mc{B}_{\a(n)\ad(m)}(h)$ is of the form $\mc{B}_{\a(n)\ad(m)}(h)=(\cA h)_{\a(n)\ad(m)}$, 
where $\cA$ is a linear differential operator of order $(m+n)$ involving 
the Lorentz covariant derivative $\mc{D}_a$, the curvature tensors
 $\big(W_{\a(4)},\bar{W}_{\ad(4)}\big)$, $R_{\a(2)\ad(2)}$ and $R$, 
and their covariant derivatives.

\item
$\mc{B}_{\a(n)\ad(m)}(h)$ has vanishing gauge variation under \eqref{GTCHSprepM4} if spacetime is Bach-flat,
\begin{align}
\d_{\xi} \mc{B}_{\a(n)\ad(m)} (h)= \mc{O}\big( B\big)~.
\label{HSBachGIM4}
\end{align}

\item
 $\mc{B}_{\a(n)\ad(m)}(h)$ is divergenceless if spacetime is Bach-flat,
\begin{align}
\mc{D}^{\b\bd} \mc{B}_{\b\a(n-1)\bd\ad(m-1)}(h) = \mc{O}\big( B\big)~.\label{HSBachDivM4}
\end{align} 
Here and in \eqref{HSBachGIM4}, 
$\mc{O}\big( B\big)$ stands for contributions involving the Bach tensor and
its covariant derivatives.
\end{enumerate}

As we discuss below, for non-conformally-flat backgrounds, a field strength  $ \mc{B}_{\a(n)\ad(m)} (h)$ satisfying these properties does not exist for general $m$ and $n$. 
It should also be noted that the above properties do not determine the higher-spin Weyl or Bach tensors uniquely in non-conformally-flat and non-Bach-flat backgrounds respectively. 

\subsection{Conformal higher-spin action }
Suppose that the background spacetime under consideration is conformally flat, 
\begin{align}
W_{\a(4)}=0~.
\label{CFM4}
\end{align}
Then from \eqref{HSWeylGIM4} it follows that the tensors $\mc{W}^{(m,n)}_{\a(m+n)}(h)$ and $\mc{W}^{(n,m)}_{\a(m+n)}(\bar{h})$
are invariant under the gauge transformations \eqref{GTCHSprepM4},
\begin{align}
 \delta_{\xi}\mc{W}^{(m,n)}_{\a(m+n)}(h)=0~,\qquad \delta_{\xi}\mc{W}^{(n,m)}_{\a(m+n)}(\bar{h})=0~. \label{HSWeylGICFM}
\end{align}
As a consequence, the property \eqref{HSWeylGICFM} and the Weyl transformation laws \eqref{HSWeylWeylT} (recall that $\Delta_{\mc{W}^{(n,m)}}=2+\frac{1}{2}(m-n)$), tell us that the following action\footnote{Of course, we could have instead used the Lagrangian $\ms{L}=  \overline{\mc{W}}_{(m,n)}^{\ad(m+n)}(h)\overline{\mc{W}}^{(n,m)}_{\ad(m+n)}(\bar{h})$. However, on account of \eqref{HSWeylCC}, this structure is already contained within the `$+$c.c.' sector of \eqref{CHSActionM4}. } 
\begin{align}
S_{\rm{CHS}}^{(m,n)} [ h,\bar{h}]  
=\frac{1}{2}\text{i}^{m+n} \int \rd^4 x \, e\, \mc{W}_{(m,n)}^{\a(m+n)}(h)\mc{W}^{(n,m)}_{\a(m+n)}(\bar{h}) +\text{c.c.}~,
\label{CHSActionM4}
\end{align}
is both gauge and Weyl invariant, 
\bea
\d_\xi S_{\rm{CHS}}^{(m,n)} [ h,\bar{h}]  =0~, \qquad 
\d^{(\text{weyl})}_\s S_{\rm{CHS}}^{(m,n)} [ h,\bar{h}]  =0~.
\eea
In \eqref{CHSActionM4} the `$+$c.c.' means to add the complex conjugate, so that the action is real.  

Suppose instead that the background spacetime under consideration is Bach-flat,
\begin{align}
B_{\a(2)\ad(2)}=0~.
\end{align}
Then the properties \eqref{HSBachGIM4} and \eqref{HSBachDivM4} imply that the higher-spin Bach tensor is both gauge invariant and transverse\footnote{The field
$\mc{D}^{\b\bd} \mc{B}_{\b\a(m-1)\bd\ad(n-1)}(h)$ is primary,
and hence \eqref{Thismaybeconserved} is Weyl invariant.}
\begin{subequations}
\begin{align}
\d_{\xi} \mc{B}_{\a(n)\ad(m)} (h)&=0~,\\
\mc{D}^{\b\bd} \mc{B}_{\b\a(n-1)\bd\ad(m-1)}(h) &=0~. \label{Thismaybeconserved}
\end{align}
\end{subequations}
This in turn implies that the action
\begin{align}
S_{\rm{CHS}}^{(m,n)} [ h,\bar{h}]  
=\frac{1}{2}\text{i}^{m+n} \int \rd^4 x \, e\, \bar{h}^{\a(n)\ad(m)}\mc{B}_{\a(n)\ad(m)}(h) +\text{c.c.}~, \label{CHSActionB4}
\end{align}
is gauge invariant on a Bach-flat background. On account of \eqref{HSBachWeylT} and \eqref{4777577398.44} it is also Weyl invariant on an arbitrary background. We note that, by virtue of the invariance of \eqref{CHSActionM4} and \eqref{CHSActionB4} under the spacetime gauge group $\mc{G}$, it follows that it is invariant under the conformal transformations \eqref{ConfT4D}, 
 \begin{align}
 \d^{(\mc{G})}_{\L[\z]}S_{\rm{CHS}}^{(m,n)} [ h,\bar{h}] =0~,
 \end{align}
where the background is held fixed, $\d^{(\mc{G})}_{\L[\z]} e_{a}{}^{m}=0$. 

The two functionals \eqref{CHSActionM4} and \eqref{CHSActionB4} represent two seemingly different  ways to formulate the conformal higher-spin action in four dimensions. On conformally-flat backgrounds we will see that the two actually coincide. On the other hand, on Bach-flat (but non-conformally-flat) backgrounds, the full gauge invariant CHS action cannot be represented in the form \eqref{CHSActionM4},\footnote{By inspection of the linearised conformal gravity action \eqref{CGquad} this fact is already apparent.} and \eqref{CHSActionB4} will be the more fundamental formulation.

 In fact, as we will see, the situation is not so simple. In particular, for all values of $m$ and $n$, the higher-spin Weyl tensors $\mc{W}^{(m,n)}_{\a(m+n)}(h)$ are guaranteed to exist.\footnote{Indeed, in section \ref{secCHSCF4} we construct explicit expressions for them. We stress that although they exist on non-conformally flat backgrounds,  they will not be gauge invariant, as dictated by property (iii). } In contrast, it will be shown through explicit examples that the higher-spin Bach tensor $\mc{B}_{\a(n)\ad(m)}(h)$ does not necessarily exist on Bach-flat backgrounds. See section \ref{secCHSBach} for further discussion.  

\section{Conformal higher-spin models in $\mb{M}^{4}$} \label{secCHSMink4}
So far our discussion of the models for $4d$ CHS fields $h_{\a(m)\ad(n)}$ has been rather abstract.
In this section we recall the concrete expressions for them in Minkowski space $\mb{M}^4$.
In the case $|m-n|=0$ and $|m-n|=1$, they were first described in \cite{FT}. The actions for mixed symmetry CHS fields with $|m-n|=\text{even}$  were given in \cite{Vasiliev2009}, and for   $|m-n|=\text{odd}$ in \cite{KMT}. 
We also present their formulation in terms of the $4d$ spin-projection operators. 

\subsection{CHS models}

Let us fix two integers $m\geq 1$ and $n\geq 1$, and consider the CHS gauge field $h_{\a(m)\ad(n)}$ and its conjugate $\bar{h}_{\a(n)\ad(m)}$. In $\mb{M}^4$, $h_{\a(m)\ad(n)}$ is defined modulo the gauge transformations 
\begin{align}
\delta_{\xi}h_{\a(m)\ad(n)}=\pa_{(\a_1(\ad_1}\xi_{\a_2\dots\a_m)\ad_2\dots\ad_n)}~. \label{CHSGTMink4}
\end{align}
Explicit expressions for the higher-spin Weyl tensors with only undotted indices are\footnote{For $|m-n|=0$, the higher-spin Weyl tensors were first written down by Weinberg \cite{WeinbergWeyl} and in the general $m\neq n$ case by 
Gates et al. \cite{GGRS, SG} (see also \cite{FL1, FL-4D} for $|m-n|=0,1$).  }
\begin{subequations} \label{HSWeylIMink4}
\begin{align}
\mc{W}_{\a(m+n)}(h)&:=\pa_{(\a_1}{}^{\bd_1}\cdots\pa_{\a_n}{}^{\bd_n}h_{\a_{n+1}\dots\a_{m+n})\bd(n)}~,\label{HSWeylIMink4a}\\
\mc{W}_{\a(m+n)}(\bar{h})&:=\pa_{(\a_1}{}^{\bd_1}\cdots\pa_{\a_m}{}^{\bd_m}\bar{h}_{\a_{m+1}\dots\a_{m+n})\bd(m)}~,\label{HSWeylIMink4b}
\end{align}
whilst those for the higher-spin Weyl tensors with only dotted indices are given by 
\begin{align}
\overline{\mc{W}}_{\ad(m+n)}(h)&:=\pa_{(\ad_1}{}^{\b_1}\cdots\pa_{\ad_m}{}^{\b_m}h_{\b(m)\ad_{m+1}\dots\ad_{m+n})}~,\label{HSWeylIIMink4a}\\
\overline{\mc{W}}_{\ad(m+n)}(\bar{h})&:=\pa_{(\ad_1}{}^{\b_1}\cdots\pa_{\ad_n}{}^{\b_n}\bar{h}_{\b(n)\ad_{n+1}\dots\ad_{m+n})}~.\label{HSWeylIIMink4b}
\end{align}
\end{subequations}
Now it is clear that they are related via complex conjugation according to eq. \eqref{HSWeylCC},
\begin{align}\label{HSWeylCCRelsMink4}
\Big(\mc{W}_{\a(m+n)}(h)\Big)^*=\overline{\mc{W}}_{\ad(m+n)}(\bar{h})~,\qquad \Big(\mc{W}_{\a(m+n)}(\bar{h})\Big)^*=\overline{\mc{W}}_{\ad(m+n)}(h)~.
\end{align}

The higher-spin Weyl tensors are invariant under the gauge transformations \eqref{CHSGTMink4},\footnote{A $4d$ analogue of the gauge completeness property was provided in \cite{DD}: A prepotential is pure gauge if and only if its Weyl tensor vanishes, $\mc{W}_{\a(m+n)}(\phi)=0 ~ \Leftrightarrow ~ \phi_{\a(m)\ad(n)}=\pa_{\a\ad}\xi_{\a(m-1)\ad(n-1)}$.}
\begin{align}
\delta_{\xi}\mc{W}_{\a(m+n)}(h)=0~,\qquad \delta_{\xi}\mc{W}_{\a(m+n)}(\bar{h})=0~.
\end{align} 
This means that the following action for the conformal higher-spin field $h_{\a(m)\ad(n)}$ 
\begin{align}
S_{\rm{CHS}}^{(m,n)} [ h,\bar{h}] 
=\frac{1}{2}\text{i}^{m+n} \int \rd^4 x \, \mc{W}^{\a(m+n)}(h)\mc{W}_{\a(m+n)}(\bar{h}) +\text{c.c.}~,
\label{CHSActionMink4}
\end{align}
is gauge invariant. It is also invariant under the conformal transformations \eqref{ConfT4D}, 
\begin{align}
\delta_{\xi}S_{\rm{CHS}}^{(m,n)} [ h,\bar{h}]  =0~,\qquad \delta^{(\mc{G})}_{\L[\z]}S_{\rm{CHS}}^{(m,n)} [ h,\bar{h}]  =0~.
\end{align}


We may obtain an alternative form of the CHS action by integrating \eqref{CHSActionMink4} by parts and moving all derivatives onto the prepotential $h_{\a(m)\ad(n)}$, with the result 
\begin{align}
S_{\rm{CHS}}^{(m,n)} [ h,\bar{h}] 
=\frac{1}{2}\text{i}^{m+n} \int \rd^4 x \, \bar{h}^{\a(n)\ad(m)}\mc{B}_{\a(n)\ad(m)}(h) +\text{c.c.} \label{CHSABachMink4a}
\end{align}
On the other hand, we might instead start with the `$+$c.c.' sector in \eqref{CHSActionMink4}, and then move all derivatives on $h_{\a(m)\ad(n)}$, with the result
\begin{align}
S_{\rm{CHS}}^{(m,n)} [ h,\bar{h}] 
=\frac{1}{2}\text{i}^{m+n} \int \rd^4 x \, \bar{h}^{\a(n)\ad(m)}\widehat{\mc{B}}_{\a(n)\ad(m)}(h) +\text{c.c.}\label{CHSABachMink4b}
\end{align}
In eqs. \eqref{CHSABachMink4a} and \eqref{CHSABachMink4b} we have defined the higher-spin linearised Bach tensors
\begin{subequations}\label{HSBachPrepMink}
\begin{align}
\mc{B}_{\a(n)\ad(m)}(h)&:=\pa_{(\ad_1}{}^{\b_1}\cdots \pa_{\ad_m)}{}^{\b_m}\pa_{(\a_1}{}^{\bd_1}\cdots\pa_{\a_n}{}^{\bd_n}h_{\b_1\dots\b_m)\bd(n)}~,\label{HSBachPrepMinka}\\
\widehat{\mc{B}}_{\a(n)\ad(m)}(h)&:=\pa_{(\a_1}{}^{\bd_1}\cdots \pa_{\a_n)}{}^{\bd_n}\pa_{(\ad_1}{}^{\b_1}\cdots\pa_{\ad_m}{}^{\b_m}h_{\b(m)\bd_1\dots\bd_n)}~.\label{HSBachPrepMinkb}
\end{align}
\end{subequations}
The two types of Bach tensors turn out to be equivalent to each other,\footnote{In view of \eqref{HSBachHSWeyl}, this may be interpreted as a Bianchi identity on the HS Weyl tensor, see e.g. \cite{KRDuality}. }
\begin{align}
\mc{B}_{\a(n)\ad(m)}(h)=\widehat{\mc{B}}_{\a(n)\ad(m)}(h)~. \label{HSBachTypeEquiv}
\end{align}
To prove this one can assume, without loss of generality, that $m\geq n$. It may then be shown that both sides of the equality \eqref{HSBachTypeEquiv} evaluate to 
\begin{align}
\frac{m!n!}{(m+n)!}\sum_{j=0}^{n}\binom{m}{j}\binom{n}{j}\Box^j\Big(\pa_{\a}{}^{\bd}\Big)^{n-j}\Big(\pa_{\ad}{}^{\b}\Big)^{m-j} h_{\a(j)\b(m-j)\ad(j)\bd(n-j)}~. 
\end{align}
As a consequence of \eqref{HSBachTypeEquiv}, we have the following relation\footnote{In the special case when $m=n=s$, identities \eqref{HSBachTypeEquiv} and \eqref{HSBachCCMink} tell us that $\mc{B}_{\a(s)\ad(s)}(h)$ is real. } 
\begin{align}
\mc{B}_{\a(m)\ad(n)}(\bar{h})=\Big(\mc{B}_{\a(n)\ad(m)}(h)\Big)^*\equiv \overline{\mc{B}}_{\a(m)\ad(n)}(\bar{h}) ~.\label{HSBachCCMink}
\end{align}

One can show that $\mc{B}_{\a(n)\ad(m)}(h)$ is transverse and gauge invariant
\begin{subequations}\label{HSBachMink4Prop}
\begin{align}
\pa^{\b\bd}\mc{B}_{\a(n-1)\b\ad(m-1)\bd}(h)&=0~,\label{HSBachMink4TT}\\
\delta_{\xi}\mc{B}_{\a(n)\ad(m)}(h)&=0~.\label{HSBachMink4GI}
\end{align}
\end{subequations}
 Property \eqref{HSBachMink4TT} follows from equivalence of \eqref{CHSActionMink4} and \eqref{CHSABachMink4a}, and the gauge invariance of \eqref{CHSActionMink4}.
Property \eqref{HSBachMink4GI} is manifest when written in terms of the Weyl tensors
\begin{subequations}\label{HSBachHSWeyl}
\begin{align}
\mc{B}_{\a(n)\ad(m)}(h)&=\pa_{(\ad_1}{}^{\b_1}\cdots \pa_{\ad_m)}{}^{\b_m}\mc{W}_{\a(n)\b(m)}(h)~,\\
\widehat{\mc{B}}_{\a(n)\ad(m)}(h)&=\pa_{(\a_1}{}^{\bd_1}\cdots \pa_{\a_n)}{}^{\bd_n}\overline{\mc{W}}_{\ad(m)\bd(n)}(h)~.
\end{align}
\end{subequations}
The equation of motion for the CHS action is the vanishing of the higher-spin Bach tensor.

The overall normalisation of $S_{\rm{CHS}}^{(m,n)} [ h,\bar{h}]$ can be explained as follows. The factor of i$^{m+n}$ is chosen over, say, i$^{m+n+1}$ on account of the identity (valid modulo a total derivative)
\begin{align}
\text{i}^{m+n+1}\int \rd^4 x \,  \mc{W}^{\a(m+n)}(h)\mc{W}_{\a(m+n)}(\bar{h})
+{\rm c.c.} = 0~,
\end{align}
which follows from \eqref{HSBachCCMink}. Using \eqref{HSBachCCMink} one can also show that
\begin{align}
S_{\rm{CHS}}^{(m,n)} [ h,\bar{h}] 
=\text{i}^{m+n} \int \rd^4 x \, \mc{W}^{\a(m+n)}(h)\mc{W}_{\a(m+n)}(\bar{h})~,
\end{align}
which explains the factor of $\frac{1}{2}$ in \eqref{CHSActionMink4}. Therefore, the following are all equivalent expressions for one and  the same conformal invariant 
\begin{align} 
\int \rd^4 x \, &\mc{W}^{\a(m+n)}(h)\mc{W}_{\a(m+n)}(\bar{h})  = \int \rd^4 x \, \bar{h}^{\a(n)\ad(m)}\mc{B}_{\a(n)\ad(m)}(h) \non\\
&=  \int \rd^4 x \, \bar{h}^{\a(n)\ad(m)}\widehat{\mc{B}}_{\a(n)\ad(m)}(h)=\int \rd^4 x \,\overline{\mc{W}}^{\ad(m+n)}(h)\overline{\mc{W}}_{\ad(m+n)}(\bar{h})~. 
\end{align}

\subsection{Transverse projectors} \label{secTTProjMink4}

It is instructive to recall the $4d$ spin $s=\frac{1}{2}(m+n)$ transverse projectors $\Pi_{\perp}^{(m,n)}$ and $\widehat{\Pi}_{\perp}^{(m,n)}$. Their action on 
an arbitrary field $\phi_{\a(m)\ad(n)} \in V_{(m,n)}$ is defined as 
\begin{subequations}\label{TTprojMink}
\begin{align}
\Pi_{\perp}^{(m,n)}\phi_{\a(m)\ad(n)}&:=\Delta_{(\ad_1}{}^{\b_1}\cdots\Delta_{\ad_n)}{}^{\b_n}\Delta_{(\b_1}{}^{\bd_1}\cdots\Delta_{\b_n}{}^{\bd_n}\phi_{\a_1\dots\a_m)\bd(n)} ~,\label{TTprojMinka}\\
\widehat{\Pi}_{\perp}^{(m,n)}\phi_{\a(m)\ad(n)}&:=\Delta_{(\a_1}{}^{\bd_1}\cdots\Delta_{\a_m)}{}^{\bd_m}\Delta_{(\bd_1}{}^{\b_1}\cdots\Delta_{\bd_m}{}^{\b_m}\phi_{\b(m)\ad_1\dots\ad_n)}\label{TTprojMinkb} ~.
\end{align}
\end{subequations}
Here we have made use of the non-local\footnote{In the transverse projectors \eqref{TTprojMink} there is always an even integer power of $\Box^{-\frac{1}{2}}$.} involutive operator
\begin{align}
\Delta_{\a}{}^{\ad}=\frac{\pa_{\a}{}^{\ad}}{\sqrt{\Box}}~,\qquad \Delta_{\a}{}^{\bd}\Delta_{\bd}{}^{\b}=\delta_{\a}{}^{\b}~,\qquad \Delta_{\ad}{}^{\b}\Delta_{\b}{}^{\bd}=\delta_{\ad}{}^{\bd}~.\label{TTProjDeltaOp}
\end{align}
The operators $\Pi_{\perp}^{(m,n)}$ and $\widehat{\Pi}_{\perp}^{(m,n)}$ satisfy the properties
\begin{subequations}\label{TTproj4prop}
\begin{align}
\Pi_{\perp}^{(m,n)}\Pi_{\perp}^{(m,n)}=\Pi_{\perp}^{(m,n)}~&,\qquad \widehat{\Pi}_{\perp}^{(m,n)}=\widehat{\Pi}_{\perp}^{(m,n)}\widehat{\Pi}_{\perp}^{(m,n)}~,\label{TTproj4propIdem}\\
\pa^{\b\bd}\Pi_{\perp}^{(m,n)}\phi_{\b\a(m-1)\bd\ad(n-1)}=0~&,\qquad 0=\pa^{\b\bd}\widehat{\Pi}_{\perp}^{(m,n)}\phi_{\b\a(m-1)\bd\ad(n-1)}~\label{TTproj4propTrans}.
\end{align} 
\end{subequations}
In a fashion similar to the proof of \eqref{HSBachTypeEquiv}, it may be shown that both projectors are equal,
\begin{align}
&\Pi_{\perp}^{(m,n)}\phi_{\a(m)\ad(n)}=\widehat{\Pi}_{\perp}^{(m,n)}\phi_{\a(m)\ad(n)}~.
\end{align}
Thus, the operator $\Pi_{\perp}^{(m,n)}$ is the unique rank-$(m,n)$ transverse projector in $\mb{M}^4$. When restricted to the subspace $V^{\perp}_{(m,n)}$ of transverse fields, $\Pi_{\perp}^{(m,n)}$ acts as the identity operator.  

Let us define the projector $\Pi_{\parallel}^{(m,n)}$ which is the orthogonal complement of $\Pi_{\perp}^{(m,n)}$, 
\begin{align}
 \Pi_{\parallel}^{(m,n)}:= \mathds{1} -\Pi_{\perp}^{(m,n)}~,\qquad \Pi_{\parallel}^{(m,n)}\Pi_{\parallel}^{(m,n)}=\Pi_{\parallel}^{(m,n)}~.
\end{align} 
By definition the two resolve the identity, $\mathds{1}=\Pi_{\perp}^{(m,n)}+\Pi_{\parallel}^{(m,n)}$, and are orthogonal, 
\begin{align}
\Pi_{\perp}^{(m,n)}\Pi_{\parallel}^{(m,n)}=0~. \label{TTProjOrthMink4}
\end{align}
 Moreover, it may be shown that $\Pi_{\parallel}^{(m,n)}$ projects $\phi_{\a(m)\ad(n)}$ onto its longitudinal component,
\begin{align}
\Pi_{\parallel}^{(m,n)}\phi_{\a(m)\ad(n)}=\pa_{\a\ad}\phi_{\a(m-1)\ad(n-1)}~,
\end{align}
for some unconstrained $\phi_{\a(m-1)\ad(n-1)}$. It immediately follows that any $\phi_{\a(m)\ad(n)}$ may be decomposed into irreducible parts according to the rule
\begin{align}
\phi_{\a(m)\ad(n)}=\phi^{\perp}_{\a(m)\ad(n)}+\sum_{t=1}^{n-1}\big(\pa_{\a\ad}\big)^t\phi^{\perp}_{\a(m-t)\ad(n-t)}+\big(\pa_{\a\ad}\big)^{n}\phi_{\a(m-n)} \label{DecompMink4}
\end{align}
where we have assumed, without loss of generality, that $m\geq n$. In eq. \eqref{DecompMink4} each of the fields $\phi^{\perp}_{\a(m-t)\ad(n-t)}$, with $0 \leq t \leq n-1$, are transverse.

The rank-$(m,n)$ spin projection operator $\Pi^{(m,n)}_{\perp}$ selects the first term in the decomposition \eqref{DecompMink4}, i.e. the transverse component of an unconstrained $\phi_{\a(m)\ad(n)}$ with the maximal spin of $s=\frac{1}{2}(m+n)$. For two special cases, $m=n$ and $m=n+1$,
the projection operators are equivalent to 
the Behrends-Fronsdal projectors  \cite{BF,Fronsdal58}. The spin projectors for fields with generic $m\neq n $ mixed symmetry were first constructed in \cite{SG,GGRS}, where they introduced the operator $\D_{\a\ad}$ and used it to construct various superprojectors. 

Let us now consider the case when the field $\phi_{\a(m)\ad(n)}$ is a CHS field, $\phi_{\a(m)\ad(n)}\equiv h_{\a(m)\ad(n)}$.
We can express the higher-spin Weyl tensors of $h_{\a(m)\ad(n)}$ in terms of $\Pi_{\perp}^{(m,n)}$,
\begin{subequations}\label{HSWeylProj}
\begin{align}
\mc{W}_{\a(m+n)}(h)&=\Box^{\frac n2}\big(\Delta_{\a}{}^{\bd}\big)^n\Pi_{\perp}^{(m,n)}h_{\a(m)\bd(n)}~,\label{HSWeylProja}\\
\mc{W}_{\a(m+n)}(\bar{h})&=\Box^{\frac m2}\big(\Delta_{\a}{}^{\bd}\big)^m\Pi_{\perp}^{(n,m)}\bar{h}_{\a(n)\bd(m)}~\label{HSWeylProjb}.
\end{align}
\end{subequations}
Similarly, the higher-spin Bach tensor may also be rewritten in the form 
\begin{subequations}
\begin{align}
\mc{B}_{\a(n)\ad(m)}(h)&= \Box^{\frac{1}{2}(m+n)}\big(\Delta_{\ad}{}^{\b}\big)^{m-n}\Pi^{(m,n)}_{\perp}h_{\b(m-n)\a(n)\ad(n)}~,\\
\mc{B}_{\a(n)\ad(m)}(h)&=\Box^{\frac{1}{2}(m+n)}\big(\Delta_{\a}{}^{\bd}\big)^{n-m}\Pi^{(m,n)}_{\perp}h_{\a(m)\ad(m)\bd(n-m)} ~.
\end{align}
\end{subequations}
In the first and second expressions we have assumed that $m\geq n$ and $n\geq m$ respectively.
This allows us to recast the CHS actions \eqref{CHSABachMink4a} into the form (below we assume $m\geq n$)
\begin{align}
S_{\rm{CHS}}^{(m,n)} [ h,\bar{h}] 
=\text{i}^{m+n} \int \rd^4 x \, \bar{h}^{\a(n)\ad(m)}\Box^{n}\big(\pa_{\ad}{}^{\b}\big)^{m-n}\Pi^{(m,n)}_{\perp}h_{\b(m-n)\a(n)\ad(n)}~. \label{CHSATTProjMink}
\end{align}
In the special cases $m=n=s$ 
and $m=n+1=s$ respectively, this reads
\begin{subequations}
\begin{align}
S_{\rm{CHS}}^{(s,s)} [ h ] 
&=(-1)^s\int \rd^4 x \, h^{\a(s)\ad(s)}\Box^{s}\Pi^{(s,s)}_{\perp}h_{\a(s)\ad(s)}~, \label{FTCHSB}\\
S_{\rm{CHS}}^{(s,s-1)} [ h,\bar{h}] 
&=(-1)^{s} \text{i}\int \rd^4 x \, \bar{h}^{\a(s-1)\ad(s)}\Box^{s-1}\pa_{\ad}{}^{\b}\Pi^{(s,s-1)}_{\perp}h_{\b\a(s-1)\ad(s-1)}~.\label{FTCHSF}
\end{align}
\end{subequations}
Actions \eqref{FTCHSB} and \eqref{FTCHSF} are equivalent to the original bosonic and fermionic CHS actions proposed by Fradkin and Tseytlin (though the projectors in \cite{FT} were implicit).

Since the CHS action \eqref{CHSATTProjMink} is gauge invariant, we may impose the gauge condition
\begin{align}
h_{\a(m)\ad(n)}\equiv h^{\perp}_{\a(m)\ad(n)}~, \qquad \pa^{\b\bd}h^{\perp}_{\a(m-1)\b\ad(n-1)\bd}=0~. \label{TTGF}
\end{align} 
In this case, the projector $\Pi^{(m,n)}_{\perp}$ acts as the identity on $h^{\perp}_{\a(m)\ad(n)}$ and \eqref{CHSATTProjMink} becomes\footnote{Equivalently, one can arrive at \eqref{CHSATTProjMinkGF} by using the decomposition \eqref{DecompMink4} in \eqref{CHSATTProjMink}, and noting that all longitudinal components 
drop out due to gauge invariance.} 
\begin{align}
S_{\rm{CHS}}^{(m,n)} [ h,\bar{h}] 
=\text{i}^{m+n} \int \rd^4 x \, \bar{h}_{\perp}^{\a(n)\ad(m)}\Box^{n}\big(\pa_{\ad}{}^{\b}\big)^{m-n}h^{\perp}_{\b(m-n)\a(n)\ad(n)}~. \label{CHSATTProjMinkGF}
\end{align}
It should be noted that the gauge condition \eqref{TTGF} is not preserved by conformal transformations,\footnote{The condition that a primary field be transverse is only preserved under conformal transformations if it has  weight $2+\frac{1}{2}(m+n)$. This is why the  weight of the HS Bach tensor is fixed to \eqref{HSBachWeylT}. } 
and hence, in contrast to \eqref{CHSATTProjMink},  the action \eqref{CHSATTProjMinkGF} is not conformally invariant. Finally, we point out that although $h_{\a(m)\ad(n)}$, $\mc{W}_{\a(m+n)}(h)$ and $\mc{B}_{\a(n)\ad(m)}(h)$ are conformal primary fields, $\Pi^{(m,n)}_{\perp}h_{\a(m)\ad(n)}$ is not.

It is possible to recast the spin-projection operators into a form which depends solely on Casimir operators of the $4d$ Poincar\'e algebra $\mf{iso}(3,1)$.
In particular, let us define the square of the Pauli-Lubankski vector,
\begin{align}\label{PauliLubankskiSquare}
\mathbb{W}^2=\mathbb{W}^a\mathbb{W}_a~,\qquad \mathbb{W}_a:=-\frac{1}{2}\ve_{abcd}M^{bc}\pa^d~. 
\end{align}
On the field $\phi_{\a(m)\ad(n)}$, it may be shown that $\mathbb{W}^2$ assumes the form (see e.g. \cite{BK}) 
\begin{align}
\mathbb{W}^2\f_{\a(m)\ad(n)}=s(s+1)\Box\f_{\a(m)\ad(n)} +mn \pa_{\a\ad}\pa^{\b\bd}\f_{\a(m-1)\b\ad(n-1)\bd}~,
\end{align}  
where we have defined $s:=\frac{1}{2}(m+n)$. On any transverse field $\psi_{\a(m)\ad(n)}$ this reduces to $\big(\mathbb{W}^2-s(s+1)\Box\big)\psi_{\a(m)\ad(n)}=0$.
One can express $\Pi_{\perp}^{(m,n)}$ purely in terms of the Casimir operators $\mathbb{W}^2$ and $\Box$ of $\mf{iso}(3,1)$ as follows
\begin{subequations} \label{CasimirProjectorsMink4}
\begin{align}
\Pi_{\perp}^{(m,n)}\f_{\a(m)\ad(n)}=\frac{m!}{(m+n)!n!}\frac{1}{\Box^n}&\prod_{j=0}^{n-1}\Big(\mathbb{W}^2-(s-j)(s-j-1)\Box\Big)\f_{\a(m)\ad(n)}\\
=\frac{n!}{(m+n)!m!}\frac{1}{\Box^m}&\prod_{j=0}^{m-1}\Big(\mathbb{W}^2-(s-j)(s-j-1)\Box\Big)\f_{\a(m)\ad(n)}~.
\end{align}
\end{subequations}
The operators \eqref{CasimirProjectorsMink4} satisfy the two-component spinor version of the properties \eqref{TTProjPropDDim}.
In the special case $m=n=s$ we have $\Pi^{(s,s)}_{\perp} \equiv \Pi^{(s)}_{\perp}$ with
\begin{align}
\Pi^{(s)}_{\perp}=\frac{1}{\Box^s(2s)!}\prod_{j=0}^{s-1}\Big(\mb{W}^2-j(j+1)\Box\Big)~,
\end{align}
and it may be shown that $\Pi^{(s)}_{\perp}\phi_{\a(s')\ad(s')}=0$ for $s'<s$. Therefore, given a symmetric but traceful field $\bm h_{a(s)}$, $\Pi^{(s)}_{\perp}$ projects $\bm h_{a(s)}$ onto its transverse and traceless (TT) component. 

\section{Conformal higher-spin models in AdS$_{4}$ } \label{secCHSAdS4}
In this section we derive the gauge invariant actions for arbitrary rank CHS fields $h_{\a(m)\ad(n)}$ in AdS$_4$,\footnote{Gauge-invariant actions for integer spin-$s$ CHS fields $h_{a(s)}$ on AdS$_d$ were first constructed by Metsaev \cite{Metsaev2014} (see also \cite{NTCHS})  via the ordinary derivative formulation.} and use them to construct the corresponding AdS$_4$ transverse projectors. 
Various applications of these results are investigated. In particular, a connection between the poles of the projectors and (partially-)massless fields is established. This relation allows us to easily demonstrate that the kinetic operator of the AdS$_4$ CHS action factorises into partially massless wave operators. This section is based on our paper \cite{AdSprojectors}.

 For the algebra of AdS$_4$ covariant derivatives we adopt the convention\footnote{In \cite{AdSprojectors} we used the convention $\big[\mc{D}_a,\mc{D}_b\big]=\frac{1}{2}R_{ab}{}^{cd}M_{cd}$ and $R=-12\mu\mub$. In order to maintain the same commutator algebra, $\big[\mc{D}_{a},\mc{D}_{b}\big]=-\mu\mub M_{ab}$, in this thesis we take the AdS$_4$ scalar curvature to be positive, $R=12\mu\mub$. } 
 \begin{align}
 \big[\mathcal{D}_a,\mathcal{D}_b\big]=-\mu\mub M_{ab}\qquad \Longleftrightarrow \qquad \big[\mathcal{D}_{\a\ad},\mathcal{D}_{\b\bd}\big]=-2\mu\mub\big(\ve_{\a\b}\bar{M}_{\ad\bd}+\ve_{\ad\bd}M_{\a\b}\big)~. \label{LCDalgebraAdS4}
 \end{align} 
Here the parameter $\mu\mub$ is related to the scalar curvature via $R=12\mu\mub$. 
We parametrise the curvature in terms of 
$\mu\mub$ in order for the notation to be consistent with that used in $\cN=1$ 
AdS superspace \eqref{algebraAdS4|4}. An analysis similar to that given below applies in the case of de Sitter space, one just needs to replace all occurrences of $\mu\mub$ with $-\mu\mub$.

Of crucial importance to our subsequent analysis is the quadratic Casimir operator of the AdS$_4$ isometry group, whose realisation on tensor fields is
 \begin{align}
\mathcal{Q}:=\Box-\mu\mub\big(M^{\g\d}M_{\g\d}+\bar{M}^{\gd\dd}\bar{M}_{\gd\dd}\big)~,\qquad \big[\mathcal{Q},\mathcal{D}_{\a\ad}\big]=0~,
\label{QCasimirAdS4}
\end{align}
where $\Box = \cD^a \cD_a = - \hf \cD^{\a\ad} \cD_{\a\ad}$. 


%


\subsection{Transverse projectors}\label{secTTprojAdS4}

 To inspire an ansatz for the operators which project a field defined on AdS$_4$ onto its transverse component, we look towards the CHS models on the same background.
 In section \ref{secCHSCF4} we will show that on any conformally-flat Einstein background, the corresponding CHS models are simply the minimal lift of the ones defined on $\mb{M}^4$, obtained by the replacement $\pa_{\a\ad}\rightarrow \mc{D}_{\a\ad}$. Thus, in particular, the HS Bach tensors on AdS$_4$ are 
\begin{subequations} \label{HSBachAdS4}
\begin{align}
\mc{B}_{\a(n)\ad(m)}(h)&=\mc{D}_{(\ad_1}{}^{\b_1}\cdots \mc{D}_{\ad_m)}{}^{\b_m}\mc{D}_{(\a_1}{}^{\bd_1}\cdots\mc{D}_{\a_n}{}^{\bd_n}h_{\b_1\dots\b_m)\bd(n)}~,\label{HSBachAdS4a}\\
\widehat{\mc{B}}_{\a(n)\ad(m)}(h)&=\mc{D}_{(\a_1}{}^{\bd_1}\cdots \mc{D}_{\a_n)}{}^{\bd_n}\mc{D}_{(\ad_1}{}^{\b_1}\cdots\mc{D}_{\ad_m}{}^{\b_m}h_{\b(m)\bd_1\dots\bd_n)}~.\label{HSBachAdS4b}
\end{align}
\end{subequations}
They are invariant under the gauge transformations \eqref{GTCHSprepM4},
\begin{align}
\delta_{\xi}\mc{B}_{\a(n)\ad(m)}(h)=0~,\qquad \delta_{\xi}\widehat{\mc{B}}_{\a(n)\ad(m)}(h)=0~.\label{HSBachAdS4GI}
\end{align}
For our current purpose, more important is the fact that they are transverse 
\begin{align}
\mc{D}^{\b\bd}\mc{B}_{\a(n-1)\b\ad(m-1)\bd}(h)=0~,\qquad \mc{D}^{\b\bd}\widehat{\mc{B}}_{\a(n-1)\b\ad(m-1)\bd}(h)=0~.\label{HSBachAdS4TT}
\end{align}

 We can use the differential operator which appears in the HS Bach tensors as a foundation for the transverse projectors. However, it is clear that any projector must preserve the rank of the tensor field on which it acts. This can be rectified by appropriately removing or inserting vector derivatives in \eqref{HSBachAdS4} to convert the indices. With these remarks in mind, we define the following two differential operators by their action on an unconstrained field $\phi_{\a(m)\ad(n)}\in V_{(m,n)}$ (which need not be a CHS field),
\begin{subequations}\label{ProjectorsStripped}
\begin{align}
\mc{P}^{(m,n)}\phi_{\a(m)\ad(n)} \equiv \mc{P}_{\a(m)\ad(n)}(\phi)&=\mathcal{D}_{(\ad_1}{}^{\b_1}\cdots\mathcal{D}_{\ad_n)}{}^{\b_n}\mathcal{D}_{(\b_1}{}^{\bd_1}\cdots\mathcal{D}_{\b_n}{}^{\bd_n}\phi_{\a_1\dots\a_m)\bd(n)}~, \label{ProjectorsStrippeda}\\[6pt]
\widehat{\mc{P}}^{(m,n)}\phi_{\a(m)\ad(n)}\equiv \widehat{\mc{P}}_{\a(m)\ad(n)}(\phi)&=\mathcal{D}_{(\a_1}{}^{\bd_1}\cdots\mathcal{D}_{\a_m)}{}^{\bd_m}\mathcal{D}_{(\bd_1}{}^{\b_1}\cdots\mathcal{D}_{\bd_m}{}^{\b_m}\phi_{\b(m)\ad_1\dots\ad_n)}~. \label{ProjectorsStrippedb}
\end{align}
\end{subequations}
Both operators \eqref{ProjectorsStripped} project out the transverse component of the field $\phi_{\a(m)\ad(n)}$,
\begin{subequations}\label{ProjectorsStrippedT}
\begin{align}
\mathcal{D}^{\b\bd}\mc{P}_{\b\a(m-1)\bd\ad(n-1)}(\phi)&=0~,\\
\mathcal{D}^{\b\bd}\widehat{\mc{P}}_{\b\a(m-1)\bd\ad(n-1)}(\phi)&=0~.
\end{align} 
\end{subequations}
However they are not projectors in the sense that they do not square to themselves. In fact, one may show that they instead satisfy
\begin{subequations}\label{ProjectorsStrippedSq}
\begin{align}
\mc{P}^{(m,n)}\mc{P}^{(m,n)}\phi_{\a(m)\ad(n)}=\prod_{t=1}^{n}(\mathcal{Q}-\t_{(t,m,n)}\mu\mub)\mc{P}^{(m,n)}\phi_{\a(m)\ad(n)}~, \label{ProjectorsStrippedSqa}\\
\widehat{\mc{P}}^{(m,n)}\widehat{\mc{P}}^{(m,n)}\phi_{\a(m)\ad(n)}=\prod_{t=1}^{m}(\mathcal{Q}-\t_{(t,m,n)}\mu\mub)\widehat{\mc{P}}^{(m,n)}\phi_{\a(m)\ad(n)}~, \label{ProjectorsStrippedSqb}
\end{align}
\end{subequations}
where the parameters $\t_{(t,m,n)}$ are defined by
\begin{align}
 \t_{(t,m,n)}&:= \frac{1}{2}\Big[(m+n-t+3)(m+n-t-1)+(t-1)(t+1)\Big]~.\label{PMvalAdS4}
\end{align}
From eq. \eqref{2.5} it follows that the two operators $\Pi_{\perp}^{(m,n)}$ and $\widehat{\Pi}_{\perp}^{(m,n)}$, where
\begin{subequations}\label{TTprojectorsAdS4}
\begin{align}
\Pi_{\perp}^{(m,n)}\phi_{\a(m)\ad(n)}\equiv \Pi^{\perp}_{\a(m)\ad(n)}(\phi)=\bigg[\prod_{t=1}^{n}(\mathcal{Q}-\t_{(t,m,n)}\mu\mub)\bigg]^{-1}\mc{P}_{\a(m)\ad(n)}(\phi)~, \label{TTprojectorsAdS4a}\\
\widehat{\Pi}_{\perp}^{(m,n)}\phi_{\a(m)\ad(n)}\equiv \widehat{\Pi}^{\perp}_{\a(m)\ad(n)}(\phi)=\bigg[\prod_{t=1}^{m}(\mathcal{Q}-\t_{(t,m,n)}\mu\mub)\bigg]^{-1}\widehat{\mc{P}}_{\a(m)\ad(n)}(\phi)~, \label{TTprojectorsAdS4b}
\end{align}
\end{subequations}
 square to themselves and project out the transverse subspace of $V_{(m,n)}$,
 \begin{subequations}\label{TTprojectorsAdS4T}
\begin{align}
\Pi^{(m,n)}\Pi^{(m,n)}\phi_{\a(m)\ad(n)}&=\Pi^{(m,n)}\phi_{\a(m)\ad(n)}~,\qquad \mathcal{D}^{\b\bd}\Pi_{\b\a(m-1)\bd\ad(n-1)}(\phi)=0~,\\
\widehat{\Pi}^{(m,n)}\widehat{\Pi}^{(m,n)}\phi_{\a(m)\ad(n)}&=\widehat{\Pi}^{(m,n)}\phi_{\a(m)\ad(n)}~,\qquad \mathcal{D}^{\b\bd}\widehat{\Pi}_{\b\a(m-1)\bd\ad(n-1)}(\phi)=0~.
\end{align}
\end{subequations}
 Actually, the two types of projectors prove to coincide,
\begin{align}\label{2.9}
\Pi_{\a(m)\ad(n)}(\phi)=\widehat{\Pi}_{\a(m)\ad(n)}(\phi)~,
\end{align}
and so it suffices to consider only the first, \eqref{TTprojectorsAdS4a}. Therefore the operator $\Pi_{\perp}^{(m,n)}$ is the unique rank-$(m,n)$ transverse projector in AdS$_4$. When restricted to the subspace $V^{\perp}_{(m,n)}$ of transverse fields, $\Pi_{\perp}^{(m,n)}$ acts as the identity. 

We would like to point out that, in a fashion similar to \eqref{CasimirProjectorsMink4}, it should be possible to obtain new realisations for the  AdS$_4$ spin projection operators  \eqref{TTprojectorsAdS4} in terms of the Casimir operators of the algebra $\mf{so}(3,2)$. In this case, $\pa^a\pa_a$ should be replaced with the quadratic Casimir $\mc{Q}$,  eq. \eqref{Q}.
 Furthermore, the role of the square of the Pauli-Lubankski vector $\mb{W}^2$, eq. \eqref{PauliLubankskiSquare}, should be played by the quartic Casimir operator $\mb{W}^2_{\text{AdS}}$,
\bea
\mb{W}^2_{\text{AdS}}&:=&-\frac{1}{2}\big(\mb{Q}+2\mu\mub\big)\big(M^2+\bar{M}^2\big)+\mc{D}^{\a\ad}\mc{D}^{\b\bd}M_{\a\b}\bar{M}_{\ad\bd} \non \\
&&
-\frac{1}{4}\mu\mub\big(M^2M^2+\bar{M}^2\bar{M}^2+6M^2\bar{M}^2\big)~,
\eea
which satisfies $\big[\mb{W}^2_{\text{AdS}},\mc{D}_{\a\ad}\big]=0~$, and where we have defined
\bea
M^2:=M^{\a\b}M_{\a\b}~,
\qquad  \bar{M}^2:=\bar{M}^{\ad\bd}\bar{M}_{\ad\bd}~. 
 \eea

Let us define the projector $\Pi_{\parallel}^{(m,n)}$ which is the orthogonal complement of $\Pi_{\perp}^{(m,n)}$, 
\begin{align}
 \Pi_{\parallel}^{(m,n)}:= \mathds{1} -\Pi_{\perp}^{(m,n)}~,\qquad \Pi_{\parallel}^{(m,n)}\Pi_{\parallel}^{(m,n)}=\Pi_{\parallel}^{(m,n)}~.
\end{align} 
By definition the two resolve the identity, $\mathds{1}=\Pi_{\perp}^{(m,n)}+\Pi_{\parallel}^{(m,n)}$, and are orthogonal, 
\begin{align}
\Pi_{\perp}^{(m,n)}\Pi_{\parallel}^{(m,n)}=0~. \label{TTProjOrthAdS4}
\end{align}
 Moreover, it may be shown that $\Pi_{\parallel}^{(m,n)}$ projects $\phi_{\a(m)\ad(n)}$ onto its longitudinal component, 
\begin{align}
\Pi_{\parallel}^{(m,n)}\phi_{\a(m)\ad(n)}=\mc{D}_{\a\ad}\phi_{\a(m-1)\ad(n-1)}~,
\end{align}
for some unconstrained field $\phi_{\a(m-1)\ad(n-1)}$. It immediately follows that any $\phi_{\a(m)\ad(n)}$ may be decomposed into irreducible parts according to the rule
\begin{align}
\phi_{\a(m)\ad(n)}=\phi^{\perp}_{\a(m)\ad(n)}+\sum_{t=1}^{n-1}\big(\mc{D}_{\a\ad}\big)^t\phi^{\perp}_{\a(m-t)\ad(n-t)}+\big(\mc{D}_{\a\ad}\big)^{n}\phi_{\a(m-n)} \label{DecompAdS4}
\end{align}
where we have assumed, without loss of generality, that $m\geq n$. In eq. \eqref{DecompAdS4} each of the fields $\phi^{\perp}_{\a(m-t)\ad(n-t)}$, with $0 \leq t \leq n-1$, are transverse.



\subsection{On-shell massive and (partially-)massless fields}

It is of interest to understand the physical significance of the parameters \eqref{PMvalAdS4} which appear in the definition of the projectors \eqref{TTprojectorsAdS4}. With this in mind we now introduce on-shell fields.
Given two positive integers $m$ and $n$, a tensor field $\phi_{\a(m)\ad(n)}$ of Lorentz type $(m/2,n/2)$ in AdS$_4$  is said to be on-shell if it satisfies the equations
\begin{subequations}\label{OSFAdS4}
\begin{align}
0~&=\big(\mathcal{Q}-\rho^2\big)\phi_{\a(m)\ad(n)}~,\label{OSC1AdS4}\\
0~&=\mc{D}^{\b\bd}\phi_{\a(m-1)\b\ad(n-1)\bd}~.\label{OSC2AdS4}
\end{align}
\end{subequations}
We say that such a field describes a spin $s=\frac{1}{2}(m+n)$ particle with pseudo-mass $\rho$.

Unitary irreducible representations of the Lie algebra $\mf{so}(3,2)$ of the AdS$_4$ isometry group may be realised in terms of the on-shell fields \eqref{OSFAdS4} for certain values of $\r$.  In brief, these representations may be labelled by the pseudo-mass $\rho$ (which is related to the more conventional minimal energy $E_0$ via eq. \eqref{PseudoMassEnergy}) and the spin $s=\frac{1}{2}(m+n)$, and are denoted by $\mf{D}(\rho,s)$. Unitary massive representations may be realised by those on-shell fields whose pseudo mass satisfies the unitarity bound $\rho^2 \geq \tau_{(1,m,n)}\mu\mub $. These facts are discussed in more detail in appendix \ref{AppGenForm4}.

It is typical to choose $m=n=s$ for spin-$s$ bosonic fields whereas the usual choice for spin-$\big(s+\frac 12\big)$ fermionic fields is $m=n+1=s$. By now it is well known that in these cases, the system of equations \eqref{OSFAdS4} is invariant under gauge transformations of depth $t$
\begin{subequations}
\begin{align}
\delta_{\xi}\phi_{\a(s)\ad(s)}&=\mathcal{D}_{(\a_1(\ad_1}\dots\mathcal{D}_{\a_t\ad_t}\xi_{\a_{t+1}\dots\a_s)\ad_{t+1}\dots\ad_{s})}\equiv \big(\mc{D}_{\a\ad}\big)^t\xi_{\a(s-t)\ad(s-t)}~,\\
\delta_{\xi}\phi_{\a(s+1)\ad(s)}&=\mathcal{D}_{(\a_1(\ad_1}\dots\mathcal{D}_{\a_t\ad_t}\xi_{\a_{t+1}\dots\a_s)\ad_{t+1}\dots\ad_{s+1})}\equiv \big(\mc{D}_{\a\ad}\big)^t\xi_{\a(s-t+1)\ad(s-t)}~,
\end{align} 
\end{subequations}
for an on-shell gauge parameter when the pseudo-mass takes the values\footnote{Due to our definition \eqref{OSC1AdS4}, the mass values \eqref{PMValsBosFerm} are shifted with respect to the usual ones.}\cite{DeserW4, Zinoviev, Metsaev2006}
\begin{subequations}\label{PMValsBosFerm}
\begin{align}
\rho^2_{(t,s)}&=\frac{1}{2}\big[(2s-t+3)(2s-t-1)+(t-1)(t+1)\big]\mu\mub~,\label{PMValsBos}\\
\rho^2_{(t,s+\frac 12)}&=\frac{1}{2}\big[(2s-t+4)(2s-t)+(t-1)(t+1)\big]\mu\mub~, \label{PMValsFerm}
\end{align}
\end{subequations}
where $1\leq t \leq s$. Strictly massless fields correspond to $t=1$ whilst all other values of $t$ correspond to partially massless fields. 

Remarkably, we see that for these values of $m$ and $n$, the partially massless values coincide with the parameters in the projectors,
\begin{align}
\r^2_{(t,s)}=\t_{(t,s,s)}\mu\mub^2~,\qquad \r^2_{(t,s+\frac 12)}=\t_{(t,s,s+1)}\mu\mub~.
\end{align} 
Therefore, we can extend this notion and say that a field $\phi_{\a(m)\ad(n)}$ is partially massless when it satisfies the on-shell conditions \eqref{OSFAdS4} with\footnote{Such a pseudo-mass violates the unitarity bound (see appendix \ref{AppGenForm4}), and hence the partially-massless representations are non-unitary. } 
\begin{align}
\r^2_{(t,m,n)}=\t_{(t,m,n)}\mu\mub~,\qquad 1\leq t \leq \text{min}(m,n)~.\label{PMPointsAdS4}
\end{align}
Indeed, as shown in appendix \eqref{secAppPMGAdS4}, for these values a gauge invariance with depth $t$ emerges in the system of equations \eqref{OSFAdS4},
\begin{align}
\delta_{\xi}\phi_{\a(m)\ad(n)}=\mathcal{D}_{(\a_1(\ad_1}\dots\mathcal{D}_{\a_t\ad_t}\xi_{\a_{t+1}\dots\a_{m})\ad_{t+1}\dots\ad_n)}\equiv \big(\mc{D}_{\a\ad}\big)^t\xi_{\a(m-t)\ad(n-t)}~. \label{PMGSAdS4}
\end{align}
This is true as long as the gauge parameter $\xi_{\a(m-t)\ad(n-t)}$ is also on-shell with the same pseudo-mass,
\begin{subequations}\label{GPP}
\begin{align}
0~&=\big(\mathcal{Q}-\tau_{(t,m,n)}\mu\mub\big)\xi^{(t)}_{\a(m-t)\ad(n-t)}~,\label{GPPa}\\
0~&=\mc{D}^{\b\bd}\xi^{(t)}_{\a(m-t-1)\b\ad(n-t-1)\bd}~.\label{GPPb}
\end{align}
\end{subequations}
This gauge symmetry\footnote{Gauge-invariant actions for partially massless fields in (A)dS were given in 
\cite{Zinoviev, Metsaev2006,SV}.}
 is our main motivation for choosing the upper bound of min$(m,n)$ for $t$ in the definition \eqref{PMPointsAdS4}.
 
 In summary, the parameters  $\tau_{(t,m,n)}$ determine the pole structure of the off-shell transverse projection operators \eqref{TTprojectorsAdS4} for  fields
of Lorentz type $(m/2,n/2)$ in AdS$_4$.
 This observation leads to a new understanding of the depth-$t$ partially massless fields $\phi^{(t)}_{\a(m)\ad(n)}$. 
Specifically, the gauge symmetry \eqref{PMGSAdS4} of the field $\phi^{(t)}_{\a(m)\ad(n)}$
 is associated with the pole $\rho^2=\tau_{(t,m,n)}\mu\mub$
of the corresponding spin projection operator in AdS$_4$.
In appendix \ref{secAppPMGAdS4} we provide a systematic derivation of this claim using the spin projection operators \eqref{TTprojectorsAdS4}.




\subsection{Factorisation of the conformal higher-spin action} \label{secFactorCHS}

As an application of our analysis, we would like to discuss the factorisation of the kinetic operator (aka the `higher-spin Bach operator') associated with the action for a conformal higher-spin field in AdS$_4$. Accordingly, we now consider the specific case when $\phi_{\a(m)\ad(n)}$ is a CHS field, $\phi_{\a(m)\ad(n)}\equiv h_{\a(m)\ad(n)}$.
We recall that the corresponding CHS action in AdS$_4$ takes the form (see eq. \eqref{HSBachAdS4} and the discussion therein)
\begin{align}
S_{\text{CHS}}^{(m,n)}[h, \bar{h}]&=\frac{1}{2}\text{i}^{m+n}\int \text{d}^4x \, e \, \bar{h}^{\a(n)\ad(m)}\mc{B}_{\a(n)\ad(m)}(h)+{\rm c.c.}  \label{CHSActAdS4}
\end{align}
where the linearised higher-spin Bach tensor was defined in eq. \eqref{HSBachAdS4},
\begin{align}
\big(\mc{B}^{(m,n)}h\big)_{\a(n)\ad(m)}\equiv \mc{B}_{\a(n)\ad(m)}(h)=\mc{D}_{(\ad_1}{}^{\b_1}\cdots \mc{D}_{\ad_m)}{}^{\b_m}\mc{D}_{(\a_1}{}^{\bd_1}\cdots\mc{D}_{\a_n}{}^{\bd_n}h_{\b_1\dots\b_m)\bd(n)} \label{NewBachAdS}
\end{align}
The latter is transverse \eqref{HSBachAdS4TT} and invariant \eqref{HSBachAdS4GI} under the gauge transformations 
\begin{align}
\delta_{\xi}h_{\a(m)\ad(n)}=\mathcal{D}_{(\a_1(\ad_1}\xi_{\a_2\dots\a_m)\ad_2\dots\ad_n)} 
~.
\label{GT456654}
\end{align}
This means that the action \eqref{CHSActAdS4} is also gauge invariant under \eqref{GT456654}, in addition to being invariant under the conformal transformations \eqref{ConfT4D}.
The equation of motion for $\bar{h}_{\a(n)\ad(m)}$ is the vanishing of the  higher-spin Bach tensor
\begin{align}
\mc{B}_{\a(n)\ad(m)}(h)=0~. \label{EOMCHSAdS4}
\end{align}
The gauge freedom \eqref{3.99} allows us to impose the transverse gauge
\begin{align}
h_{\a(m)\ad(n)}\equiv h^{\perp}_{\a(m)\ad(n)}~,\qquad \mathcal{D}^{\b\bd}h^{\perp}_{\b\a(m-1)\bd\ad(n-1)}=0~. \label{TTGFAdS4}
\end{align}

There are three separate scenarios that we should consider, the first of which occurs when $m=n=s$. In this case the bosonic higher-spin Bach operator $\mc{B}^{(s,s)}$ coincides with the operator $\mc{P}^{(s,s)}$ defined in eq. \eqref{ProjectorsStrippeda},
\begin{align}
\big(\mc{B}^{(s,s)}h\big)_{\a(s)\ad(s)}=\mc{P}^{(s,s)}h_{\a(s)\ad(s)}=\prod_{t=1}^{s}\big(\mathcal{Q}-\t_{(t,s,s)}\mu\mub\big)\Pi_{\perp}^{(s,s)}h_{\a(s)\ad(s)}~.\label{BosBachOp}
\end{align}
Consequently, the corresponding CHS action may be expressed through the projector as\footnote{This is the AdS$_4$ version of the bosonic action originally proposed by Fradkin and Tseytlin \cite{FT}. }
\begin{align} \label{FTCHSGen1}
S_{\text{CHS}}^{(s,s)}[h]&=(-1)^s\int \text{d}^4x \, e \, h^{\a(s)\ad(s)}\prod_{t=1}^{s}\big(\mathcal{Q}-\t_{(t,s,s)}\mu\mub\big)\Pi_{\perp}^{(s,s)}h_{\a(s)\ad(s)}~.
\end{align}
The projector $\Pi_{\perp}^{(s,s)}$ acts as the identity on any transverse field, which means that in the gauge \eqref{TTGFAdS4} the Bach operator \eqref{BosBachOp} factorises,
\begin{align}
\big(\mc{B}^{(s,s)}h^{\perp}\big)_{\a(s)\ad(s)}=\prod_{t=1}^{s}(\mathcal{Q}-\t_{(t,s,s)}\mu\mub)h^{\perp}_{\a(s)\ad(s)} \label{factor185}
\end{align}
and hence so too does the gauge fixed action. 

In the introduction we mentioned that the action $S_{\text{CHS}}^{(s,s)}[h]$ describes $s(s+1)$ propagating degrees of freedom (dof). 
This result was established in \cite{FT}, here we provide an interesting alternative derivation. 
Using \eqref{factor185} the equation of motion \eqref{EOMCHSAdS4} reads $0=\prod_{t=1}^{s}(\mathcal{Q}-\t_{(t,s,s)}\mu\mub)h^{\perp}_{\a(s)\ad(s)}$. The solution to this is a linear combination of depth-$t$ (partially-)massless fields with $1\leq t \leq s$. A spin-$s$ depth-$t$ (partially-)massless field carries $2t$ dof, resulting from the $2s+1$ for transverse $h^{\perp}_{\a(s)\ad(s)}$ subtract $2s-2t+1$ from the transverse gauge parameter $\xi^{\perp}_{\a(s-t)\ad(s-t)}$ (the system is on-shell \eqref{OSFAdS4} and \eqref{GPP}). Hence the total dof is $\sum_{t=1}^{s}2t=s(s+1)$.  

Next, let us consider the case when $n > m$. By taking appropriate derivatives of the Bach tensor \eqref{NewBachAdS} one arrives at the following relation\footnote{Due to the transversality of the HS Bach tensor, the left hand side of \eqref{355.19} is automatically totally symmetric in its dotted indices.}
\begin{align}
\big(\mc{D}_{\ad}{}^{\b}\big)^{n-m}\mc{B}_{\b(n-m)\a(m)\ad(m)}(h)=\mc{P}_{\a(m)\ad(n)}(h)~, \label{355.19}
\end{align}
which may be inverted to give
\begin{align}
\big(\mc{B}^{(m,n)}h\big)_{\a(n)\ad(m)}&=\bigg[\prod_{t=m+1}^{n}\big(\mathcal{Q}-\t_{(t,m,n)}\mu\mub\big)\bigg]^{-1}\big(\mc{D}_{\a}{}^{\bd}\big)^{n-m}\mc{P}^{(m,n)}h_{\a(m)\ad(m)\bd(n-m)}~\non\\
&=\prod_{t=1}^{m}\big(\mathcal{Q}-\t_{(t,m,n)}\mu\mub\big)\big(\mc{D}_{\a}{}^{\bd}\big)^{n-m}\Pi_{\perp}^{(m,n)}h_{\a(m)\ad(m)\bd(n-m)}~.
\end{align}
Consequently, the corresponding CHS action may be expressed through the projector as
\begin{align}
S_{\text{CHS}}^{(m,n)}[h,\bar{h}]=\frac{1}{2}\text{i}^{m+n}\int \text{d}^4x \, e \, &\bar{h}^{\a(n)\ad(m)}\prod_{t=1}^{m}\big(\mathcal{Q}-\t_{(t,m,n)}\mu\mub\big)\non\\
&\times\big(\mc{D}_{\a}{}^{\bd}\big)^{n-m}\Pi_{\perp}^{(m,n)}h_{\a(m)\ad(m)\bd(n-m)}+\text{c.c.}
\end{align}
It follows that in the gauge \eqref{TTGFAdS4}, the Bach operator factorises as
\begin{align}
\big(\mc{B}^{(m,n)}h^{\perp}\big)_{\a(n)\ad(m)}=\prod_{t=1}^{m}\big(\mathcal{Q}-\t_{(t,m,n)}\mu\mub\big)\big(\mc{D}_{\a}{}^{\bd}\big)^{n-m}h^{\perp}_{\a(m)\ad(m)\bd(n-m)}~. \label{factor2}
\end{align}
We see that due to the mismatch of $m$ and $n$, the conformal operator $\mc{B}^{(m,n)}$ does not factorise wholly into products of second-order operators. However, using \eqref{355.19} it is easy to see that for transverse fields the following  equation can be derived from \eqref{EOMCHSAdS4},
\begin{align}
\prod_{t=1}^{n}(\mathcal{Q}-\t_{(t,m,n)}\mu\mub)h^{\perp}_{\a(m)\ad(n)}=0~. \label{FullySickMate}
\end{align}

If on the other hand $m > n$, then the Bach tensor may be written in terms of $\mc{P}^{(m,n)}$ 
\begin{align}
\big(\mc{B}^{(m,n)}h\big)_{\a(n)\ad(m)}&=\big(\mc{D}_{\ad}{}^{\b}\big)^{m-n}\mc{P}^{(m,n)}h_{\b(m-n)\a(n)\ad(n)}~\non\\
&=\prod_{t=1}^{n}\big(\mathcal{Q}-\t_{(t,m,n)}\mu\mub\big)\big(\mc{D}_{\ad}{}^{\b}\big)^{m-n}\Pi_{\perp}^{(m,n)}h_{\b(m-n)\a(n)\ad(n)}~.
\end{align}
The corresponding CHS action takes the form 
\begin{align} \label{FTCHSGen2}
S_{\text{CHS}}^{(m,n)}[h,\bar{h}]=\frac{1}{2}\ri^{m+n}\int \text{d}^4x \, e \, &\bar{h}^{\a(m)\ad(n)}\prod_{t=1}^{n}\big(\mathcal{Q}-\t_{(t,m,n)}\mu\mub\big)\non\\
&\times\big(\cD_{\ad}{}^{\b}\big)^{m-n}\Pi_{\perp}^{(m,n)}h_{\a(n)\b(m-n)\ad(n)}+\text{c.c.}~.
\end{align}
Once again,  it follows that in the transverse gauge $\mc{B}^{(m,n)}$ factorises as
\begin{align}
\big(\mc{B}^{(m,n)}h^{\perp}\big)_{\a(n)\ad(m)}=\prod_{t=1}^{n}\big(\mathcal{Q}-\t_{(t,m,n)}\mu\mub\big)\big(\mc{D}_{\ad}{}^{\b}\big)^{m-n}h^{\perp}_{\b(m-n)\a(n)\ad(n)}~, \label{factor3}
\end{align}
and so too does \eqref{FTCHSGen2}. This time the higher-derivative equation derivable from \eqref{EOMCHSAdS4} is 
\begin{align}
\prod_{t=1}^{m}(\mathcal{Q}-\t_{(t,m,n)}\mu\mub)h^{\perp}_{\a(m)\ad(n)}=0~. \label{31123.20}
\end{align}

An interesting observation is that according to the definition \eqref{PMPointsAdS4}, when $m\neq n$ there is a discrete set of (non-unitary) mass values corresponding to the range $\text{min}(m,n)<t\leq \text{max}(m,n)$ which are not partially massless but which enter the spectrum of the higher-order wave equations \eqref{FullySickMate} and \eqref{31123.20}.

As a side remark, we note that to obtain the effective actions corresponding to the CHS models with $m\neq n$,
it is convenient to make use of the method of squaring which is always applied in the spinor theory.
In this method we have to deal with the operator
\begin{align}
\mc{B}^{(n,m)}\mc{B}^{(m,n)}h_{\a(m)\ad(n)}=\prod_{t=1}^{m}\big(\mathcal{Q}-\t_{(t,m,n)}\mu\mub\big)\mc{P}^{(m,n)}h_{\a(m)\ad(n)}
\end{align}
which for a transverse field becomes
\begin{align}
 \mc{B}^{(n,m)}\mc{B}^{(m,n)}h^{\perp}_{\a(m)\ad(n)}=\prod_{t=1}^{m}\big(\mathcal{Q}-\t_{(t,m,n)}\mu\mub\big)\prod_{k=1}^{n}\big(\mathcal{Q}-\t_{(k,m,n)}\mu\mub\big)h^{\perp}_{\a(m)\ad(n)}~. 
 \end{align}

For lower-spin values $s=3/2 $ and $s= 2,$ corresponding to $m=n-1=1$ and $m=n=2$ respectively, the factorisation of the conformal Bach operators \eqref{NewBachAdS} was observed long ago in \cite{DeserN1,Tseytlin5, Tseytlin6}. 
This factorisation was conjectured, based on lower-spin examples, by Tseytlin
\cite{Tseytlin13} for the bosonic $m=n$ and fermionic $m=n-1 $ cases, 
and also by Joung and Mkrtchyan  \cite{Karapet1,Karapet2} 
for certain bosonic CHS models.
More recently, the factorisation was proved by several groups 
 \cite{ Metsaev2014, NTCHS,GrigorievH} for those bosonic CHS models
on AdS${}_d$, with even $d$, 
 which are described by completely symmetric arbitrary spin conformal fields 
(the $m=n$ case in four dimensions). 

Using the formalism of spin projector operators in AdS${}_4$  developed here, 
the known factorisation properties for $m=n$ follow  immediately, 
 and are captured through the expression \eqref{factor185}. We have also provided the first derivation of the factorisation for conformal operators of arbitrary Lorentz type $(m/2,n/2)$, which is encapsulated by expressions \eqref{factor2} and \eqref{factor3}. This encompasses the case of arbitrary fermionic spin, $m=n-1$, 
 which to our knowledge was not covered previously in the literature. As in the bosonic case, we find that the spectrum of \eqref{factor2} and \eqref{factor3} consists of all partial masses. In contrast however, we find that the spectrum of the wave equations \eqref{FullySickMate} and \eqref{31123.20} contain a discrete set of (non-unitary) massive values.

\section{Conformal higher-spin models on conformally-flat backgrounds} \label{secCHSCF4}
 In this section we make use of the framework of conformal space\footnote{For discussion on the utility of conformal space to the problem at hand, see the beginning of sec. \ref{sec3DCHSCF}.}
 to construct gauge and Weyl invariant models for conformal higher-spin fields on arbitrary conformally-flat backgrounds. We also derive analogous  models for the `generalised' conformal higher-spin fields, which have higher-depth gauge transformations. 
 This section is based on our paper \cite{Confgeo}. 


We recall that in $d=4$, upon imposing appropriate constraints, the algebra of conformally covariant derivatives $\nabla_{a}$ is determined by the Weyl tensor: 
\begin{align}\label{CCDal4Vec}
[\nabla_a,\nabla_b]=-\frac{1}{2}W_{abcd}M^{cd}-\frac{1}{2}\nabla^dW_{abcd}K^c~.
\end{align}
In the two-component spinor notation, this  
takes the form\footnote{This algebra simplifies on any Einstein background \eqref{EinsteinSpace} since, using the Bianchi identity \eqref{BianchiIBP} (see also \eqref{BianchiIBPspinor}), it may be shown that $\nabla_{\ad}{}^{\b}W_{\a(3)\b}=0$ and $\nabla_{\a}{}^{\bd}\bar{W}_{\ad(3)\bd}=0$.}
\begin{align}
-\big[\nabla_{\a\ad},\nabla_{\b\bd} \big] = \ve_{\ad\bd}&\Big(W_{\a\b}{}^{\g\d}M_{\g\d}+\frac{1}{4}\nabla^{\d\gd}W_{\a\b\d}{}^{\g}K_{\g\gd}\Big) ~~~~~~~~~~~~~~~\non \\
+\ve_{\a\b}&\Big(\bar{W}_{\ad\bd}{}^{\gd\dd}\bar{M}_{\gd\dd}+\frac{1}{4}\nabla^{\g\dd}\bar{W}_{\ad\bd\dd}{}^{\gd}K_{\g\gd}\Big) ~.\label{CCDA4D}
\end{align}
From this one may derive the following two very useful identities
\begin{subequations}
\begin{align}
\nabla_{\a}{}^{\bd}\nabla_{\b\bd}&=-\ve_{\a\b}\Box_c-W_{\a\b}{}^{\g(2)}M_{\g(2)}-\frac{1}{4}\nabla^{\d\gd}W_{\a\b\d}{}^{\g}K_{\g\gd} ~,\\
\nabla_{\ad}{}^{\b}\nabla_{\bd\b}&=-\ve_{\ad\bd}\Box_c-\bar{W}_{\ad\bd}{}^{\gd(2)}\bar{M}_{\gd(2)}-\frac{1}{4}\nabla^{\g\dd}\bar{W}_{\ad\bd\dd}{}^{\gd}K_{\g\gd} ~.
\end{align}
\end{subequations}
The commutation relations of the remaining generators of $\mf{so}(4,2)$ with 
$\nabla_{\b\bd} $ 
are given by
\begin{subequations}\label{confalM4}
\bea
\big[M_{\a\g},\nabla_{\b\bd} \big]&=&
\ve_{\b(\a}\nabla_{\g)\bd}~,\qquad 
\big[ \bar{M}_{\ad\gd},\nabla_{\b\bd} \big]=\ve_{\bd(\ad}\nabla_{\b\gd)}~,~~~~~\\
&&~~\big[\mathbb{D},\nabla_{\b\bd} \big]=\nabla_{\b\bd}~,\\
\big[K_{\a\ad},\nabla_{\b\bd}\big] &=& 4\Big(\ve_{\ad\bd}M_{\a\b}+\ve_{\a\b}\bar{M}_{\ad\bd}-\ve_{\a\b}\ve_{\ad\bd}\mathbb{D}\Big)~.
\eea
\end{subequations}
For future calculations, it is also helpful to have identities for the following commutators 
\begin{subequations}
\begin{align}
\big[K_{\a\ad},\Box_c\big]&=-4\Big(\nabla_{\a}{}^{\bd}\bar{M}_{\ad\bd}+\nabla_{\ad}{}^{\b}M_{\a\b}-\nabla_{\a\ad}\big(\mb{D}-1\big)\Big)~,\\
\big[\nabla_{\a\ad},\Box_c\big]&=\Big(\nabla_{\ad}{}^{\b}W_{\a\b}{}^{\g(2)}+W_{\a\b}{}^{\g(2)}\nabla_{\ad}{}^{\b}\Big)M_{\g(2)}
+\Big(\nabla_{\a}{}^{\bd}\bar{W}_{\ad\bd}{}^{\gd(2)}+\bar{W}_{\ad\bd}{}^{\gd(2)}\nabla_{\a}{}^{\bd}\Big)\bar{M}_{\gd(2)}\non\\
&+\frac{1}{4}\Big(\nabla_{\b}{}^{\gd}W_{\a}{}^{\b(2)\g}\nabla_{\b\ad}+\nabla_{\bd}{}^{\g}\bar{W}_{\ad}{}^{\bd(2)\gd}\nabla_{\a\bd}+\frac{1}{2}B_{\a}{}^{\g\gd}{}_{\ad}\Big)K_{\g\gd}~,
\end{align}
\end{subequations}
where $B_{\a(2)\ad(2)} $ is the Bach tensor \eqref{BachCS}.

From \eqref{CCDA4D} it follows that on any conformally-flat background the covariant derivatives of conformal space commute, 
\begin{align}
W_{\a(4)}=0 \qquad \implies \qquad  \big[\nabla_{\a\ad},\nabla_{\b\bd}\big]=0~. \label{FlatCCDCF4}
\end{align}
The following discussions apply on an arbitrary background, unless otherwise stated. 

\subsection{Conformal higher-spin gauge fields} \label{secCHSPrepCS4}

Within the setting of conformal space, a spin-$\frac{1}{2}(m+n)$ conformal gauge field is also realised by the totally symmetric rank-$(m,n)$ spinor field $h_{\a(m)\ad(n)}$ and its conjugate $\bar{h}_{\a(n)\ad(m)}$, with $m\geq 1$ and $n\geq 1$. However, in place of those listed in section \ref{secCHS4prep}, we now define $h_{\a(m)\ad(n)}$  to satisfy the following properties: 
\begin{enumerate}[label=(\roman*)]

\item $h_{\a(m)\ad(n)}$ is a primary tensor field
 with conformal weight $\Delta_{h_{(m,n)}}=\big(2-\frac{1}{2}(m+n)\big)$
 \begin{align}
 K_{\b\bd}h_{\a(m)\ad(n)}=0~,\qquad \mathbb{D}h_{\a(m)\ad(n)}=\big(2-\frac{1}{2}(m+n)\big)h_{\a(m)\ad(n)}~. \label{CHS4Dprepprop}
 \end{align}
 
 \item $h_{\a(m)\ad(n)}$ is defined modulo gauge transformations of the form
 \begin{align}
\d_\x h_{\a(m)\ad(n) } = \nabla_{(\a_1 (\ad_1 } \x_{\a_2 \dots \a_m)\ad_2\dots\ad_n) }~, \label{CHSprepGT4}
\end{align}
where the gauge parameter $\xi_{\a(m-1)\ad(n-1)}$ is also a primary field, 
\end{enumerate}
\vspace{-12pt}
\begin{align}
 K_{\b\bd}\x_{\a(m-1)\ad(n-1)}=0~,\qquad \mathbb{D}\xi_{\a(m-1)\ad(n-1)}=\big(1-\frac{1}{2}(m+n)\big)\xi_{\a(m-1)\ad(n-1)}~.
\end{align}

In this context, the conformal weight of $h_{\a(m)\ad(n)}$ is fixed by consistency of the above properties with the conformal algebra \eqref{confalM4}. From \eqref{GtransPhi}, it follows that under the conformal gravity gauge group $\mc{G}$, the prepotential transforms according to the rule 
\begin{align}
\delta_{\L}^{(\mc{G})}h_{\a(m)\ad(n)}=
\Big(\xi^a\nabla_a +\frac{1}{2}K^{ab}M_{ab}+\Delta_{h_{(m,n)}}\s\Big)h_{\a(m)\ad(n)}~, \label{transrule679}
\end{align}
where we have used the properties \eqref{CHS4Dprepprop}.
We note that after degauging, i.e. imposing the gauge condition \eqref{degauge}, the transformation rules \eqref{CHSprepGT4} and \eqref{transrule679} reduce to \eqref{GTCHSprepM4} and \eqref{transruleM444} respectively. Therefore, we see that the definition of a conformal gauge field  in conformal space is equivalent to the one given in the vielbein approach. 

\subsection{Linearised higher-spin Weyl and Bach tensors}\label{secNSFieldStr}

In conformal space we can construct the higher-spin Weyl and Bach tensors of $h_{\a(m)\ad(n)}$ in complete analogy with the guiding principles set out in section \ref{secLinWeylBachVA}. In particular, the analogue of the defining features of the higher-spin Weyl tensors are as follows:
\begin{enumerate}[label=(\roman*)]
\item
$\mc{W}_{\a(m+n)}(h)$ and $\overline{\mc{W}}_{\ad(m+n)}(h)$ are primary tensors with conformal weights $\Delta_{\mc{W}^{(m,n)}}=\big(2-\frac{1}{2}(m-n)\big)$ and $\Delta_{\overline{\mc{W}}^{(m,n)}}=\big(2+\frac{1}{2}(m-n)\big)$ respectively, 
\end{enumerate}
\vspace{-12pt}
\begin{subequations}\label{HSWeylCP}
\begin{align}
K_{\b\bd}\mc{W}_{\a(m+n)}(h)&=0~,\qquad 
\mathbb{D}\mc{W}_{\a(m+n)}(h)=\Big(2-\frac{1}{2}(m-n)\Big)\mc{W}_{\a(m+n)}(h)~,\\
K_{\b\bd}\overline{\mc{W}}_{\ad(m+n)}(h)&=0~,\qquad \mathbb{D}\overline{\mc{W}}_{\ad(m+n)}(h)=\Big(2+\frac{1}{2}(m-n)\Big)\overline{\mc{W}}_{\ad(m+n)}(h)~.
\end{align}
\end{subequations}
\begin{enumerate} [label=(\roman*)] \setcounter{enumi}{1}
\item  
$\mc{W}_{\a(m+n)}(h)$ and $\overline{\mc{W}}_{\ad(m+n)}(h)$  are descendants of the form $\mc{W}_{\a(m+n)}(h)=(\hat{\cA} h)_{\a(m+n)}$ and $\overline{\mc{W}}_{\ad(m+n)}(h)=(\check{\cA} h)_{\ad(m+n)}$. Here $\hat{\cA}$ and $\check{\cA}$ are some linear differential operators of order $n$ and $m$ respectively, each involving 
the conformally covariant derivative $\nabla_{\a\ad}$, the Weyl tensor $\big(W_{\a(4)},\bar{W}_{\ad(4)}\big)$ 
and its covariant derivatives.

\item
$\mc{W}_{\a(m+n)}(h)$ and $\overline{\mc{W}}_{\ad(m+n)}(h)$ have vanishing gauge variation under \eqref{CHSprepGT4} if spacetime is conformally-flat,
\begin{subequations} \label{HSWeylGICS4}
\begin{align}
\d_{\xi} \mc{W}_{\a(m+n)} (h)&= \mc{O}\big( W\big)~,\\
\d_{\xi} \overline{\mc{W}}_{\ad(m+n)} (h)&= \mc{O}\big( W\big)~.
\end{align}
\end{subequations}

\end{enumerate}

Similarly, in conformal space the defining features of the linearised higher-spin Bach tensor $\mc{B}_{\a(n)\ad(m)}(h)$ of $h_{\a(m)\ad(n)}$ are as follows:
\begin{enumerate}[label=(\roman*)]
\item
$\mc{B}_{\a(n)\ad(m)}(h)$ is a primary tensor with conformal weight $\Delta_{\mc{B}^{(m,n)}}=\big(2+\frac{1}{2}(m+n)\big)$, 
\begin{align}
K_{\b\bd}\mc{B}_{\a(n)\ad(m)}(h)=0~,\qquad \mb{D} \mc{B}_{\a(n)\ad(m)}(h) = \Big(2+\frac{1}{2}(m+n) \Big) \mc{B}_{\a(n)\ad(m)}(h)~.
\label{HSBachCP}
\end{align}

\item  
$\mc{B}_{\a(n)\ad(m)}(h)$ is of the form $\mc{B}_{\a(n)\ad(m)}(h)=(\cA h)_{\a(m)\ad(n)}$, 
where $\cA$ is a linear differential operator of order $(m+n)$ involving 
the conformally covariant derivative $\nabla_{\a\ad}$, the Weyl tensor
 $\big(W_{\a(4)},\bar{W}_{\ad(4)}\big)$ and its covariant derivatives.

\item
$\mc{B}_{\a(n)\ad(m)}(h)$ has vanishing gauge variation under \eqref{CHSprepGT4} if spacetime is Bach-flat,
\begin{align}
\d_{\xi} \mc{B}_{\a(n)\ad(m)} (h)= \mc{O}\big( B\big)~.
\label{HSBachGICS4}
\end{align}

\item
 $\mc{B}_{\a(n)\ad(m)}(h)$ is transverse if spacetime is Bach-flat,
\begin{align}
\nabla^{\b\bd} \mc{B}_{\b\a(n-1)\bd\ad(m-1)}(h) = \mc{O}\big( B\big)~.\label{HSBachDivCS4}
\end{align} 
\end{enumerate}

Properties \eqref{HSBachGICS4}  and \eqref{HSBachDivCS4} cannot be satisfied for generic $m$ and $n$; see section \ref{secCHSBach}.
Let us, for the moment, consider only those spacetimes with vanishing Weyl tensor,
\begin{align}
W_{abcd}=0~.
\end{align} 
Due to property (ii), the only possible expressions for $\mc{W}_{\a(m+n)}(h)$ and $\overline{\mc{W}}_{\ad(m+n)}(h)$, which reduce to \eqref{HSWeylIMink4} in the flat limit, is the minimal lift, $\pa_{\a\ad} \rightarrow \nabla_{\a\ad}$, of \eqref{HSWeylIMink4}: 
\begin{subequations}\label{HSWeylCS}
\begin{align}
\mc{W}_{\a(m+n)}(h)&=\nabla_{(\a_1}{}^{\bd_1}\dots\nabla_{\a_n}{}^{\bd_n}h_{\a_{n+1}\dots\a_{n+m})\bd(n)}~,\label{HSWeylCSa}\\
\overline{\mc{W}}_{\ad(m+n)}(h)&=\nabla_{(\ad_1}{}^{\b_1}\dots\nabla_{\ad_m}{}^{\b_m}h_{\b(m)\ad_{m+1}\dots\ad_{m+n})}~.\label{HSWeylCSb}
\end{align}
\end{subequations}
Since spacetime is conformally-flat, the covariant derivatives commute, \eqref{FlatCCDCF4}.
Therefore, gauge invariance of \eqref{HSWeylCS} under \eqref{CHSprepGT4} follows immediately from that of \eqref{HSWeylIMink4},
\begin{align}
\delta_{\xi}\mc{W}_{\a(m+n)}(h)=0~,\qquad \delta_{\xi}\overline{\mc{W}}_{\ad(m+n)}(h)=0~. \label{HSWeylCFGI}
\end{align}
It is clear that both \eqref{HSWeylCSa} and \eqref{HSWeylCSb} are of the form specified in property (ii), and also that they have the correct Weyl weights as prescribed in property (i). It remains to show that they are primary, to which we now turn. 
 
Using the algebra \eqref{confalM4} it is possible to prove, via induction on $j$, that the following identity holds 
\begin{align}
\big[K_{\b\bd},\big(\nabla_{\a\ad}\big)^j\big]&= 4j\big(\nabla_{\a\ad}\big)^{j-1}\Big[ 
\ve_{\bd\ad}M_{\b\a}+\ve_{\b\a}\bar{M}_{\bd\ad} -\ve_{\b\a}\ve_{\bd\ad}\big(\mb{D}+j-1\big)\Big]~. \label{CSFundaId1}
\end{align}
It is important to note that in the above identity we are using the convention \eqref{SymCon2}.
Using \eqref{CSFundaId1} it is easy to check that the special conformal generators annihilate the higher-spin Weyl tensors \eqref{HSWeylCS}, which are therefore primary. In fact, in deriving \eqref{CSFundaId1}, no assumptions on the background spacetime need to be made. Hence it applies equally well on all spacetimes, from which it follows that \eqref{HSWeylCP} are primary on a generic background.
 
 In a similar manner, we deduce that in a conformally-flat background, the correct expressions for the higher-spin Bach tensors is  the minimal lift of \eqref{HSBachPrepMink}:
\begin{subequations} \label{HSBachCSCF}
\begin{align}
\mc{B}_{\a(n)\ad(m)}(h)&=\nabla_{(\ad_1}{}^{\b_1}\cdots\nabla_{\ad_m)}{}^{\b_m}\mc{W}_{\a(n)\b(m)}(h)~,\label{HSBachCSCFa}\\
\widehat{\mc{B}}_{\a(n)\ad(m)}(h)&=\nabla_{(\a_1}{}^{\bd_1}\cdots\nabla_{\a_m)}{}^{\bd_m}\overline{\mc{W}}_{\ad(m)\bd(n)}(h)~.\label{HSBachCSCFb}
\end{align} 
\end{subequations}
They possess the same features as their $\mb{M}^4$ analogues. In particular, they are both equal to one another
\begin{align}
\mc{B}_{\a(n)\ad(m)}(h)=\widehat{\mc{B}}_{\a(n)\ad(m)}(h)~, \label{HSBachTypeEquivCF}
\end{align}
in addition to being transverse and gauge invariant,
\begin{subequations}\label{HSBachCS4Prop}
\begin{align}
\nabla^{\b\bd}\mc{B}_{\a(n-1)\b\ad(m-1)\bd}(h)&=0~,\label{HSBachCS4TT}\\
\delta_{\xi}\mc{B}_{\a(n)\ad(m)}(h)&=0~.\label{HSBachCS4GI}
\end{align}
\end{subequations}
Finally, the identity \eqref{CSFundaId1} may also be used to show that \eqref{HSBachCSCFa} is primary in a generic curved spacetime, i.e. satisfies \eqref{HSBachCP}.

  
  \subsection{Conformal higher-spin action}
  
The property \eqref{HSWeylCFGI} ensures that the action for the conformal gauge field $h_{\a(m)\ad(n)}$ 
\begin{align}
S_{\rm{CHS}}^{(m,n)} [ h,\bar{h}] 
&=\frac{1}{2}\text{i}^{m+n} \int \rd^4 x \, e \, \mc{W}^{\a(m+n)}(h)\mc{W}_{\a(m+n)}(\bar{h}) +\text{c.c.}~,
\label{CHSActionCF4}
\end{align}
is invariant under the gauge transformations \eqref{CHSprepGT4} on any conformally-flat background,
\begin{align}
W_{\a(4)}=0 \qquad \implies \qquad \delta_{\xi}S_{\text{CHS}}^{(m,n)}[h,\bar{h}]=0~.
\end{align}
On a generic background, this action is also invariant under the conformal gravity gauge group $\mc{G}$, $\delta_{\L}^{(\mc{G})}S_{\rm{CHS}}^{(m,n)} [ h,\bar{h}]=0$, on account of \eqref{HSWeylCP}. Hence it is a primary functional, which means that it is Weyl invariant upon degauging. For a discussion on lifting gauge invariance of this action to more general backgrounds, see section \ref{secCHSBach}.

 In a conformally-flat spacetime the conformally covariant derivatives obey the flat algebra \eqref{FlatCCDCF4}. This means that on such backgrounds, all of the CHS actions and their constituents (the higher-spin Bach and Weyl tensors)  inherit the nice features belonging to their Minkowski counterparts which were derived in section \ref{secCHSMink4}. For example, an equivalent form of the CHS action \eqref{CHSActionCF4} is
\begin{align}
S_{\rm{CHS}}^{(m,n)} [ h,\bar{h}] &=\frac{1}{2}\text{i}^{m+n} \int \rd^4 x \, e \, \bar{h}^{\a(n)\ad(m)}\mc{B}_{\a(n)\ad(m)}(h) +\text{c.c.}\non\\
&=\text{i}^{m+n} \int \rd^4 x \, e \,  \bar{h}^{\a(n)\ad(m)}\mc{B}_{\a(n)\ad(m)}(h)~.\label{CHSABachCF4}
\end{align}
 The only non-trivial part in proving their equivalence is the technical issue of integration by parts (IBP) in conformal space. However, in this case IBP is valid since the above functionals satisfy the prerequisites required for the IBP rule \eqref{Z.111} to hold.

Finally, we would like to point out that it is possible to extend the transverse spin projection operators \eqref{TTprojMink} of Minkowski space to arbitrary conformally-flat backgrounds. This is achieved by minimally lifting the operator \eqref{TTProjDeltaOp}
\begin{align}
\Delta_{\a}{}^{\ad}=\frac{\nabla_{\a}{}^{\ad}}{\sqrt{\Box_c}}~,\qquad \Delta_{\a}{}^{\bd}\Delta_{\bd}{}^{\b}=\delta_{\a}{}^{\b}~,\qquad \Delta_{\ad}{}^{\b}\Delta_{\b}{}^{\bd}=\delta_{\ad}{}^{\bd}~,
\end{align}
whereupon the action of the projectors on some field $\phi_{\a(m)\ad(n)}$ takes the form
\begin{subequations}\label{TTprojCF4}
\begin{align}
\Pi_{\perp}^{(m,n)}\phi_{\a(m)\ad(n)}&=\Delta_{(\ad_1}{}^{\b_1}\cdots\Delta_{\ad_n)}{}^{\b_n}\Delta_{(\b_1}{}^{\bd_1}\cdots\Delta_{\b_n}{}^{\bd_n}\phi_{\a_1\dots\a_m)\bd(n)} ~,\label{TTprojCF4a}\\
\widehat{\Pi}_{\perp}^{(m,n)}\phi_{\a(m)\ad(n)}&=\Delta_{(\a_1}{}^{\bd_1}\cdots\Delta_{\a_m)}{}^{\bd_m}\Delta_{(\bd_1}{}^{\b_1}\cdots\Delta_{\bd_m}{}^{\b_m}\phi_{\b(m)\ad_1\dots\ad_n)}\label{TTprojCF4b} ~.
\end{align}
\end{subequations}
This is of course valid only for tensors fields $\phi_{\a(m)\ad(n)}$ transforming under some representation of the conformal gravity gauge group $\mc{G}$, such as a CHS gauge field. In this case, the operators $\Pi_{\perp}^{(m,n)}$ and $\widehat{\Pi}_{\perp}^{(m,n)}$ satisfy all of the properties derived in section \eqref{secTTProjMink4}. 
In particular, in the spirit of Fradkin and Tseytlin, by minimally lifting \eqref{CHSATTProjMink}, they may be used to provide alternative realisations for the CHS action \eqref{CHSActionCF4} in terms of the transverse projectors.




 \subsection{Generalised CHS models}

As a simple extension of the above models, we now consider  
the generalised CHS gauge field $h^{(t)}_{\a(m)\ad(n)}$ of depth $t$, with $t$ a positive integer taking values in the range $1\leq t \leq \text{min}(m,n)$.
Gauge-invariant actions for arbitrary bosonic mixed symmetry generalised CHS fields in $\mb{M}^d$ were derived by Vasiliev \cite{Vasiliev2009}. 
Below we extend them to $4d$ conformally-flat backgrounds, and we also discuss fermionic generalised CHS fields for the first time. The field $h^{(t)}_{\a(m)\ad(n)}$ is defined to possess the following properties:
\begin{enumerate}[label=(\roman*)]

\item $h^{(t)}_{\a(m)\ad(n)}$ is a primary field
 with conformal weight $\Delta_{h_{(m,n,t)}}=\big((t+1)-\frac{1}{2}(m+n)\big)$
 \begin{align}
 K_{\b\bd}h^{(t)}_{\a(m)\ad(n)}=0~,\qquad \mathbb{D}h^{(t)}_{\a(m)\ad(n)}=\big((t+1)-\frac{1}{2}(m+n)\big)h^{(t)}_{\a(m)\ad(n)}~. \label{CHS4Dgenprepprop}
 \end{align}
 
 \item $h^{(t)}_{\a(m)\ad(n)}$ is defined modulo depth-$t$ gauge transformations of the form\footnote{Although the form of the gauge transformations \eqref{CHSgenprepGT4} is similar to that of a depth-$t$ partially massless field (cf. eq. \eqref{PMGSAdS4}), they are two different types of fields (the latter are not conformal). }
 \begin{align}
\delta_{\xi}h^{(t)}_{\a(m)\ad(n)}=\nabla_{(\a_1(\ad_1}\cdots\nabla_{\a_t\ad_t}\xi_{\a_{t+1}\dots\a_m)\ad_{t+1}\dots\ad_n)}\equiv \big(\nabla_{\a\ad}\big)^t\xi_{\a(m-t)\ad(n-t)}~, \label{CHSgenprepGT4}
\end{align}
where the gauge parameter $\xi_{\a(m-t)\ad(n-t)}$ is also a primary field, 
\end{enumerate}
\vspace{-12pt}
\begin{align}
 K_{\b\bd}\x_{\a(m-t)\ad(n-t)}=0~,\qquad \mathbb{D}\xi_{\a(m-t)\ad(n-t)}=\big(1-\frac{1}{2}(m+n)\big)\xi_{\a(m-1)\ad(n-1)}~.
\end{align}

The conformal weight of $h^{(t)}_{\a(m)\ad(n)}$ is fixed by consistency of the above properties with the conformal algebra \eqref{confalM4}, for which one should use the identity \eqref{CSFundaId1}. Through comparison with the discussion in section \ref{secCHSPrepCS4}, it is clear that the usual CHS gauge fields $h_{\a(m)\ad(n)}$ correspond to generalised CHS fields with the minimal depth of $t=1$.

As was done in the case $t=1$, from $h^{(t)}_{\a(m)\ad(n)}$ we may construct generalised higher-spin Weyl tensors $\mc{W}^{(t)}_{\a(m+n-t+1)\ad(t-1)}(h)$ and $\overline{\mc{W}}^{(t)}_{\a(t-1)\ad(m+n-t+1)}(h)$ which are primary,
\begin{align}
K_{\b\bd}\mc{W}^{(t)}_{\a(m+n-t+1)\ad(t-1)}(h)&=0~,\qquad K_{\b\bd}\overline{\mc{W}}^{(t)}_{\a(t-1)\ad(m+n-t+1)}(h)=0~,
\end{align}
and which possess the following conformal weights,
\begin{subequations}\label{GenHSWeylWeights}
\begin{align}
\mathbb{D}\mc{W}^{(t)}_{\a(m+n-t+1)\ad(t-1)}(h)&=\Big(2-\frac{1}{2}(m-n)\Big)\mc{W}^{(t)}_{\a(m+n-t+1)\ad(t-1)}(h)~,\\
\mathbb{D}\overline{\mc{W}}^{(t)}_{\a(t-1)\ad(m+n-t+1)}(h)&=\Big(2-\frac{1}{2}(n-m)\Big)\overline{\mc{W}}^{(t)}_{\a(t-1)\ad(m+n-t+1)}(h)~.
\end{align}
\end{subequations}

In a general curved spacetime one may show, using the identity \eqref{CSFundaId1}, 
that these properties are satisfied by the following expressions:
\begin{subequations}\label{GenHSWeylCF4}
\begin{align}
\mc{W}^{(t)}_{\a(m+n-t+1)\ad(t-1)}(h)&= \big(\nabla_{\a}{}^{\bd}\big)^{n-t+1}h^{(t)}_{\a(m)\ad(t-1)\bd(n-t+1)}~,\label{GenHSWeylCF4a}\\
\overline{\mc{W}}^{(t)}_{\a(t-1)\ad(m+n-t+1)}(h)&=\big(\nabla_{\ad}{}^{\b}\big)^{m-t+1}h^{(t)}_{\b(m-t+1)\a(t-1)\ad(n)}~.\label{GenHSWeylCF4b}
\end{align}
\end{subequations}
The action of the generalised HS Weyl operators on the conjugate field is given by
\begin{subequations}\label{GenHSWeylCCCF4}
\begin{align}
\mc{W}^{(t)}_{\a(m+n-t+1)\ad(t-1)}(\bar{h})&= \big(\nabla_{\a}{}^{\bd}\big)^{m-t+1}\bar{h}^{(t)}_{\a(n)\ad(t-1)\bd(m-t+1)}~,\\
\overline{\mc{W}}^{(t)}_{\a(t-1)\ad(m+n-t+1)}(\bar{h})&=\big(\nabla_{\ad}{}^{\b}\big)^{n-t+1}\bar{h}^{(t)}_{\b(n-t+1)\a(t-1)\ad(m)}~.
\end{align}
\end{subequations}
Once again, we see that the two types of generalised higher-spin Weyl tensors are related through complex conjugation according to the rule 
\begin{subequations}
\begin{align}
\Big(\mc{W}^{(t)}_{\a(m+n+t-1)\ad(t-1)}(h)\Big)^*&=\overline{\mc{W}}^{(t)}_{\a(t-1)\ad(m+n+t-1)}(\bar{h})~,\\
 \Big(\mc{W}^{(t)}_{\a(m+n-t+1)\ad(t-1)}(\bar{h})\Big)^*&=\overline{\mc{W}}^{(t)}_{\a(t-1)\ad(m+n-t+1)}(h)~.
\end{align}
\end{subequations}
of which \eqref{HSWeylCCRelsMink4} is a special case. Associated with these field strengths is the generalised CHS action
\begin{align}
S_{\text{CHS}}^{(m,n,t)}[h,\bar{h}]=\frac{1}{2}{\rm i}^{m+n}\int \text{d}^4x \, e \,\mc{W}_{(t)}^{\a(m+n-t+1)\ad(t-1)}(h)\mc{W}^{(t)}_{\a(m+n-t+1)\ad(t-1)}(\bar{h}) +\text{c.c.}~,\label{GenCHSActionCF4}
\end{align}
which is invariant under the conformal gravity gauge group $\mathcal{G}$, $\delta_{\L}^{(\mc{G})}S_{\rm{CHS}}^{(m,n,t)} [ h,\bar{h}]=0$ .
In any conformally-flat spacetime the generalised higher-spin Weyl tensors 
\eqref{GenHSWeylCF4} prove to be
invariant under the gauge transformations \eqref{CHSgenprepGT4}
\begin{align}
W_{abcd}=0\quad\implies\quad \delta_{\xi}\mc{W}^{(t)}_{\a(m+n-t+1)\ad(t-1)}(h)=\delta_{\xi}\overline{\mc{W}}^{(t)}_{\a(t-1)\ad(m+n-t+1)}(h) =0~,
\end{align}
and hence so too is the action \eqref{GenCHSActionCF4},
\begin{align}
W_{abcd}=0\quad\implies\quad \delta_{\xi}S_{\text{CHS}}^{(m,n,t)}[h,\bar{h}]=0~.
\end{align}

 The equations of motion that follow from \eqref{GenCHSActionCF4} are the vanishing of the generalised higher-spin Bach tensors, which are defined as follows
\begin{subequations}\label{GenHSBach}
\begin{align}
\mc{B}^{(t)}_{\a(n)\ad(m)}(h)&=\big(\nabla_{\ad}{}^{\b}\big)^{m-t+1}\mc{W}^{(t)}_{\b(m-t+1)\a(n)\ad(t-1)}(h)~,\label{GenHSBacha}\\ 
\widehat{\mc{B}}^{(t)}_{\a(n)\ad(m)}(h)&=\big(\nabla_{\a}{}^{\bd}\big)^{n-t+1}\overline{\mc{W}}^{(t)}_{\a(t-1)\ad(m)\bd(n-t+1)}(h)~.\label{GenHSBachb}
\end{align}
\end{subequations}
Both \eqref{GenHSBacha} and \eqref{GenHSBachb} are weight $3-t+\frac{1}{2}(m+n)$ primary descendents  in a general curved spacetime. In any conformally-flat spacetime, they are equal to one another,
\begin{align}
\mc{B}^{(t)}_{\a(n)\ad(m)}(h)=\widehat{\mc{B}}^{(t)}_{\a(n)\ad(m)}(h)~,
\end{align}
the proof of which proceeds in a fashion similar to that of \eqref{HSBachTypeEquiv}. Furthermore, $\mc{B}^{(t)}_{\a(n)\ad(m)}(h)$ satisfies the $t$-extended versions of the properties \eqref{HSBachCS4Prop}. Namely, it is partially conserved and invariant under the gauge transformations \eqref{CHSgenprepGT4},
\begin{subequations}\label{GenHSBachCS4Prop}
\begin{align} 
\big(\nabla^{\b\bd}\big)^t\mc{B}^{(t)}_{\a(n-t)\b(t)\ad(m-t)\bd(t)}(h)=0~,\label{GenHSBachCS4PropGI}\\
\delta_{\xi}\mc{B}^{(t)}_{\a(n)\ad(m)}(h)=0~.\label{GenHSBachCS4PropTT} 
\end{align} 
\end{subequations}

As was done in the $d=3$ case, we can once again ask which values of $t$ yield 
second-order Lagrangians in the action \eqref{GenCHSActionCF4}.\footnote{First-order models in this context are not well defined.} A similar analysis reveals that second order models exist only for bosonic spin. These models possess maximal depth with $t=m=n=s$ and were first discussed in \cite{Sachs,EO} (however, only Weyl symmetry was considered in \cite{Sachs,EO}, gauge symmetry was not discussed).


 \subsection{Degauging}
 
 To conclude this section, we discuss the $d=4$ degauging procedure, which turns out to be much more tractable than the $d=3$ one. 
In the gauge \eqref{degauge}, the conformal covariant derivative reads
\begin{align}
\nabla_{\a\ad}=\mathcal{D}_{\a\ad}-\frac{1}{4}S_{\a\ad,}{}^{\b\bd}K_{\b\bd}~. \label{DGCCD4}
\end{align}
 The Schouten tensor may be decomposed into irreducible ${\sSL} (2, {\mathbb C})$ components as
 \begin{align}
S_{\a\ad,\b\bd}=(\s^a)_{\a\ad}(\s^b)_{\b\bd}S_{ab}=\frac{1}{2}R_{\a\b\ad\bd}-\frac{1}{12}\ve_{\a\b}\ve_{\ad\bd}R~,
 \end{align}
 whereupon the conformally covariant derivative takes the form,
\begin{align}
\nabla_{\a\ad}=\mathcal{D}_{\a\ad}-\frac{1}{8}R_{\a}{}^{\b\bd}{}_{\ad}K_{\b\bd}+\frac{1}{48}RK_{\a\ad}~.\label{FullyDegaugedMate}
\end{align} 
 The aim is then to express the higher-spin Weyl and Bach tensors
 in terms of the torsion-free Lorentz covariant derivative, the curvature and the prepotential.
 
Let $\Phi$ be a weight $\Delta_{\Phi}$ primary tensor field (with Lorentz indices suppressed).  Using \eqref{CSFundaId1} and the degauged covariant derivative \eqref{FullyDegaugedMate}, it is possible to show that the following identity holds true
\begin{align}
\big(\mc{D}_{\a\ad}\big)^j\big(\nabla_{\a\ad}\big)^k\Phi&=\big(\mc{D}_{\a\ad}\big)^{j+1}\big(\nabla_{\a\ad}\big)^{k-1}\Phi+\frac{1}{2}(k-1)\big(\mc{D}_{\a\ad}\big)^{j}\Big\{R^{\b}{}_{\a\ad(2)}\big(\nabla_{\a\ad}\big)^{k-2}M_{\b\a}\non\\
&+R_{\a(2)\ad}{}^{\bd}\big(\nabla_{\a\ad}\big)^{k-2}\bar{M}_{\bd\ad}+R_{\a(2)\ad(2)}\big(\nabla_{\a\ad}\big)^{k-2}\big(\Delta_{\Phi}+k-2\big)\Big\}\Phi~.
 \end{align}
Here $j$ and $k$ are two arbitrary positive integers. Therefore, in any background spacetime with a vanishing traceless Ricci tensor, or in other words an Einstein space, 
 \begin{align}
R_{\a\b\ad\bd}=0 \qquad \Longleftrightarrow \qquad R_{ab} = \l \eta_{ab}~, \label{EinsteinSpace}
 \end{align}
 the degauging procedure is trivial and we obtain
\begin{subequations}
\begin{align}
\mc{W}^{(t)}_{\a(m+n-t+1)\ad(t-1)}(h)&= \big(\mc{D}_{\a}{}^{\bd}\big)^{n-t+1}h^{(t)}_{\a(m)\ad(t-1)\bd(n-t+1)}~,\label{HSWeylaDG}\\
\overline{\mc{W}}^{(t)}_{\a(t-1)\ad(m+n-t+1)}(h)&=\big(\mc{D}_{\ad}{}^{\b}\big)^{m-t+1}\bar{h}^{(t)}_{\b(m-t+1)\a(t-1)\ad(n)}~,\label{HSWeylbDG}\\
\mc{B}^{(t)}_{\a(n)\ad(m)}(h)&=\big(\mc{D}_{\ad}{}^{\b}\big)^{m-t+1}\mc{W}^{(t)}_{\b(m-t+1)\a(n)\ad(t-1)}(h)~,\label{HSBachaDG}\\ 
\widehat{\mc{B}}^{(t)}_{\a(n)\ad(m)}(h)&=\big(\mc{D}_{\a}{}^{\bd}\big)^{n-t+1}\overline{\mc{W}}^{(t)}_{\a(t-1)\ad(m)\bd(n-t+1)}(h)~.\label{HSBachbDG}
\end{align}
 \end{subequations}
We emphasise that these degauged field strengths are generally only gauge invariant if, in addition to \eqref{EinsteinSpace}, spacetime is restricted to be conformally-flat.  
  
 In the case where $R_{\a(2)\ad(2)}$ does not vanish we were not able to obtain a closed-form expression for the degauged version of any of the above tensors, for arbitrary $m$ and $n$. 
 However, the expressions for the bosonic higher-spin Weyl tensor $\mc{W}^{(t)}_{\a(m+n-t+1)\ad(t-1)}(h)$ with $t=1$ and $m=n=s$ for $s=2,3,4,5$ are as follows:
\begin{subequations} 
 \begin{align}
 \mc{W}_{\a(4)}(h)&=\big(\mc{D}_{\a}{}^{\bd}\big)^2h_{\a(2)\bd(2)} -\frac{1}{2}R_{\a(2)}{}^{\bd(2)}h_{\a(2)\bd(2)}~,\label{LinWeylDG}\\[6pt]
 \mc{W}_{\a(6)}(h)&=\big(\mc{D}_{\a}{}^{\bd}\big)^3h_{\a(3)\bd(3)}-\big(\mathcal{D}_{\a}{}^{\bd}R_{\a(2)}{}^{\bd(2)}\big)h_{\a(3)\bd(3)}-2R_{\a(2)}{}^{\bd(2)}\mathcal{D}_{\a}{}^{\bd}h_{\a(3)\bd(3)}~,\\[6pt]
  \mc{W}_{\a(8)}(h)&=\big(\mc{D}_{\a}{}^{\bd}\big)^4h_{\a(4)\bd(4)} -\frac{3}{2}\Big(\big(\mc{D}_{\a}{}^{\bd}\big)^2R_{\a(2)}{}^{\bd(2)}\Big)h_{\a(4)\bd(4)}-5\Big(\mathcal{D}_{\a}{}^{\bd}R_{\a(2)}{}^{\bd(2)}\Big)\mathcal{D}_{\a}{}^{\bd}h_{\a(4)\bd(4)} \notag\\
 &-5R_{\a(2)}{}^{\bd(2)}\big(\mc{D}_{\a}{}^{\bd}\big)^2h_{\a(4)\bd(4)} +\frac{9}{4}R_{\a(2)}{}^{\bd(2)}R_{\a(2)}{}^{\bd(2)}h_{\a(4)\bd(4)} ~,\\[6pt]
 \mc{W}_{\a(10)}(h)&=\big(\mc{D}_{\a}{}^{\bd}\big)^5h_{\a(5)\bd(5)} -2\Big(\big(\mc{D}_{\a}{}^{\bd}\big)^3R_{\a(2)}{}^{\bd(2)}\Big)h_{\a(5)\bd(5)}-9\Big(\big(\mc{D}_{\a}{}^{\bd}\big)^2R_{\a(2)}{}^{\bd(2)}\Big)\mathcal{D}_{\a}{}^{\bd}h_{\a(5)\bd(5)}\notag\\
 &-15\Big(\mc{D}_{\a}{}^{\bd}R_{\a(2)}{}^{\bd(2)}\Big)\big(\mc{D}_{\a}{}^{\bd}\big)^2h_{\a(5)\bd(5)} +16R_{\a(2)}{}^{\bd(2)}\Big(\mc{D}_{\a}{}^{\bd}R_{\a(2)}{}^{\bd(2)}\Big)h_{\a(5)\bd(5)}  \notag\\
 &-10R_{\a(2)}{}^{\bd(2)}\big(\mc{D}_{\a}{}^{\bd}\big)^3h_{\a(5)\bd(5)}+16R_{\a(2)}{}^{\bd(2)}R_{\a(2)}{}^{\bd(2)}\mathcal{D}_{\a}{}^{\bd}h_{\a(5)\bd(5)}~.
 \end{align}
 \end{subequations}
 Modulo terms involving the background Weyl tensor, eq. \eqref{LinWeylDG} proves to coincide with the linearised anti-self-dual  Weyl tensor $W_{\a(4)}$.
 
The expressions for the fermionic higher-spin Weyl tensor $\mc{W}^{(t)}_{\a(m+n-t+1)\ad(t-1)}(h)$ with $t=1$ and $m-1=n=s$ for $s=1,2,3,4$ are as follows:
\begin{subequations} 
 \begin{align}
 \mc{W}_{\a(3)}(h)&=\mathcal{D}_{\a}{}^{\bd}h_{\a(2)\bd} ~,\\[6pt]
 \mc{W}_{\a(5)}(h)&=\big(\mc{D}_{\a}{}^{\bd}\big)^2h_{\a(3)\bd(2)}-\frac 12 R_{\a(2)}{}^{\bd(2)}h_{\a(3)\bd(2)}~, \\[6pt]
  \mc{W}_{\a(7)}(h)&=\big(\mc{D}_{\a}{}^{\bd}\big)^3h_{\a(4)\bd(3)} -\Big(\mc{D}_{\a}{}^{\bd}R_{\a(2)}{}^{\bd(2)}\Big)h_{\a(4)\bd(3)} -2R_{\a(2)}{}^{\bd(2)}\mathcal{D}_{\a}{}^{\bd}h_{\a(4)\bd(3)}~,\\[6pt]
 \mc{W}_{\a(9)}(h)&=\big(\mc{D}_{\a}{}^{\bd}\big)^4h_{\a(5)\bd(4)} -\frac 32 \Big(\big(\mc{D}_{\a}{}^{\bd}\big)^2R_{\a(2)}{}^{\bd(2)}\Big)h_{\a(5)\bd(4)}-5\Big(\mc{D}_{\a}{}^{\bd}R_{\a(2)}{}^{\bd(2)}\Big)\mc{D}_{\a}{}^{\bd}h_{\a(5)\bd(4)}\notag\\
 &-5R_{(\a(2)}{}^{\bd(2)}\big(\mc{D}_{\a}{}^{\bd}\big)^2h_{\a(5)\bd(4)} +\frac 94 R_{\a(2)}{}^{\bd(2)}R_{\a(2)}{}^{\bd(2)}h_{\a(5)\bd(4)}~,
 \end{align}
 \end{subequations}
 whilst the corresponding expressions for $\mc{W}^{(t)}_{\a(m+n-t+1)\ad(t-1)}(\bar{h})$ are:
\begin{subequations} 
 \begin{align}
 \mc{W}_{\a(3)}(\bar{h})&=\big(\mc{D}_{\a}{}^{\bd}\big)^2\bar{h}_{\a\bd(2)} -\frac 12 R_{\a(2)}{}^{\bd(2)}\bar{h}_{\a\bd(2)}~,\\[6pt]
  \mc{W}_{\a(5)}(\bar{h})&=\big(\mc{D}_{\a}{}^{\bd}\big)^3\bar{h}_{\a(2)\bd(3)} -\Big(\mc{D}_{\a}{}^{\bd}R_{\a(2)}{}^{\bd(2)}\Big)\bar{h}_{\a(2)\bd(3)}-2R_{\a(2)}{}^{\bd(2)}\mc{D}_{\a}{}^{\bd}\bar{h}_{\a(2)\bd(3)}~,\\[6pt]
\mc{W}_{\a(7)}(\bar{h})&=\big(\mc{D}_{\a}{}^{\bd}\big)^4\bar{h}_{\a(3)\bd(4)}  -\frac 32 \Big(\big(\mc{D}_{\a}{}^{\bd}\big)^2R_{\a(2)}{}^{\bd(2)}\Big)\bar{h}_{\a(3)\bd(4)}-5\Big(\mc{D}_{\a}{}^{\bd}R_{\a(2)}{}^{\bd(2)}\Big)\mc{D}_{\a}{}^{\bd}\bar{h}_{\a(3)\bd(4)} \notag\\
 & -5R_{\a(2)}{}^{\bd(2)}\big(\mc{D}_{\a}{}^{\bd}\big)^2\bar{h}_{\a(3)\bd(4)} +\frac 94 R_{\a(2)}{}^{\bd(2)}R_{\a(2)}{}^{\bd(2)}\bar{h}_{\a(3)\bd(4)}~,\\[6pt]
 \mc{W}_{\a(9)}(\bar{h})&=\big(\mc{D}_{\a}{}^{\bd}\big)^5\bar{h}_{\a(4)\bd(5)} -2\Big(\big(\mc{D}_{\a}{}^{\bd}\big)^3R_{\a(2)}{}^{\bd(2)}\Big)\bar{h}_{\a(4)\bd(5)}-9\Big(\big(\mc{D}_{\a}{}^{\bd}
 \big)^2R_{\a(2)}{}^{\bd(2)}\Big)\mc{D}_{\a}{}^{\bd}\bar{h}_{\a(4)\bd(5)}\notag\\
 &-15\Big(\mc{D}_{\a}{}^{\bd}R_{\a(2)}{}^{\bd(2)}\Big)\big(\mc{D}_{\a}{}^{\bd}\big)^2\bar{h}_{\a(4)\bd(5)}+16R_{\a(2)}{}^{\bd(2)}\Big(\mc{D}_{\a}{}^{\bd}R_{\a(2)}{}^{\bd(2)}\Big)\bar{h}_{\a(4)\bd(5)}   \notag\\
 &-10R_{\a(2)}{}^{\bd(2)}\big(\mc{D}_{\a}{}^{\bd}\big)^3\bar{h}_{\a(4)\bd(5)}+16R_{\a(2)}{}^{\bd(2)}R_{\a(2)}{}^{\bd(2)}\mc{D}_{\a}{}^{\bd}\bar{h}_{\a(4)\bd(5)}~.
 \end{align}
 \end{subequations}


\section{Conformal higher-spin models on Bachgrounds} \label{secCHSBach}

Although the (generalised) CHS models derived in the previous section are Weyl invariant on arbitrarily curved backgrounds, they are gauge invariant only on conformally-flat ones. The purpose of this section is to investigate the conditions under which it is possible to lift the gauge invariance of the action \eqref{CHSActionCF4} to less restrictive backgrounds. 

A natural question to ask is; what are the minimal restrictions that should be placed on the metric in order for gauge invariant CHS models to exist? As is well known, consistent models for conformal spin-$3/2$ and spin-$2$ fields may be formulated at most on Bach-flat backgrounds (or `Bachgrounds'),
\begin{align}
B_{\a(2)\ad(2)}=0~. \label{CSBachFlat}
\end{align} 
In terms of the conformally covariant derivative $\nabla_{\a\ad}$, the Bach tensor takes the form\footnote{It may be checked that \eqref{BachCS} is equivalent to \eqref{BachSpinor} upon degauging. The reality of $B_{\a(2)\ad(2)}$ is not obvious, but follows from the Bianchi identity \eqref{BianchiCCDCS} upon projecting along the $K_{\a\ad}$ direction.} 
\begin{align}
B_{\a(2)\ad(2)}=\nabla_{(\ad_1}{}^{\b}\nabla_{\ad_2)}{}^{\b}W_{\a(2)\b(2)}=\nabla_{(\a_1}{}^{\bd}\nabla_{\a_2)}{}^{\bd}\bar{W}_{\ad(2)\bd(2)}=\bar{B}_{\a(2)\ad(2)}~. \label{BachCS}
\end{align}
It is therefore natural to expect that, for CHS models with spin greater than two, the vanishing of the background Bach tensor is also a necessary condition for gauge invariance.  
 This is the point of view that we adopt in this thesis, and we hereby restrict our attention to backgrounds satisfying \eqref{CSBachFlat}. 
 
Let us consider the depth-$t$ CHS gauge field $h_{\a(m)\ad(n)}^{(t)}$. In any non-conformally-flat background, the model which we seek must reduce to \eqref{GenCHSActionCF4} in the conformally-flat limit. In this sense the minimal lift of \eqref{GenCHSActionCF4}, which we denote by $S_{\rm{Skeleton}}^{(m,n,t)} [ h,\bar{h}]$,
\begin{align}
S_{\text{Skeleton}}^{(m,n,t)}[h,\bar{h}]=\frac{1}{2}{\rm i}^{m+n}\int \text{d}^4x \, e \,\mc{W}_{(t)}^{\a(m+n-t+1)\ad(t-1)}(h)\mc{W}^{(t)}_{\a(m+n-t+1)\ad(t-1)}(\bar{h}) +\text{c.c.}~,\label{GenCHSSkeleton}
\end{align} 
 constitutes the skeleton of the full gauge invariant CHS action.
This action is not gauge invariant, but is invariant under the conformal gravity gauge group $\mc{G}$. The gauge variation of the skeleton is directly proportional to the background Weyl tensor. Therefore, in order to restore gauge invariance to \eqref{GenCHSSkeleton}, we will need to supplement it with a non-minimal\footnote{Within the setting of conformal space, by non-minimal we mean that the subject has explicit dependence on the background Weyl tensor, so that it vanishes in the conformally-flat limit. } sector which is quadratic in $h_{\a(m)\ad(n)}$ and which explicitly depends on the Weyl tensor,
\begin{align}
S_{\rm{CHS}}^{(m,n,t)} [ h,\bar{h}]=S_{\rm{Skeleton}}^{(m,n,t)} [ h,\bar{h}]+S_{\text{NM}}^{(m,n,t)}[h,\bar{h}] ~,\qquad \delta_{\xi}S_{\rm{CHS}}^{(m,n,t)} [ h,\bar{h}]\approx 0~. \label{CHSBachActionTrunc}
\end{align}
 Here the symbol $\approx$ represents equality modulo terms involving the Bach tensor, $=\big|_{B_{ab}=0}$. 
 
 In eq. \eqref{CHSBachActionTrunc} we have denoted the non-minimal functional by $S_{\text{NM}}^{(m,n,t)}[h,\bar{h}]$, and it takes the generic form
\begin{align}
S_{\text{NM}}^{(m,n,t)}[h,\bar{h}]=\frac{1}{2}\text{i}^{m+n} \int \rd^4 x \, e \, \bar{h}_{(t)}^{\a(n)\ad(m)}\mc{J}^{(t)}_{\a(n)\ad(m)}(h)+\text{c.c.} \label{HSNMFunc}
\end{align}
The tensor $\mc{J}^{(t)}_{\a(n)\ad(m)}(h)$ is some non-minimal primary descendent of $h^{(t)}_{\a(m)\ad(n)}$ whose weight coincides with the Bach-tensor of $h^{(t)}_{\a(m)\ad(n)}$,
\begin{align}
K_{\b\bd}\mc{J}^{(t)}_{\a(n)\ad(m)}(h)=0~,\qquad \mb{D} \mc{J}^{(t)}_{\a(n)\ad(m)}(h) = \Big(3-t+\frac{1}{2}(m+n) \Big) \mc{J}^{(t)}_{\a(n)\ad(m)}(h)~. \label{HSNMBachCP}
\end{align}
This ensures that  \eqref{HSNMFunc}, and consequently \eqref{CHSBachActionTrunc}, is also invariant under $\mc{G}$.

Naively, one might expect that for CHS fields with spin $s>2$, the condition \eqref{CSBachFlat} is necessary and sufficient to ensure consistent propagation, as was true for $s\leq 2 $.  
However, recent studies of the (incomplete) minimal depth conformal spin-3 theory
\cite{NTCHS,GrigorievT,BeccariaT,Manvelyan2} indicate that this is not the case. 
More specifically, for a generic Bach-flat background, Nutma and Taronna \cite{NTCHS} constructed the conformal spin-3 action up to and including terms linear in the background curvature (this action was primary and gauge invariant up to, but not including, second order in curvature). A few years later Grigoriev and Tseytlin \cite{GrigorievT} demonstrated that for the pure conformal spin-3 theory,
 this is the best that one can do.\footnote{The authors also argued that the pure spin-$s$ bosonic CHS theory can always be made gauge invariant at least to first order in the Weyl tensor.} They conjectured that it might be possible to restore gauge invariance to all orders in the curvature by switching on a coupling to a conformal spin-1 field.\footnote{We point out that in section \ref{secSpin3Recipe}, using superspace techniques, we deduce that in addition to a conformal spin $1-3$ coupling, there should also be a conformal spin $2-3$ coupling present.} The details of the proposed spin 1$-$3 coupling were then worked out by Beccaria and Tseytlin \cite{BeccariaT}, though its specific relation to the pure spin 3 sector has not yet been determined, since the latter is known only to first order in the curvature. It should also be mentioned that Manvelyan and Poghosyan \cite{Manvelyan2} recently employed a geometric approach (in the sense of \cite{deWitFreedman,DD}) to explicitly construct an exact spin-3 non-minimal primary tensor field $\mc{J}^{(1)}_{\a(3)\ad(3)}(h)$ (see \eqref{HSNMBachCP} with $m=n=t+2=3$) which is at least first order in the curvature.\footnote{Such $\mc{J}^{(1)}_{\a(3)\ad(3)}(h)$ are the most difficult ones to construct. In contrast, one may easily construct ones that are third order in curvature, such as $\mc{J}^{(1)}_{\a(3)\ad(3)}(h)=W_{\a(2)}{}^{\b(2)}W_{\b(2)}{}^{\g(2)}W_{\g(2)}{}^{\d(2)}h^{(1)}_{\d(2)\a\ad(3)}$. } 



It is then natural to expect that this trend continues for fields of arbitrary rank, and that in general, full gauge invariance of the action for a CHS field is only attainable if one allows for mixing with `lower-spin' or `subsidiary' fields. 
In this case, for generic $m$ and $n$, the full CHS action will not be of the diagonal form \eqref{CHSBachActionTrunc}, but of the non-diagonal form
\begin{align}
S_{\rm{CHS}}^{(m,n,t)} [ h,\bar{h},\dots]=S_{\rm{Skeleton}}^{(m,n,t)} [ h,\bar{h}]+S_{\text{NM}}^{(m,n,t)}[h,\bar{h}] +\cdots~, \label{CHSBachAction}
\end{align}
where the dots represent any other conformal fields whose presence is necessary for gauge invariance. At present it seems that the only way to determine which particular extra fields are necessary, if any, is by explicit construction on a case-by-case basis. 

  
On account of the technical difficulty, in this thesis we do not tackle the problem of constructing a consistent model for the minimal depth conformal spin-3 field.\footnote{Though this is a work in progress.} 
However, it appears that new insights into the problem under consideration may be obtained by studying somewhat simpler dynamical systems -- generalised CHS fields
   in a gravitational background. The point is that one can decrease the number of derivatives appearing in the action by increasing the depth of the gauge transformations. This bypasses some of the technical difficulties associated with higher-derivative models
 such as the conformal spin-3 one. 
Therefore, in subsections \ref{secCMDspin3} and \ref{secCMDspin52} we construct gauge invariant models for the conformal maximal depth (CMD) spin-3 and spin-5/2 fields on arbitrary Bachgrounds. They were presented in \cite{spin3depth3}, and are the first examples of complete gauge invariant CHS models which exhibit the lower-spin coupling mechanism. In particular, we will see that in order to achieve gauge invariance it is necessary to couple these fields to so called `conformal non-gauge' fields, which are described in appendix \ref{AppCNGM}. In subsection  \ref{secConformalHook}, we present a gauge invariant model for the hooked conformal graviton on a Bachground. This model was presented in \cite{SCHS} and also requires a coupling to conformal non-gauge fields. It is the first example of a minimal depth CHS model with this property. Another novel feature of the hooked conformal graviton is that the gauge invariant model is not unique, and there is in fact a one parameter family of such models \cite{SCHSgen}.
 
  
 Prior to discussing these, we will first present models for CHS fields which meet our naive expectations; that gauge invariance may be achieved on Bachgrounds without the use of subsidiary fields. This includes models for the conformal 
(i) gravitino  \ref{secCGravitino}; (ii) graviton \ref{secCGraviton}; and (iii) maximal depth graviton  \ref{secCMDgraviton}. Both (i) and (ii) are not new, whilst (iii) first appeared in \cite{Confgeo}.
 For these examples, the full model takes the form \eqref{CHSBachActionTrunc}, and it is interesting to note that they may be recast into the form 
 \begin{align}
S_{\rm{CHS}}^{(m,n,t)} [ h,\bar{h}]&=\frac{1}{2}\text{i}^{m+n} \int \rd^4 x \, e \, \bar{h}_{(t)}^{\a(n)\ad(m)}\mc{B}^{(t)}_{\a(n)\ad(m)}(h) +\text{c.c.} \label{BachBachChicken}
\end{align}
 Here we have defined the full linearised higher-spin Bach-tensor $\mc{B}^{(t)}_{\a(n)\ad(m)}(h)$,
 \begin{align}
\mc{B}^{(t)}_{\a(n)\ad(m)}(h):=\widetilde{\mc{B}}^{(t)}_{\a(n)\ad(m)}(h)+\mc{J}^{(t)}_{\a(n)\ad(m)}(h)~,
\end{align} 
 where $\widetilde{\mc{B}}^{(t)}_{\a(n)\ad(m)}(h)$ is the minimally lifted higher-spin Bach tensor \eqref{GenHSBach}. 
 The descendent $\mc{B}^{(t)}_{\a(n)\ad(m)}(h)$ is gauge invariant and transverse on a Bach-flat background,
 \begin{align}
 \delta_{\xi}\mc{B}_{\a(n)\ad(m)}(h)\approx 0~, \qquad \nabla^{\b\bd}\mc{B}_{\a(n-1)\b\ad(m-1)\bd}(h)\approx 0~.
 \end{align}
Models for CHS fields which require lower-spin couplings cannot be recast into this form.

\subsection{The conformal gravitino}\label{secCGravitino}

The field with $m=n+1=2$ and $t=1$ corresponds to the conformal gravitino, and the gauge-invariant model can be extracted from the action for $\cN=1$ conformal supergravity \cite{KTvN1,KTvN2} by linearising it around a Bach-flat background. Below we review this model, but from a constructive perspective. 

The conformal gravitino is described by a complex primary field $h_{\a(2)\ad}$ and its conjugate each with Weyl weight $+1/2$ , 
\begin{align}
K_{\b\bd}h_{\a(2)\ad}=0~,\qquad\mathbb{D}h_{\a(2)\ad}=\frac{1}{2}h_{\a(2)\ad}~,
\end{align}
It is defined modulo gauge transformations of the type
 \begin{align}\label{GravitinoGT}
 \delta_{\xi}h_{\a(2)\ad}=\nabla_{\a\ad}\xi_{\a}~.
 \end{align}
 where the complex gauge parameter $\xi_{\a}$ is primary of dimension $-1/2$.
 
 
  Associated with the gravitino are the two field strengths 
 \begin{align}
 \mc{W}_{\a(3)}(h)=\nabla_{\a}{}^{\bd}h_{\a(2)\bd}~,\qquad \mc{W}_{\a(3)}(\bar{h})=\nabla_{\a}{}^{\bd}\nabla_{\a}{}^{\bd}\bar{h}_{\a\bd(2)}~,
 \end{align} 
  which are primary descendants of dimensions $+3/2$ and $+5/2$ respectively.
Under \eqref{GravitinoGT}, their variations are proportional to the background Weyl tensor,
 \begin{align}
 \d_{\xi}\mc{W}_{\a(3)}(h)=W_{\a(3)\d}\xi^{\d}~,\qquad \d_{\xi}\mc{W}_{\a(3)}(\bar{h})=\frac{1}{2}W_{\a(3)\d}\nabla^{\d\dd}\bar{\xi}_{\dd}-\nabla^{\d\dd}W_{\a(3)\d}\bar{\xi}_{\dd}~.
 \end{align} 
 Consequently, the skeleton action \eqref{GenCHSSkeleton} has a non-zero gauge variation given by 
 \begin{align}
 \delta_{\xi}S_{\text{Skeleton}}^{(2,1)}[h,\bar{h}]=-\frac{1}{2}\text{i}\int\text{d}^4x \, e \,&\bigg\{ \frac{1}{2}\bar{\xi}^{\ad}\bigg[W^{\b(3)\d}\nabla_{\d\ad}\mc{W}_{\b(3)}(h)+3\nabla_{\d\ad}W^{\b(3)\d}\mc{W}_{\b(3)}(h)\bigg]\notag\\
& -\xi^{\a}W_{\a}{}^{\b(3)}\mc{W}_{\b(3)}(\bar{h})\bigg\}+\text{c.c.}
 \end{align}
However, if the skeleton is supplemented by the non-minimal primary correction
\begin{subequations}
\begin{align}
 S_{\text{NM}}^{(2,1)}[h,\bar{h}]&=
 -\frac{1}{2}\ri\int\text{d}^4x\,e\, \bar{h}^{\a\ad(2)}\mc{J}_{\a\ad(2)}(h)+\text{c.c.}~,\\ \mc{J}_{\a\ad(2)}(h)&=W_{\a}{}^{\b(3)}\nabla_{\b\ad}h_{\b(2)\ad}+2\nabla_{\b\ad}W_{\a}{}^{\b(3)}h_{\b(2)\ad}~,
 \end{align}
 \end{subequations}
where $\mc{J}_{\a\ad(2)}$ is primary with weight $+7/2$, then the sum of the two sectors
 \begin{align}
 S_{\text{CHS}}^{(2,1)}[h,\bar{h}]=S_{\text{Skeleton}}^{(2,1)}[h,\bar{h}]+S_{\text{NM}}^{(2,1)}[h,\bar{h}]
  \label{CgravitinoAct}
 \end{align}
has gauge variation that is strictly proportional to the Bach tensor,
\begin{align}
\delta_{\xi}S_{\text{CHS}}^{(2,1)}[h,\bar{h}]=-\frac{1}{2}\text{i}\int\text{d}^4x\,e\,\bigg\{\xi^{\alpha}B_{\a}{}^{\b\bd(2)}\bar{h}_{\b\bd(2)}+\bar{\xi}^{\ad}B^{\b(2)\bd}{}_{\ad}h_{\b(2)\bd}\bigg\}+\text{c.c.}
\end{align}
Therefore, the action \eqref{CgravitinoAct} is gauge invariant when restricted to Bach-flat backgrounds
 \bea
 \delta_{\xi}S_{\text{CHS}}^{(2,1)}[h,\bar{h}]\approx 0~.
 \eea

We may recast the full gauge invariant action \eqref{CgravitinoAct} into the form \eqref{BachBachChicken}, with linearised Bach tensor
\begin{align}
\mc{B}_{\a\ad(2)}(h)&=\widetilde{\mc{B}}_{\a\ad(2)}(h)+\mc{J}_{\a\ad(2)}(h)~.
\end{align}
Here $\mc{B}_{\a\ad(2)}(h)$ is gauge invariant and transverse on any Bachground,
\begin{align}
\delta_{\xi}\mc{B}_{\a\ad(2)}(h)\approx 0~, \qquad \nabla^{\b\bd}\mc{B}_{\b\bd\ad}(h)\approx 0~.
\end{align}

\subsection{The conformal graviton} \label{secCGraviton}

The CHS field with $m=n=2$ and $t=1$ corresponds to the conformal graviton.  
Below we construct a conformally invariant model describing its dynamics that is gauge invariant on any Bach-flat background. This model may of course be obtained by linearising the action of $d=4$ conformal gravity around a Bach-flat background, as was done in section \eqref{SectionCGpeasant}.
 
In accordance with section \ref{secCHSPrepCS4}, the conformal graviton is described by the real  field $h_{\a(2)\ad(2)}=\bar{h}_{\a(2)\ad(2)}$ which is primary with zero Weyl weight,
\begin{align}
K_{\b\bd}h_{\a(2)\ad(2)}=0~,\qquad \mb{D}h_{\a(2)\ad(2)}=0~.
\end{align}
It is defined modulo the gauge transformations
\begin{align}
\delta_{\xi}h_{\a(2)\ad(2)}=\nabla_{\a\ad}\xi_{\a\ad}~. \label{CGravitonGT}
\end{align} 
where the real gauge parameter $\xi_{\a\ad}$ is a primary field with conformal weight $-1$ . 

Associated with the conformal graviton is the linearised Weyl tensor,
\begin{align}
\mc{W}_{\a(4)}(h)=\nabla_{\a}{}^{\bd}\nabla_{\a}{}^{\bd}h_{\a(2)\bd(2)}~, \label{LinWeylTensor}
\end{align}
which is a primary field of dimension 2. Under the gauge transformation \eqref{CGravitonGT}, its variation is given by
\begin{align}
\delta_{\xi}\mc{W}_{\a(4)}(h)=\frac{1}{2}W_{\a(4)}\nabla^{\b\bd}\xi_{\b\bd}-\xi_{\b\bd}\nabla^{\b\bd}W_{\a(4)}-2W^{\b}{}_{\a(3)}\nabla_{\a)}{}^{\bd}\xi_{\b\bd}~,
\end{align}
and hence the gauge transformation of the corresponding skeleton \eqref{GenCHSSkeleton} is
\begin{align}
\delta_{\xi}S_{\text{skeleton}}^{(2,2)}[h]=\frac{1}{2}\int\text{d}^4x \, e \, \xi^{\a\ad}&\bigg\{4\mc{W}^{\b(4)}(h)\nabla_{\b\ad}W_{\b(3)\a}+4W_{\b(3)\a}\nabla_{\b\ad}\mc{W}^{\b(4)}(h) \notag\\
&-W_{\b(4)}\nabla_{\a\ad}\mc{W}^{\b(4)}(h)-3\mc{W}^{\b(4)}(h)\nabla_{\a\ad}W_{\b(4)}
\bigg\}+\text{c.c.}
\end{align}
  
To restore gauge invariance to this model, we seek a weight $+4$ non-minimal primary correction $\mc{J}_{\a(2)\ad(2)}(h)$ to the minimal linearised Bach tensor, 
\begin{align}
K_{\b\bd}\mc{J}_{\a(2)\ad(2)}(h)=0~,\qquad \mathbb{D}\mc{J}_{\a(2)\ad(2)}(h)=4\mc{J}_{\a(2)\ad(2)}(h)~. \label{Spin2BachCorrGen}
\end{align}
Restricting our attention to the construction of tensors with the  properties of $\mc{J}_{\a(2)\ad(2)}(h)$ greatly lightens the workload. In fact, beginning with the most general weight $+4$ tensor with this index structure, the condition of being primary is so strong that one may show that there are only three (up to complex conjugation) such linearly independent tensors that are linear in the Weyl tensor. They are given by
\begin{subequations}\label{Spin2BachCorr}
\begin{align}
\mc{J}^{(1)}_{\a(2)\ad(2)}(h)&= B_{\a}{}^{\b\bd}{}_{\ad}h_{\a\b\bd\ad}~,\label{Spin2BachCorra}\\
\mc{J}^{(2)}_{\a(2)\ad(2)}(h)&=-2W_{\a(2)}{}^{\b(2)}\nabla_{\ad}{}^{\g}\nabla_{\b}{}^{\gd}h_{\g\b\ad\gd}-\nabla_{\ad}{}^{\g}W_{\a(2)}{}^{\b(2)}\nabla_{\g}{}^{\gd}h_{\b(2)\ad\gd} \notag\\
&\quad+2\nabla_{\ad}{}^{\g}W_{\a(2)}{}^{\b(2)}\nabla_{\b}{}^{\gd}h_{\b\g\ad\gd} -\nabla_{\a\ad}W_{\a}{}^{\b(3)}\nabla_{\b}{}^{\gd}h_{\b(2)\ad\gd}\notag\\
&\quad-\nabla_{\a}{}^{\gd}W_{\a}{}^{\b(3)}\nabla_{\b\ad}h_{\b(2)\ad\gd} +3\nabla_{\b}{}^{\gd}W_{\a(2)}{}^{\b(2)}\nabla_{\gd}{}^{\g}h_{\g\b\ad(2)}\notag\\
&\quad+\frac 12\nabla^{\g\gd}W_{\a(2)}{}^{\b(2)}\nabla_{\g\gd}h_{\b(2)\ad(2)} +h_{\b(2)\ad(2)}\Box_c W_{\a(2)}{}^{\b(2)}~,\label{Spin2BachCorrb}\\ 
\mc{J}^{(3)}_{\a(2)\ad(2)}(h)&=-W_{\a}{}^{\b(3)}\nabla_{\b\ad}\nabla_{\b}{}^{\gd}h_{\b\a\ad\gd}+W_{\a(2)}{}^{\b(2)}\Box_c h_{\b(2)\ad(2)}\notag\\
&\quad-2\nabla_{\b\ad}W_{\a}{}^{\b(3)}\nabla_{\a}{}^{\gd}h_{\b(2)\ad\gd}-\nabla_{\b}{}^{\gd}W_{\a(2)}{}^{\b(2)}\nabla_{\gd}{}^{\g}h_{\b\g\ad(2)}\notag\\
&\quad-\frac 12\nabla^{\g\gd}W_{\a(2)}{}^{\b(2)}\nabla_{\g\gd}h_{\b(2)\ad(2)} -\nabla_{\ad}{}^{\g}\nabla_{\b}{}^{\gd}W_{\a(2)}{}^{\b(2)}h_{\g\b\gd\ad}\notag\\
&\quad+\Box_c W_{\a(2)}{}^{\b(2)}h_{\b(2)\ad(2)}~. \label{Spin2BachCorrc}
\end{align}
We remind the reader that the convention \eqref{SymCon2} is in play. 

There are precisely three (up to complex conjugation) linearly independent structures that are quadratic in the Weyl tensor and which satisfy the properties \eqref{Spin2BachCorrGen}. They are given by
\begin{align}
	\mc{J}^{(4)}_{\a(2)\ad(2)}(h)&=W_{\a(2)}{}^{\g(2)}W_{\g(2)}{}^{\b(2)}h_{\b(2)\ad(2)} ~,\label{Spin2BachCorrd}\\ 
	\mc{J}^{(5)}_{\a(2)\ad(2)}(h)&=W_{\a\g(2)}{}^{\b}W_{\a}{}^{\b\g(2)}h_{\b(2)\ad(2)}~,\label{Spin2BachCorre}\\ 
	\mc{J}^{(6)}_{\a(2)\ad(2)}(h)&=W_{\a(2)}{}^{\b(2)}\bar{W}_{\ad(2)}{}^{\bd(2)}h_{\b(2)\bd(2)} ~. \label{Spin2BachCorrf}
\end{align}
\end{subequations}

The tensors \eqref{Spin2BachCorr} span all primary structures of the type $\mc{J}_{\a(2)\ad(2)}(h)$ and, in particular, any linear combination will also satisfy \eqref{Spin2BachCorr}. Furthermore, if we express them in the form
\begin{align}
\mc{J}^{(i)}_{\a(2)\ad(2)}(h)=\mathcal{A}_{(i)}h_{\a(2)\ad(2)}
\end{align}
where $\mathcal{A}_{(i)}$ is the corresponding linear differential operator then, with the exception of $\mathcal{A}_{(2)}$, it may be shown that each operator is symmetric in the sense $\mathcal{A}_{(i)}=\mathcal{A}_{(i)}^T$ (see \eqref{Z.1} for the definition of the transpose operator $\mathcal{A}_{(i)}^T$). This property reduces the amount of work required to compute the gauge variation of each of the functionals associated with $\mc{J}^{(i)}_{\a(2)\ad(2)}(h)$.

Any operator $\mathcal{A}$ may be decomposed into symmetric and antisymmetric parts, $\mathcal{A}=\mathcal{A}_{\text{S}}+\mathcal{A}_{\text{A}}$ with $\mathcal{A}_{\text{S}}=\frac 12 (\mathcal{A}+\mathcal{A}^T)$ and $\mathcal{A}_{\text{A}}=\frac 12 (\mathcal{A}-\mathcal{A}^T)$. It follows that the antisymmetric part of $\mathcal{A}$ vanishes identically in any integral of the form 
\begin{align}
\int\text{d}^4 x \, e \, h^J\mathcal{A}h_J=\int\text{d}^4 x \, e \, h^J\mathcal{A}_{\text{S}}h_J~. \label{DeathOfAsymm}
\end{align}

Using \eqref{DeathOfAsymm} it is possible to show that at the level of actions, the following correspondence between $\mc{J}^{(2)}$ and the remaining primary structures holds
\begin{align}
\int\text{d}^4x\, e \,h^{\a(2)\ad(2)}\mc{J}^{(2)}_{\a(2)\ad(2)}(h)= \int\text{d}^4x\, e \,h^{\a(2)\ad(2)}&\Big(2\mc{A}_{(1)}-\mc{A}_{(3)} \non\\
&+2\mc{A}_{(4)}+\mc{A}_{(5)}+\mc{A}_{(6)}\Big)h_{\a(2)\ad(2)}~.
\end{align}
Additionally, the structure $\mc{J}^{(1)}$ vanishes in any Bach-flat spacetime and will be of no use. Therefore, it suffices to consider only one of the primary structures that is linear in the Weyl tensor, say $\mc{J}^{(3)}$, and its associated functional
\begin{align}
S_{\text{NM},W^1}^{(2,2)}[h]=\frac{1}{2}\int\text{d}^4x\, e \,h^{\a(2)\ad(2)}\mc{J}^{(3)}_{\a(2)\ad(2)}(h)+\text{c.c.} \label{GravitonLinearNMStructure}
\end{align}
One can then show that under the gauge transformation \eqref{CGravitonGT} and upon integrating by parts, the action \eqref{GravitonLinearNMStructure} transforms as
\begin{align}
\d_{\xi}S_{\text{NM},W^1}^{(2,2)}[h]&=\frac{1}{4}\delta_{\xi}S_{\text{Skeleton}}^{(2,2)}[h]+\bigg(\int\text{d}^4x\, e \, \xi^{\a\ad}\bigg\{\nabla_{\d}{}^{\dd}\big[W^{\b(2)}{}_{\g(2)}W_{\a}{}^{\g(2)\d}h_{\b(2)\ad\dd}\big]\notag\\
&+\nabla_{\b}{}^{\dd}\big[W^{\b(2)}{}_{\g(2)}W_{\a}{}^{\g(2)\d}h_{\b\d\ad\dd}\big] -\nabla_{\b\bd}\big[W_{\a}{}^{\b(3)}\bar{W}_{\ad}{}^{\bd(3)}h_{\b(2)\bd(2)}\big]\bigg\}\notag\\
&-\frac{1}{2}\int \text{d}^4x\, e \, \xi^{\a\ad}\bigg\{\nabla^{\b\bd}B_{\a}{}^{\b\bd}{}_{\ad}h_{\b(2)\bd(2)}+B_{\a}{}^{\b\bd(2)}\nabla_{\ad}{}^{\b}h_{\b(2)\bd(2)}\bigg\} +\text{c.c.}\bigg)~. \non
\end{align}
To annihilate the terms quadratic in the Weyl tensor, we define the functional
\begin{align}
S_{\text{NM},W^2}^{(2,2)}[h]=\frac{1}{2}\int\text{d}^4x \, e \, h^{\a(2)\ad(2)}\bigg\{\mc{J}^{(4)}_{\a(2)\ad(2)}(h)+\mc{J}^{(5)}_{\a(2)\ad(2)}(h)+\mc{J}^{(6)}_{\a(2)\ad(2)}(h)\bigg\}+\text{c.c.}
\end{align}
which is quadratic in the Weyl tensor. It may then be shown that the action
\begin{align}
S_{\text{CHS}}^{(2,2)}[h]&=S_{\text{Skeleton}}^{(2,2)}[h]-2S_{\text{NM},W^1}^{(2,2)}[h]+2S_{\text{NM},W^2}^{(2,2)}[h] \notag\\
&=\frac{1}{2}\int\text{d}^4x \, e \, h^{\a(2)\ad(2)}\bigg\{\widetilde{\mc{B}}_{\a(2)\ad(2)}(h)+2W_{\a}{}^{\b(3)}\nabla_{\b\ad}\nabla_{\b}{}^{\gd}h_{\b\a\ad\gd}-2W_{\a(2)}{}^{\b(2)}\Box_c h_{\b(2)\ad(2)}\notag\\
&+4\nabla_{\b\ad}W_{\a}{}^{\b(3)}\nabla_{\a}{}^{\gd}h_{\b\b\ad\gd}+2\nabla_{\b}{}^{\gd}W_{\a(2)}{}^{\b(2)}\nabla_{\gd}{}^{\g}h_{\b\g\ad(2)}-2h_{\b(2)\ad(2)}\Box_c W_{\a(2)}{}^{\b(2)}\notag\\
&+\nabla^{\g\gd}W_{\a(2)}{}^{\b(2)}\nabla_{\g\gd}h_{\b(2)\ad(2)} +2h_{\g\b\gd\ad}\nabla_{\ad}{}^{\g}\nabla_{\b}{}^{\gd}W_{\a(2)}{}^{\b(2)}+2W_{\a(2)}{}^{\g(2)}W_{\g(2)}{}^{\b(2)}h_{\b(2)\ad(2)}\notag\\
& +2W_{\a\g(2)}{}^{\b}W_{\a}{}^{\b\g(2)}h_{\b(2)\ad(2)}+2W_{\a(2)}{}^{\b(2)}\bar{W}_{\ad(2)}{}^{\bd(2)}h_{\b(2)\bd(2)} \bigg\}+\text{c.c.}~, \label{CGravitonA}
\end{align}
whose variation under \eqref{CGravitonGT} is given by
\begin{align}
\d_{\xi}S_{\text{CHS}}^{(2,2)}[h]=\int \text{d}^4x\, e \, \xi^{\a\ad}\bigg\{\nabla^{\b\bd}B_{\a}{}^{\b\bd}{}_{\ad}h_{\b(2)\bd(2)}+B_{\a}{}^{\b\bd(2)}\nabla_{\ad}{}^{\b}h_{\b(2)\bd(2)}\bigg\}+\text{c.c.}~,
\end{align}
is the unique model describing the graviton that is both conformally invariant in a general curved background and gauge invariant on any Bach-flat background,
\begin{align}
 \d^{(\mc{G})}_{\L}S_{\text{CHS}}^{(2,2)}[h]=0~, \qquad \d_{\xi}S_{\text{CHS}}^{(2,2)}[h] \approx 0~.
\end{align}

The action \eqref{CGravitonA} may be rewritten in the form \eqref{BachBachChicken} by defining the following non-minimal extension of the minimal Bach-tensor
\begin{align}
\mc{B}_{\a(2)\ad(2)}(h)=\widetilde{\mc{B}}_{\a(2)\ad(2)}(h)-2\Big(\mc{A}_{(3)}-\mc{A}_{(4)}-\mc{A}_{(5)}-\mc{A}_{(6)}\Big)h_{\a(2)\ad(2)}~.
\end{align}
This tensor is transverse and gauge invariant,
\begin{align}
 \nabla^{\b\bd}\mc{B}_{\a\b\ad\bd}(h)\approx 0~, \qquad \d_{\xi}\mc{B}_{\a(2)\ad(2)}(h)\approx 0~,
\end{align}
on any Bach-flat background.

This model was analysed in \cite{Manvelyan2} using a similar methodology, the main differences being that their analysis was performed in the gauge $\mathfrak{b}_a=0$ and the conformal graviton field was not traceless. This means that their model contains an extra algebraic gauge symmetry which may be used to gauge away the trace.  The authors found two inequivalent primary Lagrangians that were linear in the Weyl tensor and used both in the construction of their gauge invariant action. Upon eliminating the trace of the graviton field, one of these structures vanishes and the other must be proportional to \eqref{GravitonLinearNMStructure} modulo terms involving the Bach tensor and also terms quadratic in the Weyl tensor.




\subsection{Maximal depth conformal graviton} \label{secCMDgraviton}

The next model that we study corresponds to the case $m=n=t=2$, which is known as the maximal depth conformal graviton. This is our last example of a CHS model which meets our naive expectations outlined at the beginning of this section. 
Upon degauging, its action coincides with the one studied earlier
in Ref. \cite{BT2015}, where it was suggested that gauge invariance could only be upheld in an Einstein space. See below for further discussion on this.


 The maximal depth conformal graviton is described by the real field $h^{(2)}_{\a(2)\ad(2)}$ which is defined modulo the depth 2 gauge transformations\footnote{Since we will be dealing exclusively with conformal maximal depth fields in the next three sub sections, we will usually drop all labels that refer to $t$ when its value is clear from the context.}
\begin{align} 
\delta_{\xi}h_{\a(2)\ad(2)}=\nabla_{\a\ad}\nabla_{\a\ad}\xi~. \label{CMDspin2GT} 
\end{align}
Here both $h_{\a(2)\ad(2)}$ and $\xi$ are primary and have Weyl weights $1$ and $-1$ respectively,
\begin{align}
\mathbb{D}h_{\a(2)\ad(2)}=h_{\a(2)\ad(2)}~,\qquad  \mathbb{D}\xi=-\xi~.
\end{align}
 The skeleton action \eqref{GenCHSSkeleton} is second order in derivatives and takes the form
\begin{align}
S_{\text{Skeleton}}^{(2,2,2)}[h]=\frac{1}{2}\int \text{d}^4x \, e \, \mc{W}^{\a(3)\ad}(h)\mc{W}_{\a(3)\ad}(h)+\text{c.c.}~,\qquad \mc{W}_{\a(3)\ad}(h)=\nabla_{\a}{}^{\bd}h_{\a(2)\ad\bd}~.\label{CMDspin2Skeleton}
\end{align}
This functional is not gauge invariant if the background Weyl tensor is non-vanishing, 
 and one can show that its variation under \eqref{CMDspin2GT} is equal to 
\begin{align}
\delta_{\xi} S_{\text{Skeleton}}^{(2,2,2)}[h]=\int\text{d}^4x\, e \, \xi\bigg\{W_{\a(3)}{}^{\d}\nabla_{\d\ad}\mc{W}^{\a(3)\ad}(h)+2\mc{W}^{\a(3)\ad}(h)\nabla_{\d\ad}W^{\d}{}_{\a(3)}\bigg\}+\text{c.c.}
\end{align}

To restore gauge invariance to the skeleton, we seek a non-minimal correction of the form $\mc{J}_{\a(2)\ad(2)}(h)$ with the conformal properties  
\begin{align}
K_{\b\bd}\mc{J}_{\a(2)\ad(2)}(h)=0~,\qquad \mathbb{D}\mc{J}_{\a(2)\ad(2)}(h)=3\mc{J}_{\a(2)\ad(2)}(h)~. \label{CMDSpin2BachCorrGen}
\end{align}
In contrast to the minimal depth conformal graviton, there is only a single structure (up to conjugation) satisfying these constraints, and it takes the form 
\begin{align}
\mc{J}_{\a(2)\ad(2)}(h)=-W_{\a(2)}{}^{\b(2)}h_{\b(2)\ad(2)}~.
\end{align}
It may then be shown that the corresponding non-minimal action,
\begin{align}
S_{\text{NM}}^{(2,2,2)}[h]=\frac{1}{2}\int\text{d}^4x &\, e \, h^{\a(2)\ad(2)}\mc{J}_{\a(2)\ad(2)}(h) +{\rm c.c.}~,  \label{CMDSpin2NMAct}
\end{align}
 has the following gauge variation under \eqref{CMDspin2GT} 
\begin{align} 
\delta_{\xi}S_{\text{NM}}^{(2,2,2)}[h]=-\delta_{\xi} S_{\text{Skeleton}}^{(2,2,2)}[h]-\bigg( \int\text{d}^4x\, e \, \xi B^{\a(2)\ad(2)}h_{\a(2)\ad(2)}+\text{c.c.}\bigg)~, \label{2.14}
\end{align}
It follows that the primary action
\begin{align}
S_{\text{CHS}}^{(2,2,2)}[h]&=S_{\text{Skeleton}}^{(2,2,2)}[h]+S_{\text{NM}}^{(2,2,2)}[h]\notag\\
&=\frac{1}{2}\int \text{d}^4x \, e \, \bigg\{\mc{W}^{\a(3)\ad}(h)\mc{W}_{\a(3)\ad}(h)-  h^{\a(2)\ad(2)}W_{\a(2)}{}^{\b(2)}h_{\b(2)\ad(2)}\bigg\}+\text{c.c.} \label{CMDspin2FullAct}
\end{align}
is gauge invariant on any Bach-flat background,
\begin{align}
\delta_{\xi}S_{\text{CHS}}^{(2,2,2)}[h]\approx 0~.
\end{align}
The corresponding gauge invariant and doubly-transverse Bach tensor takes the form 
\begin{align}
\mc{B}_{\a(2)\ad(2)}(h)=\widetilde{\mc{B}}_{\a(2)\ad(2)}(h)+\mc{J}_{\a(2)\ad(2)}(h)~,
\end{align} 
where $\widetilde{\mc{B}}_{\a(2)\ad(2)}(h)$ is the minimal lift of \eqref{GenHSBacha}.
 
 To make contact with the existing literature, it is useful to present the degauged version of this model.  Upon imposing the gauge \eqref{degauge}, implementing the usual degauging procedure process and converting to vector notation, the action \eqref{CMDspin2FullAct} takes the form
 \begin{align}
 S_{\text{CHS}}^{(2,2,2)}[h]=-4\int \text{d}^4x \, e \,\bigg\{&\mathcal{D}^ah^{bc}\mathcal{D}_{a}h_{bc}-\frac{4}{3}\mathcal{D}_{a}h^{ab}\mathcal{D}^{c}h_{bc}-2R_{ab}h^{ac}h_{c}{}^{b}+\frac{1}{6}Rh^{ab}h_{ab}\notag\\
 &+2W_{abcd}h^{ac}h^{bd}\bigg\}  \label{CMDspin2FullActVec}
 \end{align}
 where we have made use of \eqref{SpinorWeylTensor} and the definition $h_{\a(2)\ad(2)}:=(\s^a)_{\a\ad}(\s^{b})_{\a\ad}h_{ab}$ for symmetric and traceless $h_{ab}$. It is invariant under the degauged transformations \eqref{CMDspin2GT}, 
\begin{align}
\delta_{\xi}h_{\a(2)\ad(2)}=\mathcal{D}_{\a\ad}\mathcal{D}_{\a\ad}\xi-\frac{1}{2}R_{\a(2)\ad(2)}\xi ~.\label{CMDspin2GTDG}
\end{align} 
 which in vector notation reads
\begin{align}
\delta_{\xi}h_{ab}=\big(\mathcal{D}_a\mathcal{D}_b-\frac{1}{2}R_{ab}\big)\xi-\frac{1}{4}\eta_{ab}\big(\Box-\frac 12 R\big)\xi~. \label{CMDspin2GTDGVec}
\end{align} 
 
  The action \eqref{CMDspin2FullActVec} consists of two sectors that are independently invariant under Weyl transformations (the first and second lines). Various combinations of these functionals were studied earlier in \cite{DeserN1, DeserN2, LN, Sachs, EO } when trying to construct Weyl invariant second-order models for a symmetric traceless rank two tensor. However the question of gauge invariance was first raised in \cite{DeserN1, DeserN2}, but only in the case of an (A)dS$_4$ background, where the last term in \eqref{CMDspin2FullActVec} is not present. Much later, the correct action \eqref{CMDspin2FullActVec} was proposed in \cite{BT2015}, but the authors considered only gauge transformations of the type\footnote{For a general background, the right-hand side of \eqref{CMDspin2GTDGVecES} does not preserve its form under Weyl transformations.}
  \begin{align}
  \delta_{\xi}h_{ab}=\big(\mathcal{D}_a\mathcal{D}_b-\frac{1}{4}\eta_{ab}\Box\big)\xi~.\label{CMDspin2GTDGVecES}
  \end{align}
  Consequently, it was concluded that gauge invariance could only be upheld in Einstein spaces, where \eqref{CMDspin2GTDGVec} and \eqref{CMDspin2GTDGVecES} coincide. 
It is important to emphasise that the equation of motion resulting from \eqref{CMDspin2FullActVec} was observed in \cite{DeserW6} to be invariant under the gauge transformations \eqref{CMDspin2GTDGVec} in an arbitrary Bach-flat background. However the authors of \cite{DeserW6} were interested in coupling the model to conformal gravity, which lead to the conclusion that the system was inconsistent.


\subsection{Maximal depth spin-3} \label{secCMDspin3}

The next case that we would like to analyse is the CMD
spin-3 field $h_{\a(3)\ad(3)}$, with $m=n=t=3$. As mentioned earlier, this is the first complete example of a gauge invariant CHS model whose action requires non-diagonal mixing terms between the parent field $h_{\a(3)\ad(3)}$ and lower-spin fields. 

The relevant gauge freedom in this case is
\begin{align}
\delta_{\xi}h_{\a(3)\ad(3)}=\nabla_{(\a_1(\ad_1}\nabla_{\a_2\ad_2}\nabla_{\a_3)\ad_3)}\xi~.\label{CMDspin3GT}
\end{align}
Both $h_{\a(3)\ad(3)}$ and $\xi$ are primary and have Weyl weights
\begin{align}
\mathbb{D}h_{\a(3)\ad(3)}=h_{\a(3)\ad(3)}~,\qquad  \mathbb{D}\xi=-2\xi~.
\end{align}
 As is the case for all four-dimensional bosonic CHS models with maximal depth, the CMD spin-3  skeleton action is second order in derivatives.  It takes the specific form
 \begin{subequations}
\begin{align}
S_{\text{Skeleton}}^{(3,3,3)}[h]=-&\frac{1}{2}\int\text{d}^4x \, e \,\mc{W}^{\a(4)\ad(2)}(h)\mc{W}_{\a(4)\ad(2)}(h)+ {\rm c.c.} ~, \\
& \mc{W}_{\a(4)\ad(2)}(h)=\nabla_{\a}{}^{\bd}h_{\a(3)\ad(2)\bd}~,\label{CMDspin3Skeleton}
\end{align}
\end{subequations}
and has gauge variation equal to
\begin{align}
&\delta_{\xi}S_{\text{Skeleton}}^{(3,3,3)}[h]=2\int\text{d}^4x \, e \, \xi \bigg\{2\mc{W}^{\a(4)\ad(2)}(h)\nabla_{\a\ad}\nabla_{\g\ad}W_{\a(3)}{}^{\g}+\nabla_{\ad}{}^{\g}W_{
\a(4)}\nabla_{\g\ad}\mc{W}^{\a(4)\ad(2)} (h)\notag\\
&+4\nabla_{\a\ad}\mc{W}^{\a(4)\ad(2)}(h)\nabla_{\g\ad}W_{\a(3)}{}^{\g}+\frac{4}{3}W_{\a(3)}{}^{\g}\nabla_{\a\ad}\nabla_{\g\ad}\mc{W}^{\a(4)\ad(2)}(h)\bigg\}+\text{c.c.}
\end{align}

We seek non-minimal descendants $\mc{J}_{\a(3)\ad(3)}(h)$ with the properties
\begin{align}
k_{\b\bd}\mc{J}_{\a(3)\ad(3)}(h)=0~,\qquad \mb{D}\mc{J}_{\a(3)\ad(3)}(h)=3\mc{J}_{\a(3)\ad(3)}(h)~.
\end{align}
Once again, there is only one possible structure of this type,\footnote{This will be a general feature of maximal depth bosonic CHS models. For the field $h^{(s)}_{\a(s)\ad(s)}$ with $m=n=t=s$, the only non-minimal correction is $\mc{J}_{\a(s)\ad(s)}(h)=W_{\a(2)}{}^{\b(2)}h^{(s)}_{\a(s-2)\b(2)\ad(s)}$. } and the associated functional takes the form
\begin{subequations} \label{CMDspin3NMFunc}
\begin{align}
S_{\text{NM}}^{(3,3,3)}[h]&=\frac{1}{2}\int  \text{d}^4x \, e \, h^{\a(3)\ad(3)}\mc{J}_{\a(3)\ad(3)}(h)+\text{c.c.}~, \\
 &\mc{J}_{\a(3)\ad(3)}(h)=W_{\a(2)}{}^{\b(2)}h_{\b(2)\a\ad(3)}\label{CMDspin3NM}
\end{align}
\end{subequations}
and its gauge variation proves to be equal to
\begin{align}
\delta_{\xi}S_{\text{NM}}^{(3,3,3)}[h]&=-\frac{1}{2}\delta_{\xi}S_{\text{Skeleton}}^{(3,3,3)}[h] +\bigg(\int \text{d}^4x \, e \, \xi\bigg\{-\nabla^{\a\ad}B^{\a(2)\ad(2)}h_{\a(3)\ad(3)}\notag\\
&-\frac{3}{2}B^{\a(2)\ad(2)}\nabla^{\a\ad}h_{\a(3)\ad(3)} +\frac{4}{3}\nabla_{\b\bd}\big[\bar{W}^{\ad(3)\bd}W^{\a(3)\b}\big]h_{\a(3)\ad(3)}\notag\\
&+\frac{2}{3}W^{\a(3)\b}\bar{W}^{\ad(3)\bd}\nabla_{\b\bd}h_{\a(3)\ad(3)}\bigg\}+\text{c.c.}\bigg)~.\label{CMDspin3NMVar}
\end{align}
We point out that in deriving \eqref{CMDspin3NMVar}, there is a nontrivial contribution arising from integration by parts, we discuss this technicality in more detail in appendix \ref{AppIBPS3}. 

It follows that, in a Bach-flat background, the action consisting of purely the maximal depth spin-3 field 
\begin{align}
S_{hh}=S_{\text{Skeleton}}^{(3,3,3)}[h]+2S_{\text{NM}}^{(3,3,3)}[h]~, \label{CMDspin3PureAct}
\end{align}
is gauge invariant only to first order in the background Weyl tensor, since 
\begin{align}
\delta_{\xi}S_{hh} \approx \frac{4}{3}\int \text{d}^4x \, e \, \xi &\bigg\{2\bar{W}^{\ad(3)\bd}\nabla_{\b\bd}W^{\a(3)\b}h_{\a(3)\ad(3)}+2W^{\a(3)\b}\nabla_{\b\bd}\bar{W}^{\ad(3)\bd}h_{\a(3)\ad(3)}\notag\\
&+W^{\a(3)\b}\bar{W}^{\ad(3)\bd}\nabla_{\b\bd}h_{\a(3)\ad(3)}\bigg\}+\text{c.c.}
\label{CMDspin3PureActVar}
\end{align}
This is the best that one can achieve without making use of any extra fields.  

The result \eqref{CMDspin3PureActVar} is analogous to the conclusion of Ref. \cite{GrigorievT}
that the pure spin-3 action cannot be made gauge invariant beyond the first order in curvature. 
It was also conjectured in \cite{GrigorievT} that it might be possible to restore the spin-3
gauge invariance by introducing  a coupling   to a conformal spin-1 field. 
For the CMD spin-3 field, we are going to demonstrate 
that gauge invariance can indeed be restored by switching on a coupling to 
certain lower-spin fields.
 
To this end, we introduce two `lower-spin'\footnote{In this context, by lower-spin field we mean any field whose total number of indices is less than that of the parent field.} fields
$\chi_{\a(3)\ad}$ and $\vf_{\a(4)}$, along with their complex conjugates $\bar{\chi}_{\a\ad(3)}$ and $\bar{\vf}_{\ad(4)}$. They are each particular cases of the type I conformal non-gauge fields introduced in appendix \ref{AppCNGM}.\footnote{For ease of distinction, in this section we denote the two non-gauge fields by different symbols.} They are said to be non-gauge because they are defined to carry the following conformal properties
\begin{subequations}\label{CMDspin3NGprops}
\begin{align}
\mathbb{D}\chi_{\a(3)\ad}&=\chi_{\a(3)\ad}~,\qquad K_{\b\bd}\chi_{\a(3)\ad}=0~,\\
\mathbb{D}\vf_{\a(4)}&=0~,\qquad~~~~~~~~ K_{\b\bd}\vf_{\a(4)}=0~,
\end{align}
\end{subequations}
which are incompatible with the usual CHS gauge transformations involving only derivatives (of any depth). However, in non-trivial backgrounds we can consistently endow them with the following background dependent gauge transformations 
\begin{subequations}\label{CMDspin3NGGT}
\begin{align}
\delta_{\xi}\chi_{\a(3)\ad}&=W_{\a(3)}{}^{\b}\nabla_{\b\ad}\xi-2\nabla_{\b\ad}W_{\a(3)}{}^{\b}\xi~,\\
\delta_{\xi}\vf_{\a(4)}&=W_{\a(4)}\xi~.
\end{align}
\end{subequations}
The right hand sides of \eqref{CMDspin3NGGT} are fixed by the conformal properties \eqref{CMDspin3NGprops}. To cancel the variation \eqref{CMDspin3PureActVar} we introduce the following couplings between the two lower-spin fields and $h$,
\begin{subequations}\label{CMDSpin3Mixing}
\begin{align}
S_{h\chi}&=\int\text{d}^4x\, e \, h^{\a(3)\ad(3)}\bar{W}_{\ad(3)}{}^{\bd}\chi_{\a(3)\bd}+\text{c.c.}~,\\
S_{h\vf}&=\int\text{d}^4x\, e \, h^{\a(3)\ad(3)}\bigg\{W_{\a(3)}{}^{\b}\nabla_{\b\bd}\bar{\vf}_{\ad(3)}{}^{\bd}-3\nabla_{\b\bd}W_{\a(3)}{}^{\b}\bar{\vf}_{\ad(3)}{}^{\bd}\bigg\}+\text{c.c.}~,
\end{align} 
\end{subequations}
both of which are primary. Under \eqref{CMDspin3NGGT} the functionals \eqref{CMDSpin3Mixing} transform according to
\begin{subequations}\label{CMDSpin3MixingVar}
\begin{align}
\delta_{\xi}S_{h\chi}&= \int\text{d}^4x \, e \,\bigg\{\bar{W}^{\ad(3)\bd}\chi^{\a(3)}{}_{\bd}\delta_{\xi}h_{\a(3)\ad(3)} -\xi\bigg[ 3\bar{W}^{\ad(3)\bd}\nabla_{\b\bd}W^{\a(3)\b}h_{\a(3)\ad(3)}\notag\\
&+W^{\a(3)\b}\nabla_{\b\bd}\bar{W}^{\ad(3)\bd}h_{\a(3)\ad(3)}+W^{\a(3)\b}\bar{W}^{\ad(3)\bd}\nabla_{\b\bd}h_{\a(3)\ad(3)}\bigg]\bigg\}+\text{c.c.} ~,\label{CMDSpin3MixingVara}\\
\delta_{\xi}S_{h\vf}&= \int\text{d}^4x \, e \,\bigg\{\delta_{\xi}h^{\a(3)\ad(3)}\bigg[W_{\a(3)}{}^{\b}\nabla_{\b\bd}\bar{\vf}_{\ad(3)}{}^{\bd}-3\nabla_{\b\bd}W_{\a(3)}{}^{\b}\bar{\vf}_{\ad(3)}{}^{\bd}\bigg] \notag\\
&-\xi\bigg[4\bar{W}^{\ad(3)\bd}\nabla_{\b\bd}W^{\a(3)\b}h_{\a(3)\ad(3)}+W^{\a(3)\b}\bar{W}^{\ad(3)\bd}\nabla_{\b\bd}h_{\a(3)\ad(3)}\bigg]\bigg\}+\text{c.c.}~ \label{CMDSpin3MixingVarb}
\end{align}
\end{subequations}

Of course, the presence of the non-diagonal sector \eqref{CMDSpin3Mixing} forces us to introduce diagonal actions for each non-gauge field so that we may cancel the first term in each of the variations \eqref{CMDSpin3MixingVara} and \eqref{CMDSpin3MixingVarb}. The appropriate kinetic actions are those derived in appendix \ref{AppCNGM}
\begin{subequations}\label{CMDspin3NGKinetic}
\begin{align}
S_{\chi\bar{\chi}}=&~\frac{1}{2}\int\text{d}^4x\, e \, \chi^{\a(3)\ad}\nabla_{\a}{}^{\ad}\nabla_{\a}{}^{\ad}\bar{\chi}_{\a\ad(3)}+\text{c.c.}~,\label{CMDspin3NGKinetica}\\
S_{\vf\bar{\vf}}=&~\frac{1}{2}\int\text{d}^4x\, e \,\bar{\vf}^{\ad(4)}\nabla_{\ad}{}^{\a}\nabla_{\ad}{}^{\a}\nabla_{\ad}{}^{\a}\nabla_{\ad}{}^{\a}\vf_{\a(4)}+\text{c.c.}\label{CMDspin3NGKineticb}
\end{align}
\end{subequations}
They are both primary and prove to have the following gauge variations
\begin{subequations}\label{CMDspin3NGKineticVar}
\begin{align}
\delta_{\xi}S_{\chi\bar{\chi}}=-&\int\text{d}^4x\, e \, \bigg\{ \bar{W}^{\ad(3)\bd}\chi^{\a(3)}{}_{\bd}\delta_{\xi}h_{\a(3)\ad(3)}\notag\\
&+\xi\bigg[\chi^{\a(3)\ad}\nabla_{\a}{}^{\ad}B_{\a(2)\ad(2)}+3B_{\a(2)\ad(2)}\nabla_{\a}{}^{\ad}\chi^{\a(3)\ad}\bigg]\bigg\}+\text{c.c.}~,\\
\delta_{\xi}S_{\vf\bar{\vf}}=\phantom{-}&\int\text{d}^4x\, e \,\bigg\{-\delta_{\xi}h^{\a(3)\ad(3)}\bigg[W_{\a(3)}{}^{\b}\nabla_{\b\bd}\bar{\vf}_{\ad(3)}{}^{\bd}-3\nabla_{\b\bd}W_{\a(3)}{}^{\b}\bar{\vf}_{\ad(3)}{}^{\bd}\bigg]\notag\\
&+\xi\bigg[6B^{\a(2)\ad(2)}\nabla_{\a}{}^{\ad}\nabla_{\a}{}^{\ad}\bar{\vf}_{\ad(4)}+8\nabla_{\a}{}^{\ad}B^{\a(2)\ad(2)}\nabla_{\a}{}^{\ad}\bar{\vf}_{\ad(4)}\non\\
&+3\bar{\vf}_{\ad(4)}\nabla_{\a}{}^{\ad}\nabla_{\a}{}^{\ad}B^{\a(2)\ad(2)}\bigg]\bigg\}+\text{c.c.}
\end{align}
\end{subequations}
From \eqref{CMDSpin3MixingVar} and \eqref{CMDspin3NGKineticVar}, it follows that the conformal action
\begin{align}
&S^{(3,3,3)}_{\text{CHS}}[h,\chi,\vf]=S_{hh}+\frac{8}{3}S_{\chi\bar{\chi}}-\frac{4}{3}S_{\vf\bar{\vf}}+\frac{8}{3}S_{h\chi}-\frac{4}{3}S_{h\vf}~\label{CMDspin3FullAction} \\
&=\frac{1}{2}\int\text{d}^4x \, e \,\bigg\{-\mc{W}^{\a(4)\ad(2)}(h)\mc{W}_{\a(4)\ad(2)}(h)+2h^{\g\a(2)\ad(3)}W_{\a(2)}{}^{\b(2)}h_{\b(2)\g\ad(3)} \notag\\
&+\frac{8}{3}\chi^{\a(3)\ad}\nabla_{\a}{}^{\ad}\nabla_{\a}{}^{\ad}\bar{\chi}_{\a\ad(3)}-\frac{4}{3}\bar{\vf}^{\ad(4)}\nabla_{\ad}{}^{\a}\nabla_{\ad}{}^{\a}\nabla_{\ad}{}^{\a}\nabla_{\ad}{}^{\a}\vf_{\a(4)} +\frac{16}{3}h^{\a(3)\ad(3)}\bar{W}_{\ad(3)}{}^{\bd}\chi_{\a(3)\bd}\notag\\
&-\frac{8}{3}h^{\a(3)\ad(3)}\bigg[W_{\a(3)}{}^{\b}\nabla_{\b\bd}\bar{\vf}_{\ad(3)}{}^{\bd}-3\nabla_{\b\bd}W_{\a(3)}{}^{\b}\bar{\vf}_{\ad(3)}{}^{\bd}\bigg]\bigg\}+\text{c.c.}~, 
\end{align}
has gauge variation that is strictly proportional to the Bach tensor,
\begin{align}
\delta_{\xi}&S^{(3,3,3)}_{\text{CHS}}[h,\chi,\vf]=-\frac{1}{2}\int\text{d}^4x\, e \, \xi\bigg\{4\nabla^{\a\ad}B^{\a(2)\ad(2)}h_{\a(3)\ad(3)}+6B^{\a(2)\ad(2)}\nabla^{\a\ad}h_{\a(3)\ad(3)}\notag\\
&+16B_{\a(2)\ad(2)}\nabla_{\a}{}^{\ad}\chi^{\a(3)\ad}+\frac{16}{3}\chi^{\a(3)\ad}\nabla_{\a}{}^{\ad}B_{\a(2)\ad(2)}+16B^{\a(2)\ad(2)}\nabla_{\a}{}^{\ad}\nabla_{\a}{}^{\ad}\bar{\vf}_{\ad(4)}\notag\\
&+\frac{64}{3}\nabla_{\a}{}^{\ad}B^{\a(2)\ad(2)}\nabla_{\a}{}^{\ad}\bar{\vf}_{\ad(4)}+8\bar{\vf}_{\ad(4)}\nabla_{\a}{}^{\ad}\nabla_{\a}{}^{\ad}B^{\a(2)\ad(2)}\bigg\}+\text{c.c.}
\end{align}
It is therefore gauge invariant when restricted to a Bach-flat background,
\begin{align}
\delta_{\xi}S^{(3,3,3)}_{\text{CHS}}[h,\chi,\vf]\approx 0~.
\end{align}
It is also invariant under the conformal gravity gauge group $\mc{G}$ on a generic background. 
Clearly it is not possible to recast this action in terms of some linearised Bach-tensor which is (triply-)transverse and gauge-invariant.

Due to the presence of the non-gauge kinetic terms, the action \eqref{CMDspin3FullAction} does not reduce to \eqref{GenCHSActionCF4} in the conformally flat limit, but rather to 
\begin{align}
W_{\a(4)}=0 \quad \implies \quad S^{(3,3,3)}_{\text{CHS}}[h,\chi,\vf]=S_{\text{CHS}}^{(3,3,3)}[h]+\frac{8}{3}S_{\chi\bar{\chi}}-\frac{4}{3}S_{\vf\bar{\vf}}~.
\end{align}
However, it is clear that the spin-3 field decouples from the non-gauge fields in this limit. 

Finally, it is of interest to provide the degauged version of the pure spin-3 sector \eqref{CMDspin3PureAct} in vector notation. It may be shown to be equivalent to
\begin{align}
 S_{hh}=4\int \text{d}^4x \, e \,\bigg\{&\mathcal{D}^ah_{acd}\mathcal{D}^{b}h_{b}{}^{cd}-2\mathcal{D}_{a}h_{bcd}\mathcal{D}^{a}h^{bcd}+6R_{ab}h^{acd}h_{cd}{}^{b}-\frac{2}{3}Rh^{abc}h_{abc}\notag\\
 &-8W_{abcd}h^{acf}h_{f}{}^{bd}\bigg\}  \label{CMDspin3Vector}
 \end{align}
where we have made use of the definition $h_{\a(3)\ad(3)}:=(\s^a)_{\a\ad}(\s^b)_{\a\ad}(\s^c)_{\a\ad}h_{abc}$ for symmetric and traceless $h_{abc}$. The gauge transformations \eqref{CMDspin2GT} are then equivalent to
\begin{align}
\delta_{\xi}h_{abc}=\bigg(\mathcal{D}_{(a}\mathcal{D}_{b}\mathcal{D}_{c)}&-2R_{(ab}\mathcal{D}_{c)}-\mathcal{D}_{(a}R_{bc)}\bigg)\xi\non\\
&+\eta_{(ab}\bigg(R_{c)}{}^{d}\mathcal{D}_{d}+\frac{1}{3}R\mathcal{D}_{c)}+\frac{1}{3}\mathcal{D}_{c)}R-\frac{1}{2}\mathcal{D}_{c)}\Box\bigg)\xi~.
\end{align}
The conversion of the lower-spin sectors in \eqref{CMDspin3FullAction} is a straightforward but tedious matter and will be omitted as the final expressions are not illuminating.


\subsection{Maximal depth spin-5/2} \label{secCMDspin52}

Maximal depth fermionic models differ from  their bosonic counterparts in that their skeletons \eqref{GenCHSSkeleton} are all third order in derivatives rather than second order. This makes 
extending them to Bach-flat backgrounds technically more challenging, but conceptually there is no difference. In particular, as we show below, lower-spin fields must also be introduced to render the maximal depth spin-$5/2$ system gauge invariant beyond first order in the Weyl curvature.

The maximal depth spin-5/2 CHS models corresponds to the case $m-1=n=t=2$, and is described by the complex field $h_{\a(3)\ad(2)}$ defined modulo depth two gauge transformations
\begin{align}
\delta_{\xi}h_{\a(3)\ad(2)}=\nabla_{\a\ad}\nabla_{\a\ad}\xi_{\a}~.\label{CMDspin52GT}
\end{align}
Both $h_{\a(3)\ad(2)}$ and $\xi_{\a}$ are primary and carry Weyl weights
\begin{align}
\mathbb{D}h_{\a(3)\ad(2)}=\frac{1}{2}h_{\a(3)\ad(2)}~,\qquad \mathbb{D}\xi_{\a}=-\frac{3}{2}\xi_{\a}~.
\end{align}

The skeleton sector \eqref{GenCHSSkeleton},
\begin{align}
S_{\text{Skeleton}}^{(3,2,2)}[h,\bar{h}]=\frac{{\rm i}}{2}\int \text{d}^4x \, e \,\mc{W}^{\a(4)\ad}(h)\mc{W}_{\a(4)\ad}(\bar{h}) +\text{c.c.}~,\label{CMDspin52Skeleton}
\end{align}
is composed of the generalised Weyl tensor of both $h_{\a(3)\ad(2)}$ and $\bar{h}_{\a(2)\ad(3)}$
\begin{align}
\mc{W}_{\a(4)\ad}(h)=\nabla_{\a}{}^{\bd}h_{\a(3)\ad\bd}~,\qquad \mc{W}_{\a(4)\ad}(\bar{h})=\nabla_{\a}{}^{\bd}\nabla_{\a}{}^{\bd}\bar{h}_{\a(2)\ad\bd(2)}~.
\end{align}
Under the transformation \eqref{CMDspin52GT} it varies as
\begin{align}
&\delta_{\xi}S_{\text{Skeleton}}^{(3,2,2)}[h,\bar{h}]=\frac{\ri}{2}\int\text{d}^4x\, e \, \bigg\{\xi^{\a}\bigg[\frac{5}{2}W^{\g\b(3)}\nabla_{\g}{}^{\bd}\mc{W}_{\a\b(3)\bd}(\bar{h})+3\nabla_{\g}{}^{\bd}W^{\g\b(3)}\mc{W}_{\a\b(3)\bd}(\bar{h})\notag\\
&-\frac{3}{2}W^{\b(4)}\nabla_{\a}{}^{\bd}\mc{W}_{\b(4)\bd}(\bar{h})-\nabla_{\a}{}^{\bd}W^{\b(4)}\mc{W}_{\b(4)\bd}(\bar{h})\bigg] - \frac{1}{3}\bar\xi_{\ad}\bigg[-\nabla^{\d\dd}W^{\b(4)}\nabla_{\d\dd}\mc{W}_{\b(4)}{}^{\ad}(h)\notag\\
&+6\nabla_{\g}{}^{\dd}W^{\g\b(3)}\nabla_{\dd}{}^{\b}\mc{W}_{\b(4)}{}^{\ad}(h)+2\nabla^{\b\bd}W^{\b(3)\g}\nabla_{\g}{}^{\ad}\mc{W}_{\b(4)\bd}(h)+5\nabla^{\b\bd}\nabla_{\g}{}^{\ad}W^{\g\b(3)}\mc{W}_{\b(4)\bd}(h)\notag\\
&+10\nabla_{\g}{}^{\ad}W^{\g\b(3)}\nabla^{\b\bd}\mc{W}_{\b(4)\bd}(h)+4 W^{\b(4)}\Box_c\mc{W}_{\b(4)}{}^{\ad}(h)+4W^{\b(3)\g}\nabla_{\g}{}^{\ad}\nabla^{\b\bd}\mc{W}_{\b(4)\bd}(h)\notag\\
&-15W^{\b(2)\d(2)}W_{\d(2)}{}^{\b(2)}\mc{W}_{\b(4)}{}^{\ad}(h)+\Box_c W^{\b(4)}\mc{W}_{\b(4)}{}^{\ad}(h)\bigg] \bigg\} +\text{c.c.}
\end{align}
Unlike the previous two maximal depth models we have looked at, for spin-$5/2$ there is a family of non-minimal primary counter-terms, which is generated by the following two functionals\footnote{There are also two more functionals of the form ${\rm i}\int \text{d}^4x \, e \, h^{\a(3)\ad(2)}\mc{J}_{\a(3)\ad(2)}(\bar{h})+\text{c.c.}$, where $\mc{J}_{\a(3)\ad(2)}(\bar{h})$ is a composite primary field depending on $\bar{W}_{\ad(4)}$ and $\bar{h}_{\a(2)\ad(3)}$. However they prove to be equivalent to \eqref{CMDspin52NM} modulo total derivatives. }
\begin{subequations}\label{CMDspin52NM}
\begin{align}
S_{\text{NM}}^{(3,2,2)}[h,\bar{h}]&=\frac{\ri}{2}\int\text{d}^4x\, e \, h^{\a(3)\ad(2)}\bigg\{-\frac{5}{4}W_{\a(3)}{}^{\b}\nabla^{\b\bd}\bar{h}_{\b(2)\bd\ad(2)}+\nabla^{\b\bd}W_{\a(3)}{}^{\b}\bar{h}_{\b(2)\bd\ad(2)}\notag \\
&\phantom{={\rm i}\int\text{d}^4x\, e \, \psi^{\a(3)\ad(2)}\bigg\{ }~+3W_{\a(2)}{}^{\b(2)}\nabla_{\a}{}^{\bd}\bar{h}_{\b(2)\bd\ad(2)} \bigg\} +\text{c.c.}~, \label{CMDspin52NMa}\\
\widetilde{S}_{\text{NM}}^{(3,2,2)}[h,\bar{h}]&=\frac{{\rm i}}{2}\int\text{d}^4x\, e \, h^{\a(3)\ad(2)}\bigg\{W_{\a(3)}{}^{\b}\nabla^{\b\bd}\bar{h}_{\b(2)\bd\ad(2)}-2\nabla^{\b\bd}W_{\a(3)}{}^{\b}\bar{h}_{\b(2)\bd\ad(2)}\notag\\
&\phantom{={\rm i}\int\text{d}^4x\, e \, h^{\a(3)\ad(2)}\bigg\{ }~+3\nabla_{\a}{}^{\bd}W_{\a(2)}{}^{\b(2)}\bar{h}_{\b(2)\bd\ad(2)} \bigg\} +\text{c.c.}
\end{align}
\end{subequations}
They are characterised by weight $+7/2$ non-minimal primary descendants $\mc{J}_{\a(3)\ad(2)}(\bar{h})$. The overall factors of i in \eqref{CMDspin52NM} are chosen so that their gauge variations may cancel that of \eqref{CMDspin52Skeleton}.

To first order in the Weyl tensor, it may be shown that any linear combination of the above two functionals, $a_1S_{\text{NM}}^{(3,2,2)}+a_2\widetilde{S}_{\text{NM}}^{(3,2,2)}$, with $a_2\neq 0$ will have gauge variation not proportional to that of $S_{\text{Skeleton}}^{(3,2,2)}$. Therefore it suffices to consider only the first structure \eqref{CMDspin52NMa}. Indeed, its gauge variation may be shown to be
\begin{align}
&\delta_{\xi}S_{\text{NM}}^{(3,2,2)}=\delta_{\xi}S_{\text{Skeleton}}^{(3,2,2)}+\bigg(\frac{\text{i}}{2}\int\text{d}^4x\, e \, \bigg\{ \xi^{\a}\bigg[\frac{3}{4}B_{\a}{}^{\b\bd(2)}\nabla^{\b\bd}\bar{h}_{\b(2)\bd(3)}+\nabla^{\b\bd}B_{\a}{}^{\b\bd(2)}\bar{h}_{\b(2)\bd(3)}\notag\\
&+B^{\b(2)\bd(2)}\nabla_{\a}{}^{\bd}\bar{h}_{\b(2)\bd(3)}-\frac{7}{4}\bar W^{\gd\bd(3)}\nabla_{\g\gd}W_{\a}{}^{\g\b(2)}\bar h_{\b(2)\bd(3)}-\frac{13}{8}W_{\a}{}^{\g\b(2)}\nabla_{\g\gd}\bar W^{\gd\bd(3)}\bar h_{\b(2)\bd(3)}\notag\\
&-\frac{9}{8}\bar W^{\gd\bd(3)}W_{\a}{}^{\g\b(2)}\nabla_{\g\gd}\bar h_{\b(2)\bd(3)}\bigg]-\bar\xi_{\ad}\bigg[\nabla^{\b\bd}B^{\b(2)\bd\ad}h_{\b(3)\bd(2)}+\frac{3}{4}B^{\b(2)\bd\ad}\nabla^{\b\bd}h_{\b(3)\bd(2)} \notag\\
&+B^{\b(2)\bd(2)}\nabla^{\b\ad}h_{\b(3)\bd(2)}-\frac{49}{24}\bar W^{\ad\gd\bd(2)}\nabla_{\g\gd}W^{\g\b(3)}h_{\b(3)\bd(2)}-\frac{4}{3} W^{\g\b(3)}\nabla_{\g\gd}\bar W^{\ad\gd\bd(2)}h_{\b(3)\bd(2)}\notag\\
&-\frac{9}{8}\bar W^{\ad\gd\bd(2)}W^{\g\b(3)}\nabla_{\g\gd}h_{\b(3)\bd(2)}\bigg]\bigg\}+\text{c.c.}\bigg)~.
\end{align}

We see that once again, using just the spin-$5/2$ field, gauge invariance can only be controlled to first order in the Weyl tensor,
\begin{align}
S_{h\bar{h}}=S_{\text{Skeleton}}^{(3,2,2)}[h,\bar{h}]-S_{\text{NM}}^{(3,2,2)}[h,\bar{h}]~,\qquad \delta_{\xi}S_{h\bar{h}}=\mc{O}\big(W^2\big)
\end{align}
 To go beyond this order we need to introduce two lower-spin non-gauge fields\footnote{In principle one could also consider the field $\rho_{\a(4)\ad}$, which has the same conformal properties as $\vf_{\a(3)}$ but has the gauge transformation $\delta_{\lambda}\rho_{\a(4)\ad}=W_{\a(4)}\bar\lambda_{\ad}$. However this field turns out to be unnecessary in the construction. This could be a signal that this model is not unique, in the same sense that the conformal hook model is not (see section \ref{secConformalHook}). } 
$\chi_{\a(2)\ad}$
and $\vf_{\a(3)}$.
They possess the conformal properties
\begin{subequations}
\begin{align}
\mathbb{D}\chi_{\a(2)\ad}&=\frac{3}{2}\chi_{\a(2)\ad}~,\qquad K_{\b\bd}\chi_{\a(2)\ad}=0~,\\
\mathbb{D}\vf_{\a(3)}&=\frac{1}{2}\vf_{\a(3)}~,\qquad~~~ K_{\b\bd}\vf_{\a(3)}=0~,
\end{align}
\end{subequations}
and are defined modulo the gauge transformations 
\begin{subequations}
\begin{align}
\delta_{\xi}\chi_{\a(2)\ad}&=W_{\a(2)}{}^{\b(2)}\nabla_{\b\ad}\xi_{\b}-\nabla_{\b\ad}W_{\a(2)}{}^{\b(2)}\xi_{\b}~,\\
\delta_{\xi}\vf_{\a(3)}&=W_{\a(3)}{}^{\b}\xi_{\b}~.
\end{align}
\end{subequations}
For more details on the conformal non-gauge fields we refer the reader to appendix \ref{AppCNGM}.

The primary couplings between these fields and the spin-$5/2$ field take the form
\begin{subequations}\label{CMDspin52Mixing}
\begin{align}
S_{h\bar{\chi}}&= \text{i}\int\text{d}^4x\, e \,\bar h^{\a(2)\ad(3)}\bar W_{\ad(3)}{}^{\bd}\chi_{\a(2)\bd}+\text{c.c.}~, \\
S_{h\bar\vf}&= \text{i}\int\text{d}^4x\, e \,h^{\a(3)\ad(2)}\bigg\{W_{\a(3)}{}^{\g}\nabla_{\g}{}^{\bd}\bar\vf_{\ad(2)\bd}-2\nabla_{\g}{}^{\bd}W_{\a(3)}{}^{\g}\bar\vf_{\ad(2)\bd}\bigg\}+\text{c.c.}
\end{align}
\end{subequations}
Their gauge variations may be shown to be 
\begin{subequations}\label{CMDspin52MixingGVar}
\begin{align}
\delta_{\xi}S_{h\bar{\chi}}&= \text{i}\int\text{d}^4x\, e \, \bigg\{ \xi^{\a}\bigg[W_{\a}{}^{\g\b(2)}\bar W^{\gd\bd(3)}\nabla_{\g\gd}\bar h_{\b(2)\bd(3)}+W_{\a}{}^{\g\b(2)}\nabla_{\g\gd}\bar W^{\gd\bd(3)}\bar h_{\b(2)\bd(3)}\notag\\ 
&\phantom{= \text{i}\int\text{d}^4x\, e \, \bigg\{}+2\bar W^{\gd\bd(3)}\nabla_{\g\gd}W_{\a}{}^{\g\b(2)}\bar h_{\b(2)\bd(3)}\bigg]-\bar\xi_{\ad}\bigg[\bar{W}^{\ad\bd\gd(2)}\nabla_{\gd}{}^{\b}\nabla_{\gd}{}^{\b}\chi_{\b(2)\bd} \notag\\
&\phantom{= \text{i}\int\text{d}^4x\, e \, \bigg\{}+2\nabla_{\gd}{}^{\b}\bar{W}^{\ad\bd\gd(2)}\nabla_{\gd}{}^{\b}\chi_{\b(2)\bd} +B^{\b(2)\ad\bd}\chi_{\b(2)\bd}\bigg]\bigg\}+\text{c.c.}~,\\
\delta_{\xi}S_{h\bar\vf}&= \text{i}\int\text{d}^4x\, e \, \bigg\{\bar\xi_{\ad}\bigg[W^{\g\b(3)}\bar W^{\ad\gd\bd(2)}\nabla_{\g\gd}h_{\b(3)\bd(2)}+3\bar W^{\ad\gd\bd(2)}\nabla_{\g\gd}W^{\g\b(3)}h_{\b(3)\bd(2)}\bigg]\notag\\
&\phantom{= \text{i}\int\text{d}^4x\, e \, \bigg\{}+\xi^{\a}\bigg[W_{\a}{}^{\b(3)}\nabla_{\b}{}^{\bd}\nabla_{\b}{}^{\bd}\nabla_{\b}{}^{\bd}\bar\vf_{\bd(3)}-3B_{\a}{}^{\g\bd(2)}\nabla_{\g}{}^{\bd}\bar\vf_{\bd(3)}\notag\\
&\phantom{= \text{i}\int\text{d}^4x\, e \, \bigg\{}-2\nabla_{\g}{}^{\bd}B_{\a}{}^{\g\bd(2)}\bar\vf_{\bd(3)}\bigg]\bigg\} +\text{c.c.}
\end{align}
\end{subequations}
The kinetic actions required to cancel the variations in \eqref{CMDspin52MixingGVar} which are proportional to the non-gauge fields are
\begin{subequations}
\begin{align}
S_{\chi\bar\chi}&= \frac{\text{i}}{2}\int\text{d}^4x\, e \,\chi^{\a(2)\ad}\nabla_{\a}{}^{\ad}\bar\chi_{\a\ad(2)}+\text{c.c.}~, \\
S_{\vf\bar{\vf}}&=\frac{\text{i}}{2}\int\text{d}^4x\, e \,\bar{\vf}^{\ad(3)}\nabla_{\ad}{}^{\a}\nabla_{\ad}{}^{\a}\nabla_{\ad}{}^{\a}\vf_{\a(3)}+\text{c.c.}
\end{align}
\end{subequations} 
They are both primary and prove to have the gauge transformations
\begin{subequations}
\begin{align}
\delta_{\xi}S_{\chi\bar\chi}&=-\text{i}\int\text{d}^4x\, e \,\bar\xi_{\ad}\bigg\{2\nabla_{\gd}{}^{\b}\bar W^{\ad\bd\gd(2)}\nabla_{\gd}{}^{\b}\chi_{\b(2)\bd}+\bar W^{\ad\bd\gd(2)}\nabla_{\gd}{}^{\b}\nabla_{\gd}{}^{\b}\chi_{\b(2)\bd}\bigg\}+\text{c.c.}~,\\
\delta_{\xi}S_{\vf\bar{\vf}}&=-\text{i}\int\text{d}^4x\, e \,\xi^{\a}W_{\a}{}^{\b(3)}\nabla_{\b}{}^{\bd}\nabla_{\b}{}^{\bd}\nabla_{\b}{}^{\bd}\bar\vf_{\bd(3)}+\text{c.c.}
\end{align}
\end{subequations} 
It follows that the primary action
\begin{align}
&S_{\text{CHS}}^{(3,2,2)}[h,\chi,\vf]=S_{h\bar{h}}-\frac{37}{48}S_{h\bar{\chi}}+\frac{17}{48}S_{h\bar\vf}+\frac{37}{48}S_{\chi\bar\chi}+\frac{17}{48}S_{\vf\bar\vf} \label{3.4343}\\[2ex]
&=\frac{\text{i}}{2}\int\text{d}^4x\, e \,\bigg\{\mc{W}^{\a(4)\ad}(h)\mc{W}_{\a(4)\ad}(\bar h)+ h^{\a(3)\ad(2)}\bigg[\frac{5}{4}W_{\a(3)}{}^{\b}\nabla^{\b\bd}\bar{h}_{\b(2)\bd\ad(2)}\notag\\
&-\nabla^{\b\bd}W_{\a(3)}{}^{\b}\bar{h}_{\b(2)\bd\ad(2)}-3W_{\a(2)}{}^{\b(2)}\nabla_{\a}{}^{\bd}\bar h_{\b(2)\bd\ad(2)}\bigg]+h^{\a(3)\ad(2)}\bigg[\frac{17}{24}W_{\a(3)}{}^{\g}\nabla_{\g}{}^{\bd}\bar\vf_{\ad(2)\bd}\notag\\
&-\frac{17}{12}\nabla_{\g}{}^{\bd}W_{\a(3)}{}^{\g}\bar\vf_{\ad(2)\bd} \bigg]-\frac{37}{24}\bar h^{\a(2)\ad(3)}\bar W_{\ad(3)}{}^{\bd}\chi_{\a(2)\bd}+\frac{37}{48}\chi^{\a(2)\ad}\nabla_{\a}{}^{\ad}\bar\chi_{\a\ad(2)}\notag\\
&+\frac{17}{48}\bar{\vf}^{\ad(3)}\nabla_{\ad}{}^{\a}\nabla_{\ad}{}^{\a}\nabla_{\ad}{}^{\a}\vf_{\a(3)}\bigg\}+\text{c.c.}~,
\end{align}
has gauge variation that is strictly proportional to the Bach tensor
\begin{align}
\delta_{\xi}S^{(3,2,2)}_{\text{CHS}}&=-\frac{\text{i}}{2}\int\text{d}^4x\, e \,\bigg\{ \xi^{\a}\bigg[\frac{3}{4}B_{\a}{}^{\b\bd(2)}\nabla^{\b\bd}\bar h_{\b(2)\bd(3)}+\nabla^{\b\bd}B_{\a}{}^{\b\bd(2)}\bar h_{\b(2)\bd(3)}\notag\\
&+B^{\b(2)\bd(2)}\nabla_{\a}{}^{\bd}\bar h_{\b(2)\bd(3)}+\frac{17}{8}B_{\a}{}^{\g\bd(2)}\nabla_{\g}{}^{\bd}\bar\vf_{\bd(3)}+\frac{17}{12}\nabla_{\g}{}^{\bd}B_{\a}{}^{\g\bd(2)}\bar\vf_{\bd(3)}\bigg]\notag\\
&-\bar\xi_{\ad}\bigg[\nabla^{\b\bd}B^{\b(2)\bd\ad} h_{\b(3)\bd(2)}+\frac{3}{4}B^{\b(2)\bd\ad}\nabla^{\b\bd} h_{\b(3)\bd(2)} \notag\\
&+B^{\b(2)\bd(2)}\nabla^{\b\ad} h_{\b(3)\bd(2)}+\frac{37}{24}B^{\b(2)\bd\ad}\chi_{\b(2)\bd}\bigg]\bigg\}+\text{c.c.}
\end{align}
It is therefore gauge invariant on any Bach-flat background
\begin{align}
\delta_{\xi}S^{(3,2,2)}_{\text{CHS}}[h,\chi,\vf]\approx 0~,
\end{align}
and has the conformally flat limit
\begin{align}
W_{\a(4)}=0\quad \implies \quad S_{\text{CHS}}^{(3,2,2)}[h,\chi,\vf]=S^{(3,2,2)}_{\text{CHS}}[h,\bar{h}]+\frac{37}{48}S_{\chi\bar\chi}+\frac{17}{48}S_{\vf\bar\vf}~.
\end{align}

\subsection{The hooked conformal graviton} \label{secConformalHook}

We will refer to the conformal gauge field with $m=n+2=3$ and $t=1$ as the pseudo-graviton, 
for it is described by a field $h_{\a(3) \ad}$ (and its conjugate $\bar{h}_{\a \ad(3)}$)
with the same weight and total number of spinor indices
as the conformal graviton $h_{\a(2) \ad(2)}$. Another feature it shares with the latter is that it has minimal depth gauge transformations. This is the first complete model of a minimal depth CHS field which needs to be coupled to auxiliary fields in order to achieve gauge invariance on Bach-flat backgrounds. As we will see, it is also turns out non-unique, which is the first example with this property. 

The fields $h_{\a(3)\ad}$ and $\bar{h}_{\a \ad(3)}$ may be combined into a single real traceless\footnote{Two-derivative non-conformal theories for a traceful 
hook field were first studied in $d$-dimensional Minkowski and anti-de Sitter (AdS$_d$) spaces in \cite{Curtright1, Curtright2} and \cite{BMVpm} respectively.}  
tensor field $h_{ [ ab ]c}$ satisfying the algebraic relations associated with the Young diagram {\scalebox{0.4}{
\begin{ytableau}
~ & ~ \\
~ \\
\end{ytableau}
}}. The latter is known in the literature  as a hook field.
Previously such fields were used to describe four-derivative gauge theories
in AdS$_d$ \cite{Karapet2}. The hook terminology used in \cite{Curtright1, Curtright2, BMVpm, Karapet2} (see also \cite{Vasiliev2009})
is inconvenient in the supersymmetric case, which will be studied in section \ref{secPGSBach},  
since the hook field $h_{[ab]c}$
is contained in a gauge superfield $H_{[ab]} $ which is a column and not a hook.
Nevertheless, we may sometimes refer to $h_{\a(3)\ad}$ as the conformal hook or hooked conformal graviton. For further comments on the conformal hook and some of its incarnations throughout the literature, see appendix \ref{AppHook}.

In accordance with the discussion in section \ref{secCHSPrepCS4}, 
the pseudo-graviton $h_{\a(3)\ad}$ has the conformal properties
\begin{align}
K_{\b\bd}h_{\a(3)\ad}=0~,\qquad \mathbb{D}h_{\a(3)\ad}=0~,
\end{align}
and is defined modulo gauge transformations
\begin{align}
\label{3.2}
\delta_{\xi}h_{\a(3)\ad}=\nabla_{\a\ad}\xi_{\a(2)}~.
\end{align}
The two linearised Weyl tensors $\mc{W}_{\a(4)}(h)$ and 
$\mc{W}_{\a(4)}(\bar{h})$ defined by
\begin{align}
\mc{W}_{\a(4)}(h)=\nabla_{\a}{}^{\bd}h_{\a(3)\bd}~,\qquad \mc{W}_{\a(4)}(h)=\nabla_{\a}{}^{\bd}\nabla_{\a}{}^{\bd}\nabla_{\a}{}^{\bd}\bar{h}_{\a\bd(3)}~,
\end{align}
 are no longer gauge invariant, consequently one may show that the variation of the skeleton action \eqref{GenCHSSkeleton} is given by
\begin{align}
\delta_{\xi}S^{(3,1)}_{\text{Skeleton}}=\frac{1}{2}\int \text{d}^4x\, e \, &\bigg\{\xi^{\a(2)}\bigg[2W_{\a}{}^{\g(3)}\mc{W}_{\a\g(3)}(\bar{h})\bigg]-\frac{1}{3}\bar{\xi}^{\ad(2)}\bigg[16\nabla_{\g\ad}W^{\g\b(3)}\nabla_{\ad}{}^{\b}\mc{W}_{\b(4)}(h)\notag\\
&+4W^{\g\b(3)}\nabla_{\g\ad}\nabla_{\ad}{}^{\b}\mc{W}_{\b(4)}(h)+6\nabla_{\ad}{}^{\b}\nabla_{\g\ad}W^{\g\b(3)}\mc{W}_{\b(4)}(h)\notag\\
&+2\nabla_{\ad}{}^{\d}W^{\b(4)}\nabla_{\d\ad}\mc{W}_{\b(4)}(h)\bigg] \bigg\}+\text{c.c.}\label{hhjk}
\end{align}

To counter the variation \eqref{hhjk} we need to construct all possible non-minimal corrections $\mc{J}_{\a\ad(3)}(\bar{h})$ with the conformal properties
\begin{align}
k_{\b\bd}\mc{J}_{\a\ad(3)}(\bar{h})=0~,\qquad \mb{D} \mc{J}_{\a\ad(3)}(\bar{h})= 4\mc{J}_{\a\ad(3)}(\bar{h})
\end{align}
 Up to terms proportional to the Bach tensor, there are only three such structures:
 \begin{subequations}
\begin{align}
\mc{J}^{(1)}_{\a(3)\ad}(\bar{h})=&\phantom{+}W_{\a(3)}{}^{\b}\bar{W}_{\ad}{}^{\bd(3)}\bar{h}_{\b\bd(3)}~,\\[8pt]
\mc{J}^{(2)}_{\a(3)\ad}(\bar{h})=&\phantom{+}5W_{\a(3)}{}^{\g}\nabla_{\g}{}^{\bd}\nabla^{\b\bd}\bar{h}_{\b\bd(2)\ad}+6W_{\a(2)}{}^{\g(2)}\nabla_{\g}{}^{\bd}\nabla_{\g}{}^{\bd}\bar{h}_{\a\ad\bd(2)}\notag\\
&+\nabla_{\g}{}^{\bd}W_{\a(3)}{}^{\g}\nabla^{\b\bd}\bar{h}_{\b\bd(2)\ad} -6\nabla_{\g}{}^{\bd}W_{\a(2)}{}^{\b\g}\nabla_{\a}{}^{\bd}\bar{h}_{\b\bd(2)\ad}\notag\\
&+2\nabla^{\d\bd}W_{\a(3)}{}^{\b}\nabla_{\d}{}^{\bd}\bar{h}_{\b\bd(2)\ad}-4\nabla^{\b\bd}\nabla_{\g}{}^{\bd}W_{\a(3)}{}^{\g}\bar{h}_{\b\bd(2)\ad}~,\\[8pt]
\mc{J}^{(3)}_{\a(3)\ad}(\bar{h})=&\phantom{+}5\bar{W}_{\ad}{}^{\gd\bd(2)}\nabla_{\a\gd}\nabla_{\a}{}^{\bd}\bar{h}_{\a\bd(3)}-\bar{W}^{\gd(2)\bd(2)}\nabla_{\a\gd}\nabla_{\a\gd}\bar{h}_{\a\ad\bd(2)}~\notag\\
&+18\nabla_{\a\gd}\bar{W}_{\ad}{}^{\gd\bd(2)}\nabla_{\a}{}^{\bd}\bar{h}_{\a\bd(3)}-3\nabla_{\a\gd}\bar{W}^{\gd\bd(3)}\nabla_{\a\ad}\bar{h}_{\a\bd(3)}\notag\\
&+3\nabla_{\a}{}^{\dd}\bar{W}_{\ad}{}^{\bd(3)}\nabla_{\a\dd}\bar{h}_{\a\bd(3)}+6\nabla_{\a}{}^{\bd}\nabla_{\a\gd}\bar{W}_{\ad}{}^{\gd\bd(2)}\bar{h}_{\a\bd(3)}~.
\end{align}
\end{subequations}
We denote the associated non-minimal functional by 
\begin{align}
S_{\text{NM},i}^{(3,1)}[h,\bar{h}]=\frac{1}{2}\int \text{d}^4x \, e \, h^{\a(3)\ad}\mc{J}^{(i)}_{\a(3)\ad}(\bar{h})+\text{c.c.} \label{CHookNMAct}
\end{align} 
It may be shown that, up to a total derivative and terms involving the Bach tensor, the functional \eqref{CHookNMAct} with $i=3$ is a linear combination of the other two,
\begin{align}
\int \text{d}^4x \, e& \, h^{\a(3)\ad}\mc{J}^{(3)}_{\a(3)\ad}(\bar{h}) + \text{c.c.}=\int \text{d}^{4}x \, e \, \bar{h}^{\a\ad(3)} \bar{\mc{J}}^{(3)}_{\a\ad(3)}(h) + \text{c.c.}\notag\\
&=\int \text{d}^{4}x \, e \, h^{\a(3)\ad}\bigg\{-\mc{J}^{(2)}_{\a(3)\ad}(\bar{h})+5\mc{J}^{(1)}_{\a(3)\ad}(\bar{h})+6B_{\a(2)}{}^{\bd(2)}\bar{h}_{\a\ad\bd(2)}\bigg\} +\text{c.c.} \label{8875}
\end{align}
 Here we have denoted $\bar{\mc{J}}_{\a\ad(3)}(h):=\big(\mc{J}^{(3)}_{\a(3)\ad}(\bar{h})\big)^*$. Thus it suffices to consider only the first two independent functionals. Their gauge variations may be shown to be 
\begin{subequations}
\begin{align}
\delta_{\xi}S_{\text{NM},1}^{(3,1)}=&-\frac{1}{2}\int\text{d}^4x\, e \, \bigg\{ \xi^{\a(2)}\bigg[W_{\a(2)}{}^{\b\g}\nabla_{\g\gd}\bar{W}^{\gd\bd(3)}\bar{h}_{\b\bd(3)}+\bar{W}^{\gd\bd(3)}\nabla_{\g\gd}W_{\a(2)}{}^{\b\g}\bar{h}_{\b\bd(3)}\notag\\
&+W_{\a(2)}{}^{\b\g}\bar{W}^{\gd\bd(3)}\nabla_{\g\gd}\bar{h}_{\b\bd(3)}\bigg]+\bar{\xi}^{\ad(2)}\bigg[\bar{W}_{\ad(2)}{}^{\bd\gd}\nabla_{\g\gd}W^{\g\b(3)}h_{\b(3)\bd}\notag\\
&+W^{\g\b(3)}\nabla_{\g\gd}\bar{W}_{\ad(2)}{}^{\bd\gd}h_{\b(3)\bd}+\bar{W}_{\ad(2)}{}^{\bd\gd}W^{\g\b(3)}\nabla_{\g\gd}h_{\b(3)\bd}\bigg]\bigg\}+\text{c.c.}~,\\
\delta_{\xi}S_{\text{NM},2}^{(3,1)}=&\phantom{+}2\delta_{\xi}S_{\text{Skeleton}}^{(3,1)}-\bigg(\int\text{d}^4x\, e \, \bigg\{ \xi^{\a(2)}\bigg[3W_{\a(2)}{}^{\b\g}\nabla_{\g\gd}\bar{W}^{\gd\bd(3)}\bar{h}_{\b\bd(3)}\notag\\
&+\bar{W}^{\gd\bd(3)}\nabla_{\g\gd}W_{\a(2)}{}^{\b\g}\bar{h}_{\b\bd(3)}+2W_{\a(2)}{}^{\b\g}\bar{W}^{\gd\bd(3)}\nabla_{\g\gd}\bar{h}_{\b\bd(3)}+\frac 32 B_{\a(2)}{}^{\bd(2)}\nabla^{\b\bd}\bar{h}_{\b\bd(3)}\notag\\
&+2B_{\a}{}^{\b\bd(2)}\nabla_{\a}{}^{\bd}\bar{h}_{\b\bd(3)}+2\nabla^{\b\bd}B_{\a(2)}{}^{\bd(2)}\bar{h}_{\b\bd(3)}\bigg]+\bar{\xi}^{\ad(2)}\bigg[W^{\g\b(3)}\nabla_{\g\gd}\bar{W}_{\ad(2)}{}^{\bd\gd}h_{\b(3)\bd}\notag\\
&+3\bar{W}_{\ad(2)}{}^{\bd\gd}\nabla_{\g\gd}W^{\g\b(3)}h_{\b(3)\bd}+2\bar{W}_{\ad(2)}{}^{\bd\gd}W^{\g\b(3)}\nabla_{\g\gd}h_{\b(3)\bd}+\frac 32 B^{\b(2)}{}_{\ad(2)}\nabla^{\b\bd}h_{\b(3)\bd}\notag\\
&+2B^{\b(2)\bd}{}_{\ad}\nabla_{\ad}{}^{\b}h_{\b(3)\bd}+2\nabla^{\b\bd}B^{\b(2)}{}_{\ad(2)}h_{\b(3)\bd}\bigg]\bigg\}+\text{c.c.}\bigg)~.
\end{align}
\end{subequations}

One can see that the following primary action, which consists purely of the pseudo-graviton, has gauge variation that is strictly second order in the Weyl tensor 
\begin{align}
S^{(\Gamma)}_{h\bar{h}}=S_{\text{Skeleton}}^{(3,1)}[h,\bar{h}]+\Gamma S_{\text{NM},1}^{(3,1)}[h,\bar{h}]-\frac 12 S_{\text{NM},2}^{(3,1)}[h,\bar{h}]~, \qquad \delta_{\xi}S_{h\bar{h}}=\mc{O}\big(W^2\big)~.
\end{align}
Here  $\Gamma\in\mathbb{R}$ is a free parameter whose significance will become apparent shortly. 
 At this point, we have exhausted all possible primary terms that are purely quadratic in the field $h_{\a(3)\ad}$. The only option left is to introduce extra lower-spin fields to compensate for the second order terms. It is sufficient to introduce only
one such field, the conformal non-gauge field $\chi_{\a(2)}$ and its conjugate $\bar{\c}_{\ad(2)}$.

The non-gauge field $\chi_{\a(2)}$ is primary with weight one,
\begin{align}
K_{\b\bd}\chi_{\a(2)}=0~, \qquad \mathbb{D}\chi_{\a(2)}=\chi_{\a(2)}~.
\end{align}
It is defined to have a gauge transformation rule involving the gauge parameter of the conformal hook field
\begin{align}
\delta_{\xi}\chi_{\a(2)}=W_{\a(2)}{}^{\b(2)}\xi_{\b(2)}~. \label{chiGT}
\end{align}
The only suitable primary coupling between the fields $h_{\a(3)\ad}$ and $\chi_{\a(2)}$ is given by 
\begin{align}
S_{h\bar{\c}}=\int\text{d}^4x\, e \, h^{\a(3)\ad}\bigg\{W_{\a(3)}{}^{\g}\nabla_{\g}{}^{\bd}\bar{\chi}_{\bd\ad}-\nabla_{\g}{}^{\bd}W_{\a(3)}{}^{\g}\bar{\chi}_{\bd\ad}\bigg\}+\text{c.c.}~,
\end{align}
and it has the following gauge variation:
\begin{align}
\delta_{\xi}S_{h\bar{\c}}&=\int\text{d}^4x\, e \,\bigg\{\xi^{\a(2)}\bigg[W_{\a(2)}{}^{\g(2)}\nabla_{\g}{}^{\bd}\nabla_{\g}{}^{\bd}\bar{\chi}_{\bd(2)}-B_{\a(2)}{}^{\bd(2)}\bar{\chi}_{\bd(2)}\bigg]\notag\\
&+\bar{\xi}^{\ad(2)}\bigg[2\bar{W}_{\ad(2)}{}^{\bd\gd}\nabla_{\g\gd}W^{\g\b(3)}h_{\b(3)\bd}+\bar{W}_{\ad(2)}{}^{\bd\gd}W^{\g\b(3)}\nabla_{\g\gd}h_{\b(3)\bd}\bigg]\bigg\}+\text{c.c.}\label{CHookMixingGVar}
\end{align}
To cancel the part of \eqref{CHookMixingGVar} that is proportional to 
$\bar{\chi}_{\ad(2)}$ and  $\c_{\a(2)}$, we introduce the corresponding primary kinetic action 
\begin{align}
S_{\chi\bar{\chi}}=\frac{1}{2}\int\text{d}^4x\, e \, \bar{\chi}^{\ad(2)}\nabla_{\ad}{}^{\a}\nabla_{\ad}{}^{\a}\chi_{\a(2)}+\text{c.c.} \label{CHookKineticX}
\end{align}
Its gauge variation is given by 
\begin{align}
\delta_{\xi}S_{\chi\bar{\chi}}=\frac{1}{2}\int\text{d}^4x\, e \, \bigg\{\xi^{\a(2)}W_{\a(2)}{}^{\b(2)}\nabla_{\b}{}^{\bd}\nabla_{\b}{}^{\bd}\bar{\chi}_{\bd(2)}+\bar{\xi}^{\ad(2)}\bar{W}_{\ad(2)}{}^{\bd(2)}\nabla_{\bd}{}^{\b}\nabla_{\bd}{}^{\b}\chi_{\b(2)}\bigg\}+\text{c.c.} \label{CHookKineticXGVar}
\end{align}

From \eqref{CHookMixingGVar} and \eqref{CHookKineticXGVar}, it may be shown that the following conformal action
\begin{subequations} 
\begin{align}
&S_{\text{CHS},\Gamma=1}^{(3,1)}[h,\c]~= ~S^{(\Gamma=1)}_{h\bar{h}}-S_{h\bar{\c}}
+S_{\chi\bar{\chi}}\label{CHookGInvX} \\
&=\frac{1}{2}\int\text{d}^4x\, e \, \bigg\{  \mc{W}^{\a(4)}(h)\mc{W}_{\a(4)}(\bar{h})+h^{\a(3)\ad}W_{\a(3)}{}^{\b}\bar{W}_{\ad}{}^{\bd(3)}\bar{h}_{\b\bd(3)}\notag\\ 
&-\frac{1}{2}h^{\a(3)\ad}\bigg[5W_{\a(3)}{}^{\g}\nabla_{\g}{}^{\bd}\nabla^{\b\bd}\bar{h}_{\b\bd(2)\ad}+6W_{\a(2)}{}^{\g(2)}\nabla_{\g}{}^{\bd}\nabla_{\g}{}^{\bd}\bar{h}_{\a\ad\bd(2)}+\nabla_{\g}{}^{\bd}W_{\a(3)}{}^{\g}\nabla^{\b\bd}\bar{h}_{\b\bd(2)\ad}\notag\\
& -6\nabla_{\g}{}^{\bd}W_{\a(2)}{}^{\b\g}\nabla_{\a}{}^{\bd}\bar{h}_{\b\bd(2)\ad}+2\nabla^{\d\bd}W_{\a(3)}{}^{\b}\nabla_{\d}{}^{\bd}\bar{h}_{\b\bd(2)\ad}-4\nabla^{\b\bd}\nabla_{\g}{}^{\bd}W_{\a(3)}{}^{\g}\bar{h}_{\b\bd(2)\ad}\bigg]\notag\\
&-2h^{\a(3)\ad}\bigg[W_{\a(3)}{}^{\g}\nabla_{\g}{}^{\bd}\bar{\chi}_{\bd\ad}-\nabla_{\g}{}^{\bd}W_{\a(3)}{}^{\g}\bar{\chi}_{\bd\ad}\bigg]+\bar{\chi}^{\ad(2)}\nabla_{\ad}{}^{\a}\nabla_{\ad}{}^{\a}\chi_{\a(2)}\bigg\}+\text{c.c.}
\end{align}
\end{subequations}
has gauge variation that is strictly proportional to the Bach tensor
\begin{align}
\delta_{\xi}S_{\text{CHS},\Gamma=1}^{(3,1)}=&\phantom{+}\int\text{d}^4x\, e \, \bigg\{ \xi^{\a(2)}\bigg[3 B_{\a(2)}{}^{\bd(2)}\nabla^{\b\bd}\bar{h}_{\b\bd(3)}+4B_{\a}{}^{\b\bd(2)}\nabla_{\a}{}^{\bd}\bar{h}_{\b\bd(3)}+4\nabla^{\b\bd}B_{\a(2)}{}^{\bd(2)}\bar{h}_{\b\bd(3)}\notag\\
&+2B_{\a(2)}{}^{\bd(2)}{}\bar{\chi}_{\bd(2)}\bigg]+\bar{\xi}^{\ad(2)}\bigg[3 B^{\b(2)}{}_{\ad(2)}\nabla^{\b\bd}h_{\b(3)\bd}+4B^{\b(2)\bd}{}_{\ad}\nabla_{\ad}{}^{\b}h_{\b(3)\bd}\notag\\
&+4\nabla^{\b\bd}B^{\b(2)}{}_{\ad(2)}h_{\b(3)\bd}+2B^{\b(2)}{}_{\ad(2)}\chi_{\b(2)}\bigg]\bigg\}+\text{c.c.}
\end{align}
It is therefore gauge invariant when restricted to a Bach-flat background
\begin{align}
\delta_{\xi}S_{\text{CHS},\Gamma=1}^{(3,1)}\approx 0~.
\end{align}

Although it is sufficient to couple the conformal graviton to the non-gauge field $\chi_{\a(2)}$ to achieve gauge invariance, it is not necessary. Indeed, there exists another gauge invariant model for $h_{\a(3)\ad}$ which makes use of a different conformal non-gauge field $\vf_{\a(4)\ad(2)}$.\footnote{In a sense $\vf_{\a(4)\ad(2)}$ has higher spin relative to $h_{\a(3)\ad}$, due to the larger total number of indices.} The latter has the conformal properties
\begin{subequations}
\begin{align}
K_{\b\bd}\varphi_{\a(4)\ad(2)}=0~,& \qquad \mathbb{D}\varphi_{\a(4)\ad(2)}=\varphi_{\a(4)\ad(2)}~,  \label{CHookNGF}
\end{align}
\end{subequations} 
as prescribed in appendix \ref{AppCNGM}.  Under the pseudo-graviton gauge transformations it transforms non-trivially according to the rule 
\begin{align}
\delta_{\xi}\vf_{\a(4)\ad(2)}=W_{\a(4)}\bar{\xi}_{\ad(2)}~. \label{NGGThihello}
\end{align}

In this case, the gauge invariant action may be shown to take the form 
\begin{align}
S_{\text{CHS},\Gamma=3}^{(3,1)}[h,\vf]=S_{h\bar{h}}^{(\Gamma=3)}+S_{h\vf}-S_{\vf\bar{\vf}}~, \label{CHookFullGInvF}
\end{align}
where we have defined the primary functionals 
\begin{subequations}
\begin{align}
S_{h\vf}&= \int\text{d}^4x\, e \, h^{\a(3)\ad}\bigg\{\bar{W}_{\ad}{}^{\gd\bd(2)}\nabla_{\gd}{}^{\b}\vf_{\a(3)\b\bd(2)}-\nabla_{\gd}{}^{\b}\bar{W}_{\ad}{}^{\gd\bd(2)}\vf_{\a(3)\b\bd(2)}\bigg\}+\text{c.c.}~, \\
S_{\vf\bar{\vf}}&=\frac{1}{2}\int\text{d}^4x\, e \, \bar{\vf}^{\a(2)\ad(4)}\nabla_{\ad}{}^{\a}\nabla_{\ad}{}^{\a}\vf_{\a(4)\ad(2)}+\text{c.c.}
\end{align}
\end{subequations}

In fact, by using both of the fields $\chi_{\a(2)}$ and $\vf_{\a(4)\ad(2)}$, one can construct a one-parameter family of actions for the pseudo-graviton described by  
\begin{align}
S^{(3,1)}_{\text{CHS},\Gamma}[h,\chi,\vf]=S_{h\bar{h}}^{(\Gamma)}~+~&\frac{1}{2}(\Gamma-3)S_{h\bar{\c}}-\frac{1}{2}(\Gamma-3)S_{\c\bar{\c}} \notag\\
~+~&\frac{1}{2}(\Gamma-1)S_{h\vf}-\frac{1}{2}(\Gamma-1)S_{\vf\bar{\vf}}~,~~~~~~~~~ \label{CHookFullGInvFam}
\end{align}
which are gauge invariant on Bach-flat backgrounds for all values of $\Gamma$,
\begin{align}
\delta_{\xi}S^{(3,1)}_{\text{CHS},\Gamma}[h,\chi,\vf]\approx 0~.
\end{align}
The parameter $\Gamma$ reflects the non-uniqueness of this model. 
The actions \eqref{CHookGInvX} and \eqref{CHookFullGInvF} may be recovered by setting $\Gamma=1$ and $\Gamma=3$ respectively. 

Finally, we note that in the conformally-flat limit, the action \eqref{CHookFullGInvFam} reduces to 
\begin{align}
\label{OneParFmaCFlim}
S_{\text{CHS},\Gamma}^{(3,1)}[h,\chi,\vf]&=S_{\text{CHS}}^{(3,1)}[h,\bar{h}]-\frac{1}{2}(\Gamma-3)S_{\c\bar{\c}}-\frac{1}{2}(\Gamma-1)S_{\vf\bar{\vf}}~,
\end{align}
and each of the fields $h_{\a(3)\ad}$, $\chi_{\a(2)}$ and $\vf_{\a(4)\ad(2)}$ decouple. 

\section{Summary of results} \label{sec4DCHSdis}

This chapter was dedicated to the construction of CHS models in curved $4d$ spacetimes and applications thereof. We began in section \ref{secCHSM4} by describing the main features of the mixed symmetry CHS field $h_{\a(m)\ad(n)}$, and its higher-spin Bach tensor $\mc{B}_{\a(n)\ad(m)}(h)$, on a generic background.\footnote{We also discussed its HS Weyl tensor $\mc{W}_{\a(m+n)}(h)$. However, in the authors opinion, the descendent $\mc{B}_{\a(n)\ad(m)}(h)$ is in some sense more fundamental. For example, the latter is the direct analogue of the $3d$ HS Cotton tensor, since both are transverse. Moreover, the ability to express the CHS Lagrangian in the form $\mc{W}^{\a(m+n)}(h)\mc{W}_{\a(m+n)}(\bar{h})$ appears to be a special feature of $4d$ conformally-flat spacetimes. } 
The definition of $\mc{B}_{\a(n)\ad(m)}(h)$ is such that it has the characteristic features of a covariantly conserved conformal current if spacetime is conformally-flat, leading to the gauge and Weyl invariant CHS action \eqref{CHSActionB4}. 

In section \ref{secCHSMink4} we reviewed the explicit realisation of the conformal and gauge invariant models for $h_{\a(m)\ad(n)}$ in $\mb{M}^4$, in terms of field strengths and also spin-projection operators. We provided a novel and simple form \eqref{CasimirProjectorsMink4} for the rank-$(m,n)$ transverse-traceless projectors, which appeared recently in \cite{AdS3(super)projectors}. In section \ref{secCHSAdS4} we derived explicit gauge and conformally invariant actions for $h_{\a(m)\ad(n)}$ on AdS$_4$ \cite{Confgeo, AdSprojectors}. We used these models as motivation to construct novel rank-$(m,n)$ spin-projection operators \eqref{TTprojectorsAdS4} on AdS$_4$ \cite{AdSprojectors}. The latter contain multiple poles which were shown to be correspond to partially-massless fields of various depths. These results allowed us to demonstrate that the rank-$(m,n)$ CHS kinetic operator factorises into second order operators associated with partially-massless fields of all depths \cite{AdSprojectors}, see e.g. \eqref{factor2}. This was the first derivation of this fact for $|m-n|\neq 0$.

In section \eqref{secCHSCF4} we constructed manifestly conformal and gauge invariant CHS actions on arbitrary conformally-flat backgrounds \cite{Confgeo}. This was done for generalised CHS gauge fields $h^{(t)}_{\a(m)\ad(n)}$ of all depths $1\leq t \leq \text{min}(m,n)$ (we recall that ordinary CHS fields have minimal depth $t=1$). In section \ref{secCHSBach} we extended gauge invariance to Bach-flat backgrounds for the following special cases: (i) maximal depth conformal graviton \cite{Confgeo} with $m=n=t=2$, see eq. \eqref{CMDspin2FullAct}; (ii) maximal depth spin-3 \cite{spin3depth3} with $m=n=t=3$, see eq. \eqref{CMDspin3FullAction}; (iii) maximal depth spin-5/2 \cite{spin3depth3} with $m-1=n=t=2$, see eq. \eqref{3.4343}; and (iv) hooked conformal graviton \cite{SCHS, SCHSgen} with $m-2=n=t=1$, see eq. \eqref{CHookFullGInvFam}. Each of these are new results. Models (ii) -- (iv) all required a coupling to subsidiary conformal fields in order to restore gauge invariance, supporting the conjecture made in \cite{GrigorievT} for minimal depth spin 3. An interesting observation is the non-uniqueness of model (iv) \cite{SCHSgen}.

It is worth mentioning that each of the subsidiary conformal fields necessary for gauge invariance of (ii) -- (iv) are specific examples of what we have called non-gauge conformal fields $\chi_{\a(m)\ad(n)}$. Their Weyl invariant actions on arbitrary backgrounds were described in appendix \ref{AppCNGM}, which first appeared in \cite{SCHS}.

\begin{subappendices}

\section{Technical results in AdS$_4$} \label{AppAdS4}

In this appendix we first give a systematic discussion
of how to realise the unitary and the partially massless representations of the
AdS$_4$ isometry algebra $\mf{so}(3,2)$ in terms of on-shell tensor fields. In \ref{AppGenForm4} we develop some computational tools which aid  in proving the results of appendix \ref{secAppPMGAdS4}, in which we discuss the partially massless gauge symmetry. 

\subsection{Field theoretic representations of $\mf{so}(3,2)$} \label{AppAdS4Reps}

 The unitary irreducible representations (UIRs) of the Lie algebra $\mf{so}(3,2)$ of the AdS$_4$ isometry group were studied in detail in~\cite{Dirac:1935zz, Dirac:1963ta, Fronsdal:1965zzb, Fronsdal:1974ew, Fronsdal:1975eq, Fronsdal:1975ac, Evans,Angelopoulos,AFFS}  (see also~\cite{Nicolai:1984hb, deWit:1999ui} for a comprehensive review). They are specified by the lowest value $E_0$ of the energy $E$ and spin $s$,
and are traditionally denoted $D (E_0, s)$.\footnote{The parameters $E_0$ and $s$ determine the values of the quadratic and quartic Casimir operators of  $\mathfrak{so}(3, 2)$, as shown in  \cite{Fronsdal:1974ew,Evans}. The energy $E$ is chosen to be dimensionless.
To restore dimensionful  energy, one has to rescale  $E \to |\mu| \,E$, where $\m\bar \m$ determines the AdS curvature \eqref{LCDalgebraAdS4}.}
The allowed  spin values
are $s = 0, \frac{1}{2}, 1, \cdots $, the same as in $\mb{M}^4$. However, unlike in $\mb{M}^4$, unitarity imposes a bound on the allowed values of energy. According to the theorems proved in \cite{Evans,Angelopoulos}, $D(E_0, s)$ is unitary iff one of the following conditions holds: (i) $s=0$, $E_0 \geq \hf$; (ii) $s=\hf$,  $E_0\geq 1$, 
and (iii) $s\geq 1$, $E_0 \geq s+1$. 

The representations $D\big(\hf , 0\big) = {\rm Rac} $ and $D\big( 1, \hf \big) = {\rm Di} $ 
are known as the Dirac singletons \cite{Dirac:1963ta}.\footnote{It was found by Flato and Fronsdal \cite{Flato:1980zk,FF78} that the singletons are the square roots of massless particles in the sense that all two-singleton states are massless.} 
The representations 
$D(s+1, s) $ for $s>0$ and $D( 1, 0)  \oplus D (2, 0)$ are called massless since they contract  to the massless discrete helicity representations of the Poincar\'e group 
\cite{AFFS}. These representations prove to be restrictions of certain unitary representations of the conformal algebra   
$\mathfrak{so}(4, 2)$  to  $\mathfrak{so}(3, 2)$ \cite{AFFS,Barut:1970kp}. The remaining representations $D(E_0, s)$  are usually referred to as the massive AdS$_4$ representations. Below we will be concerned only with those representations carrying spin $s\geq 1$.

UIRs of the Lie algebra $\mf{so}(3,2)$ of the AdS$_4$ isometry group may be realised in terms of the on-shell fields $\phi_{\a(m)\ad(n)}$ \eqref{OSFAdS4}, 
\begin{subequations}\label{OSFAppdS4}
\begin{align}
0~&=\big(\mathcal{Q}-\rho^2\big)\phi_{\a(m)\ad(n)}~,\label{OSC1AppdS4}\\
0~&=\mc{D}^{\b\bd}\phi_{\a(m-1)\b\ad(n-1)\bd}~,\label{OSC2AppdS4}
\end{align}
\end{subequations}
 for certain values of pseudo-mass $\r$. The spin $s$ of the representation proves to be equal to $s=\frac{1}{2}(m+n)$. The following relation between the pseudo-mass $\rho$ and the minimal energy $E_0$ holds:
\bea
\rho^2 =\big[E_0(E_0-3) +s(s+1)\big]\mu\mub~,\label{PseudoMassEnergy}
\eea
see \cite{Nicolai:1984hb, deWit:1999ui} for pedagogical derivations.

As the name suggests, the pseudo-mass does not coincide with what is usually considered to be the physical mass $\rho_{\text{phys}}$. Rather, the two are related through
\begin{align}
\rho^2_{\text{phys}}=\rho^2-\tau_{(1,m,n)}\mu\mub~, \qquad
\tau_{(1,m,n)} = \hf (m+n+2)(m+n-2)  
~,
\label{Mphys}
\end{align}
 where $\t_{(1,m,n)}$ is one of the parameters \eqref{PMvalAdS4} (we will see that $\t_{(1,m,n)}$ is associated with a massless field). 
 The on-shell field $\phi_{\a(m)\ad(n)}$ corresponds to the irreducible representation $D\big(E_0,\frac{1}{2}(m+n)\big)$ where the minimal energy $E_0$ is related to the physical mass  through
\begin{align}
\rho_{\text{phys}}^2 
&=\big[E_0(E_0-3)-\frac{1}{4}(m+n+2)(m+n-4)\big]\mu\mub~. \label{Energy-Mass}
\end{align}
The unitarity bound for $m+n > 1$
 is $E_0\geq  \frac{1}{2}(m+n+2)$, which in terms of the masses is
\begin{align}
\rho^2_{\text{phys}}\geq 0 \quad \implies \quad \rho^2 \geq \tau_{(1,m,n)}\mu\mub~, \qquad \frac{1}{2}(m+n)\geq 1~.
\label{Ubound}
\end{align}

With these relations in mind, we usually prefer to use the pseudo-mass as a representation label in place of $E_0$. As a caveat we note that there are two distinct values of $E_0$ leading to the same value of $\rho_{\text{phys}}^2$, 
\begin{align}
\big(E_0\big)_{\pm}=\frac{3}{2}\pm \frac{1}{2}
\sqrt{
4\frac{\rho^2_{\text{phys}}}{\mu\mub}+(m+n-1)^2
}
~.
\label{2.12AdS4}
\end{align}
However, the solution $(E_0)_-$ always violates the unitarity bound for $\frac{1}{2}(m+n)\geq 1$. Thus when referring to a unitary representation with $s \geq 1$ and
pseudo-mass $\rho^2$, we are implicitly referring to the representation corresponding to $(E_0)_+$,
\begin{align}
E_0 =\frac{3}{2} + \frac{1}{2}
\sqrt{
4\frac{\rho^2_{\text{phys}}}{\mu\mub}+(m+n-1)^2}
~, \qquad \frac{1}{2}(m+n)\geq 1~.
\end{align}
For $s= 0, \hf $ (or, equivalently, $m+n=0,1$),  the unitary bound is $E_0\geq s +\hf$, 
and the solution  $(E_0)_-$ in \eqref{2.12AdS4}  does not violate the unitarity bound for certain values of $\rho^2_{\text{phys}}$. For $s=0$ the allowed values of $\rho^2_{\text{phys}}$ are  restricted by
the condition
$- \frac 14 |\m|^2 \leq \rho^2_{\text{phys}} \leq  \frac 34 |\m|^2 $, which is known as  
the Breitenlohner-Freedman bound \cite{BreitenF}. 

The massive representation of $\mathfrak{so}(3, 2)$ with spin $s=\frac{1}{2}(m+n)$
 may be realised on the space of fields $\phi_{\a(m)\ad(n)}$  satisfying the equations \eqref{OSFAppdS4}
 in which  $\r$ is constrained by 
 \begin{align}
 \rho^2>\tau_{(1,m,n)}\mu\mub \quad\implies\quad \rho_{\text{phys}}^2>0~,
 \end{align}
but is otherwise arbitrary. This restriction ensures the unitarity of the representation. 

Given two positive integers $m$ and $n$, the tensor field $\phi_{\a(m)\ad(n)}\equiv \phi^{(t)}_{\a(m)\ad(n)}$ is said to be partially massless with depth-$t$ if it satisfies the on-shell conditions \eqref{OSFAdS4} such that its pseudo-mass takes the special value \cite{DeserW4, Zinoviev, Metsaev2006, AdSprojectors}
\begin{align}
\rho^2=\tau_{(t,m,n)}\mu\mub~,\qquad 1 \leq t \leq \text{min}(m,n)~, \label{PM00}
\end{align}
where the dimensionless constants $\tau_{(t,m,n)}$
are defined in eq. \eqref{PMvalAdS4},
\begin{align}
 \t_{(t,m,n)}= \frac{1}{2}\Big[(m+n-t+3)(m+n-t-1)+(t-1)(t+1)\Big]~.\label{PMvalAdS4App}
\end{align}
The specific feature of partially massless fields is that,
for a fixed $t$, the system of equations \eqref{OSFAppdS4} and \eqref{PM00} admits a depth-$t$ gauge symmetry
\begin{align}
\delta_{\xi}\phi^{(t)}_{\a(m)\ad(n)}= \big(\mc{D}_{\a\ad}\big)^t\xi^{(t)}_{\a(m-t)\ad(n-t)}~. \label{GTNSPMApp}
\end{align}
This is true as long as the gauge parameter $\xi^{(t)}_{\a(m-t)\ad(n-t)}$ is also on-shell with the same pseudo-mass,
\begin{subequations}\label{GPPApp}
\begin{align}
0~&=\big(\mathcal{Q}-\tau_{(t,m,n)}\mu\mub\big)\xi^{(t)}_{\a(m-t)\ad(n-t)}~,\label{GPPAppa}\\
0~&=\mc{D}^{\b\bd}\xi^{(t)}_{\a(m-t-1)\b\ad(n-t-1)\bd}~.\label{GPPAppb}
\end{align}
\end{subequations}

Strictly massless fields carry depth $t=1$ and therefore have mass given by 
\begin{align}
\rho^2=\tau_{(1,m,n)}\mu\mub \quad \implies \quad \rho_{\text{phys}}^2=0~.
\label{2.17massless}
\end{align}
This saturates the bound \eqref{Ubound} and hence defines a unitary representation of $\mathfrak{so}(3, 2)$. However, on account of the inequality 
\begin{align}
\tau_{(1,m,n)}> \tau_{(t,m,n)}~,\qquad 2 \leq t \leq \text{min}(m,n)~,
\end{align}
 the true partially massless representations 
are non-unitary. In particular, this means that there are two minimal energy values, 
\begin{align}
 (E_0)_{\pm}=\frac{3}{2}\pm\frac{1}{2}(m+n-2t+1)~,
 \end{align}
which are equally valid since they both violate the unitarity bound. In this work, whenever this ambiguity arises, we always implicitly choose the positive branch, $E_0\equiv (E_0)_+$.    
To distinguish the true partially-massless representations with depth $t$ and Lorentz type $(m/2,n/2)$, we will employ the notation  
\begin{align}
P\big(t,m,n\big)~,  \qquad2\leq t \leq \text{min}(m,n)~. 
 \label{NSPMrep}
\end{align}
Such a representation carries minimal energy $E_0=\frac{1}{2}(m+n)-t+2$.

\subsection{Generating function identities}  \label{AppGenForm4}

For many practical calculations in AdS$_4$, it is useful to 
 introduce the auxiliary commuting spinor variables $\U^{\a}$ and $\bar{\U}^{\ad}$. 
Associated with a tensor field $\phi_{\a(m)\ad(n)}$ of Lorentz type $(\frac{m}{2},\frac{n}{2})$ is  a homogeneous polynomial $\phi_{(m,n)}(\U, \bar \U)$ of degree $(m,n)$ defined by 
  \bea
  \phi_{(m,n)}:=\U^{\a_1}\cdots\U^{\a_m}\bar{\U}^{\ad_1}\cdots\bar{\U}^{\ad_n}\phi_{\a_1\dots\a_m\ad_1\dots\ad_n}~.
  \eea
  We denote the linear space of such homogeneous polynomials as $\mathcal{H}_{(m,n)}$. The correspondence $\phi_{\a(m)\ad(n)} \rightarrow \phi_{(m,n)}$ is one-to-one.  We note that since, for example, the $\U^{\a}$ are commuting, we have $\U^{\a}\U_{\a}:=\U^{\a}\ve_{\a\b}\U^{\b}=0$.

Next we introduce the two AdS$_4$ differential operators 
\begin{align}
\mathcal{D}_{(1,1)}:= \U^{\a}\bar{\U}^{\ad}\mathcal{D}_{\a\ad}~,\qquad \mathcal{D}_{(-1,-1)}:=\mathcal{D}^{\a\ad}\frac{\partial}{\partial\U^{\a}}\frac{\partial}{\partial \bar{\U}^{\ad}}~\equiv \mathcal{D}^{\a\ad}\partial_{\a}\bar{\partial}_{\ad}~,\label{AdS4GenForm1}
\end{align}
which increase and decrease
the degree of homogeneity by $(1,1)$ and $(-1,-1)$ respectively.
They may be shown to satisfy the algebra
\begin{align}
\big[\mathcal{D}_{(1,1)},\mathcal{D}_{(-1,-1)}\big]=(\bm\U+1)(\Box+2\mu\mub\bar{\bm M})+(\bar{\bm \U}+1)(\Box+2\mu\mub\bm M)~, \label{AdS4GenForm2}
\end{align}
where we have defined\footnote{The auxiliary variables $\U^{\a}$ and $\bar{\U}^{\ad}$ are defined to be inert with respect to the Lorentz generators $M_{\a\b}$ and $\bar{M}_{\ad\bd}$. Alternatively, one could define their action on any $\phi_{(m,n)}\in\mc{H}_{(m,n)}$ to be given by $M_{\a\b}\phi_{(m,n)}:=-\U_{(\a}\pa_{\b)}\phi_{(m,n)}$ and $\bar{M}_{\ad\bd}\phi_{(m,n)}:=-\bar{\U}_{(\ad}\bar{\pa}_{\bd)}\phi_{(m,n)}$.}
\begin{subequations}
\begin{align}
\bm\U&=\U^{\a}\partial_{\a}~,\qquad \qquad  \qquad~ \bm\U\phi_{(m,n)}=m\phi_{(m,n)}~,\\
\bar{\bm \U}&=\bar{\U}^{\ad}\bar{\partial}_{\ad}~,\qquad\qquad \qquad~\bar{\bm \U}\phi_{(m,n)}=n\phi_{(m,n)}~,\\
\bm M&=\U^{\a}\partial^{\b}M_{\a\b}~,\qquad\qquad \bm M\phi_{(m,n)}=-\frac{1}{2}m(m+2)\phi_{(m,n)}~,\\
\bar{\bm M}&=\bar{\U}^{\ad}\bar{\partial}^{\bd}\bar{M}_{\ad\bd}~,\qquad\qquad \bar{\bm M}\phi_{(m,n)}=-\frac{1}{2}n(n+2)\phi_{(m,n)}~.\label{AdS4GenForm3}
\end{align}
\end{subequations}
Then, via induction on $k$ it is possible to show that for any $\phi_{(m,n)}\in \mathcal{H}_{(m,n)}$, the following crucial identity holds true
\begin{align}
\big[\mathcal{D}_{(-1,-1)},\mc{D}^k_{(1,1)}\big]\phi_{(m,n)}=-k(m+n+k+1)\big(\mathcal{Q}-\t_{(k,m+k,n+k)}\mu\mub\big)\mc{D}^{k-1}_{(1,1)}\phi_{(m,n)}~.\label{AdS4GenForm4}
\end{align}
We recall that $\t_{(t,m,n)}$ are the partially-massless values \eqref{PMvalAdS4App}, and $\mc{Q}$ is the AdS$_4$ quadratic Casimir operator \eqref{QCasimirAdS4}.

\subsection{Partially massless gauge symmetry} \label{secAppPMGAdS4}

Given two integers $m$ and $n$ such that $m\geq n>0$, 
let $\phi_{\a(m)\ad(n)}$  be an on-shell field,
\begin{subequations}\label{AdS4B.1}
\begin{align}
0&=\big(\mathcal{Q}- \rho^2
\big)\phi_{\a(m)\ad(n)}~,\label{AdS4B.1a}\\
0&= \mc{D}^{\b\bd}\phi_{\a(m-1)\b\ad(n-1)\bd}~.\label{AdS4B.1b}
\end{align}
\end{subequations}
We would like to determine those values of $\rho$ for which the above system of equations
is compatible with a gauge symmetry. 

We begin by positing a gauge transformation of the form 
\begin{align}
\delta_{\xi}\phi_{\a(m)\ad(n)}=\mc{D}_{\a\ad}\xi_{\a(m-1)\ad(n-1)}~ \qquad \Longleftrightarrow \qquad \delta_{\xi}\phi_{(m,n)}=\mc{D}_{(1,1)}\xi_{(m-1,n-1)} \label{AdS4BGT}
\end{align}
and look for a gauge parameter  $\xi_{(m-1,n-1)}$ such that 
$\delta_{\xi}\phi_{(m,n)}$ is a solution to the equations \eqref{AdS4B.1}. 
Clearly, for gauge invariance of \eqref{AdS4B.1a}, the gauge parameter must satisfy
\begin{align}
0&=\big(\mathcal{Q}- \rho^2
\big)\xi_{(m-1,n-1)}~. \label{AdS4Btk}
\end{align}
Next we require \eqref{AdS4B.1b} to be gauge invariant, 
\begin{align}
0=\mc{D}^{\b\bd} \d_\xi \phi_{\a(m-1)\b\ad(n-1)\bd} \qquad \Longleftrightarrow \qquad 0=\mc{D}_{(-1,-1)}\delta_{\xi}\phi_{(m,n)}~. \label{AdS4BHello}
\end{align}
To solve this problem, let us recall that,
using the spin projection operators, it was shown that any unconstrained tensor field can be decomposed into irreducible (i.e. transverse) parts \eqref{DecompAdS4}.  For the gauge parameter $\xi_{(m-1,n-1)}$, this decomposition takes the form
\begin{align}
\xi_{(m-1,n-1)}=~&\xi_{(m-1,n-1)}^{\perp}+\sum_{t=1}^{n-2}\mc{D}^{\phantom{.}t}_{(1,1)}\xi^{\perp}_{(m-t-1,n-t-1)}+\mc{D}^{n-1}_{(1,1)}\xi_{(m-n,0)}~,
\label{AdS4Bty}
\end{align}
where $\xi_{(m-n,0)}$ is unconstrained, whilst the other fields are transverse,
\begin{align}
0=\mc{D}_{(-1,-1)}\xi^{\perp}_{(m-t,n-t)}~,\qquad 1 \leq t \leq n-1~. \label{AdS4Bth}
\end{align}
Inserting this expansion into the condition \eqref{AdS4BHello}, and making use of identity \eqref{AdS4GenForm4}, one arrives at the following equation
\begin{align}
0=\sum_{k=1}^{n}k(m+n-k+1)&\Big(
\rho^2
-\tau_{(k,m,n)} \mu\mub\Big)\mc{D}_{(1,1)}^{k-1}\xi^{\perp}_{(m-k,n-k)}~.\label{AdS4Btx}
\end{align}
The right-hand side of \eqref{AdS4Btx} is the decomposition into irreducible parts.
The whole expression may vanish only  if  there exists an integer $t$ such that 
\bea
\rho^2 =  \tau_{(t,m,n)}\mu\mub~, \qquad  1 \leq t \leq n~.
\label{AdS4B.7}
\eea
In addition,
each $\xi^{\perp}_{(m-k,n-k)}$ in \eqref{AdS4Btx} except for $\xi^{\perp}_{(m-t,n-t)}$ must vanish identically. Hence the decomposition \eqref{AdS4Bty} reduces to
\begin{align}
\xi_{(m-1,n-1)}&=\mc{D}_{(1,1)}^{t-1}\xi^{\perp}_{(m-t,n-t)}~, 
\end{align}
and the gauge transformation \eqref{AdS4BGT} becomes the  well-known one for a partially-massless field with depth $t$,
\begin{align}
\delta_{\xi}\phi_{(m,n)}&=\mc{D}_{(1,1)}^{t}\xi^{\perp}_{(m-t,n-t)}~,
\end{align}
with $\xi^{\perp}_{(m-t,n-t)}$ satisfying \eqref{AdS4Btk}, \eqref{AdS4Bth} 
and \eqref{AdS4B.7}.

The method used above is quite general in that it deduces all types of gauge symmetry compatible with the on-shell conditions \eqref{AdS4B.1}, and at which mass values they appear. 
Below we provide a more direct proof, albeit less general. Specifically, we show  that the system of equations
\eqref{AdS4B.1}, with $\rho^2=\t_{(t,m,n)}\mu\mub$,  is invariant under depth $t$ gauge transformations, $\delta_{\xi}\phi^{(t)}_{(m,n)}=\mc{D}_{(1,1)}^{\phantom{.}t}\xi^{(t)}_{(m-t,n-t)}$, with $\xi^{(t)}_{(m-t,n-t)}$ on-shell \eqref{GPPApp}.

To begin with, it is clear that \eqref{OSC1AppdS4} is gauge invariant only if the gauge parameter is also on-shell with the same pseudo-mass $\rho$
\begin{align}
\big(\mathcal{Q}-\r^2\big)\xi^{(t)}_{(m-t,n-t)}=0~. \label{App4PMG1}
\end{align}
We also need to ensure that the  transverse condition \eqref{OSC2AppdS4} is invariant, $0=\mc{D}_{(-1,-1)}\delta_{\xi}\phi^{(t)}_{(m,n)}$.
Using the identity \eqref{AdS4GenForm4}, one may show that this is equivalent to
\begin{align}
0=\Big[\mc{D}^{\phantom{.}t}_{(1,1)}\mc{D}_{(-1,-1)} -t(m+n-t+1)\big(\mathcal{Q}-\t_{(t,m,n)}\mu\mub\big)\mc{D}^{t-1}_{(1,1)}\Big]\xi^{(t)}_{(m-t,n-t)} ~.
\end{align}
The first term vanishes if $\xi^{(t)}_{(m-t,n-t)}$ is transverse,
$0=\mc{D}_{(-1,-1)}\xi^{(t)}_{(m-t,n-t)}$, 
whilst the second term vanishes if $\r$ in \eqref{App4PMG1} satisfies $\r^2=\t_{(t,m,n)}\mu\mub$.

\section{Maximal depth spin-3: Integration by parts} \label{AppIBPS3}
In general, integrating by parts in conformal space is nontrivial because the conformal covariant derivative carries extra connections that give non-vanishing contributions from total derivatives. However, under special conditions, which in practice are usually met, we may follow the usual procedure and ignore any total derivatives that arise. See section \ref{secIBPCS} for more details. These conditions are not met for the non-minimal action \eqref{CMDspin3NMFunc} and so below we elaborate on how integration by parts works in this case. 

In what follows we drop the `$+$ c.c.' for simplicity. The gauge variation of \eqref{CMDspin3NMFunc} is 
\begin{align}
\delta_{\xi}S^{(3,3,3)}_{\text{NM}}&=\int\text{d}^4x \, e \, \nabla_{(\a_1\ad_1}\nabla_{\a_2\ad_2}\nabla_{\g)\ad_3}\xi W^{\a(2)\b(2)}h_{\b(2)}{}^{\g\ad(3)} \notag\\
&=\frac{1}{3}\int\text{d}^4x \, e \, 
\bigg[3\nabla_{\g\ad_1}\nabla_{\a_1\ad_2}\nabla_{\a_2\ad_3}\xi-2\ve_{\a_1\g}\bar{W}_{\ad(3)}{}^{\gd}\nabla_{\a_2\gd}\xi-4\ve_{\a_1\g}\nabla_{\a_2\gd}\bar{W}_{\ad(3)}{}^{\gd}\xi\bigg]\notag\\
&
\qquad \qquad \qquad 
\times W^{\a(2)\b(2)}h_{\b(2)}{}^{\g\ad(3)}
\notag\\
&= \mathcal{I}_{\text{Total}}+\frac{1}{3}\int\text{d}^4x \, e \, \xi\bigg\{-3\nabla_{\a_1\ad_1}\nabla_{\a_2\ad_2}\nabla_{\g\ad_3}\big[W^{\a(2)\b(2)}h_{\b(2)}{}^{\g\ad(3)}\big]\notag\\
&+2\nabla_{\g\gd}\big[\bar{W}_{\ad(3)}{}^{\gd}W^{\g\b(3)}h_{\b(3)}{}^{\ad(3)}\big]-4\nabla_{\g\gd}\bar{W}_{\ad(3)}{}^{\gd}W^{\g\b(3)}h_{\b(3)}{}^{\ad(3)}\bigg\}~. \label{TransposedSpin3op}
\end{align}
Here $\mathcal{I}_{\text{Total}}$ is the total derivative that arises in moving from the second to third line,
\begin{align}
\mathcal{I}_{\text{Total}}=\frac{1}{6}\int\text{d}^4x \, e \,\nabla^{\a\ad}\mathcal{Z}_{\a\ad}    ~,\label{ItotalSpin3zz}
\end{align}
with 
\begin{align}
\mathcal{Z}_{\a\ad}=&~6W^{\g(2)\d(2)}h_{\d(2)\a\ad\bd(2)}\nabla_{\g}{}^{\bd}\nabla_{\g}{}^{\bd}\xi+4\xi W_{\a}{}^{\g(3)}\bar{W}_{\ad}{}^{\gd(3)}h_{\g(3)\gd(3)}\notag\\
&+6\nabla_{\g}{}^{\bd}\xi\nabla^{\b\bd}\big[W_{\a}{}^{\g(3)}h_{\g(2)\b\bd(2)\ad}\big]-6\xi\nabla_{\g}{}^{\bd}\nabla^{\b\bd}\big[W_{\a}{}^{\g(3)}h_{\g(2)\b\bd(2)\ad}\big]~.
\end{align}
Typically, the action that we begin with is primary (as it is here), which means that all conformal covariant derivatives in the action take the form \eqref{DGCCD4}
\begin{align}
\nabla_{\a\ad}=\mathcal{D}_{\a\ad}-\frac{1}{4}S_{\a\ad,}{}^{\b\bd}K_{\b\bd}
\end{align}
where $\mathcal{D}_{\a\ad}$ is the torsion-free Lorentz covariant derivative and $S_{\a\ad,\b\bd}$ is the Schouten tensor. Since we can always ignore total derivatives from the former, this allows us to rewrite \eqref{ItotalSpin3zz} as 
\begin{align}
\mathcal{I}_{\text{Total}}=-\frac{1}{24}\int\text{d}^4x \, e \,S^{\a\ad,\b\bd}K_{\b\bd}\mathcal{Z}_{\a\ad}~.\label{ItotalSpin3}
\end{align}
The above expression vanishes in most cases because $\mathcal{Z}_{\a\ad}$ turns out to be primary, however this is not true for the current example and one can instead show that \eqref{ItotalSpin3} reduces to
\begin{align}
\mathcal{I}_{\text{Total}}=\int\text{d}^4x \, e \,S^{\a\ad,\b\bd}\nabla_{\a}{}^{\gd}\bigg\{\xi W_{\b}{}^{\g(3)}h_{\g(3)\ad\bd\gd}\bigg\}~. \label{ItotalSpin3b}
\end{align}
By making use of the Bianchi identity \eqref{BianchiIBP}
\begin{align}
\nabla^dW_{abcd}=-2\mc{D}_{[a}S_{b]c}\qquad \Longleftrightarrow \qquad \nabla_{\a}{}^{\bd}\bar{W}_{\ad(3)\bd}=\mathcal{D}_{(\ad_1}{}^{\b}S_{\b\ad_2,\a\ad_3)}~, \label{BianchiIBPspinor}
\end{align}
one can show that \eqref{ItotalSpin3b} is equivalent to
\begin{align}
\mathcal{I}_{\text{Total}}=\int\text{d}^4x \, e \, \xi\bigg\{W^{\a(3)\b}\nabla_{\b\bd}\bar{W}^{\ad(3)\bd}h_{\a(3)\ad(3)}\bigg\}~.
\end{align}

One must be careful to include this term when computing the gauge variation \eqref{CMDspin3NMVar}. This subtlety regarding integration by parts does not occur elsewhere throughout this thesis. It should be emphasised that the IBP rule \eqref{Y.10} has not failed, since its prerequisite conditions were never met.  In particular, after transposition the result (i.e. the second term in \eqref{TransposedSpin3op}) is not primary.  

\section{Conformal non-gauge models} \label{AppCNGM}

In section \ref{secCHSBach} we constructed gauge invariant models for the following fields (i) conformal maximal depth spin-$5/2$; (ii) conformal maximal depth spin-$3$; and (iii) hooked conformal graviton. Common to each of these models was the necessity to couple the parent gauge field to certain auxiliary conformal fields which were dubbed `conformal non-gauge' fields. Therefore, it is of interest to elaborate on the kinetic action for a generic conformal non-gauge field, which is the subject of this appendix. 

\subsection{Conformal non-gauge models (I)}

Let $\chi_{\a(m)\ad(n)}$ be a primary tensor field,  
such that $m \geq n\geq 0$, with the properties
\begin{align}
K_{\b\bd}\chi_{\a(m)\ad(n)}=0~,\qquad \mathbb{D}\chi_{\a(m)\ad(n)}=\big[2-\frac{1}{2}(m-n)\big]\chi_{\a(m)\ad(n)}~. \label{NGTICP}
\end{align}
 Its conjugate $\bar{\chi}_{\a(n)\ad(m)}$ is also primary with the same Weyl weight.
 
  Recall that, in order for a depth-$t$ CHS field $h^{(t)}_{\a(m)\ad(n)}$ to be both primary and defined modulo depth-$t$ gauge transformations, it must possess the conformal weight $\big(t+1-\frac{1}{2}(m+n)\big)$. Since the weight of $\chi_{\a(m)\ad(n)}$ is not equal to this, it cannot be consistently defined modulo such gauge transformations. It is for this reason that we call $\chi_{\a(m)\ad(n)}$ a conformal non-gauge field.  
 
 From $\chi_{\a(m)\ad(n)}$ one can construct the descendent
\begin{align}
\label{nongaugeDescendent}
 \mathfrak{X}_{\a(n)\ad(m)}(\chi)&=\nabla_{(\ad_1}{}^{\b_1}\cdots\nabla_{\ad_{m-n}}{}^{\b_{m-n}}\chi_{\b(m-n)\a(n)\ad_{m-n+1}\dots\ad_m)} \non\\
 &\equiv \big(\nabla_{\ad}{}^{\b}\big)^{m-n}\chi_{\b(m-n)\a(n)\ad(n)}~,
 \end{align}
which is a primary tensor field of weight $\big(2+\frac{1}{2}(m-n)\big)$, 
\begin{align}
K_{\b\bd}\mathfrak{X}_{\a(n)\ad(m)}(\chi)=0~,\qquad \mathbb{D}\mathfrak{X}_{\a(n)\ad(m)}(\chi)=\big[2+\frac{1}{2}(m-n)\big]\mathfrak{X}_{\a(n)\ad(m)}(\chi)~.\label{NGTIDesCP}
\end{align}
To prove this, one can make use of the identity \eqref{CSFundaId1}. The properties \eqref{NGTICP} and \eqref{NGTIDesCP} mean that the functional 
\begin{align}
S^{(m,n)}_{\text{NG}}[\chi,\bar{\chi}]=
\hf\ri^{m+n}\int\text{d}^4x\, e \, \bar{\chi}^{\a(n)\ad(m)}\mathfrak{X}_{\a(n)\ad(m)}(\chi)+\text{c.c.} \label{NGActionArb}
\end{align}
is primary in a generic background. It should be remarked that choosing a different overall coefficient for the first term in \eqref{NGActionArb} 
leads to a total derivative,
\bea
\text{i}^{m+n+1}\int\text{d}^4x\, e \, \bar{\chi}^{\a(n)\ad(m)}\mathfrak{X}_{\a(n)\ad(m)}(\chi)+\text{c.c.}  =0~.
\eea

The action \eqref{NGActionArb} with $(m,n)=(1,0)$ describes  a conformal Weyl spinor, while the $m=n=0$ case corresponds to an auxiliary 
complex scalar that appears, at the component level,  
in the conformal Wess-Zumino model \eqref{7.254|4}.\footnote{More generally, for $m=n$ 
 the action \eqref{NGActionArb} describes an auxiliary field.}
The action \eqref{NGActionArb} with 
$(m,n)=(2,0)$ 
describes a self-dual two-form that emerges in extended conformal supergravity 
theories \cite{deWvHVP,Bergshoeff1,Bergshoeff2,BCdeWS,BCS,vMVP,HS}
 (see \cite{FT} for a review). The conformal non-gauge fields with the values $(m,n)=\{(3,0),(2,1)\}$, $(m,n)=\{(4,0),(3,1)\}$ and $(m,n)=\{(2,0),(4,2)\}$ played pivotal roles in ensuring gauge invariance of the CHS models (i), (ii) and (iii) respectively.


\subsection{Conformal non-gauge models (II)}

For the special case $n=0$ there exists another family of primary functionals. These actions may be classified by two integers $m$ and $t$ that are associated with the non-gauge field $\chi_{\a(m)}^{(t)}$, defined to have the conformal properties
\begin{align}
K_{\b\bd}\chi_{\a(m)}^{(t)}=0~, \qquad \mathbb{D}\chi_{\a(m)}^{(t)}=\Big(2-t-\frac{1}{2}m\Big)\chi_{\a(m)}^{(t)}~.
\end{align}
We refer to conformal non-gauge fields $\chi_{\a(m)}^{(t)}$ with $t=0$ as being type I, whilst those with $t>0$ will be called type II. From $\chi_{\a(m)}^{(t)}$ we can construct the descendent
\begin{align}
 \mathfrak{X}^{(t)}_{\ad(m)}(\chi)=(\Box_c)^{t}\nabla_{(\ad_1}{}^{\a_1}\cdots\nabla_{\ad_{m})}{}^{\a_m}\chi^{(t)}_{\a(m)}~. \label{NGDesc2}
 \end{align}
 Recall that $\Box_c=\nabla^a\nabla_a$ is the conformal D'alembertian.

 Upon restricting the background to be conformally flat, one may prove, via induction on $t$, that the following identity holds
 \begin{align}
 \big[K_{\a\ad},(\Box_c)^t\big]=-4t(\Box_c)^{t-1}\bigg(\nabla_{\a}{}^{\bd}\bar{M}_{\ad\bd}+\nabla_{\ad}{}^{\b}M_{\a\b}-\nabla_{\a\ad}\big(\mathbb{D}+t-2\big)\bigg)~. \label{CSFundaId2}
 \end{align}
  In such backgrounds it may then be shown, using both \eqref{CSFundaId1} and \eqref{CSFundaId2}, that the descendent \eqref{NGDesc2} is primary with Weyl weight given by 
 \begin{align}
 K_{\b\bd} \mathfrak{X}^{(t)}_{\ad(m)}(\chi)=0~,\qquad \mathbb{D} \mathfrak{X}^{(t)}_{\ad(m)}(\chi)=\Big(2+t+\frac{1}{2}m\Big) \mathfrak{X}^{(t)}_{\ad(m)}(\chi)~. \label{NGDesc2Prop}
 \end{align}
 It follows that the functional 
 \begin{align}
S^{(m,0,t)}_{\text{NG}}[\chi,\bar{\chi}]=\frac{~\ri^{m}}{2}\int\text{d}^4x\, e \, \bar{\chi}_{(t)}^{\ad(m)}\mathfrak{X}^{(t)}_{\ad(m)}(\chi)+\text{c.c.} \label{NGtype2Act}
\end{align}
is primary in all conformally flat backgrounds. When $n=t=0$, the two models \eqref{NGActionArb} and \eqref{NGtype2Act} coincide. 

In a generic background the descendent \eqref{NGDesc2} is not primary. Naively, one might expect that it is possible to rectify this by including non-minimal corrections. Indeed, when $t=1$, the first four  primary extensions to \eqref{NGDesc2} are given by 
\begin{subequations}   
\begin{align}
\mathfrak{X}^{(1)}(\chi)
&=\Box_c\chi^{(1)}~,\\
\mathfrak{X}^{(1)}_{\ad}(\chi)
&=\Box_c\nabla_{\ad}{}^{\a}\chi_{\a}^{(1)}~, \label{7.14b}\\
\mathfrak{X}^{(1)}_{\ad(2)}(\chi)&=\Box_c\nabla_{(\ad_1}{}^{\a}\nabla_{\ad_2)}{}^{\a}\chi_{\a(2)}^{(1)}+W_{\a(2)}{}^{\b(2)}\nabla_{(\ad_1}{}^{\a}\nabla_{\ad_2)}{}^{\a}\chi_{\b(2)}^{(1)} \notag\\
&\phantom{=}+4\nabla_{(\ad_1}{}^{\a}W_{\a(2)}{}^{\b(2)}\nabla_{\ad_2)}{}^{\a}\chi_{\b(2)}^{(1)}~,\\
\mathfrak{X}^{(1)}_{\ad(3)}(\chi)&=\Box_c\nabla_{(\ad_1}{}^{\a}\nabla_{\ad_2}{}^{\a}\nabla_{\ad_3)}{}^{\a}\chi_{\a(3)}^{(1)}+3W_{\a(2)}{}^{\b(2)}\nabla_{(\ad_1}{}^{\a}\nabla_{\ad_2}{}^{\a}\nabla_{\ad_3)}{}^{\g}\chi^{(1)}_{\b(2)\g}\notag\\
&\phantom{=}+13\nabla_{(\ad_1}{}^{\a}W_{\a(2)}{}^{\b(2)}\nabla_{\ad_2}{}^{\a}\nabla_{\ad_3)}{}^{\g}\chi^{(1)}_{\b(2)\g}+2\nabla_{(\ad_1}{}^{\g}W_{\a(2)}{}^{\b(2)}\nabla_{\ad_2}{}^{\a}\nabla_{\ad_3)}{}^{\a}\chi^{(1)}_{\b(2)\g}\notag\\
&\phantom{=}+7\nabla_{(\ad_1}{}^{\g}\nabla_{\ad_2}{}^{\a}W_{\a(2)}{}^{\b(2)}\nabla_{\ad_3)}{}^{\a}\chi^{(1)}_{\b(2)\g}~,
\end{align}
\end{subequations}
which correspond to the cases $m=0,1,2,3$ respectively. 
Similar completions for $t=1$ are expected to exist for any $m>3$.
The reason being that in appendix \ref{AppNGSM} we construct a family  
of supersymmetric non-gauge models which, at the component level, 
contain the non-supersymmetric non-gauge models of this subappendix with $t=0$ and $t=1$. The supersymmetric model is primary in a generic  supergravity background which means that the $t=1$ family must also exist in a generic background. 

 However, when $t>1$, there are some values of $m$ for which no primary extension of  $\mathfrak{X}^{(t)}_{\ad(m)}(\chi)$ exists. 
 This is true, in particular,  for the following cases:
 (i) $(t,0)$ with $t>2$; and (ii) $(t,1)$ with $t>1$.
 These non-existence results were derived in the mathematical literature, 
see \cite{GJMS,Gover1,Gover2,Dirac2, Dirac3} and references therein.
 
 In the scalar case, $m=0$, it is known that  
 $\mathfrak{X}^{(2)}(\chi) 
 =(\Box_c)^{2}\chi^{(2)} $ is primary on a generic background. 
 Upon degauging $(\Box_c)^{2}\chi^{(2)} \equiv  \D_0 \c$   we obtain
 \bea
 \D_0 \c
 = \Big\{ \Box^2
  -\mc{D}^a \big(2 R_{ab} \mc{D}^b 
  -\frac 23 R \mc{D}_a \big)\Big\} 
  \c~.
  \eea
This operator was  discovered by Fradkin and Tseytlin
in 1981 \cite{FT1982a,FT1982b} (see \cite{FT} for a review)
and re-discovered 
by Paneitz in 1983 \cite{Paneitz} and Riegert in 1984 \cite{Riegert}.

In the spinor case, $m=1$, it is known that 
$ \mathfrak{X}^{(1)}_{\ad}(\chi)  =\Box_c\nabla_{\ad}{}^{\a}\chi_{\a}^{(1)} $
 is primary in a generic background. 
 Upon degauging 
 $\Box_c\nabla_{\ad}{}^{\a}\chi_{\a}^{(1)} \equiv -(\D_{\hf})_{\a\ad} \c^\a  $
  we obtain
 \bea
 (\D_{\hf})_{\a \ad} \c^{\a}
 =\Big\{ \Box\mc{D}_{\a\ad}{}
 -\frac{1}{6}\mc{D}_{\a\ad} R
 -\frac{1}{12}R\mc{D}_{\a\ad}+\hf R_{\a\b\bd\ad}\mc{D}^{\b\bd}\Big\}\chi^{\a}
 ~.
  \eea
  This operator was  introduced  by Fradkin and Tseytlin
in 1981 \cite{FT1982a,FT1982b} (see \cite{FT} for a review).

The operators $\D_0$ and $\D_\hf$ are contained in the supersymmetric 
Fradkin-Tseytlin 
 operator \cite{BdeWKL} defined by 
 \bea
 {\bm \D} \bar  \f = -\frac{1}{64}  \bar{\Nabla}^2 {\Nabla}^2  \bar{\Nabla}^2 \bar \f ~, \qquad
  {\Nabla}_\a \bar \f =0 ~,
  \eea
 where $\f $ is a primary dimension-0 chiral superfield.

\section{Comments on hook field models} \label{AppHook}

The conformal pseudo-graviton fields $h_{\a(3)\ad}$ and $\bar h_{\a \ad(3)}$ are 
 in a one-to-one correspondence with a real tensor field $h_{abc}$ satisfying the algebraic constraints
\begin{subequations} \label{R.1}
\bea
h_{abc}&=&-h_{bac}~, 
\qquad h_{abc}+h_{bca}+h_{cab}=0~; \\
h_{ab}{}^{b}&=&0~. \label{C.1bbb}
\eea
\end{subequations}
In the literature, one often refers to $h_{abc}$ 
as a hook field since its algebraic symmetries are represented by the 
Young diagram {\scalebox{0.4}{
\begin{ytableau}
	~ & ~  \\
	~ \\
\end{ytableau} 
}}. If the trace condition \eqref{C.1bbb} is omitted, one speaks of a  traceful hook field; below it is denoted by $\hat{h}_{abc}$.

Two-derivative models for 
$\hat{h}_{abc}$ were first studied in \cite{Curtright1, Curtright2} in $d$-dimensional Minkowski space.\footnote{These models were later lifted to AdS$_d$ \cite{BMVpm} where it was shown that either of the gauge symmetries in \eqref{R.2} may be upheld separately, but not simultaneously.} In the massless case, such a theory  
is
invariant under two types of gauge transformations,
\begin{align}
\delta_{\xi}\hat{h}_{abc}=\partial_{c}\xi_{ab}-\partial_{[a}\xi_{b]c}~,\qquad \delta_{\theta}\hat{h}_{abc}=2\partial_{[a}\theta_{b]c}~,
 \label{R.2}
\end{align}
 generated 
 by antisymmetric ($\xi_{ab}=-\xi_{ba}$) and symmetric ($\theta_{ab}=\theta_{ba}$) 
 gauge parameters. 
 
 Four-derivative theories for $\hat{h}_{abc}$ were studied previously in AdS$_d$ \cite{Karapet2}. In addition to the gauge transformations \eqref{R.2}, these models 
 also prove to be invariant under the 
algebraic local trace-shift  
 transformation $\delta_{\alpha}\hat{h}_{abc}=\eta_{c[a}\alpha_{b]}$. 
 This third symmetry may be used to gauge away the trace of $\hat{h}_{abc}$, resulting with 
 the traceless hook field $h_{abc}$ \eqref{R.1}. 
 In terms of 
 $h_{abc}$, 
 the action proposed in \cite{Karapet2}
 may be 
 rewritten for $d=4$
 in two-component spinor notation.
 The resulting action in Minkowski space proves to be proportional to 
\begin{align}
S_{\text{Hook}}=\int\text{d}^4x \,
\mc{W}^{\a(4)}(h)\mc{W}_{\a(4)}(\bar{h})
- \int\text{d}^4x \,
\mc{W}^{\a(4)}(h)\Box\mc{W}_{\a(4)}(h)
+\text{c.c.} \label{R.3}
\end{align} 
where $\mc{W}_{\a(4)}(h)=\partial_{(\a_1}{}^{\bd}h_{\a_2\a_3\a_4)\bd}$ and $\mc{W}_{\a(4)}(\bar{h})=\partial_{(\a_1}{}^{\bd_1}\partial_{\a_2}{}^{\bd_2}\partial_{\a_3}{}^{\bd_3}\bar{h}_{\a_4)\bd(3)}$ are the linearised Weyl tensors of the conformal hook (see e.g. \eqref{HSWeylIMink4} with $(m,n)=(3,1)$). 
These field strengths,\footnote{The field strengths $\mc{W}_{\a(4)}(h)$ and $\mc{W}_{\a(4)}(\bar{h})$ differ from those used in \cite{Karapet2}.} and hence the action \eqref{R.3}, are invariant under the gauge transformations
\begin{subequations} \label{R.4}
\bea
\delta_{\xi}h_{\a(3)\ad}&=&\partial_{(\a_1\ad}\xi_{\a_2\a_3)}~, \label{R.55a}
\eea
which is a special case of \eqref{CHSGTMink4} for $(m,n)=(3,1)$. 
In addition, action \eqref{R.3} is also invariant under the second gauge symmetry 
\bea
\delta_{\theta}h_{\a(3)\ad}=\partial_{(\a_1}{}^{\bd}\theta_{\a_2\a_3)\ad\bd}~. \label{R.66b}
\eea
\end{subequations}
However, the second term in the action \eqref{R.3} is not conformal (that is, it is not 
invariant under the special conformal transformations
of $\mb{M}^4$, see e.g. \cite{Vasiliev2009}). There is no way to lift it,  and hence the gauge symmetry \eqref{R.66b}, 
 to curved space in a Weyl invariant way.\footnote{It should  also be mentioned that the gauge symmetry \eqref{R.66b} is not compatible with the conformal 
properties of the gauge field $h_{\a(3)\ad}$ which are dictated by \eqref{R.55a}.}
It is the first term in \eqref{R.3} which admits a Weyl invariant extension to curved backgrounds.

Finally, we would like to point out that the action \eqref{R.3} admits a straightforward generalisation to a certain class of higher-rank fields. Specifically, associated with the field $h_{\a(m+n)\ad(m-n)}$ for $1\leq n \leq m-1$, is the action 
\begin{align}
S[h,\bar{h}]=\int\text{d}^4x \,
\bigg\{
\mc{W}^{\a(2m)}(h)\mc{W}_{\a(2m)}(\bar{h})
-\mc{W}^{\a(2m)}(h)\Box^m\mc{W}_{\a(2m)}(h)
\bigg\}
+\text{c.c.}~, \label{R.5}
\end{align}
which is invariant under the gauge transformations
\vspace{-0.5ex}
\begin{subequations} \label{R.6}
\begin{align}
\delta_{\xi}h_{\a(m+n)\ad(m-n)}&=\partial_{(\a_1(\ad_1}\xi_{\a_2\dots\a_{m+n})\ad_2\dots\ad_{m-n})}~,\label{R.6a}\\
 \delta_{\theta}h_{\a(m+n)\ad(m-n)}&=\partial_{(\a_1}{}^{\bd_1}\dots\partial_{\a_n}{}^{\bd_n}\theta_{\a_{n+1}\dots\a_{m+n})\ad(m-n)\bd(n)}~, \label{R.6b}
\end{align}
\end{subequations}
for a real gauge parameter $\theta_{\a(m)\ad(m)}=\bar{\theta}_{\a(m)\ad(m)}$. The field strengths in \eqref{R.5} are given in \eqref{HSWeylIMink4}. In vector notation the fields $h_{\a(m+n)\ad(m-n)}$ and $\bar{h}_{\a(m-n)\ad(m+n)}$ correspond to the  non-rectangular and traceless Young diagram \scalebox{0.55}{
\begin{tikzpicture}
\draw (0,0) -- (2,0) -- (2,0.5) -- (0,0.5) -- cycle;
\draw (0,0) -- (1,0) -- (1,-0.5) -- (0,-0.5) -- cycle;
\filldraw[black] (1,0.25) circle (0pt) node {$m$};
\filldraw[black] (0.5,-0.25) circle (0pt) node {$n$};
\end{tikzpicture}
}. In the rectangular case, when $m=n=s$, there is no longer any $\xi$-type gauge symmetry, however the corresponding action \eqref{R.5} with $\mc{W}_{\a(2s)}(h)=h_{\a(2s)}$ and $\mc{W}_{\a(2s)}(\bar{h})=\partial_{(\a_1}{}^{\bd_1}\cdots\partial_{\a_{2s})}{}^{\bd_{2s}}\bar{h}_{\bd(2s)}$ is invariant under the transformations \eqref{R.6b}. 

Note that the gauge freedom \eqref{R.6} defines a reducible gauge theory, for if we take
\begin{subequations}
\begin{align}
\hat{\xi}_{\a(m+n-1)\ad(m-n-1)}&:=\partial_{(\a_1}{}^{\bd_1}\cdots\partial_{\a_n}{}^{\bd_n}V_{\a_{n+1}\dots\a_{n+m-1})\ad(m-n-1)\bd(n)}~,\\
\hat{\theta}_{\a(m)\ad(m)}&:=\frac{m}{n-m}\partial_{(\a_1(\ad_1}V_{\a_2\dots\a_m)\ad_2\dots\ad_m)}~,
\end{align} 
\end{subequations}
with $V_{\a(m-1)\ad(m-1)}$ real, then a combined $\hat{\theta}$ and $\hat{\ell}$ transformation vanishes,
\begin{align}
(\delta_{\hat{\ell}}+\delta_{\hat{\theta}})h_{\a(m+n)\ad(m-n)}=0~.
\end{align}

\end{subappendices}


\chapter{SCHS models in three dimensional $\mc{N}=1$ superspace} \label{Chapter3Dsuperspace}

In this chapter we elaborate on models describing the dynamics of superconformal higher-spin (SCHS) gauge multiplets on various curved $3d$ superspace backgrounds. These models are higher-spin generalisations of the linearised action for conformal supergravity. We will be mainly concerned with theories possessing $\mc{N}=1$ superconformal symmetry, but also provide results for the $\mc{N}$-extended case.  
In off-shell $3d$ $\cN$-extended conformal supergravity, with $1\leq \cN \leq 6$, 
the superfield Euler-Lagrange equation states that the super-Cotton tensor vanishes
\cite{BKNT-M1,BKNT-M2,KNT-M13}.
This constraint characterises a conformally-flat superspace background, which constitutes the most general type of background on which an SCHS superfield can consistently  propagate.  
Indeed,  for $1 \leq \cN \leq 6$, we show that a gauge-invariant action exists for every SCHS superfield on arbitrary conformally-flat backgrounds, and give explicit expressions for $\mc{N}=1,2$. In each instance, the linearised action is completely determined by the relevant higher-spin super-Cotton tensor. 

This chapter is based on the publications \cite{Topological, Confgeo, CottonAdS} and is organised as follows.  
 In section \ref{secSCHSM3} we give a brief overview of the geometry of 3$d$ $\mc{N}=1$ conformal supergravity, followed by a discussion on the generic features of  
  SCHS models on arbitrary superspace backgrounds. 
In section \ref{secSCHSMink3} we review the models for SCHS superfields in $\mc{N}=1$ Minkowski superspace and use them to construct topologically massive supersymmetric gauge models. Via a component analysis, these models are shown to be related to the non-supersymmetric ones derived in chapter \ref{Chapter3D}. 
Using the framework of conformal superspace, we construct $\mc{N}$-extended SCHS models on arbitrary conformally flat superspace backgrounds in section \ref{secSCHSCF3}. 
 In section \ref{secSCHSAdS3} we derive expressions for the higher-spin super-Cotton tensors in $\mc{N}=1$ anti-de Sitter superspace AdS$^{3|2}$. 
This allows us to (i) demonstrate that the SCHS kinetic operator factorises into wave operators associated with partially-massless superfields; and (ii) construct topologically massive models on AdS$^{3|2}$.
A summary of the results obtained is given in section \ref{secSCHS3dis}.

\section{Superconformal higher-spin models in $\mc{M}^{3|2}$} \label{secSCHSM3}
Conformal higher-spin gauge superfields in $3d$ $\cN=1$ Minkowski superspace 
were introduced in \cite{K16,KT17}, as a by-product of the $\cN=2$ approach of \cite{KO}.
In this section we start by  
generalising this concept to the case of $\cN=1$ supergravity, 
building on the ideas advocated in \cite{KMT}. This section is based on our paper \cite{Topological}.

\subsection{Conformal supergravity} \label{Sec3dN=1CSGGeom}

Consider a curved $\cN=1$ 
superspace, $\cM^{3|2}$, parametrised by local real coordinates
$z^{M}=(x^m,\q^{\mu})$, with $m=0,1,2$ and $\mu=1,2$,
of which $x^m$ are bosonic and $\q^{\mu}$ fermionic. 
We introduce a basis of one-forms
$\bm E^A=(\bm E^a,\bm E^\a)$ and its dual basis $\bm e_A=(\bm e_a,\bm e_\a)$, 
\bea
\bm E^A=\rd z^M \bm e_{M}{}^A~,
\qquad \bm e_A = \bm e_A{}^M  \pa_M ~,
\label{beins}
\eea
which will be referred to as the supervielbein and its inverse, respectively.
The superspace structure group is $\sSL(2,{\mathbb R})$, 
the double cover of the connected Lorentz group $\sSO_0(2,1)$. 
The covariant derivatives have the form:
\bea
\bm \cD_{A}&=& ( \bm\cD_a, \bm\cD_\a )= \bm e_{A}-\bm\o_A~,
\label{23cd}
\eea
where $\bm \o_A$ is the Lorentz super connection,
\bea
\bm\o_A=\hf\bm\o_{A}{}^{bc}M_{bc}=-\bm\o_{A}{}^b M_b=\hf\bm\o_{A}{}^{\b\g}M_{\b\g}~.
\label{2.444234}
\eea
We use boldface characters to denote geometric objects defined on superspace, so that they may be distinguished from their corresponding analogues defined on spacetime.

The covariant derivatives are characterised by the graded commutation relations 
\bea
{[}\bm \cD_{{A}},\bm\cD_{{B}}\}&=&
- \bm T_{ {A}{B} }{}^{{C}}\bm\cD_{{C}}
-\hf \bm R_{{A} {B}}{}^{{cd}}M_{{cd}}~,
\label{algebra-0}
\eea
where $\bm T_{ {A}{B} }{}^{{C}}$ and $\bm R_{{A} {B}}{}^{{cd}}$ are
the torsion and curvature tensors, respectively. 
To describe supergravity, the covariant derivatives 
have to obey certain torsion constraints \cite{GGRS} such that 
the algebra \eqref{algebra-0} takes the form \cite{KLT-M11} 
\bsubeq \label{N=1algCS}
\bea
\{\bm\cD_\a,\bm\cD_\b\}&=&
2\ri\bm\cD_{\a\b}
-4\ri \cS M_{\a\b}
~,~~~~~~~~~
\label{N=1alg-1}
\\
\left[ \bm\cD_{\a \b} , \bm\cD_\g \right] &=& 
- 2 \ve_{\g(\a} \cS \bm\cD_{\b)} + 2 \ve_{\g(\a} \cS_{\b) \d \r} M^{\d\r}
+ \frac{2}{3} \big( \bm\cD_\g \cS M_{\a\b} 
- 4 \bm\cD_{(\a} \cS M_{\b) \g} \big) 
~,~~~~~~~~~
\label{N=1alg-3/2-2}
\\
{[}\bm\cD_{a},\bm\cD_b{]}
&=&
-\frac{1}{2}\ve_{abc}(\g^c)^{\a\b}\Big{\{}
 \ri\cS_{\a\b\g}\bm\cD^\g
+\frac{4\ri}{3}\bm\cD_{\a}\cS\bm\cD_\b
-
\ri \bm\cD_{(\a}\cS_{\b\g\d)}M^{\g\d}\non\\
&&
\phantom{-\frac{1}{2}\ve_{abc}(\g^c)^{\a\b}\Big{\{}}
+\Big(
\frac{2\ri}{3}\bm\cD^2\cS
+4\cS^2\Big)M_{\a\b}
\Big{\}}
~.
~~~~~~~~~~~~
\label{N=1alg-2}
\eea
\esubeq
Here the scalar superfield $\cS$ and the rank three symmetric spinor superfield $\cS_{\a\b\g}=\cS_{(\a\b\g)}$ are real.
The dimension-2 Bianchi identities imply that 
\bea
\bm\cD_{\a}\cS_{\b\g\d}&=&
\bm\cD_{(\a}\cS_{\b\g\d)}
-\ri \ve_{\a(\b}\bm\cD_{\g\d)}\cS \quad \Longrightarrow \quad 
\bm\cD^\g \cS_{\a\b\g} = -\frac{4\ri }{3} \bm\cD_{\a\b} \cS
~.
\eea
We  use the notation $\bm\cD^2 := \bm\cD^\a \bm\cD_\a$.

The algebra of covariant derivatives is invariant under 
the following super-Weyl transformations \cite{ZP88,ZP89,LR-brane}
\bsubeq \label{SusyDerivsWeyl3}
\bea
\d^{(\text{weyl})}_\S\bm\cD_\a&=&
\hf \S\bm\cD_\a + \bm\cD^{\b}\S M_{\a\b} \label{WeylSpinorD}
~,
\\
\d^{(\text{weyl})}_\S \bm\cD_{\a\b}&=&
\S\bm\cD_{\a\b}
-\ri\bm\cD_{(\a} \S\bm{\mc{D}}_{\b)}
+\bm\cD_{(\a}{}^{\g}\S M_{\b)\g}
~,
\eea
\esubeq
with the parameter $\S$ being a real unconstrained superfield.\footnote{We will only use the boldface notation for superfields when there is room for confusion, e.g., when doing component analysis. } The corresponding transformations of the torsion superfields are
\bea
\d^{(\text{weyl})}_\S\cS&=&\S\cS-\frac{\ri}{4}  \bm \cD^2\S~,~~~~~~
\d^{(\text{weyl})}_\S \cS_{\a\b\g}=\frac{3}{2}\S \cS_{\a\b\g}-\frac{1}{2}  \bm\cD_{(\a\b}\bm\cD_{\g)}\S
~.
\label{sW}
\eea

The $\mc{N}=1$ conformal supergravity gauge group $\bm{\mc{G}}$ is generated by the following local transformations (i) supertranslations with gauge parameter $\xi^A(z)=\big(\xi^a(z), \xi^{\a}(z)\big)$; (ii) Lorentz rotations with gauge parameter $K^{ab}(z)=-K^{ba}(z)$; and (iii) super-Weyl transformations with gauge parameter $\S(z)$. Under $\bm{\mc{G}}$ the covariant derivatives of $\mc{M}^{3|2}$ transform according to the rule (cf. \eqref{PoinDer}):
\begin{align}
\delta_{\L}^{(\scriptsize\bm{\mc{G}})}\bm{\mc{D}}_A=\big[ \L, \bm{\mc{D}}_{A} \big]+\d^{(\text{weyl})}_\S \bm{\mc{D}}_A~,\qquad \L:= \xi^A\bm{\mc{D}}_{A}+\frac{1}{2}K^{ab}M_{ab}~.
\end{align}
A tensor superfield $\Phi(z)$ (with its Lorentz indices suppressed) is said to primary with superconformal weight $\Delta_{\Phi}$ if its transformation law under $\bm{\mc{G}}$ is 
\begin{align}
\delta_{\L}^{(\bm{\scriptstyle{\cG}})}\Phi=\Big(\xi^A\bm{\mc{D}}_{A}+\frac{1}{2}K^{ab}M_{ab} +\Delta_{\Phi}\S \Big)\Phi~.
\end{align}

 A real supervector field $\z= \z^B  \bm e_B$  is called conformal Killing if 
\begin{subequations} \label{CCSVF}
\bea
\delta_{\L[\z]}^{(\scriptsize\bm{\mc{G}})}\bm\cD_A=\Big[ \z^B \bm\cD_B + \hf K^{\b \g} [\z] M_{\b\g} , \bm\cD_A\Big]  + \delta^{(\text{weyl})}_{\S [\z]}  \bm\cD_{A} = 0 ~,
 \label{CCSVF.a}
\eea
for some Lorentz ($K^{\b \g}[\z]$) and super-Weyl ($\S[ \z] $) parameters. 
One can show that this equation implies $\z^{\b \g} $ is the only independent transformation parameter, 
\bea
\z^\b = \frac{\ri}{6} \bm\cD_\g \z^{\b \g} ~,\qquad K_{\b\g} [\z] = 2 \bm\cD_{(\b} \z_{\g) }-2\mc{S}\z_{\b\g} ~,
\qquad \S[\z] = \bm\cD_\b \z^\b = \frac 13 \bm\cD_b \z^b~,
 \label{CCSVF.b}
\eea
and it obeys the superconformal Killing equation
\bea
\bm\cD_{(\a} \z_{\b\g ) } =0 \quad \implies \quad \bm\cD_{(\a\b} \z_{\g \d) } =0 
~~ \Longleftrightarrow ~~  \bm\cD^a \z^b + \bm\cD^b \z^a = 2 \eta^{ab} \S[\z] ~.
 \label{CCSVF.c}
\eea
\end{subequations}
The conformal Killing supervector fields of $\cM^{3|2}$ span the conformal superalgebra of the curved superspace.\footnote{In the case of a conformally-flat superspace, see eq. \eqref{CFSM3}, this algebra is isomorphic to  $\mathfrak{osp}(1|4, {\mathbb R})$.} 
If we consider $\cM^{3|2}$ as a fixed background superspace, then the primary superfield $\F$ possesses the following rigid superconformal transformation law
\bea
\delta_{\L[\z]}^{(\bm{\scriptstyle{\cG}})} \F = \Big(\z^B \bm\cD_B +\hf K^{\b\g}[\z] M_{\b\g} + \Delta_{\F} \S[\z]\Big) \F ~.
\label{RigidSCT}
\eea

The $\cN=1$ supersymmetric extension of the 
 Cotton tensor \eqref{CottonRicci} was constructed in  \cite{KT-M12}. 
 It is given by the expression
 \bea
\bm C_{\a\b\g} = \left(\frac{\ri }{2}\bm \cD^2 +4\cS\right) \cS_{\a\b \g} 
+   \bm\cD_{(\a\b} \bm\cD_{\g)} \cS 
\label{super-Cotton}
~.
\eea
The super-Weyl transformation of $\bm C_{\a\b\g}$ proves to be
\bea
\d^{(\text{weyl})}_\S \bm C_{\a\b\g} = \frac{5}{2} \S \bm C_{\a\b\g}~.
\eea
Furthermore, $\bm C_{\a\b\g}$ obeys the conformally invariant Bianchi identity 
\begin{align}
\bm\cD^{\a}\bm C_{\a\b\g}=0~.
\end{align}
It can be shown \cite{BKNT-M1} that the curved superspace is conformally flat if and 
only if 
\begin{align}
\bm C_{\a\b\g}=0~. \label{CFSM3}
\end{align}
The superconformal curvature $\bm C_{\a\b\g}$ arises as the first functional derivative of the three dimensional conformal supergravity action with respect to the corresponding unconstrained conformal gauge prepotential. 


\subsection{SCHS prepotentials and their field strengths} \label{sec3DN=1PrepsNStrengths}

A real tensor superfield $H_{\a(n) }(z) $, for integer $n\geq 1$,  is said to be a superspin $\frac{n}{2}$ conformal gauge supermultiplet if: 
\begin{enumerate}[label=(\roman*)]

\item $H_{\a(n) } $  is a primary superfield with superconformal weight $\Delta_{H_{(n)}}=(1-{n}/{2})$, 
\bea
\d^{(\text{weyl})}_\S H_{\a(n)} = \Big( 1-\frac{n}{2}\Big) \S H_{\a(n)}~.
\label{SCHS3prepWeyl}
\eea
\item $H_{\a(n) } $  is defined modulo gauge transformations of the form\footnote{We point out that in $3d$ $\mc{N}=1$ superspace, the following complex conjugation rule for expressions involving $\bm\cD_{\a}$ holds: $\big(\bm\cD_{\a}\Phi \big)^*=(-1)^{\ve_{\F}+1}\bm\cD_{\a}\bar{\F}$, where  $\ve_{\F}$ is the Grassmann parity of $\Phi$ and $(\F)^*=\bar{\F}$. }
\bea
\d_\L H_{\a(n) } =\ri^n \bm\cD_{(\a_1} \L_{\a_2 \dots \a_n) }~,\label{SCHS3prepGT}
\eea
where the real gauge parameter $\L_{\a(n-1)}$ is a weight $\frac{1}{2}(1-n)$ primary superfield (but is otherwise unconstrained).
\end{enumerate}

The super-Weyl weight $\Delta_{H_{(n)}} $  is uniquely fixed by requiring $\L_{\a(n-1)}$ and $\d_\L H_{\a(n) }$ to be super-Weyl primary and using \eqref{WeylSpinorD}. For a fixed background superspace, $H_{\a(n)}$ possesses the rigid superconformal transformation law \eqref{RigidSCT}.

Starting with $H_{\a(n)}$ one can construct its descendant,
${\mathfrak C}_{\a(n)}(H)$, known as the linearised higher-spin super-Cotton tensor. It is defined by the following properties: 
\begin{enumerate}[label=(\roman*)]

\item
${\mathfrak C}_{\a(n)}(H)$ is a primary superfield with superconformal weight $\Delta_{\mf{C}_{(n)}}=(1+n/2)$, 
\bea
\d^{(\text{weyl})}_\S {\mathfrak C}_{\a(n)}(H) = \Big(1+\frac{n}{2} \Big)  \S {\mathfrak C}_{\a(n)}(H)~.\label{SCHScotWeyl}
\eea

\item
${\mathfrak C}_{\a(n)}(H)$ is of the form ${\mathfrak C}_{\a(n)}(H)=\cA H_{\a(n)}$, where $\cA$ is a linear differential operator of order $n$ (in vector derivatives), which involves $\bm\cD_A$,
the torsion tensors $\cS_{\a\b\g}$ and $\cS$, and their covariant derivatives.

\item
${\mathfrak C}_{\a(n)}(H)$ has vanishing gauge variation under \eqref{SCHS3prepGT} 
if $\mc{M}^{3|2}$ is conformally flat,
\bea
\d_\L {\mathfrak C}_{\a(n)}(H) = \mc{O}\big( \bm C \big)~.
\label{SCHScotGTM}
\eea

\item
 ${\mathfrak C}_{\a(n)}(H)$ is transverse if $\mc{M}^{3|2}$ is conformally flat,
\bea
\bm\cD^{\b} {\mathfrak C}_{\b \a(n-1)}(H) = \mc{O}\big( \bm C\big)~.
\label{SCHScotDivM}
\eea 
Here and in \eqref{SCHScotGTM}, $\mc{O}\big( \bm C\big)$ stands for contributions containing the super-Cotton tensor \eqref{super-Cotton} and its covariant derivatives.
\end{enumerate}

As a simple example, we consider a U(1) vector multiplet coupled to supergravity,
which corresponds to the $n=1$ case. 
This multiplet is described by a real spinor prepotential 
$H_\a $ which is super-Weyl primary of weight $1/2$ and is defined 
modulo gauge transformations $\d_\L  H_\a = \ri \bm\cD_\a \L$, 
where the gauge parameter $\L$ is an unconstrained real superfield. 
The required super-Weyl primary descendant of weight $3/2$ 
is given by 
\bea
{\mathfrak C}_\a(H) =-\frac{\ri}{2} \bm\cD^\b \bm\cD_\a H_\b
-2\cS H_\a 
\eea
which proves to be gauge invariant, 
\bea
\d_\L  {\mathfrak C}_\a(H) =0~.
\eea
The field strength obeys the Bianchi identity 
\bea
\bm\cD^\a {\mathfrak C}_\a(H) =0~.
\eea

For $n>1$ the right-hand sides of \eqref{SCHScotGTM} and \eqref{SCHScotDivM} are non-vanishing. In addition, the above properties do not determine $\mf{C}_{\a(n)}(H)$ uniquely in non-conformally-flat backgrounds. 

\subsection{Superconformal higher-spin action}

Suppose that our background curved superspace $\cM^{3|2}$
is conformally flat, 
\bea
\bm C_{\a(3)}=0~.\label{SMCF31}
\eea
Then the tensor superfield ${\mathfrak C}_{\a(n)}(H)$ 
is gauge invariant and covariantly conserved, 
\begin{subequations}
\bea
\d_\L {\mathfrak C}_{\a(n)}(H)&=&0~,\\ 
\bm\cD^{\b} {\mathfrak C}_{\b\a(n-1)} (H)&=&0~. 
\eea
\end{subequations}
These properties and the super-Weyl transformation laws
\eqref{SCHS3prepWeyl} and  \eqref{SCHScotWeyl} imply
that the following action\footnote{The super-Weyl transformation of the superspace 
integration measure is $\d^{(\text{weyl})}_\S E = -2\S E$.} 
\bea
{\mathbb S}_{\rm{SCHS}}^{(n)} [ H] 
= - \frac{\ri^n}{2^{\left \lfloor{n/2}\right \rfloor +1}}
   \int \rd^{3|2}z \, E\,
 H^{\a(n)} 
{\mathfrak C}_{\a(n) }( H) ~, \qquad E^{-1} = {\rm Ber} (\bm e_A{}^M)~,
\label{SCHSActM3}
\eea
is gauge and super-Weyl invariant, 
\bea
\d_\L {\mathbb S}_{\rm{SCHS}}^{(n)} [H] =0~, \qquad 
\d^{(\text{weyl})}_\S  {\mathbb S}_{\rm{SCHS}}^{(n)} [ H ] =0~.
\eea
 This is the superconformal higher-spin action in three dimensional $\mc{N}=1$ superspace. In \eqref{SCHSActM3} we have used the shorthand notation $\rd^{3|2}z=\rd^3x\rd^2\theta$.
The condition \eqref{SMCF31} is required to guarantee the gauge invariance 
of ${\mathbb S}_{\rm{SCHS}}^{(n)} [ H]$ for $n\geq 2$. 
The above action is actually super-Weyl invariant in an arbitrary curved superspace. 
By virtue of the invariance of \eqref{SCHSActM3} under the conformal supergravity gauge group $\bm{\mc{G}}$, it follows that ${\mathbb S}_{\rm{SCHS}}^{(n)}$ is invariant under the rigid superconformal transformations \eqref{RigidSCT}, 
 \begin{align}
 \delta_{\L[\z]}^{(\bm{\scriptstyle{\cG}})} {\mathbb S}_{\rm{SCHS}}^{(n)} [ H] =0~,
 \end{align}
when the background is held fixed.


\section{Superconformal higher-spin models in $\mb{M}^{3|2}$} \label{secSCHSMink3}

We now review the linearised higher-spin 
super-Cotton tensors ${\mathfrak C}_{\a(n)}(H)$
on $\mc{N}=1$ Minkowski superspace $\mb{M}^{3|2}$ and their associated superconformal actions, as derived in \cite{K16}. These actions will be used to construct topologically massive higher-spin supersymmetric gauge models in $\mb{M}^{3|2}$. In section \ref{secSUSYMassiveMOds} we analyse the component structure of each of these models. This section is based on our paper \cite{Topological}.

We note that the spinor covariant derivative $D_{\a}=\frac{\pa}{\pa\theta^{\a}}+\ri\theta^{\b}\pa_{\b\a}$ of ${\mathbb M}^{3|2}$ satisfies the anti-commutation relation
\begin{align}
\big\{D_{\a},D_{\b}\big\}=2\ri\pa_{\a\b}~.
\end{align}
As a result one may show that the following identities hold
\begin{subequations}\label{3DN=1FlatId}
\begin{align}
D_{\a}D_{\b}&=\ri \pa_{\a\b}+\frac{1}{2}\ve_{\a\b}D^2~,\label{3DN=1FlatIda}\\
D&^\a D_\b D_\a =0 ~,\label{3DN=1FlatIdb}\\
[D_\a &D_\b, D_\g D_\d ]=0~,\label{3DN=1FlatIdc}\\
D^2D_{\a}=&-D_{\a}D^2=2\ri\pa_{\a\b}D^{\b}~,\\
D^2&D^2=-4\Box
\end{align}
\end{subequations}
where $D^2=D^{\a}D_{\a}$ and $\Box=\pa^a\pa_a=-\frac{1}{2}\pa^{\a\b}\pa_{\a\b}$.

\subsection{Higher-spin super-Cotton tensor in Minkowski superspace} \label{sec3dN=1SCHSMink}

In Minkowski superspace, the higher-spin 
super-Cotton tensor \cite{K16,KT17} is 
\bea
\mf{C}_{\a_1\dots\a_n}(H) = 
\Big( -\frac{\ri}{2}\Big)^n
D^{\b_1} D_{\a_1} \dots D^{\b_n} D_{\a_n} H_{\b(n)}
= \mf{C}_{(\a_1 \dots \a_n )}(H)
~.
\label{HSscotMink3a}
\eea
Using the identities \eqref{3DN=1FlatId} it is not difficult to show that $\mf{C}_{\a(n)}(H)$ is invariant
\begin{align}
 \delta_{\L}\mf{C}_{\a(n)}(H)=0 ~, \label{HSscotMink3aGI}
\end{align}
 under the gauge transformations 
\bea
\d_{\L} H_{\a(n) } &=& \ri^n D_{(\a_1 } \L_{\a_2 \dots \a_n)}~,\label{SCHSprepGTMink3}
\eea
and obeys the conservation identity 
\bea
{D}^{\b} \mf{C}_{\b\a(n-1)}(H)=0 ~.\label{HSscotMink3aTT}
\eea
The fact that $\mf{C}_{\a(n)}(H)$ 
defined by \eqref{HSscotMink3a} is completely symmetric, 
is a corollary of \eqref{3DN=1FlatIdc}.

The normalisation chosen in \eqref{HSscotMink3a} can be explained as follows.
The gauge freedom \eqref{SCHSprepGTMink3} allows us to impose the gauge condition
\bea
D^\b H_{\b \a(n-1)} =0~,
 \eea
 under which
 the expression for the super-Cotton tensor 
 simplifies,
\begin{subequations} 
\bea
 D^\b H_{\b \a(n)}=0 \quad \Longrightarrow \quad 
 \mf{C}_{\a(n)}(H) = \pa_{\a_1}{}^{\b_1} \dots \pa_{\a_n}{}^{\b_n} H_{\b(n)}~.
\eea
This result can be fine-tuned to
\bea
\mf{C}_{\a(2s)}(H) &=&\Box^s H_{\a(2s)}~, \\
\mf{C}_{\a(2s+1)}(H) &=&\Box^s \pa^\b{}_{(\a_1} H_{\a_2 \dots \a_{2s+1})\b }
=\Box^s \pa^\b{}_{\a_1 } H_{\a_2 \dots \a_{2s+1} \b }~,
\eea
\end{subequations}
where $s>0$ is an integer.

For completeness, we also give another representation 
for the higher-spin super-Cotton tensor  derived in \cite{K16,KT17}:
\bea
&&\mf{C}_{\a(n)} (H) 
:= \frac{1}{2^{n}} 
\sum\limits_{j=0}^{\left \lfloor{n/2}\right \rfloor}
\bigg\{
\binom{n}{2j}  \Box^{j}\pa_{(\a_{1}}{}^{\b_{1}}
\dots
\pa_{\a_{n-2j}}{}^{\b_{n-2j}}H_{\a_{n-2j+1}\dots\a_{n})\b_1 \dots\b_{n-2j}}~~~~
\nonumber \\
&&\qquad \qquad -\frac{\ri}{2} 
\binom{n}{2j+1}D^{2}\Box^{j}\pa_{(\a_{1}}{}^{\b_{1}}
\dots\pa_{\a_{n-2j -1}}{}^{\b_{n-2j -1}}H_{\a_{n-2j}\dots\a_{n})
\b_1 \dots \b_{n-2j -1} }\bigg\}~.~~~~~
\label{HSscotMink3b}
\eea
An alternative expression for $\mf{C}_{\a(n)} (H) $ in terms of the $\mc{N}=1$ transverse superprojectors was derived in \cite{BHHK}. In \cite{BHHK} the authors also derived explicit expressions for
 the $\mc{N}$-extended higher-spin super-Cotton tensors with $2< \mc{N}\leq 6$ on $\mb{M}^{3|2\mc{N}}$. The $\mc{N}=2$ higher-spin super-Cotton tensors were found in \cite{KO}. 

The following higher-spin action \cite{K16,KT17}
\bea
{\mathbb S}^{(n)}_{\rm SCHS}[H] = 
- \frac{\ri^n}{2^{\left \lfloor{n/2}\right \rfloor +1}}
 \int \rd^{3|2}z \,H^{\a(n)} \mf{C}_{\a(n)}(H)
\label{SCHSActMink3}
\eea
is $\cN=1$ superconformal. It is  clearly invariant under the gauge transformations 
\eqref{SCHSprepGTMink3}. The content of this model at the component level is analysed in section \ref{SCHScomponentsMink3}. 


\subsection{Massive superfields} \label{secMassiveSuperfields}

Before using the superconformal higher-spin actions \eqref{SCHSActMink3} to generate models for massive superfields, we first recall the superfield realisations of the massive $\mc{N}=1$ super-Poincar\'e representations.

The quadratic Casimirs of the $\mc{N}=1$ super-Poincar\'e algebra are the momentum squared $\mc{P}^a\mc{P}_a$ and the superhelicity operator $\mb{Z}$. 
Given a unitary irreducible representation of the super-Poincar\'e group, the mass $m$ and superhelicity $\kappa$ are defined by
\begin{align}
\mc{P}^a\mc{P}_a=-m^2\mathds{1}~,\qquad \mb{Z}= m \kappa \mathds{1}~.
\end{align}
We recall that their corresponding superfield realisations are $\pa^a\pa_a=\Box$ and
\begin{align}
\mb{Z}:=\mb{W}-\frac{\ri }{8}D^2~,\qquad \mb{W}:=\frac{1}{2}\pa^{\a\b}M_{\a\b}\label{SuperhelicityOp}
\end{align}
where $\mb{W}$ is the Pauli-Lubankski pseudo-scalar \eqref{PLscalar}.

For $n>0$, a massive superfield $\F_{\a(n)}$
is defined to be a real symmetric rank-$n$ spinor,
 $\F_{\a(n)} 
= \bar \F_{\a(n)}$,
which obeys the differential conditions \cite{KNT-M15} (see also \cite{KT17})
\begin{subequations}
\label{214}
\bea
D^\b \F_{\b \a(n-1)} &=& 0 \quad \Longrightarrow \quad
\pa^{\b\g} \F_{\b\g\a(n-2)} =0
~ , 
\label{214a} 
\\
-\frac{\ri}{2} D^2  \F_{\a(n)} &=& m \s \F_{\a(n)}~, 
\qquad \s =\pm 1~.
\label{214b}
\eea
\end{subequations}
It follows from \eqref{214a} and \eqref{3DN=1FlatIda} that 
\bea
-\frac{\ri}{2} D^2  \F_{\a(n)} =\pa^\b{}_{(\a_1} \F_{\a_2 \dots \a_n)\b} ~,
\eea
and thus $\F_{\a(n)}$ is an on-shell superfield (cf. \eqref{1.1.2}), 
\bea
\pa^\b{}_{(\a_1} \F_{\a_2 \dots \a_n)\b} = m \s\F_{\a(n)}~,
\qquad \s =\pm 1~. \label{76.7.6.4.2.}
\eea
It follows from \eqref{214b} that\footnote{Equations 
\eqref{214a} and \eqref{2177} 
are the $\cN=1$ supersymmetric extension
of the Fierz-Pauli equations.} 
\bea
(\Box -m^2) \F_{\a(n)}=0~.
\label{2177}
\eea
Using equations \eqref{SuperhelicityOp}, \eqref{214b} and \eqref{76.7.6.4.2.}, for the superhelicity of $\F_{\a(n)}$  we obtain 
\bea
\mb{Z}\F_{\a(n)}=m\k \F_{\a(n)}~,\qquad \k = \hf\left( n +\hf \right) \s~.
\label{218}
\eea
We define the superspin of $\F_{\a(n)}$ to be $n/2$. 

The massive supermultiplet $\F_{\a(n)}$ contains two ordinary 
massive fields of the type \eqref{1.1.12}, each with mass $m$. They are defined according to
\bea
\f_{\a(n)} := \F_{\a(n)} \big|_{\q=0}~, \qquad 
\f_{\a(n+1)} := \ri^{n+1} D_{(\a_1} \F_{\a_2 \dots \a_{n+1})} \big|_{\q=0}~,
\eea
and carry helicity values of $\frac{n}{2} \s $ and $\frac{n+1}{2}  \s$, respectively. See section \ref{secSUSYMassiveMOds} below for an explanation of the bar notation.


\subsection{Topologically massive supersymmetric higher-spin models}\label{secTMSHSMmink}

Topologically massive $\mc{N}=1$ supersymmetric higher-spin actions in $\mb{M}^{3|2}$ are higher-spin extensions of linearised topologically massive $\mc{N}=1$ supergravity \cite{DK, Deser}. Similar to the non-supersymmetric models \eqref{HSTMGFlat}, they are obtained by coupling the SCHS action \eqref{SCHSActMink3} (which plays the role of a Chern-Simons topological term) to a massless  sector. The type of massless sector required differs depending on the value of superspin. In particular, in the case of integer $(n=2s)$ and half-integer $(n=2s+1)$ superspin, they take the respective forms
\begin{subequations} \label{SHSTMGFlat}
\bea
{\mathbb S}_{\rm TM}^{(2s)}[H,Y]
&=&{\mathbb S}_{\rm{SCHS}}^{(2s)}[H]
-m^{2s-1}{\mathbb S}_{\rm{FO}}^{(2s)} [H, Y]~,
\label{SHSTMGFlata}
\\
{\mathbb S}_{\rm TM}^{(2s+1)}[H,X]
&=&{\mathbb S}_{\rm{SCHS}}^{(2s+1)}[H]
-m^{2s-1}{\mathbb S}_{\rm{SO}}^{(2s+1)}
[H,X]~,
\label{SHSTMGFlatb}
\eea
\end{subequations}
 where $m$ is a real parameter of mass dimension one. We have denoted the first and second order massless sectors by ${\mathbb S}_{\rm{FO}}^{(2s)}$ and ${\mathbb S}_{\rm{SO}}^{(2s+1)}$ respectively, which we describe below.
 
 On-shell, the model \eqref{SHSTMGFlata} describes a massive superfield with mass $m$, superspin $s$ and superhelicity $\left( s +\frac{1}{4} \right) \s$ where $\s=m/|m|$. Similarly, the model \eqref{SHSTMGFlatb} describes a massive superfield with mass $m$, superspin $(s+\frac{1}{2})$ and superhelicity $\left( s + \frac{3}{4}\right) \s$ where $\s=m/|m|$. They were first proposed in \cite{KT17}, where these statements were proved by analysing the resulting equations of motion at the superspace level. In section \ref{secSUSYMassiveMOds} below we provide an alternative proof by analysing their component content and relating it to the non-supersymmetric topologically massive models derived in section \ref{secMassiveActFlat}. In section \ref{secSCHSAdS3} we will also lift these models to anti-de Sitter superspace.

\subsubsection{First-order massless actions}

We introduce the reducible real gauge superfield 
${\bm \cH}_{\b , \a_1 \dots \a_{n-1}} ={\bm \cH}_{\b , (\a_1 \dots \a_{n-1})}={\bm \cH}_{\b , \a(n-1)}$ defined modulo 
gauge transformations with real gauge parameter 
\bea
\d_{\L} {\bm \cH}_{\b , \a(n-1)} = \ri^n D_\b \L_{\a(n-1)}~.
\label{SMFOredPrepGT}
\eea
A supersymmetric gauge-invariant action of lowest order in derivatives is 
\bea
{\mathbb S}^{(n)}_{\rm FO} [\bm\cH]= \frac{\ri^{n+1}}{2^{\lceil (n+1)/2\rceil}}
\int \rd^{3|2} z \, 
{\bm \cH}^{\b , \a(n-1)} 
 D^\g D_\b 
{\bm  \cH}_{\g , \a(n-1)}
   ~.
\label{SMFOredAct}
\eea
The gauge invariance of ${\mathbb S}^{(n)}_{\rm FO}$ 
follows from the identity \eqref{3DN=1FlatIdb}.
Our action \eqref{SMFOredAct} is a  higher-spin extension of the model for the massless
gravitino multiplet ($n=2$) in Minkowski superspace proposed by Siegel \cite{Siegel}
(see also \cite{GGRS}).

The reducible superfield ${\bm \cH}_{\b , \a(n-1)}$ can be decomposed 
 into irreducible superfields:
\begin{align}
{\bm \cH}_{\b , \a_1\dots\a_{n-1}} &= {\bm \cH}_{(\b, \a_1\dots\a_{n-1}) }
+\frac{n-1}{n}\ve_{\b(\a_1}{\bm \cH}^{\g,}{}_{\a_2\dots\a_{n-1})\g} \non\\
&\equiv H_{\b\a(n-1)} +(n-1)\ve_{\b(\a_1}Y_{\a_2\dots\a_{n-1})}~.
\end{align}
Here we have defined the completely symmetric and real superfields  $H_{\a(n)}:={\bm \cH}_{(\a_1,\a_2\dots\a_n)}$ and $Y_{\a(n-2)}:=\frac{1}{n}{\bm \cH}^{\g,}{}_{\g\a(n-2)}$.
Then the gauge transformation \eqref{SMFOredPrepGT} turns into
\begin{subequations}\label{GTIrredFOMasslessMink3}
\bea
\d_{\L} H_{\a (n)} &=& \ri^n D_{(\a_1 } \L_{\a_2 \dots \a_n)} ~, \\
\d_{\L} Y_{\a (n-2) } &=& \frac{\ri^n}{n} D^\b \L_{\b \a(n-2)}~.
\eea
\end{subequations}
The supersymmetric gauge-invariant action \eqref{SMFOredAct} takes the form 
\begin{align}
{\mathbb S}^{(n)}_{\rm FO}&[H,Y] = 
\frac{\ri^{n+1}}{2^{\lceil (n+1)/2\rceil}}
\int \rd^{3|2} z \,\Big\{  
H^{\b  \a (n-1)} D^\g D_\b H_{\g   \a (n-1)}  
 +2\ri (n-1) Y^{\a (n-2)} \pa^{\b\g} H_{ \b \g \a (n-2) } \non \\
&
+ (n-1) \Big( Y^{\a (n-2) }D^2 Y_{\a (n-2)} 
+(-1)^n 
(n-2)D_\b Y^{\b \a (n-3) }
D^\g Y_{\g \a (n-3)} \Big) \Big\} ~.
\label{SMFOActMink3}
\end{align}
When $n$ is even, $n=2s$, this action is the massless integer superspin action of \cite{KT17}.


\subsubsection{Second-order massless actions}

The second-order massless half-integer superspin action in $\mb{M}^{3|2}$ is given by
\begin{align}
{\mathbb S}_{\rm{SO}}^{(2s+1)}&[H,X]
= \Big(- \hf \Big)^s   \int \rd^{3|2} z \,\bigg\{ 
-\frac{\ri}{2} H^{\a(2s+1)} \Box H_{\a(2s+1)}
-\frac{\ri}{8} D_\b H^{\b \a(2s)} D^2 D^{\g}  H_{\g \a(2s)} \non \\
&+ \frac{\ri}{4} s \pa_{\b\g} H^{\b\g \a(2s-1)} 
\pa^{\r\l} H_{\r\l \a(2s-1)}
-\hf (2s-1) X^{\a(2s-2)} \pa^{\b\g} D^{\d} H_{\b\g \d \a(2s-2)} \non \\
&+\frac{\ri}{2}  (2s-1)\Big[ X^{\a(2s-2)} D^2 X_{\a(2s-2)}
- \frac{s-1}{s} D_\b X^{\b\a(2s-3)} D^\g X_{\g \a(2s-3)}\Big]\bigg\}
~, \label{SMSOActMink3}
\end{align}
where $X_{\a(2s-2)}$ is a real and totally-symmetric superfield. This action is invariant under the gauge transformations
\begin{subequations} \label{SusySOMink3GF}
\bea
\d_{\L} H_{\a(2s+1)} &=& \ri D_{(\a_1} \L_{\a_2 \dots \a_{2s+1} )} ~,\\
\d_{\L} X_{\a(2s-2)} &=& \frac{s}{2s+1} \pa^{\b\g} \L_{\b\g \a(2s-2) }~.
\eea
\end{subequations}  
The action \eqref{SMSOActMink3} is the massless half-integer superspin action proposed in \cite{KT17}.

\subsubsection{New topologically massive supersymmetric higher-spin models}

Off-shell $\mc{N}=1$ supersymmetric extensions of the new topologically massive higher-spin gauge models \eqref{HSNTMG} may now be easily obtained. For any integer $n\geq 1$, the corresponding gauge-invariant action may be formulated in terms of the unconstrained prepotential $H_{\a(n)}$ and takes the form 
\begin{align}
\mb{S}_{\text{NTM}}^{(n)}[H]=- \frac{\ri^n}{2^{\left \lfloor{n/2}\right \rfloor +1}}\int \rd^{3|2}z \, H^{\a(n)} \Big(\frac{\ri}{2}D^2+ m\Big)\mf{C}_{\a(n)}(H)~. \label{SHSNTMGmink}
\end{align}
where $m \in \mb{R}$ is a real parameter with units of mass. The equation of motion obtained by varying \eqref{SHSNTMGmink} with respect to $H_{\a(n)}$ is 
\begin{align}
0= \Big(\frac{\ri}{2}D^2+ m\Big)\mf{C}_{\a(n)}(H)~. \label{SEOM1mink}
\end{align}
For generic $m$, this equation in conjunction with the off-shell conservation identity \eqref{HSscotMink3aTT}
means that the field-strength $\mf{C}_{\a(n)}(H)$ itself describes a massive superfield with mass $|m|$, superspin $n/2$ and superhelicity $\frac{1}{2}\big(n+\frac{1}{2}\big)\s$, where $\s=m/|m|$. This is in agreement with the discussion in section \ref{secMassiveSuperfields}. This action has not appeared elsewhere in the literature.

\subsection{Supersymmetric higher-spin actions in components} \label{secSUSYMassiveMOds}

In this section we will describe the component structure of the supersymmetric 
higher-spin theories introduced in the previous section.
As in \cite{KT17}, the integration measure\footnote{This 
definition implies that $ \int \rd^{3|2} z \, V = \int \rd^3 x\, F$, for any scalar 
superfield $V(x,\q)  =\dots + \ri \q^2 F(x)$.} for $\cN=1$ Minkowski superspace
is defined as follows:
\bea
\int \rd^{3|2} z \, L= \frac{\ri}{4} \int \rd^3 x \, D^2 L\big|~.
\label{3DReductionformula}
\eea
As usual, the $\theta$-independent component of a tensor superfield ${L} ={L}(x,\q)$ (with suppressed indices) is defined by ${L}|:= { L}(x,\q )\big|_{\q =0}$. The operation $|$ is called bar-projection, and it applies to the everything to its left, e.g., $ D^2 L\big|= \big(D^2 L\big)\big|$.

\subsubsection{Superconformal higher-spin action} \label{SCHScomponentsMink3}

We start by reducing the superconformal higher-spin action 
\eqref{SCHSActMink3} to components. 
The gauge freedom \eqref{SCHSprepGTMink3} can be used to impose a Wess-Zumino gauge
\bea
H_{\a(n)} |=0~,\qquad
D^\b H_{\b \a(n-1)} | =0~.
\label{WZgaugeSCHSMink3}
\eea
In this gauge, there remains two independent component fields
\bea
h_{\a(n+1)}:= \ri^{n+1} D_{(\a_1} H_{\a_2 \dots \a_{n+1} )} \big|~, \qquad
h_{\a(n)}  := -\frac{\ri}{4} D^2 H_{\a(n)}\big|~.
\eea
In the gauge \eqref{WZgaugeSCHSMink3}, the residual gauge freedom is characterised by the conditions
\bea
D_{(\a_1 } \L_{\a_2 \dots \a_n )}|=0~, \qquad 
D^2 \L_{\a(n-1)} |= -2\ri \frac{n-1}{n+1} \pa^\b{}_{(\a_1} 
\L_{\a_2 \dots \a_{n-1} )\b} |~.
\eea
At the component level, the remaining independent gauge transformations 
are 
\begin{subequations}
\begin{align}
\delta_{\xi}h_{\a(n)}&=\pa_{(\a_1\a_2}\xi_{\a_3\dots\a_n)}~,\\
\delta_{\xi}h_{\a(n+1)}&=\pa_{(\a_1\a_2}\xi_{\a_3\dots\a_{n+1})}~,
\end{align}
\end{subequations}
where we have defined $\xi_{\a(n-1)} := (-1)^{n+1}\L_{\a(n-1)}\big|$ and 
$\xi_{\a(n-2)} := \frac{n-1}{2n}\ri^n D^\b \L_{\b\a(n-2)}\big|$.

Due to the conservation equation \eqref{HSscotMink3aTT}, the higher-spin super-Cotton tensor 
\eqref{HSscotMink3a} also has two independent components, which we define as
\bea
\mc{C}_{\a(n)} (h) := \mf{C}_{\a(n)}(H)\big|~,\qquad
\mc{C}_{\a(n+1)}(h):= \ri^{n+1} D_{(\a_1} \mf{C}_{\a_2 \dots \a_{n+1} )} (H)\big|~.
\eea
The field strengths $\mc{C}_{\a(n)} (h)$ and $\mc{C}_{\a(n+1)}(h)$ are given in terms of 
the gauge potentials $h_{\a(n)}$ and $h_{\a(n+1)}$, respectively. They coincide with the corresponding non-supersymmetric higher-spin Cotton tensors given in eq. \eqref{2.31}. 
To prove this statement, one should use the form \eqref{HSscotMink3b} of $\mf{C}_{\a(n)}(H)$ and the identity 
\bea
\binom{n}{2j} +\binom{n}{2j+1} =\binom{n+1}{2j+1} ~.
\eea
Reducing the action \eqref{SCHSActMink3} to components using \eqref{3DReductionformula} gives 
\bea
{\mathbb S}^{(n)}_{\rm SCHS}[H] =
S^{(n)}_{\rm CHS}[h]  + S^{(n+1)}_{\rm CHS}[h] ~, 
\label{SCHScomponentsMink3Red}
\eea
where the gauge invariant conformal higher-spin action $S^{(n)}_{\rm CHS}[h]$ 
is defined by eq. \eqref{CHS3ActCot}.


\subsubsection{Massless first-order model}

We now turn to 
working out the component structure of the first-order model  \eqref{SMFOActMink3}.
The gauge freedom  \eqref{GTIrredFOMasslessMink3} allows us to choose a Wess-Zumino gauge
\bea
H_{\a(n)} |=0~,\qquad
D^\b H_{\b \a(n-1)} | =0~, \qquad
Y_{\a(n-2)} |= 0~.
\eea
Then, the residual gauge freedom is characterised by the conditions
\begin{align}
D_{(\a_1 } \L_{\a_2 \dots \a_n )}|&=0~, \qquad D^{\b}\L_{\b\a(n-2)}|=0~,\non\\
D^2 \L_{\a_1 \dots \a_{n-1}} |&= -2\ri \frac{n-1}{n+1} \pa^\b{}_{(\a_1} 
\L_{\a_2 \dots \a_{n-1} )\b} |~.
\end{align}
These conditions imply that there remains only one independent gauge parameter 
at the component level. We define it as 
\bea
\xi_{\a(n-1)} := (-1)^{n+1} \L_{\a(n-1)}|~.
\eea
We define the component fields as 
\begin{subequations}
\bea
h_{\a(n+1)}&:= &\ri^{n+1} D_{(\a_1} H_{\a_2 \dots \a_{n+1} )} |~, \\
h_{\a(n)} & :=& -\frac{\ri}{4} D^2 H_{\a(n)}|~, \\
y_{\a(n-1)}&:= & \frac{\ri^{n+1}}{2n} D_{(\a_1} Y_{\a_2 \dots \a_{n-1} )} |~,\\
z_{\a(n-3)}&:=& \ri^{n+1} D^\b Y_{\b\a(n-3) } |~,\\
f_{\a(n-2)} & :=& \frac{\ri}{4} D^2 Y_{\a(n-2)}|~.
\eea
\end{subequations}
Their gauge transformation laws are
\begin{subequations}
\bea
\d_{\x} h_{\a(n+1) }&=& \pa_{(\a_1 \a_2} \xi_{\a_3 \dots \a_{n+1})} ~,\\
\d_{\x} y_{\a(n-1)} &=&\frac{1}{n+1}  \pa^\b{}_{(\a_1 } \xi_{\a_2 \dots \a_{n-1} )\b} ~,\\
\d_{\x} z_{\a(n-3) }&=& \pa^{\b \g} \xi_{\b\g \a(n-3)} ~, \\
\d_{\x} h_{\a(n)}&=&0~, \\
\d_{\x} f_{\a(n-2) } &=&0~.
\eea
\end{subequations}
Direct calculations of the component action give
\bea
{\mathbb S}^{(n)}_{\rm FO}
&&=  
\frac{\text{i}^n}{2^{\lceil n/2\rceil}}
\int \rd^3 x \, \Big\{ h^{\a(n)} h_{\a(n)}
+f^{\a(n-2)} f_{\a(n-2)} \Big\} \non \\
&&+
\frac{\ri^{n+1}}{2^{\lceil (n+1)/2\rceil}}
\int \rd^3 x \, \Big\{ h^{\b \a(n)} \pa_\b{}^\g h_{\g \a(n)}
+2(n-1) y^{\a(n-1)} \pa^{\b\g} h_{\b \g \a(n-1)}  \non \\
&&  + 4(n-1) y^{\b \a(n-2)} \pa_\b{}^\g y_{\g \a(n-2)}
+ \frac{2(n-2)(n+1)}{n}z^{\a(n-3)} \pa^{\b\g} y_{\b \g \a(n-3)} \non \\
&&- \frac{(n-2)(n-3)}{n(n-1)} z^{\b \a(n-4)} \pa_\b{}^\g z_{ \g \a(n-4)}
\Big\}~.
\eea
The fields $h_{\a(n)}$ and $z_{\a(n-2)}$ appear in the action without derivatives.
This action can  be rewritten in the form 
\bea
{\mathbb S}^{(n)}_{\rm FO}
&=&  
\frac{\text{i}^n}{2^{\lceil n/2\rceil}}
\int \rd^3 x \, \Big\{ h^{\a (n)} h_{\a (n)}
+f^{\a (n-2)} f_{\a (n-2)} \Big\} 
+ S_{\rm{FF}}^{(n+1)}
[{h}_{(n+1)},{y}_{(n-1)},
{ z}_{(n-3)}]~,~~~
\label{FOcomponentsMink3Red}
\eea
where $S_{\rm{FF}}^{(n+1)}$ is the Fang-Fronsdal first order model \eqref{MasslessFOFlat},
with $n$ replaced by $(n+1)$.


\subsubsection{Massive integer superspin action}

We are now prepared to read off the component form of the massive 
integer superspin action ${\mathbb S}_{\rm TM}^{(2s)}$ \eqref{SHSTMGFlata}.
Choosing $n=2s$ in the component actions \eqref{SCHScomponentsMink3Red} and \eqref{FOcomponentsMink3Red} 
gives
\begin{align}
{\mathbb S}_{\rm TM}^{(2s)}[H,Y] &=
 S^{(2s)}_{\rm CHS}[h_{(2s)}]  
-\hf \Big( -\hf\Big)^s m^{2s-1}
\int \rd^3 x \,  h^{\a (2s)} h_{\a (2s)} \non \\
&+ S^{(2s+1)}_{\rm CHS}[h_{(2s+1)}] - 
m^{2s-1} S_{\rm{FF}}^{(2s+1)}
[{h}_{(2s+1)},{y}_{(2s-1)},{ z}_{(2s-3)}] \non \\
&-\hf \Big( -\hf \Big)^s 
m^{2s-1} \int \rd^3 x \, f^{\a (2s-2)} f_{\a (2s-2)} ~.
\label{TMcomponentIS}
\end{align}
We see that the $f_{\a(2s-2)}$ field appears only in the third line of \eqref{TMcomponentIS}
and without derivatives, and thus  $f_{\a(2s-2)}$ is an auxiliary field.
Next, the expression in the second line of \eqref{TMcomponentIS} constitutes the topologically massive 
gauge-invariant spin-$(s+\hf)$ action \eqref{HSTMGFlatF}. The two terms in the first 
line of \eqref{TMcomponentIS} involve the $h_{\a(2s)}$ field.
Unlike $ S^{(2s)}_{\rm CHS}[h_{(2s)}]$, the second mass-like term is not gauge invariant.
However, the action 
\bea
S_{\rm massive}^{(2s)} [h_{(2s)}]
=S^{(2s)}_{\rm CHS}[h_{(2s)}]  
-\hf \Big( -\hf\Big)^s m^{2s-1}
\int \rd^3 x \,  h^{\a (2s)} h_{\a (2s)} 
\label{WeirdMassiveAction}
\eea
does describe a massive  spin-$s$ field. Therefore, the action \eqref{TMcomponentIS} is equivalent to
\begin{align}
{\mathbb S}_{\rm TM}^{(2s)}[H,Y] &=S_{\rm TM}^{(2s)}[{h}_{(2s+1)},{y}_{(2s-1)},{ z}_{(2s-3)}]+S_{\rm massive}^{(2s)} [h_{(2s)}]\non\\
&+\Big( -\hf \Big)^{s+1} m^{2s-1} \int \rd^3 x \, f^{\a (2s-2)} f_{\a (2s-2)}
\end{align}

Let us show that the action \eqref{WeirdMassiveAction} does indeed describe a massive spin-$s$ field on-shell. The corresponding equation of motion is 
\bea
\mc{C}_{\a(2s)}(h) - m^{2s-1}  h_{\a (2s)}=0~.
\label{6.20}
\eea
Since $\mc{C}_{\a(2s)}(h)$ is divergenceless, eq. \eqref{CotDefPropFlat2}, the above equation of motion implies
that $h_{\a(2s)} $ is divergenceless, $\pa^{\b(2)}h_{\b(2)\a(2s-2)}=0$.
As a consequence,  $\mc{C}_{\a(2s)}(h)$ takes the simple form given by 
\eqref{3.41a}, and the equation of motion turns into 
(cf. eq. \eqref{3.40} and \eqref{3.44})
\bea
 \Box^{s-1}\partial^{\beta}{}_{\a_1}h_{\a_2\dots\a_{2s}\beta} 
- m^{2s-1}  h_{\a (2s)}=0~,
\label{6.21}
\eea
which implies 
\bea
\Big(\Box^{2s-1}-(m^2)^{2s-1}\Big) h_{\alpha(2s)}=0~.
\label{6.22}
\eea
Since the polynomial equation $z^{2s-1}-1=0$ has only one real root, 
$z=1$, we conclude that \eqref{6.22} leads to the Klein-Gordon 
equation  \eqref{KGBaby}. As a result, the higher-derivative equation \eqref{6.21}
reduces to the first-order one, eq.  \eqref{3.47}.

The action \eqref{WeirdMassiveAction} is not gauge invariant, but can be turned into a gauge-invariant one by 
making use of the St\"uckelberg trick.
An interesting feature of the model  \eqref{WeirdMassiveAction} is that it is well-defined
on an arbitrary conformally flat space.
We note that \eqref{WeirdMassiveAction} is a higher-spin analogue
of the well-known model for a massive vector field
(see \cite{TPvN,DJ} and references therein)
with Lagrangian
\bea
\cL_{\rm SD} &=& \hf f^a f_a -\frac{1}{2m} \ve^{abc} f_a \pa_b f_c~.
\eea

The above component analysis clearly demonstrates that the model \eqref{SHSTMGFlata}
describes a single massive supermultiplet subject to the equations
\eqref{214a} and \eqref{214b}  with $n=2s$ on the mass shell.
The superfield proof was provided in \cite{KT17}.


\subsubsection{Massless second-order model}

Finally, we consider the massless half-integer superspin model described by \eqref{SMSOActMink3}.
The gauge freedom  \eqref{SusySOMink3GF} allows us to choose a Wess-Zumino gauge of the form 
\bea
H_{\a(2s+1)} \big|=0~, \qquad
D^\b H_{\b \a(2s)}\big|=0~.
\label{7.3}
\eea
To preserve these conditions, the residual gauge symmetry has to be constrained by 
\bea
 D_{(\a_1} \L_{\a_2 \dots \a_{2s+1} )} \big|=0~,\qquad
 D^2 \L_{\a(2s)}\big| = -\frac{2\ri s}{s+1} \pa^\b{}_{(\a_1} \L_{\a_2 \dots \a_{2s})\b}\big|~.
 \eea
Under the gauge conditions imposed, 
the independent component fields of $H_{\a(2s+1)} $ can be chosen as 
\bea
h_{\a(2s+2)} := -D_{(\a_1} H_{\a_2 \dots \a_{2s+2})} \big|~, \qquad
h_{\a(2s+1)}:= \frac{\ri}{4} D^2 H_{\a(2s+1)} \big|~.
\eea
The remaining independent component parameters of $\L_{\a(2s)}$ can be chosen as 
\bea
\x_{\a(2s)}:= \L_{\a(2s)}\big|~, \qquad
\x_{\a(2s-1)}:= -\ri \frac{s}{2s+1} D^\b \L_{\b \a(2s-1)}\big|~.
\eea
The gauge transformation laws of $h_{\a(2s+2)}$ and $h_{\a(2s+1)}$
can be shown to be 
\begin{subequations}
\bea
\d_\x h_{\a(2s+2)} &=& \pa_{(\a_1 \a_2} \x_{\a_3 \dots \a_{2s+2} )}~, \label{7.7a} \\
\d_\x h_{\a(2s+1)} &=& \pa_{(\a_1 \a_2} \x_{\a_3 \dots \a_{2s+1} )}~. \label{7.7b} 
\eea 
\end{subequations}
We now define the component fields of $X_{\a(2s-2)}$ as follows:
\begin{subequations}
\bea
y_{\a(2s-2)}  &:=& 2X_{\a(2s-2)} \big|~ , \\
y_{\a(2s-1)}  &:=& -\frac{\ri}{2}  D_{(\a_1 } X_{\a_2 \dots \a_{2s-1} )}\big|~, \\
z_{\a(2s-3)}  &:=& -\ri  D^\b X_{\b \a (2s-3 )}\big| ~, \\
F_{\a(2s-2)}  &:=& \frac{\ri}{4}D^2 X_{\a(2s-2)}\big|~.
\eea 
\end{subequations}
The gauge transformation laws of each are as follows:
\begin{subequations}
\bea
\d_\x y_{\a (2s-2) } &=& \frac{2s}{2s+1} \pa^{\b\g } \x_{\b \g \a (2s-2) }~,  \label{7.9a}  \\
\d_\x y_{\a(2s-1)} &=&\frac{1}{2s+1}  \pa^\b{}_{(\a_1} \x_{\a_2 \dots \a_{2s-1}) \b}~,
 \label{7.9b}  \\
\d_\x z_{\a(2s-3)} &=& \pa^{\b \g} \x_{\b\g  \a (2s-3)}~,  \label{7.9c} \\
\d_\x F_{\a(2s-2)} &=& \frac{s(s-1)}{2(s+1)(2s+1)}\pa^{\b(2)}\pa^{\b}{}_{(\a_1}\x_{\a_2\dots\a_{2s-2})\b(3)}~.
\eea 
\end{subequations}
In principle, we need not derive the gauge transformation of $F_{\a(2s-2)}$ 
since this field turns out to be auxiliary.

The bosonic transformation laws \eqref{7.7a} and \eqref{7.9a} correspond to the 
massless spin-$(s+1)$ Fronsdal action $S_{\rm{F}}^{(2s+2)}$ 
defined by eq. \eqref{MasslessSOFlat}. The fermionic  transformation laws \eqref{7.7b},
 \eqref{7.9b}  and \eqref{7.9c} correspond to the 
massless spin-$(s+\hf)$ Fang-Fronsdal action $S_{\rm{FF}}^{(2s+1)}$ 
defined by eq. \eqref{MasslessFOFlat}.

The component action follows from \eqref{SMSOActMink3} by making use of
the reduction rule \eqref{3DReductionformula}.
Direct calculations lead to the following bosonic Lagrangian:
\bea
2(-2)^{s+1} L_{\rm bos} &=& 
h^{\a(2s+2)}\Box h_{\a(2s+2)}
-\hf (s+1)\partial_{\g(2)}h^{\g(2)\a(2s)}\partial^{\b(2)}h_{\a(2s)\b(2)}\notag\\
&& -\hf{(2s-1)} y^{\a(2s-2)}\partial^{\b(2)}\partial^{\g(2)}h_{\a(2s-2)\b(2)\g(2)}
-\frac{(s+1)(2s-1)}{2s}
y^{\a(2s-4)}\Box y_{\a(2s-4)}  \non \\
&& -4 s(2s-1) \Big[ (s+1) F^{\a(2s-2)} F_{\a(2s-2)}
-\frac{s-1}{2s} F^{\a(2s-2)} \pa^\b{}_{(\a_1} y_{\a_2 \dots \a_{2s-2})\b} \Big]~.
\eea
Eliminating the auxiliary field $F_{\a(2s-2)} $  leads to 
\bea
2(-2)^{s+1} L_{\rm bos} &=& 
h^{\a(2s+2)}\Box h_{\a(2s+2)}
-\hf (s+1)\partial_{\g(2)}h^{\g(2)\a(2s)}\partial^{\b(2)}h_{\a(2s)\b(2)}\notag\\
&&-\hf{(2s-1)}\Big[ y^{\a(2s-2)}\partial^{\b(2)}\partial^{\g(2)}h_{\a(2s-2)\b(2)\g(2)}
+ \frac{2}{s+1}y^{\a(2s-2)}\Box y_{\a(2s-2)}\notag\\
&&+\frac{(s-1)(2s-3)}
{4(s+1)}\partial_{\g(2)}y^{\g(2)\a(2s-4)}\partial^{\b(2)}y_{\b(2)\a(2s-4)}\Big]
~.
\eea
This Lagrangian corresponds to the massless spin-$(s+1)$ action 
$S_{\rm{F}}^{(2s+2)}$ obtained from  \eqref{MasslessSOFlat} by the replacement $s \to s+1$.
The fermionic sector of the component action proves to coincide with 
the massless spin-$(s+\hf)$ action \eqref{MasslessFOFlat}, 
$ S_{\rm{FF}}^{(2s+1)} [{h}_{(2s+1)},{y}_{(2s-1)},{ y}_{(2s-3)}]$. Therefore, we have the result
\begin{align}
{\mathbb S}_{\rm{SO}}^{(2s+1)}&[H_{(2s+1)},X_{(2s-2)}]=S_{\rm{F}}^{(2s+2)}[h_{(2s+2)},y_{(2s-2)}]+S_{\rm{FF}}^{(2s+1)}[h_{(2s+1)},y_{(2s-1)},z_{(2s-3)}]~.
\end{align}


\subsubsection{Massive  half-integer superspin action}

We now have all of the ingredients at our disposal to read off the component form of 
the massive half-integer superspin action \eqref{SHSTMGFlatb},
\begin{align}
{\mathbb S}_{\rm TM}^{(2s+1)}
&= \phantom{+}
 {\mathbb S}_{\rm{SCHS}}^{(2s+1)}[{ H}_{(2s+1)}]
-m^{2s-1}{\mathbb S}_{\rm{SO}}^{(2s+1)}
[{ H}_{(2s+1)},{ X}_{(2s-2)}] \non \\
&\approx  \phantom{+}
 S^{(2s+2)}_{\rm CHS}[h_{(2s+2)}]  
- m^{2s-1} S_{\rm{F}}^{(2s+2)}
[{h}_{(2s+2)},{y}_{(2s-2)}]
\non \\
&\phantom{=}+ ~ S^{(2s+1)}_{\rm CHS}[h_{(2s+1)}] - m^{2s-1} S_{\rm{FF}}^{(2s+1)}
[{h}_{(2s+1)},{y}_{(2s-1)},{ z}_{(2s-3)}] \non\\
&= \phantom{+}S_{\text{TM}}^{(2s+2)}[{h}_{(2s+2)},{y}_{(2s-2)}] + S_{\rm{TM}}^{(2s+1)}
[{h}_{(2s+1)},{y}_{(2s-1)},{ z}_{(2s-3)}]  \label{SusyTMComponentsHIS}
\end{align}
Here the symbol `$\approx$' indicates that the auxiliary field has been eliminated.

The explicit  structure of the component action  \eqref{SusyTMComponentsHIS}
clearly demonstrates that the model \eqref{SHSTMGFlatb} 
describes a single massive supermultiplet subject to the equations
\eqref{214a} and \eqref{214b}  with $n=2s+1$ on the mass shell.
The superfield proof was provided in \cite{KT17}.

\section{Superconformal higher-spin models on conformally-flat backgrounds} \label{secSCHSCF3}
In this section our main purpose is to derive the $\mc{N}=1$ superconformal higher-spin models in conformally-flat superspace backgrounds. Though, in the process, we will also provide the analogous $\mc{N}=2$ models and give comments on their $\mc{N}$-extended cousins. This section is based on our paper \cite{Confgeo}. 
In three dimensions,  $\cN$-extended conformal supergravity was formulated in superspace
as the gauge theory of the superconformal group  in \cite{BKNT-M1}. Upon degauging, 
this formulation reduces to the conventional one, sketched in \cite{HIPT} and fully developed in \cite{KLT-M11}, with the local structure group $\sSL(2 , \mb{R}) \times \sSO(\cN)$.
The former formulation is known as $\cN$-extended conformal superspace, while
the latter is often referred to as $\sSO(\cN)$ superspace. 
Below we only make use of the conformal 
superspace formulations for  $\cN=1,2$.
To start with, we recall the main facts about the $3d$ $\cN$-extended superconformal algebra
and primary superfields in conformal superspace following \cite{BKNT-M1}. 

The $3d$ $\cN$-extended superconformal algebra,
 ${\mathfrak{osp}}(\cN|4, {\mathbb R})$, 
 contains bosonic and fermionic generators. Its even part
 ${\mathfrak{so}}(3,2) \oplus {\mathfrak{so}}(\cN)$
 includes the generators 
of $\mathfrak{so}(\cN)$, $N_{KL}=- N_{LK}$, where 
$K,L=1,\dots, \cN$, in addition to the generators of the conformal group
described in section \ref{GCA}.
Their commutation relations are:
\begin{subequations} \label{SCA}
\begin{gather}
[N_{KL} , N^{IJ}] = 2 \d^I_{[K} N_{L]}{}^J - 2 \d^J_{[K} N_{L]}{}^I \ . \label{SCA.e}
\end{gather}
The odd part of  ${\mathfrak{osp}}(\cN|4, {\mathbb R})$ is spanned by 
 the $Q$-supersymmetry ($Q_\a^I$) and
 $S$-supersymmetry ($S_\a^I$) generators.
 In accordance with \cite{BKNT-M1}, the fermionic operators $Q_\a^I$ obey the
algebra
\begin{gather}
\{ Q_\a^I \ , Q_\b^J \} = 2 \ri \d^{IJ} (\g^c)_{\a\b} P_c 
\ , 
\quad [Q_\a^I , P_b ] = 0 \ , \label{5.1b}\\
[M_{\a\b} , Q_\g^I] = \ve_{\g(\a} Q_{\b)}^I \ , \quad [\mathbb D, Q_\a^I] = \hf Q_\a^I \ , \quad [N_{KL} , Q_\a^I] = 2 \d^I_{[K} Q_{\a L]} \ ,
\end{gather}
while the operators $S_\a^I$ obey the
algebra
\begin{gather}
\{ S_\a^I , S_\b^J \} = 2 \ri \d^{IJ} (\g^c)_{\a\b} K_c \ , \quad [S_\a^I , K_b] = 0 \ , \label{5.1d} \\
[M_{\a\b} , S_\g^I] = \ve_{\g(\a} S_{\b)}^I \ , \quad [\mathbb D, S_\a^I] = - \hf S_\a^I \ , \quad [N_{KL} , S_\a^I] = 2 \d^I_{[K} S_{\a L]} \ .
\end{gather}
In the supersymmetric case, the translation $(P_a)$ and special conformal
$(K_a)$ generators are extended to $P_A = (P_a , Q_\a^I)$ and 
$K_A = (K_a , S_\a^I)$, respectively.
The remainder of the algebra of $K_A$ with $P_A$ is given by
\begin{gather}
[K_a , Q_\a^I ] = - \ri (\g_a)_\a{}^\b S_\b^I \ , \quad [S_\a^I , P_a] = \ri (\g_a)_\a{}^{\b} Q_{\b}^I \ , \\
\{ S_\a^I , Q_\b^J \} = 2 \ve_{\a\b} \d^{I J} \mathbb D - 2 \d^{I J} M_{\a\b} - 2 \ve_{\a\b} N^{IJ} \ .
\end{gather}
\end{subequations}
 
The superspace geometry of $\cN$-extended conformal supergravity
is formulated in terms of the covariant derivatives of the form
\bea
\bm\nabla_A  =(\bm\nabla_a,\bm \nabla_\a^I)
= \bm e_A {}^M\pa_M- \hf \Omega_A{}^{bc} M_{bc} - \hf \Phi_A{}^{PQ} N_{PQ} - B_A \mathbb D - \mathfrak{F}_A{}^B K_B \ .
\eea
Here 
$\Omega_A{}^{bc}$  
is the Lorentz connection,  $\Phi_A{}^{PQ}$  the  ${\sSO }(\cN)$ connection, $B_A$
 the dilatation connection,   and $\mathfrak F_A{}^B$ the special
superconformal connection.
The graded commutation relations of $\bm\nabla_A$ with the generators 
$M_{bc}$, $N_{PQ} $, $\mathbb D $ and $K_B$ are obtained from \eqref{SCA}
by the replacement $P_A \to \bm\nabla_A$. However the relations \eqref{5.1b} turn into 
\begin{align}
[\bm \nabla_A ,\bm \nabla_B \}
	&= -\cT_{AB}{}^C \bm\nabla_C
	- \frac{1}{2} \cR(M)_{AB}{}^{cd} M_{cd}
	- \frac{1}{2} \cR(N)_{AB}{}^{PQ} N_{PQ}
	\non \\ & \quad
	- \cR({\mathbb D})_{AB} \mathbb D
	- \cR(S)_{AB}{}^\g_K S_\g^K
	- \cR(K)_{AB}{}^c K_c~.
\end{align}
To describe the off-shell conformal supergravity multiplet, the torsion and curvature 
tensors should obey certain $\cN$-dependent covariant constraints 
given in \cite{BKNT-M1}.
The complete solutions to the constraints are derived  in \cite{BKNT-M1}. We will reproduce the 
$\cN=1$ and $\cN=2$ solutions below.

The  generators $K_A = (K_a, S_\alpha^I)$ are used to define conformal {\it primary} superfields:
\be 
K_A \Phi = 0 \ .
\ee
In accordance with \eqref{5.1d}, 
 if a superfield is annihilated by the $S$-supersymmetry generator,
then it is necessarily primary,
\bea
S_\a^I \F =0 \quad \implies \quad K_a \F =0~.
\eea

\subsection{$\cN=1$ SCHS theories}

The algebra of $\cN=1$ conformal covariant derivatives \cite{BKNT-M1} is 
\begin{subequations} \label{N=1AlgebraCSS}
\begin{align} \{ \bm\nabla_\a , \bm\nabla_\b \} &= 2 \ri \bm\nabla_{\a\b} \ , \\
[ \bm\nabla_a , \bm\nabla_\b ] &= \frac{1}{4} (\g_a)_\b{}^\g \bm C_{ \g\d \s} K^{\d\s}  ~, \\
[\bm\nabla_a , \bm\nabla_b] &= - \frac{\ri}{8} \ve_{abc} (\g^c)^{\a\b} \bm\nabla_\a \bm C_{\b\g\d} K^{\g\d} - \frac{1}{4} \ve_{abc} (\g^c)^{\a\b} \bm C_{\a\b\g} S^\g \label{N=1Algebra.3}\ .
\end{align}
\end{subequations}
It is written in terms of 
 the $\cN = 1$ super Cotton tensor $\bm C_{\a\b\g}$
which is a primary superfield of dimension 5/2,
\bea 
 S_\d \bm C_{\a\b\g} = 0 \ ,\quad
\mathbb D \bm C_{\a\b\g} = \frac{5}{2}\bm C_{\a\b\g} 
\ , 
\eea
 obeying the Bianchi identity 
\bea
\bm \nabla^\a \bm C_{\a \b\g} = 0 \ . 
\eea

Upon degauging, i.e. imposing the gauge $B_A=0$, it may be shown that the $\mc{N}=1$ superspace geometry as described in section \ref{Sec3dN=1CSGGeom} is recovered.   Additionally,   
the above super Cotton tensor proves to be proportional to \eqref{super-Cotton}. The details of this procedure shall not concern us, since we will not be degauging any of our models explicitly. We refer the reader to \cite{BKNT-M1} for the relevant details.

Consider a real primary superfield $L$ of dimension $+2$,
\bea
S_\a L =0~, \qquad {\mathbb D} L  =2L~.
\eea
Then the functional 
\bea
I =   \int \rd^{3|2}z \, E\, L ~, \qquad E^{-1} = {\rm Ber} (\bm e_A{}^M)
\eea
is locally superconformal. We will use this action principle to construct $\cN=1$ locally superconformal higher-spin actions.

We now introduce SCHS gauge prepotentials by extending the definitions given 
in section \ref{sec3DN=1PrepsNStrengths} to conformal superspace. 
Given a positive integer $n\geq 1$,
a real tensor superfield $ H_{\a(n) } $ is said to be an SCHS
gauge prepotential if 
\begin{enumerate}[label=(\roman*)]
\item $ H_{\a(n) } $ is  primary and of dimension $(1-{n}/{2})$, 
\bea
S_\b H_{\a(n) } =0 ~, \qquad {\mathbb D} H_{\a(n)} = \left(1 -\frac{n}{2} \right) H_{\a(n)}~;
\label{3dN=1SCHSprepprop}
\eea
\item $ H_{\a(n) } $  is defined modulo gauge transformations of the form
\bea
\d_\L { H}_{\a(n) } =\ri^n \bm\nabla_{(\a_1} \L_{\a_2 \dots \a_n) }~,
\label{3dN=1SCHSprepGT}
\eea
\end{enumerate}
with the  gauge parameter $\L_{\a(n-1)}$
being real and primary but otherwise unconstrained. The dimension of $H_{\a(n) }$ is 
uniquely fixed by requiring  $\L_{\a(n-1)}$ 
and the right-hand side of \eqref{3dN=1SCHSprepGT} to be primary.

In conformal superspace, the $\mc{N}=1$ higher-spin super-Cotton tensor satisfies the following properties:
\begin{enumerate}[label=(\roman*)]

\item
${\mathfrak C}_{\a(n)}(H)$ is a primary superfield with weight $\Delta_{\mf{C}_{(n)}}=(1+n/2)$, 
\bea
S_{\b}{\mathfrak C}_{\a(n)}(H) =0~,\qquad \mb{D} {\mathfrak C}_{\a(n)}(H)=\Big(1+\frac{n}{2} \Big) {\mathfrak C}_{\a(n)}(H)~.\label{SCHScotPrimCSS}
\eea

\item
${\mathfrak C}_{\a(n)}(H)$ is of the form ${\mathfrak C}_{\a(n)}(H)=\cA H_{\a(n)}$, where $\cA$ is a linear differential operator of order $n$ involving $\bm\nabla_A$,
the super-Cotton tensor $\bm C_{\a\b\g}$, and its covariant derivatives.

\item
${\mathfrak C}_{\a(n)}(H)$ has vanishing gauge variation under \eqref{3dN=1SCHSprepGT} 
if $\mc{M}^{3|2}$ is conformally flat,
\bea
\d_\L {\mathfrak C}_{\a(n)}(H) = \mc{O}\big( \bm C \big)~.
\label{SCHScotGTMCSS}
\eea

\item
 ${\mathfrak C}_{\a(n)}(H)$ is transverse if $\mc{M}^{3|2}$ is conformally flat,
\bea
\bm\nabla^{\b} {\mathfrak C}_{\b \a(n-1)}(H) = \mc{O}\big( \bm C\big)~.
\label{SCHScotDivMCSS}
\eea 

\end{enumerate}

Let us first discuss the case $n=1$ corresponding to a superconformal 
vector multiplet. Associated with the prepotential $H_\a$ 
is the real spinor descendant
\bea
{\mathfrak C}_\a (H)=-\frac{\ri}{2} \bm\nabla^\b\bm\nabla_\a { H}_\b~,\label{5.12.Thesis}
\eea
which proves to be gauge invariant, 
\bea
\d_\L  {\mathfrak C}_\a(H) =0~,
\label{5.13.Thesis}
\eea
and  primary, 
\bea
S_\b {\mathfrak C}_\a(H) = 0~, \qquad {\mathbb D} {\mathfrak C}_\a(H) =\frac 32
{\mathfrak C}_\a(H)~.
\label{5.14.Thesis}
\eea
The field strength \eqref{5.12.Thesis} obeys the Bianchi identity 
\bea
\bm\nabla^\a {\mathfrak C}_\a(H) =0~.
\label{5.15.Thesis}
\eea
In general, this conservation equation is superconformal, for some primary 
spinor  $ {\mathfrak C}_\a $, if the dimension of 
${\mathfrak C}_\a $ is equal to 3/2.
The Chern-Simons action 
\bea
\mb{S}^{(1)}_{\rm SCHS}[H]
= - \frac{\ri}{2}
   \int \rd^{3|2}z \, E\,
 { H}^{\a} 
{\mathfrak C}_{\a}( H) 
\eea
has the following basic properties: (i) it
is locally superconformal;  and (ii) it is invariant under the gauge transformations 
\eqref{3dN=1SCHSprepGT} with $n=1$.

It turns out that some of the properties of the conformal vector supermultiplet ($n=1$), given by eqs. 
\eqref{5.13.Thesis}--\eqref{5.15.Thesis}, cannot  be extended  to $n>1$ in the case
of an  arbitrary curved background. So
let us first consider a conformally flat superspace, 
\bea
\bm C_{\a\b\g}=0~.
\eea
Then it follows from \eqref{N=1AlgebraCSS} that the conformally covariant derivatives
$\bm\nabla_A =(\bm\nabla_a, \bm\nabla_\a)$ obey the same graded commutation relations 
as the flat-superspace covariant derivatives. This allows us to use the flat-superspace 
results of section \ref{sec3dN=1SCHSMink} provided local
superconformal invariance can be kept under control.
We associate with the gauge prepotential $H_{\a(n)}$ 
the following linearised higher-spin super Cotton tensor
\begin{align}
{\mathfrak C}_{\a(n)} (H)
&= \frac{1}{2^{n}} 
\sum\limits_{j=0}^{\left \lfloor{n/2}\right \rfloor}
\bigg\{
\binom{n}{2j}  (\bm\Box_c)^{j}\bm\nabla_{(\a_{1}}{}^{\b_{1}}
\dots
\bm\nabla_{\a_{n-2j}}{}^{\b_{n-2j}}H_{\a_{n-2j+1}\dots\a_{n})\b(n-2j)}~~~~
\nonumber \\
&
-\frac{\ri}{2} 
\binom{n}{2j+1}\bm\nabla^{2}(\bm\Box_c)^{j}\bm\nabla_{(\a_{1}}{}^{\b_{1}}
\dots\bm\nabla_{\a_{n-2j -1}}{}^{\b_{n-2j -1}}H_{\a_{n-2j}\dots\a_{n})
\b(n-2j-1) }\bigg\}~,~~~~~\non\\
&= \Big( -\frac{\ri}{2}\Big)^n
\bm\nabla^{\b_1} \bm\nabla_{\a_1} \cdots \bm\nabla^{\b_n} \bm\nabla_{\a_n} H_{\b(n)}
\end{align}
where we have denoted $\bm\nabla^2= \bm\nabla^\a \bm\nabla_\a$ and $\bm\Box_{c}=\bm\nabla^a\bm\nabla_a$.
Making use of \eqref{SCA} it may be shown that ${\mathfrak C}_{\a(n)} (H)$ satisfies each of the properties \eqref{SCHScotPrimCSS},\eqref{SCHScotGTMCSS} and \eqref{SCHScotDivMCSS}.
This implies that the Chern-Simons-type action
\bea
\mb{S}_{\rm{SCHS}}^{(n)} [ {H}] 
= - \frac{\ri^n}{2^{\left \lfloor{n/2}\right \rfloor +1}}
   \int \rd^{3|2}z \, E\,
 {H}^{\a(n)} 
{\mathfrak C}_{\a(n) }( H) \label{SCHSact3dN=1CSS}
\eea
has the following fundamental properties: (i) it
is locally superconformal;  and (ii) it is invariant under the gauge transformations 
\eqref{3dN=1SCHSprepGT}.



\subsection{$\mc{N}$-extended SCHS theories}

In the $\cN=2$ case it is convenient to replace the real spinor covariant derivatives $\bm\nabla_\a^I$ with 
complex ones,
\bea 
\bm\nabla_\a = \frac{1}{\sqrt{2}} (\bm\nabla_\a^{1} - \ri \bm\nabla_\a^{2}) \ , \quad \bar{\bm\nabla}_\a = - \frac{1}{\sqrt{2}} (\bm\nabla_\a^{1} + \ri \bm\nabla_\a^{2}) \ ,
\eea
which are eigenvectors, 
\bea
 [J , \bm\nabla_\a] = \bm\nabla_\a \ , \quad [J , \bar{\bm\nabla}_\a] = - \bar{\bm\nabla}_\a ~,
\eea
of the $\sU(1)$ generator $J$ defined by
\bea 
J := - \frac{\ri}{2} \ve^{KL} N_{KL} \ .
\eea
It is also useful to introduce the operators
\bea 
S_\a := \frac{1}{\sqrt{2}} (S_\a^{1} + \ri S_\a^{2}) \ , \quad 
\bar{S}_\a := \frac{1}{\sqrt{2}} (S_\a^{1} - \ri S_\a^{2}) \ ,
\eea
which have the properties
\bea
[J , \bar{S}_\a] = \bar{S}_\a \ , \quad [J , S_\a] = - S_\a \ .
\eea
The graded commutation relations specific to the new basis are
\begin{subequations} \label{gensCB}
\begin{align}
\{ S_\a , S_\b \} = 0 \ , \quad \{ \bar{S}_\a , \bar{S}_\b \}& = 0 \ , 
\quad \{ S_\a , \bar{S}_\b \} = 2 \ri K_{\a\b} \ , \\
[K_a , \bm\nabla_\a ] = - \ri (\g_a)_\a{}^\b \bar{S}_\b \ ,& \quad [K_a , \bar{\bm\nabla}_\a ] = \ri (\g_a)_\a{}^\b S_\b \ , \\
[\bar{S}_\a , \bm\nabla_a] = \ri (\g_a)_\a{}^\b \bm\nabla_{\b} \ , \quad &[S_\a , \bm\nabla_a] = - \ri (\g_a)_\a{}^\b \bar{\bm\nabla}_{\b} \ , \\
\{ \bar{S}_\a , \bm\nabla_\b \} = 0 \ , &\quad \{ S_\a , \bar{\bm\nabla}_\b \} = 0 \ , \\
\{ \bar{S}_\a , \bar{\bm\nabla}_\b \} = - 2 \ve_{\a\b} {\mathbb D} + 2 M_{\a\b} - 2 \ve_{\a\b} J \ ,& 
\quad \{ S_\a , \bm\nabla_\b \} = 2 \ve_{\a\b} {\mathbb D} - 2 M_{\a\b} - 2 \ve_{\a\b} J \ . 
\end{align}
\end{subequations}

In the complex basis,
the algebra of $\cN=2$  covariant derivatives \cite{BKNT-M1} is  
\begin{subequations} 
\begin{align} 
\{ \bm\nabla_\a , \bm\nabla_\b \} &= 0 \ , \quad \{ \bar{\bm\nabla}_\a , \bar{\bm\nabla}_\b \} = 0 \ , \\
\{ \bm\nabla_\a , \bar{\bm\nabla}_\b \} &= - 2 \ri \bm\nabla_{\a\b} - \ve_{\a\b} \bm C_{\g\d} K^{\g\d}\ , \\
[\bm\nabla_a , \bm\nabla_\b] = \frac{\ri}{2} (&\g_a)_\b{}^\g \bm\nabla_\g \bm C^{\a\d} K_{\a\d} - (\g_a)_{\b\g} \bm C^{\g\d} \bar{S}_\d \ , \\
[\bm\nabla_a , \bm\nabla_b] = - \frac{\ri}{8} \ve_{abc} (\g^c)^{\g\d} \Big( \ri [\bm\nabla_\g &, \bar{\bm\nabla}_\d] \bm C_{\a\b} K^{\a\b} + 4 \bar{\bm\nabla}_\g \bm C_{\d\b} \bar{S}^\b + 4 \bm\nabla_\g \bm C_{\d \b} S^\b 
- 8 \bm C_{\g\d} J \Big) \ , 
\end{align}
\end{subequations}
where 
 the $\cN = 2$ super Cotton tensor $\bm C_{\a\b}$ is a primary real superfield,
 \bea
 S_\g \bm C_{\a\b} = 0 \quad \Longleftrightarrow \quad \bar S_\g \bm C_{\a\b} = 0 \ ,\quad
 \mathbb D \bm C_{\a\b} = 2 \bm C_{\a\b} \ , 
\eea
with the fundamental property 
\bea
\bm\nabla^\a \bm C_{\a\b} = 0 ~. 
\eea
In $\sSO(2)$ superspace \cite{KLT-M11}, the super Cotton tensor $\bm C_{\a\b}$ was introduced
originally  in \cite{Kuzenko12}. 

Given an integer $n\geq 1$, a real  tensor superfield ${H}_{\a(n) } $ 
is said to be an $\mc{N}=2$ superconformal gauge prepotential
if 
\begin{enumerate}[label=(\roman*)]

\item ${H}_{\a(n) } $ is primary and of dimension $(-{n}/{2})$, 
\bea
S_\b H_{\a(n) } =0  \quad \Longleftrightarrow \quad \bar S_\b H_{\a(n) }=0 ~,\quad
\quad {\mathbb D} H_{\a(n)}  =  -\frac{n}{2}H_{\a(n)}~;
\eea

\item  ${H}_{\a(n) } $ is defined modulo gauge transformations of the form
\bea
\d_\L { H}_{\a(n) } =\bar{\bm\nabla}_{(\a_1} \L_{\a_2 \dots \a_n) }
-(-1)^n\bm\nabla_{(\a_1} \bar \L_{\a_2 \dots \a_n) }~,
\label{55.28}
\eea

\end{enumerate}
where the  gauge parameter $\L_{\a(n-1)}$
is a primary  complex superfield of $\sU(1)$ charge $+1$, that is,
$J \L_{\a(n-1) } = \L_{\a(n-1)}$. The dimension of the gauge prepotential  is uniquely fixed by requiring 
 $H_{\a(n)}$ and  $\L_{\a(n-1)}$ to be primary.

In the remainder of this section we assume 
that the background curved superspace $\cM^{3|4}$
is conformally flat, 
\bea
\bm C_{\a \b}=0~.
\eea

The higher-spin super-Cotton tensor ${\mathfrak C}_{\a (n)}  (H)$ associated with the gauge prepotential $H_{\a(n)} $ was derived in $\mc{N}=2$ Minkowski superspace in \cite{KO}. The minimal lift of ${\mathfrak C}_{\a (n)}  (H)$ to $\mc{N}=2$ conformal superspace is
\bea
&&{\mathfrak C}_{\a (n)}  (H)
= \frac{1}{2^{n-1}} 
\sum\limits_{j=0}^{\left \lfloor{n/2}\right \rfloor}
\bigg\{
\binom{n}{2j} 
\bm\Delta ( \bm\Box_c)^{j}\bm\nabla_{(\a_{1}}{}^{\b_{1}}
\dots
\bm\nabla_{\a_{n-2j}}{}^{\b_{n-2j}}H_{\a_{n-2j+1}\dots\a_{n})\b(n-2j)}~~~~
\nonumber \\
&&\qquad \qquad +
\binom{n}{2j+1}\bm\Delta^{2}(\bm\Box_c)^{j}\bm\nabla_{(\a_{1}}{}^{\b_{1}}
\dots\bm\nabla_{\a_{n-2j -1}}{}^{\b_{n-2j -1}}H_{\a_{n-2j}\dots\a_{n})
\b(n-2j-1) }\bigg\}~,~~~~~
\label{eq:HSFSUniversal}
\eea
where 
$\bm\D = \frac{\ri}{2} \bm\nabla^\a \bar{\bm\nabla}_\a$.
This real descendant proves to be primary, 
\bea
S_\b {\mathfrak C}_{\a (n)}(H) =0 \quad \Leftrightarrow \quad 
\bar S_\b {\mathfrak C}_{\a (n)}(H) =0~,
\quad {\mathbb D} {\mathfrak C}_{\a (n)}(H)  
=  \left(1+\frac{n}{2} \right)  {\mathfrak C}_{\a (n)}(H)~,
\eea
and gauge invariant,
\bea
\d_\L {\mathfrak C}_{\a(n)}(H)&=&0~.
\eea
Moreover, it obeys the conservation equation
\bea
\bm\nabla^\b {\mathfrak C}_{\b \a(n-1)}(H) =0 
\quad \Longleftrightarrow \quad
 \bar{\bm\nabla}^\b {\mathfrak C}_{\b \a(n-1)} =0 ~.
\eea
These properties  imply
that the action
\bea
\mb{S}^{(n)}_{\rm{SCHS}}
[ {H}] 
= - \frac{\ri^n}{2^{\left \lfloor{n/2}\right \rfloor +1}}
   \int \rd^{3|4}z \, E\,
 {H}^{\a(n)} 
{\mathfrak C}_{\a(n) }( { H})  \label{N=2SCHS3act}
\eea
is $\mc{N}=2$ superconformal and invariant under the gauge transformations 
\eqref{55.28}.

To conclude our $\cN=2$ analysis, we remark that the off-shell formulations for 
massless and massive higher-spin $\cN=2$ supermultiplets in Minkowski superspace, 
as well as in the (1,1) and (2,0) anti-de Sitter 
backgrounds were constructed in \cite{KO,HKO,Hutomo:2018iqo} (see also \cite{HHK}). 
These theories are realised in terms of 
the conformal gauge prepotentials $H_{\a (n)}$ in conjunction with certain compensating 
supermultiplets.

For $2<\mc{N}\leq 6$, the corresponding linearised higher-spin super-Cotton tensors in $\mc{N}$-extended Minkowski superspace were recently derived in \cite{BHHK}. In conjunction with the off-shell formulation of $3d$  $\mc{N}$-extended conformal superspace developed in \cite{BKNT-M1}, the results of \cite{BHHK} may be readily extended to arbitrary conformally-flat superspace backgrounds.


%
%


\section{Superconformal higher-spin models in AdS$^{3|2}$} \label{secSCHSAdS3}

In the previous section we made use of $\mc{N}=1$ conformal superspace to construct the linearised higher-spin super-Cotton tensors on arbitrary conformally-flat superspace backgrounds, of which $\mc{N}=1$ AdS superspace (AdS$^{3|2}$) is a specific example. 
The super-Cotton tensors were formulated in terms of the conformally covariant derivative $\bm \nabla_A$, which suffices to describe the superconformal higher-spin action \eqref{SCHSact3dN=1CSS}. However, in order to formulate non-conformal massive actions, such as those derived in section \ref{secTMSHSMmink}, one has to express $\mf{C}_{\a(n)}(H)$ in terms of the Lorentz covariant derivative $\bm{\mc{D}}_A$ introduced in section \ref{Sec3dN=1CSGGeom}.

To transition from the first description to the second, one has to go through the procedure of degauging (see \cite{BKNT-M1} for the details).
Much like the non-supersymmetric case, this procedure is difficult for higher-derivative descendants such as $\mf{C}_{\a(n)}(H)$,  and we were not able to complete it for arbitrary $n$.  In this section we bypass this problem by deriving a closed-form expression for the higher-spin super-Cotton tensor in  AdS$^{3|2}$ directly in terms of $\bm{\mc{D}}_A$. 
We also give several applications of this result. This section is based on our papers \cite{Topological, CottonAdS}.

\subsection{$\mc{N}=1$ AdS superspace geometry}

Before turning to the main construction, 
some comments  about the geometry of AdS$^{3|2}$ are in order.
 The geometry of AdS$^{3|2}$ 
 is encoded in its covariant derivatives, $\bm\cD_A$,
with the graded commutation relations \cite{GGRS,KLT-M12}
\begin{subequations} \label{algN1}
\begin{align}
\{ \bm{\mc{D}}_\a , \bm{\mc{D}}_\b \} &= 2\ri \bm{\mc{D}}_{\a\b} - 4\ri\mc{S} M_{\a\b}~, \\
\ [ \bm{\mc{D}}_{\a \b}, \bm{\mc{D}}_\g ] &= -2\mc{S} \ve_{\g(\a}\bm{\mc{D}}_{\b)}~, \\
\ [ \bm{\mc{D}}_{\a \b}, \bm{\mc{D}}_{\g \d} ] &= 4 \mc{S}^2 \big(\ve_{\g(\a}M_{\b)\d} + \ve_{\d(\a} M_{\b)\g}\big) \label{algN1c}
~,
\end{align}
\end{subequations}
with $\cS \neq 0$ a constant real parameter, 
which 
may be positive or negative.\footnote{This corresponds to the case where the torsion superfields in  \eqref{N=1algCS} satisfy $\mc{S}_{\a\b\g}=0$ and $\bm\cD_A\mc{S}=0$. Clearly this implies that $\bm C_{\a\b\g}=0$ and hence AdS$^{3|2}$ is conformally-flat. } The two choices, $\cS=|\cS|$ and $\cS = -|\cS|$,  correspond to the so-called $(1,0)$ and $(0,1)$  
AdS superspaces \cite{KLT-M12}, which are different realisations of $\cN=1$ AdS superspace. The $(1,0)$ and $(0,1)$ AdS superspaces are naturally embedded 
in $(1,1)$ AdS superspace \cite{KLT-M12,HHK} and are related to each other by a parity transformation.\footnote{The only difference between the $(1,0)$ and $(0,1)$ AdS supersymmetry types is the fact that the mass terms in the corresponding Killing spinor equations \eqref{440} have different signs.}

Isometries of AdS${}^{3|2}$ are generated by Killing supervector fields $\z^B$
on AdS${}^{3|2}$. They are defined to be
those  conformal Killing supervector fields, eq. \eqref{CCSVF}, which obey the additional restriction $\S[\z]=0$, 
\bea
\Big[ \z^B \bm\cD_B + \hf K^{\b \g} [\z] M_{\b\g} , \bm\cD_A\Big]   = 0 ~.
 \label{421}
\eea 
The important properties of the Killing supervector fields include the following:
\begin{subequations}\label{4.21abc}
\bea
\bm\cD_\a \z_{\b\g} &=& 2\ri ( \ve_{\a\b} \z_\g + \ve_{\a\g} \z_\b) ~,\\
\bm\cD_\a \z_\b &=& \hf K_{\a\b} [\z]   + \cS \z_{\a\b}~,\\
\bm\cD_\a K_{\b\g} [\z]&=& 4\ri \cS( \ve_{\a\b} \z_\g + \ve_{\a\g} \z_\b) ~,
\eea
\end{subequations}
and their corollary
\bea
\big( \ri \bm\cD^2 + 12 \cS \big) \z_\a=0~.
\eea
The relations \eqref{4.21abc} tell us that the $\q$-dependence of the Killing superfield
parameters  $\z^B(x,\q)$ and $K^{\b\g}(x,\q) $  in \eqref{421} is determined by their values at $\q=0$. 
 Given a tensor superfield $\F$ on AdS${}^{3|2}$, 
its AdS transformation law is obtained from \eqref{RigidSCT} by setting $\S[\z]=0$, 
\bea
\d_{\z} \F = \Big(\z^B \bm\cD_B +\hf K^{\b\g}[\z] M_{\b\g} \Big) \F ~.
\label{424}
\eea
We recall that the independent component fields of $\z^B$ are 
$\z^b |$ and $\z^\b|$.  These components are, respectively,  a Killing vector field and a Killing spinor field on AdS$_3$. The equation on the Killing spinor field follows from bar projecting
\bea
\bm\cD_{\a\b} \z_\g = - \cS (\ve_{\g \a} \z_\b + \ve_{\g\b} \z_\a) \quad 
\Longleftrightarrow \quad \bm\cD_a \z_\b = \cS (\g_a)_\b{}^\g \z_\g~, 
\eea
which is a simple corollary of \eqref{4.21abc}.

There are two independent quadratic Casimir operators\footnote{For symplectic groups our notation is $\sSp (2n, {\mathbb R})  \subset \sGL (2n,{\mathbb R})$, 
hence  $\sSp (2, {\mathbb R} ) \cong  \sSL (2,{\mathbb R} ) \cong \sSU(1,1)$.}
 of the $\cN=1$ AdS isometry supergroup, $\sOSp (1|2; {\mathbb R} ) \times  \sSL (2,{\mathbb R} ) $,
 which may be chosen as follows:
\begin{subequations}
\begin{align}
\mathbb{F}&:=-\frac{\rm{i}}{2}\bm{\mc{D}}^2+2\bm{\mc{D}}^{\a\b}M_{\a\b}~,\qquad \big[\mathbb{F},\bm{\mc{D}}_A\big]=0~, \label{SF}\\
\mathbb{Q}&:=-\frac{1}{4}\bm{\mc{D}}^2\bm{\mc{D}}^2+\rm{i}\mc{S}\bm{\mc{D}}^2~,\qquad ~~\big[\mathbb{Q},\bm{\mc{D}}_A\big]=0~. \label{SQ}
\end{align}
\end{subequations}
The second of these may be expressed in terms of the AdS d'Alembertian, $\Box = \bm\cD^a \bm\cD_a$, via the relation
\begin{align}
-\frac{1}{4} \bm\cD^2 \bm\cD^2
=  \bm\Box -2\ri {\cS} \bm\cD^2 
+2 {\cS} \bm\cD^{\a \b} M_{\a\b} -2 {\cS}^2 M^{\a\b} M_{\a\b}~.
\end{align}
The fact that they are Casimir operators follows from  
the useful identities:
\begin{subequations}  \label{A8-mod}
	\bea 
	\bm{\mc{D}}_\a \bm{\mc{D}}_\b &=& \ri \bm{\mc{D}}_{\a\b} - 2\ri\mc{S}M_{\a\b}+\frac{1}{2}\ve_{\a\b}\bm{\mc{D}}^2~, \\
	\bm{\mc{D}}^{\b} \bm{\mc{D}}_\a \bm{\mc{D}}_\b &=& 4\ri \mc{S}\bm{\mc{D}}_\a~, \quad \{ \bm{\mc{D}}^2, \bm{\mc{D}}_\a  \} = 4\ri \mc{S}\bm{\mc{D}}_\a~,\\
 \bm{\mc{D}}^2 \bm{\mc{D}}_\a &=& 2\ri \mc{S}\bm{\mc{D}}_\a + 2\ri \bm{\mc{D}}_{\a\b} \bm{\mc{D}}^\b - 4\ri \mc{S} \bm{\mc{D}}^\b M_{\a\b}~, \\
\qquad \ [ \bm{\mc{D}}_{\a} \bm{\mc{D}}_{\b},\bm{\mc{D}}^2 ]
&=& 0 \quad \Longrightarrow \quad \ [\bm{\mc{D}}_{\a\b}, \bm{\mc{D}}^2 ] = 0~,\\
\ [ \bm{\mc{D}}_\a ,  \bm\Box] &=& 2\mc{S} \bm{\mc{D}}_{\a\b} \bm{\mc{D}}^\b + 3\mc{S}^2 \bm{\mc{D}}_\a~.
	\eea
\end{subequations}
It is an instructive exercise to check these identities. 

\subsection{On-shell massive and partially massless superfields in AdS$_3$} \label{secSPM3}

We define an on-shell real superfield $\F_{\a(n)}$ (which need not be an SCHS superfield), with $n\geq 1 $, to be one which satisfies the constraints
\begin{subequations}\label{OS2}
\begin{align}
0&=\bm{\mc{D}}^{\b}\F_{\a(n-1)\b}~,\label{OS2a}\\
0&=\big(\mb{F}- M\big)\F_{\a(n)} \label{OS2b}~, 
\end{align}
\end{subequations}
for some mass parameter $M$ which can take any real value, $M \in \mb{R}$. The field $\F_{\a(n)}$ is said to be transverse with pseudo-mass $M$, superspin $n/2$ and superhelicity $\frac{1}{2}\big(n+\frac{1}{2}\big)\s$, where $\s=M/|M|$. It can be shown that any superfield satisfying \eqref{OS2a} also satisfies
\begin{align}
-\frac{\rm{i}}{2}\bm{\mc{D}}^2\F_{\a(n)}=\bigg(\bm{\mc{D}}_{(\a_1}{}^{\b}+(n+2)\mc{S}\delta_{(\a_1}{}^{\b}\bigg)\F_{\a_2\dots\a_n)\b}~. \label{Dsquare}
\end{align}
Furthermore, if it satisfies both \eqref{OS2a} and  \eqref{OS2b}, then as a consequence we have
\begin{align}
-\frac{\rm{i}}{2}\bm{\mc{D}}^2\F_{\a(n)}&=\frac{1}{2n+1}\bigg( M+2n(n+2)\mc{S}\bigg)\F_{\a(n)}~.\label{OS2c}
\end{align}
From this we can deduce that the second-order wave equation which $\F_{\a(n)}$ satisfies is
\begin{align}
0=\bigg(\mb{Q}-\frac{1}{(2n+1)^2}\big[ M+2n(n+2)\mc{S}\big]\big[M+2(n-1)(n+1)\mc{S}\big]\bigg)\F_{\a(n)}~. \label{OS2d}
\end{align}
The equations \eqref{OS2a} and \eqref{OS2c} were introduced in \cite{KNT-M15}.
In the flat superspace limit, they reduce to the mass-shell equations given in \cite{KT17}.

A novel feature of three-dimensional field theories in AdS superspace is the existence of two distinct types of partially massless superfields. We will say that an on-shell superfield has type A or type B partially massless symmetry if, in addition to \eqref{OS2}, its pseudo mass satisfies 
\begin{subequations} \label{SPM333}
\begin{align}
M\equiv M^{(A)}_{(t,n)}&=-2\big[n(n-2t)-(t+1)\big]\mc{S}~,\qquad ~~~~~ 0 \leq t \leq \lceil n/2 \rceil -1~,\label{SPMA}\\
M\equiv M^{(B)}_{(t,n)}&=\phantom{-}2\big[n(n-2t+1)-(t-1)\big]\mc{S}~,\qquad 1\leq t \leq \lfloor n/2 \rfloor~, \label{SPMB}
\end{align}
\end{subequations}
respectively. We will refer to these as type A and type B pseudo masses, and we say that the corresponding superfield carries super-depth $t$. The latter is an integer whose range of allowed values for each type is specified  in \eqref{SPM333}. The relation
\begin{align}
M^{(A)}_{(\lceil n/2 \rceil +t-1,n)}=M^{(B)}_{(\lfloor n/2 \rfloor -t+1,n)} \label{ID9}
\end{align}
 holds between the  two types of masses for all $t$. The second order wave equation \eqref{OS2d} satisfied by partially massless superfields reduces to
\begin{align}
0=\big(\mb{Q}-\lambda^{(A)}_{(t,n)}\mc{S}^2\big)\F^{(t,A)}_{\a(n)}~,\qquad 0=\big(\mb{Q}-\lambda^{(B)}_{(t,n)}\mc{S}^2\big)\F^{(t,B)}_{\a(n)}~,
\end{align}
where the constants 
\begin{align}
\lambda^{(A)}_{(t,n)}=4t(t+1)~, \qquad \lambda^{(B)}_{(t,n)}=4(n-t)(n-t+1)~,
\end{align}
will be referred to as the type A and type B partially massless values.

It may be shown that the system of equations satisfied by on-shell type A and type B partially massless superfields, $\F^{(t,A)}_{\a(n)}$ and $\F^{(t,B)}_{\a(n)}$ respectively, admit the gauge symmetries
\begin{subequations} \label{SPMG}
\begin{align}
\delta_{\L}\F^{(t,A)}_{\a(n)}&=\text{i}^n\bm{\mc{D}}_{(\a_1\a_2}\cdots\bm{\mc{D}}_{\a_{2t-1}\a_{2t}}\bm{\mc{D}}_{\a_{2t+1}}\L_{\a_{2t+2}\dots\a_n)}=\ri^n\big(\bm{\mc{D}}_{\a(2)}\big)^t\bm{\mc{D}}_{\a}\L_{\a(n-2t-1)}~,\label{SPMG1}\\
\delta_{\L}\F^{(t,B)}_{\a(n)}&=\phantom{\rm{i}^n}\bm{\mc{D}}_{(\a_1\a_2}\cdots\bm{\mc{D}}_{\a_{2t-1}\a_{2t}}\L_{\a_{2t+1}\dots\a_n)}=\big(\bm{\mc{D}}_{\a(2)}\big)^t\L_{\a(n-2t)}~.\label{SPMG2}
\end{align}
\end{subequations}
Similar to the non-supersymmetric case (see eq. \eqref{GPC1}), this is true only if the gauge parameters $\L_{\a(n-2t-1)}$ and $\L_{\a(n-2t)}$ are also on-shell with the same pseudo-mass as their gauge field. We point out that strictly massless superfields $\F_{\a(n)}$, defined modulo the standard gauge transformations $\delta_{\L}\F_{\a(n)}=\ri^n\bm\cD_{\a}\L_{\a(n-1)}$,
correspond to type A partially massless superfields with the minimal super-depth of $t=0$.


\subsection{Component analysis}

In order to make the content of the two types of partially massless supermultiplets more transparent, in this section we study their component structure in more detail. 

We start with some general comments.
Every Killing supervector field $\z^B$ on AdS$^{3|2}$ can be uniquely decomposed as a sum 
\bea
\z^B= \z^B_{(\text{even})}+\z^B_{(\text{odd})}~,
\eea
 where $\z^B_{(\text{even})}$ and $\z^B_{(\text{odd})}$ are 
  even $\big( v^b:=\z^b_{(\text{even})}| \neq 0 $ and $ \z_{(\text{even})}^\b| = 0 \big)$ and odd 
 $\big( \z_{(\text{odd})}^b| = 0 $ and $ \e^{\b}:=\z_{(\text{even})}^\b| \neq 0\big)$
 Killing supervector fields, respectively. 
 Here $v^b(x) $ is a Killing vector field on AdS$_3$, 
\bea
 \cD^a v^b + \cD^b v^a = 0~,
 \eea
 and $\e^\b(x)$ is a Killing spinor field on AdS$_3$,
 \bea
\cD_{\a\b} \e_\g = - \cS (\ve_{\g \a} \e_\b + \ve_{\g\b} \e_\a) ~.
\label{440}
\eea

Let us consider a tensor superfield $\F$ on AdS${}^{3|2}$ with  
the transformation law  \eqref{424}.
Its independent component fields are contained in the set of fields 
$\vf=\mathfrak{V}|$, where  
$\mathfrak{V}:=\big\{ \F, \bm\cD_{ \a} \F, \cdots\big\}$.
Choosing $\z^B$ in \eqref{424} to be $ \z^B_{(\text{even})}$, one observes that the component fields $\vf$ transform as tensor fields on AdS$_3$,
\bea
\d_v \vf = \Big(v^b \cD_b +\hf K^{\b\g}[v] M_{\b\g} \Big) \vf ~. 
\eea
Choosing $\z^B$ in \eqref{424} to be $ \z^B_{(\text{odd})}$, we find that 
the supersymmetry transformation laws of the component fields
are
given by\footnote{To derive this transformation rule, the relation $K_{\a\b}[\z]=\frac{1}{2}\bm{\mc{D}}^{\g}{}_{(\a}\z_{\b)\g}$ may be useful. }
\bea
\d_\e \vf =  \e^\b (\bm\cD_\b \mathfrak{V})| ~.
\label{susyT}
\eea

If $\F_{\a(n)}$ is an on-shell superfield on AdS$^{3|2}$ with pseudo-mass $M$, then it has only two independent component fields, which we define according to 
\begin{align}
\f_{\a(n)}:=\F_{\a(n)}|~,\qquad \f_{\a(n+1)}:=\text{i}^{n+1}\bm{\mc{D}}_{(\a_1}\F_{\a_2\dots\a_{n+1})}| 
=\text{i}^{n+1}\bm{\mc{D}}_{\a_1}\F_{\a_2\dots\a_{n+1}}| 
~.
\label{443}
\end{align} 
As a consequence of \eqref{OS2a} and \eqref{A8-mod}, both component fields are transverse
\begin{align}
0=\mc{D}^{\b\g}\f_{\b\g\a(n-2)}~,\qquad 0=\mc{D}^{\b\g}\f_{\b\g\a(n-1)}~. \label{OS3a}
\end{align}
To deduce the other first order constraint which they satisfy, the relation 
\begin{align}
\mb{F}\Psi_{\a(n)}|=\bigg(\frac{2n+1}{n}\mc{F}+(n+2)\mc{S}\bigg)\Psi_{\a(n)}|~, \label{ID1}
\end{align}
which holds for an arbitrary transverse superfield $\Psi_{\a(n)}$, is useful. We recall that $\mc{F}$ is the quadratic Casimir operator \eqref{F} of $\mf{so}(2,2)$. Making use of \eqref{ID1}, we find that the component fields satisfy
\begin{subequations}\label{OS3b}
\begin{align}
0&=\bigg(\mc{F}-\frac{n}{2n+1}\big[M-(n+2)\mc{S}\big]\bigg)\f_{\a(n)}~,\\
0&=\bigg(\mc{F}-\frac{n+1}{2n+1}\big[M+(n-1)\mc{S}\big]\bigg)\f_{\a(n+1)}~.
\end{align}
\end{subequations}
The equations \eqref{OS3a} and \eqref{OS3b} define on-shell fields in AdS$^3$ in accordance with section \ref{section 2.2}. Finally, making use of \eqref{susyT} and \eqref{OS2}, the supersymmetry transformation laws of the component fields
 prove to be
\begin{subequations}
\begin{align}
\delta_{\e}\f_{\a(n)}&= (-\ri)^{n+1}\e^{\b}\f_{\b\a(n)}~,\\
\delta_{\e}\f_{\a(n+1)}&=-\ri^n\e^{\b}\mc{D}_{\b\a}\f_{\a(n)}+\frac{\ri^n}{(2n+1)}\Big(M+n(4n+5)\mc{S}\Big)\e_{\a}\f_{\a(n)}~.
\end{align}
\end{subequations}

Let us now consider the case when the on-shell superfield is type A or type B partially massless with super-depth $t$. Upon substituting $M=M^{(A)}_{(t,n)}$, \eqref{OS3b} reduces to
\begin{align}
0=\Big(\mc{F}-\rho^{(-)}_{(t,n)}\Big)\f_{\a(n)}~,\qquad 0=\Big(\mc{F}-\rho^{(-)}_{(t+1,n+1)}\Big)\f_{\a(n+1)}~.
\end{align}
On the other-hand, substituting $M=M^{(B)}_{(t,n)}$  we find 
\begin{align}
0=\Big(\mc{F}-\rho^{(+)}_{(t,n)}\Big)\f_{\a(n)}~,\qquad 0=\Big(\mc{F}-\rho^{(+)}_{(t,n+1)}\Big)\f_{\a(n+1)}~.~~~
\end{align}
We recall that $\rho^{(\pm)}_{(t,n)}$ are the depth-$t$ pseudo-mass values \eqref{PM1}.

We see that the type A supermultiplet $\F_{\a(n)}^{(t,A)}$ consists of two negative helicity partially massless fields: one with spin $n/2$ and depth $t$, and the other with spin $(n+1)/2$ and depth $t+1$.  In contrast, the type B supermultiplet $\F_{\a(n)}^{(t,B)}$ consists of two positive helicity depth-$t$ partially massless fields: one with spin $n/2$ and the other with spin $(n+1)/2$.

\subsection{Higher-spin super-Cotton tensors and their factorisation}


In contrast to the non-supersymmetric case, explicit expressions for the higher-spin super-Cotton tensors are easily obtained in $\mc{N}=1$ AdS superspace. In \cite{Topological}, some of the lower rank super-Cotton tensors were derived using the operator
\begin{align}
\Delta^{\a}{}_{\b}:=-\frac{\text{i}}{2}\bm{\mc{D}}^{\a}\bm{\mc{D}}_{\b}-2\mc{S}\delta^{\a}{}_{\b} \label{Del1}
\end{align}
which satisfies the relations
\begin{align}
\bm{\mc{D}}^{\b} \Delta^{\a}{}_{\b}=0~,\qquad \Delta^{\a}{}_{\b}\bm{\mc{D}}_{\a}=0~. \label{ID82}
\end{align}
To derive the higher-spin super-Cotton tensors, we make use of the following extension of \eqref{Del1}
\begin{align}
\Delta_{[j]}^{\a}{}_{\b}:=-\frac{\text{i}}{2}\bm{\mc{D}}^{\a}\bm{\mc{D}}_{\b}-2j\mc{S}\delta^{\a}{}_{\b}~, \label{Del2}
\end{align} 
for which \eqref{Del1} corresponds to the $j=1$ instance, $\Delta^{\a}{}_{\b}\equiv \Delta_{[1]}^{\a}{}_{\b}$. Using the algebra \eqref{A8-mod}, it may be shown that they possess the following important properties
\begin{subequations}\label{DP}
\begin{align}
\big[\Delta_{[j]}^{\a_1}{}_{\b_1},\Delta_{[k]}^{\a_2}{}_{\b_2}\big]&=\ve_{\b_1\b_2}\mc{S}\big(\bm{\mc{D}}^{\a(2)}-2\mc{S}M^{\a(2)}\big)-\ve^{\a_1\a_2}\mc{S}\big(\bm{\mc{D}}_{\b(2)}-2\mc{S}M_{\b(2)}\big)~,\label{DP1}\\
\Delta_{[j]}^{\a}{}_{\g}\Delta_{[j+1]}^{\b}{}_{\d}\ve^{\d\g}&=j\mc{S}\ve^{\a\b}\big(\text{i}\bm{\mc{D}}^2+4(j+1)\mc{S}\big)~,\label{DP2}\\
\Delta_{[j+1]}^{\a}{}_{\g}\Delta_{[j]}^{\b}{}_{\d}\ve_{\a\b}&=j\mc{S}\ve_{\g\d}\big(\text{i}\bm{\mc{D}}^2+4(j+1)\mc{S}\big)~, \label{DP3}
\end{align}
\end{subequations}
for arbitrary integers $j$ and $k$.

In terms of \eqref{Del2}, the higher-spin super-Cotton tensor takes the remarkably simple form
\begin{align}
\mathfrak{C}_{\a(n)}(H)&=\Delta_{[1]}^{\b_1}{}_{(\a_1} \Delta_{[2]}^{\b_2}{}_{\a_2}\cdots\Delta_{[n]}^{\b_n}{}_{\a_n)}H_{\b(n)} \non\\
&=\Delta_{[n]}^{\b_1}{}_{(\a_1}\Delta_{[n-1]}^{\b_2}{}_{\a_2}\cdots\Delta_{[1]}^{\b_n}{}_{\a_n)}H_{\b(n)} \label{Scot}
\end{align}
The equivalence of the two above expressions for $\mathfrak{C}_{\a(n)}(H)$ follows from the identity \eqref{DP1}. The defining features of the super-Cotton tensors 
follow immediately from the properties \eqref{DP} of the operators \eqref{Del2}. In particular, transversality may be shown as follows
\begin{align}
\bm{\mc{D}}^{\g}\mf{C}_{\g\a(n-1)}(H)&=\bm{\mc{D}}^{\g}\Delta_{[1]}^{\b_1}{}_{(\g}\Delta_{[2]}^{\b_2}{}_{\a_1}\cdots\Delta_{[n]}^{\b_n}{}_{\a_{n-1})}H_{\b(n)} \non\\
&=\frac{1}{n!}\bm{\mc{D}}^{\g}\big(\Delta_{[1]}^{\b_1}{}_{\g}\Delta_{[2]}^{\b_2}{}_{\a_1}\cdots\Delta_{[n]}^{\b_n}{}_{\a_{n-1}}+(n!-1) ~\text{permutations}~\big)H_{\b(n)}~. \label{scotT}
\end{align}
In the last line, all of the $(n!-1)$ permutations may be brought into the same form as the first term using \eqref{DP2}. From the first equation in \eqref{ID82}, it follows that the right hand side of  \eqref{scotT} vanishes, and hence
\begin{align}
0=\bm{\mc{D}}^{\b}\mf{C}_{\b\a(n-1)}(H)~. \label{scotTT}
\end{align} 
In a similar vein, its variation under the gauge transformation \eqref{SCHS3prepGT}
may be computed as follows
\begin{align}
\mf{C}_{\a(n)}(\delta_{\L}H)&=\ri^n\Delta_{[n]}^{\b_1}{}_{(\a_1}\Delta_{[n-1]}^{\b_2}{}_{\a_2}\cdots\Delta_{[1]}^{\b_n}{}_{\a_n)}\bm{\mc{D}}_{(\b_1}\L_{\b_2\dots\b_n)}\non\\
&=\frac{\ri^n}{n!}\Delta_{[n]}^{\b_1}{}_{(\a_1}\Delta_{[n-1]}^{\b_2}{}_{\a_2}\cdots\Delta_{[1]}^{\b_n}{}_{\a_n)}\big(\bm{\mc{D}}_{\b_n}\L_{\b_1\dots\b_{n-1}}+(n!-1)~\text{permutations}~\big)~.
\end{align}
This time using \eqref{DP3}, all of the $(n!-1)$ permutations may be brought into the same form as the first term in the second line.  From the second identity in \eqref{ID82} it follows that the right hand side vanishes, 
\begin{align}
0=\mf{C}_{\a(n)}(\delta_{\L}H)~,
\end{align}
and hence $\mf{C}_{\a(n)}(H)$ is gauge invariant. 


  
The gauge invariance and transversality of the higher-spin super-Cotton tensor means that its associated higher-spin Chern-Simons-type functional 
\begin{align}
 S^{(n)}_{\rm SCHS}[H] = 
- \frac{\ri^n}{2^{\left \lfloor{n/2}\right \rfloor +1}} \int \rd^{3|2}z \, E \, H^{\a(n)} \mf{C}_{\a(n)}(H) ~, 
\label{SCS}
\end{align}
is gauge invariant.
In the flat-superspace limit, the action \eqref{SCS} reduces to the one given in section \ref{sec3dN=1SCHSMink}.
 On account of  the two equivalent forms of $\mf{C}_{\a(n)}(H)$ in \eqref{Scot}, it is also symmetric in the sense analogous to \eqref{sym}. 
 Since $ H^{\a(n)} $ and $\mf{C}_{\a(n)}(H) $ are primary superfields, with dimensions as prescribed in
 \eqref{SCHS3prepWeyl} and \eqref{SCHScotWeyl} respectively, the action \eqref{SCS} is superconformal.

Similar to the non-supersymmetric case, we may impose the transverse gauge condition on $H_{\a(n)}$
\begin{align}
H_{\a(n)}\equiv H^{\text{T}}_{\a(n)}~,\qquad 0=\bm{\mc{D}}^{\b}H^{\text{T}}_{\b\a(n-1)}~,
\end{align}
under which the super-Cotton tensor takes the form
\begin{align}
\mf{C}_{\a(n)}(H^{\text{T}})&=\frac{1}{(2n+1)^n}\prod_{t=0}^{ n-1 }\big(\mb{F}-M^{(A)}_{(t,n)}\big) H^{\text{T}}_{\a(n)} \non\\
&=\frac{1}{(2n+1)^n}\prod_{t=0}^{ \lceil n/2 \rceil -1 }\big(\mb{F}-M^{(A)}_{(t,n)}\big) \prod_{t=1}^{ \lfloor n/2 \rfloor }\big(\mb{F}-M^{(B)}_{(t,n)}\big) H^{\text{T}}_{\a(n)}~, \label{ScotFactor}
\end{align}
where we have used \eqref{ID9}. We see that on AdS${}^{3|2}$, the superconformal higher-spin action \eqref{SCS} factorises into first-order differential operators involving all of the type A and type B partial pseudo-mass values.  
  
Finally we note that, in any conformally flat superspace, it is also true that the vanishing of the higher-spin super-Cotton tensor $\mf{C}_{\a(n)}(H)$ is a necessary and sufficient condition for 
$H_{\a(n)}$ to be pure gauge (cf. \eqref{GaugeCompletenessCF}),
\begin{align}
\mathfrak{C}_{\a(n)}(H)=0 \quad \Longleftrightarrow \quad 
H_{\a(n)}=\ri^n\bm{\mc{D}}_{\a}\L_{\a(n-1)}~,
\end{align}
 for some $\L_{\a(n-1)}$. Therefore, from \eqref{ScotFactor}, it follows that both type A and B partially-massless superfields do not contain any local propagating degrees of freedom.

\subsection{Massive $\mc{N}=1$ gauge actions}

The topologically massive supersymmetric models \eqref{SHSNTMGmink} and \eqref{SHSTMGFlat} in $\mb{M}^{3|2}$, which were presented in section \ref{secTMSHSMmink}, may now be easily extended to AdS$^{3|2}$.
 For any integer $n\geq 1$, the new topologically massive gauge-invariant action corresponding to the unconstrained prepotential $H_{\a(n)}$ may be recast into the form 
\begin{align}
\mb{S}_{\text{NTM}}^{(n)}[H]=- \frac{\ri^n}{2^{\left \lfloor{n/2}\right \rfloor +1}}\frac{1}{M}\int \rd^{3|2}z \, E \,\mf{C}^{\a(n)}(H) \big(\mb{F}-M\big)H_{\a(n)}~. \label{SHSNTMG}
\end{align}
The equation of motion obtained by varying \eqref{SHSNTMG} with respect to $H_{\a(n)}$ is 
\begin{align}
0=\big(\mb{F}-M\big)\mf{C}_{\a(n)}(H)~. \label{SEOM1}
\end{align}
For generic $M$, this equation in conjunction with the off-shell conservation identity \eqref{scotTT}
means that the field-strength $\mf{C}_{\a(n)}(H)$ itself describes a propagating mode with pseudo-mass $M$, superspin $n/2$ and superhelicity $\frac{1}{2}\big(n+\frac{1}{2}\big)\s$, where $\s=M/|M|$. 
In the case when $M$ takes on one of the type A or type B partially massless values \eqref{SPM333}, an analysis similar to that given in the non-supersymmetric case can be conducted, with the conclusion that there are no local propagating degrees of freedom. 

To construct the topologically massive supersymmetric higher-spin actions, we need to extend the massless first order \eqref{SMFOActMink3} and second order \eqref{SMSOActMink3} models to AdS$^{3|2}$. For integer superspin, the relevant massless action\footnote{There is another massless action of this type, where the superfield $H_{\a(2s)}$ appears in the action with a first-order kinetic operator. However, the gauge transformations are of type B with $t=1$, $\delta_{\L}H_{\a(2s)}=\bm{\mc{D}}_{\a(2)}\L_{\a(2s-2)}$. The corresponding action has been derived recently in \cite{HHK}.
 } is the first-order (in vector derivatives) one, and it takes the form 
\bea
{\mathbb S}^{(2s)}_{\rm FO}[H,Y] &=& 
\bigg(-\frac{1}{2}\bigg)^s\frac{\ri}{2}
\int \rd^{3|2} z \, E\,\Big\{  
H^{\b  \a (2s-1)} \bm\cD^\g \bm\cD_\b H_{\g   \a (2s-1)}  \non \\
&&
 +2\ri (2s-1) Y^{\a (2s-2)} \bm\cD^{\b(2)} H_{ \b(2) \a (2s-2) } \non \\
&&
+ (2s-1) \Big( Y^{\a (2s-2) }\bm\cD^2 Y_{\a (2s-2)} 
+ 
(2s-2)\bm\cD_\b Y^{\b \a (2s-3) }
\bm\cD^\g Y_{\g \a (2s-3)} \Big) \non \\
&& -4\cS \ri \Big( H^{\a(2s) } H_{\a(2s) }
+2s(2s-1) Y^{\a(2s-2)} Y_{\a(2s-2)} \Big)
\Big\} ~.
\label{5.44}
\eea
It is invariant under the gauge transformations
\begin{subequations}
\bea
\d_{\L} {H}_{\a (2s)} &=& 
\bm\cD_{\a } \L_{\a(2s-1)} ~, \\
\d_{\L} {Y}_{\a (2s-2) } &=& \frac{1}{2s} \bm\cD^\b \L_{\b \a(2s-2)}~.
\eea
\end{subequations}
For half-integer superspin, the relevant massless action is the second-order one and it takes the form 
\bea
{\mathbb S}_{\rm{SO}}^{(2s+1)}[H,X]
&&= \bigg(- \hf \bigg)^s   \int \rd^{3|2} z \, E\,\bigg\{ 
-\frac{\ri}{2} H^{\a(2s+1)} {\mathbb Q} H_{\a(2s+1)}
-\frac{\ri}{8} \bm\cD_\b H^{\b \a(2s)} \bm\cD^2 \bm\cD^\g  H_{\g \a(2s)} \non \\
&&+ \frac{\ri}{4} s \bm\cD_{\b\g} H^{\b\g \a(2s-1)} 
\bm\cD^{\r\l} H_{\r\l \a(2s-1)}
-\hf (2s-1) X^{\a(2s-2)} \bm\cD^{\b\g} \bm\cD^\d H_{\b\g \d \a(2s-2)} \non \\
&&+\frac{\ri}{2}  (2s-1)\Big[X^{\a(2s-2)} \bm\cD^2 X_{\a(2s-2)}
- \frac{s-1}{s} \bm\cD_\b X^{\b\a(2s-3)} \bm\cD^\g X_{\g \a(2s-3)}\Big]
\non \\
&& +\ri s \cS H^{\b \a(2s)}  \bm\cD_\b{}^\g  H_{\g \a(2s)}
+\hf (s+1) {\cS} H^{\a(2s+1)} \bm\cD^2 H_{\a(2s+1)}
 \label{5.45} \\
&&+\ri s (2s-3) {\cS}^2 H^{\a(2s+1)} H_{\a(2s+1)}
+ \frac{(2s-1)(s^2 -3s -2)}{s} {\cS} X^{\a(2s-2)} X_{\a(2s-2)}\bigg\}
~.~~~
\non
\eea
It is invariant under the gauge transformations
\begin{subequations}
\bea
\d_{\L} H_{\a(2s+1)} &=& \ri \bm\cD_{\a} \L_{\a(2s)} ~,\\
\d_{\L} X_{\a(2s-2)} &=& \frac{s}{2s+1} \bm\cD^{\b(2)} \L_{\b(2) \a(2s-2)}~.
\eea
\end{subequations} 
The corresponding topologically massive gauge-invariant actions for integer and half-integer superspin are 
\begin{subequations} \label{5.39}
\bea
{\mathbb S}_{\rm TM}^{(2s)}[H,Y]
&=& {\mathbb S}_{\rm{SCHS}}^{(2s)}[H]
+\boldsymbol{\mu}(M,s){\mathbb S}_{\rm{FO}}^{(2s)} [H,Y]~,
\label{5.39a}
\\
{\mathbb S}_{\rm TM}^{(2s+1)}[H,X]
&=&{\mathbb S}_{\rm{SCHS}}^{(2s+1)}[H]
+\boldsymbol{\nu}(M,s){\mathbb S}_{\rm{SO}}^{(2s+1)}[H,X]~.
\label{5.39b}
\eea
\end{subequations}
The coupling constants $\boldsymbol{\mu}(M,s)$ and $\boldsymbol{\nu}(M,s)$ both have mass dimension $2s-1$, and are functions of $s$ and some parameter $M$ with mass dimension one. Their explicit form in terms of these quantities may be determined by requiring that any $H_{\a(n)}$ satisfying \eqref{OS2} is a particular solution to the resulting field equations, as in the non-supersymmetric case. We expect that both $\boldsymbol{\mu}$ and $\boldsymbol{\nu}$ will vanish at the type A and type B partially massless points \eqref{SPM333}, where the models do not describe any local propagating degrees of freedom.  

The above massive gauge-invariant actions are manifestly supersymmetric, 
that is the $\cN=1$ AdS supersymmetry is realised off-shell. 
It is worth pointing out that there also exists an on-shell construction of gauge-invariant Lagrangian formulations for massive higher-spin ${\cN}=1$ supermultiplets in $\mathbb{M}^{3}$ and AdS$^{3}$ \cite{BSZ3, BSZ4, BSZ5}, extending previous works on the non-supersymmetric cases \cite{BSZ1, BSZ2}. These frame-like formulations are based on the gauge-invariant approach to the dynamics of massive higher-spin fields proposed by Zinoviev \cite{Zinoviev, Zinoviev2} and Metsaev \cite{Metsaev2006}. 

So far we have discussed $\cN=1$ topologically massive supergravity 
and its higher spin extensions.
The off-shell formulations for $\cN$-extended topologically massive supergravity theories were presented 
in \cite{DKSS2,KLRST-M} for $\cN=2$,  in \cite{KN14} for $\cN=3$,
and in \cite{KN14,KNS} for the $\cN=4$ case.  
In all of these theories, the action functional is a sum of two terms, 
one of which is the action for pure $\cN$-extended supergravity 
(Poincar\'e or anti-de Sitter)
and the other is the action for $\cN$-extended conformal supergravity.
The off-shell actions for $\cN$-extended supergravity theories 
in three dimensions were given
in \cite{GGRS} for $\cN=1$, \cite{KLT-M11,KT-M11} for $\cN=2$ and
 \cite{KLT-M11} for the cases $\cN=3, ~4$.
The off-shell actions for $\cN$-extended conformal supergravity 
were given in \cite{vN} for $\cN=1$, \cite{RvanN86} for $\cN=2$,  
  \cite{BKNT-M2} for $ \cN =3,~4,~ 5$, and in \cite{Nishimura,KNT-M13}
for  the $\cN=6$ case.
 Refs. \cite{BKNT-M2,KNT-M13} made use of
the off-shell formulation for $\cN$-extended conformal supergravity 
proposed in \cite{BKNT-M1}. The on-shell formulation for $\cN$-extended
conformal supergravity with $\cN>2$ was given in \cite{LR89}.
On-shell approaches to  $\cN$-extended topologically massive supergravity theories 
with $4 \leq \cN \leq 8$ were presented in  
  \cite{Chu:2009gi,Gran:2012mg,Nilsson:2013fya,LS1,LS2}. 
It would be interesting to formulate topologically massive higher spin supermultiplets
for $\cN>2$.


\section{Summary of results} \label{secSCHS3dis}
This chapter was dedicated to the construction of SCHS models in curved $3d$ $\mc{N}=1$ superspace and applications thereof. We began in section \ref{secSCHSM3} by describing the salient features of the SCHS field $H_{\a(n)}$, and its higher-spin super-Cotton tensor $\mf{C}_{\a(n)}(H)$, on a generic supergravity background.
The definition of $\mf{C}_{\a(n)}(H)$ is such that it has the characteristic features of a covariantly conserved conformal supercurrent if superspace is conformally-flat \cite{Topological}, leading to the gauge and Weyl invariant SCHS action \eqref{SCHSActM3}. 

In section \ref{secSCHSMink3} we reviewed the explicit realisation of SCHS models in Minkowksi superspace \cite{K16}, and used them to construct topologically massive supersymmetric higher-spin gauge models. The latter were originally proposed  and analysed in \cite{KT17} at the superspace level. We analysed these models at the component level \cite{Topological} and related them to the topologically massive models constructed in section \ref{sec3DCHSflat}. We also proposed `new topologically massive' higher-spin supersymmetric gauge actions \cite{Topological, CottonAdS}

In section \ref{secSCHSAdS3} we constructed the gauge-invariant higher-spin super-Cotton tensors  \eqref{Scot} and the associated SCHS action \eqref{SCS} in AdS$^{3|2}$ for the first time \cite{CottonAdS}.
We found that in the transverse gauge the SCHS action factorises into products of first order operators \eqref{ScotFactor} associated with partially-massless superfields \cite{CottonAdS}. Partially-massless superfields in AdS$^{3|2}$ were discussed for the first time in \cite{CottonAdS}, of which there are two types \eqref{SPMG1} and \eqref{SPMG2}. 
These results allowed us to lift the topologically massive actions of \cite{KT17} to AdS$^{3|2}$ \cite{CottonAdS}, where we also discussed their `new topologically massive' variant  \cite{CottonAdS}, see eq. \eqref{5.39} and \eqref{SHSNTMG} respectively.

Finally, in section \ref{secSCHSCF3} we formulated gauge-invariant models \eqref{SCHSact3dN=1CSS} for the $\mc{N}=1$ SCHS prepotential $H_{\a(n)}$ on arbitrary conformally-flat backgrounds \cite{Confgeo}. We also provided analogous results for $\mc{N}=2$ SCHS theories \eqref{N=2SCHS3act} and described how to obtain the $\mc{N}$-extended versions by combining the results of \cite{BHHK} and \cite{BKNT-M1}.
Models for generalised (higher-depth) SCHS gauge prepotentials were not discussed.


\chapter{SCHS models in four dimensional $\mc{N}=1$ superspace} \label{Chapter4Dsuperspace}

In this chapter we elaborate on gauge invariant models for superconformal higher-spin (SCHS) multiplets on various curved $4d$ $\mc{N}=1$ superspace backgrounds. 
When dealing with such theories, the gravitational field belongs to the so-called Weyl multiplet, which also  contains a conformal gravitino and a $\sU(1)$ gauge field. It appears that  consistent propagation of SCHS multiplets on such a background may be defined  only if the corresponding super-Bach tensor vanishes. Therefore, the background Weyl multiplet is a solution to the equations of motion for $\mc{N}=1$ conformal supergravity. 

This observation is in accordance with the pattern established thus far; that (S)CHS theories may be consistently defined only on backgrounds which solve the equation of motion for the conformal (super)gravity theory which they generalise. 
The SCHS models discussed in this chapter also embody another important principle; that any CHS theory  may be embedded in an off-shell SCHS theory.
Consequently, an effective method of studying various CHS models is to study the corresponding SCHS models which induce them (and vice versa). This idea is exemplified in the later sections of this chapter.

This chapter is based on the publications \cite{Confgeo, SCHS, SCHSgen, AdSuperprojectors} and is organised as follows. 
In section \ref{secGenM4|4} we review the geometry of $\sU(1)$ superspace, which is then used to describe (i) off-shell 4$d$ $\mc{N}=1$ conformal supergravity; and (ii) the generic features of SCHS theories on arbitrary backgrounds. In sections \ref{secMink4|4} and \ref{secAdS4|4} we specialise to $4d$ $\mc{N}=1$ Minkowski ($\mb{M}^{4|4}$) and anti-de Sitter (AdS$^{4|4}$) superspace backgrounds respectively. The SCHS action on AdS$^{4|4}$ is shown to factorise into products of second-order operators. 
Within the framework of conformal superspace, gauge invariant models for SCHS multiplets and their higher-depth cousins  are formulated on all conformally-flat backgrounds in section \ref{secCF4|4}. In some cases, gauge invariance of the SCHS action is extended to arbitrary Bach-flat backgrounds in section \ref{secBach4|4}.  
A summary of the results obtained is given in section \ref{secSCHS4dis}.

\section{Superconformal higher-spin models in $\mc{M}^{4|4}$}\label{secGenM4|4}

There are different ways to describe $4d$ $\mc{N}=1$ conformal supergravity in superspace, including (i) the Grimm-Wess-Zumino geometry \cite{GWZ1, GWZ2} (see \cite{BK} for a review); (ii) $\sU(1)$ superspace \cite{Howe1, Howe2} (see also the works \cite{GGRS,BGG}); and (iii) conformal superspace \cite{ButterN=1}. As a starting point, we will make use of the $\sU(1)$ superspace formalism, and then transition to the more efficient conformal superspace approach. We begin by reviewing (ii), which was developed in the early 80s by Howe. Our presentation follows that of \cite{BK11, KR}.



\subsection{Conformal supergravity} \label{secCSGSS4}

Consider a curved four dimensional $\cN=1$ 
superspace, denoted by $\cM^{4|4}$, and parametrised by local coordinates
$z^{M}=(x^m,\q^{\mu}, \bar{\q}_{\mud})$, with $m=0,1,2,3$ and $\mu =1,2$. Here
the $x^m$ are bosonic and real, whilst both $\q_{\mu}$ and $\qB^{\mud}$ are fermionic and related via complex conjugation, $\bar{\q}^{\mud}=(\q^{\mu})^*$. 
The structure group of $\sU(1)$ superspace is $\sSL(2,{\mathbb C})\times \sU(1)_{R}$, which means that its covariant derivatives take the form 
\bea
\bm \cD_{A}&=& \big( \bm\cD_a, \bm\cD_\a, \bm\cDB^{\ad} \big)= \bm e_{A}{}^{M}\pa_M-\frac{1}{2}\bm\o_A{}^{bc}M_{bc}-\ri\Q_A\mb{Y} ~.
\label{U(1)CD}
\eea
Here $M_{bc}$ are the generators of Lorentz transformations with corresponding connection $\bm \o_A{}^{bc}$, and $\mb{Y}$ is the generator of chiral $\sU(1)$ rotations with corresponding connection $\Q_A$. The $R$-symmetry generator $\mb{Y}$ is normalised according to 
\begin{align}
\big[\mb{Y},\bm\cD_{\a}\big]=\bm\cD_{\a}~,\qquad \big[\mb{Y},\bm\cDB_{\ad}\big]=-\bm\cDB_{\ad}~.
\end{align}

The covariant derivatives are characterised by the graded commutation relations 
\bea
{[}\bm \cD_{{A}},\bm\cD_{{B}}\}&=&
- \bm T_{ {A}{B} }{}^{{C}}\bm\cD_{{C}}
-\hf \bm R_{{A} {B}}{}^{{cd}}M_{{cd}}-\ri\bm{F}_{AB}\mb{Y}~,
\label{U(1)universalAlg}
\eea
where $\bm T_{ {A}{B} }{}^{{C}}$ and $\bm R_{{A} {B}}{}^{{cd}}$  are
the torsion and Lorentz curvature tensors respectively,  and $\bm{F}_{AB}$ is the $\sU(1)_R$ field strength. 
In order to describe conformal supergravity, the covariant derivatives 
have to obey certain torsion constraints \cite{Howe1,Howe2} such that 
the algebra \eqref{U(1)universalAlg} takes the following form 
\begin{subequations} \label{U(1)algebraRevamped}
	\bea
	&& {} \quad \{ \bm\cD_{\a} , \bm\cDB_{\ad} \} = -2{\rm i} \bm\cD_{\a \ad} ~, \\
	\{ \bm\cD_{\a}, \bm\cD_{\b} \} &=& -4{\bar R} M_{\a \b}~, \qquad
	\{\bm\cDB_{\ad}, \bm\cDB_{\bd} \} =  4R {\bar M}_{\ad \bd}~,\label{U(1)algebraRevampedb} \\
	\left[ \bm\cD_{\a} , \bm\cD_{ \b \bd } \right]
	& = &
	{\rm i}
	{\ve}_{\a \b}
	\Big({\bar R}\,\bm\cDB_\bd + G^\g{}_\bd\bm \cD_\g
	- (\bm\cD^\g G^\d{}_\bd)  M_{\g \d}
	+2{ \bm{\bar W}}_\bd{}^{\gd \dot{\d}}
	{\bar M}_{\gd \dot{\d} }  \Big) \non \\
	&&
	+ {\rm i} (\bm\cDB_{\bd} {\bar R})  M_{\a \b}
	-\frac{\ri}{3} \ve_{\a\b} \bar X^\gd \bar M_{\gd \bd} - \frac{\ri}{2} \ve_{\a\b} \bar X_\bd \mb{Y}
	~, \\
	\left[  \bm\cDB_{\ad} , \bm\cD_{\b\bd} \right]
	& = &
	- {\rm i}
	\ve_{\ad\bd}
	\Big({R}\,\bm\cD_{\b} + G_\b{}^\gd \bm\cDB_\gd
	- (\bm\cDB^{\gd} G_{\b}{}^{\dd})  \bar M_{\gd \dd}
	+2{ \bm{W}}_\b{}^{\g \d}
	{M}_{\g \d }  \Big) \non \\
	&&
	- {\rm i} (\bm\cD_\b R)  {\bar M}_{\ad \bd}
	+\frac{\ri}{3} \ve_{\ad \bd} X^{\g} M_{\g \b} - \frac{\ri}{2} \ve_{\ad\bd} X_\b \mb{Y}
	~,\\
	\left[ \bm\cD_{\a \ad} , \bm\cD_{\b \bd} \right] & = & \ve_{\a \b} \bar \psi_{\ad \bd} + \ve_{\ad \bd} \psi_{\a \b} ~.\label{U(1)VecDer}
	\eea
	In eq. \eqref{U(1)VecDer} we have defined the operators $\psi_{\a(2)}$ and $\bar{\psi}_{\ad(2)}$ as follows
	\bea
	\psi_{\a \b} & = & - \ri G_{ ( \a }{}^{\gd} \bm\cD_{\b ) \gd} + \frac{1}{2} \bm\cD_{( \a } R \bm\cD_{\b)} + \frac{1}{2} \bm\cD_{ ( \a } G_{\b )}{}^{\gd} \bm\cDB_{\gd} +  \bm{W}_{\a \b}{}^{\g} \bm\cD_{\g} \non \\
	&& + \frac{1}{6} X_{( \a}\bm \cD_{ \b)} + \frac{1}{4} (\bm\cD^{2} - 8R) {\bar R} M_{\a \b} + \bm\cD_{( \a}  \bm{W}_{ \b)}{}^{\g \d} M_{\g \d} \non \\
	&& - \frac{1}{6} \bm\cD_{( \a} X^{\g} M_{\b) \g} - \frac{1}{2} \bm\cD_{ ( \a} \bm\cDB^{\gd} G_{\b)}{}^{\dd} {\bar M}_{\gd \dd} + \frac{1}{4} \bm\cD_{( \a} X_{\b)} \mb{Y} ~, \\
	{\bar \psi}_{\ad \bd} & = & \ri G^{\g}{}_{( \ad} \bm\cD_{\g \bd)} - \frac{1}{2} \bm\cDB_{( \ad } {\bar R} \bm\cDB_{\bd)} - \frac{1}{2}\bm \cDB_{ ( \ad } G^{\g}{}_{\bd)} \bm\cD_{\g} - { \bm{\bar W}}_{\ad \bd}{}^{\gd} \bm\cDB_{\gd} \non \\
	&& - \frac{1}{6} {\bar X}_{( \ad} \bm\cDB_{ \bd)} + \frac{1}{4} (\bm\cDB^{2} - 8{\bar R}) R {\bar M}_{\ad \bd} - \bm\cDB_{( \ad} { \bm{\bar W}}_{ \bd)}{}^{\gd \dd} {\bar M}_{\gd \dd} \non \\
	&& + \frac{1}{6} \bm\cDB_{( \ad} {\bar X}^{\gd} {\bar M}_{\bd) \gd} + \frac{1}{2} \bm\cDB_{ ( \ad} \bm\cD^{\g} G^{\d}{}_{\bd)} M_{\g \d} + \frac{1}{4} \bm\cDB_{( \ad} {\bar X}_{\bd)} \mb{Y} ~.
	\eea
\end{subequations}

By solving the aforementioned constraints and the resulting Bianchi identities, the tensors  $\bm T_{ {A}{B} }{}^{{C}}$, $\bm R_{{A} {B}}{}^{{cd}}$ and $\bm{F}_{AB}$ have been expressed in terms of the irreducible 
superfields $G_{\a\ad},~R,~X_{\a}$ and $ \bm{W}_{\a\b\g}= \bm{W}_{(\a\b\g)}$. More specifically, $G_{\a\ad}=\bar{G}_{\a\ad}$ is a real vector superfield whilst the remainder are complex chiral superfields
\begin{align}
\bm\cDB_{\ad} R = 0 ~, \qquad \bm\cDB_{\ad} X_{\a} = 0 ~, \qquad \bm\cDB_{\ad}  \bm{W}_{\a\b\g}=0~.
\end{align}
They are related via the following Bianchi identities:
\begin{subequations}
	\bea
	X_{\a} &=& \bm\cD_{\a}R - \bm\cDB^{\ad}G_{\a \ad} ~,\label{Bianchi1M4U(1)}\\
	\bm\cD^{\a} X_{\a} &=& \bm\cDB_{\ad} {\bar X}^{\ad} ~, \\
	\bm\cD^{\g}  \bm{W}_{\a \b \g} &=& {\rm i} \bm\cD_{(\a}{}^{\gd} G_{\b ) \gd} - \frac{1}{3} \bm\cD_{(\a} X_{\b)} ~.
	\eea
\end{subequations}
 Furthermore, under $\sU(1)_R$ they each possess the following charges,\footnote{Complex conjugation flips the sign of the $\sU(1)_R$ charge of a superfield; if $\mb{Y}\F=q_{\F}\F$ then $\mb{Y}\bar{\F}=-q_{\F}\bar{\F}$. \label{footnoteChiral}}
 \begin{align}
  \mb{Y}G_{\a\ad}=0~, \qquad \mb{Y}R=-2R~,\qquad \mb{Y}X_{\a}=-X_{\a}~,\qquad \mb{Y} \bm{W}_{\a(3)}=- \bm{W}_{\a(3)}~. \label{TorsionU(1)Charges}
 \end{align}

The constraints imposed on the torsion and curvature tensors proposed in \cite{Howe1,Howe2} are invariant under the following super-Weyl transformation of $\bm\cD_A$, 
\begin{subequations}
\label{superWeylTfRevamped}
\bea
\delta_{\S}^{(\text{weyl})}\bm\cD_{\a} & = & \frac{1}{2} \S \bm\cD_{\a} + 2 \bm\cD^{\b} \Sigma M_{\b \a} - \frac{3}{2} \bm\cD_{\a} 
\Sigma \mb{Y} ~, \\
\delta_{\S}^{(\text{weyl})} \bm\cDB_{\ad} & = & \frac{1}{2} \S \bm\cDB_{\ad} + 2 \bm\cDB^{\bd} \S {\bar M}_{\bd \ad} +
\frac{3}{2} \bm\cDB_{\ad} \S \mb{Y} ~, \\
\delta_{\S}^{(\text{weyl})} \bm\cD_{\a \ad} & = & \S \bm\cD_{\a \ad} + {\rm i} \bm\cD_{\a} \S \bm\cDB_{\ad} 
+ {\rm i} \bm\cDB_{\ad} \S \bm\cD_{\a}  + {\rm i} \bm\cDB_{\ad} \bm\cD^{\b} \S  M_{\b \a} \non \\
&& + {\rm i} \bm\cD_{\a} \bm\cDB^{\bd} \S { \bar M}_{\bd \ad} + \frac{3}{4} {\rm i}  \left[ \bm\cD_{\a} , \bm\cDB_{\ad} \right]\S \mb{Y} ~,
\eea
\end{subequations}
provided that the torsion superfields transform as follows
\begin{subequations}\label{superWeylTfTorsionsRevamped}
\bea
\delta_{\S}^{(\text{weyl})} R & = & \S R + \frac{1}{2} \bm\cDB^{2} \S ~, \\
\delta_{\S}^{(\text{weyl})} G_{\a \ad} & = &  \S G_{\a \ad} + [ \bm\cD_{\a} , \bm\cDB_{\ad} ] \S ~, \\
\delta_{\S}^{(\text{weyl})} X_{\a} & = & \frac{3}{2} \S X_{\a} - \frac{3}{2} (\bm\cDB^{2} - 4 R) \bm\cD_{\a} \S ~,\label{U(1)FSSweyl}\\
\delta_{\S}^{(\text{weyl})}  \bm{W}_{\a\b\g} & = & \frac{3}{2}\S  \bm{W}_{\a\b\g}~. \label{superWeylsuperWeyl}
\eea
\end{subequations}
Here the parameter $\S$ is a real unconstrained scalar superfield. The geometry described above is precisely the $\sU(1)$ superspace geometry of \cite{Howe1,Howe2,GGRS} in the form (up to conventions) described in \cite{BK11,KR}.

In four dimensions, the $\mc{N}=1$ conformal supergravity gauge group $\bm{\mc{G}}$ as described above is generated by the following local transformations (i) supertranslations with gauge parameter $\xi^A(z)=\big(\xi^a(z), \xi^{\a}(z),\bar{\xi}_{\ad}(z)\big)$; (ii) Lorentz transformations  with gauge parameter $K^{ab}(z)=-K^{ba}(z)$; (iii)  chiral $\sU(1)_R$ rotations with real gauge parameter $\rho(z)$; and (iv) super-Weyl transformations with real gauge parameter $\S(z)$. Under $\bm{\mc{G}}$ the covariant derivatives transform according to the rule:
\begin{align}
\delta_{\L}^{(\scriptsize\bm{\mc{G}})}\bm{\mc{D}}_A=\big[ \L, \bm{\mc{D}}_{A} \big]+\d^{(\text{weyl})}_\S \bm{\mc{D}}_A~,\qquad \L:= \xi^A\bm{\mc{D}}_{A}+\frac{1}{2}K^{ab}M_{ab}-\ri\rho\mb{Y}~.
\end{align}

A tensor superfield $\Phi(z)$ (with its Lorentz indices suppressed) is said to be primary with superconformal weight $\Delta_{\Phi}$ and $\sU(1)_R$ charge $q_{\F}$ if its transformation law under $\bm{\mc{G}}$ is 
\begin{align}
\delta_{\L}^{(\bm{\scriptstyle{\cG}})}\Phi=\Big(\xi^A\bm{\mc{D}}_{A}+\frac{1}{2}K^{ab}M_{ab}-\ri q_{\F}\rho +\Delta_{\Phi}\S \Big)\Phi~. \label{PrimarySFM4}
\end{align}
Suppose further that $\F$ is chiral, $\bm\cDB_{\ad}\F=0$. Then, using \eqref{superWeylTfRevamped}, we see that in order for this constraint to be covariant under super-Weyl transformations, we require that the relation
\begin{align}
\bm\cDB_{\ad}\F=0\qquad \implies \qquad q_{\F}=-\frac{2}{3}\Delta_{\F} \label{CovChiralCon}
\end{align}
between its weight and charge holds, and that it carries no dotted indices $\bar{M}_{\ad\bd}\F=0$.    

 A real supervector field $\z= \z^B  \bm e_B$  is called conformal Killing if 
\bea
\delta_{\L[\z]}^{(\scriptsize\bm{\mc{G}})}\bm\cD_A=\Big[ \z^B \bm\cD_B + \hf K^{\b \g} [\z] M_{\b\g} -i\rho[\z]\mb{Y}, \bm\cD_A\Big]  + \delta^{(\text{weyl})}_{\S [\z]}  \bm\cD_{A} = 0 ~,
 \label{CCSVF.a4}
\eea
for some Lorentz ($K^{\b \g}[\z]$), $R$-symmetry ($\rho[\z]$) and super-Weyl ($\S[ \z] $) parameters whose explicit form may be found, for example, in \cite{KR}. In particular, it may be shown that the gauge parameters in \eqref{CCSVF.a4} are completely determined by the component $\z^a$ of $\z^A$, which obeys the superconformal Killing equation
\bea
\bm\cD_{(\b}\z_{\a)\ad}=0\quad \Longleftrightarrow \quad \bm\cDB_{(\bd}\z_{\a\ad)}=0~. \label{CCSVF.c4}
\eea
The conformal Killing supervector fields of $\cM^{4|4}$ span the conformal superalgebra of the curved superspace.\footnote{In the case of a conformally-flat superspace, see eq. \eqref{CFS4}, this algebra is isomorphic to the $\mc{N}=1$ superconformal algebra $\mathfrak{su}(2,2|1)$.} 
If we regard $\cM^{4|4}$ as a fixed background superspace, then the primary superfield $\F$ (see \eqref{PrimarySFM4}) possesses the following rigid superconformal transformation law 
\bea
\delta_{\L[\z]}^{(\bm{\scriptstyle{\cG}})} \F = \Big(\z^B \bm\cD_B +\hf K^{\b\g}[\z] M_{\b\g} -\ri q_{\F}\rho[\z]+ \Delta_{\F} \S[\z]\Big) \F ~.
\label{RigidSCTM4}
\eea

Of particular interest in \eqref{U(1)algebraRevamped} is the curvature tensor $ \bm{W}_{\a\b\g}$, which is a covariantly chiral and primary superfield, as can be seen from \eqref{TorsionU(1)Charges} and \eqref{superWeylsuperWeyl}.   It is the $\mc{N}=1$ super-Weyl tensor, which is a superspace generalisation of the Weyl tensor $ W_{\a\b\g\d}$. In particular, it can be shown that curved superspace $\mc{M}^{4|4}$ is conformally-flat if and only if $ \bm{W}_{\a\b\g}$ vanishes,
\begin{align}
 \bm{W}_{\a\b\g}=0~. \label{CFS4}
\end{align}
Since $ \bm{W}_{\a(3)}$ is crucial for our subsequent analysis, we collect its main properties below,
\begin{align}
\bm\cDB_{\ad} \bm{W}_{\a(3)}=0~,\qquad \delta_{\S}^{(\text{weyl})}  \bm{W}_{\a(3)} =  \frac{3}{2}\S  \bm{W}_{\a(3)}~, \qquad \mb{Y}  \bm{W}_{\a(3)}=- \bm{W}_{\a(3)}~.
\end{align}

The action for $4d$ $\mc{N}=1$ conformal supergravity\footnote{Linearised conformal supergravity  on $\mb{M}^{4|4}$ was constructed by Ferrara and Zumino \cite{FZ2,FZ78}.}  takes the form \cite{Siegel78,Zumino}
\begin{align}
\mb{S}_{\text{CSG}}= \int \rd^4x\, \rd^2\q\, \cE\,   \bm{W}^{\a\b \g} \bm{W}_{\a\b\g} 
+{\rm c.c.} ~, \label{N=1CSGA}
\end{align}
where $\cE$ is the chiral integration measure (for more details, see e.g. \cite{BK}). It may be shown that the super-Bach tensor,\footnote{The terminology ``super-Bach tensor'' was introduced in \cite{KMT}. In linearised conformal supergravity the super-Bach tensor was first computed in \cite{FZ78}.} 
\bea\label{U(1)super-Bach}
 \bm{B}_{\a \ad} & = & \ri \bm\cD^{\b}{}_{\ad} \bm\cD^{\g}  \bm{W}_{\a \b \g} + G_{\b \ad} \bm\cD_{\g}  \bm{W}_{\a}{}^{\b \g} + \bm\cD_{\b} G_{\g \ad}  \bm{W}_{\a}{}^{\b \g} =  \bm{\bar{B}}_{\a\ad}~,
\eea
introduced in \cite{BK88, BK} (see also \cite{KMT}), arises as a functional derivative of $\mb{S}_{\text{CSG}}$ with respect to the gravitational superfield $H^{\a\ad}$ \cite{Siegel78}. More specifically,
\bea
\d  \int \rd^4x \rd^2 \q \, \cE\,  \bm{W}^{\a \b \g} \bm{W}_{\a\b\g } =
\int \text{d}^{4|4}z \, E\, \delta^{(\text{cov})} H^{\a\ad}  \bm{B}_{\a\ad}~,\qquad E^{-1}:=\text{Ber}(\bm e_{A}{}^{M})
\eea
where $\text{d}^{4|4}z=\rd^4x \rd^2 \q  \rd^2 \bar \q $ and 
$\delta^{(\text{cov})} H^{\a\ad} $ denotes the covariant variation of the gravitational superfield
defined in \cite{GrisaruSiegel1,GrisaruSiegel2}. The super-Bach tensor satisfies the conservation identities 
\bea
\bm\cD^{\a}  \bm{B}_{\a \ad} = 0 ~, \qquad \bm\cDB^{\ad}  \bm{B}_{\a \ad} = 0 ~, \label{SBcons}
\eea
which expresses the gauge invariance of the conformal supergravity action. 

\subsection{Superconformal higher-spin gauge prepotentials} \label{SCHSprepSS4}

Let $m$ and $n$ be two positive integers and consider the tensor superfield $H_{\a(m)\ad(n)}$. In order for $H_{\a(m)\ad(n)}$ to be an $\mc{N}=1$ conformal superspin-$\frac{1}{2}(m+n+1)$ gauge prepotential, it must be a primary superfield with super-Weyl weight $\Delta_{H_{(m,n)}}=-\frac{1}{2}(m+n)$,
\begin{subequations}
\begin{align}
\delta_{\S}^{(\text{weyl})}H_{\a(m)\ad(n)}&=-\frac{1}{2}(m+n)\S H_{\a(m)\ad(n)}~,    
\end{align} 
and it must carry $\sU(1)_R$ charge $q_{H_{(m,n)}}=\frac{1}{3}(m-n)$,
\begin{align}
\mb{Y}H_{\a(m)\ad(n)}&=\frac{1}{3}(m-n)H_{\a(m)\ad(n)}~.
\end{align}
\end{subequations}
Furthermore, the SCHS multiplet $H_{\a(m)\ad(n)}$ must be defined modulo certain gauge transformations which vary depending on the values of $m$ and $n$ \cite{HST81, KMT, KR}:
\begin{itemize}


\item For $1\leq m \neq n \geq 1$, the SCHS supermultiplet $H_{\a(m)\ad(n)}$ is complex and is defined modulo gauge transformations\footnote{We point out that these gauge transformations are invariant under the gauge-for-gauge symmetries $\d\L_{\a(m-1)\ad(n)}= \bm\cD_{\a}\L_{\a(m-2)\ad(n)}$ and $\d\O_{\a(m)\ad(n-1)}=\bm\cDB_{\ad}\O_{\a(m)\ad(n-2)}$. Hence the corresponding gauge theory is reducible in the sense of \cite{BV}. Similar comments apply to \eqref{SCHSprepGTII} and \eqref{SCHSprepGTIII}.} of the form \cite{KMT}
\begin{subequations}\label{SCHSprepGT}   
   \begin{align}
  \delta_{\L,\O}H_{\a(m)\ad(n)}=\bm\cD_{\a}\L_{\a(m-1)\ad(n)}+\bm\cDB_{\ad}\O_{\a(m)\ad(n-1)}~, \label{SCHSprepGTI}
   \end{align}
   for unconstrained primary complex gauge parameters $\L_{\a(m-1)\ad(n)}$ and $\O_{\a(m)\ad(n-1)}$ .


\item For $m=n=s \geq 1$, the SCHS supermultiplet $H_{\a(s)\ad(s)}$ is defined to be real
\begin{align}
H_{\a(s)\ad(s)}=\bar{H}_{\a(s)\ad(s)}~,
\end{align} 
and to possess the gauge freedom \cite{HST81}
   \begin{align}
  \delta_{\L}H_{\a(s)\ad(s)}=\bm\cD_{\a}\L_{\a(s-1)\ad(s)}-\bm\cDB_{\ad}\bar{\L}_{\a(s)\ad(s-1)}~,\label{SCHSprepGTII}
   \end{align}
   for unconstrained primary complex gauge parameter $\L_{\a(s-1)\ad(s)}$ .


\item For $m>n=0$, the complex SCHS supermultiplet  $H_{\a(m)}$  is defined modulo gauge transformations of the form \cite{KR}
\begin{align}
  \delta_{\L,\l}H_{\a(m)}=\bm\cD_{\a}\L_{\a(m-1)}+\l_{\a(m)}~,\qquad \bm\cDB_{\ad}\l_{\a(m)}=0 \label{SCHSprepGTIII}
\end{align}
\end{subequations}
where the complex gauge parameter $\L_{\a(m-1)}$ is an unconstrained primary superfield, whilst the complex gauge parameter $\l_{\a(m)}$ is  primary and covariantly chiral . 

\end{itemize}


In each case, the super-Weyl weight $\Delta_{H_{(m,n)}} $ and $\sU(1)_R$ charge $q_{H_{(m,n)}}$ are uniquely fixed by requiring both the gauge variation of $H_{\a(m)\ad(n)}$, denoted generically by $\delta_{\L} H_{\a(m)\ad(n)}$, and the corresponding gauge parameters to be super-Weyl primary and using \eqref{superWeylTfRevamped}. The analogous properties for $\bar{H}_{\a(n)\ad(m)}$ may be obtained by complex conjugation of the above formulae (bearing in mind footnote \ref{footnoteChiral}).\footnote{In $4d$ $\mc{N}=1$ supergravity, the covariant spinor derivatives satisfy the following complex conjugation relations: $\big(\bm\cD_{\a}\F\big)^*=(-1)^{\ve_{\F}}\bm\cDB_{\ad}\bar{\F}$ where $\ve_{\F}$ is the Grassmann parity of the tensor superfield $\Phi$.} 
For a fixed background superspace, $H_{\a(m)\ad(n)}$ possesses the rigid superconformal transformation law \eqref{RigidSCTM4}. 

Some comments regarding several special choices of $m$ and $n$ are in order. The gauge prepotential $H_{\a(s) \ad(s)}$ describes the conformal superspin-$(s+\hf) $ multiplet, with the lowest choice $s=1$ corresponding to linearised conformal supergravity.
It is one of the dynamical variables in terms of which the off-shell massless superspin-$(s+\hf)$ multiplets in Minkowski and AdS backgrounds are formulated \cite{KPS,KS94}.

The second special case corresponds to $m= n+1 =s>1$.
The gauge prepotential $H_{\a(s) \ad(s-1)}$ and its conjugate,
along with certain compensating supermultiplets, 
are used to describe the off-shell massless superspin-$s$ multiplet
in Minkowski and AdS backgrounds, originally proposed in  \cite{KS,KS94}
and recently  reformulated in \cite{HK2,BHK}.
  
Thirdly, the family of supermultiplets with $n=0$ and $m$ arbitrary generalise the superconformal gravitino multiplet ($m=n+1=1$). The corresponding gauge transformation rule \eqref{SCHSprepGTIII} was first proposed in \cite{KR}. The latter extend the transformation 
law given by Gates and Siegel \cite{GS} who studied an off-shell formulation 
for the  massless gravitino supermultiplet in Minkowski superspace. 

\subsection{Higher-spin super-Weyl and super-Bach tensors }

From the SCHS multiplet $H_{\a(m)\ad(n)}$ one can construct its descendent $\mf{W}^{(m,n)}_{\a(m+n+1)}(H)$,\footnote{Similar to the non-supersymmetric case, we regard $\mf{W}^{(m,n)}$ as an operator acting on a rank-$(m,n)$ superfield. In particular, the super-Weyl tensor associated with $\bar{H}_{\a(n)\ad(m)}$ is $\mf{W}^{(n,m)}_{\a(m+n+1)}(\bar{H})$, and its weight and charge are $\Delta_{\mf{W}^{(n,m)}}$ and $q_{\mf{W}^{(n,m)}}$.  We drop the labels $(m,n)$ in $\mf{W}^{(m,n)}_{\a(m+n+1)}(H)$ when it is clear which prepotential we are talking about.
Furthermore, one could also construct the HS super-Weyl tensor $\bar{\mf{W}}_{\ad(m+n+1)}(H)$ with only dotted indices (similar to e.g. \eqref{ThisIsNotImportant}), but we will not do so here. } known as the higher-spin super-Weyl tensor, which satisfies the following properties:
\begin{enumerate}[label=(\roman*)]

\item
$\mf{W}_{\a(m+n+1)}(H)$ is a primary superfield with Weyl weight $\Delta_{\mf{W}^{(m,n)}}=\frac{1}{2}\big(3+n-m\big)$ and $\sU(1)_R$ charge $q_{\mf{W}^{(m,n)}}=-\frac{1}{3}\big(3+n-m\big)$, 
\begin{subequations} \label{HSSWeylprop1}
\begin{align}
\d_\S^{(\text{weyl})} \mf{W}_{\a(m+n+1)}(H) &=\phantom{-}\frac{1}{2}\big(3+n-m\big)  \S \mf{W}_{\a(m+n+1)}(H)~,\label{HSSWeylprop1a}\\
\mb{Y}\mf{W}_{\a(m+n+1)}(H) &= -\frac{1}{3}\big(3+n-m\big)\mf{W}_{\a(m+n+1)}(H)~.\label{HSSWeylprop1b}
\end{align}
\end{subequations}

\item 
$\mf{W}_{\a(m+n+1)}(H)$ is a descendent of the form $\mf{W}_{\a(m+n+1)}(H)=\big(\mc{A}H\big)_{\a(m+n+1)}$. Here $\mc{A}$ is a linear differential operator of order $2n+3$ (in spinor derivatives) and has charge $-1$. It involves the covariant derivative $\bm\cD_A$, the curvature/torsion tensors $R,X_{\a}, G_{\a\ad}$ and $\bm W_{\a\b\g}$, and their covariant derivatives.  

\item
$\mf{W}_{\a(m+n+1)}(H)$ is covariantly chiral,
\begin{align}
\bm\cDB_{\ad}\mf{W}_{\a(m+n+1)}(H)=0~,\label{HSSWeylprop2}
\end{align}
which is consistent with \eqref{CovChiralCon}.

\item
$\mf{W}_{\a(m+n+1)}(H)$ has vanishing gauge variation under the relevant gauge transformations  \eqref{SCHSprepGT} if $\mc{M}^{4|4}$ is conformally-flat,
\begin{align}\label{HSSWeylprop3}
\d_{\L}\mf{W}_{\a(m+n+1)} (H)&= \mc{O}\big( \bm W\big)~,
\end{align}
Here $\mc{O}\big( \bm W\big)$ stands for contributions involving the super-Weyl tensor and
its covariant derivatives.
\end{enumerate}

From the SCHS multiplet $H_{\a(m)\ad(n)}$ we may construct another important superconformal field strength $\mf{B}^{(m,n)}_{\a(n)\ad(m)}(H)$, known as the higher-spin super-Bach tensor, which possesses the following properties:
\begin{enumerate}[label=(\roman*)]

\item $\mf{B}_{\a(n)\ad(m)}(H)$ is a primary superfield with Weyl weight $\Delta_{\mf{B}^{(m,n)}}=2+\frac{1}{2}\big(m+n\big)$ and $\sU(1)_R$ charge $q_{\mf{B}^{(m,n)}}=\frac{1}{3}\big(m-n\big)$,
\begin{subequations} \label{HSSBachprop1}
\begin{align}
\d_\S^{(\text{weyl})} \mf{B}_{\a(n)\ad(m)}(H) &=\Big(2+\frac{1}{2}\big(m+n\big) \Big) \S \mf{B}_{\a(n)\ad(m)}(H)~,\label{HSSBachprop1a}\\
\mb{Y}\mf{B}_{\a(n)\ad(m)}(H) &= \frac{1}{3}\big(m-n\big)\mf{B}_{\a(n)\ad(m)}(H)~.\label{HSSBachprop1b}
\end{align}
\end{subequations}

\item 
$\mf{B}_{\a(n)\ad(m)}(H)$ is a descendent of the form $\mf{B}_{\a(n)\ad(m)}(H)=\big(\mc{A}H\big)_{\a(n)\ad(m)}$. Here $\mc{A}$ is a linear differential operator of order $m+n+2$ (in vector derivatives) and has charge $0$. It involves the covariant derivative $\bm\cD_A$, the curvature/torsion tensors $R,X_{\a}, G_{\a\ad}$ and $\bm W_{\a\b\g}$, and their covariant derivatives.  

\item For $m\geq 1$ and $n\geq 1 $,  $\mf{B}_{\a(n)\ad(m)}(H)$ is TLAL if $\mc{M}^{4|4}$ is Bach-flat,
\begin{subequations}\label{HSSBachprop2}
\begin{align}
\bm\cD^{\b}\mf{B}_{\b\a(n-1)\ad(m)}(H)=\mc{O}\big( \bm B\big)~,\qquad \bm\cDB^{\bd}\mf{B}_{\a(n)\ad(m-1)\bd}(H)=\mc{O}\big( \bm B\big)~. \label{HSSBachprop2a}
\end{align}
For $m>n=0$,  $\mf{B}_{\ad(m)}(H)$ is ALTL if $\mc{M}^{4|4}$ is Bach-flat,
\begin{align}
\Big(\bm\cD^2-4\bar{R}\Big) \mf{B}_{\ad(m)}(H)=\mc{O}\big( \bm B\big)~,\qquad \bm\cDB^{\bd}\mf{B}_{\bd\ad(m-1)}(H)=\mc{O}\big( \bm B\big)~.\label{HSSBachprop2b}
\end{align} 
\end{subequations}
See the discussion immediately below for an explanation of the abbreviations. 
\item
$\mf{B}_{\a(n)\ad(m)}(H)$ has vanishing gauge variation under the relevant gauge transformations  \eqref{SCHSprepGT} if $\mc{M}^{4|4}$ is Bach-flat,
\begin{align}\label{HSSBachprop3}
\d_{\L}\mf{B}_{\a(n)\ad(m)}(H)&= \mc{O}\big( \bm B\big)~,
\end{align}
Here and in \eqref{HSSBachprop2}, $\mc{O}\big( \bm B\big)$ stands for contributions involving the super-Bach tensor and its covariant derivatives.
\end{enumerate}

As we discuss in section \ref{secSCHSactGenSS} below, for non-conformally-flat backgrounds, a field strength  $ \mf{B}_{\a(n)\ad(m)} (H)$ satisfying these properties does not exist for general $m$ and $n$. 
Moreover, the above properties do not determine the higher-spin super-Weyl or super-Bach tensors uniquely in non-conformally-flat and non-Bach-flat superspace backgrounds respectively (since one may add appropriate terms involving $\bm W_{\a(3)}$ or $\bm B_{\a\ad}$). 
In the case $m=n=1$, the descendants $\mf{W}_{\a(3)}(H)$ and $\mf{B}_{\a\ad}(H)$ prove to coincide with the linearised versions of $\bm W_{\a(3)}$ and $\bm B_{\a\ad}$ respectively.


In this chapter we will encounter superfields $\F_{\a(m)}$, with $m>n=0$, which are linear, and also superfields $\F_{\ad(n)}$, with $n>m=0$, which are anti-linear:
\begin{subequations}\label{(A)LinearGenSS}
\begin{align}
\Big(\bm\cDB^2-4R\Big) \F_{\a(m)}&=0~,\label{LinearGenSS}\\
\Big(\bm\cD^2-4\bar{R}\Big) \F_{\ad(n)}&=0~.\label{ALinearGenSS}
\end{align}
\end{subequations}
We recall that a superfield $\F_{\a(m)\ad(n)}$, with $m\geq 1$ and $n\geq 1$, which is transverse with respect to either spinor derivatives satisfies a corresponding (anti-)linearity constraint 
\begin{subequations}\label{TLALGenSS}
\begin{align}
\bm\cDB^{\bd}\F_{\a(m)\ad(n-1)\bd}&=0\qquad \implies \qquad \Big(\bm\cDB^2-2(n+2)R\Big)\F_{\a(m)\ad(n)}=0~,\label{TLGenSS}\\
\bm\cD^{\b}\F_{\b\a(m-1)\ad(n)}&=0\qquad \implies \qquad \Big(\bm\cD^2-2(m+2)\bar{R}\Big)\F_{\a(m)\ad(n)}=0~,\label{TALGenSS}
\end{align}
\end{subequations}
as a consequence of \eqref{U(1)algebraRevampedb}. Here we have denoted $\bm\cD^2=\bm\cD^{\a}\bm\cD_{\a}$ and $\bm\cDB^2=\bm\cDB_{\ad}\bm\cDB^{\ad}$. 

A superfield $\F_{\a(m)}$ satisfying \eqref{LinearGenSS} is said to be linear, whilst $\F_{\ad(n)}$  satisfying \eqref{ALinearGenSS} is said to be anti-linear.
If $\F_{\a(m)\ad(n)}$ satisfies \eqref{TLGenSS} it is said to be transverse linear, whilst it is called transverse anti-linear if it satisfies \eqref{TALGenSS}. A superfield $\F_{\a(m)}$ satisfying \eqref{LinearGenSS} and \eqref{TALGenSS} is said to be simultaneously linear and transverse anti-linear, or LTAL for brevity. A superfield $\F_{\ad(n)}$ satisfying \eqref{ALinearGenSS} and \eqref{TLGenSS} is simultaneously anti-linear and transverse linear, or ALTL. If $\F_{\a(m)\ad(n)}$  satisfies both \eqref{TLGenSS} and \eqref{TALGenSS}, then it is said to be simultaneously transverse linear and transverse anti-linear, or TLAL.


If one further requires that the TLAL field $\F_{\a(m)\ad(n)}$ or ALTL field $\F_{\ad(n)}$ be primary, then, in order for the conditions \eqref{TLALGenSS} to be preserved under super-Weyl transformations, its Weyl weight and $\sU(1)_R$ charge are uniquely fixed to take the values $\Delta_{\F_{(m,n)}}=2+\frac{1}{2}(m+n)$ and $q_{\F_{(m,n)}}=\frac{1}{3}(n-m)$ respectively. The transformations rules \eqref{superWeylTfRevamped} and \eqref{superWeylTfTorsionsRevamped} should be used to prove this. Such a supermultiplet, of which the higher-spin super-Bach tensor is an example, carries the characteristic features of a conserved conformal supercurrent. 


\subsection{Superconformal higher-spin action}\label{secSCHSactGenSS}

Suppose that our curved background superspace $\mc{M}^{4|4}$ is conformally-flat,
\begin{align}
\bm W_{\a\b\g}=0~.
\end{align}
Then, given an SCHS gauge prepotential $H_{\a(m)\ad(n)}$, it follows from \eqref{HSSWeylprop3} that its super-Weyl tensor (and that of its conjugate $\bar{H}_{\a(n)\ad(m)}$) is gauge invariant
\begin{align}
\delta_{\L} \mf{W}_{\a(m+n+1)}(H)=0~,\qquad \delta_{\L}\mf{W}_{\a(m+n+1)}(\bar{H})=0~.\label{HSSWeylGISS4}
\end{align}
Since both field strengths are also chiral,\footnote{This ensures that the action \eqref{SCHSactSWeylSS4} is invariant under superdiffeomorphisms. } eq. \eqref{HSSWeylprop2}, then one can construct the action 
\begin{align}
\mb{S}_{\text{SCHS}}^{(m,n)}[H,\bar{H}]=\frac{1}{2}\ri^{m+n}\int \rd^4x\rd^2\theta \, \cE\, \mf{W}^{\a(m+n+1)}(H)\mf{W}_{\a(m+n+1)}(\bar{H})+\text{c.c.}~, \label{SCHSactSWeylSS4}
\end{align}
where the integration is restricted to the chiral submanifold, corresponding to the constant slice $\bar{\theta}_{\mud}=0$ (the `c.c.' sector is correspondingly restricted to the anti-chiral submanifold). By virtue of the properties \eqref{HSSWeylGISS4} and \eqref{HSSWeylprop1a}, this action is invariant under both gauge and super-Weyl transformations,
\begin{align}
\delta_{\L}\mb{S}_{\text{SCHS}}^{(m,n)}[H,\bar{H}] =0~, \qquad \d_\S^{(\text{weyl})}\mb{S}_{\text{SCHS}}^{(m,n)}[H,\bar{H}] =0~.
\end{align} 
Here we have used that the chiral measure transforms according to $\d_\S^{(\text{weyl})}\cE = -3\S \cE$. On account of \eqref{HSSWeylprop1b} and $\mb{Y}\cE = 2\cE$, the action \eqref{SCHSactSWeylSS4} is also invariant under local $\sU(1)_R$ transformations.

Suppose instead that the background superspace under consideration is Bach-flat,
\begin{align}
\bm B_{\a\ad}=0~.
\end{align}
Then, given an SCHS field $H_{\a(m)\ad(n)}$, the properties \eqref{HSSBachprop2} and \eqref{HSSBachprop3} imply that its super-Bach tensor is gauge invariant and TLAL (or perhaps ALTL)
\begin{subequations}
\begin{align}
\delta_{\L}\mf{B}_{\a(n)\ad(m)}(H)&=0~,\\
\bm\cD^{\b}\mf{B}_{\b\a(n-1)\ad(m)}(H)=0~,\qquad \bm\cDB^{\bd}&\mf{B}_{\a(n)\ad(m-1)\bd}(H)=0~.
\end{align}
\end{subequations}
These properties, along with the transformation rules \eqref{HSSBachprop1}, $\d_\S^{(\text{weyl})}E=-2E$ and $\mb{Y}E=0$ (and the corresponding rules for the prepotential), ensure that the action
\begin{align}
\mb{S}_{\text{SCHS}}^{(m,n)}[H,\bar{H}]=\frac{1}{2}\ri^{m+n}\int \rd^{4|4}z \, E \, \bar{H}^{\a(n)\ad(m)}\mf{B}_{\a(n)\ad(m)}(H)+\text{c.c.}~, \label{SCHSactSBachSS4}
\end{align}
is invariant under local gauge, super-Weyl and $\sU(1)_R$ transformations. It is therefore invariant under the conformal supergravity gauge group $\bm\cG$, which means that for a fixed  background, it is invariant under the rigid superconformal transformations \eqref{RigidSCTM4},
\begin{align}
\delta_{\L[\z]}^{(\bm{\scriptstyle{\cG}})}\phantom{.}\mb{S}_{\text{SCHS}}^{(m,n)}[H,\bar{H}]=0~. \label{RigidSCsymSCHS}
\end{align}

The two superconformal higher-spin actions \eqref{SCHSactSWeylSS4} and \eqref{SCHSactSBachSS4} turn out to be equivalent on conformally-flat backgrounds, but not on generic Bach-flat backgrounds. This is because both $\mf{W}_{\a(m+n+1)}(H)$ and  $\mf{B}_{\a(n)\ad(m)}(H)$ always exist on conformally-flat backgrounds, and are related to each other via simple integration by parts. To establish this relation, one would need to lift the functional \eqref{SCHSactSWeylSS4} to the full superspace using the rule
\begin{align}
\int \rd^{4|4}z \, E \, \mc{L} = -\frac{1}{4}\int \rd^4x\rd^2\theta \, \cE\, \Big(\bm\cDB^2-4R\Big)\mc{L}~, \label{U(1)chiralintegralrule}
\end{align}
for some scalar superfield $\mc{L}(z)$.  

Only $\mf{W}_{\a(m+n+1)}(H)$ is guaranteed to exist in a generic Bach-flat background, and it will not be gauge invariant, as prescribed in \eqref{HSSWeylprop3}. In contrast, the higher-spin super-Bach tensor $\mf{B}_{\a(n)\ad(m)}(H)$ does not generally exist on Bach-flat backgrounds. Indeed, similar to the non-supersymmetric case, we will see that in order to restore gauge invariance to the SCHS action, one needs to introduce 
non-minimal counter terms as well as couplings to additional superconformal multiplets. The fact that $\mf{B}_{\a(n)\ad(m)}(H)$ does not generally exist on Bach-flat backgrounds is a consequence of this last statement.


\section{Superconformal higher-spin models in $\mb{M}^{4|4}$}\label{secMink4|4}

Within the component setting, superconformal higher-spin theories on $\mb{M}^4$ were first introduced by Fradkin and Linetsky \cite{FL-4D} at the quadratic and cubic level. More recently, the off-shell gauge supermultiplet $H_{\a(m)\ad(n)}$, described in section \ref{SCHSprepSS4}, was used in \cite{KMT} to construct free SCHS actions on $\mc{N}=1$ Minkowski superspace $\mb{M}^{4|4}$. In this section we summarise the elements of SCHS models on $\mb{M}^{4|4}$ which are relevant for our purposes. 

For the covariant derivatives $D_A=\big(\pa_a, D_{\a},\DB^{\ad}\big)$  of $\mb{M}^{4|4}$ we adopt the conventions of \cite{BK}, in which $D_A$ satisfy the anti-commutation relations
\begin{subequations}
\begin{align}
\big\{D_{\a},\DB_{\ad}\big\}=&-2\ri \pa_{\a\ad}~,\\
\big\{D_{\a},D_{\b}\big\}=0~,\qquad &\big\{\DB_{\ad},\DB_{\bd}\big\}=0~.
\end{align}
\end{subequations}
Many useful identities may be readily obtained from this algebra, such as
\begin{subequations}
\begin{align}
D_{\a}D_{\b}&=\phantom{-}\frac{1}{2}\ve_{\a\b}D^2~,\qquad \big[D^2,\DB_{\ad}\big]=-4\ri \pa_{\a\ad}D^{\a}~,\\
\DB_{\ad}\DB_{\bd}&=-\frac{1}{2}\ve_{\ad\bd}\DB^2~, \qquad \big[\DB^2,D_{\a}\big]=\phantom{-}4\ri \pa_{\a\ad}\DB^{\ad}~,
\end{align}
\end{subequations}
where we have denoted $D^2=D^{\a}D_{\a}$ and $\DB^2=\DB_{\ad}\DB^{\ad}$. Other useful identities may be obtained by taking the flat limit of \eqref{A.2Blah}.

Let us first consider the SCHS supermultiplet $H_{\a(m)\ad(n)}$ and its conjugate $\bar{H}_{\a(n)\ad(m)}$ with  strictly positive integers $m\geq 1 $ and $n\geq 1$. In Minkowski superspace, $H_{\a(m)\ad(n)}$ is defined modulo the gauge transformations
\begin{align}
  \delta_{\L,\O}H_{\a(m)\ad(n)}=D_{\a}\L_{\a(m-1)\ad(n)}+\DB_{\ad}\O_{\a(m)\ad(n-1)}~. \label{SCHSprepGTImink4}
 \end{align}
The corresponding super-Weyl tensors take the form
\begin{subequations}\label{HSSWeylMink4}
\begin{align}
\mf{W}_{\a(m+n+1)}(H)&=-\frac{1}{4}\DB^2\pa_{(\a_1}{}^{\bd_1}\cdots\pa_{\a_n}{}^{\bd_n}D_{\a_{n+1}}H_{\a_{n+2}\dots\a_{m+n+1})\bd(n)}~,\label{HSSWeylMink4a}\\
\mf{W}_{\a(m+n+1)}(\bar{H})&=-\frac{1}{4}\DB^2\pa_{(\a_1}{}^{\bd_1}\cdots\pa_{\a_m}{}^{\bd_m}D_{\a_{m+1}}\bar{H}_{\a_{m+2}\dots\a_{m+n+1})\bd(m)}~.\label{HSSWeylMink4b}
\end{align}
\end{subequations}
 They are clearly chiral and gauge invariant by construction,
\begin{subequations} \label{HSSWeylpropMinkaloo}
\begin{align}
\DB_{\ad}\mf{W}_{\a(m+n+1)}(H)&=0~,\qquad \delta_{\L,\O}\mf{W}_{\a(m+n+1)}(H)=0~,\\
\DB_{\ad}\mf{W}_{\a(m+n+1)}(\bar{H})&=0~,\qquad  \delta_{\L,\O}\mf{W}_{\a(m+n+1)}(\bar{H})=0~.
\end{align}
\end{subequations}
Rather than building super-Weyl tensors with indices of only the undotted type, we could have instead formulated them with only dotted indices. In this case, one would arrive at the complex conjugates of \eqref{HSSWeylMink4},
\begin{subequations}\label{HSSWeylMink4cc}
\begin{align}
\overline{\mf{W}}_{\ad(m+n+1)}(H)&=(-1)^{m+n+1}\frac{1}{4}D^2\pa_{(\ad_1}{}^{\b_1}\cdots\pa_{\ad_m}{}^{\b_m}\DB_{\ad_{m+1}}H_{\b(m)\ad_{m+2}\dots\ad_{m+n+1})}~,\\
\overline{\mf{W}}_{\ad(m+n+1)}(\bar{H})&=(-1)^{m+n+1}\frac{1}{4}D^2\pa_{(\ad_1}{}^{\b_1}\cdots\pa_{\ad_n}{}^{\b_n}\DB_{\ad_{n+1}}\bar{H}_{\b(n)\ad_{n+2}\dots\ad_{m+n+1})}~,
\end{align}
\end{subequations}
 which are gauge invariant and anti-chiral. 
 
 From the prepotential $H_{\a(m)\ad(n)}$ one may also derive its higher-spin super-Bach tensors for which there are two types:
\begin{subequations}\label{HSSBACHmink4}
 \begin{align} 
 \mf{B}_{\a(n)\ad(m)}(H)&=-\frac{1}{4}\pa_{(\ad_1}{}^{\b_1}\cdots\pa_{\ad_m)}{}^{\b_m}D^{\g}\DB^2D_{(\g}\pa_{\a_1}{}^{\bd_1}\cdots\pa_{\a_n}{}^{\bd_n}H_{\b_1\dots\b_m)\bd(n)}~,\label{HSSBACHmink4a}\\
 \widehat{\mf{B}}_{\a(n)\ad(m)}(H)&=\phantom{-}\frac{1}{4}\pa_{(\a_1}{}^{\bd_1}\cdots\pa_{\a_n)}{}^{\bd_n}\DB^{\gd}D^2\DB_{(\gd}\pa_{\ad_1}{}^{\b_1}\cdots\pa_{\ad_m}{}^{\b_m}H_{\b(m)\bd_1\dots\bd_n)}~.\label{HSSBACHmink4b}
 \end{align}
 \end{subequations} 
 They are each simultaneously transverse linear and transverse anti-linear (TLAL),
 \begin{subequations}\label{HSSBACHTLALmink}
 \begin{align}
 D^{\b}\mathfrak{B}_{\b\a(n-1)\ad(m)}(H)&= 0~, \qquad \DB^{\bd}\mathfrak{B}_{\a(n)\ad(m-1)\bd}(H)=0~,\\
D^{\b}\widehat{\mathfrak{B}}_{\b\a(n-1)\ad(m)}(H)&= 0~, \qquad \DB^{\bd}\widehat{\mathfrak{B}}_{\a(n)\ad(m-1)\bd}(H)=0~.
 \end{align}
 \end{subequations} 
In a fashion similar to the proof of \eqref{HSBachTypeEquiv}, one may show that the two are equivalent,
\begin{align}
\mf{B}_{\a(n)\ad(m)}(H)=\widehat{\mf{B}}_{\a(n)\ad(m)}(H)~.
\end{align} 
Consequently, the following non-trivial relation holds
\begin{align}
\mf{B}_{\a(m)\ad(n)}(\bar{H})=\Big(\mf{B}_{\a(n)\ad(m)}(H)\Big)^*\equiv \overline{\mf{B}}_{\a(m)\ad(n)}(\bar{H}) ~.\label{HSSBachCCMink}
\end{align}
In addition to satisfying the conservation equations \eqref{HSSBACHTLALmink}, the tensors \eqref{HSSBACHmink4} are also gauge invariant, which is made manifest when they are expressed in terms of the higher-spin super-Weyl tensors:
\begin{subequations}\label{sBachsWeyl}
\begin{align}
\mf{B}_{\a(n)\ad(m)}(H)&=\pa_{(\ad_1}{}^{\b_1}\cdots\pa_{\ad_m)}{}^{\b_m}D^{\b_{m+1}}\mf{W}_{\a(n)\b(m+1)}(H)~,\\
\widehat{\mf{B}}_{\a(n)\ad(m)}(H)&=(-1)^{m+n+1}\pa_{(\a_1}{}^{\bd_1}\cdots\pa_{\a_n)}{}^{\bd_n}\DB^{\bd_{n+1}}\overline{\mf{W}}_{\ad(m)\bd(n+1)}(H)~.
\end{align}
\end{subequations}

When $m=n=s$, the prepotential satisfies the reality condition $H_{\a(s)\ad(s)}=\bar{H}_{\a(s)\ad(s)}$, and has the gauge transformation law \eqref{SCHSprepGTImink4} with $\O_{\a(s)\ad(s-1)}=-\bar{\L}_{\a(s)\ad(s-1)}$. In this case, all of the above formulae apply equally well. In addition, we find that \eqref{HSSWeylMink4a} and \eqref{HSSWeylMink4b} coincide and that  the identity \eqref{HSSBachCCMink} implies $\mf{B}_{\a(s)\ad(s)}(H)$ is real. 
Next we consider the  SCHS multiplet $H_{\a(m)}$ with $m>n=0$, having conjugate $\bar{H}_{\ad(m)}$, which is defined modulo the gauge transformations
\begin{align}
  \delta_{\L,\l}H_{\a(m)}=D_{\a}\L_{\a(m-1)}+\l_{\a(m)}~,\qquad \DB_{\ad}\l_{\a(m)}=0~. \label{SCHSprepGTIIImink4}
\end{align}
The above formulae are also valid in this case, provided one interprets expressions such as $\pa_{\a_1}{}^{\bd_1}\cdots\pa_{\a_0}{}^{\bd_0}$ to be equal to unity. The only change is that the higher-spin super-Bach tensor $\mf{B}_{\ad(m)}(H)$ has only dotted spinorial indices, and is ALTL rather than TLAL. 

 The above properties mean that the superspin-$\frac{1}{2}(m+n+1)$ SCHS action
 \begin{subequations}\label{SCHSactMink4|4}
 \begin{align}
 \mb{S}_{\text{SCHS}}^{(m,n)}[H,\bar{H}]&= \frac{1}{2}\ri^{m+n}\int \rd^4x\rd^2\theta  \mf{W}^{\a(m+n+1)}(H)\mf{W}_{\a(m+n+1)}(\bar{H})+\text{c.c.}~\\
 &= \frac{1}{2}(-\ri)^{m+n+2} \int \rd^{4|4}z \bar{H}^{\a(n)\ad(m)}\mf{B}_{\a(n)\ad(m)}(H) +\text{c.c.}
 \end{align}
 \end{subequations}
is manifestly gauge invariant. It is also invariant under the rigid $\mc{N}=1$ superconformal transformations of $\mb{M}^{4|4}$, which follows since \eqref{SCHSactMink4|4} admits a super-Weyl and $\sU(1)_R$ invariant extension to curved superspace, as we show in section \ref{secSCHSCF}. We note that the operator $\mf{B}^{(m,n)}$ is symmetric in the sense
\begin{align}
\int \rd^{4|4}z \bar{H}^{\a(n)\ad(m)}\mf{B}_{\a(n)\ad(m)}(H)= \int \rd^{4|4}z H^{\a(m)\ad(n)}\mf{B}_{\a(m)\ad(n)}(\bar{H})~,
\end{align}
which holds up to a total derivative. In conjunction with the identity \eqref{HSSBachCCMink}, this relation implies that the following identity holds true
\begin{align}
\ri^{m+n+1}\int \rd^{4|4}z \bar{H}^{\a(n)\ad(m)}\mf{B}_{\a(n)\ad(m)}(H) +\text{c.c.}=0~,
\end{align}
which explains the overall normalisation of \eqref{SCHSactMink4|4}.

 In section \ref{secSCHSCF} we will extend these models to arbitrary conformally-flat backgrounds using the framework of conformal superspace. There we also study the content of the action $\mb{S}_{\text{SCHS}}^{(m,n)}[H,\bar{H}]$ (on bosonic conformally-flat backgrounds) at the component level and compare it to the results of chapter \ref{Chapter4D}. For a discussion of this component analysis we refer the reader to section  \ref{secSCHStoCHS}.

\section{Superconformal higher-spin models in AdS$^{4|4}$} \label{secAdS4|4}
In this section we study the gauge invariant actions for superconformal higher-spin fields in $4d$ $\mc{N}=1$ AdS superspace, or AdS$^{4|4}$.\footnote{Originally, AdS$^{4|4}$ was  introduced \cite{Keck,Zumino77,IS} as the coset space ${\rm AdS^{4|4} } := {{\sOSp}(1|4)}/{{\sSO}(3,1)}$. The same superspace is equivalently realised as a unique maximally symmetric solution of the following two off-shell formulations for $\cN=1$ AdS supergravity:
(i) the well-known minimal theory (see, e.g., \cite{GGRS,BK} for reviews); and 
(ii)  the more recently discovered non-minimal theory \cite{BK11}.} In particular, with the help of the AdS superspin projection operators, we demonstrate that the kinetic operator of the AdS$^{4|4}$ SCHS action factorises into products of second-order operators assocatiated with (partially-)massless supermultiplets. To establish this connection, we will first need to describe on-shell partially-massless supermultiplets at the superspace level, which itself is a novel result.\footnote{The structure of on-shell $\mathcal{N}=1$ supermultiplets containing partially-massless fields was recently discussed 
 at the component level in \cite{G-SHR, BKSZ}.} This section is based on selected results from our paper \cite{AdSuperprojectors}.



\subsubsection{$\mc{N}=1$ AdS superspace geometry}\label{secAdS4|4Geometry}

The geometry of ${\rm AdS}^{4|4}$ is charcterised by its corresponding covariant derivatives\footnote{One may arrive at this geometry from the $\sU(1)$ superspace geometry described in section \ref{secCSGSS4} via the following steps. First, the super-Weyl freedom should be partially fixed by using \eqref{U(1)FSSweyl} to impose the gauge $X_{\a}=0$. In this gauge the $\sU(1)_R$ curvature vanishes, and the $\sU(1)_R$ connection $\Theta_A$ may be gauged away. There is a class of residual super-Weyl transformations preserving this gauge, see e.g. \cite{KR} for the explicit transformations rules.  After this, one arrives at the Grimm-Wess-Zumino superspace geometry \cite{GWZ1,GWZ2}, for which AdS$^{4|4}$ corresponds to the case where  $R=\mu,~G_{\a\ad}=0$ and $\bm W_{\a(3)}=0$.  }
\bea
\bm \cD_{A}&=& \big( \bm\cD_a, \bm\cD_\a, \bm\cDB^{\ad} \big)= \bm e_{A}{}^{M}\pa_M-\frac{1}{2}\bm\o_A{}^{bc}M_{bc} ~,
\label{AdS4|4CD}
\eea
which satisfy the graded commutation relations:
 \begin{subequations}\label{algebraAdS4|4}
\bea
\{\bm\cD_\a,\bm\cDB_{\ad} \} &=& -2 \ri \bm\cD_{\a\ad}~, \\
\{ \bm\cD_\a, \bm\cD_\b \} &=& -4\mub M_{\a\b}~, \qquad \{ \bm\cDB_\ad, \bm\cDB_\bd \} = 4 \m \bar{M}_{\ad \bd}~, \\
\, [\bm\cD_\a , \bm\cD_{\b\bd} ] &=& \ri \mub \varepsilon_{\a\b}\bm\cDB_{\bd},  \qquad  \, [\bm\cDB_\ad, \bm\cD_{\b\bd} ] = - \ri \m \varepsilon_{\ad\bd}\bm\cD_\b, \\
\, [\bm\cD_{\a\ad},\bm\cD_{\b\bd}] &=& - 2\mub \m (\varepsilon_{\a\b}\bar{M}_{\ad\bd} + \varepsilon_{\ad\bd}M_{\a\b} )~.
\eea
\end{subequations}
Here $\m\neq 0$ is a constant complex parameter. 

We make use of the following identities, which can be readily derived from \eqref{algebraAdS4|4}:
\begin{subequations} 
\label{A.2Blah}
\bea 
\bm\cD_\a\bm\cD_\b
\!&=&\!\frac{1}{2}\ve_{\a\b}\bm\cD^2-2{\bar \m}\,M_{\a\b}~,
\quad\qquad \,\,\,
{\bm\cDB}_\ad{\bm\cDB}_\bd
=-\frac{1}{2}\ve_{\ad\bd}{\bm\cDB}^2+2\m\,{\bar M}_{\ad\bd}~,  \label{A.2aBlah}\\
\bm\cD_\a\bm\cD^2
\!&=&\!4 \bar \m \,\bm\cD^\b M_{\a\b} + 4{\bar \m}\,\bm\cD_\a~,
\quad\qquad
\bm\cD^2\bm\cD_\a
=-4\bar \m \,\bm\cD^\b M_{\a\b} - 2\bar \m \, \bm\cD_\a~, \label{A.2bBlah} \\
{\bm\cDB}_\ad{\bm\cDB}^2
\!&=&\!4 \m \,{\bm\cDB}^\bd {\bar M}_{\ad\bd}+ 4\m\, \bm\cDB_\ad~,
\quad\qquad
{\bm\cDB}^2{\bm\cDB}_\ad
=-4 \m \,{\bm\cDB}^\bd {\bar M}_{\ad\bd}-2\m\, \bm\cDB_\ad~,  \label{A.2cBlah}\\
\left[\bm\cDB^2, \bm\cD_\a \right]
\!&=&\!4\rm i \bm\cD_{\a\bd} \bm\cDB^\bd +4 \m\,\bm\cD_\a = 
4\rm i \bm\cDB^\bd \bm\cD_{\a\bd} -4 \m\,\bm\cD_\a~,
 \label{A.2dBlah} \\
\left[\bm\cD^2,{\bm\cDB}_\ad\right]
\!&=&\!-4\rm i \bm\cD_{\b\ad}\bm\cD^\b +4\bar \m\,{\bm\cDB}_\ad = 
-4\rm i \bm\cD^\b \bm\cD_{\b\ad} -4 \bar \m\,{\bm\cDB}_\ad~,
 \label{A.2e}
\eea
\end{subequations} 
where $\bm\cD^2=\bm\cD^\a\bm\cD_\a$, and ${\bm\cDB}^2={\bm\cDB}_\ad{\bm\cDB}^\ad$. 
Other useful identities are: 
\begin{subequations}\label{A.4Blah}
	\bea
	\bm\cD_\a{}^\bd \bm\cD_\bd{}^\b &=& \d_\a{}^\b \bm\Box -2\m\mub M_\a{}^\b~, \label{A.4aBlah}\\
	\bm\cD^\a{}_\ad \bm\cD_\a{}^\bd &=& \d_\ad{}^\bd \bm\Box - 2 \mu \mub \bar{M}_\ad{}^\bd~,
	\eea
\end{subequations}
where $\bm\Box=\bm\cD^a\bm\cD_a$. Of special importance are the relations:
\begin{subequations} \label{A.5Blah}
\bea
\bm\Box +2\m \bar \m = 
- \frac 18 \bm\cD^{\a}\big(\bm\cDB^2-4\mu\big)\bm\cD_{\a} 
&+&\frac{1}{16} \big\{\bm\cD^2 -4\bar \m ,\bm\cDB^2 -4 \m\big\} ~, \label{A.5aBlah}\\
\bm\cD^{\a}\big(\bm\cDB^2-4\mu\big)\bm\cD_{\a}
&=& \bm\cDB_{\ad}\big(\bm\cD^2-4\mub\big)\bm\cDB^{\ad}~,\\
\big[\bm\cD^2,\bm\cDB^2\big]=-4\text{i}\big[\bm\cD^\b,\bm\cDB^{\bd}\big]\bm\cD_{\b\bd}&+&8\mu\bm\cD^2-8\mub\bm\cDB^2 \non \\
=-4\text{i}   \bm\cD_{\b\bd}\big[\bm\cD^\b,\bm\cDB^{\bd}\big]
&-&8\mu\bm\cD^2+8\mub\bm\cDB^2~.
\eea
\end{subequations}

In what follows, we make use of the Casimir operator $\mathbb{Q}$ of the $\mc{N}=1$ AdS$_4$ superalgebra $\mf{osp}(1|4)$, whose realisation on superfields takes the form \cite{BKS}
\bea \label{Cas1}
\mathbb{Q}:=\bm\Box+\frac{1}{4}\Big(\mu\bm\cD^2+\mub\bm\cDB^2\Big)
&-&\mu\mub\Big(M^{\a\b}M_{\a\b}+\bar{M}^{\ad\bd}\bar{M}_{\ad\bd}\Big)~,\qquad  \big[\mathbb{Q},\bm\cD_{A}\big]=0~, 
\eea
which is the $\mc{N}=1$ supersymmetric extension of \eqref{QCasimirAdS4}. We point out that the unitary irreducible representations of $\mf{osp}(1|4)$ were studied in \cite{Heidenreich:1982rz} (see also \cite{Nicolai:1984hb, deWit:1999ui}).


\subsection{On-shell supermultiplets}


Given two integers $m\geq 1$ and $n\geq 1$, a supermultiplet $\F_{\a(m)\ad(n)}$ on AdS$^{4|4}$ is said to be on-shell if it satisfies the set of equations
\begin{subequations}\label{OSAdS4|4}
\begin{align}
\big(\mb{Q}-M^2\big)&\F_{\a(m)\ad(n)}=0~,\label{OSAdS4|4,1}\\
\bm\cDB^{\bd}\F_{\a(m)\ad(n-1)\bd}=0~,&\qquad \bm\cD^{\b}\F_{\b\a(m-1)\ad(n)\bd}=0~.\label{OSAdS4|4,2}
\end{align}
\end{subequations} 
Hence the superfield $\Phi_{\a(m)\ad(n)}$ is simultaneously transverse linear and transverse anti-linear (TLAL), and is said to have super-spin $s=\frac{1}{2}(m+n+1)$ and pseudo-mass $M$. In the case when $m>n=0$ and $n>m=0$, the condition \eqref{OS2} should be modified to
\begin{subequations} \label{OSAdS4|43}
\begin{align}
\bm\cD^{\b}\F_{\a(m-1)\b}&=0~, \qquad \big(\bm\cDB^2-4\mu\big)\F_{\a(m)}=0~,  \label{OSAdS4|43.a}\\
\bm\cDB^{\bd}\F_{\ad(n-1)\bd}&=0~, \qquad \big(\bm\cD^2-4\mub\big)\F_{\ad(n)}=0~,  \label{OSAdS4|43.b}
\end{align} 
\end{subequations}
respectively, whilst \eqref{OSAdS4|4,1} remains the same.

Given an on-shell supermultiplet $\F_{\a(m)\ad(n)}$ satisfying \eqref{OSAdS4|4}, it is said to be massless if its pseudo-mass takes the value
\begin{align}
M^2=\lambda_{(1,m,n)}\mu\mub \label{masslessAdS4|4}
\end{align}
whilst it is said to be partially-massless with depth-$t$ if $M$ satisfies 
\begin{align}
M^2=\lambda_{(t,m,n)}\mu\mub~,\qquad 2 \leq t \leq  \text{min}(m+1,n+1)~.\label{DefineSPM}
\end{align}
For fixed $m$ and $n$, a strictly massless supermultiplet may be considered to be a partially-massless supermultiplet with depth $t=1$.
The supermultiplet $\F_{\a(m)\ad(n)}$ is said to be massive if $M^2>\lambda_{(1,m,n)}$.\footnote{This ensures that the supermultiplet furnishes a unitary representation of $\mf{osp}(1|4)$, see e.g. \cite{AdSuperprojectors}. The massless representations are unitary whilst the true partially-massless representations are non-unitary.}
Here we have defined a family of dimensionless parameters $\lambda_{(t,m,n)}$, called partially-massless values, according to
\begin{align}
\l_{(t,m,n)} = \hf \Big [ (m+n-t+1)(m+n-t+4) + t(t-1) \Big ]~.\label{SPM}
\end{align}
 
In the case of an on-shell massless superfield, which we denote by $\F^{(1)}_{\a(m)\ad(n)}$, the corresponding system of equations is invariant under the gauge transformations
\begin{subequations} \label{SPMgaugesymmetry}
\begin{align}
t=1~: \quad \delta_{\L,\O}\F^{(1)}_{\a(m)\ad(n)}=\bm\cDB_{\ad}\L_{\a(m)\ad(n-1)}+\bm\cD_{\a}\O_{\a(m-1)\ad(n)}~.~~~~~~~~~~~~~~~~~~~~\label{MasslessGIaAdS4|4}
\end{align} 
Similarly, the system of equations satisfied by an on-shell partially massless superfield of depth-$t$, which we denote by $\F_{\a(m)\ad(n)}^{(t)}$, is invariant under certain depth-$t$ gauge transformations. In the special case $m=n=s$,\footnote{We have not yet found the corresponding gauge transformations when $m\neq n$.} they take the form 
\begin{align}
2\leq t \leq s~&: \quad \delta_{\L}\F^{(t)}_{\a(s)\ad(s)}=\big(\bm\cD_{\a\ad}\big)^{t-1}\big(\bm\cDB_{\ad}\L_{\a(s-t+1)\ad(s-t)}-\bm\cD_{\a}\bar{\L}_{\a(s-t)\ad(s-t+1)}\big)~,\label{Blah7AdS4|4}\\[5pt]
t=s+1~&:\quad  \delta_{\s}\F^{(s+1)}_{\a(s)\ad(s)}=\big(\bm\cD_{\a\ad}\big)^{s}\big(\s+\bar{\s}\big)~. \label{Blah8AdS4|4}
\end{align}
\end{subequations}

The massless system of equations is only invariant under \eqref{MasslessGIaAdS4|4} if the gauge parameters $\L_{\a(m)\ad(n-1)}$ and 
$\O_{\a(m-1) \ad (n)}$ are TLAL and obey the constraints
\begin{subequations}
\bea 
\bm\cD_{(\a_1}{}^\bd \O_{\a_2 \dots \a_m) \bd \ad (n-1)} &=&\phantom{-} \ri (n+1) \m \L_{\a(m)\ad (n-1)}~, \label{GoodDaySirA}\\
\bm\cD^\b{}_{(\ad_1} \L_{\b \a(m-1)  \ad_2 \dots  \ad_n)} 
&=& -\ri (m+1) \bar \m \O_{\a(m-1)\ad (n)}~.\label{GoodDaySirB}
\eea
\end{subequations}
These on-shell conditions imply that $\L_{\a(m)\ad (n-1)}$ and 
$\O_{\a(m-1) \ad (n)}$ satisfy the equations  
\bea
 \big(\mathbb{Q}-\lambda_{(1,m,n)}\mu\mub\big)\L_{\a(m)\ad(n-1)}=0~, \qquad
  \big(\mathbb{Q}-\lambda_{(1,m,n)}\mu\mub\big)\O_{\a(m-1)\ad(n)}=0~,\label{Blah3AdS4|4}
\eea
which is consistent with the gauge variation of \eqref{OSAdS4|4,1}.
Likewise, the partially-massless system of equations is invariant under the transformations \eqref{Blah7AdS4|4} as long as the gauge parameters are TLAL and satisfy the reality conditions
\begin{subequations}\label{Blah5AdS4|4}
\begin{align}
\bm\cD^\b{}_{\ad} \L_{\b\a(s-t)\ad(s-t)} &= \phantom{-}\ri (s+1) \mub \bar{\L}_{\a(s-t)\ad (s-t+1)}~, \\
\bm\cD_{\a}{}^{\bd} \bar{\L}_{\a(s-t)\ad(s-t)\bd} &= -\ri (s+1) \mu \L_{\a(s-t+1)\ad (s-t)}~. 
\end{align}
\end{subequations}
The same is true for the transformations \eqref{Blah8AdS4|4} given that $\s$ is chiral, $\bm\cDB_{\ad}\s=0$, and that it satisfies the equations
\begin{subequations}\label{BLAH7AdS4|4}
\begin{align}
-\frac{1}{4}\big(\bm\cD^2-4\mub\big)\s+(s+1)\mub\bar{\s}&=0~,\\
-\frac{1}{4}\big(\bm\cDB^2-4\mu\big)\bar{\s}+(s+1)\mu\s &=0~.
\end{align}
\end{subequations}
It may be shown that \eqref{Blah5AdS4|4} and \eqref{BLAH7AdS4|4} imply that each gauge parameter satisfies the same mass-shell equation as its corresponding gauge field, as required.

\subsection{Transverse superprojectors} \label{construction}

In order to arrive at a factorised form of the SCHS action, we will make use of the superspin projection operators (or simply superprojectors\footnote{Superprojectors have been studied for over 45 years. See \cite{SalamS,Sokatchev,Sokatchev81,RS,SG,GGRS,BHHK} for superprojectors in $\mc{N}$-extended Minkowski superspace, and \cite{OS1,OS2,GS,GKP} for applications thereof.}) in AdS$^{4|4}$. The latter were constructed recently in \cite{AdSuperprojectors}, to where we refer the reader for a more thorough discussion; here we give only the final result.\footnote{The author of this thesis does not claim ownership of the AdS$^{4|4}$ superprojectors \eqref{superprojAdS4|4}. They were derived by Daniel Hutchings, a co-author of \cite{AdSuperprojectors}. Here we use them as a tool to derive the factorisation properties of the SCHS actions.  } Given an unconstrained superfield $\F_{\a(m)\ad(n)}$, with $m$ and $n$ arbitrary positive integers  (either of which may be zero, but not both), the superprojectors $\bm\Pi_{\perp}^{(m,n)}$ take the form
\begin{subequations}\label{superprojAdS4|4}
\begin{align}
\bm\Pi_{\perp}^{(m,n)}\Phi_{\a(m)\ad(n)}\equiv \bm\Pi^{\perp}_{\a(m)\ad(n)}(\Phi)&:=\bigg [\prod_{t=1}^{n+1} \big ( \mathbb{Q}-\l_{(t,m,n)}\m \mub \big ) \bigg ]^{-1}\mathbb{P}_{\a(m)\ad(n)}(\Phi)~, \label{superprojAdS4|41}\\
\bm{\widehat{\Pi}}{}^{(m,n)}_{\perp}\Phi_{\a(m)\ad(n)}\equiv \bm{\widehat{\Pi}}{}^{\perp}_{\a(m)\ad(n)}(\Phi)&:=\bigg [ \prod_{t=1}^{m+1} \big ( \mathbb{Q}-\l_{(t,m,n)}\m \mub \big ) \bigg ]^{-1} \widehat{\mathbb{P}}_{\a(m)\ad(n)}(\Phi)~. \label{superprojAdS4|42}
\end{align}
\end{subequations}
Here we have defined the two higher-derivative descendants
\begin{subequations}\label{strippedprojectors}
\begin{align}
\mathbb{P}_{\a(m)\ad(n)}(\Phi)&=-\frac{1}{8}\bm\cD_{(\ad_1}{}^{\b_1}\dots \bm\cD_{\ad_n)}{}^{\b_n}\bm\cD^{\g}(\bm\cDB^2-4\mu  )\bm\cD_{(\g}\bm\cD_{\b_1}{}^{\bd_1}\cdots\bm\cD_{\b_n}{}^{\bd_n} \Phi_{\a_1\dots\a_m)\bd(n)}~,\label{Pss1}\\
\widehat{\mathbb{P}}_{\a(m)\ad(n)}(\Phi)&=\phantom{-}\frac{1}{8}\bm\cD_{(\a_1}{}^{\bd_1}\dots \bm\cD_{\a_m)}{}^{\bd_m}\bm\cDB^{\gd} (\bm\cD^2-4\mub  )\bm\cDB_{(\gd}\bm\cD_{\bd_1}{}^{\b_1}\cdots\bm\cD_{\bd_m}{}^{\b_m} \Phi_{\b(m)\ad_1\dots\ad_n)}~,\label{Pss2}
\end{align}
\end{subequations}
and the parameters $\lambda_{(t,m,n)}$ are the partially-massless values \eqref{SPM}.

The two types of projectors prove to coincide (much like the HS super-Bach tensors)
\begin{align}
\bm\Pi_{\perp}^{(m,n)}\Phi_{\a(m)\ad(n)}=\bm{\widehat{\Pi}}{}^{(m,n)}_{\perp}\Phi_{\a(m)\ad(n)}~.\label{CoincidenceProjAdS4|4}
\end{align}
In addition, it may be shown that $\bm\Pi_{\perp}^{(m,n)}$ satisfies the following fundamental properties, which are the AdS$^{4|4}$ analogues of the properties \eqref{TTProjPropDDim}:
\begin{itemize}
\item $\bm\Pi_{\perp}^{(m,n)}$ is a projector in the sense that it squares to itself
\begin{subequations}
\begin{align}
\bm\Pi_{\perp}^{(m,n)}\bm\Pi_{\perp}^{(m,n)}\Phi_{\a(m)\ad(n)}=\bm\Pi_{\perp}^{(m,n)}\Phi_{\a(m)\ad(n)}~.
\end{align}

\item $\bm\Pi_{\perp}^{(m,n)}$ projects onto the space of TLAL superfields\footnote{For superprojectors which project onto the space of TL or TAL superfields in AdS$^{4|4}$, see \cite{IS}.} when $m\geq 1$ and $n\geq 1$,\footnote{The operators \eqref{strippedprojectors} also satisfy this property.}
\begin{align}\label{TLALproj}
\bm\cDB^\bd \bm\P^{\perp}_{\a(m)\ad(n-1)\bd}(\Phi)&=0 ~,\qquad \bm\cD^\b \bm\P^{\perp}_{\b\a(m-1)\ad(n)}(\Phi)=0 ~.
\end{align}
In the case $m>n=0$, $\bm\Pi_{\perp}^{(m,0)}$ projects onto the space of LTAL superfields.
 
 \item $\bm\Pi_{\perp}^{(m,n)}$ acts as the identity operator on the space of TLAL superfields $\Psi_{\a(m)\ad(n)}$ 
\begin{align}
\bm\cDB^\bd \Psi_{\a(m)\ad(n-1)\bd}=0~,\qquad &\bm\cD^\b \Psi_{\b\a(m-1)\ad(n)}=0 ~,\non\\
 \implies \quad  \bm\Pi_{\perp}^{(m,n)}\Psi_{\a(m)\ad(n)}&=\Psi_{\a(m)\ad(n)}~.\label{TLALprojSurjective}
\end{align}
\end{subequations}
A similar statement holds for LTAL superfields $\Psi_{\a(m)}$ and the projector $\bm\Pi_{\perp}^{(m,0)}$.
\end{itemize}


Given a superfield $\F_{\a(m)\ad(n)}$ lying on the mass-shell \eqref{OSAdS4|4,1},  it follows from the above properties that its projection $\bm\Pi_{\perp}^{(m,n)}\F_{\a(m)\ad(n)}$ is TLAL and hence satisfies the on-shell conditions \eqref{OSAdS4|4}.  Such superfields correspond to irreducible representations of $\mf{osp}(1|4)$. Similar to the non-supersymmetric case, we see that the poles of the superprojectors \eqref{superprojAdS4|4} contain information regarding the (partially-)massless supermultiplets.


\subsection{Factorisation of the superconformal higher-spin action} \label{section 4}

Let us now consider the case when $\F_{\a(m)\ad(n)}$ is an SCHS prepotential $H_{\a(m)\ad(n)}$ on AdS$^{4|4}$. When $m\geq 1$ and $n\geq 1$, the latter is defined modulo the gauge transformations 
\begin{align}
  \delta_{\L,\O}H_{\a(m)\ad(n)}=\bm\cD_{\a}\L_{\a(m-1)\ad(n)}+\bm\cDB_{\ad}\O_{\a(m)\ad(n-1)}~, \label{SCHSprepGTIAdS}
   \end{align}
whilst in the case $m>n=0$  it has the gauge transformation law
\begin{align}
  \delta_{\L,\l}H_{\a(m)}=\bm\cD_{\a}\L_{\a(m-1)}+\l_{\a(m)}~,\qquad \bm\cDB_{\ad}\l_{\a(m)}=0~. \label{SCHSprepGTIIIAdS}
\end{align}
In both cases, associated with  $H_{\a(m)\ad(n)}$ and its complex conjugate  $\bar H_{\a(n)\ad(m)}$ are the higher-spin super-Weyl tensors (see \cite{KS94} for the $m=n=s$ case)
\begin{subequations}\label{HSWAdS4|4}
\begin{align}
\mathfrak{W}_{\a(m+n+1)}(H)&= -\frac{1}{4}\big(\bm\cDB^2-4\mu\big)\bm\cD_{(\a_1}{}^{\bd_1}\cdots\bm\cD_{\a_n}{}^{\bd_n}\bm\cD_{\a_{n+1}}H_{\a_{n+2}\dots\a_{m+n+1})\bd(n)}~, \label{HSWAdS4|4a}\\
\mathfrak{W}_{\a(m+n+1)}(\bar{H})&= -\frac{1}{4}\big(\bm\cDB^2-4\mu\big)\bm\cD_{(\a_1}{}^{\bd_1}\cdots\bm\cD_{\a_m}{}^{\bd_m}\bm\cD_{\a_{m+1}}\bar{H}_{\a_{m+2}\dots\a_{m+n+1})\bd(m)}~,\label{HSWAdS4|4b}
\end{align}
\end{subequations}
which are chiral and gauge invariant,
\begin{align}
\delta_{\L}\mathfrak{W}_{\a(m+n+1)}(H)=0~,\qquad \delta_{\L}\mathfrak{W}_{\a(m+n+1)}(\bar{H})=0~.
\end{align}
The corresponding higher-spin super-Bach tensors
\begin{subequations} \label{HSBachAdS4|4}
\begin{align}
\mathfrak{B}_{\a(n)\ad(m)}(H)&= -\frac{1}{4}\bm\cD_{(\ad_1}{}^{\b_1}\cdots\bm\cD_{\ad_{m})}{}^{\b_m}\bm\cD^{\g}\big(\bm\cDB^2-4\mu\big)\bm\cD_{(\g}\bm\cD_{\a_1}{}^{\bd_1}\cdots\bm\cD_{\a_n}{}^{\bd_n}H_{\b_1\dots\b_m)\bd(n)}~, \label{HSBach1AdS4|4}\\
\widehat{\mathfrak{B}}_{\a(n)\ad(m)}(H)&=\phantom{-}\frac{1}{4}\bm\cD_{(\a_1}{}^{\bd_1}\cdots\bm\cD_{\a_{n})}{}^{\bd_n}\bm\cDB^{\gd}\big(\bm\cD^2-4\mub\big)\bm\cDB_{(\gd}\bm\cD_{\ad_1}{}^{\b_1}\cdots\bm\cD_{\ad_m}{}^{\b_m}H_{\b(m)\bd_1\dots\bd_n)}~, \label{HSBach2AdS4|4}
\end{align}
\end{subequations}
are TLAL for $m\geq 1$ and $n\geq 1$, and ALTL for $m>n=0$. They also prove to be equal
\begin{align}
\mathfrak{B}_{\a(n)\ad(m)}(H)=\widehat{\mathfrak{B}}_{\a(n)\ad(m)}(H)~. \label{CoincidenceBachAdS4|4}
\end{align}

The SCHS action is typically written as a functional over the chiral subspace of the full superspace (see e.g. \eqref{SCHSactSWeylSS4}).
A specific feature of AdS${}^{4|4}$ is the identity \cite{Siegel78}
\bea
 \int\rd^4x\rd^2\q\, \cE \,\cL_{\rm c}
= \int\rd^4x\rd^2\q\rd^2\qb\, \frac{E}{\m} \,\cL_{\rm c}~, \qquad \bm\cDB_\ad \cL_{\rm c} =0~,
\label{full-chiral}
\eea
which relates the integration over the chiral subspace to that over the full superspace.
Keeping in mind  \eqref{full-chiral}, 
the gauge-invariant action $S_{\text{SCHS}}^{(m,n)}[H,\bar{H}]$ is given by\footnote{This action first appeared in \cite{Confgeo, AdSuperprojectors}, though the higher-spin super-Weyl tensors \eqref{HSWAdS4|4} for $m=n=s$ were written down earlier in \cite{KS94}. } 
\begin{align}
\mb{S}_{\text{SCHS}}^{(m,n)}[H,\bar{H}]=\frac{1}{2}\text{i}^{m+n}\int 
\text{d}^{4|4}z \, \frac{E}{\m} \, 
 \mathfrak W^{\a(m+n+1)}(H)\mathfrak{W}_{\a(m+n+1)}(\bar{H}) +{\rm c.c.}
 ~, \label{SCHSchiral}
\end{align}
Upon integrating by parts, the action \eqref{SCHSchiral} may be written in the form
\begin{subequations} \label{AdS4|4SBaction}
\begin{align}
\mb{S}_{\text{SCHS}}^{(m,n)}[H,\bar{H}]&=\frac{1}{2}(-\ri)^{m+n+2} \int \text{d}^{4|4}z \, E \, \bar{H}^{\a(n)\ad(m)}\mathfrak{B}_{\a(n)\ad(m)}(H) +\text{c.c. } \label{SCHSAdS4|4,1}\\
&=\frac{1}{2}(-\ri)^{m+n+2} \int \text{d}^{4|4}z \, E \, \bar{H}^{\a(n)\ad(m)}\widehat{\mathfrak{B}}_{\a(n)\ad(m)}(H) +\rm{c.c.}\label{SCHSAdS4|4,2}
\end{align}
\end{subequations}
Since the action is gauge invariant, it is possible to impose the TLAL gauge on $H_{\a(m)\ad(n)}$,
\begin{align}
H_{\a(m)\ad(n)}&\equiv  H^{\text{T}}_{\a(m)\ad(n)}~,\non\\
\bm\cD^{\b}H^{\text{T}}_{\b\a(m-1)\ad(n)}=0~,&\qquad \bm\cDB^{\bd}H^{\text{T}}_{\a(m)\ad(n-1)\bd}=0~. \label{TLALgaugeCond}
\end{align}
It should be mentioned that the action $\mb{S}_{\text{SCHS}}^{(m,n)}[H,\bar{H}]$ is invariant under $\mc{N}=1$ superconformal transformations (in the sense of \eqref{RigidSCsymSCHS}), but the above gauge condition is not.

Let us now relate the higher-spin super-Bach tensors to the superprojectors \eqref{superprojAdS4|4}.
We begin with the familiar case when $m=n=s$. It is clear that for these values the higher-spin Bach tensor \eqref{HSBach1AdS4|4} and the descendant \eqref{Pss1} are proportional to each other, so that we have the relations
\begin{align}
\frac{1}{2}\mathfrak{B}_{\a(s)\ad(s)}(H)=\mathbb{P}_{\a(s)\ad(s)}(H)= \prod_{t=1}^{s+1} \big ( \mathbb{Q}-\l_{(t,s,s)}\m \mub \big )\bm\P^{\perp}_{\a(s)\ad(s)}(H)~.\label{bosbachAdS4|4}
\end{align}
Therefore the superspin-$(s+\frac{1}{2})$ SCHS action \eqref{SCHSAdS4|4,1} takes the form 
\begin{align}
\mb{S}_{\text{SCHS}}^{(s,s)}[H]&=2(-1)^{s+1} \int \text{d}^{4|4}z \, E \, H^{\a(s)\ad(s)}\prod_{t=1}^{s+1} \big ( \mathbb{Q}-\l_{(t,s,s)}\m \mub \big )\bm\P^{\perp}_{\a(s)\ad(s)}(H)~.\label{FTSCHSGen1}
\end{align}
It follows from \eqref{TLALprojSurjective} that in the TLAL gauge \eqref{TLALgaugeCond}, we obtain the following factorisation
 \begin{align}\label{ThisguyFactors}
\mb{S}_{\text{SCHS}}^{(s,s)}[H^{\text{T}}]&=2(-1)^{s+1} \int \text{d}^{4|4}z \, E \, H_{\text{T}}^{\a(s)\ad(s)}\prod_{t=1}^{s+1} \big ( \mathbb{Q}-\l_{(t,s,s)}\m \mub \big )H^{\text{T}}_{\a(s)\ad(s)}~.
\end{align}

Next we consider the complex SCHS prepotential $H_{\a(m)\ad(n)}$ with $n > m$. In contrast to \eqref{bosbachAdS4|4}, one may show that in this case the following relation holds
\begin{align}
2\mathbb{P}_{\a(m)\ad(n)}(H)=\big(\bm\cD_{\ad}{}^{\b}\big)^{n-m}\mathfrak{B}_{\a(m)\b(n-m)\ad(m)}(H)~,
\end{align}
which, upon inverting, yields
\begin{align}
\frac{1}{2}\mathfrak{B}_{\a(n)\ad(m)}(H)= \bigg [ \prod_{t=m+2}^{n+1} \big ( \mathbb{Q}-\l_{(t,m,n)}\m \mub \big ) \bigg ]^{-1}&\big(\bm\cD_{\a}{}^{\bd}\big)^{n-m}\mathbb{P}_{\a(m)\bd(n-m)\ad(m)}(H)~.
\end{align}
In terms of the TLAL projector this reads
\begin{align}
\mathfrak{B}_{\a(n)\ad(m)}(H)=\prod_{t=1}^{m+1} \big ( \mathbb{Q}-\l_{(t,m,n)}\m \mub \big ) \big(\bm\cD_{\a}{}^{\bd}\big)^{n-m}\bm\Pi^{\perp}_{\a(m)\bd(n-m)\ad(m)}(H) ~.
\end{align}
The corresponding SCHS action \eqref{HSBach1AdS4|4} is 
\begin{align}
\mb{S}_{\text{SCHS}}^{(m,n)}[H,\bar{H}]=(-\ri)^{m+n+2} \int \text{d}^{4|4}z \, E \, &\bar{H}^{\a(n)\ad(m)}\prod_{t=1}^{m+1} \big ( \mathbb{Q}-\l_{(t,m,n)}\m \mub \big )\non\\
&\times \big(\bm\cD_{\a}{}^{\bd}\big)^{n-m}\bm\Pi^{\perp}_{\a(m)\bd(n-m)\ad(m)}(H) +\text{c.c. }~,
\end{align}
which in the TLAL gauge factorises according to
\begin{align}
\mb{S}_{\text{SCHS}}^{(m,n)}[H^{\text{T}},\bar{H}^{\text{T}}]=(-\ri)^{m+n+2} \int \text{d}^{4|4}z \, E \, &\bar{H}_{\text{T}}^{\a(n)\ad(m)}\prod_{t=1}^{m+1} \big ( \mathbb{Q}-\l_{(t,m,n)}\m \mub \big )~\non\\
&\times \big(\bm\cD_{\a}{}^{\bd}\big)^{n-m}H^{\text{T}}_{\a(m)\ad(m)\bd(n-m)} +\text{c.c. }~~~~~~
\end{align}
 We see that, due to the mismatch of $m$ and $n$, the SCHS action does not factorise wholly into products of second-order operators. However, upon taking appropriate derivatives of the equation of motion resulting from varying the prepotential, one arrives at the following fully factorised on-shell equation 
\begin{align}
0=\prod_{t=1}^{n+1} \big ( \mathbb{Q}-\l_{(t,m,n)}\m \mub \big )H^{\text{T}}_{\a(m)\ad(n)}~. \label{eom1AdS4|4}
\end{align}
According to our definition \eqref{DefineSPM}, we see that some extra (non-unitary) massive modes, corresponding to the values of $\lambda_{(t,m,n)}$ with $m+1< t \leq n+1$, enter the spectrum of the wave equation \eqref{eom1AdS4|4}.

Finally, if $m > n$, then one may show that the following relation holds
\begin{align}
2\widehat{\mathbb{P}}_{\a(m)\ad(n)}(H)=\big(\bm\cD_{\ad}{}^{\b}\big)^{m-n}\widehat{\mathfrak{B}}_{\a(n)\b(m-n)\ad(n)}(H)~.
\end{align}
Upon inverting and expressing in terms of the TLAL projector, we find
\begin{align}
\frac{1}{2}\mathfrak{B}_{\a(n)\ad(m)}(H) 
=\prod_{t=1}^{n+1} \big ( \mathbb{Q}-\l_{(t,m,n)}\m \mub \big ) \big(\bm\cD_{\ad}{}^{\b}\big)^{m-n}\bm\Pi^{\perp}_{\a(n)\b(m-n)\ad(n)}(H) ~,
\end{align}
where we have used the identities \eqref{CoincidenceProjAdS4|4} and \eqref{CoincidenceBachAdS4|4}. Consequently, the corresponding SCHS action is 
\begin{align}
\mb{S}_{\text{SCHS}}^{(m,n)}[H,\bar{H}]=(-\ri)^{m+n+2} \int \text{d}^{4|4}z \, E \, &\bar{H}^{\a(n)\ad(m)}\prod_{t=1}^{n+1} \big ( \mathbb{Q}-\l_{(t,m,n)}\m \mub \big )\non\\
&\times \big(\bm\cD_{\ad}{}^{\b}\big)^{m-n}\bm\Pi^{\perp}_{\a(n)\b(m-n)\ad(n)}(H) +\text{c.c. }~,\label{FTSCHSGen2}
\end{align}
which factorises according to
\begin{align}
\mb{S}_{\text{SCHS}}^{(m,n)}[H^{\text{T}},\bar{H}^{\text{T}}]=(-\ri)^{m+n+2} \int \text{d}^{4|4}z \, E \, &\bar{H}_{\text{T}}^{\a(n)\ad(m)}\prod_{t=1}^{n+1} \big ( \mathbb{Q}-\l_{(t,m,n)}\m \mub \big )\non\\
&\times \big(\bm\cD_{\ad}{}^{\b}\big)^{m-n} H^{\text{T}}_{\a(n)\b(m-n)\ad(n)}+\text{c.c. }
\end{align}
The analogous on-shell equation is
\begin{align}
0=\prod_{t=1}^{m+1} \big ( \mathbb{Q}-\l_{(t,m,n)}\m \mub \big )H^{\text{T}}_{\a(m)\ad(n)}~. \label{eom2AdS4|4}
\end{align}
Once again, (non-unitary) massive modes appear in the resulting wave equation. 

In the non-supersymmetric case, the factorisation of the conformal operators was demonstrated in section \ref{secFactorCHS}. In this section we have provided the first derivation of the factorisation of the superconformal higher-spin actions in AdS$^{4|4}$ superspace. Just as in the non-supersymmetric case, the latter factor into products of minimal second-order differential operators involving all partial mass values. 
Moreover, the SCHS actions \eqref{FTSCHSGen1} and \eqref{FTSCHSGen2} (in the case $m=n+1=s$), written in terms of the superprojectors, are analogous to those expressions provided for the conformal higher-spin models in Minkowski space  \cite{FT} and in AdS$_4$ \eqref{FTCHSGen1} and \eqref{FTCHSGen2}. 

\section{Superconformal higher-spin models on conformally-flat backgrounds}\label{secCF4|4}
In this section we employ the framework of $\mc{N}=1$ conformal superspace, developed by Butter\cite{ButterN=1}, to construct gauge invariant models for the superconformal multiplets $H_{\a(m)\ad(n)}$ in arbitrary conformally-flat superspace backgrounds. We also describe models for the so-called generalised superconformal multiplets, which have gauge transformations of higher-depth relative to $H_{\a(m)\ad(n)}$. This section is based on our papers \cite{Confgeo, SCHS, SCHSgen}. We begin by reviewing the main relevant features of  $\mc{N}=1$ conformal superspace.

\subsection{$\mc{N}=1$ conformal superspace} \label{secN=1CSS}

In $\mc{N}=1$ conformal superspace, the structure group of the curved superspace $\mc{M}^{4|4}$ is chosen to be $\sSU(2,2|1)$; the $\mathcal{N}=1$ superconformal group. 
In particular, this means that the corresponding covariant derivatives  $\Nabla_A = (\Nabla_a, \Nabla_\alpha, \bar\Nabla^\ad)$ take the form
\begin{align}
\Nabla_A &= \bm e_A{}^M \pa_M - \frac{1}{2}\Omega_A{}^{bc} M_{bc} - \ri \Q_A \mb{Y}
- B_A \mb{D} - \mf{F}_{A}{}^B K_B ~.\label{CSS4covder} 
\end{align}
Here
$\Omega_A{}^{bc}$  
denotes the Lorentz connection,  $\Q_A$  the  $\rm U(1)_R$ connection, $B_A$
the dilatation connection, and  $\mathfrak F_A{}^B$ the special
superconformal connection.

Below we list the graded commutation relations for the $\cN=1$ superconformal 
algebra $\mathfrak{su}(2,2|1)$  following the conventions
adopted in \cite{ButterN,BKN}, keeping in mind that (i) the translation generators 
$P_A = (P_a, Q_\a ,\bar Q^\ad)$ are replaced with $\Nabla_A$; and (ii) the graded commutator 
$[\Nabla_A , \Nabla_B\}$ differs to that obtained from $[P_A , P_B\}$ by torsion and curvature dependent terms, 
\bea
\label{BasicAlgebraConfSS}
[\Nabla_A, \Nabla_B\} 
& = & -\cT_{AB}{}^C \Nabla_C - \hf \cR_{AB}{}^{cd} (M)M_{cd} 
- \ri \cR_{AB}(\mb{Y}) \mb{Y} - \cR_{AB} (\mathbb{D}) \mathbb{D} \non \\
&& - \cR_{AB}{}^C (K)K_C 
~.
\eea

\begin{subequations} \label{MotherBoardAlgebra}
	The Lorentz generators act on vectors and Weyl spinors as follows:
	\bea
	M_{ab} V_{c} = 2 \eta_{c[a} V_{b]} ~, \qquad 
	M_{\a \b} \j_{\g} = \ve_{\g (\a} \j_{\b)} ~, \qquad \bar{M}_{\ad \bd} \bar \j_{\gd} = \ve_{\gd ( \ad} \bar \j_{\bd )} ~.
	\eea
	The  $\rm U(1)_R$ and dilatation generators $\mb{Y}$ and $\mb{D}$ obey
	\begin{align}
	[\mb{Y}, \Nabla_\a] &= \Nabla_\a ~,\quad [\mb{Y}, \bar\Nabla^\ad] = - \bar\Nabla^\ad~,   \\
	[\mathbb{D}, \Nabla_a] = \Nabla_a ~, \quad
	&[\mathbb{D}, \Nabla_\a] = \hf \Nabla_\a ~, \quad
	[\mathbb{D}, \bar\Nabla^\ad ] = \hf \bar\Nabla^\ad ~.
	\end{align}
	The special superconformal generators $K^A = (K^a, S^\alpha, \bar S_\ad)$
	carry opposite $\rm U(1)_R$ and dilatation weight to $\Nabla_A$:
	\begin{align}
	[\mb{Y}, S^\a] &= - S^\a ~, \quad
	[\mb{Y}, \bar{S}_\ad] = \bar{S}_\ad~,  \\
	[\mathbb{D}, K_a] = - K_a ~, \quad
	&[\mathbb{D}, S^\a] = - \hf S^\a~, \quad
	[\mathbb{D}, \bar{S}_\ad ] = - \hf \bar{S}_\ad ~.
	\end{align}
Among themselves, these obey the algebra
\begin{align}
\{ S_\a , \bar{S}_\ad \} &= 2 \ri  K_{\aa}~,\label{ACommSSBK}
\end{align}
with all the other (anti-)commutators vanishing. Finally, the algebra of $K^A$ and $\Nabla_B$ is
\begin{align}
[K_\aa, \Nabla_\bb] &= 4 \big(\ve_{\ad \bd} M_{\a \b} +  \ve_{\a \b} \bar{M}_{\ad \bd} -  \ve_{\a \b} \ve_{\ad \bd} \mathbb{D} \big) ~, \\
\{ S_\a , \Nabla_\b \} &= \ve_{\a \b} \big( 2 \mathbb{D} - 3 \mb{Y} \big) - 4 M_{\a \b} ~,\label{SnablaComm} \\
\{ \bar{S}_\ad , \bar{\Nabla}_\bd \} &= - \ve_{\ad \bd} \big( 2 \mathbb{D} + 3 \mb{Y}) + 4 \bar{M}_{\ad \bd}  ~, \\
[K_{\a \ad}, \Nabla_\b] &= - 2 \ri \ve_{\a \b} \bar{S}_{\ad} \ , \qquad \qquad \qquad[K_\aa, \bar{\Nabla}_\bd] =
2 \ri  \ve_{\ad \bd} S_{\a} ~,  \\
[S_\a , \Nabla_\bb] &= 2 \ri \ve_{\a \b} \bar{\Nabla}_{\bd} \ , \qquad \qquad \quad \qquad[\bar{S}_\ad , \Nabla_\bb] =
- 2 \ri \ve_{\ad \bd} \Nabla_{\b} \ ,
\end{align}
\end{subequations}
where all other graded commutators vanish.

In this setting, the $\mc{N}=1$ conformal supergravity gauge group $\bm\cG$ includes local $\L$-transformations of the form
\bea
\delta_{\L}^{(\scriptsize\bm{\mc{G}})} \Nabla_{A} = \big[ \L , \Nabla_{A} \big] ~, \qquad \L := \xi^{B} \Nabla_{B} + \frac{1}{2}K^{bc} M_{bc} + \ri \rho \mb{Y} + \S \mathbb{D} + \tau^{B} K_{B} ~.~~~
\label{SGGauge}
\eea
Here the gauge parameter $\L$ incorporates several parameters describing 
the general coordinate  ($\xi^{B}$), local Lorentz ($K^{ab}$), chiral ($\r$), scaling ($\S$), and special superconformal ($\tau^{B}$) transformations.  A tensor superfield $\F$ (with Lorentz indices suppressed) is said to be $\bm\cG$-covariant if
 it transforms under $\bm\cG$ according to the rule
\begin{align}
\delta_{\L}^{(\bm{\scriptstyle{\cG}})}\Phi = \L  \F ~.\label{TensorGaugeTf}
\end{align}
A $\bm\cG$-covariant superfield $\F$ is said to be primary with super-Weyl weight $\Delta$ and $\sU(1)_R$ charge $q$ if it satisfies
\begin{align}
K_A\F=0~,\qquad \mb{D}\F=\Delta\F~,\qquad \mb{Y}\F=q\F~.
\end{align}
We note that, due to the anticommutator \eqref{ACommSSBK}, the following (one way) relation holds
\begin{align}
S_{\a}\F=\bar{S}_{\ad}\F=0~\qquad \implies \qquad K_{\a\ad}\F=0~.
\end{align}
If we further specify that the primary superfield $\F$ is covariantly chiral, then consistency of this constraint with e.g. the anti-commutator \eqref{SnablaComm} fixes the charge-to-weight ratio:
\begin{align}
\bNabla_{\ad}\F=0 \qquad \implies \qquad q=-\frac{2}{3}\Delta~. \label{CovChiralConCSS}
\end{align}

In conformal superspace, 
the torsion and  curvature tensors in \eqref{BasicAlgebraConfSS} are subject to covariant constraints such that 
$[\Nabla_A, \Nabla_B\}$ is expressed solely in terms of the super-Weyl tensor
$\bm W_{\alpha \beta \gamma}$, its conjugate $\bm{ \bar W}_{\ad \bd \gd}$ and their covariant derivatives (cf. \eqref{U(1)algebraRevamped}). The latter is a primary and chiral 
$\bm\cG$-covariant superfield with weight 3/2 and charge -1, 
\bea
K_B \bm W_{\a(3)} =0~, \quad \bar \Nabla_\bd \bm W_{\a(3)}=0 ~, \quad 
{\mathbb D} \bm W_{\a(3)} = \frac 32 \bm W_{\a(3)}~,\quad \mb{Y}\bm W_{\a(3)}=-\bm W_{\a(3)}
\eea
The solutions to the aforementioned constraints are given by \cite{ButterN=1}
\begin{subequations}
	\label{CSSAlgebra}
	\bea
	\{ \Nabla_{\a} , \Nabla_{\b} \} & = & 0 ~, \quad \{ \bar{\Nabla}_{\ad} , \bar{\Nabla}_{\bd} \} = 0 ~, \quad \{\Nabla_{\a} , \bar{\Nabla}_{\ad} \} = - 2 \ri \Nabla_{\a \ad} ~, \\
	\big[ \Nabla_{\a} , \Nabla_{\b \bd} \big] & = & \ri \ve_{\a \b} \Big( 2 \bm{\bar{W}}_{\bd}{}^{\gd \dd} \bar{M}_{\gd \dd} - \frac{1}{2} \bar{\Nabla}^{\ad} \bm{\bar{W}}_{\ad \bd \gd} \bar{S}^{\gd} + \frac{1}{2} \Nabla^{\g \ad} \bm{\bar{W}}_{\ad \bd}{}^{\gd} K_{\g \gd} \Big) ~, \\
	\big[ \bar{\Nabla}_{\ad} , \Nabla_{\b \bd} \big] & = & - \ri \ve_{\ad \bd} \Big( 2 \bm W_{\b}{}^{\g \d} M_{\g \d} + \frac{1}{2} \Nabla^{\a}  \bm W_{\a \b \g} S^{\g} + \frac{1}{2} \Nabla^{\a \gd} \bm W_{\a \b}{}^{\g} K_{\g \gd} \Big) ~,\\
	\big[ \Nabla_{\a \ad} , \Nabla_{\b \bd} \big] & = & \ve_{\ad \bd} \psi_{\a \b} + \ve_{\a \b} \bar{\psi}_{\ad \bd} ~. \label{2.8d}
	\eea
	In eq. \eqref{2.8d} we have defined the operators $\psi_{\a(2)} $ and $\bar{\psi}_{\ad(2)}$ as follows
	\bea
	\psi_{\a \b} & = & \bm W_{\a \b}{}^{\g} \Nabla_{\g} + \Nabla^{\g} \bm W_{\a \b}{}^{\d} M_{\g \d} - \frac{1}{8} \Nabla^{2}\bm  W_{\a \b \g} S^{\g} + \frac{\ri}{2} \Nabla^{\g \gd}\bm  W_{\a \b \g} \bar{S}_{\gd} \non \\
	&& + \frac{1}{4} \Nabla^{\g \dd} \Nabla_{(\a}\bm W_{\b) \g}{}^{\d} K_{\d \dd} + \frac{1}{4} \Nabla^{\g}\bm W_{\a \b \g} \big(2\mathbb{D} - 3\mb{Y}\big) ~, \\
	\bar{\psi}_{\ad \bd} & = & - \bm{\bar{W}}_{\ad \bd}{}^{\gd} \bar{\Nabla}_{\gd} - \bar{\Nabla}^{\gd} \bm{\bar{W}}_{\ad \bd}{}^{\dd} \bar{M}_{\gd \dd} + \frac{1}{8} \bar{\Nabla}^{2} \bm{\bar{W}}_{\ad \bd \gd} \bar{S}^{\gd} + \frac{\ri}{2} \Nabla^{\g \gd} \bm{\bar{W}}_{\ad \bd \gd} S_{\g} \non \\
	&& - \frac{1}{4} \Nabla^{\d \gd} \bar{\Nabla}_{(\ad} \bm{\bar{W}}_{\bd) \gd}{}^{\dd} K_{\d \dd} - \frac{1}{4} \bar{\Nabla}^{\gd} \bm{\bar{W}}_{\ad \bd \gd} \big( 2\mathbb{D} +3\mb{Y}\big) ~.
	\eea
\end{subequations}
Therefore, we see that if $\mc{M}^{4|4}$ is conformally-flat, then the superconformal covariant derivatives obey the same algebra as the covariant derivatives of $\mb{M}^{4|4}$,
\begin{align}
\bm W_{\a(3)}=0 \qquad \implies \qquad \big[\Nabla_{A},\Nabla_{\b\bd}\big]=0~.\label{FlatCSSalgebra}
\end{align}

We also find that $\bm W_{\a \b \g}$ obeys the Bianchi identity
\bea
\bm B_{\a\ad} :=  \ri \Nabla^\b{}_{\ad} \Nabla^\g \bm W_{\a\b\g}
=\ri \Nabla_{\a}{}^{ \bd} \bar \Nabla^\gd \bm{\bar W}_{\ad\bd\gd}
= \bm{\bar B}_{\a\ad}~,
\label{super-Bach}
\eea
where $\bm B_{\a\ad}$   is
the super-Bach tensor.\footnote{It is instructive to check that \eqref{U(1)super-Bach} and \eqref{super-Bach} coincide upon degauging to $\sU(1)$ superspace.  }
It is a primary superfield with weight 3, 
\bea
K_B \bm B_{\a\ad} &=&0~, 
\qquad {\mathbb D} \bm B_{\a\ad} = 3 \bm B_{\a\ad} ~,
\eea
which obeys the conservation equation 
\bea
\Nabla^\a \bm B_{\a\ad}=0 \quad & \Longleftrightarrow &\quad  
\bar \Nabla^\ad \bm B_{\a\ad} =0~.
\label{616}
\eea

In appendix \ref{AppCSS4} we review how the geometry of $\mc{N}=1$ conformal superspace may be degauged to recover the $\sU(1)$ superspace geometry described in section \ref{secCSGSS4}. Additionally, in the context of conformal superspace, we discuss the issues of (i) integration by parts; and (ii) transitioning between integrations over the full superspace and its chiral subspace. 

Finally, we note that if we restrict our attention to bosonic backgrounds, defined by
\begin{subequations}\label{Bbackground}
\begin{align}
\Nabla_{a}| = \nabla_a~, \quad \bm W_{\a(3)}| = 0~, \quad  \Nabla^2& \bm W_{\a(3)}| = 0 ~, \quad \Nabla^{\b}\bm W_{\b\a(2)}|=0~,\\[5pt]
\Nabla_{\a} \bm W_{\b(3)}| &= - W_{\a \b(3)}~, \label{ComponentWeyl}
\end{align}
\end{subequations}
then the resulting geometry describes conformal gravity (see section \ref{secCGfromCS}). For such backgrounds, the only surviving component field of $\bm B_{\a \ad}$ is the Bach tensor $B_{\a(2)\ad(2)}$
\bea
\label{Bach}
-\frac{1}{2} \big[ \Nabla_{(\a_1} , \bar{\Nabla}_{(\ad_1}\big] \bm B_{\a_2) \ad_2)} | =  \nabla_{(\ad_1}{}^{\b_1}\nabla_{\ad_2)}{}^{\b_2}W_{\a(2)\b(2)}=B_{\a(2) \ad(2)} ~,
\eea
where  $W_{\a(4)}$ is the Weyl tensor. The gauge choice \eqref{Bbackground} means that we are switching off the axial gauge field and all covariant fermionic fields (such as the gravitino), which comes with a loss of local supersymmetry and $\sU(1)_R$ symmetry; only rigid versions remain. 

\subsection{SCHS models} \label{secSCHSCF}

We define SCHS prepotentials $H_{\a(m)\ad(n)}$ in conformal superspace
by lifting the properties of the corresponding SCHS prepotentials living in $\sU(1)$ superspace. The latter were described in section \ref{SCHSprepSS4}, and were based on the earlier constructions  \cite{HST81, KMT, KR}. In particular, for all allowed values of $m$ and $n$, the SCHS prepotential $H_{\a(m)\ad(n)}$ is primary
\begin{subequations} \label{SCHSpreppropCSS}
\begin{align}
K_A H_{\a(m)\ad(n)}=0~,
\end{align}  
and has the following super-Weyl weight and $\sU(1)_R$ charge
\begin{align}
\mb{D}H_{\a(m)\ad(n)}&=-\frac{1}{2}(m+n)H_{\a(m)\ad(n)}~,\\
\mb{Y}H_{\a(m)\ad(n)}&=\phantom{-}\frac{1}{3}(m-n)H_{\a(m)\ad(n)}~.
\end{align} 
\end{subequations}
The weight and charge of $H_{\a(m)\ad(n)}$ is fixed by requiring consistency of the algebra \eqref{MotherBoardAlgebra} with its corresponding gauge transformations, which vary depending on $m$ and $n$: 

\begin{itemize}

\item For 
 $1\leq m \neq n \geq 1$,  $H_{\a(m)\ad(n)}$ is defined modulo gauge transformations
\begin{subequations}\label{SCHSprepGTCSS}
   \begin{align}
  \delta_{\L,\O}H_{\a(m)\ad(n)}=\Nabla_{\a}\L_{\a(m-1)\ad(n)}+\bar \Nabla_{\ad}\O_{\a(m)\ad(n-1)}~, \label{SCHSprepGTICSS}
   \end{align}
for unconstrained primary complex gauge parameters $\L_{\a(m-1)\ad(n)}$ and $\O_{\a(m)\ad(n-1)}$.

\item For $m=n=s\geq 1$, $H_{\a(s)\ad(s)}$ is real and defined modulo the gauge transformations 
\begin{align}
  \delta_{\L}H_{\a(s)\ad(s)}=\Nabla_{\a}\L_{\a(s-1)\ad(s)}-\bar\Nabla_{\ad}\bar{\L}_{\a(s)\ad(s-1)}~,\label{SCHSprepGTIICSS}
   \end{align}
   for unconstrained primary complex gauge parameter $\L_{\a(s-1)\ad(s)}$.

\item For $m>n=0$, $H_{\a(m)}$ is defined modulo the gauge transformations 
\begin{align}
  \delta_{\L,\l}H_{\a(m)}=\Nabla_{\a}\L_{\a(m-1)}+\l_{\a(m)}~,\qquad \bar\Nabla_{\ad}\l_{\a(m)}=0 ~.\label{SCHSprepGTIIICSS}
\end{align}
\end{subequations}
The complex gauge parameter $\L_{\a(m-1)}$ is primary (but otherwise unconstrained), whilst the complex gauge parameter $\l_{\a(m)}$ is primary and covariantly chiral. 

\end{itemize}

In $\mc{N}=1$ conformal superspace, the higher-spin super-Weyl tensor of the prepotential $H_{\a(m)\ad(n)}$ and its conjugate $\bar{H}_{\a(n)\ad(m)}$ is the minimal lift of \eqref{HSSWeylMink4},
\begin{subequations}\label{HSSWeylCSS4}
\begin{align}
\mf{W}_{\a(m+n+1)}(H)&=-\frac{1}{4}\bar\Nabla^2\Nabla_{(\a_1}{}^{\bd_1}\cdots\Nabla_{\a_n}{}^{\bd_n}\Nabla_{\a_{n+1}}H_{\a_{n+2}\dots\a_{m+n+1})\bd(n)}~,\label{HSSWeylCSSa}\\
\mf{W}_{\a(m+n+1)}(\bar{H})&=-\frac{1}{4}\bar\Nabla^2\Nabla_{(\a_1}{}^{\bd_1}\cdots\Nabla_{\a_m}{}^{\bd_m}\Nabla_{\a_{m+1}}\bar{H}_{\a_{m+2}\dots\a_{m+n+1})\bd(m)}~.\label{HSSWeylCSS4b}
\end{align}
\end{subequations}
The former satisfies the following properties:
\begin{enumerate}[label=(\roman*)]

\item
$\mf{W}_{\a(m+n+1)}(H)$ is a primary superfield,
\begin{subequations} \label{HSSWeylprop1CSS}
\begin{align}
K_A\mf{W}_{\a(m+n+1)}(H)&=0~, \label{HSSWeylprop1abcCSS}
\end{align}
 with super-Weyl weight and $\sU(1)_R$ charge given by
\begin{align}
\mb{D} \mf{W}_{\a(m+n+1)}(H) &=\phantom{-}\frac{1}{2}\big(3+n-m\big)  \mf{W}_{\a(m+n+1)}(H)~,\label{HSSWeylprop1aCSS}\\
\mb{Y}\mf{W}_{\a(m+n+1)}(H) &= -\frac{1}{3}\big(3+n-m\big)\mf{W}_{\a(m+n+1)}(H)~.\label{HSSWeylprop1bCSS}
\end{align}
\end{subequations}


\item
$\mf{W}_{\a(m+n+1)}(H)$ is covariantly chiral (which is consistent with \eqref{CovChiralConCSS}),
\begin{align}
\bar\Nabla_{\ad}\mf{W}_{\a(m+n+1)}(H)=0~.\label{HSSWeylprop2CSS}
\end{align}

\item
$\mf{W}_{\a(m+n+1)}(H)$ is gauge invariant under \eqref{SCHSprepGTCSS} if $\mc{M}^{4|4}$ is conformally-flat,
\begin{align}\label{HSSWeylprop3CSS}
\bm W_{\a(3)}=0 \quad \implies \quad \d_{\L}\mf{W}_{\a(m+n+1)} (H)&= 0~. 
\end{align}

\end{enumerate}

The descendent $\mf{W}_{\a(m+n+1)}(\bar{H})$ satisfies the same properties as $\mf{W}_{\a(m+n+1)}(H)$, but with $m$ and $n$ interchanged in \eqref{HSSWeylprop1CSS}.
In order to prove the validity of \eqref{HSSWeylprop1abcCSS}, the following identities are useful
\begin{subequations}
\begin{align}
\big[S_{\b},\big(\Nabla_{\a\ad}\big)^k\big]&= \phantom{-}2k\ri \ve_{\b\a}\bar\Nabla_{\ad}\big(\Nabla_{\a\ad}\big)^{k-1}~,\\
\big[\bar{S}_{\bd},\big(\Nabla_{\a\ad}\big)^k\big]&=-2k\ri \ve_{\bd\ad}\Nabla_{\a}\big(\Nabla_{\a\ad}\big)^{k-1}~,\\
\big[S_{\a}, \Nabla^2\big]&=-2\Nabla_{\a}\big(2\mb{D}-3\mb{Y}-4\big)+8\Nabla^{\b}M_{\a\b}~,\\
\big[\bar{S}_{\ad}, \bar\Nabla^2\big]&=-2\bar\Nabla_{\ad}\big(2\mb{D}+3\mb{Y}-4\big)+8\bar\Nabla^{\bd}\bar{M}_{\ad\bd}~.
\end{align}
\end{subequations}
They hold for an arbitrary positive integer $k$ and on a generic background. 
The property \eqref{HSSWeylprop3CSS} is a simple consequence of \eqref{HSSWeylpropMinkaloo} and \eqref{FlatCSSalgebra}.


Similarly, the minimal lifts of the higher-spin super-Bach tensors \eqref{HSSBACHmink4a} and \eqref{HSSBACHmink4b} to conformal superspace are 
\begin{subequations}\label{HSSBACHCSS4CF}
 \begin{align} 
 \mf{B}_{\a(n)\ad(m)}(H)&=-\frac{1}{4}\Nabla_{(\ad_1}{}^{\b_1}\cdots\Nabla_{\ad_m)}{}^{\b_m}\Nabla^{\g}\bar \Nabla^2 \Nabla_{(\g}\Nabla_{\a_1}{}^{\bd_1}\cdots\Nabla_{\a_n}{}^{\bd_n}H_{\b_1\dots\b_m)\bd(n)}~,\label{HSSBACHCSS4CF1}\\
 \widehat{\mf{B}}_{\a(n)\ad(m)}(H)&=\phantom{-}\frac{1}{4}\Nabla_{(\a_1}{}^{\bd_1}\cdots\Nabla_{\a_n)}{}^{\bd_n}\bar\Nabla^{\gd}\Nabla^2\bar\Nabla_{(\gd}\Nabla_{\ad_1}{}^{\b_1}\cdots\Nabla_{\ad_m}{}^{\b_m}H_{\b(m)\bd_1\dots\bd_n)}~.\label{HSSBACHCSS4CF2}
 \end{align}
 \end{subequations} 
 They may also be expressed in terms of the higher-spin super-Weyl tensors, as in \eqref{sBachsWeyl}.
On a generic background, the descendent $\mf{B}_{\a(n)\ad(m)}(H)$ is a primary superfield,
\begin{subequations} \label{HSSBachprop1CSS}
\begin{align}
K_{A}\mf{B}_{\a(n)\ad(m)}(H)=0~,
\end{align}
 with Weyl weight and $\sU(1)_R$ charge given by
\begin{align}
\mb{D}\mf{B}_{\a(n)\ad(m)}(H) &=\Big(2+\frac{1}{2}\big(m+n\big) \Big) \mf{B}_{\a(n)\ad(m)}(H)~,\label{HSSBachprop1aCSS}\\
\mb{Y}\mf{B}_{\a(n)\ad(m)}(H) &= \frac{1}{3}\big(m-n\big)\mf{B}_{\a(n)\ad(m)}(H)~.\label{HSSBachprop1bCSS}
\end{align}
\end{subequations}
The other descendent $\widehat{\mf{B}}_{\a(n)\ad(m)}(H)$ carries the same superconformal properties. On any conformally-flat background, 
\begin{align}
\bm W_{\a(3)}=0~, \label{CFCSS}
\end{align}
they possess the following additional properties:
\begin{enumerate}[label=(\roman*)]
\item $\mf{B}_{\a(n)\ad(m)}(H)$ and $\widehat{\mf{B}}_{\a(n)\ad(m)}(H)$ are equal to one another
\begin{align}
\mf{B}_{\a(n)\ad(m)}(H)=\widehat{\mf{B}}_{\a(n)\ad(m)}(H)~,
\end{align}
which leads to the following reality relation
\begin{align}
\mf{B}_{\a(m)\ad(n)}(\bar{H})=\Big(\mf{B}_{\a(n)\ad(m)}(H)\Big)^*\equiv \overline{\mf{B}}_{\a(m)\ad(n)}(\bar{H})~.
\end{align}
\item For $m\geq 1$ and $n\geq 1 $  $\mf{B}_{\a(n)\ad(m)}(H)$ is TLAL,
\begin{subequations}\label{HSSBachprop2CSS}
\begin{align}
\Nabla^{\b}\mf{B}_{\b\a(n-1)\ad(m)}(H)&=0 \qquad \implies \qquad \Nabla^2\mf{B}_{\a(n)\ad(m)}(H)=0~,\\
\bar\Nabla^{\bd}\mf{B}_{\a(n)\ad(m-1)\bd}(H)&=0 \qquad \implies \qquad \bar\Nabla^2\mf{B}_{\a(n)\ad(m)}(H)=0~. \label{HSSBachprop2aCSS}
\end{align}
\end{subequations}
For $m>n=0$  $\mf{B}_{\ad(m)}(H)$ is ALTL, whilst for $n>m=0$ $\mf{B}_{\a(n)}(H)$ is LTAL.
\item $\mf{B}_{\a(n)\ad(m)}(H)$ is invariant under the gauge transformations \eqref{SCHSprepGTCSS} 
\begin{align}\label{HSSBachprop3CSS}
\d_{\L}\mf{B}_{\a(n)\ad(m)}(H)&= 0~.
\end{align}

\end{enumerate}

The properties \eqref{HSSWeylprop3CSS}, \eqref{HSSBachprop2CSS} and \eqref{HSSBachprop3CSS} mean that on any conformally-flat background \eqref{CFCSS},
 the minimal lift of the action \eqref{SCHSactMink4|4} to conformal superspace,\footnote{Moving between the full superspace and its (anti-)chiral subspace is non-trivial in conformal superspace, see appendix \ref{AppCSS4}. However, the two forms \eqref{SCHSactCSSCF4|4a} and \eqref{SCHSactCSSCF4|4b} of the SCHS action prove to be equivalent (at least on conformally-flat backgrounds). }
\begin{subequations}\label{SCHSactCSSCF4|4}
 \begin{align}
 \mb{S}_{\text{SCHS}}^{(m,n)}[H,\bar{H}]&= \frac{1}{2}\ri^{m+n}\int \rd^4x\rd^2\theta \, \cE \, \mf{W}^{\a(m+n+1)}(H)\mf{W}_{\a(m+n+1)}(\bar{H})+\text{c.c.}~\label{SCHSactCSSCF4|4a}\\
 &= \frac{1}{2}(-\ri)^{m+n+2} \int \rd^{4|4}z \, E \,  \bar{H}^{\a(n)\ad(m)}\mf{B}_{\a(n)\ad(m)}(H) +\text{c.c.}~,\label{SCHSactCSSCF4|4b}
 \end{align}
 \end{subequations}
is gauge invariant $\delta_{\L} \mb{S}_{\text{SCHS}}^{(m,n)}[H,\bar{H}]=0$. Due to the properties \eqref{SCHSpreppropCSS}, \eqref{HSSWeylprop1CSS} and \eqref{HSSBachprop1CSS}, the same action is invariant under the $\mc{N}=1$ conformal supergravity gauge group $\bm\cG$ on a generic supergravity background.

The action \eqref{SCHSactCSSCF4|4} ceases to be gauge invariant on non-conformally-flat backgrounds. In section \ref{secBach4|4} we will extend the gauge invariance of $\mb{S}_{\text{SCHS}}^{(m,n)}[H,\bar{H}]$ to Bach-flat backgrounds for some low values of $m$ and $n$. 

\subsection{Component reduction: SCHS model $\longrightarrow$ CHS models} \label{secSCHStoCHS}

In this section we discuss how the gauge freedom \eqref{SCHSprepGTCSS} may be partially fixed to construct a Wess-Zumino gauge on the SCHS prepotential $H_{\a(m)\ad(n)}$. We then make use of this gauge to reduce the corresponding SCHS action $\mb{S}_{\text{SCHS}}^{(m,n)}[H,\bar{H}]$ to components, and explore the resulting CHS actions which they induce. 
Since the relevant gauge freedom changes according to the values of $m$ and $n$, an analysis should proceed on a case-by-case basis. Here, we focus only on two of the three cases; $m=n=s\geq 1$ and $m>n=0$.


\subsubsection{Case $m=n=s \geq 1$}

For this case we restrict our attention to bosonic superspace backgrounds \eqref{Bbackground} which are conformally-flat \eqref{CFCSS}, i.e. those satisfying
\begin{align} 
\Nabla_a|=\nabla_a~,\qquad \bm W_{\a(3)}=0 \quad \implies \quad W_{\a(4)}=0~.
\end{align}
This means that there is only a rigid supersymmetry present (which we will not be concerned with). 

The gauge freedom \eqref{SCHSprepGTIICSS} of the real prepotential $H_{\a(s)\ad(s)}$ allows one to impose a Wess-Zumino gauge such that there are only three independent component fields:
\begin{subequations}\label{WZGaugem=n=s}
\begin{align}
h_{\a(s+1)\ad(s+1)}&:=\frac{1}{2}\big[\Nabla_{\a},\bNabla_{\ad}\big]H_{\a(s)\ad(s)}|=\bar{h}_{\a(s+1)\ad(s+1)}~,\\
\psi_{\a(s+1)\ad(s)}&:= -\frac{1}{4}\Nabla_{\a}\bNabla^2H_{\a(s)\ad(s)}|~,\\
\bar{\psi}_{\a(s)\ad(s+1)}&:= -\frac{1}{4}\bNabla_{\ad}\Nabla^2H_{\a(s)\ad(s)}|~,\\
g_{\a(s)\ad(s)}&:=\frac{1}{32}\{\Nabla^2,\bNabla^2\}H_{\a(s)\ad(s)}|=\bar{g}_{\a(s)\ad(s)}~.
\end{align}
\end{subequations}
Each of these fields are primary and have the following Weyl weights:
\begin{subequations}\label{casem=n=sCP}
\begin{align}
K_b h_{\a(s+1)\ad(s+1)}&=0~,  \qquad    \mb{D} h_{\a(s+1)\ad(s+1)}=\big(1-s\big) h_{\a(s+1)\ad(s+1)}~,\\
K_b \psi_{\a(s+1)\ad(s)}&=0~, \qquad   ~\phantom{..} \mb{D} \psi_{\a(s+1)\ad(s)}= \big(\frac{3}{2}-s\big)\psi_{\a(s+1)\ad(s)}~,\\
K_b g_{\a(s)\ad(s)} &=0~,     \qquad   ~~~~\phantom{..} \mb{D} g_{\a(s)\ad(s)}=\big(2-s\big) g_{\a(s)\ad(s)}~.
\end{align}
\end{subequations}

The residual gauge freedom is generated by the (primary) parameters
\begin{subequations}
\begin{align}
\xi_{\a(s)\ad(s)}      &:=  -\ri \Big( \Nabla_{\a}\L_{\a(s-1)\ad(s)}|+ \bNabla_{\ad}\bar{\L}_{\a(s)\ad(s-1)}|\Big)~,\\
\z_{\a(s)\ad(s-1)}     &:=  -\frac{\ri}{2}\Nabla^2\bar{\L}_{\a(s)\ad(s-1)}|~,\\
\ell_{\a(s-1)\ad(s-1)} &:=   \frac{\ri s}{8(s+1)}\Big(\bNabla^{\bd}\Nabla^2\L_{\a(s-1)\ad(s-1)\bd}|+\Nabla^{\b}\bNabla^2\bar{\L}_{\b\a(s-1)\ad(s-1)}|\Big)~,
\end{align}
\end{subequations}
under which the fields \eqref{WZGaugem=n=s} transform according to
\begin{subequations}\label{casem=n=sGT}
\begin{align}
\delta_{\xi}h_{\a(s+1)\ad(s+1)}&=\nabla_{\a\ad}\xi_{\a(s)\ad(s)}~,\\
\delta_{\z}\psi_{\a(s+1)\ad(s)}&=\nabla_{\a\ad}\z_{\a(s)\ad(s-1)}~,\\
\delta_{\ell}g_{\a(s)\ad(s)}&=\nabla_{\a\ad}\ell_{\a(s-1)\ad(s-1)}~.
\end{align}
\end{subequations}
The properties \eqref{casem=n=sCP} and \eqref{casem=n=sGT} allows us to identify the component fields $h_{\a(s+1)\ad(s+1)}$, $\psi_{\a(s+1)\ad(s)}$ and $ g_{\a(s)\ad(s)}$ as spin-$(s+1)$, spin-$(s+\frac{1}{2})$ and spin-$s$ conformal depth $t=1$ fields respectively (see section \ref{secCHSPrepCS4}). 

By making use of the above definitions, one can show that the only non-zero component fields of the higher-spin super-Weyl tensor \eqref{HSSWeylCSSa} of $H_{\a(s)\ad(s)}$ are
\begin{subequations}
\begin{align}
\mf{W}_{\a(2s+1)}(H)|&=\mc{W}_{\a(2s+1)}(\psi)~,\quad~\Nabla^{\b}\mf{W}_{\b\a(2s)}(H)|=-2\frac{(2s+2)}{(2s+1)}\mc{W}_{\a(2s)}(g)~, \\
 \Nabla_{\a}\mf{W}_{\a(2s+1)}(H)|&=\ri\mc{W}_{\a(2s+2)}(h)~,\quad \Nabla^2 \mf{W}_{\a(2s+1)}(H)|=-4\ri \mc{W}_{\a(2s+1)}(\bar{\psi})~.
\end{align}
\end{subequations}
Using this data, one can readily reduce the action \eqref{SCHSactCSSCF4|4a} to components as follows: 
\begin{align}\label{DecompActionSCHS}
\mb{S}_{\text{SCHS}}^{(s,s)}[H]&=(-1)^s\int \rd^4 x \, e \, \Big\{-\frac{1}{4}\mc{W}^{\a(2s+2)}(h)\mc{W}_{\a(2s+2)}(h)~\non\\
&+\ri \mc{W}^{\a(2s+1)}(\bar\psi)\mc{W}_{\a(2s+1)}(\psi) +\frac{(s+1)}{(2s+1)}\mc{W}^{\a(2s)}(g)\mc{W}_{\a(2s)}(g)
\Big\} +\text{c.c.} \non\\
&= \frac{1}{2}S_{\text{CHS}}^{(s+1,s+1)}[h]+2S_{\text{CHS}}^{(s+1,s)}[\psi,\bar{\psi}]+\frac{(2s+2)}{(2s+1)}S_{\text{CHS}}^{(s,s)}[g]~.
\end{align}
Here we have made use of the notation introduced in chapter \ref{Chapter4D}.

From the decomposition \eqref{DecompActionSCHS} of the action $\mb{S}_{\text{SCHS}}^{(s,s)}[H]$, it is clear
that every gauge invariant model of a superconformal higher-spin superfield gives rise to a multiplet of gauge invariant models of conformal higher-spin fields. This fact may be exploited to deduce non-trivial information regarding various properties of the induced CHS actions, even if we do not know the explicit form of the SCHS action. For example, in section \ref{secSpin3Recipe} we are able to deduce (some of) the spectrum of the conformal spin-3 model on a Bach-flat background simply by supposing the existence of the SCHS action with $s=2$. 


\subsubsection{Case $m>n=0$}

To begin this analysis, we restrict our attention to bosonic superspace backgrounds \eqref{Bbackground} which may not necessarily be conformally-flat. Only, when we reduce the action to components do we restrict to conformally-flat backgrounds. 

The gauge freedom \eqref{SCHSprepGTIIICSS} of the prepotential $H_{\a(m)}$ allows one to impose a Wess-Zumino gauge such that there are only four non-zero component fields:
\begin{subequations}\label{WZschs}
\bea
h_{\a(m+1) \ad} &=& \frac{1}{2} \big[\Nabla_{(\a_1} , \bNabla_{\ad} \big] H_{\a_2 \dots \a_{m+1})} | ~, \label{2.34a}\\
\psi_{\a(m) \ad} &=& - \frac{1}{4} \bNabla_{\ad} \Nabla^{2} H_{\a(m)} | ~, \\
\varphi_{\a(m+1)} &=& - \frac{1}{4} \Nabla_{(\a_1} \bNabla^{2} H_{\a_2 \dots \a_{m+1})} | ~, \\
\rho_{\a(m)} &=& \frac{1}{16} \Nabla^{\b} \bNabla^{2} \Nabla_{\b} H_{\a(m)} | -  \frac{\ri m}{4(m+2)} \nabla^{\b \bd} h_{\b \a(m) \bd} ~\label{2.34d}.
\eea
\end{subequations}
Here we have defined $\rho_{\a(m)}$ in such a way that it is annihilated by $K_{a}$. Likewise, by a routine computation, one can show that each component field is primary
\bea
K_{\bb} h_{\a(m+1) \ad} = 0 ~, \quad K_{\bb} \psi_{\a(m) \ad} = 0 ~, \quad K_{\bb} \varphi_{\a(m+1)} = 0 ~, \quad K_{\bb} \rho_{\a(m)} = 0 ~.
\eea
Their Weyl weights may be easily deduced from \eqref{WZschs}.
Underlying this gauge fixing are the following constraints on the gauge parameters:
\begin{subequations}
\bea
\label{WZFixingCondtions}
\ell_{\a(m)} := 2 \ri \l_{\a(m)}| &=& - 2 \ri \bm{\nabla}_{(\a_1} \L_{\a_2\dots\a_{m})} | ~, \\
\mu_{\a(m-1)} := - \frac{\ri m}{m+1} \Nabla^{\b} \l_{\b \a(m-1)} | & = & \frac{\ri}{2} \Nabla^{2} \L_{\a(m-1)}| ~, \\
\Nabla_{(\a_1} \l_{\a_2\dots\a_{m+1})} | & = & 0  ~, \\
\Nabla^{2} \l_{\a(m)} | &=& 0 ~, \\
\big[ \Nabla_{(\a_1} , \bNabla_{\ad} \big] \L_{\a_2\dots\a_m)} | & = & - 2 \ri \nabla_{(\a_1 \ad} \L_{\a_2\dots\a_m)} | ~, \\
\Nabla_{(\a_1} \bNabla^{2} \L_{\a_2\dots\a_m)} | & = & - 4 \ri \nabla_{(\a_1 \ad} \bNabla^{\ad} \L_{\a_2\dots\a_m)} | ~, \\
\{ \Nabla^{2} , \bNabla^{2} \} \L_{\a(m-1)} | & = & - 16 \Box \L_{\a(m-1)} | ~, \\
\bNabla_{\ad} \Nabla^{2} \L_{\a(m-1)}| & = & \frac{2 m}{m+1} \nabla^{\b}{}_{\ad} \ell_{\b \a(m-1)} ~.
\eea 
\end{subequations}

The residual transformations associated with this gauge are generated by the fields $\ell_{\a(m)}$ and $\mu_{\a(m-1)}$. These transform $h_{\a(m+1) \ad}$ and $\psi_{\a(m) \ad}$ independently of the background geometry
\begin{subequations}
\bea
\label{WZResidual}
\d_{\ell} h_{\a(m+1) \ad} = \nabla_{\a \ad} \ell_{\a(m)} ~,\qquad \d_{\mu} \psi_{\a(m) \ad} = \nabla_{\a \ad} \mu_{\a(m-1)} ~.
\eea
The two fields $\vf_{\a(m+1)}$ and $\rho_{\a(m)}$ are both conformal non-gauge fields (see appendix \ref{AppCNGM}), which do not have their own gauge freedom. However, the $\ell$-transformation also acts on $\rho_{\a(m)} $
when the background Weyl tensor is non-vanishing
\bea
\d_{\ell} \rho_{\a(m)} = \ri \frac{m(m-1)}{2(m+2)} W^{\b(2)}{}_{\a(2)} \ell_{\a(m-2)\b(2)} ~. \label{rhoGT}
\eea
\end{subequations}
From this transformation law, it follows that $\r_{\a(m)}$ should play an important role in ensuring gauge invariance of the model describing $h_{\a(m+1)\ad}$. Since \eqref{rhoGT} is non-zero for $m \geq 2$, we therefore expect that any  gauge-invariant (non-supersymmetric) model describing $h_{\a(m+1)\ad}$ in a Bach-flat background should couple to a lower-spin field.

Upon further restricting the geometry to be conformally flat, the action \eqref{SCHSactCSSCF4|4} may be readily reduced to components. One finds
\bea
(-1)^m\mb{S}^{(m,0)}_{\text{SCHS}} &=& -\frac{\ri^{m+1}}{2}\int \rd^4 x \, e\, \bigg\{
- \mc{W}^{\a(m+1)}(\psi) \mc{W}_{\a(m+1)}(\bar{\psi}) - \frac{\ri}{2} \mc{W}^{\a(m+2)}(h) \mc{W}_{\a(m+2)}(\bar{h}) \non \\
&& + \bar{\varphi}^{\ad(m+1)} \mathfrak{X}_{\ad(m+1)}(\varphi) + 2 \ri \frac{m+2}{m+1} \bar{\rho}^{\ad(m)} \mathfrak{X}_{\ad(m)} (\rho)
\bigg\} + \text{c.c.}  \label{2.364|4}\\
&=& \frac{1}{2}S_{\text{CHS}}^{(m+1,1)}[h,\bar{h}]+ S_{\text{CHS}}^{(m,1)}[\psi,\bar \psi]-S_{\text{NG}}^{(m+1,0)}[\vf,\bar{\vf}]+\frac{(m+2)}{(m+1)}S_{\text{NG}}^{(m,0)}[\rho,\bar{\rho}]~,\non
\eea
which has been presented in a manifestly gauge-invariant form and where we have made use of the notation introduced in section \ref{secNSFieldStr} and appendix \ref{AppCNGM}. We would like to point out that upon fixing $m=2$ (and making the appropriate rescalings) in \eqref{2.364|4}, the relative coefficient between the pseudo-graviton sector and the non-gauge sector (the  $\r_{\a(2)}$ field) does not agree with that of \eqref{CHookGInvX}. This fact will play an important role in our analysis of the superconformal pseudo-graviton multiplet in section \ref{secPGSBach}.

This analysis also leads to non-trivial information concerning the $H_{\a(3)}$ supermultiplet in a generic background. For instance, at the component level, \eqref{WZResidual} implies that $\psi_{\a(3)\ad}$ is identifiable with the conformal pseudo-graviton field and the remaining component fields are inert under its gauge transformation. However, it was shown in section \ref{secConformalHook} that $\psi_{\a(3)\ad}$ must couple to another field to ensure gauge invariance. Thus, $H_{\a(3)}$ must also couple to a suitable supermultiplet containing this extra field. As we will see in section \ref{secPGSBach}, a lower-spin coupling is also required in the superconformal pseudo-graviton multiplet. Therefore, it seems reasonable to expect that similar couplings are necessary for all $H_{\a(m)}$, where $m \geq 2$.

\subsection{Generalised SCHS models} \label{secGenSCHSCF}

A specific feature of $H_{\a (m) \ad (n)} $ with $ m \geq 1$ and $n\geq1$ is the presence of only a single spinor derivative in its gauge transformations \eqref{SCHSprepGTICSS}.
In this section we generalise these  SCHS multiplets by increasing the number of spinor derivatives appearing in their gauge transformations. The resulting prepotentials will be called generalised SCHS multiplets.

A depth-$t$ SCHS
multiplet $H^{(t)}_{\a(m)\ad(n)}$, with $m\geq 1$ and $n \geq 1$, is defined modulo  gauge transformations of the form
\begin{align}
\delta_{\L}H^{(t)}_{\a(m)\ad(n)}&=\big[\Nabla_{(\a_1},\bar{\Nabla}_{(\ad_1}\big]\Nabla_{\a_2\ad_2}\cdots\Nabla_{\a_t\ad_t}\L^{(t)}_{\a_{t+1}\dots\a_m)\ad_{t+1}\dots\ad_{n})}\non\\
&\equiv \big[\Nabla_{\a},\bar{\Nabla}_{\ad}\big]\big(\Nabla_{\a\ad}\big)^{t-1}\L^{(t)}_{\a(m-t)\ad(n-t)}~,\label{SCHSgenGT}
\end{align}
where the gauge parameter $\L^{(t)}_{\a(m-t)\ad(n-t)}$ is unconstrained and $1\leq t \leq \text{min}(m,n)$. 
If we require that both the prepotential and its gauge parameter are primary,
\begin{subequations}\label{SCHSgenPrepprop1}
\begin{align}
K_BH^{(t)}_{\a(m)\ad(n)}=0~,\qquad K_B\L^{(t)}_{\a(m-t)\ad(n-t)}=0~, \label{SCHSgenPrepprop1b}
\end{align}
 then the Weyl weight and $\sU(1)_{R}$ charge carried by $H^{(t)}_{\a(m)\ad(n)}$ must take the values
\begin{align}
\mb{D}H^{(t)}_{\a(m)\ad(n)}&=\big(t-\frac{1}{2}(m+n)\big)H^{(t)}_{\a(m)\ad(n)}~,\\
\mb{Y}H^{(t)}_{\a(m)\ad(n)}&=\frac{1}{3}(m-n)H^{(t)}_{\a(m)\ad(n)}~. \label{SCHSgenPrepprop1a}
\end{align}
\end{subequations}
For $m=n=s$, this allows us to choose both $H^{(t)}_{\a(s)\ad(s)}$ and $\L^{(t)}_{\a(s-t)\ad(s-t)}$ to be real,  
\begin{align}
H^{(t)}_{\a(s) \ad(s)} =\bar{H}^{(t)}_{\a(s)\ad(s)}~,\qquad  \L^{(t)}_{\a(s-t)\ad(s-t)}=\bar{\L}^{(t)}_{\a(s-t)\ad(s-t)}~.\label{SCHSgenReality}
\end{align}

Unlike the non-supersymmetric case, minimal depth SCHS superfields (i.e. those with $t=1$) do not correspond to the ordinary SCHS superfields described in the previous section. 
Accordingly, in order to collectively discuss all types of SCHS gauge multiplets, we will refer to ordinary SCHS multiplets $H_{\a(m)\ad(n)}$, defined modulo the gauge transformations \eqref{SCHSprepGTCSS}, as having depth $t=0$; $H_{\a(m)\ad(n)}\equiv H^{(0)}_{\a(m)\ad(n)}$ .

In principle one can instead consider supermultiplets $\hat{H}^{(t)}_{\a(m)\ad(n)}$ with the gauge freedom
\begin{align}
\delta_{\hat{\L}}\hat{H}^{(t)}_{\a(m)\ad(n)}=\Nabla_{\a}\bNabla_{\ad}\big(\Nabla_{\a\ad}\big)^{t-1}\hat{\L}^{(t)}_{\a(m-t)\ad(n-t)}~.
\end{align}
In order for $\hat{H}^{(t)}_{\a(m)\ad(n)}$ to be primary it must have weight $\big(t-\frac{1}{2}(m+n)\big)$ and charge $\frac{1}{3}\big(m-n+2t\big)$. In this case one may consistently define $\hat{H}^{(t)}_{\a(m)\ad(n)}$ to be longitudinal anti-linear, $\Nabla_{\a}\hat{H}^{(t)}_{\a(m)\ad(n)}=0$. However, these supermultiplets possess several undesirable features. In particular, it is not possible to impose a reality condition similar to \eqref{SCHSgenReality}, nor does it seem possible to construct gauge and super-Weyl invariant actions describing their dynamics. It is for this reason that we will focus only on the supermultiplets $H^{(t)}_{\a(m)\ad(n)}$.  

From the prepotential $H^{(t)}_{\a(m)\ad(n)}$ and its conjugate one may construct the following higher-derivative descendants 
\begin{subequations}\label{genSWeyl}
\begin{align}
\mathfrak{W}_{\a(m+n-t+1)\ad(t-1)}^{(t)}(H)&=\big(\bar{\Nabla}^{\bd}\Nabla_{\a}-\frac{\text{i}t}{n-t+1}\Nabla_{\a}{}^{\bd}\big)\big(\Nabla_{\a}{}^{\bd}\big)^{n-t}H^{(t)}_{\a(m)\ad(t-1)\bd(n-t+1)}~, \label{genSWeyla}\\
\mathfrak{W}_{\a(m+n-t+1)\ad(t-1)}^{(t)}(\bar{H})&=\big(\bar{\Nabla}^{\bd}\Nabla_{\a}-\frac{\text{i}t}{m-t+1}\Nabla_{\a}{}^{\bd}\big)\big(\Nabla_{\a}{}^{\bd}\big)^{m-t}\bar{H}^{(t)}_{\a(n)\ad(t-1)\bd(m-t+1)}~.\label{genSWeylb}
\end{align}
\end{subequations}
They are the higher-depth analogues of the higher-spin super-Weyl tensors \eqref{HSSWeylCSS4}. In particular, \eqref{genSWeyla} proves to be a primary superfield in a generic background
\begin{subequations}
\begin{align}
K_B\mathfrak{W}_{\a(m+n-t+1)\ad(t-1)}^{(t)}(H)=0~, 
\end{align}
which carries the following super-Weyl weight and $\sU(1)_R$ charge
\begin{align}
\mb{D}\mathfrak{W}_{\a(m+n-t+1)\ad(t-1)}^{(t)}(H)&= \big(1-\frac{1}{2}(m-n)\big)\mathfrak{W}_{\a(m+n-t+1)\ad(t-1)}^{(t)}(H)~,\\
\mb{Y}\mathfrak{W}_{\a(m+n-t+1)\ad(t-1)}^{(t)}(H)&=\frac{1}{3}(m-n)\mathfrak{W}_{\a(m+n-t+1)\ad(t-1)}^{(t)}(H)~.
\end{align}
\end{subequations}
The descendent \eqref{genSWeylb} possesses the same superconformal properties but with $m$ interchanged with $n$. 
In addition, when restricted to conformally-flat backgrounds,
\begin{align}
\bm W_{\a(3)}=0~, \label{ConfFBBG}
\end{align}
one can show they are invariant under the higher-depth gauge transformations \eqref{SCHSgenGT},
\begin{align}
\delta_{\L}\mathfrak{W}_{\a(m+n-t+1)\ad(t-1)}^{(t)}(H)=0~, \qquad \delta_{\L} \mathfrak{W}_{\a(m+n-t+1)\ad(t-1)}^{(t)}(\bar{H})=0~.
\end{align}

Using the field strengths \eqref{genSWeyl}, one may construct the action functional
\begin{align} 
\mb{S}_{\text{SCHS}}^{(m,n,t)}[H,\bar{H}]=\frac{1}{2}\text{i}^{m+n+2}\int\text{d}^{4|4}z\, E \, \mathfrak{W}^{\a(m+n-t+1)\ad(t-1)}_{(t)}(H)\mathfrak{W}_{\a(m+n-t+1)\ad(t-1)}^{(t)}(\bar{H}) +\text{c.c.}~, \label{SCHSgenAction}
\end{align}
which is locally $\mc{N}=1$ superconformal (on any background) and gauge invariant on conformally-flat backgrounds. The overall coefficient of $\text{i}^{m+n+2}$ in \eqref{SCHSgenAction} has been chosen because of the identity
\begin{align}
\text{i}^{m+n+1}\int\text{d}^{4|4}z\, E \, \mathfrak{W}^{\a(m+n-t+1)\ad(t-1)}_{(t)}(H)\mathfrak{W}_{\a(m+n-t+1)\ad(t-1)}^{(t)}(\bar{H}) +\text{c.c.} ~=~ 0~,
\end{align}
 which holds on conformally-flat backgrounds up to an irrelevant total derivative.
 

The action \eqref{SCHSgenAction} may be recast into the alternative forms
\begin{align}
\mb{S}_{\text{SCHS}}^{(m,n,t)}[H,\bar{H}]&=\frac{1}{2}\text{i}^{m+n+2}\int\text{d}^{4|4}z\, E \, \bar{H}^{\a(n)\ad(m)}_{(t)}\mathfrak{B}_{\a(n)\ad(m)}^{(t)}(H) +\text{c.c.} \notag\\
&= \frac{1}{2}\text{i}^{m+n+2}\int\text{d}^{4|4}z\, E \, \bar{H}^{\a(n)\ad(m)}_{(t)}\widehat{\mathfrak{B}}_{\a(n)\ad(m)}^{(t)}(H) +\text{c.c.}
\end{align}
where we have made use of the definitions 
\begin{subequations}\label{genSBach}
\begin{align}
\mathfrak{B}_{\a(n)\ad(m)}^{(t)}(H)&= \phantom{..}\big(\Nabla^{\b}\bar{\Nabla}_{\ad}-\frac{\text{i}t}{m-t+1}\Nabla_{\ad}{}^{\b}\big)\big(\Nabla_{\ad}{}^{\b}\big)^{m-t}\mathfrak{W}^{(t)}_{\b(m-t+1)\a(n)\ad(t-1)}(H)~, \label{genSBach1}\\
\widehat{\mathfrak{B}}_{\a(n)\ad(m)}^{(t)}(H)&=-\big(\bNabla^{\bd}\Nabla_{\a}-\frac{\text{i}t}{n-t+1}\Nabla_{\a}{}^{\bd}\big)\big(\Nabla_{\a}{}^{\bd}\big)^{n-t}\overline{\mathfrak{W}}^{(t)}_{\a(t-1)\ad(m)\bd(n-t+1)}(H)~, \label{genSBach2}
\end{align}
\end{subequations}
and where $\overline{\mathfrak{W}}^{(t)}_{\a(t-1)\ad(m+n-t+1)}(H)\equiv \big(\mathfrak{W}_{\a(m+n-t+1)\ad(t-1)}^{(t)}(\bar{H})\big)^*$.

 The descendent \eqref{genSBach1} is primary in a generic background,
 \begin{subequations}
\begin{align}
 K_B \mathfrak{B}_{\a(n)\ad(m)}^{(t)}(H)=0~,
\end{align} 
and carries the following weight and charge
\begin{align}
\mb{D}\mathfrak{B}_{\a(n)\ad(m)}^{(t)}(H)&=\big(2-t+\frac{1}{2}(m+n)\big)\mathfrak{B}_{\a(n)\ad(m)}^{(t)}(H) ~,\\
\mb{Y}\mathfrak{B}_{\a(n)\ad(m)}^{(t)}(H)&=\frac{1}{3}(m-n)\mathfrak{B}_{\a(n)\ad(m)}^{(t)}(H)~.
\end{align} 
\end{subequations}
The other descendent \eqref{genSBach2} carries the same superconformal attributes as \eqref{genSBach1}. 
 On any conformally-flat background, they possess the additional properties:
\begin{enumerate}[label=(\roman*)]
\item $\mf{B}_{\a(n)\ad(m)}(H)$ and $\widehat{\mf{B}}_{\a(n)\ad(m)}(H)$ are equal to one another
\begin{align}
\mf{B}^{(t)}_{\a(n)\ad(m)}(H)=\widehat{\mf{B}}^{(t)}_{\a(n)\ad(m)}(H)~,
\end{align}
which leads to the following reality relation
\begin{align}
\mf{B}^{(t)}_{\a(m)\ad(n)}(\bar{H})=\Big(\mf{B}^{(t)}_{\a(n)\ad(m)}(H)\Big)^*\equiv \overline{\mf{B}}^{(t)}_{\a(m)\ad(n)}(\bar{H})~.
\end{align}
\item $\mathfrak{B}_{\a(n)\ad(m)}^{(t)}$ is `partially conserved',
\begin{align}
0&=\big[\Nabla^{\b},\bar{\Nabla}^{\bd}\big]\big(\Nabla^{\b\bd}\big)^{t-1}\mathfrak{B}_{\b(t)\a(m-t)\bd(t)\ad(n-t)}^{(t)}(H)~.
\end{align} 
\item $\mathfrak{B}_{\a(n)\ad(m)}^{(t)}(H)$ is invariant under the gauge transformations \eqref{SCHSgenGT}
\begin{align} 
\d_{\L}\mf{B}^{(t)}_{\a(n)\ad(m)}(H)&= 0~.
\end{align}
\end{enumerate}
On account of these properties, both \eqref{genSBach1} and \eqref{genSBach2} may be interpreted as the higher-depth analogues of the higher-spin super-Bach tensors \eqref{HSSBACHCSS4CF1} and \eqref{HSSBACHCSS4CF2}.

Finally, we note that for the supermultiplets with depth $t>1$, the presence of vector derivatives in the gauge transformations \eqref{SCHSgenGT} prevents one from constructing a Wess-Zumino gauge. Consequently, all possible component fields are present in this case, some of which are depth $t-1$, $t$ and $t+1$ CHS gauge fields (there are also conformal non-gauge fields present). However in the special case $t=1$, when the depth assumes its minimal value, this is no longer an obstruction and the multiplet shortens. 
 For an analysis of the corresponding Wess-Zumino gauge with $m$ and $n$ arbitrary, we refer the reader to \cite{SCHSgen}, where the component reduction of $\mb{S}_{\text{SCHS}}^{(m,n,t)}[H,\bar{H}]$ with $t=1$ was also given. In section \ref{secCMDgravitonSusy} we give an in-depth analysis of the component content for the case $m=n=t=1$.

\section{Superconformal higher-spin models on Bachgrounds} \label{secBach4|4}

At this point we extend our considerations to superspace backgrounds with non-vanishing super-Weyl tensor, $\bm W_{\a \b \g} \neq 0$, and focus our attention on generic superspace backgrounds which solve the equation of motion for $\mc{N}=1$ conformal supergravity \eqref{N=1CSGA}. Such backgrounds are said to be Bach-flat since they have vanishing super-Bach tensor,
\begin{align}
\bm{B}_{\a\ad}=0~, \label{super-Bach-flat}
\end{align} 
where $\bm{B}_{\a\ad}$ is defined in eq. \eqref{super-Bach}.
In this case it turns out that the SCHS actions \eqref{SCHSactCSSCF4|4} and \eqref{SCHSgenAction}, which we now denote by $\mb{S}_{\text{skeleton}}^{(m,n,t)}[H,\bar{H}]$, are no longer gauge invariant. 
The ensuing situation is very similar to the situation which occurs in the non-supersymmetric case, described in section \ref{secCHSBach}, to which we refer the reader for a thorough discussion. 

In short, to restore gauge invariance it is necessary, at the very least, to add primary non-minimal (i.e. vanishing in the conformally-flat limit $\bm W_{\a \b \g} = 0$) counter-terms. Such terms are quadratic in the prepotential $H_{\a(m)\ad(n)}^{(t)}$ and take the form 
\begin{align}
\mb{S}_{\text{NM}}^{(m,n,t)}[H,\bar{H}]=\frac{1}{2}\text{i}^{m+n} \int \rd^{4|4} z \, E \, H_{(t)}^{\a(m)\ad(n)}\mf{J}^{(t)}_{\a(m)\ad(n)}(\bar{H})+\text{c.c.}~ \label{HSNMFuncSusy}
\end{align}  
The tensor $\mf{J}^{(t)}_{\a(m)\ad(n)}(\bar{H})$ is some non-minimal primary descendant of $\bar{H}^{(t)}_{\a(n)\ad(m)}$ which shares the same superconformal properties as the super-Bach tensor $\mf{B}^{(t)}_{\a(m)\ad(n)}(\bar{H})$, see \eqref{HSSBachprop1CSS} and \eqref{genSBach}. The would-be gauge invariant SCHS action then takes the form 
\begin{align}
\mb{S}_{\text{SCHS}}^{(m,n,t)}[H,\bar{H}]= \mb{S}_{\text{skeleton}}^{(m,n,t)}[H,\bar{H}]+\mb{S}_{\text{NM}}^{(m,n,t)}[H,\bar{H}]~,\qquad \delta_{\L}\mb{S}_{\text{SCHS}}^{(m,n,t)}[H,\bar{H}]\approx0~.\label{NaiveSCHS}
\end{align}
Throughout this section the symbol $\approx$ represents equality modulo terms involving the super-Bach tensor. 
 In section \ref{secSCHSgravitino} and \ref{secCMDgravitonSusy} below we will see two explicit examples of where this is valid, namely, the:\footnote{Recall that we regard ordinary SCHS multiplets $H_{\a(m)\ad(n)}\equiv H^{(0)}_{\a(m)\ad(n)}$ as having depth $t=0$.} (i) conformal gravitino supermultiplet with $m=n+1=t+1=1$; and (ii) maximal depth conformal graviton supermultiplet with $m=n=t=1$.\footnote{It should also be possible to express the gauge invariant action for the conformal supergravity multiplet, with $m=n=t+1=1$, in the form \eqref{NaiveSCHS}. This is because the corresponding action must be attainable by linearising the action for $\mc{N}=1$ conformal supergravity about a Bach-flat background.  } 

However, we saw in the non-supersymmetric case that simply including non-minimal terms was not sufficient to ensure gauge invariance of the CHS models with higher spin. This must also be the case for SCHS models since, for bosonic Bach-flat superspace backgrounds satisfying \eqref{Bbackground} and \eqref{super-Bach-flat}, a gauge invariant
SCHS action decomposes into a multiplet of gauge invariant 
CHS actions at the component level. Indeed, in section \ref{secPGSBach} we will see this explicitly for the pseudo-graviton supermultiplet with $m-2=n=t=0$. 
Finally, in section \ref{secSpin3Recipe}, by supposing the existence of a gauge invariant SCHS model with $m=n=t+2=2$, we deduce non-trivial information regarding the ingredients necessary to construct a gauge invariant CHS model for the conformal spin-3 field.


\subsection{The conformal gravitino supermultiplet}\label{secSCHSgravitino}

The gauge-invariant model describing the superconformal gravitino multiplet was constructed in the earlier work \cite{KMT} within the framework of the Grimm-Wess-Zumino 
geometry \cite{GWZ1, GWZ2}.\footnote{See \cite{BK} for a review of the Grimm-Wess-Zumino geometry.} 
Below we review this model, but in the setting of conformal superspace. 

The conformal gravitino supermultiplet is described by the prepotential $H_{\a}$, which is a primary superfield with Weyl weight and $\sU(1)_{R}$ charge given by
\begin{align}
K_{B}H_{\a}=0~,\qquad \mb{D}H_{\a}=-\frac{1}{2}H_{\a}~,\qquad \mb{Y}H_{\a}=\frac{1}{3}H_{\a}~.
\end{align}
It corresponds to the case $m-1=n=t=0$, and possesses the gauge freedom
\bea
\d_{\L,\l} H_{\a} = \Nabla_{\a} \L + \lambda_{\a} ~, \qquad \bar{\Nabla}_{\ad} \l_{\a} = 0 ~,
\label{5.111}
\eea
where both $\L$ and $\l_{\a}$ are complex primary superfields, and $\l_{\a}$ is covariantly chiral. Under $\L$ and $\lambda$ gauge transformations, the skeleton action
\bea
\mb{S}^{(1,0,0)}_{\text{skeleton}}[H,\bar{H}]=\frac{1}{2}\ri  \int \rd^4x \rd^2 \q \, \cE\, \mathfrak{W}^{\a (2)}(H)
\mathfrak{W}_{\a (2)}(\bar{H}) + {\rm c.c.} ~,
\eea
which is composed of the super-Weyl tensor for $H_{\a}$ and $\bar{H}_{\ad}$
\begin{align}
\mf{W}_{\a (2)}(H)=-\frac{1}{4}\bNabla^2\Nabla_{(\a_1}H_{\a_2)}~,\qquad 
\mf{W}_{\a (2)}(\bar{H})=-\frac{1}{4}\bNabla^2\Nabla_{(\a_1}{}^{\bd}\Nabla_{\a_2)}\bar{H}_{\bd}~,
\end{align}
 has the following variations 
\begin{subequations}
\begin{align}
\delta_{\L}\mb{S}^{(1,0,0)}_{\text{skeleton}}&=-\frac{1}{4}\int\text{d}^{4|4}z\, E \, \L\bigg\{\bm{\bar{W}}^{\bd(3)}\bar{\Nabla}_{\bd}\Nabla^2\bar{\Nabla}_{\bd}\bar{H}_{\bd}-2\bar{\Nabla}_{\bd}\bm{\bar{W}}^{\bd(3)}\Nabla^2\bar{\Nabla}_{\bd}\bar{H}_{\bd}\bigg\} +\text{c.c.}~,\label{tinozeta}\\
\delta_{\lambda}\mb{S}^{(1,0,0)}_{\text{skeleton}}&=-\text{i}\int
 \rd^4x \rd^2 \q 
\, \cE \, \lambda^{\alpha}\bm W_{\a}{}^{\b(2)} \mathfrak{W}_{\b(2)}(\bar{H})+\text{c.c.}
\end{align}
\end{subequations}

To compensate for the non-zero variation of the skeleton, we need to introduce a non-minimal primary action of the form
\begin{align}
\mb{S}_{\text{NM},i}^{(1,0,0)}[H,\bar{H}]=\frac{1}{2}\text{i}\int \text{d}^{4|4}z \, E \, H^{\a}\mathfrak{J}^{(i)}_{\alpha}(\bar{H}) + \text{c.c.} ~,\label{3.777}
\end{align}
where $\mathfrak{J}_{\alpha}^{(i)}(\bar{H})$ is a primary spinor superfield
of Weyl weight $5/2$ and $\sU(1)_{R}$ charge $-1/3$
 that depends explicitly on the super-Weyl tensor. 
 It may be shown that there are two structures which satisfy these requirements, and they take the form
\begin{subequations}
\begin{align}
\mathfrak{J}^{(1)}_{\alpha}(\bar{H})&=2\bm{\bar{W}}^{\bd(3)}\Nabla_{\a\bd}\bar{\Nabla}_{\bd}\bar{H}_{\bd}-\text{i}\bar{\Nabla}_{\bd}\bm{\bar{W}}^{\bd(3)}\Nabla_{\a}\bar{\Nabla}_{\bd}\bar{H}_{\bd}+2\Nabla_{\a\bd}\bm{\bar{W}}^{\bd(3)}\bar{\Nabla}_{\bd}\bar{H}_{\bd}~,\\
\mathfrak{J}^{(2)}_{\alpha}(\bar{H})&=2\bm W_{\a}{}^{\b(2)}\Nabla_{\b}{}^{\bd}\Nabla_{\b}\bar{H}_{\bd}-\ri\Nabla_{\b}\bm{W}_{\a}{}^{\b(2)}\bar{\Nabla}^{\bd}\Nabla_{\b}\bar{H}_{\bd}-2\Nabla_{\b}{}^{\bd}\bm{W}_{\a}{}^{\b(2)}\Nabla_{\b}\bar{H}_{\bd}~.
\end{align}
\end{subequations}
However, the corresponding actions \eqref{3.777} are not independent of one another. Modulo a total derivative one may show that they are related via\footnote{See appendix \ref{AppIBPCSS4|4} for a discussion on the technical issue of integration by parts (IBP) in conformal superspace. There, an IBP rule is proposed and all instances of IBP in this subsection are justified. }
\begin{align}
\text{i}\int \text{d}^{4|4}z \, E \, H^{\alpha}\mathfrak{J}^{(1)}_{\alpha}(\bar{H}) + \text{c.c.}&=\text{i}\int \text{d}^{4|4}z \, E \, \bar{H}^{\ad}\bar{\mathfrak{J}}^{(1)}_{\ad}(H) + \text{c.c.}\notag\\
&=\text{i}\int \text{d}^{4|4}z \, E \, H^{\alpha}\bigg\{\mathfrak{J}^{(2)}_{\alpha}(\bar{H})+2\ri \bm B_{\a\ad}\bar{H}^{\ad}\bigg\} +\text{c.c.}  \label{2020}
\end{align}
Therefore, it suffices to consider only the first functional.

Using the identity \eqref{2020}, it can be shown that a generic (infinitesimal) variation of the action \eqref{3.777} (i.e. with respect to both $H_{\a}$ and $\bar{H}_{\ad}$), with $i=1$, takes the form
\begin{align}
\delta \mb{S}_{\text{NM},1}^{(1,0,0)}&=\frac{\ri}{2} \int \text{d}^{4|4}z \, E \, \delta H^{\a}\bigg\{\mathfrak{J}^{(1)}_{\alpha}(\bar{H})+\mathfrak{J}^{(2)}_{\alpha}(\bar{H})+2\ri \bm{B}_{\a\ad}\bar{H}^{\ad}\bigg\}+\text{c.c.} \label{9.87}
\end{align}
In addition to the aforementioned properties, the superfields $\mathfrak{J}^{(i)}_{\alpha}(\bar{H})$ may be shown to satisfy the following useful identities
\begin{subequations}\label{3.88}
\begin{align}
\Nabla^{\alpha}\mathfrak{J}^{(1)}_{\alpha}(\bar{H})&=\frac{1}{2}\bm{\bar{W}}^{\bd(3)}\bar{\Nabla}_{\bd}\Nabla^2\bar{\Nabla}_{\bd}\bar{H}_{\bd}-\bar{\Nabla}_{\bd}\bm{\bar{W}}^{\bd(3)}\Nabla^2\bar{\Nabla}_{\bd}\bar{H}_{\bd}~,\qquad \label{3.88a}\\
\Nabla^{\alpha}\mathfrak{J}^{(2)}_{\alpha}(\bar{H})&=-2\ri \bm{B}_{\a\ad}\Nabla^{\a}\bar{H}^{\ad}~,\label{3.88b}\\
\bar{\Nabla}^2\mathfrak{J}^{(1)}_{\alpha}(\bar{H})&=\phantom{-}2\text{i}\bar{\Nabla}^2\big(\bm{B}_{\a\ad}\bar{H}^{\ad}\big)~,\label{3.88c}\\
 \bar{\Nabla}^2\mathfrak{J}^{(2)}_{\alpha}(\bar{H})&=-8\bm{W}_{\a}{}^{\b(2)}\mathfrak{W}_{\b(2)}(\bar{H})~.\label{3.88d}
\end{align}
\end{subequations}
The relations \eqref{3.88a} and \eqref{3.88b} are useful when computing the $\L$-gauge variation of \eqref{3.777}. For technical reasons, the $\lambda$-gauge variation is best done in the chiral subspace (see appendix \ref{AppChiralSubspace} for a discussion on this), for which \eqref{3.88c} and \eqref{3.88d} are crucial. 

By virtue of \eqref{9.87} and \eqref{3.88}, one can see that the gauge variation of the functional $\mb{S}_{\text{NM}}^{(1,0,0)}$ is given by
\begin{align}
\delta_{\L, \lambda}\mb{S}_{\text{NM}}^{(1,0,0)}=-\delta_{\L,\lambda}\mb{S}_{\text{Skeleton}}^{(1,0,0)}+\bigg(\frac{1}{2}\int
 \rd^4x \rd^2 \q \,
\cE \, \bar{\Nabla}^2\big(\lambda^{\a}\bm B_{\a\ad}\bar{H}^{\ad}\big) +\text{c.c.}\bigg)~.
\end{align}
It follows that the action
\begin{subequations}
\begin{align}
\mb{S}_{\text{SCHS}}^{(1,0,0)}[H,\bar{H}]&=\mb{S}^{(1,0,0)}_{\text{Skeleton}}[H,\bar{H}]+\mb{S}^{(1,0,0)}_{\text{NM},1}[H,\bar{H}]~\\
&=\frac{\ri}{2} \int \rd^4x \rd^2 \q \,\cE \, 
\mathfrak{W}^{\a(2)}(H)\mathfrak{W}_{\a(2)}(\bar{H})+\ri\int\text{d}^{4|4}z \, E \, H^{\a}\bigg\{\bm{\bar{W}}^{\bd(3)}\Nabla_{\a\bd}\bar{\Nabla}_{\bd}\bar{H}_{\bd}\notag~~~~~~~~~~~~~~~~~~~\\
&~~~-\frac{\text{i}}{2}\bar{\Nabla}_{\bd}\bm{\bar{W}}^{\bd(3)}\Nabla_{\a}\bar{\Nabla}_{\bd}\bar{H}_{\bd}+\Nabla_{\a\bd}\bm{\bar{W}}^{\bd(3)}\bar{\Nabla}_{\bd}\bar{H}_{\bd}\bigg\} +\text{c.c.}
\end{align}
\end{subequations}
has gauge variation that is strictly proportional to the super-Bach tensor,
\begin{align}
\delta_{\L,\lambda} \mb{S}_{\text{SCHS}}^{(1,0,0)}= -2\int \text{d}^{4|4}z \, E \, \lambda^{\a}\bm B_{\a\ad}\bar{H}^{\ad}+\text{c.c.}~,
\end{align}
and is hence gauge invariant when restricted to a super-Bach-flat background
\begin{align}
\delta_{\L,\lambda} \mb{S}_{\text{SCHS}}^{(1,0,0)}\approx 0~.
\end{align}
One may check that upon using \eqref{2020} and degauging to Grimm-Wess-Zumino superspace, the above action coincides with the one given in \cite{KMT} modulo a term proportional to the super-Bach tensor. 

\subsection{The maximal-depth conformal graviton supermultiplet} \label{secCMDgravitonSusy}

The maximal-depth conformal graviton supermultiplet is described by the weightless real primary superfield\footnote{In this section we drop all labels referring to the depth since we deal only with $t=1$.} $H_{\a\ad}$ which is inert under $\sU(1)_{R}$ transformations,
\begin{subequations}\label{gengrav}
\begin{align}
K_{B}H_{\a\ad}=0~,\qquad \mathbb{D}H_{\a\ad}=0~,\qquad \mb{Y}H_{\a\ad}=0~,\qquad  H_{\a\ad}=\bar{H}_{\a\ad}~,
\end{align}
 and which is defined modulo the depth-1 gauge transformations
\begin{align}
\delta_{\L}H_{\a\ad}=\big[\Nabla_{\a},\bar{\Nabla}_{\ad}\big]\L~. \label{gt24|4}
\end{align}
\end{subequations}
Here the real gauge parameter $\L$ is primary and unconstrained. 
This corresponds to the generalised supermultiplet of section \ref{secGenSCHSCF} with $m=n=t=1$ and is called the maximal-depth conformal graviton supermultiplet because, as will be shown shortly, it contains the maximal-depth conformal graviton at the component level.


\subsubsection{Gauge invariant action in Bach-flat background}

The linearised super-Weyl tensor associated with $H_{\a\ad}$, and its corresponding gauge variation under \eqref{gt24|4}, is given by
\begin{subequations}
\begin{align}
\mathfrak{W}_{\a(2)}(H)&=\Big(\bar{\Nabla}^{\bd}\Nabla_{(\a_1}-\text{i}\Nabla_{(\a_1}{}^{\bd}\Big)H_{\a_2)\bd} ~,\\
\delta_{\L}\mathfrak{W}_{\a(2)}(H)&= 3\Big(2\bm W_{\a(2)}{}^{\b}\Nabla_{\b}\L+\Nabla_{\b}\bm W_{\a(2)}{}^{\b}\L\Big)~.
\end{align}
\end{subequations}
 It follows that the skeleton action
\begin{align}
\label{HSkeleton}
\mb{S}^{(1,1,1)}_{\text{skeleton}}[H]=\frac{1}{3}\int \text{d}^{4|4} z\, E \, \mathfrak{W}^{\a(2)}(H)\mathfrak{W}_{\a(2)}(H) +\text{c.c.}
\end{align}
(here we have chosen a different overall normalisation as compared to \eqref{SCHSgenAction}) has gauge variation proportional to the background super-Weyl tensor 
\begin{align}
\delta_{\L}S^{(1,1,1)}_{\text{skeleton}}[H]=2\int\text{d}^{4|4}z\, E \, \L \bigg\{3\Nabla_{\g}\bm W^{\g\a(2)}\mathfrak{W}_{\a(2)}(H)-2\bm W^{\g\a(2)}\Nabla_{\g}\mathfrak{W}_{\a(2)}(H)\bigg\} +\text{c.c.}
\end{align}

It is possible to restore gauge invariance to the skeleton by supplementing it with the non-minimal primary action
\begin{align}
\mb{S}^{(1,1,1)}_{\text{NM}}[H]=\int\text{d}^{4|4}z\, E \, H^{\a\ad}\bm W_{\a}{}^{\b(2)}\Nabla_{\b}H_{\b\ad}+\text{c.c.}
\end{align}  
The action which is gauge invariant in a Bach-flat background may then be shown to be
\begin{align}
\mb{S}^{(1,1,1)}_{\text{SCHS}}[H]=\mb{S}^{(1,1,1)}_{\text{skeleton}}[H]-3\mb{S}^{(1,1,1)}_{\text{NM}}[H]~,\qquad \delta_{\L}\mb{S}^{(1,1,1)}_{\text{SCHS}}[H]\approx 0~. \label{3.774|4}
\end{align}


\subsubsection{The component action}
\label{WZMDSpin2}

Let us study the component content of the model \eqref{3.774|4} when restricted to a bosonic Bach-flat superspace background, i.e. one satisfying \eqref{Bbackground} and \eqref{super-Bach-flat}.
By making use of the gauge freedom \eqref{gt24|4}, one may choose a Wess-Zumino gauge such that the only non-vanishing component fields contained within $H_{\a\ad}$ are 
\begin{subequations}
\label{HComponents}
\bea
\psi_{\a(2)\ad} &=& \Nabla_{(\a_1} H_{\a_2) \ad}| ~, \\
A_{\aa} &=& - \frac{1}{4} \Nabla^2 H_{\aa} | ~, \\
h_{\a(2) \ad(2)} &=& \frac{1}{2} \big[ \Nabla_{(\a_1} , \bNabla_{(\ad_1} \big] H_{\a_2) \ad_2)} | ~, \\
\chi_{\a(2)} &=& \frac{1}{2} \big[\Nabla_{(\a_1} , \bNabla^{\ad} \big] H_{\a_2) \ad} | ~, \\
\varphi_{\a(2)\ad} &=& - \frac{1}{4} \Nabla_{(\a_1} \bNabla^2 H_{\a_2) \ad} | + \frac{3 \ri}{2} \nabla_{(\a_1}{}^{\bd} \bar{\psi}_{\a_2) \ad \bd} ~, \\
\phi_\a &=& - \frac{1}{4} \bNabla^{\ad} \Nabla^2 H_{\aa}| - \frac{\ri}{2} \nabla^{\bb} \psi_{\a \b \bd} ~, \\
V_{\aa} &=& \frac{1}{32} \{ \Nabla^2 , \bNabla^2 \} H_{\aa} | + \frac{\ri}{4} \nabla^{\b}{}_{\ad} \chi_{\a \b} - \frac{\ri}{4} \nabla_{\a}{}^{\bd} \bar{\chi}_{\ad \bd} ~.
\eea
\end{subequations}
Both $h_{\a(2)\ad(2)}$ and $V_{\a\ad}$ are real whilst all other component fields are complex. All the fields are primary, and their Weyl weights may be deduced from \eqref{gengrav}. Associated with \eqref{HComponents} are the following gauge fixing conditions
\begin{align} \label{Hgfc}
\big[ \Nabla_{\a} , \bNabla_{\ad} \big] \L | = 0 ~, \qquad 
\bNabla_{\ad} \Nabla^2 \L | = 3 \ri \Nabla_{\aa} \Nabla^{\a} \L | ~, \qquad
\{ \Nabla^2 , \bNabla^2 \} \L | = - 4 \Box \L |  ~.
\end{align}
The residual gauge transformations are generated by the fields
\begin{align}
\l := - \frac{1}{4} \Nabla^2 \L|~,\qquad \e_{\a}:=2\ri \Nabla_{\a}\L|~,\qquad \xi:= -2\L|~.
\end{align}
 Consequently, the fields $A_\aa, \psi_{\a(2) \ad}$ and $h_{\a(2) \ad(2)}$ have the gauge transformation laws
\begin{subequations}
	\bea
	\d_{\l} A_\aa &=& \nabla_{\aa} \l ~, \\
	\d_{\e} \psi_{\a(2) \ad} &=& \nabla_{\a\ad} \e_{\a} ~, \\
	\d_\xi h_{\a(2) \ad(2)} &=& \nabla_{\a\ad} \nabla_{\a\ad} \xi ~. \label{333}
	\eea
\end{subequations}
On the other hand, the three fields $\chi_{\a(2)}, \vf_{\a(2)\ad}$ and $\phi_{\a}$ are type I conformal non-gauge fields (see appendix \ref{AppCNGM}) and do not have any gauge freedom. The field $V_{\a\ad}$ is auxiliary. 

Using the above definitions, the action \eqref{3.774|4} may be readily reduced to components:
\bea
\label{MaxDepthSpin2ComponentAction}
\mb{S}^{(1,1,1)}_{\text{SCHS}}[H] &=& - \int \rd^4 x \, e\, \bigg\{ \frac{3}{4} \bigg[\mc{W}^{\a(3) \ad}(h) \mc{W}_{\a(3) \ad}(h) -h^{\a(2)\ad(2)}W_{\a(2)}{}^{\b(2)}h_{\b(2)\ad(2)}\bigg] \non\\
&+& \frac{3 \ri}{2} \bigg[\mc{W}^{\a(3)} (\psi)\mc{W}_{\a(3)} (\bar{\psi})-\psi^{\a(2)\ad}\Big(W_{\a(2)}{}^{\b(2)}\nabla_{\b}{}^{\bd}\bar{\psi}_{\b\ad\bd}-\nabla_{\b}{}^{\bd}W_{\a(2)}{}^{\b(2)}\bar{\psi}_{\b\ad\bd}\Big)\bigg]  \non\\
&+& 2 \mc{W}^{\a(2)} (A) \mc{W}_{\a(2)} (\bar{A}) - 2 \ri \bar{\varphi}^{\a \ad(2)} \mathfrak{X}_{\a \ad(2)}(\varphi) + \frac{1}{4} \bar{\chi}^{\ad(2)} \mathfrak{X}_{\ad(2)} (\chi) - \frac{ \ri}{2} \bar{\phi}^{\ad} \mathfrak{X}_\ad(\phi)\non \\
&-& 2 V^{\aa} V_{\aa} \bigg\} + \text{c.c.} \\
&\equiv& - \frac{3}{2} S^{(2,2,2)}_{\text{CHS}}[h] + 3 S^{(2,1,1)}_{\text{CHS}}[\psi, \bar{\psi}] + 4 S^{(1,1,1)}_{\text{CHS}}[A,\bar{A}] \non\\
&\phantom{-}&- 4  S^{(2,1)}_{\text{NG}}[\varphi , \bar{\varphi}] + \frac{1}{2} S^{(2,0)}_{\text{NG}}[\chi,\bar{\chi}] +  S^{(1,0)}_{\text{NG}}[\phi, \bar{\phi}] - 4 S_{\text{NG}}^{(2,2)}[V]  ~. 
\eea
This action has been expressed in a manifestly conformal and gauge invariant (in a Bach-flat background) form by using the field strengths associated with each field,
\begin{subequations}
\begin{align}
\mc{W}_{\a(3)} (\psi)&=\nabla_{(\a_1}{}^{\bd}\psi_{\a_2\a_3)\bd}~, \qquad\qquad\quad \mc{W}_{\a(3)\ad}(h)=\nabla_{(\a_1}{}^{\bd}h_{\a_2\a_3)\ad\bd} ~,\\
\mc{W}_{\a(3)} (\bar{\psi})&=\nabla_{(\a_1}{}^{\bd_1}\nabla_{\a_2}{}^{\bd_2}\bar{\psi}_{\a_3)\bd(2)}~,\qquad ~~\mathfrak{X}_{\ad(2)}(\chi)=\nabla_{(\ad_1}{}^{\b_1}\nabla_{\ad_2)}{}^{\b_2}\chi_{\b(2)} ~,\\
\mc{W}_{\a(2)} (A)&=\nabla_{(\a_1}{}^{\bd}A_{\a_2)\bd}~, \qquad\qquad\quad \phantom{..}~\mathfrak{X}_{\a \ad(2)}(\varphi)=\nabla_{(\ad_1}{}^{\b}\vf_{\b\a\ad_2)}~,\\
\mc{W}_{\a(2)} (\bar{A})&=\nabla_{(\a_1}{}^{\bd}\bar{A}_{\a_2)\bd}~, \qquad \qquad \qquad ~\phantom{..}~ \mathfrak{X}_\ad(\phi)=\nabla_{\ad}{}^{\b}\phi_{\b}~, 
\end{align}
\end{subequations}
along with the non-minimal counter terms necessary for gauge invariance.

The analysis above indicates that the component action decomposes into a (diagonal) sum of gauge invariant  CHS actions -- denoted $S^{(2,2,2)}_{\text{CHS}}[h], S^{(2,1,1)}_{\text{CHS}}[\psi, \bar{\psi}]$ and $S^{(1,1,1)}_{\text{CHS}}[A,\bar{A}]$ -- describing a maximal-depth conformal graviton $h_{\a(2)\ad(2)}$ \eqref{CMDspin2FullAct}, a conformal gravitino $\psi_{\a(2)\ad}$ \eqref{CgravitinoAct} and a complex Maxwell field $A_{\a\ad}$ respectively.  In addition, there are also several non-gauge fields $\chi_{\a(2)}, \vf_{\a(2)\ad}$ and $\phi_{\a}$ present, the latter of which describes a massless Weyl spinor. Their corresponding actions may be found in eq. \eqref{NGActionArb}. The vector field $V_\aa$ is auxiliary since it enters the action with no derivatives.


The overall sign of the action \eqref{SCHSgenAction} has been chosen so that the Maxwell action 
\begin{align}
S^{(1,1,1)}_{\text{CHS}}[A,\bar{A}]=- \frac{1}{2} \int \text{d}^4x \, e \, F^{ab}(A)F_{ab}(\bar{A})~,\qquad F_{ab}(A):=\mathcal{D}_{a}A_{b}-\mathcal{D}_{b}A_{a}~,
\end{align}
in \eqref{MaxDepthSpin2ComponentAction} comes with canonical sign. Finally, we would like to point out that the action \eqref{3.774|4} may be recast into the form
\begin{align}
\mb{S}^{(1,1,1)}_{\text{SCHS}}[H] =\frac{3}{2}\int \text{d}^{4|4}z H^{\a\ad}&\bigg\{\frac{3}{2}D^{\b}\bar{D}^2D_{\b}H_{\a\ad}-\frac{1}{2}\big[D_{\a},\bar{D}_{\ad}\big]\big[D_{\b},\bar{D}_{\bd}\big]H^{\b\bd} \non\\
&~-4\partial_{\a\ad}\partial^{\b\bd}H_{\b\bd}-\frac{1}{4}\big\{D^2,\bar{D}^2\big\} H_{\a\ad}\bigg\}
\label{3.154|4}
\end{align}
 in Minkowski superspace and defines a superconformal field theory.
The action is invariant under the gauge transformation 
$\delta_{\L}H_{\a\ad}=\big[D_{\a},\bar{D}_{\ad}\big]\L$.
 There exists a model for linearised supergravity constructed in \cite{BGLP} with a larger gauge freedom
\bea
\delta H_{\a\ad}=\big[D_{\a},\bar{D}_{\ad}\big]\L + \l_{\a\ad} + \bar \l_{\a\ad}~, 
\qquad \bar D_\bd \l_{\a\ad} =0~,
\eea
than that which the action \eqref{3.154|4} possesses. However, the $\l$ gauge symmetry proves to be incompatible with 
the superconformal invariance. 

\subsection{The conformal pseudo-graviton supermultiplet} \label{secPGSBach}

Let us now examine a curved superspace extension of the theory described by 
the prepotential $H_{\a(2)}$, with $m-2=n=t=0$. We recall that when restricted to a bosonic background, the gauge freedom \eqref{SCHSprepGTIIICSS} allows us to adopt a Wess-Zumino gauge where the component of $H_{\a(2)}$ defined by \eqref{2.34a} is identifiable with the conformal pseudo-graviton field examined in section \ref{secConformalHook}. 
In addition, the prepotential $H_{\a(2)}$ has the same dimension and total number of spinor indices as the  prepotential $H_{\a\ad}$ of conformal supergravity \cite{OS1,FZ2}.
Hence we will refer to this model as the superconformal pseudo-graviton multiplet.

Conformal superspace is a powerful formalism for the construction of superconformal invariants.
However, its use also brings in certain
technical subtleties associated with integration by parts and with the transition between integrals over the full superspace and the chiral subspace.\footnote{Unlike conformal superspace, such subtleties do not occur within the $\sU(1) $ superspace setting. However, $\sU(1)$ superspace is much less 
powerful for constructing higher-derivative superconformal invariants. It is fair to say that conformal superspace and $\sU(1)$ superspace are complementary.} 
Without developing efficient rules to perform these operations (and such rules are absent at the moment)
conformal superspace becomes
impractical to do various field theoretic calculations for higher-derivative models such as the ones under consideration. In appendix \ref{AppCSS4} we give a detailed discussion of these issues and in some cases propose such a rule. 

Fortunately, for the current model it can be shown that any contributions arising from these subtleties are at least second order in the super-Weyl tensor. In fact, it turns out that there is a shortcut which will allow us to deduce the most important properties of the full gauge-invariant model, such as the presence of a lower-spin superfield. It is for this reason that parts of our subsequent analysis will be restricted to first order in $\bm W_{\a(3)}$. 

The superfield $H_{\a(2)}$ possesses the superconformal properties
\begin{align}
K_{B}H_{\a(2)}=0~,\qquad \mathbb{D}H_{\a(2)}=-H_{\a(2)}~,\qquad \mb{Y}H_{\a(2)}=\frac{2}{3}H_{\a(2)}~,
\end{align}
and is defined modulo the gauge transformations
\bea
\d_{\L,\l} H_{\a(2)} = \Nabla_{(\a_1} \L_{\a_2)} + \lambda_{\a(2)} ~, 
\qquad \bar{\Nabla}_{\bd} \l_{\a(2)} = 0 ~.\label{PGGT}
\eea 
Just as in the previous cases, we begin with the skeleton action
\bea
\mb{S}^{(2,0,0)}_{\text{skeleton}}[H,\bar{H}]= -\frac{1}{2}  \int \rd^4x \rd^2 \q \, \cE\, \mathfrak W^{\a (3)}(H)
\mathfrak W_{\a (3)}(\bar{H}) +{\rm c.c.} 
\eea
It has the following variations under the gauge transformations \eqref{PGGT}
\begin{subequations}
\begin{align}
\delta_{\L} \mb{S}_{\text{skeleton}}^{(2,0,0)}&=\frac{1}{2}\int\text{d}^{4|4}z\, E \, \L^{\a} \bigg\{ \non\\
&~~~- \ri \Nabla_{\a}{}^{\ad} \bar{\Nabla}^{\bd} \bar{\bm{W}}_{\bd}{}^{\ad(2)} \Nabla^2 \bar{\Nabla}_{(\ad_1} \bar{H}_{\ad_2 \ad_3)} - \ri \Nabla_{\a}{}^{\bd} \bar{\Nabla}^{\ad} \bar{\bm{W}}_{\bd}{}^{\ad(2)} \Nabla^2 \bar{\Nabla}_{(\ad_1} \bar{H}_{\ad_2 \ad_3)}\non \\
&~~~ - \frac{\ri}{2} \Nabla_{\a}{}^{\ad} \bar{\bm{W}}^{\ad(2) \bd} \bar{\Nabla}_{\bd} \Nabla^2 \bar{\Nabla}_{(\ad_1} \bar{H}_{\ad_2 \ad_3)} + \frac{7 \ri}{12} \Nabla_{\a}{}^{\bd} \bar{\bm{W}}_{\bd}{}^{\ad(2)} \bar{\Nabla}^{\gd} \Nabla^2 \bar{\Nabla}_\gd \bar{H}_{\ad(2)} \non \\
&~~~ + \frac{7 \ri}{6} \Nabla_{\a}{}^{\bd}\bar{\bm{W}}_{\bd}{}^{\ad(2)} \bar{\Nabla}^\gd \bar{\Nabla}_{(\ad_1} \bar{H}_{\ad_2) \gd} - \frac{9 \ri}{4} \Nabla^{\bd} \bar{\bm{W}}_{\bd}{}^{\ad(2)} \Nabla_{\a}{}^{\ad} \Nabla^2 \bar{\Nabla}_{(\ad_1} \bar{H}_{\ad_2 \ad_3)} \non \\
&~~~
+ \frac{\ri}{2} \bar{\Nabla}^{\bd} \bm{W}^{\ad(3)} \Nabla_{\a \bd} \Nabla^2 \bar{\Nabla}_{(\ad_1} \bar{H}_{\ad_2 \ad_3)} - \frac{3 \ri}{4} \bar{\bm{W}}^{\ad(2) \bd} \Nabla_{\a}{}^{\ad} \bar{\Nabla}_{\bd} \Nabla^2 \bar{\Nabla}_{(\ad_1} \bar{H}_{\ad_2 \ad_3)} \non \\
&~~~ - \frac{3 \ri}{4} \bar{\bm{W}}^{\ad(2) \bd} \Nabla_{\a \bd} \bar{\Nabla}^{\ad} \Nabla^2 \bar{\Nabla}_{(\ad_1} \bar{H}_{\ad_2 \ad_3)} \bigg\}  +\text{c.c.}~, \\
\delta_{\lambda}\mb{S}_{\text{skeleton}}^{(2,0,0)}&=- \int
 \rd^4x \rd^2 \q 
\, \cE \, \lambda^{\alpha \beta }\bm{W}^{\a(2)}{}_{\b} \mathfrak{W}_{\a(3)}(\bar{H})+\text{c.c.} 
\end{align}
\end{subequations}

Following the usual philosophy, in order to cancel the parts of the variation that are linear in the super-Weyl tensor we must introduce primary non-minimal corrections.  In general, they will take the form
\begin{align}
\mb{S}_{\text{NM},i}^{(2,0,0)}[H,\bar{H}] = \frac{1}{2}\int \text{d}^{4|4}z \, E \, H^{\alpha(2)}\mathfrak{J}^{(i)}_{\alpha(2)}(\bar{H}) + \text{c.c.} \label{6.34|4}
\end{align}
where $\mathfrak{J}_{\alpha(2)}^{(i)}(\bar{H})$ is a primary superfield
of Weyl weight $3$ and $\sU(1)_{R}$ charge $-2/3$
 that depends explicitly on the super-Weyl tensor. 
It may be shown that to linear order in $\bm W_{\a(3)}$, there are exactly two such structures, and they are given by 
\begin{subequations}\label{66.5}
\begin{align}
\mathfrak{J}^{(1)}_{\alpha(2)}(\bar{H})&= \Nabla^{\g \ad} \Nabla^{\b \ad} \bm{W}_{\b \a(2)} \Nabla_{\g} \bar{H}_{\ad(2)} + \frac{\ri}{4} \Nabla^{\b \ad} \Nabla_{\b} \bm{W}_{\a(2)}{}^{\g} \bar{\Nabla}^{\ad} \Nabla_{\g} \bar{H}_{\ad(2)} \non \\
&~~~ + \frac{\ri}{2} \Nabla^{\b \ad} \Nabla^{\g} \bm{W}_{\g \a(2)} \bar{\Nabla}^{\ad} \Nabla_{\b} \bar{H}_{\ad(2)} + \frac{5 \ri}{24} \Nabla^{\b \ad} \bm{W}_{\b \a(2)} \bar{\Nabla}^{\ad} \Nabla^2 \bar{H}_{\ad(2)} \non \\
&~~~ - \frac{1}{3} \Nabla^{\b \ad} \bm{W}_{\b \a(2)} \Nabla^{\g \ad} \Nabla_{\g} \bar{H}_{\ad(2)} + \frac{3}{2} \Nabla^{\b \ad} \bm{W}_{\a(2)}{}^{\g} \Nabla_{\b}{}^{\ad} \Nabla_{\g} \bar{H}_{\ad(2)} \non \\
&~~~ - \Nabla_{\a}{}^{\ad} \bm{W}_{\a}{}^{\b(2)} \Nabla_{\b}{}^{\ad} \Nabla_{\b} \bar{H}_{\ad(2)} + \frac{7 \ri}{12} \Nabla^{\b} \bm{W}_{\b \a(2)} \Nabla^{\g \ad} \bar{\Nabla}^{\ad} \Nabla_{\g} \bar{H}_{\ad(2)} \non \\
&~~~ + \frac{\ri}{4} \Nabla^{\b} \bm{W}_{\a(2)}{}^{\g} \Nabla_{\g}{}^{\ad} \bar{\Nabla}^{\ad} \Nabla_{\b} \bar{H}_{\ad(2)} + \frac{\ri}{2} \Nabla^{\b} \bm{W}_{\a \b}{}^{\g} \Nabla_{\g}{}^{\ad} \bar{\Nabla}^{\ad} \Nabla_{\a} \bar{H}_{\ad(2)} \non \\
&~~~ + \frac{2}{3} \bm{W}_{\a(2)}{}^{\b} \Nabla_{\b}{}^{\ad} \Nabla^{\g \ad} \Nabla_{\g} \bar{H}_{\ad(2)} + \frac{5 \ri}{24} \bm{W}_{\a(2)}{}^{\b} \Nabla_{\b}{}^{\ad} \Nabla^{\ad} \Nabla^{2} \bar{H}_{\ad(2)} \non \\
&~~~ + \bm{W}_{\a}{}^{\b(2)} \Nabla_{\b}{}^{\ad} \Nabla_{\b}{}^{\ad} \Nabla_{\a} \bar{H}_{\ad(2)} - \frac{2}{3} \bm{W}_{\a(2)}{}^{ \b} \bar{\bm{W}}^{\ad(2) \bd} \bar{\Nabla}_{\bd} \Nabla_{\b} \bar{H}_{\ad(2)}
~, \label{66.5a}\\
\mathfrak{J}^{(2)}_{\alpha(2)}(\bar{H})&= - \frac{2}{3} \Nabla_{\a}{}^{\ad} \Nabla_{\a}{}^{ \bd} \bar{\bm{W}}^{\ad}{}_{\bd}{}^{\gd} \bar{\Nabla}_{\gd} \bar{H}_{\ad(2)} + \frac{\ri}{3} \Nabla_{\a}{}^{\bd} \bar{\Nabla}_{\bd} \bar{\bm{W}}^{\ad(2) \gd} \bar{\Nabla}_{\gd} \Nabla_{\a} \bar{H}_{\ad(2)} \non \\
&~~~ - \frac{2 \ri}{3} \bar{\Nabla}^{\bd} \Nabla_{\a}{}^{\gd} \bar{\bm{W}}_{\gd}{}^{\ad(2)} \bar{\Nabla}_{\bd} \Nabla_{\a} \bar{H}_{\ad(2)} + \frac{4}{3} \bar{\Nabla}^{\bd} \Nabla_{\a}{}^{\ad} \bar{\bm{W}}^{\ad}{}_{\bd}{}^{\gd} \Nabla_{\a\gd} \bar{H}_{\ad(2)} \non \\
&~~~ - \frac{2}{3} \bar{\Nabla}^{\bd} \Nabla_{\a\bd} \bar{\bm{W}}^{\ad(2) \gd} \Nabla_{\a\gd} \bar{H}_{\ad(2)} - \frac{5 \ri}{12} \Nabla_{\a}{}^{ \bd} \bar{\bm{W}}_{\bd}{}^{\ad(2)} \bar{\Nabla}^2 \Nabla_{\a} \bar{H}_{\ad(2)} \non \\
&~~~ + \frac{1}{3} \Nabla_{\a}{}^{\bd} \bar{\bm{W}}_{\bd}{}^{\ad(2)} \bar{\Nabla}^{\gd} \Nabla_{\a\gd} \bar{H}_{\ad(2)} - \frac{1}{3} \Nabla_{\a}{}^{ \bd} \bar{\bm{W}}^{\ad(2) \gd} \bar{\Nabla}_{\gd} \Nabla_{\a\bd} \bar{H}_{\ad(2)} \non \\
&~~~ + \Nabla_{\a}{}^{ \ad} \bar{\bm{W}}^{\ad \bd(2)} \bar{\Nabla}_{\bd} \Nabla_{\a\bd} \bar{H}_{\ad(2)} - \Nabla_{\a}{}^{\bd} \bar{\bm{W}}^{\ad}{}_{\bd}{}^{\gd} \bar{\Nabla}^{\ad} \Nabla_{\a\gd} \bar{H}_{\ad(2)} \non \\
&~~~ - \frac{\ri}{2} \bar{\Nabla}^{\bd} \bar{\bm{W}}^{\ad}{}_{\bd}{}^{\gd} \bar{\Nabla}^{\ad} \Nabla_{\a} \Nabla_{\a\gd} \bar{H}_{\ad(2)} + \frac{2}{3} \bar{\Nabla}^{\bd} \bar{\bm{W}}^{\ad(2) \gd} \Nabla_{\a\bd} \Nabla_{\a\gd} \bar{H}_{\ad(2)} \non \\
&~~~ + \frac{7 \ri}{12} \bar{\Nabla}^{\bd} \bar{\bm{W}}_{\bd}{}^{\ad(2)} \bar{\Nabla}^{\gd} \Nabla_{\a} \Nabla_{\a\gd} \bar{H}_{\ad(2)} - \bar{\Nabla}^{\ad} \bar{\bm{W}}^{\ad \bd(2)} \Nabla_{\a\bd} \Nabla_{\a\bd} \bar{H}_{\ad(2)} \non \\
&~~~ + \frac{\ri}{6} \bar{\Nabla}^{\bd} \bar{\bm{W}}^{\ad(2) \gd} \bar{\Nabla}_{\bd} \Nabla_{\a} \Nabla_{\a\gd} \bar{H}_{\ad(2)} - \frac{2}{3} \bar{\bm{W}}^{\ad(2) \bd} \bar{\Nabla}^{\gd} \Nabla_{\a\gd} \Nabla_{\a\bd} \bar{H}_{\ad(2)} \label{66.5b}\\
&~~~
+ \frac{5 \ri}{24} \bar{\bm{W}}^{\ad(2) \bd} \bar{\Nabla}^2 \Nabla_{\a} \Nabla_{\a\bd} \bar{H}_{\ad(2)} + \bar{\bm{W}}^{\ad \bd(2)} \bar{\Nabla}^{\ad} \Nabla_{\a\bd} \Nabla_{\a\bd} \bar{H}_{\ad(2)} + \mathcal{O}(\bm{W}^2)
~. \non 
\end{align}
\end{subequations}
We remind the reader that we use the convention whereby all indices denoted by indistinguishable characters are to be symmetrised over. 

A few comments regarding the general structure of the primary superfields \eqref{66.5} are in order. Firstly, the superfield \eqref{66.5a} is an exact primary (i.e. it is primary to all orders in the super-Weyl tensor). There are terms quadratic in $\bm W_{\a(3)}$ which ensure this property. Their presence is new and did not appear in the first order structures for previous models.
Secondly, the superfield \eqref{66.5b} is primary only to first order in the super-Weyl tensor. This does not present a problem since it may be shown that \eqref{66.5b} and \eqref{66.5a} are not independent of one another and are related via the identity
\begin{align}
&\int \text{d}^{4|4}z \, E \, H^{\alpha(2)}\mathfrak{J}^{(1)}_{\alpha(2)}(\bar{H}) + \text{c.c.}= \int \text{d}^{4|4}z \, E \, \bar{H}^{\ad(2)}\bar{\mathfrak{J}}^{(1)}_{\ad(2)}(H) + \text{c.c.} \non\\
&=\int \text{d}^{4|4}z \, E \, H^{\alpha(2)}\bigg\{\mathfrak{J}^{(2)}_{\alpha(2)}(\bar{H}) + \ri \Nabla_{\a}{}^{\ad} \bm{B}_{\a}{}^{\ad} \bar{H}_{\ad(2)} + \frac{1}{6} \Nabla_{\a} \bar{\Nabla}^{\ad} \bm{B}_{\a}{}^{\ad} \bar{H}_{\ad(2)} + \frac{1}{3} \bar{\Nabla}^{\ad} \bm{B}_{\a}{}^{\ad} \Nabla_{\a} \bar{H}_{\ad(2)} \non \\
&~~~ + \frac{1}{6} \Nabla_{\a} \bm{B}_{\a}{}^{\ad} \bar{\Nabla}^{\ad} \bar{H}_{\ad(2)} + \frac{1}{2} \bm{B}_{\a}{}^{\ad} \bar{\Nabla}^{\ad} \Nabla_{\a} \bar{H}_{\ad(2)} + \frac{2 \ri}{3} \bm{B}_{\a}{}^{\ad} \Nabla_{\a}{}^{\ad} \bar{H}_{\ad(2)} \bigg\}+\text{c.c.}  \label{PGrel}
\end{align}
Hence, for the purpose of restoring gauge invariance it suffices to consider only $\mathfrak{J}^{(1)}_{\a(2)} (\bar{H})$. 

By making use of the identity \eqref{PGrel}, one can show that a generic (infinitesimal) variation of the action \eqref{6.34|4}, with $i=1$, takes the form
\begin{align}
\delta \mb{S}_{\text{NM},1}^{(2,0,0)}&= \frac{1}{2}\int \text{d}^{4|4}z \, E \, \delta H^{\a(2)}\bigg\{\mathfrak{J}^{(1)}_{\alpha(2)}(\bar{H})+\mathfrak{J}^{(2)}_{\alpha(2)}(\bar{H})+ \ri \Nabla_{\a}{}^{\ad} \bm{B}_{\a}{}^{\ad} \bar{H}_{\ad(2)} + \frac{1}{6} \Nabla_{\a} \bar{\Nabla}^{\ad}\bm{B}_{\a}{}^{\ad} \bar{H}_{\ad(2)} \non \\
&~~~ + \frac{1}{3} \bar{\Nabla}^{\ad} \bm{B}_{\a}{}^{\ad} \Nabla_{\a} \bar{H}_{\ad(2)}  + \frac{1}{6} \Nabla_{\a} \bm{B}_{\a}{}^{\ad} \bar{\Nabla}^{\ad} \bar{H}_{\ad(2)} + \frac{1}{2} \bm{B}_{\a}{}^{\ad} \bar{\Nabla}^{\ad} \Nabla_{\a} \bar{H}_{\ad(2)} \non \\
&~~~ + \frac{2 \ri}{3} \bm{B}_{\a}{}^{\ad} \Nabla_{\a}{}^{\ad} \bar{H}_{\ad(2)} \bigg\}+\text{c.c.} \label{PGvar}
\end{align}
Furthermore, to first order it is possible to show that the superfields  \eqref{66.5} satisfy the following identities
\begin{subequations}\label{divcur}
\begin{align}
\Nabla^{\beta}\mathfrak{J}^{(1)}_{\alpha \beta}(\bar{H})&= - \Nabla^{\b \ad} \bm{B}_{\a}{}^{\ad} \Nabla_\b \bar{H}_{\ad(2)} - \frac{1}{4} \bar{\Nabla}^{\ad} \bm{B}_{\a}{}^{\ad} \Nabla^2 \bar{H}_{\ad(2)} - \frac{1}{4} \Nabla^{\b} \bm{B}_{\a}{}^{\ad} \bNabla^{\ad} \Nabla_{\b} \bar{H}_{\ad(2)} \non \\
&~~~ - \frac{\ri}{2} \bm{B}^{\b \ad} \Nabla_{\a}{}^{\ad} \Nabla_\b \bar{H}_{\ad(2)} + \frac{7 \ri}{12} \bm{B}_{\a}{}^{\ad} \Nabla^{\b \ad} \Nabla_{\b}
\bar{H}_{\ad(2)} + \frac{1}{24} \bm{B}_{\a}{}^{\ad} \bar{\Nabla}^{\ad} \Nabla^{2} \bar{H}_{\ad(2)}~, \\
\Nabla^{\beta}\mathfrak{J}^{(2)}_{\alpha \beta}(\bar{H})&= \frac{1}{4}\bigg\{ \ri \Nabla_{\a}{}^{\ad} \Nabla^{\bd} 
\bar{\bm{W}}_{\bd}{}^{\ad(2)} \Nabla^{2} \bar{\Nabla}_{(\ad_1} 
\bar{H}_{\ad_2 \ad_3)} + \ri \Nabla_{\a}{}^{\bd} 
\bar{\Nabla}^{\ad} \bar{\bm{W}}_{\bd}{}^{\ad(2)} \Nabla^2 
\bar{\Nabla}_{(\ad_1} \bar{H}_{\ad_2 \ad_3)} \non \\
&~~~ + \frac{5 \ri}{4} \Nabla_{\a}{}^{\bd} \bar{\bm{W}}_{\bd}{}^{\ad(2)} \bar{\Nabla}^{2} \Nabla^2 \bar{H}_{\ad(2)} - 2 \Nabla_{\a}{}^{\gd} \bar{\bm{W}}^{\ad(2) \bd} \Nabla^{\b}{}_{\gd} \bar{\Nabla}_{\bd} \Nabla_{\b} \bar{H}_{\ad(2)} \non \\
&~~~ + 6 \Nabla_{\a}{}^{\bd} \bar{\bm{W}}^{\ad}{}_{\bd}{}^{\gd} \Nabla^{\b}{}_{\gd} \bar{\Nabla}^{\ad} \Nabla_{\b} \bar{H}_{\ad(2)} - \Nabla_{\a}{}^{\bd} \bar{\bm{W}}_{\bd}{}^{\ad(2)} \Nabla^{\b \gd} \bar{\Nabla}_{\gd} \Nabla_{\b} \bar{H}_{\ad(2)} \non \\
&~~~ + \frac{9 \ri}{4} \bar{\Nabla}^{\bd} \bar{\bm{W}}_{\bd}{}^{\ad(2)} \Nabla_{\a}{}^{\ad} \Nabla^{2} \bar{\Nabla}_{(\ad_1} \bar{H}_{\ad_2 \ad_3 )} - \frac{\ri}{2} \bar{\Nabla}^{\bd} \bar{\bm{W}}^{\ad(3)} \Nabla_{\a \bd} \Nabla^2 \bar{\Nabla}_{\ad} \bar{H}_{\ad(2)} \non \\
&~~~ - \bar{\bm{W}}^{\ad(2) \bd} \Nabla_{\a \bd} \Nabla^{\b \gd} \bar{\Nabla}_{\gd} \Nabla_{\b} \bar{H}_{\ad(2)} - 4 \bar{\bm{W}}^{\ad \bd(2)} \Nabla_{\a \bd} \Nabla^{\b}{}_{\bd} \bar{\Nabla}^{\ad} \Nabla_{\b} \bar{H}_{\ad(2)}  \non \\
&~~~ + \frac{1}{3} \Nabla_{(\a} \bm{B}_{\b)}{}^{\ad} \bar{\Nabla}^{\ad} \Nabla^{\b} \bar{H}_{\ad(2)} + \frac{4 \ri}{3}\bm{B}_{(\a}{}^{\ad} \Nabla^{\b} \Nabla_{\b)}{}^{\ad} \bar{H}_{\ad(2)}\bigg\}
~.
\end{align}
\end{subequations}

Both relations \eqref{PGvar} and \eqref{divcur} are fundamental in computing the gauge variation of the superconformal action $\mb{S}_{\text{NM},1}^{(2,0,0)}$ and to first order it can be shown that the $\L$ gauge variation is given by
\begin{align}
\delta_{\L}\mb{S}_{\text{NM,1}}^{(2,0,0)}&=-\frac{1}{4}\delta_{\L}\mb{S}_{\text{skeleton}}^{(2,0,0)} + \int \rd^{4|4}z \, E \, \L^{\a} \bigg \{ - \Nabla^{\b \ad} \bm{B}_{\a}{}^{\ad} \Nabla_\b \bar{H}_{\ad(2)} - \frac{1}{8} \bar{\Nabla}^{\ad}\bm{B}_{\a}{}^{\ad} \Nabla^2 \bar{H}_{\ad(2)} \non \\
&~~~ - \frac{1}{6} \Nabla^{\b} \bm{B}_{\a}{}^{\ad} \Nabla^{\ad} \Nabla_{\b} \bar{H}_{\ad(2)} - \frac{15 \ri}{16} \bm{B}^{\b \ad} \Nabla_{\a}{}^{\ad} \Nabla_\b \bar{H}_{\ad(2)} + \frac{47 \ri}{48} \bm{B}_{\a}{}^{\ad} \Nabla^{\b \ad} \Nabla_{\b} \bar{H}_{\ad(2)} \non \\
&~~~ -\frac{\ri}{12} \bm{B}_{\a}{}^{\ad} \Nabla^{\b \ad} \Nabla_{\b} \bar{H}_{\ad(2)} - \frac{1}{6} \bm{B}_{\a}{}^{\ad} \bar{\Nabla}^{\ad} \Nabla^2 \bar{H}_{\ad(2)} \bigg\} ~.
\end{align}
This implies that the superconformal action
\begin{align}
\mb{S}_{H\bar{H}}^{(2,0,0)}[H,\bar{H}] = \mb{S}^{(2,0,0)}_\text{skeleton}[H,\bar{H}] + 4 \mb{S}_\text{NM,1}^{(2,0,0)}[H,\bar{H}]  \label{PGM}
\end{align}
is invariant under $\L$-gauge transformations to leading order in the super-Weyl tensor (when restricted to a Bach-flat background)
\begin{align}
\delta_{\L,\l} \mb{S}_{H\bar{H}}^{(2,0,0)} \approx  \mathcal{O}(\bm W^2) ~.
\end{align}
It may also be checked that under a chiral gauge transformation, the action \eqref{PGM} is proportional to terms quadratic in $\bm W_{\a(3)}$ or linear in $\bm B_{\a\ad}$.

Due to the problems outlined in the beginning of this section, we have so far restricted our analysis of gauge invariance to be of first order in the super-Weyl tensor. However, in section \ref{secConformalHook} we constructed the non-supersymmetric model for the conformal pseudo-graviton which was gauge invariant to all orders (i.e. second order) in the background curvature. Actually, in section \ref{secConformalHook} we constructed a family (parametrised by $\Gamma$) of gauge invariant models for the conformal hook, described by the action $S^{(3,1)}_{\text{CHS},\Gamma}[h,\chi,\vf]$, see eq. \eqref{CHookFullGInvFam}.
It is instructive to compare the model with $\Gamma=1$ to the supersymmetric model considered in this section, and see what conclusions we can draw regarding the second order completion of the latter. Comments regarding the other values of $\Gamma$ will be given at the end of this section. 

In the conformally-flat limit, we know from \eqref{OneParFmaCFlim}  with $\Gamma =1 $ and \eqref{2.364|4} that the non-supersymmetric model and the (bosonic sector of the)
supersymmetric skeleton (in the Wess-Zumino gauge) are given by the actions
\begin{subequations}\label{8.14|4}
\begin{align}
S_{\text{CHS}}^{(3,1)}&=\frac{1}{2}\int\text{d}^4x\, e \, \bigg\{\mc{W}^{\a(4)}(h)\mc{W}_{\a(4)}(\bar{h})+\bar{\chi}^{\ad(2)}\mathfrak{X}_{\ad(2)}(\chi)\bigg\}+\text{c.c.}~,\label{8.14|4a}\\
\mb{S}_{{H\bar{H}}}^{(2,0,0)}\Big|_{\rm Bosonic}&=\frac{1}{4}\int \text{d}^4x \, e \, \bigg\{\mc{W}^{\a(4)}(h)\mc{W}_{\a(4)}(\bar{h})-\frac{16}{3}\bar{\rho}^{\ad(2)}\mathfrak{X}_{\ad(2)}(\rho)\bigg\} +\text{c.c.}~\label{8.14|4b}
\end{align}
\end{subequations}

From \eqref{8.14|4} one can see that the relative coefficients between the pseudo-graviton $h_{\a(3)\ad}$ and the non-gauge fields $\chi_{\a(2)}$ and $\rho_{\a(2)}$ do not agree. The only way for these models to be consistent with one another is if there is an additional lower-spin superfield present in the supersymmetric pseudo-graviton model. By construction, the model \eqref{PGM} is gauge invariant to first order in the super-Weyl tensor, therefore the purpose of this lower-spin superfield must be to ensure gauge invariance to second order. 

It turns out that the only possible candidate for such a field is the chiral non-gauge superfield $\Omega_{\a}$, which is characterised by the superconformal properties
\begin{align}
K_{B}\Omega_{\a}=0~,\qquad \mathbb{D}\Omega_{\a}=\frac{1}{2}\Omega_{\a}~,\qquad \mb{Y}\Omega_{\a}=-\frac{1}{3}\Omega_{\a}~,\qquad \bNabla_{\bd}\Omega_{\a}=0~.\label{hithere}
\end{align}
In appendix \ref{AppNGSM} we discuss this model and its higher-rank extensions. In particular, the superconformal kinetic action for $\O_{\a}$ is given by
\begin{align}
\mb{S}_{\O\bar{\O}}[\O,\bar{\O}]=-\frac{\ri}{4} \int \text{d}^4 x \rd^2\q \, \cE \, \O^{\a}\bNabla^2\Nabla_{\a\ad}\bar{\O}^{\ad} ~,
\end{align}
see appendix \ref{AppNGSM} for more details. 

 In order to provide corrections to the overall gauge variation, it is clear that $\Omega_{\a}$ must transform non-trivially and couple to the pseudo-graviton $H_{\a(2)}$. The only transformation which preserves all of the properties \eqref{hithere} is\footnote{Given the chirality of $\Omega_{\a}$ and the transformations \eqref{chiralGT}, we can further conclude that the purpose of $\Omega_{\a}$ is to ensure second-order invariance under chiral ($\l$) gauge transformations. } 
\begin{align}
\delta_{\lambda}\Omega_{\a}=\bm W_{\a}{}^{\b(2)}\lambda_{\b(2)}~.\label{chiralGT}
\end{align}
There is only a single possible quadratic non-minimal primary coupling between $H$ and $\Omega$, and it takes the form
\begin{subequations}
\begin{align}
\mb{S}_{H\bar{\O}}&[H,\Omega]=\ri \int \text{d}^{4|4}z\, E\,
H^{\a(2)}\mathfrak{J}_{\a(2)}(\bar{\Omega})+\text{c.c.}~,\\
 \mathfrak{J}_{\a(2)}(\bar{\Omega})=\Nabla_{\b\bd}& \bm  W_{\a(2)}{}^{\b}\bar{\Omega}^{\bd}-\bm W_{\a(2)}{}^{\b}\Nabla_{\b\bd}\bar{\Omega}^{\bd}+\frac{\ri}{2}\Nabla_{\b}\bm W_{\a(2)}{}^{\b}\bar{\Nabla}_{\bd}\bar{\Omega}^{\bd}~.
\end{align}
\end{subequations}
It follows that the total action for the pseudo-graviton multiplet is
\begin{align}
\mb{S}_{\text{SCHS}}^{(2,0,0)}=\mb{S}_{H\bar{H}}^{(2,0,0)}[H,\bar{H}]+\mu \mb{S}_{H\bar{\O}}[H,\Omega]+\mu \mb{S}_{\O\bar{\O}}[\Omega,\bar{\Omega}] +\mathcal{O}(\bm W^2)\label{8.4|4}
\end{align}
for some constant $\mu\in \mathbb{R}$ and where $\mathcal{O}(\bm W^2)$ represent terms that are quadratic in both $H$ and the super-Weyl tensor, and are primary. All possible structures of this type may be shown to take the form \eqref{6.34|4} with
\begin{subequations} \label{PGquadratic}
\begin{align}
\mathfrak{J}^{(3)}_{\alpha(2)}(\bar{H})&= \Nabla_\b \bm{W}^{\b}{}_{\a(2)} \bar{\Nabla}_{\bd} \bar{\bm{W}}^{\bd \ad(2)} \bar{H}_{\ad(2)} - 2 \ri \Nabla_{\bb} \bm{W}^{\b}{}_{\a(2)} \bar{\bm{W}}^{\bd \ad(2)} \bar{H}_{\ad(2)} \non \\
&~~~ + 2 \ri \bm{W}_{\a(2)}{}^{\b} \Nabla_{\bb} \bar{\bm{W}}^{\bd \ad(2)} \bar{H}_{\ad(2)}~,  \\
\mathfrak{J}^{(4)}_{\alpha(2)}(\bar{H})&= \Nabla_\b \bm{W}^{\b}{}_{\a(2)} \bar{\bm{W}}^{\ad(2) \bd} \bar{\Nabla}_{\bd} \bar{H}_{\ad(2)} + \bm{W}_{\a(2)}{}^{\b} \bar{\bm{W}}^{\ad(2) \bd} \Nabla_\b \bar{\Nabla}_\bd \bar{H}_{\ad(2)}~, \\
\mathfrak{J}^{(5)}_{\alpha(2)}(\bar{H})&= \bm{W}_{\a(2)}{}^{\b} \bar{\Nabla}_{\bd} \bar{\bm{W}}^{\bd \ad(2)} \Nabla_{\b} \bar{H}_{\ad(2)} - \bm{W}_{\a(2)}{}^{\b} \bar{\bm{W}}^{\bd \ad(2)} \bar{\Nabla}_{\bd} \Nabla_{\b} \bar{H}_{\ad(2)}~, 
\end{align}
\end{subequations}
or are expressible as linear combinations thereof. We note that the lower-spin sector in \eqref{8.4|4} is $\L$-gauge invariant in a Bach-flat background whilst their relative coefficient of unity between $\mb{S}_{H\bar{\O}}$ and $\mb{S}_{\O\bar{\O}}$ is fixed by $\lambda$-invariance. 

We can actually go a step further and deduce the value of $\mu$, but to do so we must take a closer look at the component structure of \eqref{8.4|4}.\footnote{According to our general procedure, $\mu$ would typically be determined by second order $\lambda$-invariance. } For the component fields we use the definitions \eqref{WZschs}, \eqref{WZFixingCondtions} and \eqref{7.18}. Using \eqref{8.14|4b} and the results of appendix \ref{AppNGSM}, one can show that in the conformally-flat limit the bosonic sector of \eqref{8.4|4} is
\begin{align}
\mb{S}_{\text{SCHS}}^{(2,0,0)}\Big|_{\rm Bosonic}
&= \frac{1}{4}\int \text{d}^4x \, e \, \bigg\{\mc{W}^{\a(4)}(h)\mc{W}_{\a(4)}(\bar{h})-\frac{16}{3}\bar{\rho}^{\ad(2)}\mathfrak{X}_{\ad(2)}(\rho)\notag\\
&\phantom{SPACE SPACE~~}-\mu\bigg(\bar{U}^{\ad(2)}\mathfrak{X}_{\ad(2)}(U) +\frac{1}{2}\bar{V}\Box V\bigg)\bigg\} +\text{c.c.}  \label{8.5}
\end{align} 
In a generic background we find that the fields in \eqref{8.14|4a} and \eqref{8.5} are defined modulo the gauge transformations
(see \eqref{chiGT} and \eqref{rhoGT} )
\begin{align}
&\delta_{\ell}h_{\a(3)\ad}=\nabla_{\a\ad}\ell_{\a(2)}~, \quad\delta_{\ell}V=0~,\\[4pt]
\delta_{\ell}\chi_{\a(2)}=W_{\a(2)}{}^{\b(2)}\ell_{\b(2)}~&,\quad \delta_{\ell}\rho_{\a(2)}=\frac{\ri}{4}W_{\a(2)}{}^{\b(2)}\ell_{\b(2)}~, \quad \delta_{\ell}U_{\a(2)}=\frac{\ri}{2}W_{\a(2)}{}^{\b(2)}\ell_{\b(2)}~. \non
\end{align}

In \eqref{8.5} there are clearly two distinct non-gauge fields, $\rho_{\a(2)}$ and $U_{\a(2)}$, which are of the same tensor type as the non-gauge field $\chi_{\a(2)}$ in the non-supersymmetric model and which have non-trivial gauge transformations (and so play a role in ensuring gauge invariance). However, if we make the following field redefinitions
\begin{align}
X_{\a(2)}=\frac{4\ri}{3}\big(\rho_{\a(2)}-2U_{\a(2)}\big)~,\qquad Y_{\a(2)}=\frac{4\ri}{3}\big(-2\rho_{\a(2)}+U_{\a(2)}\big)~,
\end{align}
then the fields $X_{\a(2)}$ and $Y_{\a(2)}$ transform in the manner
\begin{align}
\delta_{\ell}X_{\a(2)}=W_{\a(2)}{}^{\b(2)}\ell_{\b(2)}~, \qquad \delta_{\ell}Y_{\a(2)}=0~.
\end{align}
If in addition we choose $\mu=-16/3$ then the action \eqref{8.5} becomes 
\begin{align}
\mb{S}_{\text{SCHS}}^{(2,0,0)}\Big|_{\rm Bosonic}
= &\frac{1}{4}\int \text{d}^4x \, e \, \bigg\{\mc{W}^{\a(4)}(h)\mc{W}_{\a(4)}(\bar{h})+\bar{X}^{\ad(2)}\mathfrak{X}_{\ad(2)}(X)\notag\\
&\phantom{SPACE SPACE~~}-\bar{Y}^{\ad(2)}\mathfrak{X}_{\ad(2)}(Y) +\frac{8}{3}\bar{V}\Box V\bigg\} +\text{c.c.}  ~~~~~~\label{8.7}
\end{align}

The fields $Y_{\a(2)}$ and $V$ do not transform under $\ell_{\a(2)}$. Consequently, they do not play a role in establishing gauge invariance and are present only to ensure supersymmetry. Since the non-gauge fields $\chi_{\a(2)}$ and $X_{\a(2)}$ share the same gauge transformations and relative coefficient between their own kinetic sector and that of the pseudo-graviton, we may identify them: $\chi_{\a(2)}\equiv X_{\a(2)}$. Thus we conclude that the lower-spin sectors in the supersymmetric model \eqref{8.4|4} have coefficients $\mu=-16/3$. This means that the coefficients of all sectors in the pseudo-graviton multiplet, except for the second order ones \eqref{PGquadratic}, have been determined.

 \subsubsection{Other superspace realisations of the conformal pseudo-graviton} 
 
In section \ref{secConformalHook} we derived a one-parameter family of gauge invariant models for the conformal hook field, which was parameterised by $\Gamma$. 
 By analysing the resulting component structure, it was shown above that the model \eqref{CHookFullGInvFam} with $\Gamma=1$ (i.e. \eqref{CHookGInvX}) could be embedded within a supersymmetric gauge invariant action for a superfield $H_{\a(2)}$ coupled to a chiral non-gauge superfield $\Omega_{\a}$. In this model, the latter two transformed according to
\begin{subequations} 
\begin{align}
\delta_{\L,\lambda}H_{\a(2)}=\Nabla_{(\a_1}&\L_{\a_2)}+\lambda_{\a(2)}~,\qquad \bar{\Nabla}_{\ad}\lambda_{\a(2)}=0~, \label{d4|4}\\
&\delta_{\lambda} \Omega_{\a}=\bm W_{\a}{}^{\b(2)}\lambda_{\b(2)}~,
\end{align} 
\end{subequations}
 with $\L_{\a}$ unconstrained and $\lambda_{\a(2)}$ covariantly chiral. 
 
For $\Gamma\neq 3$, it may be possible to embed \eqref{CHookFullGInvFam} within a one-parameter family of supersymmetric actions by introducing, in addition to $H_{\a(2)}$ and $\Omega_{\a}$, the longitudinal linear non-gauge superfield $\Omega_{\a(3)\ad(2)}$ (see appendix \ref{AppNGSM} for a discussion of this type of non gauge field). The latter should be entangled with $H_{\a(2)}$ through the transformations
 \begin{align}
 \delta_{\L}\Omega_{\a(3)\ad(2)}=\bm W_{\a(3)}\bar{\Nabla}_{(\ad_1}\bar{\L}_{\ad_2)}~.\label{a4|4} 
 \end{align}
 Then, the component field of $\Omega_{\a(3)\ad(2)}$ defined by \eqref{4.13c} corresponds to $\vf_{\a(4)\ad(2)}$ and has the correct gauge transformations \eqref{NGGThihello} (in an appropriate Wess-Zumino gauge).
 
When $\Gamma=3$, the non-gauge field $\chi_{\a(2)}$ is no longer present in the action \eqref{CHookFullGInvFam}. However, as explained previously, $\chi_{\a(2)}$ appears as a component field of the gauge superfield $H_{\a(2)}$. Hence, at the component level, any gauge invariant model for $H_{\a(2)}$ will necessarily contain a coupling between $h_{\a(3)\ad}$ and $\chi_{\a(2)}$. Therefore  $H_{\a(2)}$ is not a suitable supermultiplet in providing a supersymmetric embedding for the action \eqref{CHookFullGInvF}. 

A possible candidate for this role\footnote{One can rule out the superfield $H_{\a(3)}$ (which also contains $h_{\a(3)\ad}$) as there is no possible coupling to a non-gauge superfield which contains $\vf_{\a(4)\ad(2)}$ at the component level and which preserves all symmetries.} is the spin-5/2 supermultiplet $H_{\a(2)\ad}$ coupled to a longitudinal non-gauge field $\Omega_{\a(4)\ad(2)}$ according to\footnote{We would like to point out that the gauge transformations \eqref{a4|4} and \eqref{b4|4} preserve the gauge-for-gauge symmetries $\zeta_{\a}\rightarrow \zeta_{\a}+\Nabla_{\a}\sigma$ and $\Lambda_{\a\ad}\rightarrow \Lambda_{\a\ad}+\Nabla_{\a}\bar{\eta}_{\ad}$ of \eqref{d4|4} and \eqref{c4|4}, respectively.}
\begin{subequations}
\begin{align}
\delta_{\L,\O}H_{\a(2)\ad}&=\Nabla_{\a}\L_{\a\ad}+\bar{\Nabla}_{\ad}\z_{\a(2)}\label{c4|4} ~,\\
\delta_{\Lambda}\Omega_{\a(4)\ad(2)}&=\bm W_{\a(3)}\bar{\Nabla}_{\ad}\bar{\Lambda}_{\a\ad}~,\label{b4|4}
\end{align}
\end{subequations}
with unconstrained and complex gauge parameters $\L_{\a\ad}$ and $\z_{\a(2)}$. Here, it is the component field \eqref{4.13e} of $\Omega_{\a(4)\ad(2)}$ which may be identified with the non-gauge field $\vf_{\a(4)\ad(2)}$.

\subsection{The conformal spin-3 supermultiplet} \label{secSpin3Recipe}

Our model for the superconformal pseudo-graviton multiplet provides an example of a  universal pattern, that general CHS theories possess supersymmetric embeddings. 
This gives the rationale to study SCHS theories as they may uncover interesting features of non-supersymmetric CHS models. 
As an example, 
let us assume the existence of a gauge-invariant model for the conformal spin-3 supermultiplet 
in a Bach-flat background
and see what can be deduced regarding its non-supersymmetric counterpart. The former is described by the real prepotential $H_{\a(2)\ad(2)}$ which is a   primary superfield of weight $-2$ with the gauge freedom
\begin{align}
\delta_{\L}H_{\a(2)\ad(2)}=\bNabla_{(\ad_1}\L_{\a(2)\ad_2)}
- \Nabla_{(\a_1} \bar \L_{\a_2)\ad(2)}~.
\end{align}  
This transformation law defines a reducible gauge theory (in the terminology of \cite{BV})
since the unconstrained gauge parameter
$\L_{\a(2)\ad}$ is defined modulo arbitrary local shifts
\bea
\L_{\a(2)\ad} \to \L'_{\a(2)\ad} = \L_{\a(2)\ad} + \bNabla_\ad \s_{\a(2)}
\label{8.2}
\eea
such that both parameters $\L_{\a(2)\ad} $ and $ \L'_{\a(2)\ad} $ generate the same transformation of the prepotential,
$\delta_{\L'}H_{\a(2)\ad(2)} = \delta_{\L}H_{\a(2)\ad(2)}$.
 
A Wess-Zumino gauge may be chosen 
such that the only non-vanishing bosonic component fields of $H_{\a(2)\ad(2)}$ are
\begin{subequations}
\begin{align}
h_{\a(3)\ad(3)}&:=\frac{1}{2}\big[\Nabla_{(\a_1},\bNabla_{(\ad_1}\big]H_{\a_2\a_3)\ad_2\ad_3)}|~,\label{8.2a}\\
h_{\a(2)\ad(2)}&:=\frac{1}{32}\big\{\Nabla^2,\bNabla^2\big\}H_{\a(2)\ad(2)}|~.\label{8.2b}
\end{align}
\end{subequations}
The residual gauge freedom is given by 
\begin{subequations}
\begin{align}
\delta_{\ell}h_{\a(3)\ad(3)}&=\nabla_{(\a_1(\ad_1}\ell_{\a_2\a_3)\ad_2\ad_3)}~, \label{8.3a}\\
\delta_{\ell}h_{\a(2)\ad(2)}&=\nabla_{(\a_1(\ad_1}\ell_{\a_2)\ad_2)} +\bigg[ \frac{\ri}{6} W_{\a(2)}{}^{\b(2)} \ell_{\b(2) \ad(2)} + \text{c.c.} \bigg] ~. \label{8.3b}
\end{align}
\end{subequations}
Eq. \eqref{8.3a} is the standard gauge transformation of the conformal spin-3 field. 
Similarly, the first term on the right of \eqref{8.3b} is the usual gauge transformation of the conformal spin-2 field. However, the second term in \eqref{8.3b} 
is a new feature. Its presence means that the spin-2 field also varies under the spin-3 gauge transformation.

In principle, there is a possibility  that the kinetic action for $H_{\a(2)\ad(2)}$ can be made gauge invariant without introducing any extra supermultiplet. However,
the analysis in this thesis indicates that, 
most likely, $H_{\a(2)\ad(2)}$ is to be accompanied by a lower-spin supermultiplet in order to ensure gauge invariance of the action on Bach-flat backgrounds. In addition, the results of \cite{GrigorievT, BeccariaT}
also indicate that for a consistent description of 
the conformal spin-3 field in such backgrounds, it should be accompanied by a spin-1 gauge field which gets shifted under the spin-3 gauge transformation. Thus 
our theory should involve a lower-spin supermultiplet containing a gauge vector field. 
There are only two options for such a gauge prepotential: (i) a real scalar  $V$; 
and (ii) a real vector $H_{\a\ad}$. Now we will consider both of them in turn.

The gauge prepotential $V$ is a primary superfield with weight $0$ and describes a vector multiplet. In addition to the standard gauge transformation 
\begin{subequations}
\bea
\d_\l V &=& \l + \bar \l~, \qquad  \bar{\Nabla}_{\ad} \l = 0~,
\eea
there is a unique  entanglement with the gauge parameter
$ \L_{\a(2) \ad} $ of $H_{\a(2) \ad(2)}$, given by
\bea
\d_{\L} V &=&   4 \ri \Nabla_{\a}{}^{\ad} \bm W^{\a(3)} \L_{\a(2) \ad} - \Nabla_{\a} \bm W^{\a(3)} \bar{\Nabla}^{\ad} \L_{\a(2) \ad} - 2 \ri \bm W^{\a(3)} \Nabla_{\a}{}^{\ad} \L_{\a(2) \ad} \non \\
&& - \frac{1}{2}\bm  W^{\a(3)} \Nabla_{\a} \bar{\Nabla}^{\ad} \L_{\a(2) \ad} + \text{c.c.} ~ \label{pencilcase}
\eea
\end{subequations}
The local shift \eqref{8.2} leaves the variation $\d_{\L} V +\d_{\l} V$ invariant provided the parameter 
$\l$ is also shifted in the following way
\bea
\l \rightarrow \l' = \l 
- \bar{\Nabla}^2 \Big( \Nabla_{\a} \bm W^{\a(3)} \s_{\a(2)} + \frac{1}{2}\bm W^{\a(3)} \Nabla_{\a} \s_{\a(2)} \Big) ~.
\eea

We choose a Wess-Zumino gauge such that the only bosonic fields of $V$ are
\bea
h_{\a \ad} = \frac{1}{2}\big[ \Nabla_{\a} , \bar{\Nabla}_{\ad} \big] V |~, \qquad 
D:=\frac{1}{32}\big\{\Nabla^2,\bNabla^2\big\} V|~.
\eea
The auxiliary field $D$ is irrelevant to our discussion.
For the complete gauge transformation of $h_{\a\ad}$ we obtain
\bea
\d_{\ell} h_{\a \ad} = \nabla_{\aa} \ell + \frac{2}{3} \bigg[ W_{\a}{}^{\b(3)} \nabla_{\b}{}^{\bd} \ell_{\b(2)\ad \bd} - 3 \nabla_{\b}{}^{\bd} W_{\a}{}^{\b(3)} \ell_{\b(2) \ad \bd} +\text{c.c.} \bigg] ~. \label{kk}
\eea
The first term on the right of \eqref{kk} is the standard gauge transformation of the spin-1 field.  The second term in \eqref{kk} tells us 
that the spin-1 field also varies under the spin-3 gauge transformation.
The complete variation \eqref{kk} is equivalent to  the gauge transformation law postulated 
by Grigoriev and Tseytlin \cite{GrigorievT}.\footnote{Indeed, the overall coefficient of \eqref{pencilcase} was chosen so that this is the case.}

Next, let us consider the case where a coupling between $H_{\a(2) \ad(2)}$ and a superconformal spin-2 multiplet $H_{\aa}$ is switched on. In addition to the standard law \eqref{SCHSprepGTIICSS}, the gauge transformation of $H_{\a\ad}$ may be entangled with the spin-3 gauge parameter as follows
\begin{align}
\delta_{\L}H_{\a\ad}=\bNabla_{\ad}\L_{\a} - \Nabla_{\a}\bar\L_{\ad}
+\bm W_{\a}{}^{\b(2)}\L_{\b(2)\ad} - \bar{\bm{W}}_{\ad}{}^{\bd(2)}\bar \L_{\a\bd(2)}~.
\end{align}
The local shift \eqref{8.2} does not have any effect on $ \delta_{\L}H_{\a\ad}$
provided the parameter $\L_\a$ also gets shifted as follows
\bea
\L_\a \to \L'_\a = \L_\a + \bm W_{\a}{}^{\b(2)}\s_{\b(2)}~.
\eea

A Wess-Zumino gauge may also be constructed for the field $H_{\a\ad}$ such that its only non-vanishing bosonic component fields are
\begin{subequations}
\begin{align}
\tilde{h}_{\a(2)\ad(2)}&:=\frac{1}{2}\big[\Nabla_{(\a_1},\bNabla_{(\ad_1}\big]H_{\a_2)\ad_2)}|~, \label{8.4a}\\
\tilde{h}_{\a\ad}&:=\frac{1}{32}\big\{\Nabla^2,\bNabla^2\big\}H_{\a\ad}|~, \label{8.4b}
\end{align}
\end{subequations}
with gauge transformation laws
\begin{subequations}
\begin{align}
\delta_{\ell,\tilde{\ell}}\tilde{h}_{\a(2)\ad(2)}&=\nabla_{(\a_1(\ad_1}\tilde{\ell}_{\a_2)\ad_2)}+ \bigg[  \frac{\ri}{2} W_{\a(2)}{}^{\b(2)}\ell_{\b(2)\ad(2)}+\text{c.c.} \bigg] ~, \label{8.5a}\\
\delta_{\ell,\tilde{\ell}}\tilde{h}_{\a\ad}&=\nabla_{\a\ad}\tilde{\ell} - \frac{1}{24} \bigg[ W_{\a}{}^{\b(3)}\nabla_{\b}{}^{\bd}\ell_{\b(2)\bd\ad}-3\nabla_{\b}{}^{\bd}W_{\a}{}^{\b(3)}\ell_{\b(2)\bd\ad} +{\rm c.c.} \bigg] \label{8.5b}
\end{align}
\end{subequations}
These transformations are analogous to \eqref{8.3b} and \eqref{kk}.

It is important to note that $h_{\a(2)\ad(2)}$ and $\tilde{h}_{\a(2) \ad(2)}$, defined by \eqref{8.2b} and \eqref{8.4a}, describe two independent conformal spin-2 fields,\footnote{This is reminiscent of the situation in section \ref{secPGSBach} where there were two non-gauge fields, $\rho_{\a(2)}$ and $U_{\a(2)}$, of the same tensor type.} both of which possess their own gauge transformations and are shifted under spin-3 gauge transformations. However, upon performing the field redefinitions
\begin{subequations}
\bea
{\bm h}_{\a(2) \ad(2)} = h_{\a(2) \ad(2)} - \frac{1}{3} \tilde{h}_{\a(2) \ad(2)}~, &\quad& \tilde{\bm h}_{\a(2) \ad(2)} = h_{\a(2) \ad(2)} + \frac{1}{3} \tilde{h}_{\a(2) \ad(2)}~, \label{8.9a}\\
\l_{\a \ad} = \ell_{\a \ad} - \frac{1}{3} \tilde{\ell}_{\a \ad}~, &\quad& \tilde{\l}_{\a \ad} = \ell_{\a \ad} + \frac{1}{3} \tilde{\ell}_{\a \ad}~, \label{8.9b}
\eea
the fields ${\bm h}_{\a(2) \ad(2)}$ and $\tilde{\bm h}_{\a(2) \ad(2)}$ transform according to
\bea
\d_{\l} {\bm h}_{\a(2) \ad(2)} &=& \nabla_{(\a_1 (\ad_1} \l_{\a_2) \ad_2)} ~, \\
\d_{\tilde{\l},\ell} \tilde{\bm h}_{\a(2) \ad(2)} &=& \nabla_{(\a_1 (\ad_1} \tilde{\l}_{\a_2) \ad_2)} + \bigg[ \frac{\ri}{3} W_{\a(2)}{}^{\b(2)} \ell_{\b(2) \ad(2)} + \text{c.c.} \bigg]. \label{8.9d}
\eea
\end{subequations}
In particular, we see that ${\bm h}_{\a(2) \ad(2)}$ transforms independently of the spin-3 gauge parameter $\ell_{\a(2) \ad(2)}$ and so it must decouple from the other fields in the action. It is present only to ensure supersymmetry.
As it is not possible to eliminate the shift transformation present in \eqref{8.9d}, $\tilde{\bm h}_{\a(2) \ad(2)}$ and the spin-3 field $h_{\a(3) \ad(3)}$ cannot be decoupled.

The need for a vector field, which  accompanies the spin-3 field and transforms according to \eqref{kk},  
was proposed in \cite{GrigorievT} and detailed calculations were carried out in \cite{BeccariaT}. However, the coupling between the spin-3 and spin-2 fields, described by \eqref{8.9d}, was not considered. Existence of the gauge invariant supersymmetric model dictates that any gauge invariant model of the non-supersymmetric conformal spin-3 field must necessarily possess this coupling. Similar considerations regarding the fermionic component fields of $H_{\a(2)\ad(2)}$, $H_{\aa}$ and $V$ also allows us to conclude that a gauge invariant model of the conformal spin-$5/2$ field must necessarily include a coupling between the spin-$5/2$ field, the conformal gravitino 
and/or the conformal Weyl spinor.

The analyses in \cite{GrigorievT,BeccariaT} were based on the use of the interacting bosonic CHS theory developed in \cite{Segal} (see also \cite{Tseytlin,BJM1,BJM2,Bonezzi} ).
This theory involves a single instance of each conformal spin-$s$ field 
$h_{\a(s) \ad(s)}$ (with $s=1,2\dots$).
There also exists a superconformal scenario briefly discussed  in 
\cite{KMT}, which involves a single instance of each gauge supermultiplet
$H_{\a(s)\ad(s)}$ (with $s=0,1\dots$).  At the component level,  such a superconformal
theory contains two copies of each conformal spin-$s$ gauge field 
$h_{\a(s) \ad(s)}$. This is exactly the situation in our analysis above.

\section{Summary of results} \label{secSCHS4dis}
This chapter was dedicated to the construction of SCHS models in curved $4d$ $\mc{N}=1$ superspace and applications thereof. We began in section \ref{secGenM4|4} by describing the general features of the mixed symmetry SCHS field $H_{\a(m)\ad(n)}$, and its higher-spin super-Bach tensor $\mf{B}_{\a(n)\ad(m)}(H)$, on a generic supergravity background.
The definition of $\mf{B}_{\a(n)\ad(m)}(H)$ is such that it has the characteristic features of a covariantly conserved conformal supercurrent if superspace is conformally-flat, leading to the gauge and super-Weyl invariant SCHS action \eqref{SCHSactSBachSS4}. 

In section \ref{secMink4|4} we reviewed the corresponding SCHS models in $\mc{N}=1$ Minkowski superspace $\mb{M}^{4|4}$. In section \ref{secAdS4|4} we discussed on-shell partially-massless $\mc{N}=1$ superfields in AdS$^{4|4}$ and elaborated on their gauge freedom for the first time \cite{AdSuperprojectors}, see eq. \eqref{SPMgaugesymmetry}. 
We then constructed gauge invariant actions \eqref{AdS4|4SBaction} for the SCHS prepotential $H_{\a(m)\ad(n)}$ on AdS$^{4|4}$ \cite{AdSuperprojectors}.
We found that the corresponding SCHS kinetic operator factorises into products of minimal second order operators associated with partially-massless supermultiplets of all (super)depths, see e.g.   \eqref{ThisguyFactors}. To facilitate this derivation, we made use of the TLAL (i.e. simultaneously transverse linear and transverse anti-linear) superprojectors which were constructed in \cite{AdSuperprojectors} (the latter are not strictly part of this thesis). 

In section \ref{secCF4|4} we provided a review of $\mc{N}=1$ conformal superspace. This was used to derive closed-form models for the SCHS field $H_{\a(m)\ad(n)}$ which are gauge invariant on all conformally-flat backgrounds \cite{Confgeo, SCHS},  see eq. \eqref{SCHSactCSSCF4|4}. Their corresponding content at the component level  was discussed for some special cases on bosonic backgrounds, see e.g. \eqref{DecompActionSCHS}. We also found the superconformal multiplets $H^{(t)}_{\a(m)\ad(n)}$ which contain generalised (i.e. higher-depth) CHS fields at the component level \cite{SCHSgen}. They also have gauge transformations of depth $t$ (in vector derivatives), see eq. \eqref{SCHSgenGT}. Their super-Weyl and gauge invariant actions \eqref{SCHSgenAction} on conformally-flat backgrounds were derived \cite{SCHSgen}. 

In section \ref{secBach4|4} we extended gauge invariance of the action for the maximal-depth conformal graviton supermultiplet ($m=n=t=1$) to Bach-flat backgrounds \cite{SCHSgen}, see \eqref{3.774|4}. It was shown that this model gives rise to a collection \eqref{MaxDepthSpin2ComponentAction} of gauge invariant (diagonal) CHS actions at the component level, as one would expect. We then constructed a model \eqref{8.4|4} for the conformal pseudo-graviton supermultiplet $H_{\a(2)}$ ($m-2=n=t=0$) gauge invariant to second order in the Weyl-tensor on a Bach-flat background \cite{SCHS}. We used our knowledge of the (non-supersymmetric) pseudo-graviton model \eqref{CHookFullGInvFam} to deduce that $H_{\a(2)}$ must be non-minimally coupled to a certain subsidiary superfield $\O_{\a}$ to achieve gauge invariance \cite{SCHS}. The relative coefficient between the two was determined. 

The subsidiary superfield $\O_{\a}$ belongs to a family of so-called superconformal chiral non-gauge fields $\O_{\a(m)}$, whose super-Weyl invariant actions were given \cite{SCHS} in appendix \ref{AppNGSM}. We also provided models for their higher-rank extensions, the so-called superconformal longitudinal non-gauge fields $\O_{\a(m)\ad(n)}$ \cite{SCHSgen}.
Finally, in section \ref{secSpin3Recipe}, we deduced the necessity of a non-minimal coupling between the conformal spin-3 and conformal spin-2 fields in any gauge invariant model for the former on a Bach-flat background \cite{SCHS}. Our only assumption was the existence of a gauge invariant-model for the SCHS multiplet $H_{\a(2)\ad(2)}$ on a Bach-flat background.

\begin{subappendices}

 
\section{Technical issues in conformal superspace} \label{AppCSS4}

In this appendix we touch upon various technical aspects of $4d$ $\mc{N}=1$ conformal superspace, which was reviewed in section \ref{secN=1CSS}.

\subsection{Degauging to the $\sU(1)$ superspace geometry} \label{AppDegaugeCSS4N=1}

It is well known that $\sU(1)$ superspace, described in section \ref{secCSGSS4}, is a gauge-fixed version of the $\mc{N}=1$ conformal superspace geometry. Here, we briefly outline the procedure to `degauge' from the latter to the former.

According to \eqref{SGGauge}, under an infinitesimal special superconformal gauge transformation with $\L = \tau^{B} K_{B}$, the covariant derivative transforms as follows
\bea
\d^{(\text{sct})}_{\tau} \Nabla_{A}\big|_{\mathbb{D}} = 2 (-1)^{A} \tau_{A} \quad \implies \quad \d^{(\text{sct})}_{\tau} B_{A} = - 2 (-1)^{A}  \tau_{A} ~.
\eea
Thus, it is possible to construct a gauge where the dilatation connection vanishes 
\begin{align}
B_{A} = 0~.
\end{align}
 Associated with this is a loss of unconstrained special superconformal gauge freedom.\footnote{There is class of residual gauge transformations preserving the gauge $B_{A}=0$. These generate the super-Weyl transformations of $\sU(1)$ superspace; see below.} As a result, the corresponding connection becomes auxiliary and must be manually extracted from $\Nabla_{A}$,
\bea
\Nabla_{A} &=& \mathscr{D}_{A} - \mathfrak{F}_{A}{}^{B} K_{B} ~, \label{ND}
\eea
which should be treated as the definition of $\mathscr{D}_{A}$. After identifying $\mf{F}_{AB}$ with the various $\sU(1)$ torsion superfields (see below), one finds 
\begin{subequations}
\label{A.16}
\bea
\mathscr{D}_{\a \ad} = \bm\cD_{\a \ad} + \frac{\ri}{2} G^{\b}{}_{\ad} M_{\a \b} &-& \frac{\ri}{2} G_{\a}{
}^{\bd} \bar{M}_{\ad \bd} - \frac{3 \ri}{4} G_{\a \ad} \mb{Y} ~,  \\
\mathscr{D}_{\a} = \bm\cD_{\a} ~, \, && \, \bar{\mathscr{D}}^{\ad} = \bm\cDB^{\ad} ~, 
\eea
\end{subequations}
where $\bm\cD_{A}$ is the $\sU(1)$ superspace covariant derivative \eqref{U(1)CD}\footnote{Technically we should use a different notation for the $\sU(1)_R$ connection  residing within $\Nabla_A$ and the other one in $\bm\cD_A$.}
\bea
\bm \cD_{A}&=&  \bm e_{A}{}^{M}\pa_M-\frac{1}{2}\bm\o_A{}^{bc}M_{bc}-\ri\Q_A\mb{Y} ~.
\label{U(1)CDapp}
\eea
It should also be noted that the super-Weyl tensor $\bm W_{\a \b \g}$ satisfies the new covariant chirality constraint
\bea
\bm\cDB_{\ad} \bm W_{\a \b \g} = \big( \bar{\Nabla}_{\ad} + \bar{\mathfrak{F}}_{\ad}{}^{B} K_{B} \big) \bm W_{\a \b \g} = 0 ~.
\eea

To relate the superfields $\mathfrak{F}_{A}{}^{B}$ to the torsion superfields of $\sU(1)$ superspace, it is necessary to make use of the result
\bea
[ \Nabla_{A} , \Nabla_{B} \} &=& [ \mathscr{D}_{A} , \mathscr{D}_{B} \} - \big(\mathscr{D}_{A} \mathfrak{F}_B{}^C - (-1)^{AB} \mathscr{D}_{B} \mathfrak{F}_A{}^C \big) K_C - \mathfrak{F}_A{}^C [ K_{C} , \Nabla_B \} \non \\
&& + (-1)^{AB} \mathfrak{F}_B{}^C [ K_{C} , \Nabla_A \} + (-1)^{BC} \mathfrak{F}_A{}^C \mathfrak{F}_B{}^D [K_D , K_C \} ~.
\eea
This allows one to solve for $\mathfrak{F}_{AB}$ by making use of the defining constraints of $\sU(1)$ superspace \cite{Howe1,Howe2} in addition to \eqref{CSSAlgebra}. We will not provide a detailed analysis for this step and instead refer the reader to the proof in \cite{ButterN=1}. The result is:
	\bea \label{connections}
	\mathfrak{F}_{\a \b} & = & - \frac{1}{2} \ve_{\a \b} \bar{R} ~, \quad \bar{\mathfrak{F}}_{\ad \bd} = \frac{1}{2} \ve_{\ad \bd} R ~, \quad
	\mathfrak{F}_{\a \bd} = - \frac{1}{4} G_{\a \bd} ~, \quad \bar{\mathfrak{F}}_{\ad \b} = \frac{1}{4} G_{\b \ad} ~,\non \\
	\mathfrak{F}_{\a , \b \bd} & = & - \frac{\ri}{4} \bm\cD_{\a} G_{\b \bd} - \frac{\ri}{6} \ve_{\a \b} \bar{X}_{\bd}  ~, \quad \bar{\mathfrak{F}}_{\ad , \b \bd} = \frac{\ri}{4} \bm\cDB_{\ad} G_{\b \bd} + \frac{\ri}{6} \ve_{\ad \bd} X_{\b} ~,\\
	\mathfrak{F}_{\b \bd , \a} & = & \phantom{-}\frac{\ri}{4} \bm\cD_{\a} G_{\b \bd} + \frac{\ri}{6} \ve_{\a \b} \bar{X}_{\bd}  ~, \quad \mathfrak{F}_{\b \bd , \ad} = - \frac{\ri}{4} \bm\cDB_{\ad} G_{\b \bd} - \frac{\ri}{6} \ve_{\ad \bd} X_{\b} ~,\non\\
	\mathfrak{F}_{\a \ad , \b \bd} & = & - \frac{1}{8} \big[\bm \cD_{\a} , \bm\cDB_{\ad} \big] G_{\b \bd} - \frac{1}{12} \ve_{\ad \bd} \bm\cD_{\a} X_{\b} + \frac{1}{12} \ve_{\a \b} \bm\cDB_{\ad} \bar{X}_{\bd} + \frac{1}{2} \ve_{\a \b} \ve_{\ad \bd} \bar{R} R + \frac{1}{8} G_{\a \bd} G_{\b \ad} ~, \non
	\eea
where $R, X_{\a}$ and $G_{\a\ad}$ are the torsion superfields defined in section \ref{secCSGSS4}.

In section \ref{SectionDegauging} we saw that upon imposing the gauge $\mf{b}_a=0$ and  degauging conformal space, the Weyl transformations arose as the local dilatation transformations preserving this gauge. In a similar manner, the super-Weyl transformations of $\sU(1)$ superspace may be understood as those local dilatation transformations which preserve the gauge $B_A=0$. To see how this works in detail, we refer the reader to \cite{SCHS}.


\subsection{Integration by parts in conformal superspace}\label{AppIBPCSS4|4}

Integration by parts (IBP) in conformal superspace is more complicated than in the non-supersymmetric case, which was discussed in section \ref{secIBPCS}. This is because the superfield counterpart $V^A=(V^{\a},\bar{U}_{\ad},V^{a})$ of the vector field $V^a$ in \eqref{Z.1} is generally not primary and the argument leading to the IBP rule \eqref{Y.10} breaks down. 

To see this, we consider an integral of the form
\begin{align}
I=\int\text{d}^{4|4}z\, E \, \mathcal{L}+\text{c.c.}~,\quad \mathbb{D}\mathcal{L}=2\mathcal{L}~,\quad Y\mathcal{L}=0~,\quad K_A\mathcal{L}=0~ \label{G.-1}
\end{align}
and suppose that $\mathcal{L}$ takes the form $\mathcal{L}=g^J\mathcal{A}_J(h)$. Here $g_J$ and $h_J$ are primary fields with abstract index structure and $\mathcal{A}$ is a linear differential operator such that $\mathcal{A}_J(h)\equiv \mathcal{A}_J{}^{I}h_I$ is also primary.
 Any total superconformal derivative which arises in moving to the transposed operator $\mathcal{A}^T$ (see \eqref{Z.1}) may be expressed as 
\begin{align}
I_{\text{Total}}=\int\text{d}^{4|4}z\, E \, \Omega+\text{c.c.}~,\quad \Omega=\Nabla_AV^A=\Nabla_{\a}V^{\a}+\bar{\Nabla}^{\ad}\bar{U}_{\ad}-\frac{1}{2}\Nabla_{\a\ad}V^{\a\ad}~ \label{G.0}
\end{align} 
for some set of complex composite superfields $V^A=V^A(g,h)$ whose weight and charge may be deduced from \eqref{G.-1}. 

Once again, if $\mathcal{A}^{T}_J(g)$ is primary then so too is $\Omega$; $K_A\Omega=0$. This means that all dependence on the dilatation connection $B_A$ vanishes and the superconformal covariant derivative takes the form\footnote{This is equivalent to imposing the gauge $B_A=0$ and degauging to $\sU(1)$ superspace.} \eqref{ND} where the $\mathfrak{F}_{A}{}^{B}$ are given by \eqref{connections}. Since we can always ignore total $\sU(1)$ derivatives $\bm{\mathcal{D}}_A$ (and will do so liberally in the sequel), the integral \eqref{G.0} is equivalent to
\begin{align}
I_{\text{Total}}=-\int\text{d}^{4|4}z\, E \, \mathfrak{F}_{A}{}^{B}K_{B}V^{A}+\text{c.c.} \label{G.14|4}
\end{align}

Unfortunately we are not able to argue, in a fashion similar to section \ref{secIBPCS}, that \eqref{G.14|4} vanishes on account of $V^A$ being primary. This is because $V^A$ is generally not primary. The argument given in the non-supersymmetric case has no merit here because the condition $K_A\Omega=0$ yields three constraints on the components of $V^A$
\begin{subequations}
\begin{align}
0&=\text{i}\bar{\Nabla}^{\ad}V_{\a\ad}+(-1)^{\ve_{B}}\Nabla_{B}S_{\a}V^B~,\\
0&=\text{i}\Nabla^{\a}V_{\a\ad}-(-1)^{\ve_{B}}\Nabla_{B}\bar{S}_{\ad}V^B~,\\
0&=2\text{i}\bar{S}_{\ad}V_{\a}+2\text{i}S_{\a}\bar{U}_{\ad}+4V_{\a\ad}-\Nabla_{B}K_{\a\ad}V^{B}~,
\end{align}
\end{subequations}
 from which it cannot be concluded that $V^A$ is primary. 
 
 Despite this we still believe that an integration by parts rule similar to \eqref{Y.10} exists which allows us to conclude that \eqref{G.14|4} vanishes.  In the spirit of the non-supersymmetric case, we propose the following rule:
 \begin{align}
\int\text{d}^{4|4}z\, E \, g^J\mathcal{A}_J(h)=\int\text{d}^{4|4}z\, E \, h^J\mathcal{A}^T_J(g) \label{G.2}
\end{align}
if $K_Ag_I=K_Ah_I=K_A\big(\mathcal{A}_I(h)\big)=K_A\big(\mathcal{A}^T_I(g)\big)=0$. 
 
 We would now like to give two examples, originating from the superconformal gravitino model in section \ref{secSCHSgravitino}, where the above conditions are met and non-trivial cancellations ensure that the corresponding total derivatives vanish and \eqref{G.2} holds.
 
%

 \subsubsection{Example 1}
 
 As our first example, we take a closer look at the total derivative which arises when proving the relation \eqref{2020}. By construction, the left hand side of \eqref{2020} is primary and so too is the second line on the right hand side (here $\mathfrak{J}^{(2)}$ is the conjugate transpose of  $\mathfrak{J}^{(1)}$). Thus the conditions of the rule \eqref{G.2} are met. 
 
 It remains to show that the corresponding total derivative, which may be expressed in the form \eqref{G.14|4} with
 \begin{subequations}
 \begin{align}
 V^{\a}&=2\text{i}H_{\g}\Nabla_{\g\gd}\bm W^{\a\g(2)}\bar{H}^{\gd}-2\text{i}H_{\g}\bm W^{\a\g(2)}\Nabla_{\g\gd}\bar{H}^{\gd}-H_{\g}\Nabla_{\g}\bm W^{\a\g(2)}\bar{\Nabla}_{\gd}\bar{H}^{\gd}~,\\
 \bar{U}_{\ad}&=-\Nabla_{\g}H_{\g}\Nabla_{\g}\bm W^{\g(3)}\bar{H}_{\ad}~,\\
 V^{\a\ad}&=-4\text{i}\Nabla_{\g}H_{\g}\bm W^{\a\g(2)}\bar{H}^{\ad}~,
 \end{align}
 \end{subequations} 
 vanishes. In this case it turns out that $V_{\a\ad}$ is primary and hence $\Nabla_{\a\ad}V^{\a\ad}\approx 0$ (here the symbol $\approx$ means equality modulo a total $\sU(1)$ derivative), but the same is not true of the spinor components. Indeed, using \eqref{ND} one can show that 
 \begin{align}\label{G.3}
 \bar{\Nabla}^{\ad}\bar{U}_{\ad}&\approx  G_{\a}{}^{\ad}\Nabla_{\g}H_{\g}\bm W^{\a\g(2)}\bar{H}_{\ad}~,\\
 \Nabla_{\a}V^{\a}& \approx\bm\cD_{\g}G_{\a}{}^{\ad}H_{\g}\bm W^{\a\g(2)}\bar{H}_{\ad}-G_{\a}{}^{\ad}H_{\g}\Nabla_{\g}\bm W^{\a\g(2)}\bar{H}_{\ad}+G_{\a}{}^{\ad}H_{\g}\bm W^{\a\g(2)}\Nabla_{\g}\bar{H}_{\ad}~.\non
 \end{align}
Since both $\bm W_{\a(3)}$ and $H_{\a}$ transform trivially under $K_A$, we can use \eqref{ND} to replace each occurrence of $\Nabla_{\g}$ on the right hand side of \eqref{G.3} with $\bm\cD_{\g}$. Consequently, the total superconformal derivative reduces to a total $\sU(1)$ derivative:
 \begin{align}
 \int\text{d}^{4|4}z\, E \, \Omega =  \int\text{d}^{4|4}z\, E \,\bm\cD_{\g}\big(G_{\a}{}^{\ad}H_{\g}\bm W^{\a\g(2)}\bar{H}_{\ad}\big) \approx 0~.
 \end{align}

 \subsubsection{Example 2}

Our next example is less trivial than the last, and concerns the total derivative which arises when computing the $\L$ gauge variation \eqref{tinozeta} of the gravitino skeleton. Under a $\L$ gauge transformation, the skeleton has the transformation law
\begin{subequations}\label{G.4}
\begin{align}
\delta_{\L} \mb{S}_{\text{skeleton}}^{(1,0,0)}&=\text{i}\int\text{d}^{4|4}z\, E \,\bigg\{\Nabla^{\a}\delta_{\L}H^{\a}\mf{W}_{\a(2)}(\bar{H}) +\Nabla^{\a}H^{\a}\mf{W}_{\a(2)}(\delta_{\L}\bar{H})\bigg\} +\text{c.c.} \label{G.4a}\\
&=\text{i}\int\text{d}^{4|4}z\, E \, \bigg\{\Nabla^{\a}\delta_{\L}H^{\a}\mf{W}_{\a(2)}(\bar{H})-\frac{1}{4}\delta_{\L}\bar{H}^{\ad}\Nabla^{\a}\Nabla_{\ad}{}^{\b}\bar{\Nabla}^2\Nabla_{(\a}H_{\b)}\bigg\} +\text{c.c.}\label{G.4b}
\end{align}
\end{subequations}
It may be checked that the second term on the right hand side of \eqref{G.4b} is primary, and so the conditions of the rule \eqref{G.2} are met. In this case, the total derivative which arises in moving from \eqref{G.4a} to \eqref{G.4b} takes the form \eqref{G.0} with 
\begin{subequations}
\begin{align}
V^{\a}&=-\frac{\text{i}}{4}\bar{\psi}^{\bd}\Nabla_{\b\bd}\bar{\Nabla}^2\Nabla^{(\a}H^{\b)}~,\\
 \bar{U}_{\ad}&=\frac{\text{i}}{4}\Nabla_{\b\bd}\Nabla_{\a}\bar{\psi}^{\bd}\bar{\Nabla}_{\ad}\Nabla^{(\a}H^{\b)}-\frac{\text{i}}{4}\bar{\Nabla}_{\ad}\Nabla_{\b\bd}\Nabla_{\a}\bar{\psi}^{\bd}\Nabla^{(\a}H^{\b)}~,\\
 V^{\a\ad}&=-\frac{\text{i}}{2}\Nabla_{\b}\bar{\psi}^{\ad}\bar{\Nabla}^2\Nabla^{(\a}H^{\b)}~,
\end{align}
\end{subequations}
where $\bar{\psi}_{\ad}:=\delta_{\L}\bar{H}_{\ad}$. Once again $V_{\a\ad}$ turns out to be primary (this is not the case for, e.g., the pseudo-graviton multiplet) and so $\Nabla_{\a\ad}V^{\a\ad}\approx 0$. On the otherhand, modulo a total $\sU(1)$ derivative, the spinor parts of $\Omega$ may be shown to take the form
\begin{subequations}\label{G.5}
\begin{align}
\bar{\Nabla}^{\ad}\bar{U}_{\ad}&=  \frac{1}{8}G_{\a\ad}\bar{\Nabla}_{\bd}\Nabla_{\b}\bar{\psi}^{\bd}\bar{\Nabla}^{\ad}\Nabla^{(\a}H^{\b)}+\frac{1}{8}\bm\cDB_{\bd}G_{\a\ad}\Nabla_{\b}\bar{\psi}^{\ad}\bar{\Nabla}^{\bd}\Nabla^{(\a}H^{\b)}\notag\\
&-\frac{1}{8}\bm\cDB_{\bd}G_{\a\ad}\bar{\Nabla}^{\ad}\Nabla_{\b}\bar{\psi}^{\bd}\Nabla^{(\a}H^{\b)}-\frac{1}{16}G_{\a\ad}\bar{\Nabla}^2\Nabla_{\b}\bar{\psi}^{\ad}\Nabla^{(\a}H^{\b)} \notag\\
&-\frac{1}{12}X_{\a}\Nabla_{\b}\bar{\psi}^{\ad}\bar{\Nabla}_{\ad}\Nabla^{(\a}H^{\b)} -\frac{1}{12}X_{\a}\bar{\Nabla}_{\ad}\Nabla_{\b}\bar{\psi}^{\ad}\Nabla^{(\a}H^{\b)}~,\label{G.5a}\\[8pt]
\Nabla_{\a}V^{\a}&= \frac{1}{8}G_{\a\ad}\Nabla_{\b}\bar{\psi}^{\ad}\bar{\Nabla}^2\Nabla^{(\a}H^{\b)}~. \label{G.5b}
\end{align}
\end{subequations}

The next step is to make use of \eqref{ND} and the fact that both $\Nabla_{\a}\bar{\psi}_{\ad}$ and $\Nabla_{(\a}H_{\b)}$ are primary to rewrite \eqref{G.5} as
\begin{subequations}\label{G.6}
\begin{align}
\bar{\Nabla}^{\ad}\bar{U}_{\ad}&=  \frac{1}{8}G_{\a\ad}\bm\cDB_{\bd}\bm\cD_{\b}\bar{\psi}^{\bd}\bm\cDB^{\ad}\bm\cD^{(\a}H^{\b)}+\frac{1}{8}\bm\cDB_{\bd}G_{\a\ad}\bm\cD_{\b}\bar{\psi}^{\ad}\bm\cDB^{\bd}\bm\cD^{(\a}H^{\b)}\notag\\
&-\frac{1}{8}\bm\cDB_{\bd}G_{\a\ad}\bm\cDB^{\ad}\bm\cD_{\b}\bar{\psi}^{\bd}\bm\cD^{(\a}H^{\b)}-\frac{1}{16}G_{\a\ad}\big(\bm\cDB^2-4R\big)\bm\cD_{\b}\bar{\psi}^{\ad}\bm\cD^{(\a}H^{\b)} \notag\\
&-\frac{1}{12}X_{\a}\bm\cD_{\b}\bar{\psi}^{\ad}\bm\cDB_{\ad}\bm\cD^{(\a}H^{\b)} -\frac{1}{12}X_{\a}\bm\cDB_{\ad}\bm\cD_{\b}\bar{\psi}^{\ad}\bm\cD^{(\a}H^{\b)}~,\label{G.6a}\\[8pt]
\Nabla_{\a}V^{\a}&= \frac{1}{8}G_{\a\ad}\bm\cD_{\b}\bar{\psi}^{\ad}\big(\bm\cDB^2-4R\big)\bm\cD^{(\a}H^{\b)}~. \label{G.6b}
\end{align}
\end{subequations}
Now that everything is in terms of the $\sU(1)$ covariant derivative we may freely integrate by parts. By making use of the algebra \eqref{U(1)algebraRevamped}, the identity $\bm\cDB^2G_{\a\ad}=6G_{\a\ad}R-4\text{i}\bm\cDB_{\a\ad}R$ and the Bianchi identity \eqref{Bianchi1} one may show, modulo a total $\sU(1)$ derivative, that \eqref{G.6a} and \eqref{G.6b} are given by
\begin{align}
\bar{\Nabla}^{\ad}\bar{U}_{\ad}&= -\frac{1}{4}G_{\a\ad}R\bm\cD_{\b}\bar{\psi}^{\ad}\bm\cD^{(\a}H^{\b)}+\frac{\text{i}}{2}\bm\cD_{\a\ad}R\bm\cD_{\b}\bar{\psi}^{\ad}\bm\cD^{(\a}H^{\b)}\notag\\
&\phantom{L~}-\frac{1}{4}\bm\cDB_{\bd}G_{\a\ad}\bm\cDB^{\bd}\bm\cD_{\b}\bar{\psi}^{\ad}\bm\cD^{(\a}H^{\b)}-\frac{1}{8}G_{\a\ad}\bm\cDB^2\bm\cD_{\b}\bar{\psi}^{\ad}\bm\cD^{(\a}H^{\b)}~,\\
\Nabla_{\a}V^{\a}&=-\bar{\Nabla}^{\ad}\bar{U}_{\ad}~.
\end{align}
Hence $\Omega\approx 0$ and the total conformal derivative has been reduced to a total $\sU(1)$ derivative. 

\subsection{Integration over the chiral subspace} \label{AppChiralSubspace}

Here we will investigate some subtleties associated with integrals defined over the chiral subspace. In particular, a careless application of the usual formulae from the GWZ or $\sU(1)$ geometries can lead to inconsistencies. 

We consider a locally superconformal integral defined over the chiral subspace of the following form
\begin{align}
	\mathcal{J} = -\frac{1}{4} \int \rd^4x \rd^2 \q 
	\, \cE \, \bar{\Nabla}^2 U ~,\quad \mathbb{D}U=2U~,\quad \mb{Y}U=0~,\quad K_A\bar{\Nabla}^2 U=0 \label{C.1} ~.
\end{align}
Once again, we emphasise that since the integrand is primary, all dependence on $B_{A}$ drops out, allowing us to make use of \eqref{ND}. Naively, we might expect that this integral may be uplifted to the full superspace by simply extracting the chiral projector, however this fails in general. An obvious justification for this is because the full superspace integral can only be superconformal if $U$ itself is primary, thus we postulate the following
\begin{align}
	\mathcal{J} = -\frac{1}{4} \int \rd^4x \rd^2 \q 
	\, \cE \, \bar{\Nabla}^2 U = \int \rd^{4|4}z 
	\, E \, U \quad \iff \quad K_A U = 0 \label{C.24|4} ~.
\end{align}

If this chiral integration rule works for the superconformal integral $\mathcal{J}$, then the requirement that $U$ must be primary follows trivially.
On the other hand, if we assume that $U$ is primary, then we may make use of the identity
\bea
K_A U=0 \quad \implies \quad -\frac{1}{4} \bar{\Nabla}^{2} U = - \frac{1}{4} ( \bm\cDB^2 - 4 R) U ~, \label{C.34|4}
\eea
along with the $\sU(1)$ superspace chiral integration rule \eqref{U(1)chiralintegralrule} to prove \eqref{C.24|4},
\bea
-\frac{1}{4} \int \rd^4x \rd^2 \q 
\, \cE \, \bar{\Nabla}^2 U = -\frac{1}{4} \int \rd^4x \rd^2 \q 
\, \cE \, (\bm\cDB^2 - 4 R) U = \int \rd^{4|4}z 
\, E \, U \label{C.4} ~.
\eea
Thus we have demonstrated the validity of the rule \eqref{C.24|4}. 

 Although this is a valid rule to transition between the chiral and full superspaces, it has limited applicability. This is because quite often one encounters scenarios where $U$ is not primary.
For example, let us consider the SCHS model described by the supermultiplet $H_{\a(m)}$, with $m>n=0$.
 We recall that the corresponding action \eqref{SCHSactCSSCF4|4} is formulated in terms of the two chiral primary superfields \eqref{HSSWeylCSS4}; the $\mathcal{N}=1$ higher-spin super-Weyl tensors. As a consequence, there are two obvious forms in which this integral may be presented to resemble \eqref{C.1}:
\begin{subequations}
	\bea
	\mb{S}^{(m,0)}_{\text{skeleton}}&=&-\frac{\ri^m}{4} \int \rd^4x \rd^2 \q 
	\, \cE \, \bar{\Nabla}^2 \bigg[ \mf{W}^{ \a (m+1)} ( H ) \Nabla_{\a_1}{}^{\bd_1}
	\cdots  \Nabla_{\a_{m}}{}^{\bd_{m}}
	\Nabla_{\a_{m+1}} \bar{H}_{\bd(m)}  \bigg] + \text{c.c.}~~~~~~~~~~~\\
	&=&
	-\frac{\ri^m}{4} \int \rd^4x \rd^2 \q 
	\, \cE \, \bar{\Nabla}^2 \bigg[ \Nabla^{\a} H^{\a(m) } \mf{W}_{\a (m+1)} (\bar{H})\bigg] + \text{c.c.}\label{chiralsubspace1}
	\eea
\end{subequations}
In this case it turns out that only the  $U$ corresponding to \eqref{chiralsubspace1} is primary,
\bea
K_{A} \Nabla_{(\a_1}{}^{\bd_1}
\cdots  \Nabla_{\a_{m}}{}^{\bd_{m}}
\Nabla_{\a_{m+1})} \bar H_{\bd(m)} \neq 0 ~, \qquad K_{A} \Nabla^{(\a_{1}} H^{\a_{2} \dots \a_{m+1} )} = 0 ~,
\eea
and hence the rule \eqref{C.24|4} works only for \eqref{chiralsubspace1}.

At this point, we have deduced that there is a single correct pathway to lifting these actions. Unfortunately, when the background superspace is not conformally flat, this introduces some complications in the analysis of chiral gauge transformations $\delta_{\l}H_{\a(m)}$. These are given by
\bea
\d_{\l} \mb{S}^{(m,0)}_{\text{skeleton}} & = & 
-\frac{\ri^m}{4} \int \rd^4x \rd^2 \q 
\, \cE \, [\bar{\Nabla}^2 , \Nabla^{\a_{1}}] \l^{\a_{2} \dots \a_{m+1} } \mf{W}_{\a (m+1)} (\bar{H}) + \text{c.c.} \non \\
&=& - \frac{m \ri^{m}}{2} \int \rd^4x \rd^2 \q 
\, \cE \, \bar{\Nabla}^2 \bigg[ \bm W^{\a(2)}{}_{\b} \l^{\b \a(m-1)} \Nabla_{\a_1}{}^{\bd_1}
\cdots  \Nabla_{\a_{m}}{}^{\bd_{m}}
\Nabla_{\a_{m+1}} \bar H_{\bd(m)} \bigg] ~~~~~~~\non \\
&& + \, \text{c.c.}
\eea
which cannot be trivially lifted. It is not obvious if this variation may be written in a form explicitly dependent on the super-Weyl tensor when integrating over the full superspace (given that one can not carelessly integrate by parts). Hence, the chiral gauge variation of any non-minimal sector must be reduced to the chiral subspace in order to compare them with that of the skeleton - see section \ref{secSCHSgravitino} for an in depth calculation for the $m=1$ case. Fortunately, the $\L$ gauge transformations do not suffer from such issues and the relevant calculations may be performed in the usual manner; within the full superspace.

The approach outlined above is crucial in analysing the class of models described by $H_{\a(m)}$. 
However, for the generic SCHS supermultiplets  $H_{\a(m)\ad(n)}$, with $m\geq 1$ and $n\geq 1$, there is no obvious way to lift the action \eqref{SCHSactCSSCF4|4a} to the full superspace by invoking the rule \eqref{C.24|4}.
Nevertheless, one can circumvent this problem altogether by working only with the SCHS action in the second form \eqref{SCHSactCSSCF4|4b} rather than \eqref{SCHSactCSSCF4|4a}, i.e. in terms of higher-spin super-Bach tensors instead of higher-spin super-Weyl tensors. Then, all calculations take place within the full superspace. 

\section{Superconformal non-gauge models} \label{AppNGSM}

Conformal non-gauge fields $\chi_{\a(m)\ad(n)}$ were discussed in appendix \ref{AppCNGM}.
On Bach-flat backgrounds, these fields  play an essential role in ensuring gauge invariance in models for the following three CHS fields: (i) conformal maximal-depth spin-3, section \ref{secCMDspin3}; (ii) conformal maximal-depth spin-5/2, section \ref{secCMDspin52}; and (iii) conformal (minimal-depth) hook field, section \ref{secConformalHook}. 

Common to all of these models is the presence of non-gauge fields $\chi_{\a(m)}$ with $n=0$. Such fields may be found sitting within the `chiral non-gauge' supermultiplets $\Omega_{\a(m)}$ discussed below.
A specific example of the latter was used in section \ref{secPGSBach}, where $\Omega_{\a}$ played an important role in ensuring gauge invariance of the supersymmetric extension of (iii). However, non-gauge fields with $n> 0$ were also important in models for (i) and (ii), and these are not contained within $\Omega_{\a(m)}$ at the component level. 
They may be found residing in the `longitudinal linear non-gauge' supermultiplets, which we also discuss below. 

\subsection{Chiral non-gauge superconformal multiplets} \label{secCNGS}

We define $\Omega_{\a(m)}$, with $m\geq 1$, to be a primary chiral superfield,
\begin{subequations}\label{7.11+12}
\begin{align}
K_B\Omega_{\a(m)}=0~,\qquad \bar{\Nabla}_{\ad}\Omega_{\a(m)}=0~, \label{7.11}
\end{align}
 which has Weyl weight and $\sU(1)_{R}$ charge given by 
\begin{align}
\mathbb{D}\Omega_{\a(m)}=\frac{1}{2}(2-m)\Omega_{\a(n)} \quad 
\implies \quad
\mb{Y}\Omega_{\a(m)}=\frac{1}{3}(m-2)\Omega_{\a(m)}~. \label{7.12}
\end{align}
\end{subequations}
It is possible to show that the composite scalar superfield defined by
\begin{align}
\mathcal{F}^{(m)}\big(\Omega,\bar{\Omega}\big)=&\sum_{k=0}^{m}(-1)^k\big(\Nabla_{\a\ad}\big)^{k}\Omega^{\a(m)}\big(\Nabla_{\a\ad}\big)^{m-k}\bar{\Omega}^{\ad(m)} \notag\\
-\frac{\text{i}}{2}&\sum_{k=1}^{m}(-1)^{m+k}\Nabla_{\a}\big(\Nabla_{\a\ad}\big)^{k-1}\Omega^{\a(m)}\bar{\Nabla}_{\ad}\big(\Nabla_{\a\ad}\big)^{m-k}\bar{\Omega}^{\ad(m)}
\label{F-Lagrangian}
\end{align}
is primary on a generic background. The superconformal properties of $\mathcal{F}^{(m)}$ may therefore be summarised as follows
\begin{align}
K_{A}\mathcal{F}^{(m)}=0~,\qquad \mathbb{D}\mathcal{F}^{(m)}=2\mathcal{F}^{(m)}~,\qquad \mb{Y} \mathcal{F}^{(m)}=0~.
\end{align} 
Furthermore, one can show that it satisfies the complex conjugation property 
\begin{align}
\mathcal{F}^{(m)}=(-1)^m\bar{\mathcal{F}}^{(m)}~. 
\end{align}
It follows that the action functional
\begin{align}
S_{\text{NG}}^{(m)}[\Omega,\bar{\Omega}]=\text{i}^{m}\int \text{d}^{4|4}z\, E \, \mathcal{F}^{(m)}\big(\Omega,\bar{\Omega}\big) \label{7.16}
\end{align}
is real and invariant under the conformal supergravity gauge group $\bm \cG$. When written as an integral over the chiral subspace, this action simplifies to 
\begin{align}
S_{\text{NG}}^{(m)}[\Omega,\bar{\Omega}]=-\frac{~\text{i}^{m}}{4}\int \rd^4x \rd^2 \q \, \cE\, \O^{\a(m)}\bar{\Nabla}^2\big(\Nabla_{\a\ad}\big)^{m}\bar{\O}^{\ad(m)}~.
\end{align}

Let us briefly comment on the models \eqref{7.16} for small values of $m$. Firstly, the models \eqref{7.16} were motivated by our analysis in section \ref{secPGSBach}, which made use of the case $m=1$,
\bea
\mathcal{F}^{(1)}\big(\Omega,\bar{\Omega}\big)
= \O^\a \Nabla_{\a\ad} \bar \O^\ad
-\Nabla_{\a\ad} \O^\a \bar \O^\ad  
-\frac{\ri}{2} \Nabla_{\a} \O^\a \bNabla_{\ad} \bar \O^\ad~.
\label{7.24}
\eea

Secondly, we note that the relations \eqref{7.11+12} are also well defined in the $m=0$ case, which corresponds to the conformal scalar supermultiplet $\O$. In this case the Lagrangian \eqref{F-Lagrangian} turns into 
the  Wess-Zumino kinetic term 
\bea
\mathcal{F}^{(0)} \big(\Omega,\bar{\Omega}\big)= \O \bar \O~.
\label{7.254|4}
\eea

Thirdly, the equation of motion for $\O^\a$ 
in the model \eqref{7.24} is 
\bea
\bar\Nabla^2 \Nabla_{\a\ad} \bar \O^\ad =0~.
\label{7.26}
\eea
Here the left-hand side is a primary chiral spinor superfield. Its lowest component is proportional to \eqref{7.14b}. More generally, it may be shown that the equation of motion for the model defined by eqs. \eqref{F-Lagrangian} and \eqref{7.16} is
\bea
\bar\Nabla^2 \big(\Nabla_{\a \ad} \big)^{m} \bar \O^{\ad(m)} =0~.
\label{7.27} 
\eea
Here the left-hand side, 
\bea
\Xi_{\a(m)} (\bar \O) 
:= \bar\Nabla^2\big(\Nabla_{\a \ad} \big)^{m} 
\bar \O^{\ad(m)} ~,\qquad \Nabla_\b \bar \O^{\ad (m)} =0
\label{7.28} 
\eea
 is a primary chiral tensor superfield. 
$\Xi_{\a(m)}$ is a new superconformal operator for $m>0$.

Finally, upon degauging to $\sU(1)$ superspace, the actions \eqref{7.16} (with $m=1,2,3$)  may be shown to take the form
\begin{subequations}
\begin{align}
S_{\text{NG}}^{(1)}[\Omega,\bar{\Omega}]=~&\text{i}\int\text{d}^{4|4}z\, E \, \O^{\a}\bigg\{\bm\cD_{\a\ad}+\text{i}G_{\a\ad}\bigg\}\bar{\O}^{\ad} ~,\\
S_{\text{NG}}^{(2)}[\Omega,\bar{\Omega}]=~&-\int\text{d}^{4|4}z\, E \, \O^{\a(2)}\bigg\{\bm\cD_{\a\ad}\bm\cD_{\a\ad}+3\text{i}G_{\a\ad}\bm\cD_{\a\ad}+\frac{3\text{i}}{2}\big(\bm\cD_{\a\ad}G_{\a\ad}\big)-2G_{\a\ad}G_{\a\ad} ~~~~~~~~~~~~~\notag\\
& -\frac{1}{4}\big(\big[\bm\cD_{\a},\bm\cDB_{\ad}\big]G_{\a\ad}\big)\bigg\}\bar{\O}^{\ad(2)} ~,\\
S_{\text{NG}}^{(3)}[\Omega,\bar{\Omega}]=~& -\text{i}\int\text{d}^{4|4}z\, E \, \O^{\a(3)}\bigg\{\bm\cD_{\a\ad}\bm\cD_{\a\ad}\bm\cD_{\a\ad}+6\text{i}G_{\a\ad}\bm\cD_{\a\ad}\bm\cD_{\a\ad}+6\text{i}\big(\bm\cD_{\a\ad}G_{\a\ad}\big)\bm\cD_{\a\ad} \notag\\ 
&-11G_{\a\ad}G_{\a\ad}\bm\cD_{\a\ad}-11G_{\a\ad}\big(\bm\cD_{\a\ad}G_{\a\ad}\big)-6\text{i}G_{\a\ad}G_{\a\ad}G_{\a\ad}\notag\\
&+\frac{\text{i}}{2}\big(\bm\cD_{\a}G_{\a\ad}\big)\big(\bm\cDB_{\ad}G_{\a\ad}\big)-\big(\big[\bm\cD_{\a},\bm\cDB_{\ad}\big]G_{\a\ad}\big)\bm\cD_{\a\ad}-\frac{1}{2}\big(\bm\cD_{\a\ad}\big[\bm\cD_{\a},\bm\cDB_{\ad}\big]G_{\a\ad}\big)\notag\\
&-8\text{i}G_{\a\ad}\big(\big[\bm\cD_{\a},\bm\cDB_{\ad}\big]G_{\a\ad}\big)\bigg\}\bar{\O}^{\ad(3)}~.
\end{align}
\end{subequations}
For further details on the degauging procedure, we refer the reader to appendix \ref{AppDegaugeCSS4N=1}.

It is instructive to analyse the component structure of the model described by \eqref{7.16}. To this end, we restrict ourselves to the bosonic backgrounds \eqref{Bbackground} of conformally flat superspaces, which are characterised by
\begin{align}
\Nabla_{a} | = \nabla_{a} ~, \qquad  \bm W_{\a\b\g}=0~.
\end{align}
 On account of \eqref{Bbackground} this means that the Weyl tensor vanishes, $W_{\a(4)}=0$. Since the superfield $\Omega_{\a(m)}$ is chiral it has four independent component fields (all of which are $K_a$ primary), which we define as
\begin{subequations} \label{7.18}
 \begin{align}
 A_{\a(m)}&:=\Omega_{\a(m)}\big| ~,\\
 U_{\a(m+1)}&:= \Nabla_{(\a_1}\Omega_{\a_2\dots\a_{m+1})}\big|~,\\
 V_{\a(m-1)}&:= \Nabla^{\g}\Omega_{\g\a(m-1)}\big|~,\\
 D_{\a(m)}&:=-\frac{1}{4}\Nabla^2\Omega_{\a(m)}\big|~.
 \end{align}
 \end{subequations}
Using these definitions, one may show that the functional \eqref{7.16} is equivalent to
\begin{align}
&(-1)^mS_{\text{NG}}^{(m)}[\Omega,\bar{\Omega}]=~\big(-\text{i})^{m}\int \text{d}^{4}x\, e \,\bigg\{\bar{A}^{\ad(m)}\Box\big(\nabla_{\a\ad}\big)^mA^{\a(m)}+\bar{D}^{\ad(m)}\big(\nabla_{\a\ad}\big)^mD^{\a(m)} \notag\\
&-\frac{\text{i}}{2}\bar{U}^{\ad(m+1)}\big(\nabla_{\a\ad}\big)^{m+1}U^{\a(m+1)} +\frac{\text{i}}{2}\frac{m}{m+1}\bar{V}^{\ad(m-1)} \Box \big(\nabla_{\a\ad}\big)^{m-1}V^{\a(m-1)}\bigg\} \notag\\
&=~ \text{i}^{m}\int \text{d}^{4}x\, e \,\bigg\{\bar{A}_{(1)}^{\ad(m)}\mathfrak{X}^{(1)}_{\ad(m)}(A)
+\bar{D}_{(0)}^{\ad(m)}\mathfrak{X}^{(0)}_{\ad(m)}(D)-\frac{\text{i}}{2}\bar{U}_{(0)}^{\ad(m+1)}\mathfrak{X}^{(0)}_{\ad(m+1)}(U)~~~~~~~~~~~~~~~\notag\\
&\phantom{(-\text{i})^{m}\int \text{d}^{4}x\, e \,\bigg\{}-\frac{\text{i}}{2}\frac{m}{m+1}\bar{V}_{(1)}^{\ad(m-1)}\mathfrak{X}^{(1)}_{\ad(m-1)}(V)\bigg\}~\non\\
&= 2S^{(m,0,1)}_{\text{NG}}[A,\bar{A}]+2S^{(m,0,0)}_{\text{NG}}[D,\bar{D}]-S^{(m+1,0,0)}_{\text{NG}}[U,\bar{U}]+\frac{m}{m+1}S^{(m-1,0,1)}_{\text{NG}}[V,\bar{V}]~.\label{mmm}
\end{align}
In the last line we have adopted the notation from appendix \ref{AppCNGM}, since both type I and type II conformal non-gauge fields are present.

\subsection{Longitudinal linear non-gauge superconformal multiplets}

A superfield $\Omega_{\a(m)\ad(n)}$, with $m \geq n$, is said to be longitudinal linear if it obeys the constraint
\begin{align}
 \bar{\Nabla}_{(\ad_1}\Omega_{\a(m)\ad_2\dots\ad_{n+1})}=0~\quad \implies \quad \bNabla^2\Omega_{\a(m)\ad(n)}=0~.
\end{align}

Similar to the chiral case, requiring $\Omega_{\a(m)\ad(n)}$ to be primary fixes its $\sU(1)_{R}$ charge in terms of its conformal weight as follows
\begin{align}
\mathbb{D}\Omega_{\a(m)\ad(n)}=\Delta \Omega_{\a(m)\ad(n)}~, \qquad \mb{Y}\Omega_{\a(m)\ad(n)}=-\frac{2}{3}(\Delta+n)\Omega_{\a(m)\ad(n)}~.
\end{align}
Choosing $\Delta=1-\frac{1}{2}(m-n)$ allows one to construct the following superconformal action
\begin{align}
\mb{S}_{~||}^{(m,n)}[\Omega,\bar{\Omega}]=\text{i}^{m+n}\int\text{d}^{4|4}z\, E \, \mathcal{F}^{(m,n)}(\Omega,\bar{\Omega})~,\label{long}
\end{align}
where $\mathcal{F}^{(m,n)}(\Omega,\bar{\Omega})$ is the composite scalar superfield
\begin{align}
&\mathcal{F}^{(m,n)}(\Omega,\bar{\Omega})=\sum_{k=0}^{m-n}(-1)^k\big(\Nabla_{\ad}{}^{\b}\big)^k\bar{\Omega}^{\a(n)\ad(m)}\big(\Nabla_{\ad}{}^{\b}\big)^{m-n-k}\Omega_{\a(n)\b(m-n)\ad(n)} \notag\\
-\frac{\text{i}}{2}&\sum_{k=1}^{m-n}(-1)^{m+n+k}\bar{\Nabla}_{\ad}\big(\Nabla_{\ad}{}^{\b}\big)^{k-1}\bar{\Omega}^{\a(n)\ad(m)}\Nabla^{\b}\big(\Nabla_{\ad}{}^{\b}\big)^{m-n-k}\Omega_{\a(n)\b(m-n)\ad(n)}~,
\end{align}
possessing the properties
\begin{subequations}
\begin{align}
K_{A}\mathcal{F}^{(m,n)}=0~,&\qquad \mathbb{D}\mathcal{F}^{(n,m)}=2\mathcal{F}^{(n,m)}~,\qquad \mb{Y}\mathcal{F}^{(n,m)}=0~, \\[5pt]
&~~~ \overline{\mathcal{F}^{(n,m)}}=(-1)^{n+m}\mathcal{F}^{(n,m)}~. 
\end{align}
\end{subequations}
 When written in the chiral subspace the action \eqref{long} takes the form
\begin{align}
\mb{S}_{~||}^{(m,n)}&[\Omega,\bar{\Omega}]=-\frac{\text{i}^{m+n}}{4}\int\text{d}^{4}x\text{d}^2\theta \,  \cE \, \bigg\{ \Omega_{\a(m)\b(n-m)\bd(m)}\bar{\Nabla}^2\big(\Nabla_{\ad}{}^{\b}\big)^{m-n}\bar{\Omega}^{\a(n)\ad(m-n)\bd(n)} \notag\\
&+\frac{2n}{n+1}\big(\Nabla_{\ad}{}^{\b}\big)^{m-n}\bar{\Nabla}_{\dd}\bar{\Omega}^{\a(n)\ad(m-n)\dd\bd(n-1)}\bar{\Nabla}^{\gd}\Omega_{\a(n)\b(m-n)\bd(n-1)\gd} \bigg\}~. 
\end{align}

The non-vanishing independent component fields of $\Omega_{\a(n)\ad(m)}$ are defined as
\begin{subequations}\label{4.13}
\begin{align}
A_{\a(m)\ad(n)}&:=\Omega_{\a(m)\ad(n)}| ~,\label{4.13a}\\
B_{\a(m-1)\ad(n)}&:= \Nabla^{\b}\Omega_{\a(m-1)\b\ad(n)}|~, \label{4.13b}\\
C_{\a(m+1)\ad(n)}&:= \Nabla_{(\a_1}\Omega_{\a_2\dots\a_{m+1})\ad(n)}|~,\label{4.13c}\\
D_{\a(m)\ad(n-1)}&:= \bNabla^{\bd}\Omega_{\a(m)\ad(n-1)\bd}|~,\label{4.13d}\\
E_{\a(m)\ad(n)}&:=-\frac{1}{4}\Nabla^2\Omega_{\a(m)\ad(n)}|~,\label{4.13e}\\
F_{\a(m-1)\ad(n-1)}&:=\frac{1}{2}\big[\Nabla^{\b},\bNabla^{\bd}\big]\Omega_{\a(m-1)\b\ad(n-1)\bd}|+\text{i}\frac{n+1}{m+1}\Nabla^{\b\bd}\Omega_{\a(m-1)\b\ad(n-1)\bd}|~,\label{4.13f}\\
G_{\a(m+1)\ad(n-1)}&:=\frac{1}{2}\big[\Nabla_{(\a_1},\bNabla^{\bd}\big]\Omega_{\a_2\dots\a_{m+1})\ad(n-1)\bd}|~,\label{4.13g}\\
H_{\a(m)\ad(n-1)}&:= -\frac{1}{4}\bNabla^{\bd}\Nabla^2\Omega_{\a(m)\ad(n-1)\bd}|+\text{i}\frac{n-m}{n}\Nabla^{\b\bd}\Nabla_{(\b}\Omega_{\a_1\dots\a_{m})\ad(n-1)\bd}|~.\label{4.13h}
\end{align}
\end{subequations}
 
 The first two fields, \eqref{4.13a} and \eqref{4.13b}, are primary and have the same conformal weight as that of a maximal depth CHS field of the same rank (though the former do not have any gauge symmetry). The next four fields \eqref{4.13c} -- \eqref{4.13f}
 are all primary and are conformal non-gauge. However, the last two fields \eqref{4.13g} and \eqref{4.13h} are not able to be defined so that they are primary. Instead they transform non-trivially under a $K$-transformation, 
\begin{subequations}
\begin{align}
K_{\b\bd}G_{\a(m+1)\ad(n-1)}&=\phantom{-}8\text{i}(n+1)\ve_{\b(\a_1}A_{\a_2\dots\a_{m+1})\ad(n-1)\bd}~, \\
K_{\b\bd}H_{\a(m)\ad(n-1)}&=-4\text{i}m\frac{n+1}{m+1}\ve_{\b(\a_1}B_{\a_2\dots\a_m)\ad(n-1)\bd}~,
\end{align} 
\end{subequations} 
and do not correspond to typical (i.e. generalised CHS or non-gauge) conformal fields.\footnote{See, however, \cite{MetsaevOrd1, MetsaevOrd2} where various conformal fields were defined to transform non-trivially under special conformal transformations.}

Here we do not give the corresponding component action, since it is not illuminating. Rather it suffices to give a few comments regarding its structure. First, by setting $n=0$ in the above models, one recovers the rank-$m$ chiral non-gauge models from the previous section. Being chiral, the component content of these supermultiplets is simple and there are only four non-vanishing fields, all of which turn out to be conformal non-gauge (see previous section). The component action is also simple in the sense that it consists only of the kinetic terms for the four non-gauge fields and is diagonal. However, the longitudinal linear supermultiplets have twice as many component fields, and not all of them are primary but instead transform into one another under Weyl transformations. Thus, in order to maintain Weyl invariance, the component action necessarily becomes non-diagonal, resulting in a much more complicated structure.

\end{subappendices}


\chapter{Conclusion and outlook} \label{ChapterFinito}


In this thesis we have provided a systematic study of models describing the dynamics of (super)conformal higher-spin fields on various curved backgrounds in three and four dimensions.
The purpose of this chapter is to summarise our key findings, place them in a broader context, and comment on some open problems.
For a detailed survey on the original results obtained in this thesis, we refer the reader to the last section of each of the chapters  \ref{Chapter3D} -- \ref{Chapter4Dsuperspace}.  
In order to outline the salient features in a universal way, our discussion will focus on the non-supersymmetric models for conformal fields with integer spin $s$.  
Extensions to models for conformal fields of any type (i.e. fermionic, mixed symmetry, higher-depth) and their supersymmetric analogues are straightforward.

Free conformal higher-spin actions are generalisations of the linearised action for conformal gravity.
Let us denote by $E_{ab}$ the primary descendant of the curvature tensor which defines the equation of motion for conformal gravity in $d$ dimensions. For $d=3$ this is the Cotton tensor $C_{ab}$, whilst for $d=4$ it is the Bach tensor $B_{ab}$.
It appears natural to expect that if one restricts their attention to background spacetimes satisfying
\begin{align}
E_{ab}=0~, \label{conc.0}
\end{align}
then a gauge and Weyl invariant linearised action for the conformal spin-$s$ field $h_{a(s)}$ is guaranteed to exist. 
Indeed, all of the models constructed in this thesis indicate that this statement is true. 

In three dimensions, the equation \eqref{conc.0} defines a conformally-flat background. 
For such backgrounds we have been able to obtain closed-form expressions for CHS actions which are manifestly gauge and Weyl invariant. 
They are completely characterised by the higher-spin Cotton tensor $\mc{C}_{a(s)}(h)$, which is a primary descendant of $h_{a(s)}$ that is traceless, transverse and gauge invariant.
For some specific cases, we showed that Weyl covariance can be extended to more general backgrounds, but gauge invariance cannot. 

In four dimensions, we also derived closed-form expressions for gauge and Weyl invariant CHS actions on conformally-flat backgrounds.
They are completely determined by the higher-spin Bach tensor $\mc{B}_{a(s)}(h)$, which is a primary descendant of $h_{a(s)}$ that is traceless, transverse and gauge invariant.
However, in $d=4$ the equation \eqref{conc.0} defines a Bach-flat background, which is not necessarily conformally-flat.
In such backgrounds the situation becomes significantly more complicated. In particular, it has been conjectured \cite{GrigorievT} that in order to ensure gauge invariance of the conformal spin-$s$ field on Bach-flat backgrounds, it is necessary to non-minimally couple it to lower-spin conformal fields.  
In support of this proposal we have derived the first few examples of complete gauge-invariant models on Bach-flat backgrounds for which a coupling between the parent field and certain subsidiary fields are required.

Each of these subsidiary fields belong to a family which we have called conformal non-gauge fields. 
They are not standard conformal gauge fields because they do not possess the correct Weyl weight required in order to have their own gauge freedom. 
Their necessity  appears to be an artefact of either the higher-depth or the mixed symmetry nature of the parent field in the CHS models that we have considered.
Although it is conceivable that non-gauge fields are present in complete models for minimal depth conformal spin-$s$ fields, it would seem unnatural. This is because the corresponding action should be attainable via linearisation of the fully interacting CHS model of \cite{Segal}, which consists of an infinite tower of CHS fields $h_{a(s)}$ with $s\geq 0$.
In this setting, minimal depth CHS fields with spin $s'<s$ are natural candidates for the role of subsidiary fields in a Bach-flat completion of the linearised spin-$s$ model. 
Our results indicate that the higher the spin of the parent conformal field, the higher the number of subsidiary fields required in its Bach-flat completion. 
In fact, by reiterating the supersymmetric argument given in section \ref{secSpin3Recipe}, one is led to conclude that conformal fields of all spins $s'<s$ must participate in a consistent Bach-flat completion of the linearised spin-$s$ model.

 Another possibility is that the gauge-invariant linearised action on a Bach-flat background is not unique, as was the case for the hooked conformal graviton. 
Currently,  for a given CHS field, the only way to determine which subsidiary field(s) will be necessary for gauge invariance is through explicit calculation. 
In order to perform such computations in finite time, new insight or techniques are required. 
For example, one could attempt to combine the conformal (super)space approach employed in this thesis, with that of the ordinary derivative formulation of \cite{MetsaevOrd1,MetsaevOrd2}.
In such an approach, one would have to work out how to efficiently deal with the numerous non-primary auxiliary fields required in this formulation.\footnote{In particular, their presence will complicate the process of integration by parts in conformal space. } 
 Alternatively, one could try to generate CHS models from a single hyper-action defined on a higher-dimensional `tensorial' or `hyper' space. 
There has been much attraction in developing hyperspace descriptions of massless higher-spin (super)fields \cite{HS1,HS2,HS3,HS4,HS5,HS6,HS7,HS8,HS9,HS10,HS11,HS12,HS13,HS14,HS15}, see also \cite{Sorokin:2017irs} for a review.
 It would be interesting to study whether the methods of conformal (super)space can be accommodated within this approach.


One of the simplifying features of three and four dimensional spacetimes is the existence of  the two component spinor formalism. We have taken advantage of this fact and practically all models in this thesis have been presented in this setting. However, it is interesting to note that there exists a simple way to formulate the models for a CHS field in vector notation. In particular, on any conformally-flat background, the gauge- and Weyl-invariant action for the conformal spin-$s$ field $h_{a(s)}$ can be expressed in the form\footnote{This may be easily extended to fermionic and mixed symmetry conformal fields by making use of the spin-projection operators given in \cite{AdS3(super)projectors} (see also \eqref{CasimirProjectorsMink4} for $d=4$). }
\begin{subequations}\label{FinalAction}
\begin{align}
S_{\text{CHS}}^{(d=3)}[h_{(s)}] &\propto \int \text{d}^3x \, e \, h^{a(s)}\mb{W}\prod_{j=1}^{s-1}\Big(\mb{W}^2-j^2\Box_c\Big)h_{a(s)}~,\label{FinalActiona}\\
S_{\text{CHS}}^{(d=4)}[h_{(s)}] &\propto \int \text{d}^4x \, e \, h^{a(s)}\prod_{j=0}^{s-1}\Big(\mb{W}^a\mb{W}_a-j(j+1)\Box_c\Big)h_{a(s)}~.\label{FinalActionb}
\end{align}
\end{subequations}
Here $\Box_c=\nabla^a\nabla_a$ and we have made use of the Pauli-Lubankski pseudo-scalar $\mb{W}$ and pseudo-vector $\mb{W}^a$, which have been minimally lifted to conformal space:
\begin{align}
\mb{W}:=-\frac{1}{2}\ve^{abc}\nabla_aM_{bc}~,\qquad \mb{W}^a:=-\frac{1}{2}\ve^{abcd}\nabla_{b}M_{cd}~.
\end{align}


The kinetic operators in \eqref{FinalAction} are manifestly factorised into products of (non-minimal) second-order operators, without assuming a transverse gauge condition. 
By making use of the $d$-dimensional Pauli-Lubankski pseudo-tensor $\mb{W}^{a_1 \dots a_{d-3}}:=-\frac{1}{2}\ve^{a_1\dots a_{d-3} bfg}\nabla_bM_{fg}$ (see e.g. \cite{KuzenkoPindur}), it should be possible to obtain $d$-dimensional analogues of the conformal actions \eqref{FinalActiona} and \eqref{FinalActionb}.
The field $h_{a(s)}$ in \eqref{FinalAction} is totally symmetric and traceless. 
However, upon replacing $h_{a(s)}$ with the traceful field ${\bm h}_{a(s)}$, the actions \eqref{FinalAction}  are also invariant under the gauge transformations \eqref{FlatDGT} and \eqref{FlatAGT}. This is because the kinetic operators appearing in \eqref{FinalAction} are directly related to the new spin projection operators (aka traceless and transverse projectors) derived in
 \cite{AdS3(super)projectors} for $\mb{M}^3$ and $\mb{M}^4$ (see also section \ref{secCHSMink4} for $d=4$).

Given a maximally symmetric spacetime, the unitary irreducible representations (irreps) of its
isometry algebra may be realised on the space of tensor fields satisfying certain differential
constraints. 
The purpose of the spin projection operator is to take an unconstrained field, which describes a multiplet of irreps off the mass-shell, and return the component corresponding to the irrep with maximal spin. 
In a series of papers \cite{AdSprojectors, AdSuperprojectors, AdS3(super)projectors} we have constructed (super)spin projection operators in (A)dS$^d$ space, and AdS$^d$ superspace for $d=3,4$.
In each case we have found that the poles of the (super)projectors are intimately related with partially-massless (super)fields.
It would be interesting to construct (super)projectors in $d>4$ and investigate whether this property persists. 
 In this thesis have reformulated all (S)CHS actions in AdS$^d$ (super)space in terms of the (super)projectors.
 This allowed us to show that the corresponding (S)CHS kinetic operator factorises into products of (minimal) second-order operators, each associated with a partially-massless (super)field.



Another interesting problem would be to construct spin projection operators on more general curved spacetimes. 
Since the work of Fradkin and Tseytlin \cite{FT}, it has been evident that spin projection operators are inextricably linked with linearised CHS models. 
Indeed, in \cite{FT} such operators played a fundamental role in the formulation of their CHS action, see eq. \eqref{DdimCHSA}. 
Given this relationship, and the results of this thesis, it is tempting to conclude that such projectors can be formulated at most on conformally-flat backgrounds. This is because the property of transversality of the higher-spin Cotton and Bach tensors cannot be extended to more general backgrounds.\footnote{In the case of $4d$ Bach-flat backgrounds, the necessity of subsidiary fields for gauge invariance implies that, for a single conformal field of spin $s>2$, a transverse operator (i.e. higher-spin Bach operator) does not exist.  }

In this thesis we have also constructed free models for topologically massive higher-spin gauge  (super)fields (of both the old and new variant) in $\mb{M}^3$ and AdS$^{3}$ (super)space. These are higher-spin extensions of linearised topologically massive (super)gravity \cite{DJT1, DJT2, DK} which consist of two sectors; one conformal and the other massless. An intriguing question is: Do higher-spin non-linear extensions exist?
Within the approach initiated in
\cite{Nilsson1,Nilsson2}, Linander and Nilsson \cite{LN}
constructed the full nonlinear spin-3 Cotton equation coupled to 
spin-2. 
The construction of the non-linear spin-3 Cotton tensor \cite{LN} requires an elimination 
of certain auxiliary fields, a procedure that becomes extremely difficult
for $s>3$.  However, so far this is unexplored territory.
There exist non-linear formulations for the massless spin-3 theory
\cite{HR,CFPT}, and the generalisation from $s=3$ to $s>3$ 
is shown in \cite{CFPT} to be trivial within the formulation developed. 
These results indicate that it is possible to construct a non-linear 
topologically massive higher-spin field theory. 


\begin{thebibliography}{66}
\begin{footnotesize}








\bibitem{Topological} 
S.~M.~Kuzenko and M.~Ponds,  
``Topologically massive higher spin gauge theories,''
JHEP {\bf 1810}, 160 (2018)
\href{https://arxiv.org/abs/1806.06643}{[arXiv:1806.06643 [hep-th]]} .



\bibitem{3Dprojectors} 
E. I. Buchbinder, S. M. Kuzenko, J. La Fontaine and M. Ponds, 
``Spin projection operators and  higher-spin 
Cotton tensors in three dimensions,'' 
Phys.\ Lett.\ B {\bf 790}, 389 (2019)
 \href{https://arxiv.org/abs/1812.05331}{[arXiv:1812.05331 [hep-th]]} .
  
  
\bibitem{Confgeo}
S.~M.~Kuzenko and M.~Ponds,
``Conformal geometry and (super)conformal higher-spin gauge theories,''
JHEP {\bf 1905},  113 (2019) 
\href{https://arxiv.org/abs/1902.08010}{[arXiv:1902.08010 [hep-th]]}.


 \bibitem{AdSprojectors}
S.~M.~Kuzenko and M.~Ponds,
``Spin projection operators in (A)dS and partial masslessness,''
Phys. Lett. B \textbf{800}, 135128 (2020)
\href{https://arxiv.org/abs/1910.10440}{[arXiv:1910.10440 [hep-th]]}.


\bibitem{spin3depth3} 
  S.~M.~Kuzenko and M.~Ponds,
  ``Generalised conformal higher-spin fields in curved backgrounds,''
  JHEP {\bf 2004}, 021 (2020)
  \href{https://arxiv.org/abs/1912.00652}{[arXiv:1912.00652 [hep-th]]}.


\bibitem{SCHS}
S.~M.~Kuzenko, M.~Ponds and E.~S.~N. Raptakis,
``New locally (super)conformal gauge models in Bach-flat backgrounds,''
JHEP \textbf{2008}, 068 (2020)
\href{https://arxiv.org/abs/2005.08657}{[arXiv:2005.08657 [hep-th]]}.


 \bibitem{SCHSgen}
S.~M.~Kuzenko, M.~Ponds and E.~S.~N.~Raptakis,
``Generalised superconformal higher-spin multiplets,''
JHEP \textbf{03}, 183 (2021)
\href{https://arxiv.org/abs/2011.11300}{[arXiv:2011.11300 [hep-th]]}.


\bibitem{CottonAdS}
S.~M.~Kuzenko and M.~Ponds,
``Higher-spin Cotton tensors and massive gauge-invariant actions in AdS$_3$,'' 
JHEP \textbf{05}, 275 (2021)
\href{https://arxiv.org/abs/2103.11673}{[arXiv:2103.11673 [hep-th]]}.


\bibitem{AdSuperprojectors}
E.~I.~Buchbinder, D.~Hutchings, S.~M.~Kuzenko and M.~Ponds, 
``AdS superprojectors,'' JHEP {\bf 2104}, 074 (2021)
\href{https://arxiv.org/abs/2101.05524}{[arXiv:2101.05524 [hep-th]]}.


\bibitem{AdS3(super)projectors}
 D.~Hutchings, S.~M.~Kuzenko and M.~Ponds, 
``AdS (super)projectors in three dimensions and partial masslessness,'' 
\href{https://arxiv.org/abs/2107.12201}{[arXiv:2107.12201 [hep-th]]}.





\bibitem{Wigner} 
E.~P.~Wigner,
``On unitary representations of the inhomogeneous Lorentz group,''
Annals Math.\  {\bf 40}, 149 (1939)
[Nucl.\ Phys.\ Proc.\ Suppl.\  {\bf 6}, 9 (1989)].

\bibitem{VasilievReview1}
M.~A.~Vasiliev,
``Higher spin gauge theories in various dimensions,''
PoS  JHW {\bf 003} (2003)
 \href{https://arxiv.org/abs/hep-th/0401177}{[arXiv:0401177 [hep-th]]}.

\bibitem{SorokinHS}
D.~Sorokin,
``Introduction to the classical theory of higher spins,''
AIP Conf. Proc. \textbf{767}, no.1, 172-202 (2005)
 \href{https://arxiv.org/abs/hep-th/0405069}{[arXiv:hep-th/0405069 [hep-th]]}.

\bibitem{BekaertVasilievReview}
X.~Bekaert, S.~Cnockaert, C.~Iazeolla and M.~A.~Vasiliev,
``Nonlinear higher spin theories in various dimensions,''
\href{https://arxiv.org/abs/hep-th/0503128}{[arXiv:hep-th/0503128 [hep-th]]}.

\bibitem{VasilievReview2}
M.~A.~Vasiliev,
``V L Ginzburg and higher-spin fields,''
Phys. Usp. \textbf{54}, 641-648 (2011).

\bibitem{BekaertReview}
X.~Bekaert, N.~Boulanger and P.~Sundell,
``How higher-spin gravity surpasses the spin two barrier: no-go theorems versus yes-go examples,''
Rev. Mod. Phys. \textbf{84}, 987-1009 (2012)
 \href{https://arxiv.org/abs/1007.0435}{[arXiv:1007.0435 [hep-th]]}.

\bibitem{DidenkoSkvortsov}
V.~E.~Didenko and E.~D.~Skvortsov,
``Elements of Vasiliev theory,''
 \href{https://arxiv.org/abs/1401.2975}{[arXiv:1401.2975 [hep-th]]}.
 
 \bibitem{RahmanTaronna}
R.~Rahman and M.~Taronna,
``From Higher Spins to Strings: A Primer,''
 \href{https://arxiv.org/abs/1512.07932}{[arXiv:1512.07932 [hep-th]]}.

\bibitem{Fronsdal1}
C.~Fronsdal,
``Massless fields with integer spin,''
Phys.\ Rev.\  D {\bf18},   3624 (1978).
   
\bibitem{FF}
J.~Fang and C.~Fronsdal,
``Massless fields with half-integral spin,''
Phys.\ Rev.\  D {\bf 18}, 3630 (1978).
  
\bibitem{Fronsdal2}
C.~Fronsdal,
``Singletons and massless, integral-spin fields on de Sitter space,''
Phys.\ Rev.\  D {\bf 20},  848 (1979). 

\bibitem{FF2}
J.~Fang and C.~Fronsdal,
``Massless, half-integer-spin fields in de Sitter space,''
Phys.\ Rev.\  D {\bf 22},  1361 (1980).

\bibitem{KS94}
S.~M.~Kuzenko and A.~G.~Sibiryakov,
``Free massless higher-superspin superfields on the anti-de Sitter superspace"
Phys.\ Atom.\ Nucl.\  {\bf 57}, 1257 (1994) [Yad.\ Fiz.\  {\bf 57}, 1326 (1994)]
\href{https://arxiv.org/abs/1112.4612}{[arXiv:1112.4612 [hep-th]]}.

\bibitem{KPS}
S.~M.~Kuzenko, A.~G.~Sibiryakov and V.~V.~Postnikov,
``Massless gauge superfields of higher half integer superspins,''
JETP Lett.\  {\bf 57},  534 (1993)  [Pisma Zh.\ Eksp.\ Teor.\ Fiz.\  {\bf 57},  521 (1993)].

\bibitem{KS}
S.~M.~Kuzenko and A.~G.~Sibiryakov,
``Massless gauge superfields of higher integer superspins,''
JETP Lett.\  {\bf 57},   539 (1993)  
[Pisma Zh.\ Eksp.\ Teor.\ Fiz.\  {\bf 57}, 526 (1993)].

\bibitem{Curtright:1979uz}
T.~Curtright,
``Massless Field Supermultiplets With Arbitrary Spin,''
Phys. Lett. B \textbf{85}, 219-224 (1979).

\bibitem{Vasiliev:1980as}
M.~A.~Vasiliev,
``'Gauge' form of description of massless fields with arbitrary spin. (In Russian),''
Yad. Fiz. \textbf{32}, 855-861 (1980).

\bibitem{SHSI1}
I.~L.~Buchbinder, S.~J.~Gates and K.~Koutrolikos,
``Higher Spin Superfield interactions with the Chiral Supermultiplet: Conserved Supercurrents and Cubic Vertices,''
Universe \textbf{4}, no.1, 6 (2018)
\href{https://arxiv.org/abs/1708.06262}{[arXiv:1708.06262 [hep-th]]}.

\bibitem{SHSI2}
J.~Hutomo and S.~M.~Kuzenko,
``Non-conformal higher spin supercurrents,''
Phys. Lett. B \textbf{778}, 242-246 (2018)
\href{https://arxiv.org/abs/1710.10837}{[arXiv:1710.10837 [hep-th]]}.

\bibitem{SHSI3}
K.~Koutrolikos, P.~Koci and R.~von Unge,
``Higher Spin Superfield interactions with Complex linear Supermultiplet: Conserved Supercurrents and Cubic Vertices,''
JHEP \textbf{03}, 119 (2018)
\href{https://arxiv.org/abs/1712.05150}{[arXiv:1712.05150 [hep-th]]}.


\bibitem{BHK} 
E.~I.~Buchbinder, J.~Hutomo and S.~M.~Kuzenko,
``Higher spin supercurrents in anti-de Sitter space,''
JHEP {\bf 1809}, 027 (2018)
\href{https://arxiv.org/abs/1805.08055}{[arXiv:1805.08055 [hep-th]]}.

\bibitem{SHSI4}
R.~R.~Metsaev,
``Cubic interaction vertices for N=1 arbitrary spin massless supermultiplets in flat space,''
JHEP \textbf{08}, 130 (2019)
\href{https://arxiv.org/abs/1905.11357}{[arXiv:1905.11357 [hep-th]]}.

\bibitem{SHSI5}
M.~V.~Khabarov and Y.~M.~Zinoviev,
``Cubic interaction vertices for massless higher spin supermultiplets in $d$ = 4,''
JHEP \textbf{02}, 167 (2021)
\href{https://arxiv.org/abs/2012.00482}{[arXiv:2012.00482 [hep-th]]}.
  
\bibitem{AragoneDeser}
C.~Aragone and S.~Deser,
``Consistency Problems of Hypergravity,''
Phys. Lett. B \textbf{86}, 161-163 (1979).

\bibitem{FV-vertices87a} 
E.~S.~Fradkin and M.~A.~Vasiliev,
``On the gravitational interaction of massless higher spin fields,''
Phys.\ Lett.\ B {\bf 189}, 89 (1987).

\bibitem{FV-vertices87b} 
E.~S.~Fradkin and M.~A.~Vasiliev,
``Cubic interaction in extended theories of massless higher spin fields,''
Nucl.\ Phys.\ B {\bf 291}, 141 (1987).

\bibitem{VasilievFull}
M.~A.~Vasiliev,
``Consistent equation for interacting gauge fields of all spins in (3+1)-dimensions,''
Phys. Lett. B \textbf{243}, 378-382 (1990).

\bibitem{VasilievFulld}
M.~A.~Vasiliev,
``Nonlinear equations for symmetric massless higher spin fields in (A)dS(d),''
Phys. Lett. B \textbf{567}, 139-151 (2003)
\href{https://arxiv.org/abs/hep-th/0304049}{[arXiv:hep-th/0304049 [hep-th]]}.

\bibitem{FT} 
E.~S.~Fradkin and A.~A.~Tseytlin,  
``Conformal supergravity,''
Phys.\ Rept.\  {\bf 119}, 233 (1985).	

\bibitem{Tseytlin} 
A.~A.~Tseytlin,
``On limits of superstring in AdS(5) x S**5,''
Theor.\ Math.\ Phys.\  {\bf 133}, 1376 (2002) [Teor.\ Mat.\ Fiz.\  {\bf 133}, 69 (2002)]
\href{https://arxiv.org/abs/hep-th/0201112}{[hep-th/0201112]}.

\bibitem{Segal} 
A.~Y.~Segal,
``Conformal higher spin theory,''
Nucl.\ Phys.\ B {\bf 664}, 59 (2003)
\href{https://arxiv.org/abs/hep-th/0207212}{[hep-th/0207212]}. 

\bibitem{BJM2} 
X.~Bekaert, E.~Joung and J.~Mourad,
``Effective action in a higher-spin background,''
JHEP {\bf 1102}, 048 (2011)
\href{https://arxiv.org/abs/1012.2103}{[arXiv:1012.2103 [hep-th]]}.
  
\bibitem{BJM1}
X.~Bekaert, E.~Joung and J.~Mourad,
``On higher spin interactions with matter,''
JHEP \textbf{05}, 126 (2009)
\href{https://arxiv.org/abs/0903.3338}{[arXiv:0903.3338 [hep-th]]}.  

\bibitem{Bonezzi} 
R.~Bonezzi,
``Induced action for conformal higher spins from worldline path integrals,''
Universe {\bf 3}, no. 3, 64 (2017)
\href{https://arxiv.org/abs/1709.00850}{[arXiv:1709.00850 [hep-th]]}.
  
\bibitem{Tseytlin13} 
A.~A.~Tseytlin,
``On partition function and Weyl anomaly of conformal higher spin fields,''
Nucl.\ Phys.\ B {\bf 877}, 598 (2013)
\href{https://arxiv.org/abs/1309.0785}{[arXiv:1309.0785 [hep-th]]}.  

\bibitem{TseytlinSixSphere}
A.~A.~Tseytlin,
``Weyl anomaly of conformal higher spins on six-sphere,''
Nucl. Phys. B \textbf{877}, 632-646 (2013)
 \href{https://arxiv.org/abs/1310.1795}{[arXiv:1310.1795 [hep-th]]}.

\bibitem{Beccaria:2014jxa} 
M.~Beccaria, X.~Bekaert and A.~A.~Tseytlin,
``Partition function of free conformal higher spin theory,''
JHEP {\bf 1408}, 113 (2014)
\href{https://arxiv.org/abs/1406.3542}{[arXiv:1406.3542 [hep-th]]}.
 
\bibitem{BT2015} 
M.~Beccaria and A.~A.~Tseytlin,
``On higher spin partition functions,''
J.\ Phys.\ A {\bf 48}, no. 27, 275401 (2015)
\href{https://arxiv.org/abs/1503.08143}{[arXiv:1503.08143 [hep-th]]}.

\bibitem{Joung:2015eny}
E.~Joung, S.~Nakach and A.~A.~Tseytlin,
``Scalar scattering via conformal higher spin exchange,''
JHEP \textbf{02}, 125 (2016)
\href{https://arxiv.org/abs/1512.08896}{[arXiv:1512.08896 [hep-th]]}.

\bibitem{Beccaria:2016syk} 
M.~Beccaria, S.~Nakach and A.~A.~Tseytlin,
``On triviality of S-matrix in conformal higher spin theory,''
JHEP {\bf 1609}, 034 (2016)
\href{https://arxiv.org/abs/1607.06379}{[arXiv:1607.06379 [hep-th]]}. 

\bibitem{Adamo1}
T.~Adamo, P.~H\"ahnel and T.~McLoughlin,
``Conformal higher spin scattering amplitudes from twistor space,''
JHEP \textbf{04}, 021 (2017)
 \href{https://arxiv.org/abs/1611.06200}{[arXiv:1611.06200 [hep-th]]}.

\bibitem{BeccariaTseytlin9000}
M.~Beccaria and A.~A.~Tseytlin,
``C$_{T}$ for higher derivative conformal fields and anomalies of (1, 0) superconformal 6d theories,''
JHEP \textbf{06}, 002 (2017)
 \href{https://arxiv.org/abs/1705.00305}{[arXiv:1705.00305 [hep-th]]}.

\bibitem{BeccariaTseyltinSphere}
M.~Beccaria and A.~A.~Tseytlin,
``C$_{T}$ for conformal higher spin fields from partition function on conically deformed sphere,''
JHEP \textbf{09}, 123 (2017)
 \href{https://arxiv.org/abs/1707.02456}{[arXiv:1707.02456 [hep-th]]}.

\bibitem{AABD}
S.~Acevedo, R.~Aros, F.~Bugini and D.~E.~Diaz,
``On the Weyl anomaly of 4D Conformal Higher Spins: a holographic approach,''
JHEP \textbf{11}, 082 (2017)
 \href{https://arxiv.org/abs/1710.03779}{[arXiv:1710.03779 [hep-th]]}.

\bibitem{Adamo2}
 T.~Adamo, S.~Nakach and A.~A.~Tseytlin,
 ``Scattering of conformal higher spin fields,''
 JHEP {\bf 1807}, 016 (2018)
 \href{https://arxiv.org/abs/1805.00394}{[arXiv:1805.00394 [hep-th]]}.

\bibitem{Beccaria:2018rxp}
M.~Beccaria and A.~A.~Tseytlin,
``Superconformal index of higher derivative $\mathcal N=1$ multiplets in four dimensions,''
JHEP {\bf 1810}, 087 (2018)
\href{https://arxiv.org/abs/1807.05911}{[arXiv:1807.05911 [hep-th]]}.  

\bibitem{ABD}
R.~Aros, F.~Bugini and D.~E.~Diaz,
``A calculation of the Weyl anomaly for 6D Conformal Higher Spins,''
JHEP \textbf{05}, 241 (2021)
 \href{https://arxiv.org/abs/2104.05673}{[arXiv:2104.05673 [hep-th]]}.

\bibitem{Metsaev10} 
R.~R.~Metsaev,
``Gauge invariant two-point vertices of shadow fields, AdS/CFT, and conformal fields,''
Phys.\ Rev.\ D {\bf 81}, 106002 (2010)
\href{https://arxiv.org/abs/0907.4678}{[arXiv:0907.4678 [hep-th]]}.
  
\bibitem{AdS/CFT3}
S.~Giombi, I.~R.~Klebanov, S.~S.~Pufu, B.~R.~Safdi and G.~Tarnopolsky,
``AdS Description of Induced Higher-Spin Gauge Theory,''
JHEP \textbf{10}, 016 (2013)
\href{https://arxiv.org/abs/1306.5242}{[arXiv:1306.5242 [hep-th]]}.
   
\bibitem{AdS/CFT4}
S.~Giombi, I.~R.~Klebanov and B.~R.~Safdi,
``Higher Spin AdS$_{d+1}$/CFT$_d$ at One Loop,''
Phys. Rev. D \textbf{89}, no.8, 084004 (2014)
\href{https://arxiv.org/abs/1401.0825}{[arXiv:1401.0825 [hep-th]]}.
  
\bibitem{AdS/CFT5}
M.~Beccaria and A.~A.~Tseytlin,
``Iterating free-field AdS/CFT: higher spin partition function relations,''
J. Phys. A \textbf{49}, no.29, 295401 (2016)
\href{https://arxiv.org/abs/1602.00948}{[arXiv:1602.00948 [hep-th]]}.  

\bibitem{BF} 
R.~E.~Behrends and C.~Fronsdal,
``Fermi decay of higher spin particles,''
Phys.\ Rev.\  {\bf 106}, no. 2, 345 (1957).

\bibitem{Fronsdal58}
C. Fronsdal, ``On the theory of higher spin fields,'' 
Nuovo Cim. {\bf  9}, 416 (1958).

\bibitem{MetsaevOrd1}
R.~R.~Metsaev,
``Ordinary-derivative formulation of conformal low spin fields,''
JHEP \textbf{1201}, 064 (2012)
\href{https://arxiv.org/abs/0707.4437}{[arXiv:0707.4437 [hep-th]]}.

\bibitem{MetsaevOrd2}
R.~R.~Metsaev,
``Ordinary-derivative formulation of conformal totally symmetric arbitrary spin bosonic fields,''
JHEP \textbf{06}, 062 (2012)
\href{https://arxiv.org/abs/0709.4392}{[arXiv:0709.4392 [hep-th]]}.

\bibitem{Marnelius} 
R.~Marnelius,
``Lagrangian conformal higher spin theory,''
\href{https://arxiv.org/abs/0805.4686}{[arXiv:0805.4686 [hep-th]]}.

\bibitem{Marnelius2}
R.~Marnelius,
``Lagrangian higher spin field theories from the O(N) extended supersymmetric particle,''
\href{https://arxiv.org/abs/0906.2084}{[arXiv:0906.2084 [hep-th]]}.

\bibitem{Vasiliev2009} 
M.~A.~Vasiliev,
``Bosonic conformal higher-spin fields of any symmetry,''
Nucl.\ Phys.\ B {\bf 829}, 176 (2010)
 \href{https://arxiv.org/abs/0909.5226}{[arXiv:0909.5226 [hep-th]]}.
  
 
\bibitem{Metsaev:2012hr}
R.~R.~Metsaev,
``Conformal totally symmetric arbitrary spin fermionic fields,''
Proc. Steklov Inst. Math. \textbf{309}, 202-218 (2020)
\href{https://arxiv.org/abs/1211.4498}{[arXiv:1211.4498 [hep-th]]}.  
   
\bibitem{PopeTownsend} 
C.~N.~Pope and P.~K.~Townsend,
``Conformal higher spin in (2+1) dimensions,''
Phys.\ Lett.\ B {\bf 225}, 245 (1989).
  
\bibitem{FL-3D} 
E.~S.~Fradkin and V.~Y.~Linetsky,
``A superconformal theory of massless higher spin fields in $D = (2+1)$,''
 Annals Phys.\  {\bf 198}, 293 (1990).

\bibitem{K16} 
S.~M.~Kuzenko,
``Higher spin super-Cotton tensors and generalisations of the linear-chiral duality in three dimensions,'' 
Phys.\ Lett.\ B {\bf 763}, 308 (2016) 
\href{https://arxiv.org/abs/1606.08624}{[arXiv:1606.08624 [hep-th]]}.

\bibitem{DD} 
T.~Damour and S.~Deser,
``'Geometry' of spin 3 gauge theories,''
Ann.\ Inst.\ H.\ Poincare Phys.\ Theor.\  {\bf 47}, 277 (1987).   

\bibitem{BHT}
E.~A.~Bergshoeff, O.~Hohm and P.~K.~Townsend,
``On higher derivatives in 3D gravity and higher spin gauge theories,''
Annals Phys.\  {\bf 325},  1118 (2010) 
\href{https://arxiv.org/abs/0911.3061}{[arXiv:0911.3061 [hep-th]]}.

\bibitem{BKRTY} 
E.~A.~Bergshoeff, M.~Kovacevic, J.~Rosseel, P.~K.~Townsend and Y.~Yin,
``A spin-4 analog of 3D massive gravity,''
Class.\ Quant.\ Grav.\  {\bf 28}, 245007 (2011)
\href{https://arxiv.org/abs/1109.0382}{[arXiv:1109.0382 [hep-th]]}.

\bibitem{Nilsson1} 
B.~E.~W.~Nilsson,
``Towards an exact frame formulation of conformal higher spins in three dimensions,''
JHEP {\bf 1509}, 078 (2015)
\href{https://arxiv.org/abs/1312.5883}{[arXiv:1312.5883 [hep-th]]} .

\bibitem{Nilsson2} 
B.~E.~W.~Nilsson,
``On the conformal higher spin unfolded equation for a three-dimensional self-interacting scalar field,''
JHEP {\bf 1608}, 142 (2016)
\href{https://arxiv.org/abs/1506.03328}{[arXiv:1506.03328 [hep-th]]}.

\bibitem{HHL} 
M.~Henneaux, S.~H\"ortner and A.~Leonard,
``Higher spin conformal geometry in three dimensions and prepotentials for higher spin gauge fields,''
JHEP {\bf 1601}, 073 (2016)
\href{https://arxiv.org/abs/1511.07389}{[arXiv:1511.07389 [hep-th]]}.  

\bibitem{LN} 
H.~Linander and B.~E.~W.~Nilsson,
``The non-linear coupled spin 2 - spin 3 Cotton equation in three dimensions,''
JHEP {\bf 1607}, 024 (2016)
\href{https://arxiv.org/abs/1602.01682}{[arXiv:1602.01682 [hep-th]]}.

\bibitem{KO} 
S.~M.~Kuzenko and D.~X.~Ogburn,
``Off-shell higher spin N=2 supermultiplets in three dimensions,''
Phys.\ Rev.\ D {\bf 94}, no. 10, 106010 (2016)
\href{https://arxiv.org/abs/1603.04668}{[arXiv:1603.04668 [hep-th]]} .

\bibitem{KT17} 
S.~M.~Kuzenko and M.~Tsulaia,
``Off-shell massive N=1 supermultiplets in three dimensions,''
Nucl.\ Phys.\ B {\bf 914}, 160 (2017)
\href{https://arxiv.org/abs/1609.06910}{[arXiv:1609.06910 [hep-th]]}.
 
\bibitem{BBB} 
T.~Basile, R.~Bonezzi and N.~Boulanger,
``The Schouten tensor as a connection in the unfolding of 
3D conformal higher-spin fields,''
JHEP {\bf 1704}, 054 (2017)
\href{https://arxiv.org/abs/1701.08645}{[arXiv:1701.08645 [hep-th]]}.

\bibitem{HKO} 
J.~Hutomo, S.~M.~Kuzenko and D.~Ogburn,
``${\cal N}=2$ supersymmetric higher spin gauge theories and current multiplets in three dimensions,''
 Phys.\ Rev.\ D {\bf 98}, no. 12, 125004 (2018)
  \href{https://arxiv.org/abs/1807.09098}{[arXiv:1807.09098 [hep-th]]}.

\bibitem{HLLMP} 
M.~Henneaux, V.~Lekeu, A.~Leonard, J.~Matulich and S.~Prohazka,
``Three-dimensional conformal geometry and prepotentials for four-dimensional fermionic higher-spin fields,''
JHEP {\bf 1811}, 156 (2018)
\href{https://arxiv.org/abs/1810.04457}{[arXiv:1810.04457 [hep-th]]}.

\bibitem{BHHK}
E.~I.~Buchbinder, D.~Hutchings, J.~Hutomo and S.~M.~Kuzenko,
``Linearised actions for $ \mathcal{N} $-extended (higher-spin) superconformal gravity,''
JHEP \textbf{08} (2019) 077
\href{https://arxiv.org/abs/1905.12476}{[arXiv:1905.12476 [hep-th]]}.

\bibitem{Grigoriev:2019xmp}
M.~Grigoriev, I.~Lovrekovic and E.~Skvortsov,
``New conformal higher spin gravities in $3d$,''
JHEP \textbf{01}, 059 (2020)
\href{https://arxiv.org/abs/1909.13305 }{[arXiv:1909.13305 [hep-th]]}.

\bibitem{FL-algebras} 
E.~S.~Fradkin and V.~Y.~Linetsky,
``Conformal superalgebras of higher spins,''
Annals Phys.\  {\bf 198}, 252 (1990).
  
\bibitem{FL1} 
E.~S.~Fradkin and V.~Y.~Linetsky,
``Cubic interaction in conformal theory of integer higher-spin fields in 
four dimensional space-time,''  
Phys.\ Lett.\ B {\bf 231}, 97 (1989).

\bibitem{FV1} 
E.~S.~Fradkin and M.~A.~Vasiliev,
``Candidate to the role of higher spin symmetry,''
Annals Phys.\  {\bf 177}, 63 (1987).

\bibitem{FV2} 
E.~S.~Fradkin and M.~A.~Vasiliev,
``Superalgebra of higher spins and auxiliary fields,''
Int.\ J.\ Mod.\ Phys.\ A {\bf 3}, 2983 (1988).

\bibitem{Vasiliev88} 
M.~A.~Vasiliev,
``Extended higher spin superalgebras and their realizations in terms of quantum operators,''
Fortsch.\ Phys.\  {\bf 36}, 33 (1988).
    
\bibitem{ChiralHS1}
R.~R.~Metsaev,
``Poincare invariant dynamics of massless higher spins: Fourth order analysis on mass shell,''
Mod. Phys. Lett. A \textbf{6}, 359-367 (1991). 
 
\bibitem{ChiralHS2}
R.~R.~Metsaev,
``S matrix approach to massless higher spins theory. 2: The Case of internal symmetry,''
Mod. Phys. Lett. A \textbf{6}, 2411-2421 (1991).


 \bibitem{ChiralHS3}
D.~Ponomarev and E.~D.~Skvortsov,
``Light-Front Higher-Spin Theories in Flat Space,''
J. Phys. A \textbf{50}, no.9, 095401 (2017)
 \href{https://arxiv.org/abs/1609.04655}{[arXiv:1609.04655 [hep-th]]}.

\bibitem{ChiralHS4}
D.~Ponomarev,
``Chiral Higher Spin Theories and Self-Duality,''
JHEP \textbf{12}, 141 (2017)
 \href{https://arxiv.org/abs/1710.00270}{[arXiv:1710.00270 [hep-th]]}.

\bibitem{ChiralHS5}
E.~D.~Skvortsov, T.~Tran and M.~Tsulaia,
``Quantum Chiral Higher Spin Gravity,''
Phys. Rev. Lett. \textbf{121}, no.3, 031601 (2018)
 \href{https://arxiv.org/abs/1805.00048}[arXiv:1805.00048 [hep-th]]{}.

\bibitem{ChiralHS6}
D.~Ponomarev, E.~Sezgin and E.~Skvortsov,
``On one loop corrections in higher spin gravity,''
JHEP \textbf{11}, 138 (2019)
 \href{https://arxiv.org/abs/1904.01042}{[arXiv:1904.01042 [hep-th]]}.

\bibitem{ChiralHS7}
E.~Skvortsov, T.~Tran and M.~Tsulaia,
``More on Quantum Chiral Higher Spin Gravity,''
Phys. Rev. D \textbf{101}, no.10, 106001 (2020)
 \href{https://arxiv.org/abs/2002.08487}{[arXiv:2002.08487 [hep-th]]}.

\bibitem{ChiralHS8}
E.~Skvortsov and T.~Tran,
``One-loop Finiteness of Chiral Higher Spin Gravity,''
JHEP \textbf{07}, 021 (2020)
 \href{https://arxiv.org/abs/2004.10797}{[arXiv:2004.10797 [hep-th]]}.

\bibitem{ChiralHS9}
K.~Krasnov, E.~Skvortsov and T.~Tran,
``Actions for Self-dual Higher Spin Gravities,''
 \href{https://arxiv.org/abs/2105.12782}{[arXiv:2105.12782 [hep-th]]}.    

\bibitem{HST81}
P.~S.~Howe, K.~S.~Stelle and P.~K.~Townsend,
``Supercurrents,'' 
 Nucl.\ Phys.\  B {\bf 192}, 332 (1981).
 
 \bibitem{KMT} 
S.~M.~Kuzenko, R.~Manvelyan and S.~Theisen,
``Off-shell superconformal higher spin multiplets in four dimensions,''
JHEP {\bf 1707}, 034 (2017)
\href{https://arxiv.org/abs/1701.00682}{[arXiv:1701.00682 [hep-th]]} .


\bibitem{KRN=2SCHS}
S.~M.~Kuzenko and E.~S.~N.~Raptakis,
``$\mathcal{N} = 2$ superconformal higher-spin gauge theories in four dimensions,''
 \href{https://arxiv.org/abs/2104.10416}{[arXiv:2104.10416 [hep-th]]}.   
   
\bibitem{KRDuality}
S.~M.~Kuzenko and E.~S.~N.~Raptakis,
``Duality-invariant (super)conformal higher-spin models,''
\href{https://arxiv.org/abs/2107.02001}{[arXiv:2107.02001 [hep-th]]}. 

\bibitem{FL-4D} 
E.~S.~Fradkin and V.~Y.~Linetsky,
``Superconformal higher spin theory in the cubic approximation,''
Nucl.\ Phys.\ B {\bf 350}, 274 (1991).

\bibitem{Metsaev2014} 
R.~R.~Metsaev,
``Arbitrary spin conformal fields in (A)dS,''
Nucl.\ Phys.\ B {\bf 885}, 734 (2014)
\href{https://arxiv.org/abs/1404.3712}{[arXiv:1404.3712 [hep-th]]}. 

\bibitem{NTCHS} 
T.~Nutma and M.~Taronna,
``On conformal higher spin wave operators,''
JHEP {\bf 1406}, 066 (2014)
\href{https://arxiv.org/abs/1404.7452}{[arXiv:1404.7452 [hep-th]]}.  
 
\bibitem{Eisen} 
L. P. Eisenhart, {\it Riemannian Geometry}, 
Princeton University Press, Princeton, 1926.      
 
\bibitem{DJT1}
S.~Deser, R.~Jackiw and S.~Templeton,
``Three-dimensional massive gauge theories,''
Phys.\ Rev.\ Lett.\  {\bf 48}, 975 (1982).

\bibitem{DJT2}
S.~Deser, R.~Jackiw and S.~Templeton,
``Topologically massive gauge theories,''
Annals Phys.\  {\bf 140}, 372 (1982)
[Erratum-ibid.\  {\bf 185}, 406 (1988)].

\bibitem{vN}
P.~van Nieuwenhuizen,
``D = 3 conformal supergravity and Chern-Simons terms,''
Phys.\ Rev.\  D {\bf 32}, 872 (1985).

\bibitem{HW} 
J.~H.~Horne and E.~Witten,
``Conformal gravity in three dimensions as a gauge theory,''
Phys.\ Rev.\ Lett.\  {\bf 62}, 501 (1989).
  
\bibitem{BKNT-M2} 
D.~Butter, S.~M.~Kuzenko, J.~Novak and G.~Tartaglino-Mazzucchelli,
``Conformal supergravity in three dimensions: Off-shell actions,''
JHEP {\bf 1310}, 073 (2013)
 \href{https://arxiv.org/abs/1306.1205}{[arXiv:1306.1205 [hep-th]]}.
  
\bibitem{KNT-M13} 
S.~M.~Kuzenko, J.~Novak and G.~Tartaglino-Mazzucchelli,
``$\mc{N}=6$ superconformal gravity in three dimensions from superspace,''
JHEP {\bf 1401}, 121 (2014)
\href{https://arxiv.org/abs/1308.5552}{[arXiv:1308.5552 [hep-th]]}.
 
\bibitem{KTvN1} 
M.~Kaku, P.~K.~Townsend and P.~van Nieuwenhuizen,
``Gauge theory of the conformal and superconformal group,''
Phys.\ Lett.\  {\bf 69B}, 304 (1977).
  
\bibitem{KTvN2} 
M.~Kaku, P.~K.~Townsend and P.~van Nieuwenhuizen,
``Properties of conformal supergravity,''
Phys.\ Rev.\ D {\bf 17}, 3179 (1978).

\bibitem{BK88}
I.~L.~Buchbinder and S.~M.~Kuzenko,
``Quantization of the classically equivalent theories in the superspace of simple supergravity and quantum equivalence,''
Nucl.\ Phys.\ B {\bf 308}, 162 (1988).  
  
\bibitem{BK} 
I.~L. Buchbinder and S.~M. Kuzenko, {\it Ideas and Methods of Supersymmetry and
Supergravity, Or a Walk Through Superspace},
IOP, Bristol, 1995 (Revised Edition 1998). 

\bibitem{GrigorievT} 
M.~Grigoriev and A.~A.~Tseytlin,
``On conformal higher spins in curved background,''
J.\ Phys.\ A {\bf 50}, no. 12, 125401 (2017)
\href{https://arxiv.org/abs/1609.09381}{[arXiv:1609.09381 [hep-th]]}.  
  
\bibitem{BeccariaT} 
M.~Beccaria and A.~A.~Tseytlin,
``On induced action for conformal higher spins in curved background,''
Nucl.\ Phys.\ B {\bf 919}, 359 (2017)
\href{https://arxiv.org/abs/1702.00222}{[arXiv:1702.00222 [hep-th]]}.  
  
\bibitem{Manvelyan2} 
R.~Manvelyan and G.~Poghosyan,
``Geometrical structure of Weyl invariants for spin three gauge field in general gravitational background in $d=4$,''
Nucl.\ Phys.\ B {\bf 937}, 1 (2018)
\href{https://arxiv.org/abs/1804.10779}{[arXiv:1804.10779 [hep-th]]}. 
  
\bibitem{MacDowellMansouri}
S.~W.~MacDowell and F.~Mansouri,
``Unified Geometric Theory of Gravity and Supergravity,''
Phys. Rev. Lett. \textbf{38}, 739 (1977)
[erratum: Phys. Rev. Lett. \textbf{38}, 1376 (1977)].


\bibitem{ButterN=1} 
D.~Butter,
``N=1 conformal superspace in four dimensions,''
Annals Phys.\  {\bf 325}, 1026 (2010)
\href{https://arxiv.org/abs/0906.4399}{[arXiv:0906.4399 [hep-th]]}. 
  
\bibitem{ButterN=2} 
D.~Butter,
``N=2 conformal superspace in four dimensions,''
JHEP {\bf 1110}, 030 (2011)
\href{https://arxiv.org/abs/1103.5914}{[arXiv:1103.5914 [hep-th]]}.  
  
\bibitem{BKNT-M1} 
D.~Butter, S.~M.~Kuzenko, J.~Novak and G.~Tartaglino-Mazzucchelli,
``Conformal supergravity in three dimensions: New off-shell formulation,''
JHEP {\bf 1309}, 072 (2013)
\href{https://arxiv.org/abs/1305.3132}{[arXiv:1305.3132 [hep-th]]}.  
  
\bibitem{BKNT-M5D} 
D.~Butter, S.~M.~Kuzenko, J.~Novak and G.~Tartaglino-Mazzucchelli,
``Conformal supergravity in five dimensions: New approach and applications,''
JHEP {\bf 1502}, 111 (2015)
\href{https://arxiv.org/abs/1410.8682}{[arXiv:1410.8682 [hep-th]]}. 

\bibitem{BKNT17}
D.~Butter, S.~M.~Kuzenko, J.~Novak and S.~Theisen,
``Invariants for minimal conformal supergravity in six dimensions,''
JHEP \textbf{12}, 072 (2016)
 \href{https://arxiv.org/abs/1606.02921}{[arXiv:1606.02921 [hep-th]]}.

\bibitem{BEG}
T. N.  Bailey, M. G.  Eastwood, A. R.  Gover, 
``Thomas's structure bundle for conformal, projective and related structures,''
Rocky Mt. J. Math. {\bf 24},1191 (1994).

\bibitem{Gover} 
A. R. Gover, 
``Invariant theory and calculus for conformal geometries,''
Adv. Math. {\bf 163}, 206, 2001. 

\bibitem{Thomas} 
T.~Y.~Thomas,
{\it The Differential Invariants of Generalized Spaces},  
Cambridge University Press 1934.

\bibitem{Gover:2008sw} 
A.~R.~Gover, A.~Shaukat and A.~Waldron,
``Tractors, mass and Weyl invariance,''
Nucl.\ Phys.\ B {\bf 812}, 424 (2009)
\href{https://arxiv.org/abs/0810.2867}{[arXiv:0810.2867 [hep-th]]} .

\bibitem{Gover:2008pt}  
A.~R.~Gover, A.~Shaukat and A.~Waldron,
``Weyl invariance and the origins of mass,''
Phys.\ Lett.\ B {\bf 675}, 93 (2009)
\href{https://arxiv.org/abs/0812.3364}{[arXiv:0812.3364 [hep-th]]}. 

\bibitem{Bonezzi:2010jr} 
R.~Bonezzi, E.~Latini and A.~Waldron,
``Gravity, two times, tractors, Weyl invariance and six-dimensional quantum mechanics,''
Phys.\ Rev.\ D {\bf 82}, 064037 (2010)
\href{https://arxiv.org/abs/1007.1724}{[arXiv:1007.1724 [hep-th]]}. 

\bibitem{Grigoriev:2011gp} 
M.~Grigoriev and A.~Waldron,
``Massive higher spins from BRST and tractors,''
Nucl.\ Phys.\ B {\bf 853}, 291 (2011)
\href{https://arxiv.org/abs/1104.4994}{[arXiv:1104.4994 [hep-th]]} .

\bibitem{Joung:2013doa} 
E.~Joung, M.~Taronna and A.~Waldron,
``A Calculus for higher spin interactions,''
JHEP {\bf 1307}, 186 (2013)
\href{https://arxiv.org/abs/1305.5809}{[arXiv:1305.5809 [hep-th]]}.

\bibitem{JoungConfgeo}
E.~Joung, M.~g.~Kim and Y.~Kim,
``Unfolding Conformal Geometry,''
\href{https://arxiv.org/abs/2108.05535}{[arXiv:2108.05535 [hep-th]]}.
  
\bibitem{DeserN1} 
S.~Deser and R.~I.~Nepomechie,
``Anomalous propagation of gauge fields in conformally flat spaces,''
Phys.\ Lett.\  {\bf 132B}, 321 (1983).

\bibitem{DeserN2}
S.~Deser and R.~I.~Nepomechie,
``Gauge invariance versus masslessness in de Sitter space,''
Annals Phys.\  {\bf 154} (1984) 396.

\bibitem{BG2013} 
X.~Bekaert and M.~Grigoriev,
``Higher order singletons, partially massless fields and their boundary values in the ambient approach,''
Nucl.\ Phys.\ B {\bf 876}, 667 (2013)
\href{https://arxiv.org/abs/1305.0162}{[arXiv:1305.0162 [hep-th]]}. 

\bibitem{Barnich}
G.~Barnich, X.~Bekaert and M.~Grigoriev,
``Notes on conformal invariance of gauge fields,''
J. Phys. A \textbf{48}, no.50, 505402 (2015)
 \href{https://arxiv.org/abs/1506.00595}{[arXiv:1506.00595 [hep-th]]}.

\bibitem{GrigorievH} 
M.~Grigoriev and A.~Hancharuk,
``On the structure of the conformal higher-spin wave operators,''
JHEP {\bf 1812}, 033 (2018)
\href{https://arxiv.org/abs/1808.04320}{[arXiv:1808.04320 [hep-th]]}.
  
\bibitem{Grigoriev:2020lzu}
M.~Grigoriev, K.~Mkrtchyan and E.~Skvortsov,
``Matter-free higher spin gravities in 3D: Partially-massless fields and general structure,''
Phys. Rev. D \textbf{102}, no.6, 066003 (2020)
\href{https://arxiv.org/abs/2005.05931}{[arXiv:2005.05931 [hep-th]]}.

\bibitem{Higuchi1} 
A.~Higuchi,
``Forbidden mass range for spin-2 field theory in de Sitter space-time,''
Nucl.\ Phys.\ B {\bf 282}, 397 (1987).
  
\bibitem{Higuchi2} 
A.~Higuchi,
``Symmetric tensor spherical harmonics on the $N$ sphere and their application to the de Sitter group SO($N$,1),''
J.\ Math.\ Phys.\  {\bf 28}, 1553 (1987).

\bibitem{Higuchi3} 
A.~Higuchi,
``Massive symmetric tensor field in space-times with a positive cosmological constant,''
Nucl.\ Phys.\ B {\bf 325}, 745 (1989).
  
\bibitem{BMVpm}
L.~Brink, R.~R.~Metsaev and M.~A.~Vasiliev,
``How massless are massless fields in AdS(d),''
Nucl. Phys. B \textbf{586}, 183 (2000),
\href{https://arxiv.org/abs/hep-th/0005136}{[arXiv:hep-th/0005136 [hep-th]]}. 

\bibitem{DeserW1} 
S.~Deser and A.~Waldron,
``Gauge invariances and phases of massive higher spins in (A)dS,''
Phys.\ Rev.\ Lett.\  {\bf 87}, 031601 (2001)
\href{https://arxiv.org/abs/hep-th/0102166}{[hep-th/0102166]}.
 
\bibitem{DeserW2} 
S.~Deser and A.~Waldron,
``Partial masslessness of higher spins in (A)dS,''
Nucl.\ Phys.\ B {\bf 607}, 577 (2001)
\href{https://arxiv.org/abs/hep-th/0103198}{[hep-th/0103198]}.  
 

\bibitem{DeserW3} 
S.~Deser and A.~Waldron,
``Stability of massive cosmological gravitons,''
Phys.\ Lett.\ B {\bf 508}, 347 (2001)
\href{https://arxiv.org/abs/hep-th/0103255}{[hep-th/0103255]}. 

\bibitem{DeserW4} 
S.~Deser and A.~Waldron,
``Null propagation of partially massless higher spins in (A)dS and cosmological constant speculations,''
Phys.\ Lett.\ B {\bf 513}, 137 (2001)
\href{https://arxiv.org/abs/hep-th/0105181}{[hep-th/0105181]}.


\bibitem{Zinoviev} 
Y.~M.~Zinoviev,  
``On massive high spin particles in AdS,''  
\href{https://arxiv.org/abs/hep-th/0108192}{[hep-th/0108192]}. 
 
\bibitem{DNW} 
L.~Dolan, C.~R.~Nappi and E.~Witten,
``Conformal operators for partially massless states,''
JHEP {\bf 0110}, 016 (2001)
\href{https://arxiv.org/abs/hep-th/0109096}{[hep-th/0109096]}.  

\bibitem{DeserW5} 
S.~Deser and A.~Waldron,
``Conformal invariance of partially massless higher spins,''
Phys.\ Lett.\ B {\bf 603}, 30 (2004)
\href{https://arxiv.org/abs/hep-th/0408155}{[hep-th/0408155]}.

\bibitem{SV}
E.~D.~Skvortsov and M.~A.~Vasiliev,
``Geometric formulation for partially massless fields,''
Nucl. Phys. B \textbf{756}, 117 (2006)
\href{https://arxiv.org/abs/hep-th/0601095}{[arXiv:hep-th/0601095 [hep-th]]}.

\bibitem{Metsaev2006}
R.~R.~Metsaev,
``Gauge invariant formulation of massive totally symmetric fermionic fields in (A)dS space,''
Phys.\ Lett.\ B {\bf 643} (2006) 205
\href{https://arxiv.org/abs/hep-th/0609029}{[hep-th/0609029]}.
 
\bibitem{Brust} 
C.~Brust and K.~Hinterbichler,
``Partially massless higher-spin theory,''
JHEP {\bf 1702}, 086 (2017)
\href{https://arxiv.org/abs/1610.08510}{[arXiv:1610.08510 [hep-th]]}. 

\bibitem{G-SHR}
S.~Garcia-Saenz, K.~Hinterbichler and R.~A.~Rosen,
``Supersymmetric partially massless fields and non-unitary superconformal representations,''
JHEP \textbf{11}, 166 (2018)
\href{https://arxiv.org/abs/1810.01881}{[arXiv:1810.01881 [hep-th]]}.

\bibitem{BKSZ}
I.~L.~Buchbinder, M.~V.~Khabarov, T.~V.~Snegirev and Y.~M.~Zinoviev,
``Lagrangian description of the partially massless higher spin N = 1 supermultiplets in AdS$_{4}$ space,''
JHEP \textbf{08}, 116 (2019)
\href{https://arxiv.org/abs/1904.01959}{[arXiv:1904.01959 [hep-th]]}.   

\bibitem{BG-SHR}
N.~Bittermann, S.~Garcia-Saenz, K.~Hinterbichler and R.~A.~Rosen,
``${\cal N}$ = 2 Supersymmetric Partially Massless Fields and Non-Unitary Superconformal Representations,''
JHEP \textbf{08}, 115 (2021)
 \href{https://arxiv.org/abs/2011.05994}{[arXiv:2011.05994 [hep-th]]}.
 
\bibitem{Tseytlin5}
A.~A.~Tseytlin,
``Effective action in de Sitter space and conformal supergravity,''
Yad.\ Fiz.\  {\bf 39}, 1606 (1984)
[Sov.\ J.\ Nucl.\ Phys.\  {\bf 39}, no. 6, 1018 (1984)].

\bibitem{Tseytlin6}
E.~S.~Fradkin and A.~A.~Tseytlin,
``Instanton zero modes and beta functions in supergravities. 2. Conformal supergravity,''
Phys.\ Lett.\  {\bf 134B} (1984) 307.
  
\bibitem{Karapet1}
E.~Joung and K.~Mkrtchyan,
``A note on higher-derivative actions for free higher-spin fields,''
JHEP {\bf 1211} (2012) 153
\href{https://arxiv.org/abs/1209.4864}{[arXiv:1209.4864 [hep-th]]}.  
  
\bibitem{Karapet2}
E.~Joung and K.~Mkrtchyan,
``Weyl action of two-column mixed-symmetry field and its factorization around (A)dS space,''
JHEP {\bf 1606} (2016) 135
\href{https://arxiv.org/abs/1604.05330}{[arXiv:1604.05330 [hep-th]]}.  
  
\bibitem{SKsymm}
S.~M.~Kuzenko,
``Supersymmetric Spacetimes from Curved Superspace,''
PoS \textbf{CORFU2014}, 140 (2015)
\href{https://arxiv.org/abs/1504.08114}{[arXiv:1504.08114 [hep-th]]}.
 
\bibitem{FVP} 
D.~Z.~Freedman and A.~Van Proeyen,
{\it Supergravity}, 
Cambridge University Press, 2012.
 
\bibitem{Eguchi}
T.~Eguchi, P.~B.~Gilkey and A.~J.~Hanson,
``Gravitation, Gauge Theories and Differential Geometry,''
Phys. Rept. \textbf{66}, 213 (1980).

\bibitem{Ortin}
T.~Ortin,
{\it Gravity and strings},
Cambridge University Press,  2004.
  
\bibitem{Mackey}
G.~W.~Mack,
``Infinite-dimensional group representations,''
Bull. Am. Math. Soc. \textbf{69}, 628 (1963).

\bibitem{MackSalam}
G.~W.~Mack and A.~Salam,
``Finite component field representations of the conformal group,''
Annals Phys. \textbf{53}, 174-202 (1969).

\bibitem{AT}
A.~Ach\'ucarro and P.~K.~Townsend,
``A Chern-Simons action for three-dimensional anti-de Sitter supergravity theories,''
Phys.\ Lett.\  B {\bf 180}, 89 (1986).
    
\bibitem{Witten} 
E.~Witten,
``(2+1)-dimensional gravity as an exactly soluble system,''
Nucl.\ Phys.\ B {\bf 311}, 46 (1988).
  
\bibitem{FMS}
D.~Francia, J.~Mourad and A.~Sagnotti,
``Current exchanges and unconstrained higher spins,''
Nucl. Phys. B \textbf{773}, 203-237 (2007)
\href{https://arxiv.org/abs/hep-th/0701163}{[arXiv:hep-th/0701163 [hep-th]]}.

\bibitem{PoTs}
D.~Ponomarev and A.~A.~Tseytlin,
``On quantum corrections in higher-spin theory in flat space,''
JHEP \textbf{05}, 184 (2016)
\href{https://arxiv.org/abs/1603.06273}{[arXiv:1603.06273 [hep-th]]}.

\bibitem{Isaev:2017nud}
A.~P.~Isaev and M.~A.~Podoinitsyn,
``Two-spinor description of massive particles and relativistic spin projection operators,''
Nucl.\ Phys.\ B {\bf 929} (2018) 452
\href{https://arxiv.org/abs/1712.00833}{[arXiv:1712.00833 [hep-th]]}.
  
\bibitem{Vasiliev1980}
M.~A.~Vasiliev,
````Gauge'' form of description of massless fields with arbitrary spin,''
Sov.\ J.\ Nucl.\ Phys.\ \ {\bf 32},  439 (1980)  [Yad.\ Fiz.\ \ {\bf 32}, 855 (1980)].

\bibitem{Vasiliev87} 
M.~A.~Vasiliev,
``Free massless fields of arbitrary spin in the de Sitter space and initial data for a higher spin superalgebra,''
Fortsch.\ Phys.\  {\bf 35}, 741 (1987).
    
\bibitem{Novikov} 
S. P. Novikov and I. A. Taimanov,
{\it Modern Geometric Structures and Fields}, 
American Mathematical Society, Providence, 2006.
 
\bibitem{DiracCS}
P.~A.~M.~Dirac,
``Wave equations in conformal space,''
Annals Math. \textbf{37}, 429-442 (1936).

\bibitem{VasilievCS}
C.~R.~Preitschopf and M.~A.~Vasiliev,
``Conformal field theory in conformal space,''
Nucl. Phys. B \textbf{549}, 450-480 (1999)
 \href{https://arxiv.org/abs/hep-th/9812113}{[arXiv:hep-th/9812113 [hep-th]]}.
   
\bibitem{Heyl1}
F.~W.~Heyl,
``Four lectures on Poincar\'e gauge field theory,'' in {\it Proceedings of the sixth course of the International School of Cosmology and Gravitation}, Plenum Press, New York, 1978.  
  
\bibitem{IvanovCG}
E.~A.~Ivanov and J.~Niederle,
``Gauge Formulation of Gravitation Theories. 1. The Poincare, De Sitter and Conformal Cases,''
Phys. Rev. D \textbf{25}, 976 (1982).

\bibitem{Lord1}
E.~ A.~ Lord,
``Gauge theory of a group of diffeomorphisms. II. The conformal and de Sitter groups,''
J. Math. Phys. {\bf27}, 3051 (1986).
   
\bibitem{WB} 
J.~Wess and J.~Bagger,
{\it Supersymmetry and Supergravity},
Princeton University Press, 1992. 

\bibitem{KLT-M11} 
S.~M.~Kuzenko, U.~Lindstr\"om and G.~Tartaglino-Mazzucchelli,
``Off-shell supergravity-matter couplings in three dimensions,''
JHEP {\bf 1103}, 120 (2011)
\href{https://arxiv.org/abs/1101.4013}{[arXiv:1101.4013 [hep-th]]}.
  
\bibitem{Binegar}
B.~Binegar,  
``Relativistic field theories in three dimensions,''
J.\ Math.\ Phys.\  {\bf 23},   1511 (1982).  
  
\bibitem{JN} 
R.~Jackiw and V.~P.~Nair,
``Relativistic wave equations for anyons,''
Phys.\ Rev.\ D {\bf 43}, 1933 (1991).  
  
\bibitem{GKL} 
I.~V.~Gorbunov, S.~M.~Kuzenko and S.~L.~Lyakhovich,
``On the minimal model of anyons,''
Int.\ J.\ Mod.\ Phys.\ A {\bf 12}, 4199 (1997)
\href{https://arxiv.org/abs/hep-th/9607114}{[hep-th/9607114]}.
  
\bibitem{TV}
I.~V.~Tyutin and M.~A.~Vasiliev,
``Lagrangian formulation of irreducible massive fields of arbitrary spin in (2+1) dimensions,'' 
Teor.\ Mat.\ Fiz.\  {\bf 113N1}, 45 (1997) [Theor.\ Math.\ Phys.\  {\bf 113},  1244 (1997)] 
\href{https://arxiv.org/abs/hep-th/9704132}{[hep-th/9704132]}.      
  
\bibitem{FP} 
M.~Fierz and W.~Pauli,
``On relativistic wave equations for particles of arbitrary spin in an electromagnetic field,''  Proc.\ Roy.\ Soc.\ Lond.\ A {\bf 173}, 211 (1939).  
  
\bibitem{ABdeRST} 
R.~Andringa, E.~A.~Bergshoeff, M.~de Roo, O.~Hohm, E.~Sezgin and P.~K.~Townsend,
``Massive 3D supergravity,''
Class.\ Quant.\ Grav.\  {\bf 27}, 025010 (2010)
\href{https://arxiv.org/abs/0907.4658}{[arXiv:0907.4658 [hep-th]]} .  
  
\bibitem{DK} 
S.~Deser and J.~H.~Kay,
``Topologically massive supergravity,''
Phys.\ Lett.\ B {\bf 120}, 97 (1983).

  
\bibitem{GPS} 
G.~W.~Gibbons, C.~N.~Pope and E.~Sezgin,
``The general supersymmetric solution of topologically massive supergravity,''
Class.\ Quant.\ Grav.\  {\bf 25}, 205005 (2008)
\href{https://arxiv.org/abs/0807.2613}{[arXiv:0807.2613 [hep-th]]}.

\bibitem{Siegel}
W.~Siegel,  ``Unextended superfields in extended supersymmetry,''
Nucl.\ Phys.\  B {\bf 156}, 135 (1979).

\bibitem{JT} 
R.~Jackiw and S.~Templeton,
``How super-renormalizable interactions cure their infrared divergences,''
Phys.\ Rev.\ D {\bf 23}, 2291 (1981).

\bibitem{Schonfeld} 
J.~F.~Schonfeld,
``A mass term for three-dimensional gauge fields,''
Nucl.\ Phys.\ B {\bf 185}, 157 (1981).

\bibitem{Chen1} 
B.~Chen and J.~Long,
``High spin topologically massive gravity,''
JHEP {\bf 1112}, 114 (2011)
\href{https://arxiv.org/abs/1110.5113}{[arXiv:1110.5113 [hep-th]]}. 

\bibitem{Chen2} 
B.~Chen, J.~Long and J.~d.~Zhang,
``Classical aspects of higher spin topologically massive gravity,''
Class.\ Quant.\ Grav.\  {\bf 29}, 205001 (2012)
\href{https://arxiv.org/abs/1204.3282}{[arXiv:1204.3282 [hep-th]]}.

\bibitem{FS} 
D.~Francia and A.~Sagnotti,
``On the geometry of higher spin gauge fields,''
Class.\ Quant.\ Grav.\  {\bf 20}, S473 (2003)
\href{https://arxiv.org/abs/hep-th/0212185}{[hep-th/0212185]}.

\bibitem{ST} 
A.~Sagnotti and M.~Tsulaia,
``On higher spins and the tensionless limit of string theory,''
Nucl.\ Phys.\ B {\bf 682}, 83 (2004)
\href{https://arxiv.org/abs/hep-th/0311257}{[hep-th/0311257]}.
  
\bibitem{FPT} 
A.~Fotopoulos, K.~L.~Panigrahi and M.~Tsulaia,
``Lagrangian formulation of higher spin theories on AdS space,''
Phys.\ Rev.\ D {\bf 74}, 085029 (2006)
\href{https://arxiv.org/abs/hep-th/0607248}{[hep-th/0607248]}. 

\bibitem{Sorokin:2008tf} 
D.~P.~Sorokin and M.~A.~Vasiliev,
``Reducible higher-spin multiplets in flat and AdS spaces and their geometric frame-like formulation,''
Nucl.\ Phys.\ B {\bf 809}, 110 (2009)
\href{https://arxiv.org/abs/0807.0206}{[arXiv:0807.0206 [hep-th]]}.

\bibitem{AAS} 
A.~Agugliaro, F.~Azzurli and D.~Sorokin,
``Fermionic higher-spin triplets in AdS,''
Nucl.\ Phys.\ B {\bf 907}, 633 (2016)
\href{https://arxiv.org/abs/1603.02251}{[arXiv:1603.02251 [hep-th]]} .


\bibitem{Sorokin:2018djm} 
D.~Sorokin and M.~Tsulaia,
``Supersymmetric reducible higher-spin multiplets in various dimensions,''
Nucl.\ Phys.\ B {\bf 929}, 216 (2018)
\href{https://arxiv.org/abs/1801.04615}{[arXiv:1801.04615 [hep-th]]}.

\bibitem{Dalmazi}
D.~Dalmazi and A.~L.~R.~d.~Santos,
``On higher spin analogues of linearized topologically massive gravity and linearized ``new massive gravity'',''
\href{https://arxiv.org/abs/2107.08879}{[arXiv:2107.08879 [hep-th]]}.

\bibitem{Sachs} 
A.~Iorio, L.~O'Raifeartaigh, I.~Sachs and C.~Wiesendanger,
``Weyl gauging and conformal invariance,''
Nucl.\ Phys.\ B {\bf 495}, 433 (1997)
\href{https://arxiv.org/abs/hep-th/9607110}{[hep-th/9607110]}.  

\bibitem{BPSS} 
N.~Boulanger, D.~Ponomarev, E.~Sezgin and P.~Sundell,
``New unfolded higher spin systems in $AdS_3$,''
Class.\ Quant.\ Grav.\  {\bf 32}, no. 15, 155002 (2015)
\href{https://arxiv.org/abs/1412.8209}{[arXiv:1412.8209 [hep-th]]}.

\bibitem{BHRST} 
E.~A.~Bergshoeff, O.~Hohm, J.~Rosseel, E.~Sezgin and P.~K.~Townsend,
``On critical massive (super)gravity in adS3,''
J.\ Phys.\ Conf.\ Ser.\  {\bf 314}, 012009 (2011)
\href{https://arxiv.org/abs/1011.1153}{[arXiv:1011.1153 [hep-th]]}. 

\bibitem{BSZ1} 
I.~L.~Buchbinder, T.~V.~Snegirev and Y.~M.~Zinoviev,
``Gauge invariant Lagrangian formulation of massive higher spin fields in $(A)dS_3$ space,''
Phys.\ Lett.\ B {\bf 716}, 243 (2012)
\href{https://arxiv.org/abs/1207.1215}{[arXiv:1207.1215 [hep-th]]}.  

\bibitem{BSZ2} 
I.~L.~Buchbinder, T.~V.~Snegirev and Y.~M.~Zinoviev,
``Frame-like gauge invariant Lagrangian formulation of massive fermionic higher spin fields in $AdS_3$ space,''
Phys.\ Lett.\ B {\bf 738}, 258 (2014)
\href{https://arxiv.org/abs/1407.3918}{[arXiv:1407.3918 [hep-th]]}. 

\bibitem{DKSS} 
S.~Deger, A.~Kaya, E.~Sezgin and P.~Sundell,
``Spectrum of D = 6, N=4b supergravity on AdS in three-dimensions x S**3,''
Nucl.\ Phys.\ B {\bf 536}, 110 (1998)
\href{https://arxiv.org/abs/hep-th/9804166}{[hep-th/9804166]}.

\bibitem{DeserW}
S.~Deser and A.~Waldron,
``Arbitrary spin representations in de Sitter from dS/CFT with applications to dS supergravity,''
Nucl. Phys. B \textbf{662}, 379-392 (2003)
\href{https://arxiv.org/abs/hep-th/0301068}{[arXiv:hep-th/0301068 [hep-th]]}.

\bibitem{HHK}
D.~Hutchings, J.~Hutomo and S.~M.~Kuzenko,
``Higher-spin gauge models with (1,1) supersymmetry in AdS${}_3$: Reduction to (1,0) superspace,''
\href{https://arxiv.org/abs/2011.14294}{[arXiv:2011.14294 [hep-th]]}.

\bibitem{Segal01} 
A.~Y.~Segal,
``A Generating formulation for free higher spin massless fields,''
\href{https://arxiv.org/abs/hep-th/0103028}{[hep-th/0103028]}.

\bibitem{Buchbinder:2001bs} 
I.~L.~Buchbinder, A.~Pashnev and M.~Tsulaia,
``Lagrangian formulation of the massless higher integer spin fields in the AdS background,''
Phys.\ Lett.\ B {\bf 523}, 338 (2001)
\href{https://arxiv.org/abs/hep-th/0109067}{[hep-th/0109067]} . 

\bibitem{M1}
R.~R.~Metsaev,
``Lowest eigenvalues of the energy operator for totally (anti)symmetric massless fields of the n-dimensional anti-de Sitter group,''
Class.\ Quant.\ Grav.\  {\bf 11} (1994) L141.

\bibitem{M2} 
R.~R.~Metsaev,
``Free totally (anti)symmetric massless fermionic fields in d-dimensional anti-de Sitter space,''
Class.\ Quant.\ Grav.\  {\bf 14}, L115 (1997)
\href{https://arxiv.org/abs/hep-th/9707066}{[hep-th/9707066]}.

\bibitem{M3}  
R.~R.~Metsaev,
``Fermionic fields in the d-dimensional anti-de Sitter space-time,''
Phys.\ Lett.\ B {\bf 419}, 49 (1998)
\href{https://arxiv.org/abs/hep-th/9802097}{[hep-th/9802097]}.

\bibitem{M4} 
R.~R.~Metsaev,
``Arbitrary spin massless bosonic fields in d-dimensional anti-de Sitter space,''
Lect.\ Notes Phys.\  {\bf 524}, 331 (1999)
\href{https://arxiv.org/abs/hep-th/9810231}{[hep-th/9810231]}.

\bibitem{CF} 
A.~Campoleoni and D.~Francia,
``Maxwell-like Lagrangians for higher spins,''
JHEP {\bf 1303}, 168 (2013)
\href{https://arxiv.org/abs/1206.5877}{[arXiv:1206.5877 [hep-th]]}.

\bibitem{FMM} 
D.~Francia, G.~L.~Monaco and K.~Mkrtchyan,
``Cubic interactions of Maxwell-like higher spins,''
JHEP {\bf 1704}, 068 (2017)
\href{https://arxiv.org/abs/1611.00292}{[arXiv:1611.00292 [hep-th]]}.

\bibitem{WeinbergWeyl}
S.~Weinberg,
``Photons and gravitons in perturbation theory: Derivation of Maxwell's and Einstein's equations,''
Phys. Rev. \textbf{138}, B988-B1002 (1965).

\bibitem{SG} 
W.~Siegel and S.~J.~Gates, Jr.,
``Superprojectors,''
Nucl.\ Phys.\ B {\bf 189}, 295 (1981).  

\bibitem{GGRS}
S.~J.~Gates, Jr., M.~T.~Grisaru, M.~Ro\v{c}ek and W.~Siegel,
{\it Superspace, or One Thousand and One Lessons in Supersymmetry},
Front.\ Phys.\  {\bf 58}, 1 (1983) 
\href{https://arxiv.org/abs/hep-th/0108200}{[arXiv:hep-th/0108200]}.

\bibitem{EO} 
J.~Erdmenger and H.~Osborn,
``Conformally covariant differential operators: Symmetric tensor fields,''
Class.\ Quant.\ Grav.\  {\bf 15}, 273 (1998)
\href{https://arxiv.org/abs/gr-qc/9708040}{[gr-qc/9708040]} .

\bibitem{deWitFreedman}
B.~de Wit and D.~Z.~Freedman,
``Systematics of Higher Spin Gauge Fields,''
Phys. Rev. D \textbf{21}, 358 (1980)

\bibitem{DeserW6} 
S.~Deser, E.~Joung and A.~Waldron,
``Partial masslessness and conformal gravity,''
J.\ Phys.\ A {\bf 46}, 214019 (2013)
\href{https://arxiv.org/abs/1208.1307}{[arXiv:1208.1307 [hep-th]]}.

\bibitem{Curtright1}
T.~Curtright and P.~G.~O.~Freund,
``Massive dual fields,''
Nucl. Phys. B \textbf{172},  413 (1980).

\bibitem{Curtright2}
T.~Curtright,
``Generalized gauge fields,''
Phys. Lett. B \textbf{165}, 304 (1985).
  
\bibitem{Dirac:1935zz}
P.~A.~M.~Dirac,
``The electron wave equation in De-Sitter space,''
Annals Math. \textbf{36}, 657-669 (1935).

\bibitem{Dirac:1963ta}
P.~A.~M.~Dirac,
``A remarkable representation of the 3 + 2 de Sitter group,''
J. Math. Phys. \textbf{4}, 901-909 (1963).
  
\bibitem{Fronsdal:1965zzb}
C.~Fronsdal,
``Elementary particles in a  curved space. 1.,''
Rev. Mod. Phys. \textbf{37}, 221-224  (1965).

\bibitem{Fronsdal:1974ew}
C.~Fronsdal,
``Elementary particles in a curved space. 2.,''
Phys. Rev. D \textbf{10}, 589-598 (1974).

\bibitem{Fronsdal:1975eq}
C.~Fronsdal and R.~B.~Haugen,
``Elementary particles in a curved space. 3.,''
Phys. Rev. D \textbf{12}, 3810-3818  (1975).

\bibitem{Fronsdal:1975ac}
C.~Fronsdal,
``Elementary particles in a curved space. 4. Massless particles,''
Phys. Rev. D \textbf{12}, 3819 (1975).

\bibitem{Evans} 
N.~T.~Evans, 
``Discrete series for the universal covering group of the 3 + 2 de Sitter group,"
J. Math. Phys. {\bf 8},  170 (1967). 

\bibitem{Angelopoulos} 
E. Angelopoulos,
``$\overline{\sSO}_0 (3, 2)$: Linear and unitary irreducible representations,'' in 
{\it Quantum Theory, Groups, Fields and Particles}, A. O. Barut (Ed.),
D. Reidel Publishing, 1983, pp 101--148. 

\bibitem{AFFS}
E.~Angelopoulos, M.~Flato, C.~Fronsdal and D.~Sternheimer,
``Massless particles, conformal group and de Sitter universe,''
Phys. Rev. D \textbf{23}, 1278 (1981).
	
\bibitem{Nicolai:1984hb}
H.~Nicolai,
"Representations of supersymmetry in anti-de Sitter space,"
in: {\it Supersymmetry and Supergravity '84}, Proceedings of the Trieste Spring School, eds. B. de Wit, P. Fayet, P. van Nieuwehuizen (Worlds Scientific, 1984).

\bibitem{deWit:1999ui}
B.~de Wit and I.~Herger,
``Anti-de Sitter supersymmetry,''
Lect. Notes Phys. \textbf{541}, 79--100  (2000)
\href{https://arxiv.org/abs/hep-th/9908005}{[arXiv:hep-th/9908005 [hep-th]]}.

\bibitem{FF78}
M.~Flato and C.~Fronsdal,
``One massless particle equals two Dirac singletons: 
Elementary particles in a curved space. 6,''
Lett. Math. Phys. \textbf{2}, 421-426 (1978).

\bibitem{Flato:1980zk}
M.~Flato and C.~Fronsdal,
``On Dis and Racs,''
Phys. Lett. B \textbf{97}, 236-240 (1980).

\bibitem{Barut:1970kp}
A.~O.~Barut and A.~Boehm,
``Reduction of a class of o(4,2) representations with respect to so(4,1) and so(3,2),''
J. Math. Phys. \textbf{11}, 2938-2945 (1970).

\bibitem{BreitenF}
P.~Breitenlohner and D.~Z.~Freedman,
``Stability in gauged extended supergravity,''
Annals Phys. \textbf{144}, 249 (1982).

\bibitem{Bergshoeff2}
E.~Bergshoeff, M.~de Roo and B.~de Wit,
``Extended conformal supergravity,''
Nucl.\ Phys.\  B {\bf 182}, 173 (1981).

\bibitem{deWvHVP}
B.~de Wit, J.~W.~van Holten and A.~Van Proeyen,
``Transformation rules of N=2 supergravity multiplets,''
Nucl.\ Phys.\  B {\bf 167}, 186 (1980).

\bibitem{Bergshoeff1}
E.~Bergshoeff, M.~de Roo, J.~W.~van Holten, B.~de Wit and A.~Van Proeyen,
``Extended conformal supergravity and its applications,''
in {\it Superspace and Supergravity}, S. W. Hawking and M. Ro\v{c}ek (Eds.), 
Cambridge University Press, Cambridge, 1981, pp. 237--256.  

\bibitem{BCdeWS}
D.~Butter, F.~Ciceri, B.~de Wit and B.~Sahoo,
``Construction of all N=4 conformal supergravities,''
Phys. Rev. Lett. \textbf{118}, no.8, 081602 (2017)
\href{https://arxiv.org/abs/1609.09083}{[arXiv:1609.09083 [hep-th]]}.

\bibitem{BCS}
D.~Butter, F.~Ciceri and B.~Sahoo,
``N=4 conformal supergravity: the complete actions,''
\href{https://arxiv.org/abs/1910.11874}{[arXiv:1910.11874 [hep-th]]}.
 
\bibitem{vMVP}
J.~van Muiden and A.~Van Proeyen,
``The $ \mathcal{N} $ = 3 Weyl multiplet in four dimensions,''
JHEP \textbf{01}, 167 (2019)
\href{https://arxiv.org/abs/1702.06442}{[arXiv:1702.06442 [hep-th]]}.
 
\bibitem{HS}
S.~Hegde and B.~Sahoo,
``Comment on "The N = 3 Weyl Multiplet in Four Dimensions",''
Phys. Lett. B \textbf{791}, 92 (2019)
\href{https://arxiv.org/abs/1810.05089}{[arXiv:1810.05089 [hep-th]]}. 

\bibitem{GJMS}
C.R.~Graham, R.~Jenne, L.J.~Mason and G.A.J.~Sparling, 
``Conformally invariant powers of the Laplacian, I: Existence,'' 
J. Lond. Math. Soc. {\bf s2-46}, 557 (1992).

\bibitem{Dirac2}
J.~Holland and G.~Sparling, 
``Conformally invariant powers of the ambient Dirac operator''
\href{https://arxiv.org/abs/math/0112033}{[arXiv:math/0112033 [math.DG]]} .
  
\bibitem{Gover1}
A.~Gover and L.~J.~Peterson,
``Conformally invariant powers of the Laplacian, Q-curvature, and tractor calculus,''
Commun. Math. Phys. \textbf{235}, 339 (2003)
\href{https://arxiv.org/abs/math-ph/0201030}{[arXiv:math-ph/0201030 [math-ph]]}.

\bibitem{Gover2}
A.R.~Gover and K. Hirachi, 
``Conformally invariant powers of the Laplacian: a complete non-existence theorem'' 
J. Am. Math. Soc. {\bf 17}, 389 (2004).

\bibitem{Dirac3}
M.~Fischmann,  
``On conformal powers of the Dirac operator on spin manifolds''  
\href{https://arxiv.org/abs/1311.4182}{[arXiv:1311.4182 [math.DG]]}.
   
\bibitem{FT1982a} 
E.~S.~Fradkin and A.~A.~Tseytlin,
``Asymptotic freedom in extended conformal supergravities,''
Phys.\ Lett.\ B {\bf 110}, 117 (1982).
  
\bibitem{FT1982b} 
E.~S.~Fradkin and A.~A.~Tseytlin,
``One-loop beta function in conformal supergravities,''
Nucl.\ Phys.\ B {\bf 203}, 157 (1982).
  
\bibitem{Paneitz}
S.~M.~Paneitz,
``A quartic conformally covariant differential operator for arbitrary pseudo-Riemannian manifolds,''
MIT preprint, March 1983; published posthumously in:  SIGMA {\bf 4},  036 (2008)
\href{https://arxiv.org/abs/0803.4331}{[arXiv:0803.4331 [math.DG]]}.

\bibitem{Riegert}
R.~J.~Riegert,
``A non-local action for the trace anomaly,''
Phys.\ Lett.\ B {\bf 134}, 56 (1984).

\bibitem{BdeWKL}
D.~Butter, B.~de Wit, S.~M.~Kuzenko and I.~Lodato,
``New higher-derivative invariants in N=2 supergravity and the Gauss-Bonnet term,''
JHEP {\bf 1312},  062  (2013) 
\href{https://arxiv.org/abs/1307.6546}{[arXiv:1307.6546 [hep-th]]}.

\bibitem{ZP88}
B.~M.~Zupnik and D.~G.~Pak,
``Superfield formulation of the simplest three-dimensional gauge theories and conformal supergravities,''
Theor.\ Math.\ Phys.\  {\bf 77} (1988) 1070
[Teor.\ Mat.\ Fiz.\  {\bf 77} (1988) 97]. 

\bibitem{ZP89} 
B.~M.~Zupnik and D.~G.~Pak,
``Differential and integral forms in supergauge theories and supergravity,''
Class.\ Quant.\ Grav.\  {\bf 6}, 723 (1989).
  
\bibitem{LR-brane}
U.~Lindstr\"om and M.~Ro\v{c}ek,
``A super-Weyl-invariant spinning membrane,''
Phys.\ Lett.\  B {\bf 218}, 207 (1989).

\bibitem{KT-M12} 
S.~M.~Kuzenko and G.~Tartaglino-Mazzucchelli,
``Conformal supergravities as Chern-Simons theories revisited,''
JHEP {\bf 1303}, 113 (2013)
\href{https://arxiv.org/abs/1212.6852}{[arXiv:1212.6852 [hep-th]]}.

\bibitem{KLT-M12} 
S.~M.~Kuzenko, U.~Lindstr\"om and G.~Tartaglino-Mazzucchelli,
``Three-dimensional (p,q) AdS superspaces and matter couplings,''
JHEP {\bf 1208}, 024 (2012)  
\href{https://arxiv.org/abs/1205.4622}{[arXiv:1205.4622 [hep-th]]}.

\bibitem{KNT-M15}
S.~M.~Kuzenko, J.~Novak and G.~Tartaglino-Mazzucchelli,
``Higher derivative couplings and massive supergravity in three dimensions,''
JHEP {\bf 1509}, 081 (2015) 
\href{https://arxiv.org/abs/1506.09063}{[arXiv:1506.09063 [hep-th]]}. 

\bibitem{Deser}
S.~Deser,
``Cosmological topological supergravity,''
in {\it Quantum Theory Of Gravity}, S. M. Christensen (Ed.), 
Adam Hilger, Bristol, 1984, pp. 374-381.

\bibitem{TPvN} 
P.~K.~Townsend, K.~Pilch and P.~van Nieuwenhuizen,
``Self-duality in odd dimensions,''
Phys.\ Lett.\  {\bf 136B}, 38 (1984)
Addendum: [Phys.\ Lett.\  {\bf 137B}, 443 (1984)].

\bibitem{DJ} 
S.~Deser and R.~Jackiw,
``Self-duality of topologically massive gauge theories,''
Phys.\ Lett.\  {\bf 139B}, 371 (1984).
 
\bibitem{HIPT}
P.~S.~Howe, J.~M.~Izquierdo, G.~Papadopoulos and P.~K.~Townsend,
``New supergravities with central charges and Killing spinors in 2+1 dimensions,''
Nucl.\ Phys.\  B {\bf 467}, 183 (1996)
\href{https://arxiv.org/abs/hep-th/9505032}{[arXiv:hep-th/9505032]}.

\bibitem{Kuzenko12} 
S.~M.~Kuzenko,
``Prepotentials for N=2 conformal supergravity in three dimensions,''
JHEP {\bf 1212}, 021 (2012)  
\href{https://arxiv.org/abs/1209.3894}{[arXiv:1209.3894 [hep-th]]}.

\bibitem{Hutomo:2018iqo} 
J.~Hutomo and S.~M.~Kuzenko,
``Higher spin supermultiplets in three dimensions: (2,0) AdS supersymmetry,''
Phys.\ Lett.\ B {\bf 787}, 175 (2018)
\href{https://arxiv.org/abs/1809.00802}{[arXiv:1809.00802 [hep-th]]} .

\bibitem{BSZ3} 
I.~L.~Buchbinder, T.~V.~Snegirev and Y.~M.~Zinoviev,
``Lagrangian formulation of the massive higher spin supermultiplets in three dimensional space-time,''
JHEP {\bf 1510}, 148 (2015)
\href{https://arxiv.org/abs/1508.02829}{[arXiv:1508.02829 [hep-th]]}.  
  
\bibitem{BSZ4} 
I.~L.~Buchbinder, T.~V.~Snegirev and Y.~M.~Zinoviev,
``Lagrangian description of massive higher spin supermultiplets in AdS$_{3}$ space,''
JHEP {\bf 1708}, 021 (2017)
\href{https://arxiv.org/abs/1705.06163}{[arXiv:1705.06163 [hep-th]]}.  
     
\bibitem{BSZ5}
I.~L.~Buchbinder, T.~V.~Snegirev and Y.~M.~Zinoviev,
``Supersymmetric higher spin models in three dimensional spaces,''
Symmetry \textbf{10} (2017) no.1, 9
\href{https://arxiv.org/abs/1711.11450}{[arXiv:1711.11450 [hep-th]]}.
  
\bibitem{Zinoviev2}
Y.~M.~Zinoviev,
``Frame-like gauge invariant formulation for massive high spin particles,''
Nucl. Phys. B \textbf{808} (2009), 185-204
\href{https://arxiv.org/abs/0808.1778}{[arXiv:0808.1778 [hep-th]]}.

\bibitem{DKSS2} 
N.~S.~Deger, A.~Kaya, H.~Samtleben and E.~Sezgin,
``Supersymmetric warped AdS in extended topologically massive supergravity,''
Nucl.\ Phys.\ B {\bf 884}, 106 (2014)
\href{https://arxiv.org/abs/1311.4583}{[arXiv:1311.4583 [hep-th]]}.

\bibitem{KLRST-M} 
S.~M.~Kuzenko, U.~Lindstr\"om, M.~Ro\v{c}ek, I.~Sachs and G.~Tartaglino-Mazzucchelli,
``Three-dimensional N=2 supergravity theories: From superspace to components,''
Phys.\ Rev.\ D {\bf 89}, 085028 (2014)
\href{https://arxiv.org/abs/1312.4267}{[arXiv:1312.4267 [hep-th]]}.

\bibitem{KN14} 
S.~M.~Kuzenko and J.~Novak,
``Supergravity-matter actions in three dimensions and Chern-Simons terms,''
JHEP {\bf 1405}, 093 (2014)
\href{https://arxiv.org/abs/1401.2307}{[arXiv:1401.2307 [hep-th]]}.

\bibitem{KNS} 
S.~M.~Kuzenko, J.~Novak and I.~Sachs,
``Minimal $ \mathcal{N}=4 $ topologically massive supergravity,''
JHEP {\bf 1703}, 109 (2017)
\href{https://arxiv.org/abs/1610.09895}{[arXiv:1610.09895 [hep-th]]} .

\bibitem{KT-M11}
S.~M.~Kuzenko and G.~Tartaglino-Mazzucchelli,
``Three-dimensional N=2 (AdS) supergravity and associated supercurrents,''
JHEP {\bf 1112}, 052 (2011)
\href{https://arxiv.org/abs/1109.0496}{[arXiv:1109.0496 [hep-th]]}.

\bibitem{RvanN86} 
M.~Ro\v{c}ek and P.~van Nieuwenhuizen,
``N $\geq$ 2 supersymmetric Chern-Simons terms as d = 3 extended conformal supergravity,''
Class.\ Quant.\ Grav.\  {\bf 3}, 43 (1986).
  
\bibitem{Nishimura} 
M.~Nishimura and Y.~Tanii,
``N=6 conformal supergravity in three dimensions,''
JHEP {\bf 1310}, 123 (2013)
\href{https://arxiv.org/abs/1308.3960}{[arXiv:1308.3960 [hep-th]]}. 

\bibitem{LR89}
U.~Lindstr\"om and M.~Ro\v{c}ek,
``Superconformal gravity in three dimensions as a gauge theory,''
Phys.\ Rev.\ Lett.\  {\bf 62}, 2905 (1989).

\bibitem{Chu:2009gi}
X.~Chu and B.~E.~W.~Nilsson,
``Three-dimensional topologically gauged N=6 ABJM type theories,''
JHEP {\bf 1006} (2010) 057
\href{https://arxiv.org/abs/0906.1655}{[arXiv:0906.1655 [hep-th]]}.

\bibitem{Gran:2012mg}
U.~Gran, J.~Greitz, P.~S.~Howe and B.~E.~W.~Nilsson,
``Topologically gauged superconformal Chern-Simons matter theories,''
JHEP {\bf 1212} (2012) 046
\href{https://arxiv.org/abs/1204.2521}{[arXiv:1204.2521 [hep-th]]} .  

\bibitem{Nilsson:2013fya}
B.~E.~W.~Nilsson,
``Critical solutions of topologically gauged N = 8 CFTs in three dimensions,''
JHEP {\bf 1404} (2014) 107
\href{https://arxiv.org/abs/1304.2270}{[arXiv:1304.2270 [hep-th]]} .

\bibitem{LS1} 
F.~Lauf and I.~Sachs,
``On topologically massive gravity with extended supersymmetry,''
Phys.\ Rev.\ D {\bf 94}, 065028 (2016)
\href{https://arxiv.org/abs/1605.00103}{[arXiv:1605.00103 [hep-th]]}.  

\bibitem{LS2} 
F.~Lauf and I.~Sachs,
``Complete superspace classification of three-dimensional Chern-Simons-matter theories coupled to supergravity,''
JHEP {\bf 1802}, 154 (2018)
\href{https://arxiv.org/abs/1709.01461}{[arXiv:1709.01461 [hep-th]]} .

\bibitem{GWZ1}
R.~Grimm, J.~Wess and B.~Zumino,
``Consistency checks on the superspace formulation of supergravity,''
Phys.\ Lett.\ B {\bf 73}, 415 (1978).
  
\bibitem{GWZ2}
R.~Grimm, J.~Wess and B.~Zumino,
``A complete solution of the Bianchi identities in superspace,''
Nucl.\ Phys.\ B {\bf 152},  255 (1979).

\bibitem{Howe1}
P.~S.~Howe,
``A superspace approach to extended conformal supergravity,''
Phys.\ Lett.\ B {\bf 100}, 389 (1981).

\bibitem{Howe2}
P.~S.~Howe,
``Supergravity in superspace,'' 
Nucl.\ Phys.\  B {\bf 199}, 309 (1982).
  
\bibitem{BGG}
P.~Binetruy, G.~Girardi and R.~Grimm,
``Supergravity couplings: A Geometric formulation,''
Phys. Rept. \textbf{343}, 255-462 (2001)
\href{https://arxiv.org/abs/hep-th/0005225}{[arXiv:hep-th/0005225 [hep-th]]}.

\bibitem{KR}
S.~M.~Kuzenko and E.~S.~N.~Raptakis,
``Symmetries of supergravity backgrounds and supersymmetric field theory,''
JHEP {\bf 2004}, 133 (2020)
\href{https://arxiv.org/abs/1912.08552}{[arXiv:1912.08552 [hep-th]]}. 

\bibitem{BK11} 
D.~Butter and S.~M.~Kuzenko,
``A dual formulation of supergravity-matter theories,''
Nucl.\ Phys.\ B {\bf 854}, 1 (2012)
\href{https://arxiv.org/abs/1106.3038}{[arXiv:1106.3038 [hep-th]]}.

\bibitem{FZ2}
S.~Ferrara and B.~Zumino,
``Structure of conformal supergravity,''  
Nucl.\ Phys.\  B {\bf 134}, 301 (1978).

\bibitem{FZ78} 
S.~Ferrara and B.~Zumino,
``Structure of linearized supergravity and conformal supergravity,''
Nucl.\ Phys.\ B {\bf 134}, 301 (1978).  

\bibitem{Siegel78}
W.~Siegel,
``Solution to constraints in Wess-Zumino supergravity formalism,''
Nucl.\ Phys.\  B {\bf 142}, 301 (1978). 

\bibitem{Zumino} 
B.~Zumino,
``Supergravity and superspace,''
in {\it Recent Developments in  Gravitation - Carg\`ese 1978}, 
M. L\'evy and S. Deser (Eds.), N.Y., Plenum Press, 1979, pp. 405--459.

\bibitem{GrisaruSiegel1} 
M.~T.~Grisaru and W.~Siegel,
``Supergraphity (I). Background field formalism,''
Nucl.\ Phys.\ B {\bf 187}, 149 (1981).
  
\bibitem{GrisaruSiegel2} 
M.~T.~Grisaru and W.~Siegel,
``Supergraphity (II). Manifestly covariant rules and higher loop finiteness,''
Nucl.\ Phys.\ B {\bf 201}, 292 (1982).

\bibitem{BV}
I.~A.~Batalin and G.~A.~Vilkovisky,
``Quantization of gauge theories with linearly dependent generators,''
Phys.\ Rev.\  {\bf D28},   2567 (1983).
  
\bibitem{HK2} 
J.~Hutomo and S.~M.~Kuzenko,
``The massless integer superspin multiplets revisited,''
JHEP {\bf 1802}, 137 (2018)
\href{https://arxiv.org/abs/1711.11364}{[arXiv:1711.11364 [hep-th]]}.

\bibitem{GS}
S.~J.~Gates Jr. and W.~Siegel,
``(3/2, 1) superfield of O(2) supergravity,''
Nucl.\ Phys.\ B {\bf 164}, 484 (1980).  

\bibitem{Keck}
B.~W.~Keck,
``An alternative class of supersymmetries,''
J.\ Phys.\ A  {\bf 8}, 1819 (1975).			

\bibitem{Zumino77}
B.~Zumino,
``Nonlinear realization of supersymmetry in de Sitter space,''
Nucl.\ Phys.\  B {\bf 127}, 189 (1977).

\bibitem{IS}
E.~A.~Ivanov and A.~S.~Sorin,
``Superfield formulation of OSp(1,4) supersymmetry,''
J.\ Phys.\ A  {\bf 13} (1980) 1159.

\bibitem{BKS} 
I.~L.~Buchbinder, S.~M.~Kuzenko and A.~G.~Sibiryakov,
``Quantization of higher spin superfields in the anti-De Sitter superspace,''
Phys.\ Lett.\ B {\bf 352}, 29 (1995)
\href{https://arxiv.org/abs/hep-th/9502148}{[hep-th/9502148]}.

\bibitem{Heidenreich:1982rz}
W.~Heidenreich,
``All linear unitary irreducible representations of de Sitter supersymmetry with positive energy,"
Phys. Lett. B \textbf{110}, 461-464 (1982).

\bibitem{SalamS} 
A.~Salam and J.~A.~Strathdee,
``On superfields and Fermi-Bose symmetry,''
Phys.\ Rev.\ D {\bf 11}, 1521 (1975).  

\bibitem{Sokatchev} 
E.~Sokatchev,
``Projection operators and supplementary conditions for superfields with an arbitrary spin,''
Nucl.\ Phys.\ B {\bf 99}, 96 (1975).

\bibitem{Sokatchev81}
E.~Sokatchev,
``Irreducibility conditions for extended superfields,''
Phys. Lett. B \textbf{104}, 38-40 (1981).

\bibitem{RS}
V.~Rittenberg and E.~Sokatchev,
``Decomposition of extended superfields into irreducible representations of supersymmetry,''
Nucl. Phys. B \textbf{193}, 477-501 (1981).

\bibitem{OS1} 
V.~Ogievetsky and E.~Sokatchev,
``On vector superfield generated by supercurrent,''
Nucl.\ Phys.\ B {\bf 124}, 309 (1977).

\bibitem{OS2} 
V.~I.~Ogievetsky and E.~Sokatchev,
``Superfield equations of motion,''
J.\ Phys.\ A {\bf 10}, 2021 (1977).    
  
\bibitem{GKP}
S.~J.~Gates Jr., S.~M.~Kuzenko and J.~Phillips,
``The off-shell (3/2,2) supermultiplets revisited,''
Phys.\ Lett.\  B {\bf 576}, 97 (2003)  [arXiv:hep-th/0306288].

\bibitem{ButterN} 
D.~Butter and J.~Novak,
``Component reduction in N=2 supergravity: the vector, tensor, and vector-tensor multiplets,''
JHEP {\bf 1205}, 115 (2012)
\href{https://arxiv.org/abs/1201.5431}{[arXiv:1201.5431 [hep-th]]} .

\bibitem{BKN} 
D.~Butter, S.~M.~Kuzenko and J.~Novak,
``The linear multiplet and ectoplasm,''
JHEP {\bf 1209}, 131 (2012)
\href{https://arxiv.org/abs/1205.6981}{[arXiv:1205.6981 [hep-th]]}.

\bibitem{BGLP}
I.~L.~Buchbinder, S.~J.~Gates Jr., W.~D.~Linch, III and J.~Phillips,
``New 4D, $N=1$ superfield theory: Model of free massive superspin 3/2 multiplet,''
Phys. Lett. B \textbf{535}, 280 (2002)
\href{https://arxiv.org/abs/hep-th/0201096}{[arXiv:hep-th/0201096 [hep-th]]}.

\bibitem{KuzenkoPindur}
S.~M.~Kuzenko and A.~E.~Pindur,
``Massless particles in five and higher dimensions,''
Phys. Lett. B \textbf{812}, 136020 (2021)
\href{https://arxiv.org/abs/2010.07124}{[arXiv:2010.07124 [hep-th]]}.

\bibitem{HS1} 
I.~A.~Bandos, J.~Lukierski and D.~P.~Sorokin,
``Superparticle models with tensorial central charges,''
Phys.\ Rev.\ D {\bf 61}, 045002 (2000)
\href{https://arxiv.org/abs/hep-th/9904109}{[hep-th/9904109]} . 
 
\bibitem{HS2} 
M.~A.~Vasiliev,
``Conformal higher spin symmetries of 4d massless supermultiplets and osp(L,2M) invariant equations in generalized (super)space,''
Phys.\ Rev.\ D {\bf 66}, 066006 (2002)
\href{https://arxiv.org/abs/hep-th/0106149}{[hep-th/0106149]}. 

\bibitem{HS3} 
V.~E.~Didenko and M.~A.~Vasiliev,
``Free field dynamics in the generalized AdS (super)space,''
J.\ Math.\ Phys.\  {\bf 45}, 197 (2004)
\href{https://arxiv.org/abs/hep-th/0301054}{[hep-th/0301054]}. 

\bibitem{HS4} 
O.~A.~Gelfond and M.~A.~Vasiliev,
``Higher rank conformal fields in the Sp(2M) symmetric generalized space-time,''
Theor.\ Math.\ Phys.\  {\bf 145}, 1400 (2005)
[Teor.\ Mat.\ Fiz.\  {\bf 145}, 35 (2005)]
\href{https://arxiv.org/abs/hep-th/0304020}{[hep-th/0304020]}.

\bibitem{HS5} 
M.~A.~Vasiliev and V.~N.~Zaikin,
``On Sp(2M) invariant Green functions,''
Phys.\ Lett.\ B {\bf 587}, 225 (2004)
\href{https://arxiv.org/abs/hep-th/0312244}{[hep-th/0312244]}.

\bibitem{HS6} 
I.~Bandos, P.~Pasti, D.~Sorokin and M.~Tonin,
``Superfield theories in tensorial superspaces and the dynamics of higher spin fields,''
JHEP {\bf 0411}, 023 (2004)
\href{https://arxiv.org/abs/hep-th/0407180}{[hep-th/0407180]} .

\bibitem{HS7} 
I.~Bandos, X.~Bekaert, J.~A.~de Azcarraga, D.~Sorokin and M.~Tsulaia,
``Dynamics of higher spin fields and tensorial space,''
JHEP {\bf 0505}, 031 (2005)
\href{https://arxiv.org/abs/hep-th/0501113}{[hep-th/0501113]} . 

\bibitem{HS8} 
E.~Ivanov and J.~Lukierski,
``Higher spins from nonlinear realizations of OSp(1|8),''
Phys.\ Lett.\ B {\bf 624}, 304 (2005)
\href{https://arxiv.org/abs/hep-th/0505216}{[hep-th/0505216]}.
  
\bibitem{HS9} 
O.~A.~Gelfond and M.~A.~Vasiliev,
``Sp(8) invariant higher spin theory, twistors and geometric BRST formulation of unfolded field equations,''
JHEP {\bf 0912}, 021 (2009)
\href{https://arxiv.org/abs/0901.2176}{[arXiv:0901.2176 [hep-th]]}. 
  
  
\bibitem{HS10} 
I.~A.~Bandos, J.~A.~de Azcarraga and C.~Meliveo,
``Extended supersymmetry in massless conformal higher spin theory,''
Nucl.\ Phys.\ B {\bf 853}, 760 (2011)
\href{https://arxiv.org/abs/1106.5199}{[arXiv:1106.5199 [hep-th]]}. 
  
   
\bibitem{HS11} 
I.~Florakis, D.~Sorokin and M.~Tsulaia,
``Higher spins in hyperspace,''
JHEP {\bf 1407}, 105 (2014)
\href{https://arxiv.org/abs/1401.1645}{[arXiv:1401.1645 [hep-th]]}.  
  
\bibitem{HS12} 
I.~Florakis, D.~Sorokin and M.~Tsulaia,
``Higher spins in hyper-superspace,''
Nucl.\ Phys.\ B {\bf 890}, 279 (2014)
\href{https://arxiv.org/abs/1408.6675}{[arXiv:1408.6675 [hep-th]]}.  

\bibitem{HS13} 
S.~Fedoruk and J.~Lukierski,
``New spinorial particle model in tensorial space-time and interacting higher spin fields,''
JHEP {\bf 1302}, 128 (2013)
\href{https://arxiv.org/abs/1210.1506}{[arXiv:1210.1506 [hep-th]]}.  
   
\bibitem{HS14} 
E.~Skvortsov, D.~Sorokin and M.~Tsulaia,
``Correlation functions of Sp(2n) invariant higher-spin systems,''
JHEP {\bf 1607}, 128 (2016)
\href{https://arxiv.org/abs/1605.08498}{[arXiv:1605.08498 [hep-th]]}.  
  
\bibitem{HS15} 
M.~A.~Vasiliev,
``Holography, unfolding and higher-spin theory,''
J.\ Phys.\ A {\bf 46}, 214013 (2013)
\href{https://arxiv.org/abs/1203.5554}{[arXiv:1203.5554 [hep-th]]}.

\bibitem{Sorokin:2017irs} 
D.~Sorokin and M.~Tsulaia,
``Higher Spin Fields in Hyperspace. A Review,''
Universe {\bf 4}, no. 1, 7 (2018)
\href{https://arxiv.org/abs/1710.08244}{[arXiv:1710.08244 [hep-th]]}.  

\bibitem{HR}  
M.~Henneaux and S.~J.~Rey,
``Nonlinear $W_{infinity}$ as asymptotic symmetry of three-dimensional higher spin AdS gravity,''
JHEP {\bf 1012}, 007 (2010)
\href{https://arxiv.org/abs/1008.4579}{[arXiv:1008.4579 [hep-th]]} .

\bibitem{CFPT} 
A.~Campoleoni, S.~Fredenhagen, S.~Pfenninger and S.~Theisen,
``Asymptotic symmetries of three-dimensional gravity coupled to higher-spin fields,''
JHEP {\bf 1011}, 007 (2010)
\href{https://arxiv.org/abs/1008.4744}{[arXiv:1008.4744 [hep-th]]}.

  
  \end{footnotesize}
\end{thebibliography}
\end{document}